\documentclass[10pt]{report}
\usepackage{amsmath, amsthm, amsfonts, accents, subfiles, tikz-cd, accents, parskip, subcaption, float, xcolor, enumitem, changepage}
\usepackage[normalem]{ulem}
\usepackage{hyperref}
\usepackage[utf8]{inputenc}
\usepackage[a4paper, margin=1in]{geometry}
\usepackage{cleveref}

\usepackage[sorting=none]{biblatex}
\newcommand{\blankpage}{\newpage \pagestyle{empty}\mbox{} \newpage}

\newcommand{\pleb}[0]{Pleba\'nski}
\newcommand{\urb}[0]{Urbantke}
\newcommand{\demi}[0]{Demia\'nski}
\newcommand{\derd}[0]{Derdzi\'nski}

\hypersetup{
    colorlinks      = true,
    linkbordercolor = {white},
    pdftitle        = {General Relativity via differential forms \- explorations in \pleb{}'s Formalism for GR},
    urlcolor        = [rgb]{0.18359375, 0.40234375, 0.69140625},
    citecolor       = [rgb]{0.0, 1.0, 0.0}
}

\newcommand{\utilde}[1]{\underaccent{\tilde}{#1}}

\theoremstyle{definition}
\newtheorem{definition}{Definition}[chapter]
\newtheorem{proposition}{Proposition}[chapter]
\newtheorem{corollary}{Corollary}[chapter]

\newtheorem{theorem}{Theorem}[chapter]
\newtheorem{lemma}{Lemma}[chapter]

\newcommand{\R}[0]{\mathbb{R}}
\newcommand{\C}[0]{\mathbb{C}}
\newcommand{\so}{\mathfrak{so}}
\newcommand{\su}{\mathfrak{su}}
\newcommand{\soC}{\mathfrak{so}{(3,\C)}}
\newcommand{\id}{\mathbb{I}}
\DeclareMathOperator{\Tr}{Tr}
\DeclareMathOperator{\Real}{Re}

\newcommand{\Z}{{\mathbb{Z}}}

\newcommand{\im}{i}

\newcommand{\E}[0]{\C^3}

\newcommand{\asd}[0]{\overline{\Sigma}}
\newcommand{\teps}[0]{\tilde{\epsilon}}
\newcommand{\uteps}[0]{\utilde{\epsilon}}

\newcommand{\tE}[0]{\tilde{E}}
\newcommand{\tg}[0]{\tilde{g}}
\newcommand{\tsd}[0]{\tilde{\Sigma}}

\newcommand{\dlap}[0]{\utilde{N}}
\newcommand{\N}[0]{\mathcal{N}}
\newcommand{\spasurf}[0]{\mathcal{S}}
\newcommand{\utE}[0]{\utilde{E}}

\begin{document}
    \begin{titlepage}
        \begin{center}
            \begin{figure}
                \includegraphics[width=\linewidth]{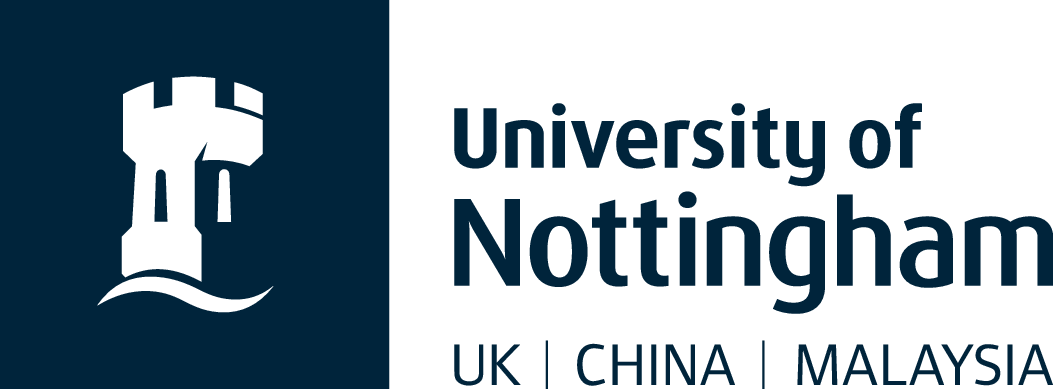}
            \end{figure}
            \vspace*{1cm}
            \fontsize{24.88}{30.5}{\textbf{General Relativity via differential forms - explorations in \pleb{}'s Formalism for GR}}
            \\
            \vspace*{2cm}
            \textbf{\LARGE Adam Grahame Shaw} \\
            \vfill
            School of Mathematical Sciences \\
            University of Nottingham \\
            \vspace*{2cm}
            A Thesis submitted for the degree of \\
            Doctor of Philosophy \\
            \vspace*{2cm}
            September, 2025
        \end{center}
    \end{titlepage}

    \blankpage
    \chapter*{Abstract}
    \setlength\parindent{24pt}
    The Lorentz group in four dimensions splits into two chiral halves, leading to a separation of the space of 2-forms into self-dual and anti-self-dual subspaces. General relativity (GR) can be encoded within one of these chiral halves, and is neatly described using \pleb{}'s chiral formulation of GR, where the main fields are a triple of 2-forms rather than a metric. Related are the chiral pure connection descriptions of gravity, where the connections are the basic fields over the 2-forms. Using the chiral formulations to write Einstein's equations (EEs) reveals a richer and more useful structure. This Thesis explores these depictions, with a particular focusing on the structure of, and generating analytic and numerical solutions to, EEs.\newline\indent The first part introduces principal bundles as the underlying mathematical framework for all subsequent work. The 2-forms arise as soldering forms on an $SO(3,\C)$ principal bundle and their derivatives define the torsion free connection. Chiral Yang-Mills theories also appear in this context and serve as a comparison throughout, due to its similarity to chiral GR.\newline\indent The second part delves into the linear structure of EEs through the chiral descriptions. We show that expanding first-order equations around a self-dual background packages them into an elliptic complex. A choice of inner products defines the adjoint and therefore the Dirac operators, the latter of which square to the Laplacian. Introducing twisting operators decomposes the complex into two truncated de Rahm complexes. Alternatively, we show that relaxing the complex requirement reveals a two parameter family of hyperbolic gauge fixings for the first-order EEs. A similar twisting operator, that separates the conformal perturbation of the metric, reveals the same pair of truncated de Rahm complexes along with their Dirac operators.\newline\indent The third part investigates the nonlinear structure of EEs. A nonlinear lift of the linear two parameter family of hyperbolic gauge fixings is found, where second-order equations have wave operators as their leading terms. Choosing a specific subfamily allows for a definition of the nonlinear twisting operators, separating the system into two sectors of size 4 and 12. Unlike the linear story where both the 4 and the 12 sectors are independent, non-linearly the 4 sector depends on the fields in the 12. This remarkable property allows part of EEs to be solved independently. Also considered in this part is an analytic method for obtaining type D Einstein metrics. In four dimensions, it is known that type D spacetimes are conformal to two distinct K\"ahler metrics. This fact, along with further assumptions on the Killing vectors, leads to a particularly rigid ansatz for the metric. We consider the implications of this for \pleb{}'s formulation and use it to reproduce the \pleb{}-\demi{} family of metrics.\newline\indent The final part of the Thesis is concentrated on the use of chiral formulations within numerical relativity. The hyperbolic gauge fixings, along with the twisting operators, are used to develop a series of evolution systems for GR. Properties of the systems are explored analytically and some basic numerical tests are performed.

    \setlength\parindent{0pt}

    \blankpage
    \section*{}
\vfill
\begin{center}
	{\bfseries\LARGE Acknowledgements}
\end{center}
\vspace{20pt}
\begin{center}
\begin{adjustwidth}{1cm}{1cm}
\it
Firstly, I would like to express my gratitude to my supervisor, Prof.~Kirill Krasnov, for his dedication, help and invaluable guidance throughout my PhD. I am also grateful to Dr.~Miguel Bezares and Prof.~Joel Fine for taking the time to carefully read and provide thoughtful comments on this Thesis.

My sincere thanks go to Niren Bhoja, Sam Close, and Thomas Laird for insightful discussions and contributions, whether direct or indirect, towards the creation of this work. 

Last, but not least, I thank my family and friends, who, throughout the past four years, have provided never-ending support and encouragement. Your combined effort has eased the journey towards this milestone that I am more than happy to have achieved.
\end{adjustwidth}
\end{center}
\vfill

    \blankpage
        
    \newpage 
    \begingroup
    \hypersetup{linkcolor=black}
    \tableofcontents
    \endgroup

    \blankpage

    \addtocontents{toc}{\protect\setcounter{tocdepth}{0}}
\chapter{Introduction}

Gravity is one of the four fundamental forces of nature that have been observed, the current best theory is general relativity (GR) which relates the matter content to the curvature of spacetime. Albert Einstein realised that certain symmetries were essential in constructing a theory of gravity. Importantly, physics should not depend on the choice of frame or coordinate system; this leads to the concepts of local Lorentz and diffeomorphism invariance. A mathematical tool that incorporates both these ideas is differential geometry, an extensive branch of mathematics that studies smooth spaces or manifolds. Using these tools, GR was first introduced by Einstein in~\cite{Die_Grundlage_d_Einste_1916,AEinsteinTheoryGravitation1915}. It is distinguished from the other forces as it is the change in the geometry of spacetime that produces the apparent gravitational force. For other fundamental forces, such as electromagnetism, the strong and weak nuclear forces, it is assumed that geometry of the spacetime is known. The forces then influence the movement of particles on this spacetime. In GR, the basic object is the metric tensor, $g_{\mu\nu}$, which in 4 dimensions contains 10 independent components. Metrics give an intuitive picture of distances, angles and curvature of the spacetime, making them an attractive view point to consider when working with manifolds. Massive objects cause deviations in the curvature of the spacetime, described by their effect on the metric. The gravitational force is then recovered by assuming that particles move along the shortest paths in spacetime, known as geodesics. Massive objects distort the surrounding geodesics causing an apparent attraction towards them. In essence, GR describes the relationship between the matter and energy content and the geometry of the spacetime. This view of gravity has become widely adopted and successful theory, over the past decade many observations of gravitational waves have been made by the LIGO and Virgo detectors~\cite{Observation_of_Collab_2016,GWTC_1_A_Gravi_Abbott_2019}. These observations have consistently agreed with the predictions of GR~\cite{Will2006}.

The equations that describe GR are a set of 10 coupled second-order nonlinear partial differential equations (PDEs) on the metric tensor, known as Einstein's field equations (EEs). In 4 dimensions, due to Lovelock's theorem~\cite{The_Four_Dimens_Lovelo_1972}, which contains some reasonable assumptions about the structure of the equations, EEs are essentially unique. Furthermore, these equations can be neatly encoded into the Einstein-Hilbert action principle~\cite{Hilbert1915}, which due to Lovelock's work can be considered a unique action for GR. Lovelock's theorem does not apply to other dimensions unless one assumes some additional structure about the differential equations. Moreover, 4 is the minimum number of dimensions in which Einstein's equations can have non-trivial solutions. These surprising properties in dimension 4 are just the beginning of many remarkable features that encourage further exploration into gravity and geometry. Usual methods of teaching GR introduce the metric as the fundamental field, with physical observables arising as derived quantities from this metric. See~\cite{Carroll_2019,Geometry_Topol_Nakaha_2018} for books that introduce GR using the metric. While this is not by any means incorrect, describing gravity in this way often misses some concepts available in differential geometry that are useful for GR. The other fundamental forces are well described by Yang-Mills (YMs) theory, they are examples of gauge theories, where there is some group that acts invariantly on the theory~\cite{Geometry_Topol_Nakaha_2018}. As with gravity, YM's can be described on a manifold using the language of differential geometry, however, the fundamental fields that arise are completely differently to the metric. This discrepancy in description between gravity and gauge theories evokes a question: why is the description of GR so different to that of YM? This is a very broad and philosophical question, for which a comprehensive discussion is beyond the scope of this Thesis. One possible answer is that they describe fundamentally different physical phenomenon, but this only leads to more questions. A different approach to this question is to consider alternative ways of describing gravity, in the hope that one description or formulation might bring it closer to gauge theories. We may not be able to tell if one description is more fundamental or correct than the other, but alternative views may, and as we see will, lead to insights into the geometry and structure of gravity in dimension 4. For these reasons we are inspired to explore and study alternative views.

\section*{Alternative Descriptions of Geometry and Gravity}

\'Elie Cartan, a pioneering figure in differential geometry, made significant contributions by developing the theory of geometric structures on manifolds. Geometric structures, or $G$-structures, are objects that are preserved under a specific Lie group action. One of the most famous example of a $G$-structure is the metric, which is associated with the Lorentz group used in Einstein's GR. More generally other objects can be considered, other examples of $G$-structures include the orientation of a manifold, almost complex and symplectic structures. Cartan's approach provides a universal framework for these concepts. His key insight was to encode $G$-structures using differential forms, which are powerful tools in differential geometry with wide-ranging applications both within and outside physics and gravity. Differential forms follow an exterior algebra and calculus that becomes useful in simplifying many standard computations. This view on geometry has not been widely adopted, at least for gravitational physicists, perhaps due to the success of the metric formulation. Nevertheless, this alternative description is still very useful and worth exploring. The group that has a Lorentzian metric as its $G$-structure is the Lorentz group, in dimension 4 it is encoded by 4 1-forms which we call the frame. Similar objects appear already in the physics literature, these are the tetrads or vielbeins, but the geometrical meaning is often ignored, and they are considered as a mathematical trick rather than something to take seriously. The metric can be recovered algebraically from the frame, we therefore take the frame as being the basic variable over the metric. Any two frame related by a Lorentz transformation describe the same metric, this is how the Lorentz invariance manifests itself. It also agrees with the physical principle that rotating ones description should not change the physical results.

The underlying structure, that $G$-structures exist on, is the principal bundle, where a group G is attached to each point in the spacetime. Yang-Mills theories are gauge theories concerning connection 1-forms on a $G$-principal bundle, the mathematical theory for which was also previously developed by Cartan. Both the frame and connections are then differential forms that exist on principal bundles, and both gauge and gravity theories can be described through this geometrical setup. This is a key step in resolving the differences between gravity and the other fundamental forces. The descriptions of all the fundamental forces has converged to be that of a collection of differential forms. Even the concept of spinors, which are usually thought of as vectors or matrices of complex numbers, can be represented as sums of mixed degree differential forms called polyforms. For further details and references see~\cite{Notes_on_Spinor_Bhoja_2022}. Gamma matrices are then mapped to operations on differential forms. Another related benefit is that one can write down a theory of GR with fermionic matter, which is not possible to do using only the metric and objects derived from it. Einstein's equations for GR can be written using the frame and its derived objects without ever needing to introduce a metric. They can also be derived from the Einstein-Cartan action, a nice property of which is that the action becomes polynomial, even when the cosmological constant is non-zero, meaning it is easier to compute. While it seems that this view on geometry is the panacea for all the issues with the metric description, there are some applications of this formulation where it is less suited and a particular example of this is in its Hamiltonian formulation.

\subsection*{Chiral Formulations of Geometry and Gravity}

The story for chiral formulations of geometry began with an aim to classify solutions to Einstein's equations using the structure of their Weyl curvature tensor. Petrov~\cite{The_Classificat_Petrov_2000} realised that once decomposed using self-dual bivectors, the Weyl tensor has a particular structure depending on the existence of certain null vectors. All metrics can then be thought of as being in one of six Petrov types; each type is associated with some physical structure present in spacetime. The most notable example is Petrov type D, where isolated massive objects such as black holes or stars appear. Since these objects are highly interesting in physics, type D solutions are particularly important. The use of self-dual objects in this classification hints at the usefulness of this description of gravity. A few years earlier, Newman and Penrose developed their null tetrad formulation or Newman-Penrose (NP) formalism~\cite{An_Approach_to_Newman_1962}. This new formulation involves complex valued 1-forms that encode the metric; it is a complex transformation of the frame formulation. Through this formulation one can decompose tensors in a granular way, constructing components of all self-dual objects but without clear connections between them. It was not until the work of Cahen, Debever and Defrise~\cite{01138a60-03bf-3425-9736-5043c4a7c399} that the link between the NP formulation, self-dual (SD) and anti-self-dual (ASD) 2-forms was established. The fact that 2-forms in 4 dimensions split into SD and ASD halves is related to another coincidence, namely that the Lie algebra for the Lorenz group splits into 
\begin{align}
	\so(1,3) = \so(3,\C) \oplus \overline{\so(3,\C)}. \label{eq:intro-lorentz-chiral-factors}
\end{align}
where $SO(3,\C)$ is group of complex 3 dimensional rotations and the bar denotes complex conjugation.  Chiral then refers to one half of the above decomposition and objects that are valued in the representations of $\so(1,3)$ can then be decomposed into these chiral halves. Although it seems like 2-forms have little to do with the metric picture of geometry, choosing a triple of 2-forms defines a unique metric up to $SO(3,\C)$ rotations and conformal transformations. Given that the 2-forms are sufficiently non-degenerate then the metric they define is such that they become self-dual with respect to it. The reason this construction is available is due to a more general realisation that a metric and a spinor, in any dimensions, can be encoded into collections of differential forms called spinorial $G$-structures. Spinorial $G$-structure in neighbouring dimensions give fruitful examples of $G$-structures, see~\cite{Aspects_of_Spin_Niren_2024,Su2StructureBhoja2024,Lorentzian_Cayl_Krasno_2023} for more examples. The usual derived quantities that appear in the metric language, such as connections and curvature, also appear naturally in the $G$-structure view and are computed using exterior calculus without the need for the metric.

Since 2-forms arise as $SO(3,\C)$ structures, using the general $G$-structure construction, we know they appear on $SO(3,\C)$ principal bundles. Again these objects appear in the same context as those that appear in gauge theories, allowing gravity to be thought of as a gauge theory itself. Further useful properties arise when considering the SD and ASD 2-form split in 4 dimensions; the Riemann curvature of the metric has a natural decomposition into SD and ASD components. Then, if one is interested in conditions on the curvature that only involve one half, the other can be ignored. One such condition is Einstein's equations wherein this simplification can be used to remove the total number of components needed to be computed, due to only needing to compute the SD/ASD part of the curvature. All chiral theories share this same feature in that the number of components needed for calculations are reduced. More generally, all gauge theories have an interpretation where connection 1-forms encode some important data; the curvature of these 1-forms is a 2-form, and it is this 2-form that can be projected onto its SD and ASD halves. Equations of motion can then be written for one half of the curvature that has the same solutions as the full theory. Remarkably this is true for both YM and GR. For both theories, the difference between the chiral and non-chiral versions of the Lagrangian is an unphysical topological term. A more detailed explanation of how the chiral formulations capture the full theory can be found in \cref{sec:chiral-yang-mills-lagrangians,sec:pleb-connection-torsion-and-curvature}.

In this Thesis, we mainly focus on Lorentzian signature metrics and the associated chiral objects. These chiral objects become complex-valued due to their nature in this context. To ensure that the metric they describe remains real-valued, certain reality conditions must be imposed. The price one pays for the simplicity gained in using chiral objects is the need to impose these conditions. In contrast, Euclidean signature metrics do not require such reality conditions. The Euclidean version of \cref{eq:intro-lorentz-chiral-factors} is 
\begin{align}
	\so(4) = \su(2) \oplus \su(2)
\end{align}
therefore the resulting 2-forms are called $SU(2)$-structure. A similar split occurs in signature $(2,2)$ metrics, but this signature is not considered here. $SU(2)$-structures appear as a triple of real 2-forms and the metric constructed from them is real-valued. All other properties of the Lorentzian 2-forms are mirrored by the Euclidean versions. It is then more natural to consider Euclidean geometry through this approach considering that no reality conditions are needed.

One of the first more widely adopted uses of the chiral formulations was Ashtekar's new variables~\cite{NewVariablesFAshtek1986}. They were discovered in an attempt to simplify the Hamiltonian formulation of GR, in the metric version the constraints are nonlinear which made calculations with them difficult. By introducing complex combinations of the spin connection components it was found that the constraints and equations of motion became polynomial. This led to various development for quantum gravity including creation of loop quantum gravity and spin foam models which are non-perturbative theories that attempt to describe quantum gravity. It was realised that the reality conditions became difficult to deal with for the quantum calculations and so alternative real formulations have become increasingly popular~\cite{i_SO_i_Alexan_2000}. Later it was realised that Ashtekar's formulations was a Legendre transformation of the covariant \pleb{} action~\cite{OnTheSeparatiPleban1977}. This action was one of the first actions to disregard the metric as a fundamental field in place of the self-dual 2-forms, for this reason it is given the name \pleb{}'s formulation. It has the structure of a BF theory with additional constraints terms that remove the unwanted components in the 2-forms, see~\cite{BF_gravity_Celada_2016} for a review. Viewing gravity through this lens has led to pure connection descriptions where main variables become the connection 1-forms instead of the 2-forms~\cite{A_pure_spin_con_Capovi_1991,Pure_Connection_Krasno_2011}. This approach brings gravity closer in form to Yang-Mills theories, where connections are also central fields. Chiral deformations of \pleb{}'s formulation have been possible by modifying only the constraints in the action~\cite{Renormalizable_Krasno_2006}, producing new families of theories that have the same degrees of freedom as GR. It is clear that the change in perspective inspired by chiral formulations has driven developments across various areas of research. In this Thesis, we aim to continue studying chiral gravity and apply it to different areas of differential geometry and GR.

\section*{The Search for Solutions}

One of the primary areas in differential geometry and gravity involves finding solutions to specific differential equations. In particular, there is significant interest in identifying Einstein or Ricci flat metrics. For physicists, such metrics represent physical spacetime on which we could potentially live, making a comprehensive study of these solutions highly desirable. From a mathematician's perspective, classifying all possible solutions remains an intriguing and rich area of research. Einstein metrics are those that satisfy Einstein's field equations, and under certain matter content restrictions are equivalent to Ricci flat metrics. The standard formulation of these equations begins with the introduction of a metric $g_{\mu\nu}$, which carries 10 real components at each point on the manifold. Einstein's equations then consist of a set of 10 coupled nonlinear second-order partial differential equations on this metric. A general solution to these equations is not known due to their complicated nonlinear structure, making it natural to consider alternative strategies for generating solutions. The two main methods we explore in this Thesis are:
\begin{itemize}
	\item Imposing symmetry or specific structures on the solutions to simplify Einstein's equations and make them analytically solvable.
	\item Exploring numerical approximations that converge to true solutions.
\end{itemize}
Both of these approaches have been extensively studied, primarily within the framework of metrics. 

Another method for generating solutions, that is not explored in the main text, is perturbation theory. Where one expands around a known solution in a Taylor series, each term is generated order by order. Two different approaches to this expansion are the parameterised post-Newtonian (PPN) formalism and the self-force approach, for reviews on PPN and self-force see~\cite{Theory_and_Expe_Will_2018} and~\cite{Motion_of_small_Pound_2015} respectively. Perturbations using the chiral formulations have a rich structure that aids most computations, advantages have been noted over their metric counterparts, for example see~\cite{ChiralPerturbaKrasno2020,Pure_connection_Krasno_2024}. Chiral formulations seem to aid computations in almost every aspect of GR. A large part of this Thesis is to use chiral formulations to assist generating solutions, through symmetry simplifications and numerically.

\subsection*{Solutions with Symmetry}

An early solution to Einstein's equation is the Kerr metric~\cite{GravitationalFKerr1963}, which describes an isolated massive body with angular momentum. This was one of the first realisations of non-trivial solutions in GR, leading to a deeper study of massive objects such as stars and black holes. The Kerr solution is generated under the assumption that spacetime is axially symmetric and static, meaning certain angular and time translations leave the solution invariant. While this simplifies the equations of motion significantly, obtaining the solution still requires considerable work; see~\cite{TheMathematicaChandr1998} for a derivation. It's natural to wonder if there are easier ways to generate solutions. Since creating a general solution is extremely challenging, some simplifying assumptions must be made. It was observed that the Kerr metric and other isolated massive bodies have a Petrov classification of type D. Given the link between chiral formulations and type D spacetimes, where self-dual 2-forms play a crucial role, it makes sense to consider how these equations in the chiral formulations behave under the type D assumption. This leads directly to the works in~\cite{SelfDualKahleDerdzi1983} and subsequently~\cite{ExplicitSelfDClaude1991}, where it was found that such metrics are completely fixed up to a single partial differential equation (PDE). This clearly indicates that chiral formulations, such as \pleb{}'s formalism, should be key ingredients in understanding solutions to Einstein's equations in four dimensions. Part of this Thesis considers in more detail the structure of Einstein's equations using the self-dual 2-form description under the type D assumption.

\subsection*{Numerical Relativity}
\hypertarget{add:added-referenecs-in-num-rel-par}
Since general solutions of Einstein's equation are seemingly impossible to find, without some simplifying assumptions, it makes sense to consider numerical methods that approximate solutions~\cite{Smarr:1979fra}.  Numerical approximations offer a practical way to represent and solve GR equations on computers by discretising spacetime into finite grids~\cite{AlcubierreMiguelIntro3+12008}. Evaluating functions on the grid provides a way to represent the solutions numerically. By formulating the equations of motion as a Cauchy problem or evolution system, initial data can be prescribed on a spatial hypersurface and evolved over time~\cite{NumericalRelatBaumga2010}. The Hamiltonian framework provides a natural way to derive suitable evolution systems. The underlying theories have a diffeomorphism and gauge invariance, which manifests itself as constraints at the level of the Hamiltonian~\cite{Dirac2001-rn}. Constraints must be imposed at each time step during the numerical evolution. Standard methods of integrating these constrained systems impose the constraints only on initial hypersurfaces and perform a free evolution~\cite{Note_on_the_pro_Fritte_1997}. However, numerical errors accumulate during evolution that drive the system away from the constraint surface~\cite{Evolution_of_th_Detwei_1987} (the surface in phase space where the constraints are satisfied). Imposing the constraints at each time step is also unfeasible due to the time complexity of this task. Early simulations that employed free evolution faced stability issues due to this problem~\cite{Three_dimension_Annino_1995}, prompting the development of artificial damping techniques that help maintain the constraints.

Structure-preserving integrators (SPIs) are alternative discretisation methods that obey certain symmetries at the discrete level, in turn they preserve the discretised versions of the corresponding constraints~\cite{HairerLubichWannerGeoNumIntStrPreAlg2010}. A large class of SPIs depend on the use of differential forms, these use discrete exterior calculus (DEC)~\cite{DiscreteExteriDesbru2005} to extend exterior calculus to discrete spaces. Previous attempts at constructing SPIs for GR have only been possible on a smaller subset of equations, see~\cite{Free_and_constr_Richte_2008,Constructing_of_Tsuchi_2016} for examples. In~\cite{NumericalRelatFarr2010} Will M. Farr was able to construct an SPI for a particular BF theory of gravity that exactly preserved the constraints generated by Lorentz transformations. This promising result inspires us further to consider these formulations based on differential forms rather than the metric formulation. Chiral GR offers a formulation that is based on differential forms and a Hamiltonian theory where the constraints are considerably simplified. These exceptional properties have already led to the development of lattice quantum gravity and spin foam models, a review of which can be found in~\cite{A_Short_Review_Ashtek_2021}. Another result,  in~\cite{Finite_element_Arnold_2009}, remarks that the existence of a differential elliptic complex implies there is a well-posed DEC problem. This provides extra motivation for the work in \cref{chap:Plebanski-Elliptic-Complex}, and should further convince the importance of studying the chiral formulation since this is the only known formulation where a suitable elliptic complex exists.

Gauge fixing plays an important role in the development of numerical schemes for GR~\cite{Gauge_condition_Alcubi_2003}. These choices can effect the well-posedness~\cite{Hyperb1+3YvonneTommaso1983}, constraint preserving properties and long term stability of numerical systems~\cite{Hyperbolic_Meth_Reula_1998}. Much effort has gone into exploring various gauge conditions that are suitable for numerical relativity in the metric language. Due to the simple structure of the evolution and constraint equations in chiral formulations of GR, it is possible that a useful gauge fixing exists that provides better performance than the metric language.

\section*{Thesis Outline}

Part \ref{part:pleb-formulation} of this Thesis is dedicated to developing the chiral formulations and describing the underlying fibre bundle constructions on which the chiral objects live. Chapter \ref{chap:fibre-bundles} introduces fibre bundles and $G$-structures explaining their general construction. The metric and frame formulations are compared in \cref{chap:metric-frame-formulation}. \pleb{}'s formulations is detailed in \cref{chap:plebanskis-formulation} along with the pure connection descriptions. Chiral YMs theory has a similar structure to the chiral formulations of GR, and is used as a comparison in various parts of the Thesis. Therefore, in \cref{chap:chiral-yang-mills} Chiral YMs is detailed and the self-dual truncations are YM and GR are compared. Part \ref{part:linearised-gravity} contains all the results that pertain to the linear structure of Einstein's equations. A new 2 parameter family of first-order hyperbolic gauge fixings is developed for chiral GR in \cref{chap:Linearised-Gravity}, these are such that the implied second-order equations of motion are the wave operators in Minkowski space. Chapter \ref{chap:Plebanski-Elliptic-Complex} is a modification of the authors paper~\cite{Plebanski_compl_Krasno_2025} where an elliptic complex is developed for the Euclidean version of Einstein's equations and is only possible due to the chiral descriptions of linearised gravity. In both the first-order gauge fixing and the elliptic complex it is shown that linearised GR splits into two sectors with the conformal components appearing in one and the remaining components in the other. This separation is a slight generalisation of the result in~\cite{ChiralPerturbaKrasno2020} to a large family of gauge fixings and using spacetime and internal indices rather than spinor. Part \ref{part:nonlinear-gravity} collects all the results relating to the full nonlinear GR equations that are not numerical approximations. Chapter \ref{chap:type-D-conformal-to-kahler} is modification of the author's paper~\cite{Kerr_metric_fro_Krasno_2024} where a new proof of \derd{}'s theorem is displayed and some new results about Killing vectors and $SU(2)$-structures are derived. Most of that chapter is then a translation of the work in~\cite{AmbitoricGeomeAposto2013}, which solves Einstein's equations for type D metric, in terms of \pleb{}'s formulation where the calculation becomes an exercise in exterior calculus. Chapter \ref{chap:nonlinear-gauge-fixing} develops the nonlinear version of the first-order hyperbolic gauge fixing found in \cref{chap:Linearised-Gravity}. Furthermore, a novel conformal separation of the gauge-fixed Einstein's equation is found that reproduces the nonlinear version of the linear separation discovered in~\cite{ChiralPerturbaKrasno2020}. Part \ref{part:numerical-relativity} contains all the numerical developments of the Thesis. Chapter \ref{chap:numerical-relativity} is an introduction to the finite difference numerical schemes in the context of GR and their historical developments. A numerical evolution system for the nonlinear hyperbolic gauge fixing is derived in \cref{chap:plebanski-numerical-relativity}. Three different systems are developed, the first corresponding to the gauge fixing in the standard Ashtekar variables. The second uses the remarkable fact that the gauge fixing conformally separates the equations, new conformal fields are introduces that follow an alternative evolution system which we call the conformal system. A subset of the conformal system can be evolved completely separately to the remaining rest, allowing for alternative ways to numerically integrate GR. The third system proposed employs a mixing of the two ideas to create what we call the partially conformal system, where certain conformal fields are used in combination with Ashtekar's variables in attempt to use some properties of the conformal system while keeping others from the original gauge fixing. Numerical results of this partial system are also displayed. Finally, in \cref{chap:pure-connection-numerical-relativity} an evolution system is developed for the pure connection formalism and is heavily based on the author's paper~\cite{WeylCurvatureKrasno2022}. A new hyperbolic gauge fixing is also developed in order to control the evolution of the constraints.

    \addtocontents{toc}{\protect\setcounter{tocdepth}{1}}

    \newpage
    \part{\pleb{} Formulation}\label{part:pleb-formulation}
    
    \newpage
    \chapter{Fibre Bundles}\label{chap:fibre-bundles}
Fibre bundles are a key concept in differential geometry and physics, most physical theories have a fibre bundle interpretation. Bundles are generalisations of the Cartesian product on topological spaces. They are used in many areas of mathematics and physics, importantly for us, they are the main tool used to describe gauge theories. Moreover, the theory of general relativity can also be formulated in the context of fibre bundles. This allows for the formal definition of many tensors, in particular the frame (or tetrad) and connection arise naturally in this setup.

In this chapter we give a brief overview of fibre bundles, the objects on which are used extensively throughout the Thesis. We begin by defining a fibre bundle through attaching fibres to a base manifold, and then continue adding structure to the fibres. Even without any additional structure one can define a connection and its curvature on the fibre bundle. We go on to define principal bundles, for which the connection defines for us a gauge covariant exterior derivative on the space of differential forms with values the Lie algebra. The soldering form is also introduced as a map between the tangent space and a reference internal space, the covariant exterior derivative of which gives rise to torsion. In a local trivialisation, we find these objects have natural counterparts that live on the base space of the bundle, these can be realised as the frame and spin connection of the first-order Einstein-Cartan gravity. Or equivalently if the soldering forms are fixed one has the setup for Yang-Mills theory. After we define $G$-structures as reduction of the structure group of a frame bundle, in this context certain tensors are preserved by the subgroup G, the main example of which is the metric. This chapter is based on the books~\cite{Kobayashi1996-ho,Krasnov2020-qr}.

\section{Fibre Bundle}
A fibre bundle is a generalisation of the Cartesian product on 2 topological spaces. 
\begin{definition}\label{def:fibre-bundle}
    A fibre bundle is a triple of topological spaces $(F,E,M)$ with the maps $F \hookrightarrow E \overset{\pi}{\rightarrow} M$,where $E$ is the total space, $M$ is the base space, $F$ are called the fibres and $\pi : E \rightarrow M$ is the projection map. On each open neighbourhood $U_i \subset M$ we require the existence of a map $\phi_i : U_i \times F \rightarrow \pi^{-1}(U_i)$ called the trivialisation. At each point $x \in U_i$ we define $\phi_i|_x : F \rightarrow \pi^{-1}(x)$, such that $\pi \circ \phi_i|_x = x$. On the intersection $U_i \cap U_j \neq \emptyset$ we define
    \begin{align*}
        t_{ij}|_x := \phi_i^{-1}|_x \circ \phi_j|_x : F \rightarrow F,
    \end{align*}
    where $t_{ij} : U_i \cap U_j \rightarrow G$ are called the transition maps and $G$ is the group of automorphisms of $F$. Furthermore, these satisfy
    \begin{align*}
        t_{ji} = t_{ij}^{-1}, \quad t_{ii} = \id, \quad t_{ik} t_{kj} = t_{ij}.
    \end{align*}
\end{definition}
A subset $\mathcal{U} \subset E$ is open if and only if $\phi_i^{-1}(\mathcal{U} \cap \pi^{-1}(U_i))$ is open in $U_i \times F$ for each $i$. Given that $U_i \times F$ has the standard product topology, this provides the topology for the fibre bundle $E$. The dimension of the base and fibre spaces, denoted $n = \dim(M)$ and $m = \dim(F)$, are not required to be equal. This definition of the fibre bundle allows for many geometrically interesting setups and is the basis for gauge theories. 

A useful notion to introduce on a general fibre bundle is the section, 
\begin{definition}
    A section is the choice of a map $s : M \rightarrow E$ that respects the projection map $\pi \circ s = id$.
\end{definition}
In general a section can only be defined locally, $s : U \rightarrow U \times F$ with $U \subset M$ being a local neighbourhood, as there can be topological obstructions to existence of global sections. In a sense the section can be defined as the ``zero'' coordinate in the total space, allowing for coordinates on the total space to be defined with respect to it.

By giving certain structures to the fibres $F$ we can define more specific fibre bundles, we list the two main examples used in this text.
\subsection{Vector Bundle} 
\begin{definition}
    A vector bundle is a fibre bundle $(V,E,M)$ where the fibres, $V$, have the structure of a vector space. In this case there exists a global zero section, $o : M \rightarrow E$, which takes each element $x \in M$ to the zero element of the vector space $\pi^{-1}(x)$.
\end{definition}
When $V = K^m$ (where $K = \mathbb{R}\ {\rm or}\ \mathbb{C}$) then the structure group is $G = GL(m,K$) as it maps vectors to vectors.

\subsection{Principal Bundle} 
A principal $G$-bundle $G \hookrightarrow P \overset{\pi}{\rightarrow} M$ is a bundle where there exists a right action of the topological group $G$ on the total space, $G: P \times G \rightarrow P$, that is free and preserves the fibres. That is given $p \in P$, such that $\pi(p) = x \in M$, and $g \in G$ then the image of the action that preserves the fibres, $p g \in P$, is in the same fibre over $x = \pi(p g)$, the free action requirement means that $g p = p \Rightarrow g = e$ where $e$ is the identity element in $G$.

Using the above two bundles we can begin adding physical interpretation to the fibres to construct more useful spaces. Our first example is that of the tangent bundle, the tangent space at a point in the manifold $M$ is defined by
\begin{align}
    T_x M = \left\{\left. \dot{\gamma}(0)\ \right|\ \forall \gamma : (-1,1) \rightarrow M, \gamma(0) = x\right\}.
\end{align}
This is a vector space spanned by the space of tangent vectors to all the curves through $x$. From this we can construct the tangent bundle 
\begin{align}
    TM = \coprod_{x \in M} T_x M
\end{align}
by attaching the tangent space to each point in the base manifold $M$.~Let $ (U_i, \phi_i) $ be an atlas for $M$, where $\phi_i : K^n \rightarrow U_i$ and $n = \dim(M)$. This induces a map, via the pushforward, $d\phi_i : K^{2n} \rightarrow \pi^{-1}(U_i)$ where $\pi : TM \rightarrow M$ is the projection map. A set $\mathcal{U} \subset TM$ is open if and only if $\phi_i^{-1}(\mathcal{U} \cap \pi^{-1}(U_i))$ is open in $K^{2n}$ for all open sets indexed by $i$. In this way the topology for $TM$ is inherited from $M$. This is the usual starting point for Riemannian geometry, except that it is assumed the existence of a metric, here we will save this assumption until later. Instead, we introduce a different object, the frame is the isomorphism
\begin{align}
    f_x : V \rightarrow T_x M
\end{align}
between a vector space $V$ ($V = \mathbb{R}^n$ or $\mathbb{C}^n$) to the tangent space $T_x M$. The vector space and $T_x M$ have the same dimension so that the inverse map can exist. A choice of a frame is equivalent to choosing a set of basis vectors for $T_x M$. Taking all possible frame maps we can define a new space,
\begin{align}
    F_x = \{f_x\ |\ \forall f_x : V \rightarrow T_x M\}.
\end{align}
These spaces can be used to construct a new fibre bundle
\begin{align}
    FM = \coprod_{x \in M} F_x
\end{align}
which we call the tangent frame bundle. Sometimes it is denoted $F(TM)$ as the frames map to the fibres of $TM$. The group $GL(n,K)$ (where $K = \R$ or $\C$) has a free and transitive natural right action on the frame bundle, that is given $f_x \in F_x$ and $g \in GL(n,K)$ then $f_x \circ g \in F_x$ is the right action. Therefore, given a preferred section in a local trivialisation $s : U \rightarrow U \times F|_U$ we can define a trivialisation of a $GL(n,K)$ principal bundle through 
\begin{align}
    G_x = \{ \left. \rho \circ s^{-1} \right| \forall \rho \in F_x \} \cong GL(n,K)
\end{align}
where $\rho \circ s^{-1} : K^n \rightarrow K^n$ and hence an element of $GL(n,K)$. This means that locally the structure of $FM$ is isomorphic to $U \times GL(n,K)$, and therefore we can view the frame bundle as a $GL(n,K)$-principal bundle. Its topology inherited in the same way as the generic fibre bundle, see the discussion below \cref{def:fibre-bundle}, constructed from the topology of $M$ and $G = GL(n,K)$.

\subsection{Associated Vector Bundles}
It is pertinent at this point to introduce the associated vector bundle to a principal $G$-bundle $P$. Given a linear representation $r : G \rightarrow GL(n,K)$ of the group $G$ there is a vector bundle 
\begin{align}
    E = P \times_G V
\end{align}
where points on $E$ are a tuple $[g,v]$ subject to the equivalence relation
\begin{align}
    [g,v] \sim [p g, r\left(g^{-1}\right) v], \quad p \in P,\ g \in G,\ v \in V.
\end{align}
It can be seen that $E$ is the vector bundle over $M$ with trivialisation $M \times V$ via the map 
\begin{align}
    [p,v] \mapsto r(p) v \in V.
\end{align}
The map is well-defined, indeed using the equivalence relations 
\begin{align}
    [p g , r\left(g^{-1}\right) v] \mapsto r(p g) r \left(g^{-1} \right) v = r(p) v
\end{align}
we see it respect the equivalence class. In the above we denote $r(p) \in GL(n,K)$ to mean the map of $p$ to its group element (on a trivialisation) and then using the representation to map to $GL(n,K)$. This shows that $E$ is the vector bundle $V \hookrightarrow E \overset{\pi_E}{\rightarrow} M$.

\subsection{Connections and Curvature}
On a fibre bundle one has the minimum geometrical setup required to define a connection. An Ehresmann connection is defined as a decomposition of the tangent bundle over the total space $TE = HE \oplus VE$ into horizontal and vertical parts respectively. The vertical vectors are defined by $VE = \ker(d\pi)$ that is the kernel of the pushforward or differential of the projection map, $d\pi : TE \rightarrow TM$. Vertical and horizontal bundles are defined to be complimentary $VE \cap HE = {0}$ and any vector $X \in TE$ can be decomposed, using the connection, into $X = X_H + X_V$ where $X_H \in HE$ and $X_V \in VE$.

This definition of the connection allows for one to define the horizontal lift of vector fields on $M$. Given a curve $\gamma(t) \in M$, such that $x = \gamma(0)$ and its tangent vector $X = \dot{\gamma}(0) \in TM$, the horizontal lift is the curve $\tilde{\gamma}(t) \in E$ such that $\dot{\tilde{\gamma}}(t) \in HE$. Then $\tilde{X} = \dot{\tilde{\gamma}}(0) \in HE$ is the horizontal lift of the vector $X$. Another important construct is the curvature of the connection, it is defined by taking the vertical projection (using the connection) of the Lie bracket of two horizontal vector fields
\begin{align}
    R(X,Y) = \left[X_H,Y_H\right]{}_V, \quad R \in \Lambda^2(E,VE) \label{eq:fibre-curvature-as-lie-bracket}
\end{align}
the failure of this to be zero measures the integrability of the subbundle $HE$. It is worth noting that despite the presence of the Lie bracket in the above definition the curvature it a only requires the definitions of the vector fields at a point and not their extension in an open set. This is because the derivatives of $X,Y$ appear only in the horizontal part of $\left[X_H,Y_H\right]$, and therefore the vertical projection removes these components. Indeed by taking $X' = X + h Z$, for $h : E \rightarrow K$ such that $h(p) = 0$, as a vector that agrees at the point $p$ with $X$ but differs elsewhere and $Z \in TE$. By computing the curvature on $X'$, and using its definition in terms of $X$ and $Z$, we find $R(X',Y)|_p = R(X,Y)|_p$, confirming that only the value of $X,Y$ at the evaluation point matters. Therefore, the curvature is a 2-form. It is convenient to package the data for a connection into a 1-form with values in the vertical tangent bundle
\begin{align}
    \omega \in \Lambda^1(E,VE)
\end{align}
such that $\omega(X) \in VE$ for all $X \in TE$. The horizontal space is then defined as the kernel of the vertical projection,
\begin{align}
    HE = \{ X \in TE\ |\ \omega(X) = 0 \}.
\end{align}
The curvature is a vertical vector field valued 2-form, constructed using the connection 1-form, 
\begin{align}
    R = d\omega + \frac{1}{2} \left[ \omega \wedge \omega \right] \in \Lambda^2(P,VE).
\end{align}
It can be checked that this definition agrees with \cref{eq:fibre-curvature-as-lie-bracket}.

\subsection{Principal Connections and Curvature}
For a principal bundle, $P$, it is natural to require that $HE$ is invariant under the right action of the group. That is $R_{g*}(H_p P) = H_{pg} P$. We call such a connection a principal connection. Moreover, if $G$ is a Lie group then the vertical bundle can be associated with the Lie algebra $\mathfrak{g} \cong V_p P$. From now on we assume that the group $G$ is always a Lie group. This allows for the 1-form connection to be valued in the Lie algebra
\begin{align}
    \omega \in \Lambda^1(P,\mathfrak{g}).
\end{align}
For the horizontal bundle to be invariant under the right action it can be shown that the connection is required to transform as
\begin{align}
    R^*_g \omega = Ad(g^{-1})(\omega).
\end{align}
Conversely, any 1-form on $P$ that is Lie algebra valued and transforms like above defines a principal connection. The horizontal vector fields are
\begin{align}
    H_p P = \{\left. X \in T_p P\ \right|\ \omega(X) = 0 \}.
\end{align}
The vertical part of a vector field can also be extracted 
\begin{align}
    X_V = \omega(X),\quad X_V \in V_p P,\quad X \in T_p P.
\end{align}
This uses the fact that the vertical tangent bundle is isomorphic to the Lie algebra.

\begin{definition}
    A tensorial q-form is a differential form with values in a vector space $V$, which is $G$-equivariant. That is, given a representation $r : G \rightarrow End(V)$, the right action on a tensorial form obeys
    \begin{align}
        R_g^* \phi = r\left(g^{-1}\right) \phi,\quad \phi \in \Lambda^q(P,V), \quad g \in G
    \end{align}
    and that 
    \begin{align}
        \phi(X_1,\ldots,X_q) = 0
    \end{align}
    if at least one of $X_i$ is vertical.
\end{definition}
Tensorial forms are named that way because they transform in the correct way under the group action, similar to tensors on a spacetime that satisfy a specific transformation rule. A useful property for tensorial forms is that they are in 1-to-1 correspondence with $E = P \times_G V$ valued q-forms on the base manifold $M$. With $\phi \in \Lambda^q(P,V)$ we have that 
\begin{align}
    \tilde{\phi}(X_1,\ldots,X_q) = [g,\phi(\tilde{X_1},\ldots,\tilde{X_q})] \in \Lambda^q(M,E)
\end{align}
where $\tilde{X}_i$ is any vector such that $d\pi_P(\tilde{X}_i) = X_i$. This map is well-defined, in that given another point $g' = g h$ we have that 
\begin{align}
    R^*_h \tilde{\phi}(X_1,\ldots,X_q) &= [g h, R^*_h \phi(\tilde{X_1},\ldots,\tilde{X_q})] \\ &= [g h, r(h^{-1}) \phi(\tilde{X_1},\ldots,\tilde{X_q})] \sim [g,\phi(\tilde{X_1},\ldots,\tilde{X_q})]
\end{align}
using the equivalence relation. This correspondence can also be constructed in the opposite direction, indeed starting with $\tilde{\phi} \in \Lambda^q(M,E)$ such that
\begin{align}
    \tilde{\phi}(\tilde{X}_1,\ldots,\tilde{X}_q)(x) = [u(x),v(\tilde{X}_1,\ldots,\tilde{X}_q)(x)], \quad u \in P, v \in \Lambda^q(M,V).
\end{align}
We can then define the 1-form on $P$ as 
\begin{align}
    \phi(d\pi_P(\tilde{X}_1),\ldots,d\pi_P(\tilde{X}_q))(p) = r(p^{-1} u(\pi_P(p))) v(\tilde{X}_1,\ldots,\tilde{X}_q)(\pi_P(p)).
\end{align}
By choosing another point $p' = pg$, and using that $\pi_P(pg)=\pi_P(p)$, we find 
\begin{align}
    R_g^* \phi(d\pi_P(\tilde{X}_1),\ldots,d\pi_P(\tilde{X}_q)) = r(g^{-1}) \phi(d\pi_P(\tilde{X}_1),\ldots,d\pi_P(\tilde{X}_q)).
\end{align}
The maps between $\Lambda^q(M,E)$ and $\Lambda^q(P,V)$ are inverses of each other, which can easily be checked by computing their composition.

The existence of a connection allows us to define the exterior covariant derivative, which is a map
\begin{align}
    D : \Lambda^q(P) \rightarrow \Lambda^{q+1}(P)
\end{align}
defined as
\begin{align}
    (D \phi)(X_1,\ldots,X_q) = (d\phi)(\tilde{X}_1,\ldots,\tilde{X}_q), \quad \phi \in \Lambda^q(P)
\end{align}
where $\tilde{X}_i$ is the horizontal part of $X_i$ with respect to $\omega$. When acting on tensorial q-forms $\phi \in \Lambda^q(P,V)$ and using the matrix representation of the Lie algebra, we have that
\begin{align}
    D = d + \omega \wedge.
\end{align}
In the matrix representation the curvature 2-form of the connection is equal to
\begin{align}
    R = D\omega = d\omega + \omega \wedge \omega \in \Lambda^2 (P,\mathfrak{g}).
\end{align}
The curvature is a tensorial 2-form and agrees with the previous definition in terms of horizontal and vertical projections. Here, the tensorial property of the curvature means that it is $G$-equivariant, or that it transforms as follows under the action of the group
\begin{align}
    R_g^* R = Ad(g^{-1})(R) = g^{-1} R g.
\end{align}

The exterior covariant derivative $D$ can be associated with a covariant derivative $D_E : \Lambda^q(M,E) \rightarrow \Lambda^{q+1}(M,E)$, it is defined by letting the following diagram commute.
\begin{equation}
\begin{tikzcd}
{\Lambda^q(M,E)} \arrow[rr, "D_E \tilde{\phi}"] \arrow[dd, "\tilde{\phi} \rightarrow \phi"] &  & {\Lambda^{q+1}(M,E)} \\
&  &  \\
{\Lambda^q(P,V)} \arrow[rr, "D\phi"] &  & {\Lambda^{q+1}(P,V)} \arrow[uu, "\phi \rightarrow \tilde{\phi}"]
\end{tikzcd}
\end{equation}
Its definition is easiest to understand in a local trivialisation, which will be computed shortly.

\subsection{Soldering Form}
One of the most important object, in the descriptions of gravity considered here, is the soldering form $\theta \in \Lambda^1(P,V)$. Given a $GL(n,K)$ principal bundle $G \hookrightarrow P \overset{\pi_P}{\rightarrow} M$ and a representation $r : G \rightarrow {\rm End}(V)$ (recall $V = \R^n$ or $\C^n$) we can construct the associated vector bundle $V \hookrightarrow E \overset{\pi_E}{\rightarrow} M$. The soldering form provides a map between the tangent space of the principal bundle and a vector space $V$. This is done by interpreting each element $p \in P$ as a particular frame $f_p \in FM$, such that at $\pi_P(p) = x \in M$ and $f_p : V \rightarrow T_{\pi_P(p)} M$. At a different point $p' =p g \in P$ the frame transforms as $f_{p'} = f_{p g} = f_p r(g)$. The soldering form is then
\begin{align}
    \theta = f_p^{-1} \circ d\pi_P \in \Lambda^1(P,V)
\end{align}
that takes vectors on the total tangent space $T_p P$ and maps them to elements in $V$. It can be seen that the soldering form is tensorial, first we can compute its transformation under the right action,
\begin{align}
    R_h^* \theta = f_{p g}^{-1} \circ d\pi_P = (f_p r(g)){}^{-1} \circ d\pi_P = r(g^{-1}) \theta
\end{align}
where we recover the correct $G$-equivariant transformation law. Secondly, vertical vectors are in the kernel of $d\pi_P$ and therefore of $\theta$ also, so the second property of tensorial forms is satisfied automatically. The soldering forms are how the tetrads appear in the context of fibre bundles, however, the usual form they are written in is the local form which we will describe later.

The torsion of a soldering form is 
\begin{align}
    T = D \theta
\end{align}
which, when $\theta$ is given its vector representation on $V$, gives
\begin{align}
    T = d \theta + \omega \wedge \theta.
\end{align}
Taking the exterior derivative on both sides gives 
\begin{align}
    dT = d \omega  \wedge \theta - \omega \wedge d \theta = d\omega \wedge \theta - \omega \wedge ( T - \omega \wedge \theta )
\end{align}
which rearranges to be 
\begin{align}
    dT + \omega \wedge T = (d\omega + \omega \wedge \omega) \wedge \theta.
\end{align}
Using $D = d + \omega \wedge$ we have the first Bianchi identity 
\begin{align}
    D T = R \wedge \theta.
\end{align}
The second Bianchi identity is 
\begin{align}
    D R = 0.
\end{align}

\section{Trivialisation of Soldering Form and Connection}
Global sections do not exist on every manifold $M$, there are subtle global restrictions that prevent this. Principle bundles only contain a global section if they are trivial, i.e. $P = M \times G$, whereas a vector bundle always has the global zero section. The Hopf fibration $S_1 \hookrightarrow S_3 \rightarrow S_2$, where $S_n$ is the $n$-sphere, is an example of a principle bundle with no global section. To overcome this we restrict ourselves to a local neighbourhood on $M$, on which these objects can always be defined. Given an open neighbourhood $U \in M$, we can define the trivialisation $P|_U \cong U \times G$ via a local section $\sigma : U \rightarrow G$,
\begin{align}
    \psi_\sigma : U \rightarrow U \times G, \quad \psi_\sigma(x) = (x,\sigma(x)) \in P.
\end{align}
We have introduced the notation $(x,g) \in P$ where $x \in M, g \in G$. At the point $(x,g) \in P$ we can define the local connection 1-form as 
\begin{align}
    \omega = g^{-1} dg + g^{-1} A g \label{eq:fibre-local-connection}
\end{align}
where $A \in \Lambda^1(U,\mathfrak{g})$ is the connection in the base. To make sense of the above formula we can introduce the Maurer-Cartan form on a Lie group, $\theta |_g = d L_{g^{-1}}|_g : T_g G \rightarrow \mathfrak{g}$, which is the pushforward of the left action of the group by the inverse of $g$. Since this maps tangent vectors on $G$ to Lie algebra elements it is clear that this is a Lie algebra valued 1-form on $G$. Now using a matrix representation $r : G \rightarrow End(V) \cong GL(V)$, which is assumed to always exist, we can construct its pushforward $dr : TG \rightarrow T GL(V) \cong GL(V)$, where we have used that the tangent space of the space of matrices is again the space of matrices. Applying this pushforward to the Maurer-Cartan form we find
\begin{align*}
    dr \circ d L_{g^{-1}} = d(r\circ L_{g^{-1}}) = d (\mathcal{L}_{r(g){}^{-1}} \circ r) = d \mathcal{L}_{r(g){}^{-1}} \circ dr.
\end{align*}
Where $\mathcal{L}_A : GL(V) \rightarrow GL(V)$, for any $A \in GL(V)$, is the left multiplication map on the matrix group, and the notation $df$ means the pushforward of the map $f$. A standard result for the matrix group is that $d\mathcal{L}_A = \mathcal{L}_A$ due to the identification of the group with its own tangent space. This allows for the interpretation
\begin{align*}
    dr \circ dL_{g^{-1}}|_g = \mathcal{L}_{r(g){}^{-1}} \circ dr |_g = r(g){}^{-1} dr |_g.
\end{align*}
The first term in \cref{eq:fibre-local-connection} follows from identifying $g^{-1}\equiv r(g){}^{-1}$ as the left multiplication by the matrix representation of $g$, and $dg \equiv dr|_g$ as the 1-form that maps the tangent vectors on $G$ to their matrix representations at the point $g$. Similarly the second term in \cref{eq:fibre-local-connection} is more accurately described as
\begin{align*}
    g^{-1} A g \equiv r(g){}^{-1} dr(A) r(g),
\end{align*}
that is each factor is mapped to its matrix representation first. All together we find that $\omega \in \Lambda^1(U \times G,GL(V))$, with the first term contributing $\Lambda^1(G,GL(V))$ and the second $\Lambda^1(U,GL(V))$, such that each term accepts a vector field from its corresponding base. Hence, the above 1-form is valued in the representation of the Lie algebra. This satisfies the correct transformation law under the right action 
\begin{align}
    R^*_h \omega = {(g h)}^{-1} d(g h) + {(gh)}^{-1} A (gh) = h^{-1} \omega h
\end{align}
for a constant $h \in G$. Using the identity trivialisation to define a local section $\sigma_0 : U \rightarrow P$, such that $\sigma_0 : x \mapsto (x,id)$ where $id$ is the identity element, allows one to define the connection on the base as the pullback of the connection in the bundle,
\begin{align}
    \sigma_0^*(\omega) = A.
\end{align}
Given another trivialisation, $\sigma'(x) = \sigma_0(x) h(x)$ and $h(x) \in G$, we find 
\begin{align}
    A' & = (s'){}^*(\omega) = (id\ h(x)){}^{-1} d(id\ h(x)) + (id\ h(x)){}^{-1} A (id\ h(x)) \\ & = h(x){}^{-1} dh(x) + h(x){}^{-1} A h(x).
\end{align}
This is the gauge transformation of the connection 1-form, the choice of section on $P$ is equivalent to a choice of gauge.

With this realisation we can give a local description of the covariant derivative on the associated vector bundle $E$. A section on $E$ is the prescription 
\begin{align}
    \tilde{s} : U \rightarrow E, \quad \tilde{s} : x \mapsto [u(x),v(x)], \quad x \in U,\ u(x) \in P,\ v(x) \in V.
\end{align} 
Moving to a section $s : P \rightarrow V$ as 
\begin{align}
    s(p) = r(p^{-1} u(x) ) v(x), \quad \textrm{where} \quad \pi_P(p) = x \in U.
\end{align}
The covariant derivative in $P$ acting on the section $s$ is given by
\begin{align}
    Ds(X) = (ds)(X-X_V) = X(s) - X_V(s)
\end{align}
where $X_V$ is the vertical part of $X$. We can now make use of the identity
\begin{align}
    X_V(\tilde{\phi}(Y_1,\ldots,Y_q)) = -[\omega(X),\tilde{\phi}(Y_1,\ldots,Y_q)] \quad \textrm{for} \quad \tilde{\phi} \in \Lambda^q(P,V),
\end{align}
where the brackets now represent the wedge product and the action of the Lie Algebra on $V$. Using this we find 
\begin{align}
    Ds(X) = X(s) + [\omega(X),s] = (d s + \omega \wedge s)(X).
\end{align}
Moving the exterior derivative back to a differential form in $\Lambda^1(M,E)$ we obtain
\begin{align}
    (D_E \tilde{s})(X) = [p(x),(d s + \omega \wedge s)(X){}_p].
\end{align}
Using the local form for the $\omega$ and $s$, at the point $(x,p) \in P$, we have
\begin{align}
    d s + \omega \wedge s & = -p^{-1} dp\ u(x) v(x) + p^{-1} d(u(x)v(x)) - p^{-1} dp p^{-1} u(x) v(x) + p^{-1} A u(x) v(x) \nonumber \\ & = p^{-1} (d(u(x)v(x)) + A u(x)v(x)) = p^{-1} d^A(u(x)v(x))
\end{align}
where the representation map is implied but not shown. The form of the covariant derivative on $E$ is  then
\begin{align}
    (D_E \tilde{s})(X) = [p(x), p(x){}^{-1}d^A(uv)(X)].
\end{align}
We now use the choice of local section $\sigma : M \rightarrow P$ to define a trivialisation of $E \cong M \times V$ via 
\begin{align}
    \psi_\sigma([p,v]) = (\pi_P(p), \sigma(x){}^{-1} r(p) v) \in M \times V.
\end{align}
It is easy to check that this is well-defined on $E$. This assigns an element in $V$ to any element in $E$, as such for the map $\tilde{s} : M \rightarrow E$ we can define 
\begin{align}
    \tilde{s}_\sigma : M \rightarrow V, \quad \tilde{s}_\sigma(x) = \sigma(x){}^{-1} u(x) v(x).
\end{align}
Applying this trivialisation to the covariant derivative on $E$ we have 
\begin{align}
    \sigma^* (D_E \tilde{s})(X) = \sigma(x){}^{-1} d^{A}(\tilde{s}_{\sigma_0})(X) = d^{A_\sigma}(\tilde{s}_{\sigma})(X).
\end{align}
Where $A_\sigma = \sigma(x){}^{-1} d\sigma(x) + \sigma(x){}^{-1} A \sigma(x)$ is the local connection gauge transformed to the point $(x,\sigma(x)) \in P$. Therefore, we can define the local gauge covariant exterior derivative on $\Lambda^q(M,V)$ as 
\begin{align}
    d^A = d + A \wedge.
\end{align}
It is not too difficult to check that this extends to all $V$-valued differential forms on $M$. The curvature of the local connection is defined as
\begin{align}
    F(A) = d^A A = dA + A \wedge A.
\end{align}
It is related to the curvature of $\omega$ through the pullback $F = \sigma^*(\Omega)$. It transforms with the adjoint action under a change of trivialisation, $F' = g^{-1} F g$. As for the soldering form at $p \in P$,
\begin{align}
    \theta = f_p^{-1} \circ d\pi_P,
\end{align}
we have that it is associated to 
\begin{align}
    \tilde{\theta} = [p,f_p^{-1}] \in \Lambda^1(M,E).
\end{align}
There is no $d\pi_P$ in the above as $\tilde{\theta}$ is a differential form on $M$ and $d\pi_P(X) = X$ for $X \in TM$. The trivialisation of $\tilde{\theta}$ given $\psi_\sigma$ is
\begin{align}
    e = f_{\sigma(x)}^{-1} \in \Lambda^1(M,V).
\end{align}
We call $e : T_x M \rightarrow V$ the frame (more correctly it is the coframe, but we ignore this subtlety and simply call it the frame), this is the tetrad that appears in physics literature. Under a gauge transformation $\sigma(x)' = \sigma(x) g(x)$ we find the frame transforms as 
\begin{align}
    e' & = f_{\sigma(x)'}^{-1} = f_{\sigma(x) g(x)}^{-1} = g(x){}^{-1} e.
\end{align}
One can also define the torsion locally using the local gauge covariant derivative 
\begin{align}
    \Theta = d^A e \in \Lambda^2(M,V).
\end{align}
It is possible to see that $\Theta = \sigma^*(T)$ is the pullback of the torsion of the soldering form.

The above shows that we can write the soldering form and the connection as 1-forms on the base $M$; at the point $(x,g) \in U \times G \subset P$ we can reconstruct the soldering form and connection via
\begin{align}
    \theta = g^{-1} e \circ d\pi_P, \quad \omega = g^{-1} dg + g^{-1} A g
\end{align}
for $e \in \Lambda^1(U,V)$ and $A \in \Lambda^1(U,\mathfrak{g})$.

This shows that to define the appropriate tensors on the frame bundle (which is soldered to the principal bundle) we only need to introduce the tensors on the base $M$ and can lift to $P$ locally through an arbitrary section. This only works when one is only interested in the local structure, one must be more careful when defining global versions of the local structures. Choosing a section in $P$ is what is known as a gauge choice in the physics literature. Physics should be invariant under gauge choices, as gauge captures the redundancies in the mathematical description. Therefore, any two gauges should produce the same equations that govern any dynamics. The correct objects to achieve this are $G$-equivariant differential forms and as we have seen all the objects in fibre bundles arise naturally in this way. We note that the connection 1-form $A$ is not $G$-equivariant, however, the curvature is, and it is this that is used to construct physical equations.

We briefly discuss the physical applications of the soldering form and connection form. As per its name, the soldering form ties the tangent manifold of the base space to the vector bundle and as such is the map $e : TM \rightarrow V$. The connection on the vector bundle is then $A$ and can be identified with the pullback of the connection on the associated principal bundle. This is a very similar setup to that of Yang-Mills theory, which also starts as a principal bundle over spacetime. The only difference between geometric setup for gravity and Yang-Mills is the introduction of a dynamical soldering form and that the equations of motion are different. Geometrically both physical theories begin with a principal bundle.

\section{\texorpdfstring{$G$}{G}-structures}
A $G$-structure, for a manifold $M$, is a reduction of the tangent frame bundle $FM$. The group $G \subset GL(n,K)$ is a subgroup of the total space of transformations on the frame. Many interesting structures on manifolds appear in this way, perhaps the most notable being the metric which appears as an $O(n)$-structure. First one fixes a local tensor, for $SO(n)$ we fix $\eta \in S^2 V^*$. The $SO(n)$-structure on $M$ is then encoded by the soldering forms or frame $e : TM \rightarrow V$, 
\begin{align}
    g = \eta(e,e) \in S^2 \Lambda^1
\end{align}
where $g$ is the metric. In this sense the $G$-structure is the pullback of the local structure on $V$ to $M$. The choice of frame is then restricted to the subset of frame that are adapted to the geometric structure, i.e.~such that a change in the choice of frame leaves the $G$-structure invariant on $M$. For the group $SO(n)$ this means that the frames $e$ are all related to each other by an $SO(n) \subset GL(n)$ rather than the full linear group  This leads to the following definition of a $G$-structure, 
\begin{definition}
    A $G$-structure on $M$ is a reduction of the principal $GL(V)$ bundle $F^*M$ of frames to a principal $G$-bundle, $G \subset GL(V)$.
\end{definition}
Depending on the subgroup there are possible global topological obstructions to  the existence of such structures, some manifold will not support all $G$-structures. Most of the constructions in this Thesis are defined locally, therefore any $G$-structure is allowed as the global existence is not needed. 

Given a subgroup $G \subset GL(V)$ and a tensor $T_G$ that is preserved under the action of $G$. The $G$-structure is then encoded into a collection of differential forms (in this case we no longer restrict to 1-forms, although higher degree forms can always be constructed from 1-forms), for which there is a canonical pullback of the local tensor $T_G$ to $M$. As such there exists an action of $G$ on the differential forms that leaves the pullback invariant also. We then have two equivalent descriptions of $G$-structures, we have $(T_G,e)$ and equivalently the pullback of $T_G$ via $e$. In this discussion the first pair of data appears more naturally than the $G$-structure itself, and it is this we assume provides the fundamental fields to any theory constructed. In the case of gravity this is taking the data $(e,\eta)$ over the metric $g$. Gravity is an important example of theories that arise through $G$-structures, in particular by encoding the metric with differential forms we see that two different formulations arise which will be described in later chapters.

    \newpage
    \chapter{Metric and Frame Formulation}\label{chap:metric-frame-formulation}
General relativity (GR) was introduced by Einstein~\cite{AEinsteinTheoryGravitation1915}, Einstein's field equations for gravity relate the curvature of the spacetime to its matter content. It was introduced using the metric formulation, the metric tensor gives a natural definition of distances and angles on a manifold. As such the geometry of spacetime is perhaps more easily understood in this language. The curvature of a metric and spacetime come with an intuitive picture, that of tangent vectors failing to commute around when parallel transported around an infinitesimal loop. Therefore, historically GR has been taught using the metric language.

However, despite the success of the metric formulation we hope to convince the reader that different formulations offer more advantageous points of view on geometry and gravity. One such formulation is the frame formulation (or tetrad formulation). In which the frame replaces the metric as the fundamental field, this is the point of view taken by \'Elie Cartan~\cite{JMPA_1922_9_1__141_0,zbMATH02603088}. It is a specific example of a $G$-structure, which more generally encodes structures on a manifold into a Lie group~\cite{The_geometry_of_Chern_1966}. Frames appear naturally in the context of $G$-structures and the metric is a derived field defined as the pullback of the canonical $SO(n)$-structure from the flat internal space to the spacetime manifold $M$. In this way the metric is encoded into a collection of differential forms, this is often the way it is thought of in the mathematical literature.

This also brings gravity in line with the other theories of nature, Yang-Mills theories are already described by connection on fibre bundles and differential forms~\cite{The_Mathematica_Marath_1992}, it then feels natural that the same should be done for gravity. The group $SO(1,3)$ naturally acts on the frames, which reveals the Lorentz symmetry of the theory. One also gains the ability to couple gravity to spinors through the frame, while this is not considered in any further details in this Thesis, it is an attractive property of the frame formulation. A possible complaint about the frame formulation is that the number of components is increased from 10 in the metric to 16. We see that this advantage for the metric is short-lived, indeed one needs to compute 40 connection components as compared to 24 for the frame. As the curvatures computed contain the same number of components it is clear that the frame formulation offers more economy during computations. This and the reduction in complexity of the computations through the use of exterior algebra is the reason that we propose the frame, and other differential form based, formulations to be a better starting point for understanding gravity.

In \cref{sec:metric-formulations} we briefly describe the Riemannian geometry associated with GR and the construction of the curvature and ultimately how Einstein's equations are imposed by relating it to the matter content. We compare this with the Einstein-Cartan frame formulations in \cref{sec:einstein-cartan-formulation} where the same objects are computed. Action principles are displayed for both that encode Einstein's equations.

More details on the different formulations of gravity can be found in~\cite{Krasnov2020-qr,PlebanskiFormuKrasno2009}. More details on Riemannian geometry and gravity can be found in~\cite{Geometry_Topol_Nakaha_2018,Carroll_2019}

\section{Metric Formulation}\label{sec:metric-formulations}

\subsection{The Metric, Levi-Civita Connection and Curvature}
We assume our spacetime is accurately described by a 4-manifold $M$. On this manifold there exists tangent vectors, that arise as derivatives of smooth curves on $M$. The metric is then an object 
\begin{align}
    g \in T^*M \otimes T^*M
\end{align}
that takes two vectors at a point $x \in M$ and maps them to a real number. This allows for the definition of angles and distances on curved spacetimes. As we are interested in the Lorentzian signature, more specifically the signature $-+++$, we consider the existence in local patches only. On such patches we introduce coordinates $x^\mu \in M$ where $\mu = 0,1,2,3$ and a basis of 1-form $dx^\mu \in \Lambda^1$. The metric, is then
\begin{align}
    g = g_{\mu\nu} dx^\mu \otimes dx^\nu 
\end{align}
where $g_{\mu\nu}$ are the components of the metric (the components will also be referred to as the metric). The metric is symmetric in its indices, $g_{\mu\nu} = g_{(\mu\nu)}$ where $g_{(\mu\nu)} = \frac{1}{2} \left( g_{\mu\nu} + g_{\nu\mu} \right)$ represents the symmetric part of the indices. The inverse of the metric is given by $g^{\mu\nu}$ such that $g^{\mu\rho} g_{\rho\nu} = \delta^\mu_\nu$ and $\delta^\mu_\nu$ is the Kronecker delta. The metric is used to raise and lower spacetime indices. It is required that the metric have non-zero determinant, such that its inverse can be defined.

\subsubsection{Levi-Civita Connection}
For the metric we then have the Levi-Civita connection or equivalently the covariant derivative $\nabla_\mu$. This is a derivation that satisfies the following properties
\begin{align}
    \nabla_\mu g_{\rho\sigma} & = 0 \\
    \nabla_{[\mu} X_{\nu]} & = \partial_{[\mu} X_{\nu]} \quad \forall X_{\mu} dx^\mu \in \Lambda^1.
\end{align}
Where $\nabla_{[\mu} X_{\nu]} = \frac{1}{2} \left( \nabla_{\mu} X_{\nu} -\nabla_{\nu} X_{\mu} \right)$ denotes the antisymmetrisation of indices. The first conditions is that the metric is preserved when parallel transported around the manifold. The second condition is the torsion free condition. When both are satisfied the covariant derivative can be expanded in terms of the uniquely defined Christoffel symbols
\begin{align}
    \Gamma^\rho_{\mu\nu} = \frac{1}{2} g^{\rho\sigma} \left( \partial_\mu g_{\nu\sigma} + \partial_\nu g_{\mu\sigma} - \partial_\sigma g_{\mu\nu} \right)
\end{align}
where $\Gamma^\rho_{[\mu\nu]} = 0$ and the expansion for a generic tensor is
\begin{align}
    \nabla_\mu T^{\mu_1\ldots\mu_q}_{\mu_{q+1}\ldots\mu_{n}} = \partial_\mu T^{\mu_1\ldots\mu_q}_{\mu_{q+1}\ldots\mu_{n}} + \Gamma^{\mu_1}_{\mu\rho} T^{\rho\ldots\mu_q}_{\mu_{q+1}\ldots\mu_{n}} + \cdots - \Gamma^{\rho}_{\mu\mu_{q+1}} T^{\mu_1\ldots\mu_q}_{\rho\ldots\mu_{n}} - \ldots
\end{align}

By applying the covariant derivative twice to a 1-form and antisymmetrising the derivatives we obtain a definition of the Riemann curvature 
\begin{align}
    R_{\mu\nu\rho}{}^\sigma X_\sigma = 2 \nabla_{[\mu} \nabla_{\nu]} X_\sigma.
\end{align}
Which implies the following definition in terms of Christoffel symbols 
\begin{align}
    R_{\mu\nu\rho}{}^\sigma = \partial_\mu \Gamma^\sigma_{\nu\rho} - \partial_\nu \Gamma^{\sigma}_{\mu\rho} + \Gamma^\alpha_{\nu\rho} \Gamma^\sigma_{\mu\alpha} -  \Gamma^\alpha_{\mu\rho} \Gamma^\sigma_{\nu\alpha}. \label{eq:metric-Riemann-curvature-tensor-definition}
\end{align}

\subsection{Einstein's Equations}
In 1915 Einstein wrote down his famous field equations describing how matter and geometry are intimately linked. They arise as algebraic conditions on the curvature of a metric $g_{\mu\nu}$. The matter content is described by a symmetric stress-energy tensor $T_{\mu\nu}$, which describes all matter in the theory. Einstein's field equations are then 
\begin{align}
    R_{\mu\nu} - \frac{1}{2} R g_{\mu\nu} + \Lambda g_{\mu\nu} = \kappa T_{\mu\nu}. \label{eq:metric-einsteins-equation}
\end{align}
where $R_{\mu\nu} = R_{\mu\rho\nu}{}^\rho$ is the Ricci tensor, $R = g^{\mu\nu} R_{\mu\nu}$ is the Ricci scalar, $\Lambda$ is the cosmological constant and $\kappa$ is the Einstein gravitational constant. They are a second order set of differential equations on the components of the metric. Solving them in general is a difficult task and analytical solutions are only known in certain situations. In dimension 4, and in Lorentzian signature, no analytical solutions without extra symmetry assumptions are known.

\subsection{Einstein-Hilbert Action}
The Einstein-Hilbert action is action principle whose equations of motion are Einstein's equations~\cite{Hilbert1915}. It is given as 
\begin{align}
    S_{EH}[g] = \int_M d^4x \sqrt{|g|} (R - 2 \Lambda) + \mathcal{L}_m.
\end{align}
Here, $\sqrt{|g|} = \sqrt{|\det(g)|}$, $R$ is the Ricci scalar, $\Lambda$ is the cosmological constant and $\mathcal{L}_m$ is the Lagrangian describing matter contribution. The action is at most second order in derivatives of the metric. By varying the action with respect to $g_{\mu\nu}$ we find 
\begin{align}
    \frac{\delta S}{\delta g_{\mu\nu}} = 0, \quad \Rightarrow R_{\mu\nu} - \frac{1}{2}g_{\mu\nu}R + \Lambda g_{\mu\nu} = T_{\mu\nu}.
\end{align}
Where the stress-energy tensor is defined through the matter Lagrangian 
\begin{align}
    T^{\mu\nu} = \frac{\delta\mathcal{L}_m}{\delta g_{\mu\nu}}.
\end{align}

Action principles are a concise way of conveying all the equations of motion. They also allow one to identify the symmetries available. In the case of gravity, the action is invariant under a diffeomorphism transformation. This is related to the idea that physics should be independent of the choice of coordinates~\cite{2018grav.book.....M}.

\section{Einstein-Cartan Formulation}~\label{sec:einstein-cartan-formulation}
Here we describe how Einstein's field equations are imposed by conditions on $SO(1,3)$-structures. We begin with a manifold $M$ of dimension $\dim(M) = 4$. We reduce the bundle of frames to an $SO(1,3)$-principal bundle so that we have an $SO(1,3)$-structure; the internal space is then $V = \R^{1,3}$. The frame that encodes the $G$-structure is then
\begin{align}
    e : T_x M \rightarrow \R^{1,3}
\end{align}
for which the soldering form at $(x,g) \in P$ is $\theta = g^{-1} e \circ d\pi_P$. Every object can be lifted from its corresponding form on the base, therefore we only need to consider differential forms on the base space $M$. Now we take a local patch so that we can introduce a basis, we call $\partial_I \in \R^{1,3}$ the basis for vectors on the internal space, where $I=0,1,2,3$ and $x^{\mu} \in M$ the coordinates for $\mu=0,1,2,3$. The coordinates on $M$ give a natural 1-form basis $dx^\mu \in \Lambda^1(M)$. This basis can be used to define the components of the frame 
\begin{align}
    e = e^I_\mu dx^\mu \otimes \partial_I \quad \textrm{or} \quad e^I = e^I_\mu dx^\mu.
\end{align}
We assume that the frame is invertible, and the inverse components are given by $e^\mu_I$, such that $e^I_\mu e^\mu_J = \delta^I_J$ and $e^I_\mu e_I^\nu = \delta_\mu^\nu$. Transformation by the group $SO(1,3)$ are chosen as to leave the tensor $\eta_{IJ} \in S^2 V^*$, where $\eta_{IJ} = {\rm diag}(-1,1,1,1)$, invariant. That is 
\begin{align}
    \eta'_{IJ} = m_I{}^K m_J{}^L \eta_{KL} = \eta_{IJ}
\end{align}
for $m^I{}_J \in SO(1,3)$. Pulling back the internal metric using the frame gives the physical metric (we often call the physical metric just the metric) on the base manifold $M$,
\begin{align}
    g_{\mu\nu} = e^I_\mu \eta_{IJ} e^J_\nu \in S^2 T^*M. \label{eq:tetrad-metric-from-tetrad}
\end{align}
Any two frames, related by an $SO(1,3)$ transformation, leave the physical metric invariant. As the frame and internal metric are defined to be invertible so is the physical metric, its inverse we denote $g^{\mu\nu}$ such that $g^{\mu\rho} g_{\rho\nu} = \delta^\mu_\nu$. Raising and lowering the internal index is done with its corresponding metric, that is $v_I = \eta_{IJ} v^J$ and $v_\mu = g_{\mu\nu} v^\nu$ and vice versa.

\subsection{Torsion-Free Spin Connection}
We introduce a metric covariant derivative $\nabla_\mu$, that is compatible with the metric and torsion free,
\begin{align}
    \nabla_\mu g_{\rho \sigma} & = 0 \\
    \nabla_{[\mu} X_{\nu]} & = \partial_{[\mu} X_{\nu]}
\end{align}
this is always possible, it exists and is unique for a given metric. The metric here should be understood as coming from \cref{eq:tetrad-metric-from-tetrad} rather than an independent object, therefore the covariant derivative is defined uniquely from a choice of frame. This is the covariant derivative that appears in the usual metric description of Riemannian geometry. The covariant derivative naturally extends to high and lower powers of the tangent bundle, as such it can act on any degree differential form. It can be expresses in terms of the Christoffel symbols $\Gamma^\rho_{\mu\nu}$ and partial derivatives, for example on the frame we have
\begin{align}
    \nabla_\mu e^I_\nu = \partial_\mu e^I_\nu - \Gamma^\rho_{\mu\nu} e^I_\rho.
\end{align}
The Christoffel symbols are defined as 
\begin{align}
    \Gamma^\rho_{\mu\nu} = \frac{1}{2} g^{\rho\sigma} \left( \partial_\mu g_{\nu\sigma} + \partial_\nu g_{\mu\sigma} - \partial_\sigma g_{\rho\sigma} \right) \label{eq:so13-metric-christoffel-symbol-definition}
\end{align}
The covariant derivative of the internal metric vanishes
\begin{align}
    \nabla_\mu \eta_{IJ} = \partial_\mu \eta_{IJ} = 0.
\end{align}
As $\eta_{IJ} = \textrm{diag}(-1,1,1,1)$ is constant. This leads to the following proposition.
\begin{proposition}
    The intrinsic torsion of the frame $e^I_\mu$ takes values in $\Lambda^1(M,\mathfrak{so}(1,3))$.
\end{proposition}
\begin{proof}
    We start by defining the intrinsic torsion, $T_\mu^I{}_J$, via 
    \begin{align}
        \nabla_\mu e^I_\nu = T_\mu^I{}_J e^J_\nu. \label{eq:so13-torsion-definition}
    \end{align}
    Then using the definition of the physical metric as the pullback of the internal metric we find 
    \begin{align}
        \nabla_\mu g_{\rho\sigma} = \nabla_\mu \left( e^I_\rho \eta_{IJ} e^J_\nu \right) = e^I_{(\rho} \eta_{IJ} \nabla_\mu e^J_{\sigma)} = T_{\mu(IJ)} e^I_\rho e^J_\sigma = 0.
    \end{align}
    As the frame is invertible we find the metric compatible condition is equivalent to
    \begin{align}
        T_\mu^{(IJ)} = 0.
    \end{align}
    The intrinsic torsion is then a 1-form with values in antisymmetric internal matrices; the matrix representation of $SO(1,3)$ Lie algebra elements are given by
    \begin{align}
        X^I{}_J \in \mathfrak{so}(1,3), \quad X^{(IJ)} = 0, \quad X^{IJ} = X^I{}_K \eta^{KJ}.
    \end{align}
    Therefore, the torsion is a 1-form with values in the Lie algebra.
\end{proof}
The fact that the intrinsic torsion is Lie algebra valued immediately gives the following result.
\begin{proposition}
    Given an $SO(1,3)$ frame $e^I$ there exists a unique connection form $A^I{}_J = A^I_\mu{}_J dx^\mu$ such that the frame is torsion free. The connection is then minus the intrinsic torsion.
\end{proposition}
\begin{proof}
    The torsion free equation for the frame is 
    \begin{align}
        d e^I + A^I{}_J \wedge e^J = 0.
    \end{align}
    Expanding in components on $M$ we find 
    \begin{align}
        \partial_{[\mu} e^I_{\nu]} + A^I_{[\mu}{}_J e^J_{\nu]} = 0
    \end{align}
    This can be solved explicitly for $A^I{}_J$ where the solution is
    \begin{align}
        A^I_\mu{}_J = e^{I\rho} e_J^\sigma \left( C_{\rho\sigma\mu} + C_{\sigma\mu\rho} - C_{\mu\rho\sigma} \right), \quad C_{\mu\rho\sigma} = e_{I\mu} \partial_{[\rho} e^I_{\sigma]}.
    \end{align}
    To see how this relates to the intrinsic torsion we note that
    \begin{align}
        \nabla_{[\mu} e^I_{\nu]} = \partial_{[\mu} e^I_{\nu]} - \Gamma^\rho_{[\mu\nu]} e^I_\rho = \partial_{[\mu} e^I_{\nu]}.
    \end{align}
    Using that the Levi-Civita connection is torsion free $\Gamma^{\mu}_{[\rho\sigma]} = 0$, as can be seen from its definition \cref{eq:so13-metric-christoffel-symbol-definition}, we find
    \begin{align}
        T^I_{[\mu}{}_J e^J_{\nu]} = \nabla_{[\mu} e^I_{\nu]} = \partial_{[\mu} e^I_{\nu]} = -A^I_{[\mu}{}_J e^J_{\nu]}
    \end{align}
    which gives $T = -A$.
\end{proof}
In the context of $SO(n)$ structures we call $A^I{}_J \in \Lambda^1$ the spin connection. This allows us to ignore the metric covariant derivative and use the torsion free connection equation to define the spin connection.

\subsection{Riemann Curvature and Spin Connection Curvature}
Given a spin connection $A^I{}_J$ we can define its curvature as 
\begin{align}
    F(A)^I{}_J = dA^I{}_J + A^I{}_K \wedge A^K{}_J \label{eq:so13-curvature-of-intrinsic-torsion-as-differential-form}
\end{align}
in components this is 
\begin{align}
    F_{\mu\nu}{}^I{}_J = \partial_{[\mu} A^I_{\nu]}{}_J + A^I_{[\mu}{}_K A^K_{\nu]}{}_J
\end{align}
or using the metric covariant derivative we have the equivalent form 
\begin{align}
    F_{\mu\nu}{}^I{}_J = \nabla_{[\mu} A^I_{\nu]}{}_J + A^I_{[\mu}{}_K A^K_{\nu]}{}_J.
\end{align}
The above form of the curvature allows us to compare it to the curvature of the metric, indeed by taking another covariant derivative of the torsion definition in \cref{eq:so13-torsion-definition} and antisymmetrising we find
\begin{align}
    \nabla_{[\mu} \nabla_{\nu]} e^I_\rho & = - (\nabla_{[\mu} A^I_{\nu]}{}_J + A^I_{[\mu}{}_K A^K_{\nu]}{}_J) e^J_\rho = -F_{\mu\nu}{}^I{}_J e^J_\rho \\
                                         & = \frac{1}{2} R_{\mu\nu\rho}{}^\sigma e^I_\rho
\end{align}
where in the first line we have used the definition of the intrinsic torsion in terms of the connection and for the second line we have used the standard result from Riemannian geometry $2 \nabla_{[\mu} \nabla_{\nu]} X_\rho = R_{\mu\nu\rho}{}^\sigma X_\sigma$ where $R_{\mu\nu\rho\sigma}$ is the Riemann curvature tensor as defined in \cref{eq:metric-Riemann-curvature-tensor-definition}. We can then relate the Riemann curvature with the curvature of the connection 
\begin{align}
    F_{\mu\nu}{}^{IJ} = \frac{1}{2} R_{\mu\nu\rho\sigma} e^{I\rho} e^{J\sigma}.
\end{align}
Hence, using \cref{eq:so13-curvature-of-intrinsic-torsion-as-differential-form} we can compute the curvature of the manifold without needing to involve the metric directly.

\subsection{Einstein's Equations}
Einstein's equations usually appear as an algebraic condition on the Riemann curvature and metric. By first defining the Ricci curvature 
\begin{align}
    R_{\mu\nu} = R_{\mu\rho\nu}{}^\rho
\end{align}
we can write the Einstein's equations, with no matter, as
\begin{align}
    R_{\mu\nu} - \frac{1}{2}g_{\mu\nu} R + \Lambda g_{\mu\nu} = 0
\end{align}

with $R = R_\mu{}^\mu$ being the Ricci scalar and $\Lambda$ being the cosmological constant. Using the frame and the curvature of its connection we find the Ricci tensor is encoded into a vector valued 1-form $R^I \in \Lambda^1(M,V)$,
\begin{align}
    R^I_\mu = R_{\mu\nu} e^{J\nu} = 2 F_{\mu\nu}{}^I{}_J e^{J\nu}.
\end{align}
The Ricci scalar is obtained from a further contraction 
\begin{align}
    R = R^I_\mu e^\mu_I.
\end{align}
Einstein's equations are then that the follow vector valued 1-form vanishes
\begin{align}
    R^I - \left(\Lambda + \frac{R}{2} \right) e^I = 0. \label{eq:tetrad-einstein-equations-as-V-1-form}
\end{align}
The metric is not needed in this computation, every object can be computed using the frame and its inverse. The above equation has 16 components, however, not all of those components are independent. Using that the curvature is related to the Riemann curvature tensor, we can show that
\begin{align}
    R^I_{[\mu} e_{I\nu]} = R_{[\mu\nu]} = 0
\end{align}
due to the symmetries of the Riemann tensor. We find that 6 of the components are not present, this leaves the 10 equations in Einstein's equations. Solutions to \cref{eq:tetrad-einstein-equations-as-V-1-form} define an Einstein metric \cref{eq:tetrad-metric-from-tetrad} that is a solution to Einstein's equation \cref{eq:metric-einsteins-equation}.

\subsection{Einstein-Cartan Action}
The differential equations of motion, in the frame formulation of gravity, that form Einstein's equations are
\begin{align}
    d e^I + A^I{}_J \wedge e^J = 0, \quad R^I - \left(\Lambda + \frac{R}{2} \right) e^I = 0. \label{eq:so13-einstein-equations-first-order-system}
\end{align}
As with the metric formulation these can be efficiently stated as an action principle
\begin{align}\label{eq:einstein-cartan-action}
    S_{EC} = \frac{1}{32\pi G}\int \epsilon_{IJKL} e^I \wedge e^J \wedge \left( F^{KL} - \frac{\Lambda}{6} e^K \wedge e^L \right)
\end{align}
where $\epsilon_{IJKL} \in \Lambda^4 V^*$ is the totally antisymmetric symbol with values in $\pm 1$. It is possible to check that this action is invariant under $SO(1,3)$ transformations of the frame and spin connection. As all the objects are differential forms or built from the exterior algebra of forms, the action is also diffeomorphism invariant. To compute the first equation in \cref{eq:so13-einstein-equations-first-order-system} we vary the action with respect to $A^{IJ}$. The variation of the curvature in this direction gives 
\begin{align}
    \delta F^{IJ} = d^A \delta A^{IJ} = d A^{IJ} + A^I{}_K \wedge \delta A^{KJ} + A^J{}_K \wedge \delta A^{IK}.
\end{align}
Applying this to the action and integrating by parts, ignoring the boundary terms, we obtain the equation of motion 
\begin{align}
    \frac{\delta S_{EC}}{\delta A^{IJ}} = \epsilon_{IJKL} d^A \left( e^K \wedge e^L \right) = 0
\end{align}
By defining a new form $T^I = d^A e^I \in \Lambda^2(M)$ we can see that this condition is equivalent too 
\begin{align}
    \epsilon_{IJKL} e^K \wedge T^L = 0.
\end{align}
Using the Hodge star, $\star : \Lambda^q \rightarrow \Lambda^{4-q}$, we find, in  components that 
\begin{align}
    (\star \epsilon_{IJKL} e^K \wedge T^L)_\mu = \frac{1}{2} \epsilon_{IJKL} \epsilon_\mu{}^{\nu\rho\sigma} e^K_\nu T^L_{\rho\sigma} = 0.
\end{align}
The frame can then be used to convert the spacetime to internal indices, allowing us to write the above condition with only internal indices
\begin{align}
     \epsilon_{I J L M} \epsilon^{K L N P} T^M_{N P} = 0
\end{align}
where $T^I_{JK} = T^I_{\mu\nu} e^\mu_J e^\nu_K$. Expanding the product of two Levi-Civita tensors into sum of Kronecker delta's and expanding we find the result is 
\begin{align}
    T^K_{IJ} + \delta_I^K T^L_{JL} - \delta_J^K T^L_{IL} = 0
\end{align}
Taking the trace with respect to $J,K$ we find that $T^J_{IJ} = 0$ and therefore,
\begin{align}
    T^I_{JK} = 0.
\end{align}
Moving back to spacetime indices and using the definition of $T^I$ we see that this equation of motion implies 
\begin{align}
    d^A e^I = 0.
\end{align}
It is clear that this equation of motion defines $A^I{}_J$ as the spin connection for the frame $e^I$. Next, we vary with respect to the frame to find 
\begin{align}
    \frac{\delta S_{EC}}{\delta e^I} = \epsilon_{IJKL} e^J \wedge \left( F^{KL} - \frac{\Lambda}{3} e^K \wedge e^L \right) = 0.
\end{align}
By raising the index $I$ the above equation is a vector valued 3-form. By employing the Hodge star map, $\star$, we can map this onto a 1-form, in components this is 
\begin{align}
    \epsilon_{\mu}{}^{\nu\rho\sigma} \epsilon^I{}_{JKL} e^J_\nu \left( F_{\rho\sigma}{}^{KL} - \frac{\Lambda}{3} e^K_\rho e^L_\sigma \right) = 0.
\end{align}
Taking advantage of the frame to translate internal and spacetime indices we can use the  product of epsilon tensors to find the above is proportional to
\begin{align}
    F^{IJ}_{\mu\nu} e^\nu_J - \frac{1}{2} F^{KL}_{\rho\sigma} e^\rho_K e^\sigma_L e^I_\mu - \Lambda e^I_\mu = 0.
\end{align}
Which is equivalent to \cref{eq:tetrad-einstein-equations-as-V-1-form}, which are themselves equivalent to Einstein's equations on the metric.

We have seen that Einstein's equations can be written as exterior differential equations on the frame and connection. The Lagrangian in this language is polynomial which is simpler than the metric case, allowing the action to be computed more easily. Another advantage is that the Einstein-Cartan formulation allows for the coupling to fermionic matter, the Dirac matrices on $M$ are the pullback of the ``flat'' internal Dirac matrices, that is $\gamma_\mu = e^I_\mu \gamma_I$ where $\gamma_I$ are the gamma matrices on the Minkowski spacetime. We can also see the storage efficiency of this formalism, the frame and connection consist of $16+24 = 40$ components whereas the metric and its Christoffel symbols constitute $10+40 = 50$ components. Therefore, the total number of components one has to compute is reduced at the price of $4$ more initial components in the frame. However, the one disadvantage to this formulation is that its Hamiltonian form contains second-class constraints~\cite{First_order_gra_Alexan_2014}, which are not present in the metric formulations, this is due to the $24$ connection momentum variables. This makes the Hamiltonian tetrad formulation more difficult deal with as one has to solve the second-class constraints manually and this is a difficult task but not impossible see~\cite{Actions_for_Gra_Peldan_1993}. A formulation that keeps all the advantages of the frame formulation and that does not contain second-class constraints, is the \pleb{} formulation of gravity.

    \newpage
    \chapter{\pleb{}'s Formulation of general relativity}\label{chap:plebanskis-formulation}
\pleb{}'s formulation is at heart of this Thesis, we use it to aid calculations in various areas of differential geometry and general relativity. This section describes the details of this formalism and how it relates to more standard formalisms. In short, it is the description of geometry in dimension $4$ through the language of self-dual 2-forms and is a chiral projection of the frame formulation. As with the metric and frame formalisms, the main fields in \pleb{}'s formulation can also be described as a $G$-structure packed into differential forms, this time it is a triple of 2-forms. Einstein's equations are then simple algebraic conditions on the curvature of the torsion free connection defined by the $G$-structure. We begin with the reverse construction, in that we assume a metric exists on the spacetime and this defines a triple of 2-forms that encodes said metric. Then we go on to show that the inverse is also possible and the Einstein's equations can be imposed in either case.

\section{Historical Remarks}

One of the first uses of self-duality in general relativity (GR) came from Petrov's classification of Einstein spacetimes~\cite{The_Classificat_Petrov_2000}, where self-dual bivectors were used to decompose the Weyl curvature into a complex $3 \times 3$ matrix. Self-dual 2-forms, in the context GR, appeared in~\cite{01138a60-03bf-3425-9736-5043c4a7c399}. In this paper the link was made between the self-dual 2-forms and the null tetrad formalism of Newmann-Penrose~\cite{A_spinor_approa_Penros_1960,An_Approach_to_Newman_1962}. It also introduces the self-dual connection defined through exterior derivatives of the self-dual 2-forms. This way the 12 complex spin coefficients of the Newmann-Penrose formalism are neatly encoded into the 3 complex 1-forms that are the self-dual connections. Lastly, this paper made clear that Einstein's equations are equivalent to the statement that the curvature of the self-dual connection is self-dual. This statement is also noted in~\cite{AtiyahHitchinSingerSelfDual1978}, along with further applications of self-dual 2-forms to chiral Yang-Mills theory.

An action principle that imposes Einstein's equations using the self-dual techniques was discovered by Jerzy \pleb{} in~\cite{OnTheSeparatiPleban1977}. There the triple of self-dual 2-forms are taken as the main variable instead of the metric, given an algebraic condition on those 2-forms that they must satisfy it is possible to show that they can be written as a product of the frame and hence encode the metric. Importantly, no metric is needed to impose Einstein's equations in this formalism, only the 2-forms need to be introduced. \pleb{}'s paper uses spinor indices to denote the triple of 2-forms, here we use an $\mathfrak{so}(3,\C)$ index which was also done in the reference~\cite{Complex_structu_Brans_1974}. This is possible due to the isomorphism between $\mathfrak{sl}(2,\C)$ and $\mathfrak{so}(3,\C)$ and allows us to ignore the more involved spinor algebra.

\section{Metric and Self-Dual 2-forms}
Although the theory of interest does not need to involve the metric for pedagogical reasons it is useful to introduce the metric and show that the self-dual 2-forms appear as a decomposition on the space of 2-forms. In this section we assume that the signature of the spacetime is $(-+++)$, the Euclidean signature version of what follow can be recovered by taking every object to be real valued and replace the imaginary unit with 1.

Let $M$ be a $4$ dimensional Lorentzian manifold with tangent bundle $TM$ and cotangent bundle $T^*M$. Let $g \in S^2 T^*M$ be the nondegenerate bilinear form we call the Lorentzian metric. Using the metric one can define the Hodge star operator $\star : \Lambda^n \rightarrow \Lambda^{4-n}$, between n-forms and $(4-n)$-forms. In this particular dimension the Hodge star is an endomorphism on the space of 2-forms,
\begin{align}
    \star : \Lambda^2 \rightarrow \Lambda^2.
\end{align}
Applying the Hodge star twice to any 2-form reveals that it is an anti-involution, $\star^2 = -\id$. The space of 2-forms can then be split into 
\begin{align}
    \Lambda^2 = \Lambda^+ \oplus \Lambda^- \label{eq:pleb-two-forms-SD-ASD-split}
\end{align}
where $\Lambda^+ $ are the self-dual (SD) and $\Lambda^-$ the anti-self-dual (ASD) spaces for eigenvalues $+i$ and $-i$ respectively. As the eigenvalues are complex, this means that each subspace contains complex 2-forms.

We can now consider the wedge product with respect to the decomposition in \cref{eq:pleb-two-forms-SD-ASD-split}. It is then known that the restriction of the wedge product metric, that is $\wedge : \Lambda^2 \otimes \Lambda^2 \rightarrow \C$, on $\Lambda^+$ (or $\Lambda^-$) is nondegenerate. Let $\Sigma^i \in \Lambda^+$ be a basis for the SD 2-forms, where $i = 1,2,3$, it can then be chosen such that
\begin{align}
    \Sigma^i \wedge \Sigma^j \sim \delta^{ij}. \label{eq:pleb-metricity-condition}
\end{align}
We call this the metricity condition for a triple of 2-forms. The tensor $\delta^{ij}$ we call the internal metric as it lives in a ``flat'' internal space. The group $SO(3,\C)$ acts on these 2-forms and leaves the metric $\delta^{ij}$ invariant under its transformation. As we are in Lorentzian signature the basis of the ASD 2-forms is related to the complex conjugate of the SD basis,
\begin{align}
    \asd^i = -{(\Sigma^i)}^* \in \Lambda^-.
\end{align}
Here ${}^*$ denotes complex conjugation. In this way one can generate the entire space of 2-forms from only the self-dual 2-forms. Self-dual 2-forms in Euclidean signature do not have such a property as they are real valued and no standard transformation exists between the SD and ASD 2-forms. A consequence of this is that instanton solutions to various theories do not exist in Lorentzian signature. In any signature the SD and ASD 2-forms are orthogonal with respect to the wedge product,
\begin{align}
    \Sigma^i \wedge \asd^j = 0. \label{eq:pleb-reality-condition}
\end{align}
Given an arbitrary 2-form $B \in \Lambda^2$ it is not true that the wedge product of itself and its complex conjugate is zero, this is only true for SD or ASD 2-forms.

This completes the construction of self-dual 2-forms starting from a metric on the manifold. Next, we take the construction in the opposite direction. We show that given a triple of 2-forms, and imposing some algebraic condition, a real valued metric is uniquely defined.
\begin{proposition}\label{prop:pleb-trip-B-defines-metric-proof}
    Let $M$ be a 4-dimensional manifold and $Q \subset \Lambda^2_\C$ be a rank 3 subspace of the complexified 2-forms that is nondegenerate with respect to the wedge product. When $Q \oplus Q^* = \Lambda^2$, with $Q^*$ being the complex conjugate of $Q$, then $Q$ defines a conformal metric on $M$.
\end{proposition}
\begin{proof}
    We present a proof given in~\cite{ModifiedGravitFreide2008} but modified for Lorentzian signature. First, we take a basis for the space of $Q$, 
    \begin{align}
        B^i \in Q \in \Lambda^2_\C, \quad i = 1,2,3.
    \end{align}
    We choose the notation 
    \begin{align}
        B^i = \frac{1}{2} B^i_{\mu\nu} dx^\mu \wedge dx^\nu
    \end{align}
    where $B^i_{\mu\nu}$ are the components of the 2-form. By using the totally antisymmetric symbol $\teps^{\mu\nu\rho\sigma} = \pm 1$ we can define the bivector dual to $B^i$ as 
    \begin{align}
        \tilde{B}^{i\mu\nu} = \frac{1}{2} \teps^{\mu\nu\rho\sigma} B^i_{\rho\sigma}.
    \end{align}
    We define a spacetime metric, built from the $B^i$'s, using \urb{}'s formula~\cite{OnIntegrabilitUrbant1984} 
    \begin{align}
        \tilde{g}_{\mu\nu} = \frac{i}{12} \epsilon_{ijk} \teps^{\rho\sigma\alpha\beta} B^i_{\mu\rho} B^j_{\nu\sigma} B^k_{\alpha\beta} = \frac{i}{6} \epsilon_{ijk} B^i_{\mu\rho} \tilde{B}^{j\rho\sigma} B^k_{\sigma\nu}. \label{eq:pleb-urbantke-BBB-formula}
    \end{align}
    As well as an internal metric
    \begin{align}
        \tilde{h}^{ij} = \frac{1}{4i} B^i_{\mu\nu}  \tilde{B}^{j\mu\nu}. \label{eq:pleb-internal-metric-BB}
    \end{align}
    When this metric is positive-definite we say that the same is true of the wedge product metric, the factor of $i$ appears in the definition due to the Lorentzian signature. As this metric is also symmetric this means that it is a real-valued matrix. Each of the above metrics are used to raise and lower the indices of their type. One can then show that 
    \begin{align}
        B^{(i}_{\mu\rho} \tilde{B}^{j)\rho\nu} & = \frac{1}{2} B^{(i}_{\mu[\rho} B^{j)}_{\alpha\beta]} \teps^{\rho\nu\alpha\beta} = \frac{1}{2} B^i_{[\mu\rho} B^j_{\alpha\beta]} \teps^{\rho\nu\alpha\beta} = \frac{1}{2} B^i_{[\mu\rho} B^j_{\alpha\beta]}  \teps^{\rho\nu\alpha\beta}  \nonumber\\ & = \frac{1}{2} B^i_{\mu' \rho'} B^j_{\alpha' \beta'} \frac{1}{4!} \delta^{[\mu' \rho' \alpha' \beta']}_{\mu \rho \alpha \beta} \teps^{\rho\nu\alpha\beta} = \frac{1}{2} B^i_{\mu' \rho'} B^j_{\alpha' \beta'} \frac{1}{4!} \left(- \teps^{\mu'\rho'\alpha'\beta'} \uteps_{\mu\rho\alpha\beta} \right) \teps^{\rho\nu\alpha\beta}  \nonumber\\ &  = -\frac{i}{3!} \left( \frac{1}{2 i} \frac{1}{4} B^i_{\mu' \rho'} B^j_{\alpha' \beta'} \teps^{\mu'\rho'\alpha'\beta'} \right) \left( -\uteps_{\mu\rho\alpha\beta} \teps^{\nu\rho\alpha\beta} \right) \nonumber \\ &= -i \tilde{h}^{ij} \delta_\mu^\nu
    \end{align}
    where we have introduced $\uteps_{\mu\nu\rho\sigma} = \pm 1$ as the totally antisymmetric tensor with indices down and $(ij)$ indicates the indices are symmetrised. The second equality in the first line uses a known result in any dimension that 
    \begin{align}
        \frac{1}{2} \left( A_{\mu[\nu} B_{\rho \sigma]} + B_{\mu[\nu} A_{\rho \sigma]} \right) = A_{[\mu\nu} B_{\rho\sigma]}, \quad A,B \in \Lambda^2
    \end{align}
    which can be checked by expanding the symmetries and comparing. A similar argument can also be used to show 
    \begin{align}
        \tilde{B}^{(i|\mu\rho} B^{|j)}_{\rho\nu} = -i \tilde{h}^{ij} \delta^\mu_\nu.
    \end{align}
    This allows us to show that 
    \begin{align}
        B^i_{\mu\rho} \tilde{B}^{j\rho\sigma} B^k_{\sigma\nu} - B^k_{\mu\rho} \tilde{B}^{j\rho\sigma} B^i_{\sigma\nu} = -2i \epsilon^{ijk} \tilde{g}_{\mu\nu}. \label{eq:pleb-BBB-prop-to_eps_g}
    \end{align}
    To see this we first note that by symmetrising $ij$ we have 
    \begin{align}
        B^{(i}_{\mu\rho} \tilde{B}^{j)\rho\sigma} B^k_{\sigma\nu} - B^k_{\mu\rho} \tilde{B}^{(j\rho\sigma} B^{i)}_{\sigma\nu} = -i \tilde{h}^{ij} B^k_{\mu\nu} + i B^k_{\mu\nu} \tilde{h}^{ji} = 0
    \end{align}
    where we have used that $\tilde{h}^{[ij]} = 0$. Using the same arguments we can also see that \cref{eq:pleb-BBB-prop-to_eps_g} is also antisymmetric in $jk$ and as it is antisymmetric in $ik$ by construction we can see that it is totally antisymmetric and hence proportional to $\epsilon^{ijk} \tilde{g}$. By inverting the internal metric we can define 
    \begin{align}
        \utilde{B}_{i\mu\nu} = \utilde{h}_{ij} B^j_{\mu\nu}, \quad \textrm{and} \quad B_i^{\mu\nu} = \utilde{h}_{ij} \tilde{B}^{j\mu\nu}
    \end{align}
    with $\utilde{h}_{ij}$ being the inverse of $\tilde{h}^{ij}$. From which we get the following identities 
    \begin{align}
        B_i^{\mu\rho} B^i_{\rho\nu} = \utilde{B}_{i\mu\rho} \tilde{B}^{i\rho\nu} & = -3i \delta^\mu_\nu \\
        B_j^{\mu\alpha} B^i_{\alpha\rho} \tilde{B}^{j\rho\nu} & = i \tilde{B}^{i\mu\nu} \\
        B^j_{\mu\rho} \tilde{B}^{i\rho\sigma} \utilde{B}_{j\sigma\nu} & = i B^i_{\mu\nu}
    \end{align}
    Using this and contracting on the left side of \cref{eq:pleb-BBB-prop-to_eps_g} we find 
    \begin{align}
        \tilde{B}^{[i\mu\rho} B^{j]}_{\rho\nu} = \epsilon^{ijk} B_k^{\mu\rho} \tilde{g}_{\rho\nu}
    \end{align}
    such that 
    \begin{align}
        \tilde{B}^{i\mu\rho} B^j_{\rho\nu} = -i \tilde{h}^{ij} \delta^\mu_\nu + \epsilon^{ijk} B_k^{\mu\rho} \tilde{g}_{\rho\nu}.
    \end{align}
    Now defining an a priori unrelated inverse metric 
    \begin{align}
        \utilde{g}^{\mu\nu} = \frac{1}{6\det(\tilde{h})} \epsilon_{ijk} \tilde{B}^{i\mu\rho} B^j_{\rho\sigma} \tilde{B}^{k\sigma\nu} \label{eq:pleb-urbantke-inverse-metric}
    \end{align}
    and reperforming the previous steps we arrive at the result 
    \begin{align}
        \tilde{B}^{[i\mu\rho} B^{j]}_{\rho\nu} = i \det(\tilde{h}) \epsilon^{ijk} \utilde{g}^{\mu\rho} \utilde{B}_{k\rho\nu}.
    \end{align}
    Therefore we conclude that 
    \begin{align}
         \tilde{B}^{i\mu\rho} \tilde{g}_{\rho\nu} = i \det(\tilde{h})\utilde{g}^{\mu\rho} B^i_{\rho\nu} \label{eq:pleb-gB-equals-Bg}
    \end{align}
    Next we can verify that $\tilde{g}_{\mu\nu}$ and $\utilde{g}^{\mu\nu}$ are inverses of each other. This is checked by computing 
    \begin{align}
        \tilde{g}_{\mu\rho} \utilde{g}^{\rho\nu} = \frac{i}{36 \det(\tilde{h})} \epsilon_{ijk} \epsilon_{lmn} B^i_{\mu\alpha} \tilde{B}^{j\alpha\beta} B^k_{\beta\rho} \tilde{B}^{l\rho\sigma} B^m_{\sigma\gamma} \tilde{B}^{n\gamma\nu}
    \end{align}
    then using 
    \begin{align}
        \epsilon_{ijk} \epsilon_{lmn} = \epsilon_{ijk} \epsilon^{l'm'n'} \det(\tilde{h}) \utilde{h}_{ll'} \utilde{h}_{mm'} \utilde{h}_{nn'}
    \end{align}
    and expanding the antisymmetric tensors into Kronecker delta's and lowering indices with $\utilde{h}_{ij}$ we are able to use the previous algebra on $B$'s to show 
    \begin{align}
        \tilde{g}_{\mu\rho} \utilde{g}^{\rho\nu} = \delta^\mu_\nu. \label{eq:pleb-metric-times-inverse-det-h}
    \end{align}
    From this we immediately see that  
    \begin{align}
        \tilde{g}_{\mu\rho} \tilde{B}^{i\rho\sigma} \tilde{g}_{\sigma\nu} = i \det(\tilde{h}) B^i_{\mu\nu}
    \end{align}
    Applying this relation to itself again reveals that 
    \begin{align}
        \det(\tilde{g}) = -\det(\tilde{h})^2.
    \end{align}
    Therefore, it is convenient to introduce a physical metric \(g_{\mu\nu} = \tilde{g}_{\mu\nu}/\sqrt{-g}\) and define $\tilde{h}^{ij} = \sqrt{-g} h^{ij}$ such that the above becomes 
    \begin{align}
        \det(h)^2 = 1.
    \end{align}
    The soon-to-be self-duality relation becomes 
    \begin{align}
        \frac{1}{2} \epsilon_{\mu\nu}{}^{\rho\sigma} B^i_{\rho\sigma} = i \det(h) B^i_{\mu\nu}
    \end{align}
    from which it is easy to see that $\det(h) = 1$ is required for $B^i$ to be self-dual, and we have used $\teps^{\mu\nu\rho\sigma} = \sqrt{-g} \epsilon^{\mu\nu\rho\sigma}$ from the definition of the Levi-Civita tensor. Finally, we see that 2-forms $B^i$ define the metric $g_{\mu\nu}$ through \urb{}'s formula that makes them self-dual. This metric only defines a conformal class as it can be multiplied by an arbitrary function and the above proof does not change.
\end{proof}

The metric in the above proof can be complex valued, this does not represent a physical situation and as such some reality conditions need to be imposed. If the physical metric defined in \cref{eq:pleb-urbantke-BBB-formula} is real-valued and the internal metric \cref{eq:pleb-internal-metric-BB} is positive-definite metric, then $\bar{B}^i$ are anti-self-dual with respect to the real-valued physical metric, that is
\begin{align}
    \star \overline{B}^i = -i \overline{B}^i
\end{align}
where $\bar{B}^i = -(B^*)^i$ are the anti-self-dual 2-forms. In this case it is easy to show that the reality condition, \cref{eq:pleb-reality-condition}, is automatically satisfied. The argument in the other direction is more involved and leads to the following proposition.

\begin{proposition}
    A triple of complex 2-forms $B^i$, that satisfy the weak metricity and reality conditions
    \begin{align}
        B^i \wedge B^j = 2i h^{ij} \nu_B \quad  B^i \wedge \overline{B}^j = 0
    \end{align} 
    where $h^{ij}$ is now a real positive-definite matrix and $\nu_B = \frac{1}{6i} B_i \wedge B^i$ is the top form, define a real metric given by
    \begin{align}
		g(u,v) \nu_B = \frac{i}{6} \epsilon_{ijk} \iota_u B^i \wedge \iota_v B^j \wedge B^k.
    \end{align}
    With which the 2-forms $B^i$ are self-dual and $\overline{B}^i$ are anti-self-dual.
\end{proposition}
\begin{proof}
    We have already seen that the metric defined by \cref{eq:pleb-urbantke-BBB-formula} makes $B^i$ self-dual, we have left to check that the condition $B^i \wedge \overline{B}^j = 0$ implies the metric is real-valued. We being by decomposing the 2-form $B^i$ into real and imaginary parts 
    \begin{align}
        B^i = S^i + i P^i
    \end{align}
    where $S^i,P^i \in \Lambda^2$. Which gives a similar decomposition for $\overline{B}^i$,
    \begin{align}
        \overline{B}^i = -S^i + i P^i
    \end{align}
    The reality and metricity conditions then imply
    \begin{align}
       S^i \wedge S^j = 0, \quad P^i \wedge P^j = 0, \quad S^i \wedge P^j = h^{ij} \nu_B.
    \end{align}
    Solutions to this can be found by introducing a frame $e^0,e^i \in \Lambda^1$, see 5.4.4 of~\cite{FormGenRelGravity2020}. The solutions are then 
    \begin{align}
        P^i = e^0 \wedge e^i, \quad S^i = -\frac{1}{2} h^{i j} \epsilon_{jkl} e^k \wedge e^l
    \end{align}
    The solutions for $P^i$ and $S^i$ can swap or change sign, this amounts to a change of sign for the top form or making $B^i$ anti-self-dual. The orientation chosen here matches the rest of the formulas in the Thesis.
    We choose this orientation and take $B^i$ to be the combination of $S^i, P^i$ above. The top form is given by 
    \begin{align}
        \frac{1}{6i} h_{ij} B^i \wedge B^j = \frac{1}{6} \epsilon_{ijk} e^{i} \wedge e^j \wedge e^k \wedge e^0 = \nu_B.
    \end{align}
    As $e^i \wedge e^j \wedge e^k$ is totally antisymmetric we can write $e^i \wedge e^j \wedge e^k \wedge e^0 = \epsilon^{ijk} \nu_B$. Computing the metric we find that in terms of the frame we have
    \begin{align}\label{eq:pleb-spacetime-metric-from-frame-and-h}
        g(u,v) = - \iota_u e^0 \iota_v e^0 + h_{ij} \iota_u e^i \iota_v e^j.
    \end{align}
    As the metric $h^{ij}$ is positive-definite so is its inverse, it being symmetric also implies that it is real-valued and therefore the metric is real-valued.
    
\end{proof} 

We now relate the arbitrary triple $B^i$ to the \pleb{} triple $\Sigma^i$. We restrict ourselves to the case where the weak metricity and reality conditions are satisfied, as such the matrix $h_{ij}$ is positive-definite and there exists a transformation matrix $b_a^i$ such that
\begin{align}
    \delta_{ab} = b_a^i b_b^j h_{ij}, \quad \textrm{or} \quad h_{ij} = b_i^a b_j^b \delta_{ab}
\end{align}
where $a,b,\ldots = 1,2,3$ and as $h_{ij}$ is positive-definite this is always possible. Moreover, as $\det(h) = 1$ this implies that $\det(b) = \pm 1$ for which we take the positive branch $\det(b) = 1$. This choice does not fully determine $b$ and there remains an $SO(3,\C)$ rotation freedom, that is $b^a_i$ and $m^a{}_b b^b_i$, where $m \in SO(3,\C)$, produce the same desired result. The transformation between the two triples of two forms is then
\begin{align}
    \Sigma^a = b^a_i B^i.
\end{align}
This is equivalent to a change of 3-frame $e^i = b^i_a e^a$ with 
\begin{align}
    \Sigma^a = i e^0 \wedge e^a - \frac{1}{2} \epsilon^{a b c} e^b \wedge e^c
\end{align}
where the raised and lowered indices $a$ do not change their meaning as their metric is $\delta^{ab}$. By choosing the positive determinant for $b$ we find that the volume forms for the two triples have the same sign, indeed computing the transformation we find 
\begin{align}
    6\nu_B = \epsilon_{ijk} e^i \wedge e^j \wedge e^k \wedge e^0 = \epsilon_{ijk} b^i_a b^j_b b^k_c e^a \wedge e^b \wedge e^c \wedge e^0 = \epsilon_{abc} e^a \wedge e^b \wedge e^c \wedge e^0 = 6\nu_\Sigma.
\end{align}
We see that the metricity condition becomes 
\begin{align}
    \Sigma^a \wedge \Sigma^b = b^a_i b^b_j B^i \wedge B^j = 2i b^a_i b^b_j h^{ij} \nu_B = 2i \delta^{ab} \nu_\Sigma.
\end{align} 
If we relabel the indices $a,b,c,\ldots \rightarrow i,j,k,\ldots$, we can ``forget'' that the metric $h^{ij}$ ever existed and work with the flat internal metric $\delta^{ij}$. In this way we recover the triple $\Sigma^i$ that is self-dual and satisfies the metricity and reality condition \cref{eq:pleb-metricity-condition,eq:pleb-reality-condition}. Much like the freedom in choosing the volume form for the physical metric we are also free to choose which triple of 2-forms we wish to make fundamental, the choice here is to make the following algebra much simpler. Seeing as this choice is always possible we define the internal metric to be orthonormal throughout the rest of the Thesis. With the internal metric being the identity matrix we have reduced the group of transformations on the 2-forms from $GL(3,\C)$ to $SO(3,\C)$. The fact that there is a group transformation on the 2-forms that leaves the metric they describe invariant suggests that these objects also arise from $G$-structure on a principal bundle.

\section{\pleb{} 2-forms as \texorpdfstring{$G$}{G}-structures}
In the frame formalism, we have seen that the metric is encoded using a collection of differential forms. Similarly, here we find it is a triple of complex 2-forms instead of a 4 real valued 1-forms that encode the metric. We also know that frames arise naturally as soldering forms on a principal bundle. It is then interesting to consider a similar construction for self-dual 2-forms.

To this end we introduce the rank 3 biframe bundle $QM$ over a manifold $M$,
\begin{align}
    QM = \coprod_{x\in M} Q_x, \quad Q_x = \{ \left. f_x \right| f_x : Q \subset \Lambda^2_\C|_x \rightarrow \E \}.
\end{align}
Where $\Lambda^2_\C |_x$ is the space of complexified 2-forms at the point $x$ and $Q$ is a rank 3 subspace of $\Lambda^2_\C |_x$. The maps $f_x$ are isomorphisms from this subspace to $\C^3$, any such map we call a biframe. The group $GL(3,\C)$ acts naturally on $QM$ by identifying two biframes. This identification allows the $QM$ to be realised as a $GL(3,\C)$ principal bundle $P$, $GL(3,\C) \hookrightarrow P \overset{\pi_P}{\rightarrow} M$. The soldering form is then defined as
\begin{align}
    \varSigma(X,Y) = f^{-1}(d\pi_P (X) \wedge d\pi_P(Y)|_Q) \in \Lambda^2(M,\E), \quad X,Y \in T_p P.
\end{align}
It is simple to check that this soldering form is $G$-equivariant by computing $R^*_g \varSigma = g^{-1} \varSigma$ and that it is Horizontal as vertical vectors are in the kernel of $d\pi_P$. This means that $\varSigma$ is a tensorial differential form just like the usual soldering form. We now fix the internal metric on $\C^3$ to be $\delta^{ij}$ and restrict our biframe bundle to contain only the biframes that are isometries between the internal metric and the wedge product metric. That is they satisfy
\begin{align}
    \varSigma(X,Y) \wedge \varSigma(V,W) \sim  (d\pi_P(X) \wedge d\pi_P(Y) \wedge d\pi_P(V) \wedge d\pi_P(W)) \delta
\end{align}
where $\delta \in S^2 \E$ is the internal metric on $\E$. This restricts the structure group to be $SO(3,\C) \subset GL(3,\C)$ and therefore we are considering $SO(3,\C)$-structures.
This defines the wedge product metric in the same way as the usual metric appears in the frame formulations. To recover the metric on the tangent space one needs to evaluate \urb{}'s formula, see \cref{eq:pleb-urbantke-metric}. A connection on the principal bundle is then a Lie algebra valued 1-form
\begin{align}
    A \in \Lambda^1(P,\mathfrak{so}(3,\C)).
\end{align}
From the general theory in \cref{chap:fibre-bundles} we know this connection defines a covariant derivative on tensorial q-forms. Indeed, the torsion of the soldering form is
\begin{align}
    \Theta = D \varSigma = d \varSigma + A \wedge \varSigma.
\end{align}
The curvature of the connection is also defined as usual
\begin{align}
    R = DA \in \Lambda^2(P,\mathfrak{so}(3,\C)).
\end{align}
We have seen that the triple of self-dual 2-forms has an interpretation of soldering forms on a principal $SO(3,\C)$-bundle. In the frame bundle it was shown that the curvature of the connection encodes the correct components of the Riemann curvature tensor needed to impose Einstein's equations. The upcoming sections explain how the same is true for the triple of self-dual 2-forms, and it is shown how Einstein's equations can be imposed elegantly in this formulation.

\section{Local Soldering form, Connection and Curvature}
In this section we summarise the move to the local versions of the objects on the principal bundle $P$. Associated to this bundle is the vector bundle $E = P \times_{SO(3,\C)} \E$ and by introducing a local section $\sigma : U \subset M \rightarrow P$ we induce a trivialisation $E|_U = U \times \E$. The pullback of the soldering forms under this section are
\begin{align}
    \Sigma^i \in Q = \Lambda^+ \subset \Lambda^2(M)
\end{align}
where the index $i = 1,2,3$ labels the components in $\E$ which we call the internal indices, Equivalently it is an $\mathfrak{so}(3,\C)$ Lie algebra index. Pulling back the isometry condition results the metricity condition on the triple of 2-forms
\begin{align}
    \Sigma^i \wedge \Sigma^j \sim \delta^{ij}.
\end{align}
For a real valued metric, as we have seen before, it is required that $\Sigma^i \wedge \overline{\Sigma}^j = 0$. The Lie algebra of $SO(3,\C)$ has its matrix representation in $3 \times 3$ antisymmetric matrices and as such we can view the connection as the following 1-form on $M$,
\begin{align}
    A^{ij} \in \Lambda^1(M,\mathfrak{so}(3,\C))
\end{align}
where $A^{ij} = A^{[ij]}$ is antisymmetric. The curvature of this connection is 
\begin{align}
    F^{ij} = dA^{ij} + A^{ik} \wedge A^{kj}.
\end{align}
The metric on $\E$ is flat and the position of the indices carries no meaning, therefore, we keep all internal indices in the raised position. In dimension 3 there is an isomorphism between vectors and antisymmetric matrices, explicitly it is captured by the map 
\begin{align}
    \phi(m)^i = \frac{1}{2} \epsilon^{ijk} m^{jk} \label{eq:pleb-isomorphism-antisym-matrices-and-vector-C3}
\end{align}
and its inverse 
\begin{align}
    \phi^{-1}(v)^{ij} = \epsilon^{ijk} v^k.
\end{align}
We take advantage of this map to introduce the connection 1-form with values in $\E$
\begin{align}
    A^i = -\frac{1}{2} \epsilon^{ijk} A^{jk} \in \Lambda^1 \otimes \E.
\end{align}
The exterior covariant derivative on the soldering form is then 
\begin{align}
    d^A \Sigma^i = d\Sigma^i + \epsilon^{ijk} A^j \wedge \Sigma^k.
\end{align}
On a tensor with an arbitrary number of internal indices the exterior covariant derivative is 
\begin{align}
    d^A \Psi^{i_1 i_2 \ldots i_n} = d \Psi^{i_1 i_2 \ldots i_n} + \epsilon^{i_1 j k} A^j \Psi^{k i_2 \ldots i_n} + \epsilon^{i_2 j k} A^j \Psi^{i_1 k i_3 \ldots i_n} + \cdots + \epsilon^{i_n \ldots j k} A^j \Psi^{i_1 i_2 \ldots k}.
\end{align}
The curvature in this representation is given by
\begin{align}
    F^i = dA^i + \frac{1}{2} \epsilon^{ijk} A^j \wedge A^k.
\end{align}
This means we can view the 2-form frame, connection 1-form and curvature form as $\E$-valued differential forms.

\section{Decomposition of \texorpdfstring{$\E$}{C3}-valued Differential forms}\label{subsec:pleb-C3-valued-forms-decomp}
As we have seen, $\E$-valued differential forms make up most of the objects of interest in the study of $SO(3,\C)$-structures. It is then meaningful to study the irreducible representations of these objects with respect to $SO(3,\C)$. This section is the Lorentzian version of the calculation done in~\cite{Su2StructureBhoja2024}, however, most of the results are identical. We have the following result for the Lorentz group that $SO(1,3) = SL(2,\C)/\Z_2 $, and since each element of $SL(2,\C)$ is complex it has a complex conjugate forming the group $\overline{SL(2,\C)}$ which both act on separate chiral halves. As such we can denote the decomposition of the Lorentz group as $SO(1,3) = SL(2,\C) \times \overline{SL(2,\C)}/\Z_2$. One of the $SL(2,\C)$ act on $\Sigma^i$ with an $SO(3,\C)$ transformation, the other leaves $\Sigma^i$ invariant. We call the first $SL{(2,\C)}_+$ for which we denote the ${\rm spin}(k/2)$ representations as $S_+^k$ and the second $SL{(2,\C)}_-$ with $S_-^k$. They are of dimension $\dim(S_\pm^k) = k+1$ and decomposes the following interesting spaces into spinors representations
\begin{align}
    \Lambda^1 = S_+ \otimes S_-, \quad \Lambda^2 = S_+^2 \oplus S_-^2, \quad \E = S_+^2.
\end{align}
First we introduce the components for the triple of 2-forms 
\begin{align}
    \Sigma^i = \frac{1}{2} \Sigma^i_{\mu\nu} dx^\mu \wedge dx^\nu
\end{align}
where $\Sigma^i_{\mu\nu}$ are the components of the 2-forms. By raising a spacetime index we can create a triple of endomorphisms of $\Lambda^1$, $\Sigma^i_\mu{}^\nu : \Lambda^1 \rightarrow \Lambda^1$. 
\subsection{Algebra and Identities}
The self-dual (SD) endomorphism, $\Sigma^i_\mu{}^\nu$, can be shown to satisfy a quaternion algebra,
\begin{align}
    \Sigma^i_\mu{}^\rho \Sigma^j_\rho{}^\nu = -\delta^{ij} \delta_\mu{}^\nu + \epsilon^{ijk} \Sigma^k_\mu{}^{\nu}. \label{eq:pleb-quaterion-algebra}
\end{align}
There are other important identities
\begin{align}
    \epsilon^{\mu\nu\rho\sigma} \Sigma^i_{\alpha\beta} & = 12 i \delta_\alpha^{[\mu} \delta_\beta^{\nu} \Sigma^{i\rho\sigma]}. \label{eq:pleb-epsilon-S-to-ddS}\\
    \Sigma^i_{\mu\nu} \Sigma^i_{\rho\sigma} & = g_{\mu\rho} g_{\nu\sigma} - g_{\mu\sigma} g_{\nu\rho} -i \epsilon_{\mu\nu\rho\sigma} \\
    \epsilon^{ijk} \Sigma^j_{\mu\nu} \Sigma^k_{\rho\sigma} & = -2\Sigma^i_{[\mu|\rho|} g_{\nu]\sigma} + 2\Sigma^i_{[\mu|\sigma|} g_{\nu]\rho} 
\end{align}
The first identity can be derived by employing the self-duality relation to replace $\Sigma^i_{\alpha\beta}$ with $\frac{-i}{2} \epsilon_{\alpha\beta}{}^{\gamma\pi} \Sigma^i_{\gamma\pi}$. Then by expanding the product of two Levi-Civita tensors we find the desired result. The second result can be found by noting that the symmetries restrict the form to be the sum of $g_{\mu\rho} g_{\nu\sigma} - g_{\mu\sigma} g_{\nu\rho}$ and $\epsilon_{\mu\nu\rho\sigma}$, the coefficients can then be found by performing contractions with the metric and Levi-Civita symbol. The last identity follows in a similar way, in that the r.h.s is the only object with the correct indices and symmetries. The overall factor is determined by comparing both sides after contractions and using the quaternion algebra. The corresponding algebra for the anti-self-dual (ASD) 2-forms $\asd^i = -(\Sigma^i)^*$ are relatively simple to derive using complex conjugate and the definition of the ASD 2-forms. Their identities are then
\begin{equation}
\begin{aligned}
    \asd^i_\mu{}^\rho \asd^j_\rho{}^\nu & = -\delta^{ij} \delta_\mu{}^\nu - \epsilon^{ijk} \asd^k_\mu{}^{\nu} \\
    \asd^i_{\mu\nu} \asd^i_{\rho\sigma} & = g_{\mu\rho} g_{\nu\sigma} - g_{\mu\sigma} g_{\nu\rho} + i \epsilon_{\mu\nu\rho\sigma} \\
    \epsilon^{ijk} \asd^j_{\mu\nu} \asd^k_{\rho\sigma} & = 2\asd^i_{[\mu|\rho|} g_{\nu]\sigma} - 2\asd^i_{[\mu|\sigma|} g_{\nu]\rho} \\
    \epsilon^{\mu\nu\rho\sigma} \asd^i_{\alpha\beta} & = -12 i \delta_\alpha^{[\mu} \delta_\beta^{\nu} \asd^{i\rho\sigma]}.
\end{aligned}
\end{equation}
The orthogonality of the SD and ASD 2-forms is equivalent to the condition
\begin{align}
    \Sigma^i_{\mu\nu} \asd^{j\mu\nu} = 0. \label{eq:pleb-reality-condition-in-components}
\end{align}
Which is another way of writing the reality condition, $\Sigma^i \wedge \asd^j = 0$. Another condition involving both the SD and ASD 2-forms is
\begin{align}
    \Sigma^i_{\mu\rho} \asd^{j\rho}{}_{\nu} = \Sigma^i_{\langle \mu|\rho|} \asd^{j\rho}{}_{\nu\rangle }. \label{eq:pleb-9-dim-basis-of-symmetric-tracefree-tensors}
\end{align}
Where $A_{\langle \mu\nu\rangle } = \frac{1}{2}(A_{\mu\nu} + A_{\nu\mu} - \frac{1}{2} g_{\mu\nu} g^{\rho\sigma} A_{\rho\sigma})$ denotes the symmetric tracefree components of the tensor $A_{\mu\nu}$. This identity can be checked by contracting the free indices, $\mu\nu$, with $\Sigma^{k\mu\nu}$ and $\asd^{k\mu\nu}$ to find that it vanishes. Next, by contracting with the metric $g^{\mu\nu}$ we find the result is also zero, hence \cref{eq:pleb-9-dim-basis-of-symmetric-tracefree-tensors} has only its symmetric and traceless components. Any symmetric tensor can be decomposed into 
\begin{align}
    S_{(\mu\nu)} = S g_{\mu\nu} + S^{ij} \Sigma^i_\mu{}^\rho \asd^j_{\rho\nu}
\end{align} 
as \cref{eq:pleb-9-dim-basis-of-symmetric-tracefree-tensors} and the metric form a basis for the symmetric $(0,2)$-tensors. With the algebra at hand we can now start deriving the decompositions of the spaces of $\E$-valued differential forms.
\begin{proposition}
    Given a triple of endomorphisms $\Sigma^i$ that satisfy the reality, metricity conditions and the quaternion algebra, there is an operator $J_1 : \E \otimes \Lambda^1 \rightarrow \E \otimes \Lambda^1$
    that decomposes the space $\E \otimes \Lambda^1$ into a 4 dimensional and 8 dimensional part
    \begin{align}
        \E \otimes \Lambda^1 = (\E \otimes \Lambda^1){}_4 \oplus (\E \otimes \Lambda^1){}_8 = (S_+ \otimes S_-) \oplus (S_+^3 \otimes S_-)
    \end{align}
    with eigenvalues $2,-1$ respectively.
\end{proposition}
\begin{proof}
    We begin by introducing the operator on $\E$-valued 1-forms
    \begin{align} \label{eq:J1-operator-definition}
        J_1 (A)^i = \epsilon^{ijk} \Sigma^j_\mu{}^\nu A^k_\nu.
    \end{align}
    Squaring this and using the quaternion algebra we find 
    \begin{align}
        J_1^2 & = \epsilon^{ikl} \Sigma^k \epsilon^{lmj} \Sigma^m = 2 \delta^{ij} \mathbb{I} + \epsilon^{ikj} \Sigma^k  \nonumber\\& = 2\mathbb{I} + J_1.
    \end{align}
    Factorising the above reveals
    \begin{align}
        (J_1 - 2)(J_1 + 1) = 0
    \end{align}
    therefore, the eigenvalues are $+2,-1$. The 4-dimensional eigenspace of the operator $J_1$ is parametrised by a 1-form $\xi_\mu \in \Lambda^1 = S_+ \otimes S_-$ and the corresponding $\E$-valued 1-form is 
    \begin{align}
        \theta^i_\mu = \Sigma^i_\mu{}^\nu \xi_\nu \in (\E \otimes \Lambda^1){}_4.
    \end{align}
    We see by computing 
    \begin{align}
        J_1(\theta)^i_\mu = \epsilon^{ijk} \Sigma^j_\mu{}^\nu \Sigma^k_\nu{}^\rho \xi_\rho = 2 \Sigma^i_\mu{}^\nu \xi_\nu = 2 \theta^i_\mu
    \end{align}
    that it has an eigenvalue of $+2$. The 8 dimensional space is then the compliment of the 12 dimensional $\E \otimes \Lambda^1$ and this 4-dimensional part. That is given a $\theta \in \E \otimes \Lambda^1$ we extract the 4-dimensional 1-forms components through 
    \begin{align}
        \theta_\mu = \Sigma^i_\mu{}^\nu \theta^i_\nu. \label{eq:pleb-4-projection-of-E-1-forms}
    \end{align}
    Using the ansatz for the $(\E \otimes \Lambda^1){}_4$ we obtain 
    \begin{align}
        (\theta^i_\mu){}_4 = \Sigma^i_\mu{}^\nu \theta_\nu = \Sigma^i_\mu{}^\nu \Sigma^j_\nu{}^\rho \theta^j_\rho = -\theta^i_\mu - \epsilon^{ikj} \Sigma^k_\mu{}^\nu \theta^j_\nu = -(\mathbb{I} + J_1)(\theta)^i_\mu. \label{eq:pleb-4-embedding-into-E-1-forms}
    \end{align}
    The compliment of this in $\E \otimes \Lambda^1$ is 
    \begin{align}
        (\theta^i_\mu){}_8 = \theta^i_\mu - (\theta^i_\mu){}_4 = (2 - J_1)(\theta)^i_\mu \in (\E \otimes \Lambda^1){}_8 = (S_+^3 \otimes S_-).
    \end{align}
    By acting with $J_1$ we recover 
    \begin{align}
        J_1(\theta_8)^i_\mu = - (\theta_8)^i_\mu
    \end{align}
    the eigenvalue to be $-1$. As such $J_1$ defines the decomposition stated.
\end{proof}
Next we have a similar result for $\E$-valued 2-forms.
\begin{proposition}
    Given a triple of endomorphisms $\Sigma^i$ that satisfy the reality, metricity conditions and the quaternion algebra, we define an operator $J_2 : \E \otimes \Lambda^2 \rightarrow \E \otimes \Lambda^2$ that decomposition the space of $\E$-valued 2-forms into 
    \begin{align}
        \E \otimes \Lambda^2 & = (\E \otimes \Lambda^2){}_1 \oplus (\E \otimes \Lambda^2){}_3 \oplus (\E \otimes \Lambda^2){}_5 \oplus (\E \otimes \Lambda^2){}_9 \label{eq:pleb-E-valued-2-form-space-decomp} \\
        & = C^\infty (M) \oplus S_+^2 \oplus S_+^4 \oplus (S_+^2 \otimes S_-^2)
    \end{align}
    with eigenvalues $0,1,-1,2$ respectively.
\end{proposition}
\begin{proof}
    First, let us define the operator $J_2$,
    \begin{align}
        J_2(B)^i_{\mu\nu} = \epsilon^{ijk} \Sigma^j_{[\mu|}{}^\rho B^k_{\rho|\nu]}, \quad B^i_{\mu\nu} \in \E \otimes \Lambda^2.
    \end{align}
    A lengthy computation gives us 
    \begin{align}
        J_2^2(B)^i_{\mu\nu} & = \frac{1}{2} B^i_{\mu\nu} - \frac{i}{2} \epsilon_{\mu\nu}{}^{\rho\sigma} B^i_{\rho\sigma} + \frac{1}{2} J_2(B)^i_{\mu\nu} + \frac{1}{2} \Sigma^i_{[\mu}{}^\alpha \Sigma^j_{\nu]}{}^\beta B^j_{\alpha\beta} \\
        J_2^3(B)^i_{\mu\nu} & = -\frac{i}{2} \epsilon_{\mu\nu}{}^{\rho\sigma} B^i_{\rho\sigma} + 2 J_2(B)^i_{\mu\nu} + \Sigma^i_{[\mu}{}^\rho \Sigma^j_{\nu]}{}^\sigma B^j_{\rho\sigma} \\
        J_2^4(B)^i_{\mu\nu} & = \frac{1}{2} B^i_{\mu\nu} - \frac{3i}{2} \epsilon_{\mu\nu}{}^{\rho\sigma} B^i_{\rho\sigma} + \frac{5}{2} J_2(B)^i_{\mu\nu} + \frac{5}{2} \Sigma^i_{[\mu}{}^\rho \Sigma^j_{\nu]}{}^\sigma B^j_{\rho\sigma}.
    \end{align}
    Taking the appropriate combinations we find 
    \begin{align}
        J_2^4 - 2 J_2^3 - J_2^2 + 2 J_2 = 0,\quad \Rightarrow J_2(J_2 - 2)(J_2 - 1)(J_2 + 1) = 0
    \end{align}
    which identifies the correct eigenvalues. As in the 1-form case we can look for parametrisations of the eigenspaces. First we compute the action of $J_2$ on $M^{ij} \Sigma^j$ where $M^{ij} = \frac{1}{3}\Tr(M) \delta^{ij} + M_{stf}^{\langle ij\rangle } + M_a^{[ij]}$ to find that 
    \begin{align}
        J_2(M^{ij} \Sigma^j_{\mu\nu}) & = \Tr(M) \Sigma^i_{\mu\nu} - M^{ji} \Sigma^j_{\mu\nu} = \frac{2}{3} \Tr(M) \Sigma^i_{\mu\nu} - M_{stf}^{ij} \Sigma^j_{\mu\nu} + M_a^{ij} \Sigma^j_{\mu\nu}.
    \end{align}
    From which we read off that multiples of $\Sigma^i$ span the $+2$ eigenspace, $-1$ is spanned by $M_{stf}^{ij} \Sigma^j_{\mu\nu}$ and $+1$ is spanned by $M_a^{ij} \Sigma^j_{\mu\nu}$. For the final eigenspace we parametrise using a tracefree tensor $h_\mu{}^\nu$ such that $h_{[\mu}{}^\rho \Sigma^i_{|\rho|\nu]} \in \E \otimes \Lambda^2$. Before acting with $J_2$ we note that $h_{\mu\nu} = \frac{1}{4}h g_{\mu\nu} + h_{[\mu\nu]} + h_{(\mu\nu)}$, the trace part has already been considered previously. The antisymmetric part can be decomposed into self-dual and anti-self-dual parts $h_{[\mu\nu]} = h^i \Sigma^i_{\mu\nu} + \bar{h}^i \asd^i_{\mu\nu}$. It can then be seen that $h^i$ is equivalent to $M_a^{ij}$ and can be ignored. For $\bar{h}^i$, the contribution vanishes due to \cref{eq:pleb-9-dim-basis-of-symmetric-tracefree-tensors}. We are left with $h_{\mu\nu} = h_{\langle \mu\nu\rangle }$ being the remaining 9 dimensional symmetric tracefree components that are not already considered in the previous components. The action of $J_2$ on these components gives
    \begin{align}
        J_2(h_{[\mu}{}^\rho \Sigma^i_{|\rho|\nu]}) = \frac{1}{2} h_\rho{}^\rho \Sigma^i_{\mu\nu} = 0.
    \end{align}
    Therefore, we see that $h_{[\mu}{}^\rho \Sigma^i_{|\rho|\nu]}$, with $h_{\mu\nu} = h_{\langle \mu\nu\rangle }$ parametrises the 9 dimensional eigenspace with eigenvalue 0.
\end{proof}

The maps in the proof above are from the irreducible spaces into $\E \otimes \Lambda^2$, we can define the maps going in the opposite direction. Given an object $B^i_{\mu\nu} \in \E \otimes \Lambda^2$ we have irreducible components 
\begin{alignat}{2}
    \Sigma^{i\mu\nu} B^i_{\mu\nu} &\in C^\infty( && M)  \\
    \epsilon^{ijk} \Sigma^{j\mu\nu} B^k_{\mu\nu} &\in \E && = S_+^2 \\
    \Sigma^{\langle j \mu\nu} B^{i\rangle }_{\mu\nu} &\in S^2_0 \E && = S_+^4 \\
    \Sigma^i_{\langle \mu}{}^\rho B^i_{|\rho|\nu\rangle } &\in S^2_0 \Lambda^1 && = S_+^2 \otimes S_-^2
\end{alignat}
for the 1,3,5 and 9 dimensional components respectively. Here we have used that $S^2_0$ denotes the symmetric tracefree part. Any higher degree differential forms are isomorphic to lower degree forms through the Hodge star operator, therefore, our analysis up to $\Lambda^2$ completes the decomposition for $\E$-valued differential forms. A corollary for the previous proposition is 
\begin{corollary}\label{col:pleb-E-Lambda2-9-is-ASD-2-forms}
    The space $(\E \otimes \Lambda^2){}_9 = \C^3 \otimes \Lambda^-$ is the space of $\C^3$-valued anti-self-dual 2-forms.
\end{corollary} 
\begin{proof}
    Given an arbitrary symmetric tracefree $h_{\mu\nu}$ we have that 
    \begin{align}
        h_{[\mu}{}^\alpha \Sigma^i_{|\alpha|\nu]} \in (\E \otimes \Lambda^2){}_9
    \end{align}
    We then compute the result of the Hodge star, using \cref{eq:pleb-epsilon-S-to-ddS}, to find
    \begin{align}
        \frac{1}{2} \epsilon_{\mu\nu}{}^{\rho\sigma} h_{[\rho}{}^\alpha \Sigma^i_{|\alpha|\sigma]} = -i h_{[\mu}{}^\alpha \Sigma^i_{|\alpha|\nu]} \in \Lambda^-
    \end{align}
    and is therefore anti-self-dual.
\end{proof}

\section{Connection, Intrinsic Torsion and Curvature} \label{sec:pleb-connection-torsion-and-curvature}
Employing the established algebra in the previous section we are able to compare, as was done with the $SO(1,3)$ frame formalism, with the standard metric definitions of curvature. We fix a representative metric in the conformal class that we choose to be our physical metric,
\begin{align}
    \sqrt{-g} g_{\mu\nu} = \frac{i}{12} \epsilon^{ijk} \teps^{\rho\sigma\alpha\beta} \Sigma^i_{\mu\rho} \Sigma^j_{\nu\sigma} \Sigma^k_{\alpha\beta}. \label{eq:pleb-urbantke-metric}
\end{align}
There is then a unique metric covariant derivative $\nabla_\mu$ that is compatible with the metric $\nabla_\mu g_{\rho\sigma} = 0$. The metric covariant derivative can be expanded as the usual partial derivative plus the Christoffel symbols $\Gamma^\rho_{\mu\nu}$, when assume no torsion for the metric we can uniquely solve for the Christoffel symbols in terms of partial derivatives of the metric. In this case the Christoffel symbols are symmetric in the lower two indices $\Gamma^\rho_{[\mu\nu]} = 0$. Using this metric covariant derivative we define the torsion of the 2-forms $\Sigma^i$ to be 
\begin{align}
    T_{\mu\rho\sigma}^i = \nabla_\mu \Sigma^i_{\rho\sigma}.
\end{align}
For which we have the following result.
\begin{proposition}
    The intrinsic torsion of a triple self-dual 2-form $\Sigma^i$, that satisfies the quaternion algebra \cref{eq:pleb-quaterion-algebra}, can be encoded into a $\E$-valued 1-form.
\end{proposition}
\begin{proof}
    Given an arbitrary vector field $X^\mu \in TM$ we can define the following $\E$-valued 2-form from the torsion 
    \begin{align}
        \nabla_X \Sigma^i_{\mu\nu} = X^\rho \nabla_\rho \Sigma^i_{\mu\nu} = T_X^i{}_{\mu\nu} \in \E \otimes \Lambda^2.
    \end{align}
    Taking the projection to the irreducible parts reveals
    \begin{align}
        \Sigma^{(j\mu\nu} \nabla_X \Sigma^{i)}_{\mu\nu} = \nabla_X (\Sigma^{j\mu\nu} \Sigma^i_{\mu\nu}) = 4 \nabla_X \delta^{ij} = 0 \\
        \epsilon^{ijk} \Sigma^{j\mu\nu} \nabla_X \Sigma^k_{\mu\nu} = \frac{1}{8} T^i_X \\
        \Sigma^i_{(\mu}{}^\rho \nabla_X \Sigma^i_{|\rho|\nu)} = \nabla_X \Sigma^i_{(\mu}{}^\rho \Sigma^i_{|\rho|\nu)} = -3 \nabla_X g_{\mu\nu} = 0.
    \end{align}
    The corresponding irreducible components of the torsion are 
    \begin{align}
        \Sigma^{(j\mu\nu} \nabla_X \Sigma^{i)}_{\mu\nu} = \Sigma^{(j\mu\nu} T_X^{i)}{}_{\mu\nu} = T_X^{(ij)} \\
        \epsilon^{ijk} \Sigma^{j\mu\nu} \nabla_X \Sigma^k_{\mu\nu} = \epsilon^{ijk} \Sigma^{j\mu\nu} T_X^k{}_{\mu\nu} = \frac{1}{8} T_X^i \\
        \Sigma^i_{(\mu}{}^\rho \nabla_X \Sigma^i_{|\rho|\nu)} = \Sigma^i_{(\mu}{}^\rho T_X^i{}_{|\rho|\nu)} = T_{X \mu\nu}
    \end{align}
    Therefore, we see that $T^i_X \in \E$ is the only non-zero component. As this was computed with an arbitrary $X^\mu$ we find the 
    \begin{align}
        \epsilon^{ijk} \Sigma^{j\rho\sigma} \nabla_\mu \Sigma^k_{\rho\sigma} = \frac{1}{8} T^i_\mu
    \end{align}
    parametrises all the active components of intrinsic torsion. Therefore, we have 
    \begin{align}
        \nabla_\mu \Sigma^i_{\rho\sigma} = \epsilon^{ijk} T^j_\mu \Sigma^k_{\rho\sigma}
    \end{align}
    where the intrinsic torsion is $T^i \in \E \otimes \Lambda^1$.
\end{proof}

We are now able to see that the intrinsic torsion is directly related to the torsion free connection 1-form.
\begin{proposition}
    Given a self-dual triple $\Sigma^i$ there exists a unique torsion free $\E$-valued connection 1-form $A^i \in \E \otimes \Lambda^1$. This connection is then minus the intrinsic torsion.
\end{proposition}
\begin{proof} \label{eq:plebanski-unique-intrinsic-torsion-proof}
    We start with the definition of torsion free
    \begin{align}
        d^A \Sigma^i = d\Sigma^i + \epsilon^{ijk} A^j \wedge \Sigma^k = 0
    \end{align}
    which is a $\E$-valued 3-form. To bring this condition into a more useful form we apply the Hodge star to both sides, in components this is 
    \begin{align}
        \epsilon_\mu{}^{\nu\rho\sigma} \partial_\nu \Sigma^i_{\rho\sigma} + \epsilon^{ijk} \epsilon_\mu{}^{\nu\rho\sigma} A^j_\nu \Sigma^k_{\rho\sigma} = 0
    \end{align}
    using the self-duality of $\Sigma^i$ we see that
    \begin{align}
        2i \epsilon^{ijk} \Sigma^j_\mu{}^\nu A^k_\nu = \epsilon_\mu{}^{\nu\rho\sigma} \partial_\nu \Sigma^i_{\rho\sigma} 
    \end{align}
    In operator notation we have 
    \begin{align}
        J_1(A)^i = -i \star d\Sigma^i.
    \end{align}
    From $J_1^2 - J_2 -2 = 0$ we see that $J_1^{-1} = \frac{1}{2}(J_1 - 1)$ is well-defined, therefore the connection is 
    \begin{align}
        A^i = -i J_1^{-1}(\star d\Sigma)^i.
    \end{align}
    Which uniquely defines the torsion free connection from $\Sigma^i$. By antisymmetrising the torsion definition we find 
    \begin{align}
        \epsilon^{ijk} T^j \wedge \Sigma^k = \nabla \Sigma^i = d \Sigma^i = - \epsilon^{ijk} A^j \Sigma^k
    \end{align}
    therefore we obtain the final statement that $A^i = -T^i$.
\end{proof}

We check that the definition of the torsion free connection in the above way is compatible with the reality conditions. Indeed, by taking the complex conjugate of the torsion free equation we find the anti-self-dual connection
\begin{align}
    \nabla_\mu \asd^i_{\rho\sigma} + \epsilon^{ijk} \bar{A}^j_\mu \asd^k_{\rho\sigma} = 0.
\end{align}
We have used that the metric is real valued so that $\nabla^* = \nabla$. It is clear then that no additional conditions are imposed by taking the derivatives of the reality condition. Applying $\nabla_\mu$ to \cref{eq:pleb-reality-condition-in-components} we see that it is zero automatically, hence no conditions need to be imposed on the connection to satisfy the reality conditions. This is perhaps unsurprising as the connections are computed using a triple of 2-forms that satisfy the reality condition.

We are now in a position to compare the curvature of the connection to that of the metric. The components of the curvature of the torsion free connection are then 
\begin{align}
    F^i_{\mu\nu} & = \partial_{[\mu} A^i_{\nu]} + \frac{1}{2}\epsilon^{ijk} A^j_{[\mu} A^k_{\nu]} \\
                 & = \nabla_{[\mu} A^i_{\nu]} + \frac{1}{2}\epsilon^{ijk} A^j_{[\mu} A^k_{\nu]}
\end{align}
By applying another covariant derivative to the definition of the intrinsic torsion and antisymmetrising we find 
\begin{align}
    \nabla_{[\mu} \nabla_{\nu]} \Sigma^i_{\rho\sigma} = R_{\mu\nu[\rho}{}^\alpha \Sigma^i_{|\alpha|\sigma]} = - \epsilon^{ijk} \left( \nabla_{[\mu} A^j_{\nu]} + \epsilon^{jlm} A^l_{[\mu} A^m_{\nu]} \right) \Sigma^k_{\rho\sigma} = -\epsilon^{ijk} F^j_{\mu\nu} \Sigma^k_{\rho\sigma}
\end{align}
By using that the commutator of two metric covariant derivatives is the curvature, we find that
\begin{align}
    R_{\mu\nu[\rho}{}^\alpha \Sigma^i_{|\alpha|\sigma]} = R_{\mu\nu}{}_{[\rho}{}^\alpha \Sigma^i_{|\alpha|\sigma]} = -\epsilon^{ijk} F^j_{\mu\nu} \Sigma^k_{\rho\sigma}.
\end{align} 
By contracting with arbitrary vectors $X^\mu$ and $Y^\nu$ we find
\begin{align}
    R_{XY}{}_{[\rho}{}^{\alpha} \Sigma^i_{|\alpha|\sigma]} = - \epsilon^{ijk} F^j_{XY} \Sigma^k_{\rho\sigma} \in \E \otimes \Lambda^2.
\end{align}
Using the knowledge of the decomposition of $\E$-valued 2-forms we know that this implies $R_{XY}{}_{[\rho}{}^\alpha \Sigma^i_{|\alpha|\sigma]} \in (\E \otimes \Lambda^2){}_3$. Therefore, it can be encoded into a $\E$ vector, the 3 components can be extracted by multiplying by with $\epsilon^{ijk} \Sigma^{j\rho\sigma}$, which gives
\begin{align}
    R_{XY}{}^{\rho\sigma} \Sigma^i_{\rho\sigma} = 2 F^i_{XY}.
\end{align}
As $X,Y$ were arbitrary we can drop them and return the indices to find that the Riemann curvature is related to the self-dual curvature by
\begin{align}
    F^i_{\mu\nu} = \frac{1}{2} R_{\mu\nu\rho\sigma} \Sigma^{i\rho\sigma}.
\end{align}
It is known that the Riemann curvature, as it is valued in $S^2 \Lambda^2$, can be decomposed into a $2 \times 2$ block form of self-dual and anti-self-dual components. We see that $F^i_{\mu\nu}$ is exactly a self-dual row of this $2 \times 2$ block decomposition.

The Ricci tensor can be extracted through 
\begin{align}
    \Sigma^i_\mu{}^\alpha R_{\alpha\nu\rho\sigma} \Sigma^{i\rho\sigma} = (\delta_\mu^\rho g^{\alpha\sigma} - \delta_\mu^\sigma g^{\alpha\rho} -i \epsilon_{\mu}{}^{\alpha\rho\sigma}) R_{\alpha\nu\rho\sigma} = - 2 R_{\mu\nu}
\end{align}
or equivalently 
\begin{align}
    R_{\mu\nu} = \Sigma^i_\mu{}^\rho F^i_{\nu\rho}. \label{eq:pleb-Ricci-tensor-from-connection-curvature}
\end{align}
To check that this tensor is symmetric we contract it with $\Sigma^{i\mu\nu}$ and $\asd^{i\mu\nu}$, which form a basis for the antisymmetric (0,2)-tensors, and we find it vanishes. Therefore, the Ricci tensor is symmetric as is expected. The Ricci scalar is then 
\begin{align}
    R = \Sigma^{i\mu\nu} F^i_{\mu\nu}.
\end{align}
The $\C$-valued 2-form $F^i$ has 18 components, we have seen that the 9 and 1 irreducible components are contained in the Ricci tensor, by computing $d^A \Sigma^i = 0$ we see that 
\begin{align}
d^A d^A \Sigma^i = \epsilon^{ijk} F^j \wedge \Sigma^k \sim \epsilon^{ijk} F^j_{\mu\nu} \Sigma^{k\mu\nu} = 0
\end{align}
which means that the $3$ dimensional irreducible components vanish. Only $5$ components remain, and they can be encoded into $\Psi^{ij} = \Psi^{\langle ij\rangle }$, a symmetric tracefree internal matrix that represents the self-dual half of the Weyl curvature. We therefore have the following decomposition of the curvature 
\begin{align}
    F^i_{\mu\nu} = \Psi^{ij} \Sigma^j_{\mu\nu} - \frac{R}{6} \Sigma^i_{\mu\nu} + R_{[\mu}{}^\alpha \Sigma^i_{|\alpha|\nu]}.
\end{align}
Introducing the tracefree Ricci curvature as $\tilde{R}_{\mu\nu} = R_{\mu\nu} - \frac{1}{4} g_{\mu\nu} R$ we can rewrite the decomposition as
\begin{align}
    F^i_{\mu\nu} = \Psi^{ij} \Sigma^j_{\mu\nu} + \frac{R}{12} \Sigma^i_{\mu\nu} + \tilde{R}_{[\mu}{}^\alpha \Sigma^i_{|\alpha|\nu]}
\end{align}
where the first two terms are self-dual and the last is anti-self-dual (as $(\E \otimes \Lambda^1){}_9 = \C^3 \otimes \Lambda^-$). The leads to the known block decomposition of the Riemann curvature; when viewed as a $\Lambda^2 \otimes \Lambda^2$ tensor it has the following self-dual anti-self-dual decomposition 
\begin{align}
R_{\mu\nu\rho\sigma} = \begin{pmatrix}
        \Psi + \frac{R}{12} && \tilde{R} \\ \tilde{\bar{R}} && \bar{\Psi} + \frac{R}{12}
    \end{pmatrix}. \label{eq:pleb-riemann-curvature-sd-asd-decomposition}    
\end{align}
With $\bar{\Psi}^{ij}$ being the anti-self-dual half of the Weyl curvature. We then have the interpretation that $F^i$ computes the top self-dual row of this block decomposition, in spirit computing half of the total Riemann curvature. However, as we are in Lorentzian signature the other half of the Riemann curvature is available through complex conjugation. Indeed, $\bar{\Psi}^{ij} = -(\Psi^*)^{ij}$ gives the relation between the two chiral halves of the Weyl curvature. In Euclidean signature it becomes a true halving of the number of components as the self-dual and ant-self-dual spaces are real-valued.

Imposing Einstein's equations is then an algebraic condition on the curvature,
\begin{align}
    \Sigma^i_\mu{}^\rho F^i_{\nu\rho} - \frac{1}{2} \Sigma^{i\rho\sigma} F^i_{\rho\sigma} g_{\mu\nu} + \Lambda g_{\mu\nu} = T_{\mu\nu}
\end{align}
or in its trace reversed form 
\begin{align}
    \Sigma^i_\mu{}^\rho F^i_{\nu\rho} = T_{\mu\nu} - \frac{1}{2} T g_{\mu\nu} + \Lambda g_{\mu\nu}.
\end{align}
This condition can also be written as
\begin{align}
    F^i_{\mu\nu} = \left( \Psi^{ij} - \frac{4\Lambda - T}{6} \delta^{ij} \right) \Sigma^j_{\mu\nu} + T_{[\mu}{}^\rho \Sigma^i_{|\rho|\nu]}.
\end{align}
When no matter is present, $T_{\mu\nu} = 0$, Einstein's equations simplify to be
\begin{align}
    F^i \in \Lambda^+, \quad \textrm{and} \quad F^i \wedge \Sigma^i \sim \Lambda.
\end{align} 
This is significantly easier to compute, especially when done by hand.

\section{\pleb{} as a Chiral Projection of the Frame Formulation}
Here we compare the frame formalism to \pleb{}'s formulation and write all the objects appearing in the latter as chiral projections of the former. First, we compare the 2-form frame and the tetrad frame. Let $e^I \in \Lambda^1$ be the frame for the tetrads, we have that the space of 2-forms is spanned by 
\begin{align}
    B^{IJ} = e^I \wedge e^J \in \Lambda^2.
\end{align}
$B^{(IJ)} = 0$ is antisymmetric with respect to its internal indices and is therefore $\mathfrak{so}(1,3)$-valued, in the sense that $B^I{}_J = B^{IK} \eta_{KJ} \in \mathfrak{so}(1,3)$. We can then use the self-dual and anti-self-dual 't Hooft symbols~\cite{Computation_of_t_Hoo_1976}, which are projectors from $\mathfrak{so}(1,3)$  to $\mathfrak{so}(3,\C)$ and $\overline{\mathfrak{so}(3,\C)}$. In components, they are
\begin{align}
    \Sigma^i_{IJ} & = i \delta^0_I \delta^i_J - i \delta^0_J \delta^i_I - \epsilon^{ijk} 
    \delta^j_I \delta^k_J \\
    \asd^i_{IJ} & = i \delta^0_I \delta^i_J - i \delta^0_J \delta^i_I + \epsilon^{ijk} \delta^j_I \delta^k_J.
\end{align}
Where $i = 1,2,3$ indexes $\C^3 = \mathfrak{so}(3,\C) = \overline{\mathfrak{so}}(3,\C)$. By raising and lowering the internal index with $\eta_{IJ} = {\rm diag}(-1,1,1,1)$ it can be shown that the 't Hooft symbols satisfy quaternion algebras 
\begin{align}
    \Sigma^i_I{}^K \Sigma^j_K{}^J = -\delta^{ij} \delta_I^J + \epsilon^{ijk} \Sigma^k_I{}^J \label{eq:pleb-internal-self-dual-2-form-quaterion-algebra} \\
    \asd^i_I{}^K \asd^j_K{}^J = -\delta^{ij} \delta_I^J - \epsilon^{ijk} \asd^k_I{}^J.
\end{align}
This offers the interpretation that $\Sigma^i_{IJ}$ are self-dual 2-forms on the internal space. We can then see the self-dual 2-forms arise as the projection of $B^{IJ}$ to $\soC$, that is 
\begin{align}
    \Sigma^i = \frac{1}{2} \Sigma^i_{IJ} B^{IJ} = \frac{1}{2} \Sigma^i_{IJ} e^I \wedge e^J = i e^0 \wedge e^i - \frac{1}{2} \epsilon^{ijk} e^j \wedge e^k. \label{eq:pleb-self-dual-2-forms-from-chiral-tetrad}
\end{align}
The anti-self-dual 2-forms are defined similarly,
\begin{align}
    \asd^i = \frac{1}{2} \asd^i_{IJ} e^I \wedge e^J = i e^0 \wedge e^i + \frac{1}{2} \epsilon^{ijk} e^j \wedge e^k. \label{eq:pleb-anti-self-dual-2-forms-from-chiral-tetrad}
\end{align}
It is clear then that any element in $\Lambda^2$ can be reached by a complex combination of $\Sigma^i$ and $\asd^i$. In particular the decomposition of $B^{IJ}$ is
\begin{align}
    B^{IJ} = \frac{1}{2} \Sigma^{iIJ} \Sigma^i + \frac{1}{2} \asd^{iIJ} \asd^i.
\end{align}

Next we consider the self-dual connection, $A^i$, as compared with the spin connection, $\omega^{IJ}$. Indeed, by taking the exterior derivative of both sides of $\Sigma^i = \frac{1}{2} \Sigma^i_{IJ}B^{IJ}$ we find that
\begin{align}
    -\epsilon^{ijk} A^j \wedge \Sigma^k & = d\Sigma^i = \frac{1}{2} \Sigma^i_{IJ} d(e^I \wedge e^J) \nonumber \\ & = \Sigma^i_{IJ} de^I \wedge e^J = - \Sigma^i_{IJ} \omega^I{}_K \wedge e^K \wedge e^J = - \Sigma^i_{IJ} \omega^I{}_K \wedge  B^{KJ} \nonumber \\ & = - \frac{1}{2}\Sigma^i_{IJ} \omega^I{}_K \left( \Sigma^{jKJ} \Sigma^j + \asd^{jKJ} \asd^j \right) = - \frac{1}{2} \epsilon^{ijk} \Sigma^{jIJ}\omega_{IJ} \wedge \Sigma^k 
\end{align}
where we have used \cref{eq:pleb-internal-self-dual-2-form-quaterion-algebra}, $\Sigma^i_{[I|K|} \asd^{jK}{}_{J]} = 0$ and that $de^I + \omega^I{}_J \wedge e^J = 0$. We can read off from the above that 
\begin{align}
    A^i = \frac{1}{2} \Sigma^i_{IJ} \omega^{IJ}.
\end{align}
Therefore, we have that the self-dual 2-forms and their connections are the self-dual projections of the 2-forms $B^{IJ}$ and the spin connection. In this sense \pleb{}'s formulation is encoding the metric into the chiral half of the tetrad formulation.

\section{Infinitesimal Transformations}
It is useful to see how infinitesimal actions of both diffeomorphism and gauge transformations affect the fields $\Sigma^i$ in $A^i$. The transformation resulting from the linearised action of the group $SO(3,\C)$ is parametrised by a Lie algebra element $\mathfrak{so}(3,\C)$ which are $3\times 3$ complex antisymmetric matrices, using the isomorphism \cref{eq:pleb-isomorphism-antisym-matrices-and-vector-C3} these are equivalent to vectors $\phi^i \in \E$. The infinitesimal $SO(3,\C)$ action is then
\begin{align}
    \delta_\phi \Sigma^i = \epsilon^{ijk} \phi^k \Sigma^j. \label{eq:pleb-infinitesimal-gauge-transformation}
\end{align}
To compute the infinitesimal action on the connection we use the torsion free definition $d^A \Sigma^i = 0$ and act the transformation on it to find 
\begin{align}
    \delta_\phi d^A \Sigma^i = d (\delta_\phi \Sigma^i) + \epsilon^{ijk} \delta_\phi A^j \wedge \Sigma^k + \epsilon^{ijk} A^j \wedge \delta_\phi \Sigma^k = 0
\end{align}
Using \cref{eq:pleb-infinitesimal-gauge-transformation} and expanding everything gives 
\begin{align}
    \epsilon^{ijk} d\phi^k \wedge \Sigma^j + \phi^j A^i \wedge \Sigma^j - \phi^i A^j \wedge \Sigma^j + \epsilon^{ijk} \delta_\phi A^j \wedge \Sigma^k = 0
\end{align}
Which by using the definition of the exterior covariant derivative we have 
\begin{align}
    \epsilon^{ijk} d^A \phi^j \wedge \Sigma^k = \epsilon^{ijk} \delta_\phi A^j \wedge \Sigma^k.
\end{align}
Taking the Hodge star and applying $J_1^{-1}$ to both sides we obtain 
\begin{align}
    \delta_\phi A^i = d^A \phi^i
\end{align}
which is the infinitesimal transformation on the gauge. Finally, we can compute its action on the curvature of $A^i$, 
\begin{align}
    \delta_\phi F^i = d^A \delta_\phi \phi^i = d^A d^A \phi^i = \epsilon^{ijk} F^j \phi^k.
\end{align}
The infinitesimal action of gauge transformations can be summarised as 
\begin{align}
    \delta_\phi \Sigma^i = \epsilon^{ijk} \phi^k \Sigma^j, \quad \delta_\phi A^i = d^A \phi^i, \quad \delta_\phi F^i = \epsilon^{ijk} F^j \phi^k.
\end{align}
For the action of linearised diffeomorphisms we can make use of Cartan's magic formula for the Lie derivative to see that along a vector $X \in TM$, which represents the infinitesimal coordinate change, we have
\begin{align}
    \mathcal{L}_X \Sigma^i = \iota_X d \Sigma^i + d \iota_X \Sigma^i = \iota d^A \Sigma^i - \epsilon^{ijk} \iota_X A^j \Sigma^k + d^A \iota_\Sigma^i = d^A \iota_X \Sigma^i - \epsilon^{ijk} \iota_X A^j \Sigma^k
\end{align}
We can see that this action induces an infinitesimal gauge transformation with gauge parameter $\phi^i = -\iota_X A^i$. To counteract this gauge transformation and extract the diffeomorphism only transformation we instead compute
\begin{align}
    \delta_X \Sigma^i = (\mathcal{L}_X + \delta_{\iota_X A})(\Sigma)^i = d^A \iota_X \Sigma^i
\end{align}
where $\delta_X$ is defined to be the action of the diffeomorphism only. The linearised diffeomorphism transformation effects the connection as
\begin{align}
    \delta_X ( d^A \Sigma^i) = d^A \delta_X \Sigma^i + \epsilon^{ijk} \delta_X A^j \wedge \Sigma^k = \epsilon^{ijk} F^j \iota_X \Sigma^k + \epsilon^{ijk} \delta_X A^j \wedge \Sigma^k = 0.
\end{align}
Applying the Hodge star to both sides and using that $d^A d^A \Sigma^i = \epsilon^{ijk} F^j \wedge \Sigma^k = 0$ we find that 
\begin{align}
    \delta_X A^i = \iota_X F^i.
\end{align}
The operation $\delta_X$ does not produce a gauge transformation for the connection, as is expected. Lastly, we have the action on the curvature 
\begin{align}
    \delta_X F^i = d^A \iota_X F^i.
\end{align}
Concluding the action of infinitesimal diffeomorphisms we have 
\begin{align}
    \delta_X \Sigma^i = d^A \iota_X \Sigma^i, \quad \delta_X A^i = \iota_X F^i,\quad \delta_X F^i = d^A \iota_X F^i.
\end{align}
An arbitrary infinitesimal action, $\delta$, on the metric can be written as a transformation on the 2-forms given by
\begin{align}
    \delta g_{\mu\nu} = \delta \Sigma^i_{(\mu|\rho|} \Sigma^i_{\nu)}{}^\rho - \frac{1}{6} g_{\mu\nu} \delta \Sigma^i_{\rho\sigma} \Sigma^{i\rho\sigma}.
\end{align}
This is derived knowing that $g_{\mu\nu} = -\frac{1}{3}\Sigma^i_{\mu\rho} g^{\rho\sigma} \Sigma^i_{\nu\sigma}$ and then using the standard linear algebra results $\delta g^{-1} = - g^{-1} \delta g g^{-1}$ to find the above equation. Computing the gauge transformation on the metric gives 
\begin{align}
    \delta_\phi g_{\mu\nu} = 2 \phi^i \Sigma^i_{(\mu\nu)} = 0
\end{align}
which is expected as the metric is invariant under rotation of the frame. For diffeomorphisms we first write its action on $\Sigma^i$ in components 
\begin{align}
    \delta_X \Sigma^i_{\mu\nu} = \partial^A_{[\mu} \left( X^\rho \Sigma^i_{\rho|\nu]} \right) = \Sigma^i_{[\mu}{}^\rho \nabla^A_{\nu]} X_\rho = \Sigma^i_{[\mu}{}^\rho \nabla_{\nu]} X_\rho
\end{align}
where we have used that $d^A = \nabla^A$ when all the indices are antisymmetrised. Substituting this into the metric formula, and using the algebra of $\Sigma$'s, we find that 
\begin{align}
    \delta_X g_{\mu\nu} = \nabla_{(\mu} X_{\nu)} = \mathcal{L}_X g_{\mu\nu}
\end{align}
which matches the standard infinitesimal action on the metric.

\section{\pleb{} Action}\label{sec:plebanski-action}
In~\cite{OnTheSeparatiPleban1977} \pleb{} first wrote down his action principal for general relativity based on a triple of self-dual 2-forms $\Sigma^i$. His idea was to write a BF-type action with $\Sigma^i$ and $A^i$ as the fundamental fields. \hypertarget{add:BF-theory-reference} BF theories are form of topological gauge theories with with actions $S_{BF} = \int B \wedge dA$ for some differential forms $B$ and $A$, with $F = dA$. For more details on BF actions see~\cite{Topological_fie_Birmin_1991}. Certain constraints need to be imposed on the self-dual 2-forms such that they encode a real valued metric, these are the metricity and reality conditions. The later reality condition involves both the self-dual and anti-self-dual 2-forms which destroys the holomorphic nature of the action, also imposing it at the level of the action is more complicated. Instead of adding the reality conditions to the action we instead impose them separately or assume they are already satisfied. In which case the \pleb{} action, without matter terms, is
\begin{align}
    S_{Pleb}[\Sigma,A,\Psi] = \frac{1}{8\pi G i} \int \Sigma^i \wedge F^i - \frac{1}{2} \left( \Psi^{ij} + \frac{\Lambda}{3}\delta^{ij} \right) \Sigma^i \wedge \Sigma^j \label{eq:pleb-action-principal}
\end{align}
where $F(A)^i = dA^i + \frac{1}{2} \epsilon^{ijk} A^j \wedge A^k$ and $\Psi^{ij} = \Psi^{\langle ij\rangle }$ are Lagrange multipliers that can be interpreted as the self-dual part of the Weyl curvature. Both $\Sigma^i$ and $A^i$ are complex fields, so this Lagrangian is complex-valued in general. When supplemented with the reality conditions 
\begin{align}
    \Sigma^i \wedge \asd^j = 0, \quad \textrm{and} \quad \Real(\Sigma^i \wedge \Sigma^i) = 0 \label{eq:pleb-both-reality-conditions}
\end{align}
it can be shown that the imaginary part of the action is the topological Holst term~\cite{Holst1996}, which is given by
\begin{align}
    S_{Holst} = \int e_I \wedge e_J \wedge F^{IJ}.
\end{align}
It can be shown that this is the integration by parts of the torsion squared term $d^\omega e^I \wedge d^\omega e_I$, which itself vanishes as we choose the torsion to be zero. Adding this term to the Einstein-Cartan actions, with a factor of the imaginary unit, produces the internal self-dual projection operator
\begin{align*}
\Sigma^i_{IJ} \Sigma^{iKL} = \frac{1}{2}\left(\delta_I^K \delta_J^L - \delta_I^L \delta_J^K - i \epsilon_{IJ}{}^{KL} \right),
\end{align*}
that is
\begin{align}
    \frac{1}{2} S_{EH} + i S_{Holst} & = \int \frac{1}{2} \epsilon_{IJKL} e^I \wedge e^J \wedge F^{KL} + i \frac{1}{2}\left( \eta_{IK} \eta_{JL} - \eta_{IL} \eta_{JK} \right) e^I \wedge e^J \wedge F^{KL} \\
                                     & = \int i \Sigma^i_{IJ} \Sigma^i_{JK} e^I e^J \wedge F^{KL} = i \int \Sigma^i \wedge F^i.
\end{align}
The result is the \pleb{} action. Since torsion is assumed to vanish this modification is equivalent to the Einstein-Cartan action, and as such does not affect the equations of motion. The real part is equivalent to the Einstein-Hilbert action \cref{eq:einstein-cartan-action}. 

Taking the variation of \cref{eq:pleb-action-principal} with respect to $\Psi^{ij}$ implies 
\begin{align}
    \frac{\delta S}{\delta \Psi^{ij}} = \Sigma^i \wedge \Sigma^j \sim \delta^{ij}
\end{align}
which recovers the metricity condition. The physical metric, which makes $\Sigma^i$ self-dual, can then be recovered using \urb{}'s formula \cref{eq:pleb-urbantke-metric}. Taking the variation with respect to $A^i$ gives 
\begin{align}
    \frac{\delta S}{\delta A^i} = d^A \Sigma^i = d\Sigma^i + \epsilon^{ijk} A^j \wedge \Sigma^k = 0
\end{align}
which is the torsion free condition. Finally, variation with $\Sigma^i$ reveals Einstein's equations 
\begin{align}
    \frac{\delta S}{\delta \Sigma^i} = F^i - \left( \Psi^{ij} + \frac{\Lambda}{3} \delta^{ij} \right) \Sigma^j = 0.
\end{align}
As we have seen the tracefree part of the Ricci tensor appears in the anti-self-dual components of $F^i$, and the Ricci scalar appears as a factor in front of the self-dual 2-forms. The above equation then imposes that the Ricci tensor or anti-self-dual components vanish and that the Ricci scalar is a constant, $\Lambda$.

A clear advantage of this formulation is that the action is cubic in fields even with $\Lambda \neq 0$, this is the only known formulation where this occurs. Counting the number of components in the self-dual 2-forms we find that in general they have $6 \times 3 = 18\C$ components, the metricity condition removes $5\C$ of these which leaves us with the $10\C + 3\C$ of a complex metric and $SO(3,\C)$ frame. The reality conditions are then $10\R = 5\C$  conditions that remove the 10 imaginary components form the metric leaving us with $10\R + 3\C$ components for a real metric and frame. In comparison, we have $16\R$ components in the frame $e^I$ which is equal the number of real components in $\Sigma^i$, we see that the self-dual 2-forms are a special repackaging of the full frame in Lorentzian signature. This clever repackaging brings with it the powerful algebra satisfied by $\Sigma$'s which simplifies many calculations, importantly the action is a holomorphic complex projection of the Einstein-Cartan action. The holomorphic nature allows us to treat the $3\C$ $SO(3,\C)$ components as $3$ real components which simplifies certain computations. We also have that $A^i$ contains $12\C$ components which equals the $24\R$ in the spin connection $\omega^{IJ}$, indeed the self-dual connection $A^i$ is the self-dual projection of spin connection into the chiral half. Again we can treat the connection holomorphically and use the connection as if it had $12$ real components. If one is interested in coupling fermions to this theory, without explicit use of the metric, certain Lagrange multipliers need to be introduced which complicated the theory. While it is possible we do not consider this matter content here.

A useful point to make clear is that all the solutions to the Einstein's equations have a corresponding solution in the case of \pleb{}'s formulation, assuming the solutions are orientable. That is no information is lost during the chiral projection and using self-dual objects. One way to view this is that for Lorentzian signature everything is complex, so once the constraints are solved it is actually the real Einstein-Cartan formulation in disguise as it contains the same number of real components. However, this is not the preferred way one should interpret this projection. In the case of Euclidean signature the objects take real values and no real and imaginary split exists to recover the lost components. To resolve this we make the assumption that the metric defined by the triple of 2-forms, with the proportionality constant fixed, is the physical metric. In this case, as we have seen in section, all the Ricci tensor components are encoded into the self-dual part of the Riemann curvature tensor. This allows us to impose all of Einstein's equations on the metric but through the use of self-dual or chiral objects.  Therefore, the number of component in the Euclidean version of \pleb{}'s formulation is truly less than the number that appears in the Einstein-Cartan formalism. The Lorentzian signature version of \pleb{}'s formalism should be thought of in the same way, in that it is a projection of the Einstein-Cartan theory, but it is holomorphic and so there are fewer components even if they are complex. A similar argument can be made about the metric the 2-forms encode being solutions to the full Einstein's equations due to the same structure of the Riemann curvature tensor. Meaning that the chiral formulation of gravity does indeed capture all of Einstein's equations and its solutions.

In summary the action for the metric for general relativity can be encoded into a tripe of self-dual 2-form $\Sigma^i$ that satisfy certain metricity and reality conditions. The intrinsic torsion of the 2-forms uniquely defines a torsion free $SO(3,\C)$ connection 1-form the curvature of which encodes the Ricci tensor and the self-dual half of the Weyl curvature. Einstein's equations are then algebraic conditions on the curvature of the connection. \pleb{}'s action, \cref{eq:pleb-action-principal}, is an action principal that encodes the metricity constraints and related equations of motion to impose Einstein's equations. To make the metric real-valued one must also impose reality conditions.

\section{Pure Connection Formalisms}\label{sec:connection-formalism}
This section will detail a related first-order description of gravity where the main fields are the self-dual connection instead of the self-dual 2-forms. This is obtained by integrating out certain fields from the action. For example, the equation $d^A \Sigma^i$ can be solved for $A^i$, this can then be substituted back into the action to recover a second-order version of \pleb{}'s action \cref{eq:pleb-action-principal} only in terms of $\Sigma^i$ and its derivatives. However, in this case the solution for $A^i$ is nonpolynomial and the resulting second-order action is unwieldy. Instead, we consider integrating out the metric to obtain the so-called pure connection formalisms.

\subsection{First-Order Pure Connection}
Here we consider integrating out the self-dual 2-forms $\Sigma^i$, to obtain a pure connection description~\cite{General_relativ_Capovi_1989,Pure_Connection_Krasno_2011}.  We will show that it results in a simple first-order description of gravity. Beginning with \cref{eq:pleb-action-principal} and taking the equation of motion from varying with respect to $\Sigma^i$ we find
\begin{align}
    F^i = (\Psi + \frac{\Lambda}{3} \delta){}^{ij} \Sigma^j
\end{align}
then assuming the matrix $M^{ij} = (\Psi + \frac{\Lambda}{3} \delta){}^{ij}$ is invertible we can solve for $\Sigma^i$,
\begin{align}
    \Sigma^i = \frac{1}{(\Psi + \frac{\Lambda}{3} \delta){}^{ij}} F^j.
\end{align}
Substituting this back into the action we find 
\begin{align}
    S_{Pleb}[A,\Psi] = \frac{1}{16\pi G i} \int \frac{1}{M^{ij}} F^i \wedge F^j \label{eq:pleb-first-order-pure-connection-aciton}
\end{align}
which is the pure connection action for general relativity. This is a diffeomorphism invariant gauge theory, where $M^{ij}$ is now a Lagrange multiplier field. Starting from the action we show its equations of motion are first-order and equivalent to those Einstein's equations in \pleb{} formalism. Taking the variation with respect to $A^i$ gives 
\begin{align}
    \frac{\delta S}{\delta A^i} = d^A \left(  \frac{1}{M^{ij}} F^j \right) = 0.
\end{align}
Using the Bianchi identity $d^A F^i = 0$ we can also write this as 
\begin{align}
    d^A M^{ij} \wedge \frac{1}{M^{jk}} F^k = 0.
\end{align}
This equation of motion is first-order in derivatives, schematically it is $\partial \Psi \partial A = 0$. Taking the variation with respect to $\Psi^{ij}$ we find 
\begin{align}
    \frac{\delta S}{\delta \Psi^{ij}} = \frac{1}{M^{ik}} \frac{1}{M^{jl}} F^k \wedge F^l \sim \delta^{ij}
\end{align}
or by multiplying with $M^2$ on both sides 
\begin{align}
    F^i \wedge F^j \sim \left(M^2\right){}^{ij}. \label{eq:pure-connection-F-F-M2-condition}
\end{align}
Again we see that the equation of motion is first-order in its derivatives on $A$, schematically this equation appears as $\partial A \partial A \sim M^2$. Therefore, we have a pair of first order equations of motion from the action \cref{eq:pleb-first-order-pure-connection-aciton}. 

To compute the metric we have to invert the steps taken to construct this action. We first introduce a 2-form 
\begin{align}
    \Sigma^i_F := \frac{1}{M^{ij}} F^j
\end{align}
built from $A^i$ and $\Psi^{ij}$. The equations of motion then imply that 
\begin{align}
    d^A \Sigma^i_F = 0, \quad \Sigma^i_F \wedge \Sigma^j_F \sim \delta^{ij}.
\end{align}
It is clear that this 2-form correspond to the triple of self-dual 2-forms seen in \pleb{}'s formulation. The metric can be recovered using the \urb{} formula 
\begin{align}
    g(u,v){}_F \nu_{\Sigma_F} = \frac{i}{6}\epsilon^{ijk} \iota_u \Sigma^i_F \wedge \iota_v \Sigma^j_F \wedge \Sigma^k_F.
\end{align}
Using the definition of $\Sigma_F$ in terms of $M,F$ we find that 
\begin{align}
    g(u,v){}_F \nu_{\Sigma_F} & = \frac{i}{6} \epsilon^{ijk} \frac{1}{M^{il}} \frac{1}{M^{jm}} \frac{1}{M^{kn}} \iota_u F^i \wedge \iota_v F^j \wedge F^k \nonumber \\ & = \frac{i}{6\det(M)} \epsilon^{ijk} \iota_u F^i \wedge \iota_v F^j \wedge F^k.
\end{align}
By changing the definition of the volume form for the connection formalism to be $\nu_F := \det(M) \nu_\Sigma = \frac{i\det(M)}{6} \frac{1}{M^{ij}} \frac{1}{M^{ik}} F^j \wedge F^k$ we fix physical metric in the conformal class to be
\begin{align}
    g(u,v) \nu_F = \frac{i}{6} \epsilon^{ijk} \iota_u F^i \wedge \iota_v F^j \wedge F^k.
\end{align}
In which case we can compute the metric using only $A^i$ and $\Psi^{ij}$. The equation defining $\Sigma^i_F$ then becomes 
\begin{align}
    F^i = \left(\Psi^{ij} + \frac{\Lambda}{3}\delta^{ij}\right) \Sigma^j_F,
\end{align}
and we can identify $\Psi^{ij}$ with the self-dual Weyl curvature and $\Lambda$ with the cosmological constant. These are Einstein's equations, and we find that the metric recovered is therefore Einstein. It is worth comparing the number of variables in this theory to that of the action \cref{eq:pleb-action-principal}. For the connection formulation one needs $12\C + 5\C = 18\C$ components for the self-dual connection and Weyl curvature. Whereas, for \pleb{}'s formulation the number of components one starts with in general is $18\C + 12\C = 30\C$ for a triple of complex 2-form and the connection. Usually the constraints are applied which leaves $10\R + 3\C + 12\C = 20\C$ components for the real metric, frame and connection. In either case the number of components is more than in the connection formulation.

It should be mentioned that a major drawback to the connection formulation is that one loses the polynomial structure of the action. Importantly, this means that when $\Lambda = 0$ this theory fails to describe Minkowski space as there $\Psi = 0$ and the $1/\Psi$ term is ill-defined. Perturbation around Minkowski have to be performed more carefully, although it is still possible to compute in this formulation. There is also the issue of reality conditions, in \pleb{}'s formulation they are algebraic conditions on $\Sigma^i$ and its complex conjugate. In these connection formulations they appear as non-holomorphic differential equations, that is 
\begin{align}
    F^i \wedge \bar{F}^j = 0,
\end{align}
these are much harder to solve in practice.

\subsection{Pure Connection Action}
Another first order pure connection action was discovered in~\cite{Pure_Connection_Krasno_2011} by noticing that when $\Lambda \neq 0$ one can integrate out the Weyl curvature and cosmological constant $M^{ij} = \Psi^{ij} + \frac{\Lambda}{3} \delta^{ij}$. Indeed, by introducing a matrix 
\begin{align}
    X^{ij} = F^i \wedge F^j / d^4x,
\end{align}
where we have divide by a top form, the equation of motion \cref{eq:pure-connection-F-F-M2-condition} can be solved for $M^{ij}$
\begin{align}
    M^{ij} = \sqrt{X}^{ij}.
\end{align}
The resulting action is 
\begin{align}
    S[A] = \int \Tr(\sqrt{F^i \wedge F^j}). \label{eq:plebanski-pure-connection-action}
\end{align}
To recover the usual formalism we notice that the combination
\begin{align}
    \Sigma^i_X = \frac{1}{\sqrt{X^{ij}}} F^j
\end{align}
satisfies 
\begin{align}
    \Sigma^i_X \wedge \Sigma^j_X \sim \delta^{ij}
\end{align}
and is therefore metric. Taking the variation of \cref{eq:plebanski-pure-connection-action} with respect to $A^i$ we find 
\begin{align}
    \frac{\delta S}{\delta A^i} = \frac{1}{2} d^A \left( \frac{1}{\sqrt{X^{ij}}} F^j \right) = d^A \Sigma^i_X = 0.
\end{align} 
Therefore, the connection $A^i$ is the intrinsic torsion for the 2-forms $\Sigma^i_X$. Einstein's equations are then hidden in the definition of the self-dual 2-forms
\begin{align}
    F^i = X^{ij} \Sigma^j_X.
\end{align}
The self-dual Weyl tensor is then encoded into $\Psi^{ij} = X^{\langle ij\rangle }$ and the Ricci scalar is its trace $R = Tr(X)$. An interesting family of modified theories also arises from this formulation, the action \cref{eq:plebanski-pure-connection-action} can be written as an arbitrary scalar function on the matrix $X$, 
\begin{align}
    S[A] = \int f(F^i \wedge F^j).
\end{align}
The modified theories all propagate the same number of degrees of freedom~\cite{A_Gauge_Theoret_Krasno_2012}, however, the metricity condition is lost and an understanding of how the metric related to the 2-forms has not yet been developed. Some developments for the linearised version of these modified theories can be found in~\cite{ModifiedGravitFreide2008}.

    \newpage
    \chapter{Chiral Yang-Mills}\label{chap:chiral-yang-mills}
The goal of this chapter is to introduce the chiral formulation of Yang-Mills. We further introduce its first-order version whose action and equations of motion closely resemble those appearing in the chiral formulations of gravity. Specifically chiral Maxwell's theory, the simplest of the Yang-Mills theories, will be used to compare with \pleb{}'s formulation throughout the Thesis.

\section{Yang-Mills Lagrangian}
The standard second order formulation of Yang-Mills starts with the same spacetime manifold $M$, over which we attach Lie groups $G$ as fibres to form a principal bundle $P$. The vector potential $A^a_\mu$ is then the pullback of a connection 1-form $\omega^a \in \Lambda^1(P,\mathfrak{g})$ to $M$ in a local trivialisation, where $a$ is an index for vector representation of the Lie algebra $\mathfrak{g}$. As such, $A^a \in \Lambda^1(M,\mathfrak{g})$, the vector potential is a Lie algebra valued 1-form on $M$. In this representation we label define the basis for the Lie algebra as $T_a \in \mathfrak{g}$, for the purposes here we consider $\mathfrak{g} = \mathfrak{su}(n)$ such that the basis satisfies
\begin{align}
    [T_a,T_b] = f_{abc} T^c, \quad \Tr(T_a T_b) = \delta^{ab}
\end{align}
where $f^{abc}$ are the structure constants and $[ \cdot, \cdot ]$ is the Lie bracket. The field strength of the vector potential then corresponds with the pullback of the curvature of the connection 1-form,
\begin{align}
    F^a = d A^a + \frac{1}{2} f^{abc} A^b \wedge A^c \in \Lambda^2(M,\mathfrak{g}).
\end{align}
The Lagrangian for Yang-Mills is given by 
\begin{align}
    L = F^a \wedge \star F^a.
\end{align}
Taking the variation with respect to $A^a$ recovers 
\begin{align}
    \frac{\delta L}{\delta A^a} = D \star F^a = 0
\end{align}
where $D \phi^a = d \phi^a + f^{abc} A^b \wedge \phi^c$ is the exterior covariant derivative. By the definition of $F^a$ we can also see that the Bianchi identity holds 
\begin{align}
    D F^a = 0.
\end{align}
The two equations above constitute all the second order equations for Yang-Mills. As we have seen in the general description of principal bundles, the curvature transforms via the adjoint under the group action of $G$. This leaves the Lagrangian invariant under the group actions. Another important action is the infinitesimal gauge transformations, given a Lie algebra element $\phi^a \in \mathfrak{g}$ we define the infinitesimal action of this on the connection as 
\begin{align}
    \delta_\phi A^a = D \phi^a.
\end{align}
Acting on the curvature this becomes 
\begin{align}
    \delta_\phi F^a= D D \phi^a = f^{abc} F^b \phi^c
\end{align}
which allows for the direct computation of its action on the Lagrangian 
\begin{align}
    \delta_\phi L = 2 \delta_\phi F^a \wedge \star F^a = 2 f^{abc} \star F^a \wedge F^b \phi^c.
\end{align}
By definition the structure constants are antisymmetric in the first two indices, that is $f^{abc} T^c = [T^a,T^b]$. Therefore, as $F^a \wedge \star F^b$ is symmetric in $ab$ the infinitesimal gauge transformation leaves the Lagrangian invariant. Gauge theories all share the same property that a gauge transformation does not affect the physical degrees of freedom, as such it is expected that the gauge action should leave the Lagrangian invariant. Furthermore, the equations of motion are also invariant under this action. Now we turn to infinitesimal diffeomorphism transformations, these are generated by a vector field $X^\mu$. We expect that these transformations should also leave the Lagrangian invariant as they amount to small changes in the coordinate description. The Lie derivative along $X^\mu$ computes the infinitesimal change of tensor along integral curves of $X^\mu$ and hence how they transform. For the connection, using Cartan's magic formula, we have 
\begin{align}
    \mathcal{L}_X A^a = \iota_X d A^a + d \iota_X A^a = \iota_X F^a + D \iota_X A^a.
\end{align}
Similarly to the case for \pleb{}'s formulation we find the diffeomorphisms also cause a gauge transformation. We can subtract this to leave only the diffeomorphism contribution
\begin{align}
    \delta_X A^a = (\mathcal{L}_{X} - \delta_{\iota_X A^a}) A^a = \iota_X F^a.
\end{align}
For the curvature we have 
\begin{align}
    \delta_X F^a = D \delta_X A^a = D \iota_X F^a.
\end{align}
The Lagrangian is then automatically invariant under diffeomorphisms as
\begin{align}
    \delta_X L = \mathcal{L}_X L = D \iota_X L = d \iota_X L
\end{align}
which is a total derivative and hence constant up to boundary terms which we can safely ignore. We have made use of the fact that $L$ is not Lie algebra valued and so the Lie algebra terms are zero when applied to it.

\section{Chiral Lagrangians}\label{sec:chiral-yang-mills-lagrangians}
Another Lagrangian exists, that is quadratic in the curvature, for the case of Yang-Mills, this is the Pontryagin Lagrangian
\begin{align}
    L_{top} = F^a \wedge F^a = d(A^a \wedge dA^a + \frac{1}{3} f^{abc} A^a \wedge A^b \wedge A^c).
\end{align}
It is a total divergence and does not affect the equations of motion. The variation with respect to $A^a$ vanishes identically due to the Bianchi identity. We are free to add this Lagrangian to the original, and by choosing a specific combination we can construct 
\begin{align}
    L = F^a \wedge (F^a - i \star F^a) = F^a \wedge F^a_+
\end{align}
where $F^a_+ \in \Lambda^+$ is the self-dual part of the curvature. We are then able to use the result that $\Lambda^2 = \Lambda^-\oplus \Lambda^+$ splits orthogonally with respect to the wedge product and write the Lagrangian as 
\begin{align}
    L = F^a_+ \wedge F^a_+
\end{align}
purely in terms of self-dual curvatures. This is the chiral Yang-Mills action, also called the Chalmers-Siegel action~\cite{The_self_dual_s_Chalme_1996}.  We can now introduce a first-order field $B^a = \phi^{ai} \Sigma^i \in \Lambda^+$, where $\Sigma^i$ are background self-dual 2-forms for Minkowski, such that the Lagrangian is 
\begin{align}
    L = \phi^{ai} \Sigma^i \wedge F^a - \frac{1}{2} \phi^{ai} \phi^{aj} \Sigma^i \wedge \Sigma^j. \label{eq:chiral-YM-action}
\end{align}
Using that $\Sigma^i$ satisfies the metricity condition we find the above action in coordinates becomes, up to an overall constant,
\begin{align}
    L = \phi^{ai} \Sigma^{i\mu\nu} F^a_{\mu\nu} - \frac{1}{2} \phi^{ai} \phi^{ai}. 
\end{align}
If one performs the same first-order reduction without the self-dual projection, we find that twice as many first-order fields are required (i.e. $\phi^{ai}$ and $\overline{\phi}^{ai}$ for $\Lambda^+$ and $\Lambda^-$). Defining $n = \dim(\mathfrak{g})$ as the dimension of the Lie group, we have that $3n + 4n = 7n$ components instead of $10n$ in the full case. As the only difference between the two action is the addition of a total divergence the equations of motion remain unchanged, and the same theory is described with both Lagrangians. The first-order chiral version of Yang-Mills is then a good starting theory for first-order perturbation theory, due to the reduced number of fields. Taking the variation with respect to $\phi^{ai}$ and $A^a$ give 
\begin{align}
    \phi^{ai} = \Sigma^{i\mu\nu} F^a_{\mu\nu}, \quad \Sigma^i_\mu{}^\nu \partial^A_\nu \phi^{ai} = 0
\end{align}
or equivalently as differential forms we have 
\begin{align}
    B^a = F{(A)}^a_+, \quad D^* B^a = 0. \label{eq:chiral-ym-chiral-differential-form-equations-of-motion}
\end{align}
Where $D^* = - \star D \star$ is the gauge covariant codifferential. It may seem like solutions to the chiral theory somehow represent ``half'' of a solution to the non-chiral theory. This is the incorrect interpretation. To see this let $F^a$ be the curvature 2-form of the connection $A^a$, then we can project $F^a$ onto its chiral half using the Hodge star of the background metric,
\begin{align}
    F{(A)}^a_+ = (i + \star) F^a
\end{align}
it is easy to check using $\star^2 = - \id$ that this is indeed self-dual. Substituting this into the chiral equations of motion \cref{eq:chiral-ym-chiral-differential-form-equations-of-motion} we find,
\begin{align}
    D^\star F^a_+ = - \star D \star (i+\star) F^a = \star D (\star i - \id) F^a = -i D^\star F^a
\end{align}
where we have used the Bianchi identity, $D F^a = 0$. This is exactly the second-order equation on the connection that corresponds to the full Yang-Mills solutions. Therefore, no information is lost in the projection to the self-dual theory. The similarities of the chiral Yang-Mills and gravity theories is worth noting and are discussed in the next section.

\section{Self-Dual Yang-Mills and Self-Dual Gravity}\label{sec:self-dual-YM-and-gravity}
Self-dual theories are important truncations of the chiral theories for Yang-Mills (YM) and general relativity (GR). They are known to be integrable as seen from many perspectives, e.g.~\cite{Self_Dual_Yang_Bardee_1996,From_2D_integra_Dunajs_1998}. They are also useful in computing quantum scattering amplitudes as the calculations take on a simple form~\cite{One_loop_self_d_Bern_1997,One_loop_i_n_i_Bern_1994,One_loop_n_poin_Bern_1998,Multi_leg_one_l_Bern_1999,Self_Dual_Gravi_Krasno_2016} and in some cases they can be used to compute amplitudes for the full theory as well~\cite{Celestial_holog_Costel_2022}. For these reasons we are motivated to show how self-dual theories are obtained from their chiral Lagrangians.

\subsection{Self-dual Yang-Mills}
For self-dual theories there solutions are instantons, these only exist in Euclidean signature as they require the self-dual and anti-self-dual components to be independent whereas in Lorentzian signature they are complex conjugates. Self-dual Yang-Mills solutions are characterised by connections whose curvatures are anti-self-dual,
\begin{align}
    F{(A)}_+ = 0.
\end{align}
A Lagrangian description of instantons is obtained from the Chalmers-Siegel Lagrangian \cref{eq:chiral-YM-action} by dropping the terms containing $\phi^2$,
\begin{align}
    L_{SDYM} = \phi^{ai} \Sigma^i \wedge F^a. \label{eq:self-dual-yang-mills-action}
\end{align}
Varying with respect to $\phi^{ai}$ reveals the instanton equations,
\begin{align}
    \Sigma^i \wedge F^a = 0, \quad \Rightarrow \quad F^a \in \Lambda^-.
\end{align}
These remain solutions to the full Yang-Mills equations, indeed we see that
\begin{align}
    D F^a = 0, \quad D \star F^a = -D F^a = 0.
\end{align}

\subsection{Self-dual Gravity}
Chiral gravity has a similar truncation whose solutions are self-dual instantons. Similarly to self-dual YM, self-dual gravity solutions are Einstein metrics with vanishing self-dual Weyl curvature and as such are only non-trivial, meaning that they are not maximally symmetric, in Euclidean signature. The first-order self-dual gravity action can be obtained by removing $\Psi^{ij}$ in \cref{eq:pleb-action-principal},
\begin{align}
    L = \Sigma^i \wedge F^i + \frac{\Lambda}{6} \Sigma^i \wedge \Sigma^i.
\end{align}
The equations of motion obtained by varying with $A^i$ and $\Sigma^i$ are
\begin{align}
    d^A \Sigma^i = 0, \quad F^i = -\frac{\Lambda}{3} \Sigma^i.
\end{align}
These have to be supplemented with $\Sigma^i \wedge \Sigma^j \sim \delta^{ij}$ to make sure that only the metric and $SO(3)$ rotation components are in the 2-forms. Solutions of this type are called instantons, and it is clear that they are also solutions to the full theory. We have seen that the 2-forms can be integrated out to leave a theory where the metric is encoded into a triple of 1-forms, $A^i$, which is the starting point of the connection formalism \cref{sec:connection-formalism}. Recall that the first-order pure connection Lagrangian is
\begin{align}
    L = \frac{1}{{(\Psi + \frac{\Lambda}{3})}^{ij}} F^i \wedge F^j.
\end{align}
The self-dual Lagrangian cannot be derived in the same way as before, by removing the self-dual Weyl curvature, $\Psi^{ij} = 0$, the only term remaining is the topological Holst term $F^i \wedge F^i$. Instead, the correct method is to expand in powers of $\Psi^{ij}$ and truncate at the first order~\cite{Self_Dual_Gravi_Krasno_2016,Pure_Connection_Krasno_2011}, for which the Lagrangian is 
\begin{align}
    L_{SDGR} = \Psi^{ij} F^i \wedge F^j. \label{eq:self-dual-connect-action}
\end{align}
Here $\Psi^{ij}$ is a symmetric tracefree tensor and is not associated with Weyl curvature any more. The variation with respect to $\Psi^{ij}$ gives
\begin{align}
    F^i \wedge F^j \sim \delta^{ij}.
\end{align}
To obtain the metric solution we define a 2-form $\Sigma^i := F^i$ and see that it satisfies the equations of motion for the full theory,
\begin{align}
    d^A \Sigma^i = 0, \quad \Sigma^i \wedge \Sigma^j \sim \delta^{ij}.
\end{align}
The final realisation is that $F^i = \Sigma^i$ implies that the curvature of the metric defined by $\Sigma^i$ has no self-dual Weyl curvature. Note that this only has solutions where $\Lambda \neq 0$. 

It was realised in~\cite{Flat_Self_dual_Krasno_2021} that there exists a self-dual action that exists on ``flat'' spacetimes, that is spacetimes where $F^+{(A)}^i = 0$. The Lagrangian for these solutions is obtained by dropping the $A \wedge A$ terms in the curvatures from the Lagrangian \cref{eq:self-dual-connect-action}, in doing so one finds
\begin{align}
    L_{fSDGR} = \Psi^{ij} dA^i \wedge dA^j
\end{align}
which is called the ``flat'' SDGR (fSDGR) Lagrangian. The equations of motion are given by 
\begin{align}
    dA^i \wedge dA^j \sim \delta^{ij}, \qquad {\rm and} \qquad d(\Psi^{ij} dA^j) = 0. \label{eq:flat-self-dual-gravity-equations-of-motion}
\end{align}
To see that this gives solution where the self-dual curvature vanishes we first define the self-dual 2-forms as $\Sigma^i = dA^i$, the $A^i$ here should not be identified with the self-dual connections, they are simply an arbitrary triple of 1-forms. From this definition of the $\Sigma^i$ we see that 
\begin{align}
    d\Sigma^i = 0, \qquad {\rm and} \qquad \Sigma^i \wedge \Sigma^j \sim \delta^{ij}
\end{align}
which implies that the self-dual connections are zero and therefore $F^i = 0$. As for the second equation in \cref{eq:flat-self-dual-gravity-equations-of-motion} this can be solved by imposing that the Lagrange multipliers $\Psi^{ij}$ are constant. 

This Lagrangian is also suitable for expanding around flat space, $dA^i \rightarrow \Sigma^i + da^i$, where $\Sigma^i$ are now the Minkowski self-dual 2-forms. Computing the linearised Lagrangian we find
\begin{align}
    \delta^2 L_{fSDGR} = \Psi^{ij} (\Sigma^i d a^j + \frac{1}{2} da^i \wedge da^j). \label{eq:flat-self-dual-gravity}
\end{align}
From here the comparison with \cref{eq:self-dual-yang-mills-action} is made explicit. To convert from YM to GR one replaces the colour index $a$ with the kinetic index $i$, this gives a realisation of the double copy~\cite{Perturbative_Qu_Bern_2010} and the colour/kinematic duality~\cite{The_Kinematic_A_Montei_2011}. Moreover, if an ansatz for the critical points of \cref{eq:self-dual-yang-mills-action,eq:flat-self-dual-gravity} are parametrised by a scalar as in~\cite{Pure_connection_Krasno_2024} then the double copy structure is realised more explicitly.

Overall, the similarities between chiral Yang-Mills and gravity give good reason to compare the two throughout the Thesis.

\section{Chiral Maxwell's Theory}\label{sec:chiral-maxwell}
By choosing $G = U(1)$ for the group in Yang-Mills we recover Maxwell's theory of electromagnetism. In this case the Lie algebra is simply $\mathfrak{g} = \R$, therefore, the connection is $A \in \Lambda^1(M)$ a normal 1-form. It is also known that the structure constants vanish, $f^{abc} = 0$. The curvature in this case is simply
\begin{align}
    F(A) = dA.
\end{align}
The first order 2-forms then expand into a basis of self-dual 2-forms, $B = \phi^i \Sigma^i$, and the equations of motion simplify to be 
\begin{align}
    \phi^i = \Sigma^{i\mu\nu} \partial_\mu A_\nu, \quad \Sigma^i_\mu{}^\nu \partial_\nu \phi^i = 0.
\end{align}
The spacetime indices are raised and lowered with a non-dynamical metric $g_{\mu\nu}$ defined by the non-dynamical 2-forms $\Sigma^i$. To see how this related to Maxwell's equations we introduce a foliation on the manifold $M$ such that $M \cong \R \times \spasurf$ where $\spasurf$ is our spatial hypersurface. To do so we introduce a distinguished coordinate $t : M \rightarrow \R$ which we call the time coordinate. The remaining coordinates we give arbitrary symbols $x^i$ for $i = 1,2,3$, this induces a basis of 1-forms,
\begin{align}
    dt, dx^i \in \Lambda^1.
\end{align}
In this basis the Minkowski self-dual 2-forms become 
\begin{align}
    \Sigma^i = i dt \wedge dx^i - \frac{1}{2} \epsilon^{ijk} dx^j \wedge dx^k
\end{align}
or in components 
\begin{align}
    \Sigma^i_{0j} = i\delta^i_j, \quad \Sigma^i_{jk} = -\epsilon^i{}_{jk}.
\end{align}
By fixing the self-dual 2-forms to those of Minkowski the metric is then also the Minkowski one $g_{\mu\nu} = \eta_{\mu\nu} = {\rm diag}(-1,1,1,1)$. The Yang-Mills connection admits a similar decomposition under this foliation
\begin{align}
    A = \theta dt + A^i dx^i.
\end{align}
Decomposing all the equations of motion into this foliation gives the following equations of motion,
\begin{align}
    \phi^i = -i (\partial_t A^i - \partial^i \theta) - \epsilon^{ijk} \partial^j A^k, \quad \partial_t \phi^i = -i \epsilon^{ijk} \partial^j \phi^k, \quad  \partial^i \phi^i = 0. \label{eq:chiral-maxwell-1+3-evolution-equations}
\end{align}
Where $\partial_t = \partial/\partial t$ and $\partial^i = \partial/\partial x^i$. To make the connection with Maxwell's equations we now equate the curvature with the Faraday 2-form, specifically the timelike and spacelike parts of the curvature are identified with the electric and magnetic field 
\begin{align}
    E^i = F_{i0} = -\partial_t A^i + \partial^i \theta, \quad B^i = \epsilon^{ijk} F_{jk} = \epsilon^{ijk} \partial^j A^k.
\end{align}
Now we see that the first field equation defines $\phi^i$ as a complex combination of the electric and magnetic field 
\begin{align}
    \phi^i = i E^i - B^i
\end{align}
which is related to the Riemann-Silberstein vector~\cite{Elektromagnetis_Silber_1907}. The real and imaginary parts of the second equation in \cref{eq:chiral-maxwell-1+3-evolution-equations} become 
\begin{align}
    \partial_t E^i = \epsilon^{ijk} \partial^j B^k, \quad \partial_t B^i = -\epsilon^{ijk} \partial^j E^k
\end{align}
which are the correct evolution equations for the electric and magnetic fields. Lastly, the final equation of motion gives 
\begin{align}
    \partial^i E^i = 0, \quad \partial^i B^i = 0
\end{align}
which are the Gauss constraints for the electric and magnetic field. We see that we have recovered all of Maxwell's equations in a vacuum. The chiral theory halves the number of first order variables needed to write down the equations of motion, in this case $\phi^i$ has 3 instead of $B \in \Lambda^2$ having 6. Although the number of fields is reduced we see that the full theory is recovered, this remains true even with the addition of source terms. One could argue that the fields introduced are complex and the extra degrees of freedom are hidden in the imaginary components, however, much like in the chiral gravity theories the fields appear holomorphically and can therefore be treated as 3 components. We then see that the chiral formulations give an efficient way to write down the equations of motion, and locally it is exactly equivalent to the usual theory.

    \newpage
    \part{Linearised Gravity}\label{part:linearised-gravity}

    \newpage
    \chapter{Linearised Gravity}\label{chap:Linearised-Gravity}
One important topic in the study of gravity is its linearisation, in general this a perturbative technique that allows one to compute approximations to nonlinear partial differential equations around a known background solution. It is a powerful technique and is used on a wide range of problems, for example a large effort is being put into studying perturbations around black hole spacetimes in order to model the gravitational waves emitted. Studying the structure of the linearised equation of motion gives an insight to the highest order structure of the nonlinear equations, and for properties such as hyperbolicity these are the only important terms. In this chapter we compute the perturbation of the \pleb{} 2-form with respect to the metric perturbation, and go on to derive the linearised Einstein's equations. It is then a simple computation to check that there are 2 propagating degrees of freedom, meaning that there are only two undetermined components in the theory once all the equations of motion have been solved. The last part of this chapter develops a hyperbolic gauge fixing for the \pleb{} formalism, this is an extension of the Harmonic gauge and implies that despite the change in the description the equations of motion remain hyperbolic and therefore well-posed. A well-posed set of differential equations has a unique solution that depends continuously on its initial data, a criterion initially proposed by Jacques Hadamard~\cite{SurLesProblemsHadamard1920}. Hyperbolicity is a property of first-order partial differential equations that is useful in analysing their well-posedness~\cite{Time_Dependent_Gustaf_2013}. Both are useful properties for our gauge-fixed equations of motion to have when we start considering numerical schemes. Before showing the linearised structure and gauge fixing for Gravity we first present the story for chiral $U(1)$ Yang-Mills. It has many similarities with linearised gravity and the gauge fixing procedure is very similar to our strategy for gravity, however, it is easier to understand for $U(1)$ Yang-Mills.

\section{Chiral Maxwell First-Order Gauge Fixing}\label{sec:first-order-maxwell-gauge-fixing}
We recall the first-order Lagrangian for $U(1)$ Yang-Mills on a Minkowski background
\begin{align}
    L = \phi^i \Sigma^{i\mu\nu} \partial_\mu A_\nu - \frac{1}{2} \phi^i \phi^i
\end{align}
where $\phi^i$ is the Riemann-Silberstein vector, $A_\mu$ is the $U(1)$ connection 1-form and $\Sigma^i_{\mu\nu}$ are the Minkowski self-dual 2-forms. Here spacetime indices are raised and lowered with the Minkowski metric $g_{\mu\nu} = \eta_{\mu\nu}$. The equations of motion for this Lagrangian are
\begin{align}
    \phi^i = \Sigma^{i\mu\nu} \partial_\mu A_{\nu}, \quad \Sigma^i_\mu{}^\nu \partial_\nu \phi^i = 0.
\end{align}
By substituting the first equation into the second find, after using the algebra of $\Sigma$'s that
\begin{align}
    \partial^\nu \partial_\nu A_\mu - \partial_\mu \partial^\nu A_\nu = 0.
\end{align}
Clearly the term $\partial^\mu A_\mu$ is obstructing wave operator from being the leading symbol of this second order equation. We can then impose a gauge fixing to remove this, the standard approach is to impose the Lorenz gauge $\partial^\mu A_\mu = 0$. As we are interested in the first-order structure we impose gauge by adding the following gauge fixing Lagrangian
\begin{align}
    L_{g.f} = \theta \partial^\mu A_\mu - \frac{1}{2} \theta^2
\end{align}
where $\theta \in C^\infty(M)$ is a new scalar field. The total Lagrangian is then 
\begin{align}
    L + L_{g.f.} = \phi^i \Sigma^{i\mu\nu} \partial_\mu A_\nu - \frac{1}{2} \phi^i \phi^i + \theta \partial^\mu A_\mu - \frac{1}{2} \theta^2.
\end{align}
Varying with respect to $\phi,\theta$ and $A_\mu$ gives 
\begin{align}
    \phi^i = \Sigma^{i\mu\nu} \partial_\mu A_\nu, \quad \theta = \partial^\mu A_\mu \label{eq:lin-maxwell-modified-phi-theta-definition} \\
    \Sigma^i_\mu{}^\nu \partial_\nu \phi^i - \partial_\mu \theta = 0. \label{eq:lin-maxwell-modified-evolution-equations}
\end{align} 
By substituting the first line of equations into the second reveals that 
\begin{align}
    \partial^\nu \partial_\nu A_\mu = 0.
\end{align}
Therefore, the solutions are solutions to wave equations. By applying the wave operator to the equations defining $\phi^i$ and $\theta$ we can see that they also follow the wave equation. We find that this gauge fixing causes every field to satisfy the wave equation. Since the wave equation is the typical example of a well-posed second order differential equation we know that this setup of PDEs is also well-posed up to gauge.

It is not clear from the equations in \cref{eq:lin-maxwell-modified-phi-theta-definition,eq:lin-maxwell-modified-evolution-equations} that this is in fact a gauge fixing, the equations of motion have been changed, and it appears that there are solutions that do not satisfy the original equations of motion. To show that the evolution under these equations is equivalent to Maxwell's equations we first decompose into the 3+1 language. By separating the coordinates into $x^\mu = t,x^i$  we find that the equations of motion become
\begin{alignat}{2}
    \phi^i & = -i(\partial_t A^i - \partial^i A_0) - \epsilon^{ijk} \partial^j A^k, \quad && \theta = -\partial_t A_0 + \partial^i A^i \\ & \partial_t \phi^i = -i \epsilon^{ijk} \partial^j \phi^k - i \partial^i \theta, \quad && \partial_t \theta = i \partial^i \phi^i.
\end{alignat}
We can see that every equation has become an evolution equation for one of $A_0,A^i,\phi^i,\theta$. For consistency with the original system of equations we require that the Gauss constraint is satisfied at all times,
\begin{align}
    C_\phi = \partial^i \phi^i = 0.
\end{align}
A physical solution to the new system will satisfy this condition. Suppose then that we prepare the system at $t = 0$ such that it is a solution to the original Maxwell equations, that is 
\begin{align}
    \partial^i \phi^i |_{t= 0} = 0 = \partial_t \theta|_{t=0}, \quad \left( \partial_t \phi^i + i \epsilon^{ijk} \partial^j \phi^k \right)|_{t=0} = 0 = \partial^i \theta|_{t=0}.
\end{align}
This implies that $\partial_\mu \theta |_{t=0} = 0$ or that $\theta |_{t=0} = \kappa$ is a constant. As $\theta$ also satisfies the wave equation we can use the result that if $\theta = \kappa$ and $\partial_\mu \theta = 0$ at $t = 0$ then $\theta = \kappa$ for all $t$. Importantly this suggests that 
\begin{align}
    C_\phi = -i \partial_t \theta = 0
\end{align}
for the entire evolution. The time component of the connection, $A_0$, acts as a Lagrange multiplier imposing the Gauss constraint, as such we are free to fix it to any value we choose. The Lorenz gauge then gives an evolution equation for $A_0$ allowing it to change during evolution without disrupting the physical degrees of freedom. We then see that given we initialise the system such that the constraint is zero then the evolution of the system remains on the constraint surface and hence corresponds to a physical Maxwell solution. Another important question to ask is why one should leave $\theta$ in the equations rather than removing it if is zero anyway? To answer this we consider our intended application, numerical study, where at each time step one needs to require that $C_\phi = 0$ is satisfied up to a tolerance. For Maxwell's equations this constraint is relatively simple to compute, however, for gravity the constraints are more complicated nonlinear differential equations. Numerical errors cause the constraint to be violated and to enforce it at each time step, by solving the constraint for a solution $\phi^i$, can require a lot of computation. Instead, a standard solution is to evolve the system in such a way that the numerical solutions approaches the constraint surface (i.e.\ the surface on which the constraint vanishes). The gauge fixing we present here gives a method of constructing the highest order terms for such equations, by imposing the correct damping terms one can force the system to the constraint surface. Damping terms are modification to equations of motion that enforce a given particular constraint by forcing the solution to decay to the desired result. These have been historically used in numerical relativity~\cite{Einstein_s_Equa_Brodbe_1998}, hence we also consider them here. The inclusion of $\theta$ allows us to measure the amount of constraint violation and modify the equations accordingly such that it is then removed. More details on this are given in \cref{sec:num-asympytotically-constrained-system} and a numerical showcase for this gauge fixing can be found in \cref{sec:numerical-chiral-maxwell}.

\section{Frame Perturbation of \texorpdfstring{$\Sigma$}{Σ}}
We now consider the perturbation of \pleb{}'s formulation of gravity with respect to the frame and metric perturbations. To begin we consider derive a convenient parametrisation of the linearised self-dual 2-forms $\Sigma^i$ in terms of the perturbation of the frame. The easiest route to this is through its definition in terms of the frame \cref{eq:pleb-self-dual-2-forms-from-chiral-tetrad}, that is 
\begin{align}
    \Sigma^i_{\mu\nu} = \frac{1}{2} \Sigma^i_{IJ} e^I_\mu e^J_\nu.
\end{align}
To linearise we then fix a background frame $e^I_\mu$ and write the first-order perturbation as $\delta e^I_\mu$ where the perturbations are assumed to be $\ll 1$ and hence higher than linear terms are ignored. The background, $e^I_\mu$, is chosen to be that of Minkowski, and using the background we have a way to convert internal and spacetime indices. The linearisation of the metric is obtained from its definition in terms of the frame,
\begin{align}
    {(\delta g)}_{\mu\nu} & = \delta( \eta_{IJ} e^I_\mu e^J_\nu ) = 2 \eta_{IJ} \delta e^I_{(\mu} e^J_{\nu)} = 2 \delta e_{(\mu\nu)}
\end{align}
in the last line we have used the background to convert the internal index to a spacetime index, $\delta e_{\mu\nu} = e_{I\mu} \delta e^I_\nu$. We see that the metric perturbation is captured by the symmetric part of $\delta e_{\mu\nu}$, the remaining antisymmetric components make up the $6$ components of the perturbation of the Lorentz frame. The antisymmetric part is a 2-form $\delta e_{[\mu\nu]} \in \Lambda^2$ and as such we can decompose it into a self-dual and anti-self-dual basis
\begin{align}
    \delta e_{[\mu\nu]} = \frac{1}{2} h^i \Sigma^i_{\mu\nu} + \frac{1}{2} \bar{h}^i \asd^i_{\mu\nu}
\end{align}
where we see the $\mathfrak{so}(3,\C)$ and $\overline{\mathfrak{so}(3,\C)}$ infinitesimal gauge transformations appearing with gauge parameters $h^i$ and $\bar{h}^i$ respectively. We relabel the symmetric part of the perturbation as $h_{\mu\nu} = \delta e_{(\mu\nu)}$ which leaves us with the parametrisation 
\begin{align}
    \delta e_{\mu\nu} = h_{\mu\nu} + \frac{1}{2} h^i \Sigma^i_{\mu\nu} + \frac{1}{2} \bar{h}^i \asd^i_{\mu\nu}. \label{eq:lin-so13-frame-perturbation-parametrisation}
\end{align} 
In this way we have parametrised the perturbation of the frame into its irreducible components. With this parametrisation in mind we can compute the linearisation of the 2-forms $\Sigma^i$,
\begin{align}
    \sigma^i_{\mu\nu} = \frac{1}{2} \Sigma^i_{IJ} \delta( e^I_\mu e^J_\nu ) = \Sigma^i_{IJ} \delta e^I_{[\mu} e^J_{\nu]} = \delta e_{[\mu}{}^\rho \Sigma^i_{|\rho|\nu]}.
\end{align}
Then using \cref{eq:lin-so13-frame-perturbation-parametrisation} we find 
\begin{align}
    \sigma^i_{\mu\nu} = h_{[\mu}{}^\rho \Sigma^i_{|\rho|\nu]} + \frac{1}{2} \epsilon^{ijk} h^k \Sigma^j_{\mu\nu}  \label{eq:lin-full-frame-Sigma-perturbation}
\end{align}
We see that $\bar{h}^i$ does not enter the linearisation, this is because only one of the chiral halves is present in the theory and the other does not act on this space of 2-forms. The $\soC$ gauge parameter can be identified with $h^i$, comparing it with \cref{eq:pleb-infinitesimal-gauge-transformation} we see it acts as an infinitesimal gauge transformation. Note that the metric perturbations appear in the $(\E \otimes \Lambda^2)_1 \oplus (\E \otimes \Lambda^2)_9$ irreducible components and the gauge transformation in $(\E \otimes \Lambda^2)_3$. One can show that conformal transformations change only the trace of the metric perturbation and as such they perturb 2-forms in the space $(\E \otimes \Lambda^2)_1$. All other metric transformations cause an anti-self-dual perturbation, due to the realisation $(\E \otimes \Lambda^2)_9 = \Lambda^-$. This parametrisation for the perturbation is also a solution of the linearised metricity and reality conditions,
\begin{align}
    \Sigma^i \wedge \sigma^j \sim \delta^{ij}, \qquad \Sigma^i \wedge \bar{\sigma}^j + \sigma^i \wedge \asd^j = 0.
\end{align}
Where therefore have that the data $h_{\mu\nu}, h^i$ defines a perturbation of the \pleb{} 2-forms. Next we can calculate the linearisation of the connection
\begin{align}
    a^i = -i J_1^{-1}(\star d \sigma^i) \label{eq:lin-connection-perturbation-in-Sigma-perturbation}
\end{align}
where $J_1$ is the operator defined in \cref{eq:J1-operator-definition} and $\star$ is the Hodge star valued on the Minkowski background. We tackle the above computation by first computing
\begin{align}
    (\star d \sigma)^i_\mu = \epsilon_{\mu}{}^{\nu\rho\sigma} \partial_\nu h_{[\rho}{}^\alpha \Sigma^i_{|\alpha|\sigma]} = i \Sigma^i_\mu{}^\nu \partial_\nu h + i \Sigma^{i\rho\sigma} \partial_\rho h_{\mu\sigma} - i \Sigma^i_\mu{}^\rho \partial^\sigma h_{\rho\sigma} + i \epsilon^{ijk} \Sigma^j_\mu{}^\nu \partial_\nu h^k.
\end{align}
Applying $J_1$ to this we find 
\begin{align}
    J_1(\star d \sigma)^i_\mu = i \Sigma^i_\mu{}^\nu \partial_\nu h - i \Sigma^{\rho\sigma} \partial_\rho h_{\mu\sigma} - i \Sigma^i_\mu{}^\rho \partial^\sigma h_{\rho\sigma} + i \epsilon^{ijk} \Sigma^j_\mu{}^\nu \partial_\nu h^k + 2i\partial_\mu h^i.
\end{align}
Therefore, we have the perturbation of the connection 1-form in terms of the metric perturbation
\begin{align}
    a^i_\mu = -\frac{i}{2} \left( J_1 - 1 \right)(\star d \sigma)^i_\mu = - \Sigma^{i\rho\sigma} \partial_\rho h_{\mu\sigma} + \partial_\mu h^i. \label{eq:lin-connection-perturbation-in-h}
\end{align}
The gauge parameter appears as an infinitesimal gauge transformation on the connection. Immediately from this we can see that the linearisation of the curvature of the connection is 
\begin{align}
    \delta F^i_{\mu\nu} = - \Sigma^{i\rho\sigma} \partial_{\rho[\mu} h_{\nu]\sigma}.
\end{align}
Where we see that the gauge parameter does not enter the curvature as it is a gauge equivariant 2-form, this can also be argued as the choice of gauge should not affect the physical curvature. The Ricci tensor can then be computed as 
\begin{align}
    R_{\mu\nu} = \Sigma^i_\mu{}^\rho F^i_{\nu\rho} = -\partial^2 h_{\mu\nu} + 2 \partial_{(\mu} \left( \partial^\rho h_{\nu)\rho} - \frac{1}{2} \partial_{\nu)} h \right).
\end{align}
This reproduces the standard linearisation result found in the metric formulation, this is to be expected as \pleb{}'s formulation is only a different way of writing the same equations. To make this even clearer we can compute the degrees of freedom for Einstein's equation with $T_{\mu\nu} = 0 = \Lambda$, for which the second order linearisation is 
\begin{align}
    \partial^2 h_{\mu\nu} - 2 \partial_{(\mu} \left( \partial^\rho h_{\nu)\rho} - \frac{1}{2} \partial_{\nu)} h \right) = 0.
\end{align}
By performing a Fourier transformation, $\partial_\mu \rightarrow k_\mu$ where $k_\mu$ is the wave-covector, and moving to frequency space the above equation becomes algebraic instead of differential,
\begin{align}
    k^2 h_{\mu\nu} - 2 k_{(\mu} \left( k^\rho h_{\nu)\rho} - \frac{1}{2} k_{\nu)} h \right) = 0.
\end{align}
To solve this we split the solution into two cases $k^2 = 0$ and $k^2 \neq 0$. First we consider $k^2 \neq 0$ for which we have 
\begin{align}
    h_{\mu\nu} = 2 \frac{k_{(\mu}}{k^2} \left( k^\rho h_{\rho)\rho} - \frac{1}{2} k_{\nu)} h \right)
\end{align}
therefore it is of the form $h_{\mu\nu} = k_{(\mu} \xi_{\nu)}$. Recalling the linear action of diffeomorphism, $\delta_\xi g_{\mu\nu} = \partial_{(\mu} \xi_{\nu)}$ we see that this solution is not physical as it can be removed by performing a diffeomorphism transformation. These linear solutions are sums of gravitational waves with wave 4-vectors $k_\mu$, it is known that gravitational waves travel at the speed of light and hence $k^2 = 0$. Therefore, we expect that solutions with $k^2 \neq 0$ are purely gauge artefacts which corroborates the solution being a diffeomorphism transformation. Next, we consider $k^2 = 0$ for which we have 
\begin{align}
    k_{(\mu} \left( k^\rho h_{\nu)\rho} - \frac{1}{2} k_{\nu)} h \right) = 0
\end{align}
which implies 
\begin{align}
    k^\rho h_{\mu\rho} - \frac{1}{2} k_{\mu} h  = 0
\end{align}
We introduce a basis on $\Lambda^1$ adapted to $k^\mu$, that is $k_\mu, l_\nu, m^A_\mu$ where $A=2,3$. The basis is chosen the orthonormal in the sense that 
\begin{align}
    k^2 = 0, \quad k^\mu l_\mu = 1, \quad k^\mu m^A_\mu = l^\mu m^A_\mu = 0, \quad m^A_\mu m^{B\mu} = \delta^{AB}.
\end{align}
Using this basis we expand the metric perturbation 
\begin{align}
    h_{\mu\nu} = k_{(\mu} \xi_{\mu)} + a l_\mu l_\nu + b_A l_{(\mu} c^A_{\nu)} + C_{AB} m^A_{\mu} m^B_{\nu}
\end{align} 
where $\xi_\mu$ is an arbitrary 1-form and $C_{AB} = C_{(AB)}$. Substituting this in we find that 
\begin{align}
    k^\nu h_{\mu\nu} = \frac{1}{2} k_\mu k^\nu \xi_\nu + \frac{1}{2} l_\mu a + \frac{1}{2} b_A m^A_\mu = \frac{1}{2} k_\mu h
\end{align}
therefore $\xi_\mu = h l_\mu$, $a = b_A = 0$. Computing the trace we obtain $h = h + C_{AB} \delta^{AB}$ which shows that $C_{AB} = C_{\langle  AB \rangle}$ should be symmetric tracefree. The solution for the metric perturbation is then 
\begin{align}
    h_{\mu\nu} = k_{(\mu} l_{\nu)} h + C_{\langle  AB \rangle} m^A_\mu m^B_\nu.
\end{align}
The first term is a diffeomorphism, leaving us with the two components in the symmetric tracefree matrix $C_{\langle  AB \rangle}$. These are the two physical degrees of freedom in the metric perturbation, and they only appear in the symmetric tracefree part of the perturbation. This result is consistent with usual metric formulation. Returning to the parametrisation of the linearised 2-forms in terms of the metric we see that the physical degrees of freedom are found in the anti-self-dual part of the perturbation. This is one of our first hints that the conformal components in gravity should be treated separately from the remaining components.

\section{Hyperbolic First-Order Gauge Fixing}
    Here we consider the problem of gauge fixing the first-order Einstein's equations (EE) using the \pleb{} formulation. We have already seen that the second order EE, when written using the metric perturbation, is equivalent to the original metric formulation. It is known that by imposing the de Donder gauge for the linearised metric, EE becomes the wave equation at highest order. Solutions to the linearised EEs can then be constructed from plane wave solutions to the wave equation. In fact the nonlinear version of this gauge fixing has been used to prove that EE's can be put in a hyperbolic and well-posed form~\cite{Hyperb1+3YvonneTommaso1983}. Much like in the chiral Maxwell case we aim to find gauge fixing that causes all the fields to satisfy the wave equation, this is a different computation to that in the metric formulation as we also account for the $SO(3,\C)$ frame perturbation.

    \subsection{First-Order Linearised Einstein's Equation}
    We begin by constructing the linearised first-order version of Einstein's equations, the gauge fixing will then be added to this in order to obtain the hyperbolic set of equations. The perturbation is done around the Minkowski spacetime, this means the background 2-forms and metric are constant, and the background connections are zero 
    \begin{align}
        \partial_\mu \Sigma^i_{\rho\sigma} = 0 = \partial_\mu g_{\rho\sigma}, \quad A^i_\mu = 0.
    \end{align}
    In the previous section we saw that the conformal perturbation appears in the trace part of the metric perturbation, and that no physical dynamics are contained in there at the linear level. For this reason we consider separating out the trace of the metric perturbation,
    \begin{align}
        h_{\mu\nu} = \hat{h}_{\mu\nu} + \frac{1}{4} g_{\mu\nu} h, \qquad g^{\mu\nu} \hat{h}_{\mu\nu} = 0.
    \end{align}
    The prescription $\hat{h}_{\mu\nu}, h, h^i$ and $h_{\mu\nu}, h^i$ are equivalent, the perturbation of $\Sigma^i$ in this conformally separated parametrisation is 
    \begin{align} \label{eq:lin-2-form-pert-in-irreducibles-frame-comps}
        \sigma^i_{\mu\nu} = \frac{1}{4} h \Sigma^i_{\mu\nu} + \frac{1}{2} \epsilon^{ijk} h^k \Sigma^j_{\mu\nu} + \hat{h}_{[\mu}{}^\rho \Sigma^i_{|\rho|\nu]}.
    \end{align}
    Similarly, for the connection we can separate the conformal term
    \begin{align} \label{eq:lin-sd-connection-pert-in-irrep-frame-comps}
        a^i_\mu = \frac{1}{4} \Sigma^i_\mu{}^\nu \partial_\nu h + \partial_\mu h^i - \Sigma^{i\rho\sigma} \partial_\rho \hat{h}_{\mu\sigma}.
    \end{align}
    The first-order Einstein condition on the curvature of the connection, $\Sigma^i_\mu{}^\rho F^i_{\nu\rho} = 0$, linearises to become
    \begin{align}
        \Sigma^i_\mu{}^\rho \partial_\nu a^i_\rho - \Sigma^i_\mu{}^\rho \partial_\rho a^i_\nu = 0.
    \end{align}
    Collecting the results we have a first-order system of equations that imposes the linearised Einstein's equations 
    \begin{align}
        a^i_\mu = \frac{1}{4} \Sigma^i_\mu{}^\nu \partial_\nu h + \partial_\mu h^i - \Sigma^{i\rho\sigma} \partial_\rho \hat{h}_{\mu\sigma}. \label{eq:lin-connection-pert-in-h-and-h-hat} \\
        \Sigma^i_\mu{}^\rho \partial_\nu a^i_\rho - \Sigma^i_\mu{}^\rho \partial_\rho a^i_\nu = 0. \label{eq:lin-einstein-equations-on-a}
    \end{align}
    Repeating the previous calculation of substituting in the equation defining $a^i_\mu$ in terms of the frame perturbation, we find that in terms of the data $\hat{h}_{\mu\nu}, h, h^i$ the second-order Einstein's equations are
    \begin{align}
        \Sigma^i_\mu{}^\rho \partial_\nu a^i_\rho - \Sigma^i_\mu{}^\rho \partial_\rho a^i_\nu = -\frac{1}{4} g_{\mu\nu} \partial^\rho \partial_\rho h - \partial^\rho \partial_\rho \hat{h}_{\mu\nu} + \partial_{(\mu} \left( 2 \partial^\rho \hat{h}_{\nu)\rho} - \frac{1}{2} \partial_{\nu)} h \right). \label{eq:lin-einstein-condition-second-order-in-h}
    \end{align}
    The first two terms here are hyperbolic as they're proportional to the linear wave equation $\partial^\mu \partial_\mu \phi = 0$. The last term in the brackets is not hyperbolic, the strategy then becomes finding a first-order gauge fixing that removes this unwanted term but keeps the same physical solutions. It also should be noted that the perturbation of the $SO(3,\C)$ frame, $h^i$, is not present and is therefore not hyperbolic. The gauge fixing developed should also give $h^i$ a hyperbolic second-order equation of motion.

        \subsection{de Donder Gauge Fixing}
        In the usual metric formalism it is known that the linearised version of the Harmonic condition, also called the de Donder gauge, is imposed to make Einstein's equation hyperbolic. We impose the de Donder gauge in its first-order form, which fixes the linear diffeomorphism freedom, by introducing the 1-form
        \begin{align}
            \Lambda^1 \ni \xi_\mu = -2 \partial^\nu \hat{h}_{\mu\nu} + \frac{1}{2}\partial_\mu h.\label{eq:lin-de-Donder-connection-from-h}
        \end{align}
        This term has exactly the same form as the non-hyperbolic term in \cref{eq:lin-einstein-condition-second-order-in-h}. It follows that to remove it from Einstein's equations we can simply modify \cref{eq:lin-einstein-equations-on-a} to include
        \begin{align}
            \Sigma^i_\mu{}^\rho \partial_\nu a^i_\rho - \Sigma^i_\mu{}^\rho \partial_\rho a^i_\nu + \partial_{(\mu} \xi_{\nu)} = 0. \label{eq:lin-modified-einstein-on-a-and-xi}
        \end{align}
        We interpret $\xi_\mu$ as being 4 extra connection component, along with the original connection $a^i_\mu$. This is because connections are, in general, derivatives of the metric variables. In fact $\xi_\mu$ can be seen as the linearisation of the 4 components of the Levi-Civita connection $\Gamma^\mu = \Gamma^\mu_{\rho\sigma} g^{\rho\sigma}$, this will help motivate the nonlinear version of this gauge in later sections. Substituting $\xi_\mu$ and $a^i_\mu$ with their definitions in terms of the metric perturbations we find
        \begin{align}
            \Sigma^i_\mu{}^\rho \partial_\nu a^i_\rho - \Sigma^i_\mu{}^\rho \partial_\rho a^i_\nu + \partial_{(\mu} \xi_{\nu)} = -\frac{1}{4} g_{\mu\nu} \partial^\rho \partial_\rho h - \partial^\rho \partial_\rho \hat{h}_{\mu\nu} = - \partial^\rho \partial_\rho h_{\mu\nu} = 0
        \end{align}
        is then hyperbolic as intended. To recover the usual de Donder gauge we take $\xi_\mu = 0$ in which case \cref{eq:lin-de-Donder-connection-from-h} is the de Donder condition and \cref{eq:lin-modified-einstein-on-a-and-xi} loses its modification and returns to the usual linearised Einstein equations. The factor chosen in the modification in \cref{eq:lin-modified-einstein-on-a-and-xi} is fixed by the definition of the 1-form $\xi_\mu$, this part of the gauge fixing is fixed and there is no freedom if one wants to obtain the wave equation.

        \subsection{Modified Lorenz Gauge}
        There still remains the $SO(3,\C)$ freedom in the theory. A common choice, and the choice used in the previous section, is to remove this gauge explicitly by setting $h^i = 0$. Here, we leave it dynamical and impose an equation of motion on $a^i$ which propagates the frame with the wave equation. The gauge choice here must be $3$ conditions on the connection, a natural choice is then the Lorenz gauge
        \begin{align}
            \partial^\mu a^i_\mu = 0.
        \end{align}
        Substituting in \cref{eq:lin-connection-pert-in-h-and-h-hat} for $a^i$, we find 
        \begin{align}
            \partial^\mu \left( \frac{1}{4} \Sigma^i_\mu{}^\nu \partial_\nu h - \partial_\mu h^i - \Sigma^{i\rho\sigma} \partial_\rho \hat{h}_{\mu\sigma} \right) = - \partial^\mu \partial_\mu h^i -\Sigma^{i\rho\sigma} \partial_\rho \partial^\mu \hat{h}_{\mu\sigma} = 0
        \end{align}
        We see that the first term is hyperbolic in $h^i$, which is the term we wish to keep, however, the second term destroys the overall hyperbolicity. To fix this we notice that the second term in the above can be written using the new connection 1-form,
        \begin{align}
            -\Sigma^{i\rho\mu} \partial_\rho \partial^\nu \hat{h}_{\mu\nu} = \Sigma^{i\rho\mu} \partial_\rho \left( -\partial^\nu \hat{h}_{\mu\nu} - \frac{1}{4} \partial_\mu h \right) = \frac{1}{2} \Sigma^{i\rho\mu} \partial_\rho \xi_\nu.
        \end{align}
        The term involving $h$ is identically zero, therefore we can change its factor from $-\frac{1}{4}$ to $+\frac{1}{2}$ so that the definition of $\xi_\mu$ appears, and realising the last equality. Clearly this produces the following modification to the Lorenz gauge
        \begin{align}
            \partial^\mu a^i_\mu - \frac{1}{2} \Sigma^{i\mu\nu} \partial_\mu \xi_\nu = -2 \partial^\mu \partial_\mu h^i = 0.
        \end{align} 
        We call this the modified Lorenz gauge and its second order equation is exactly the wave equation on the perturbation of the $SO(3,\C)$ frame, $h^i$.

        \subsection{Modified de Donder and Connections}
        The previous two gauge fixings for the diffeomorphism and $SO(3,\C)$ gauges are enough to ensure hyperbolicity for the frame sector. However, the number of components in the frame and connection sectors do not match. There are $4+12=16$ connection fields and $1+3+9=13$ in the self-dual frame. We treat real and complex components equally in this context as it is purely the number of fields that determines the structure, another way of thinking about it is that $\R \subset \C$ and hence all fields can be taken as $\C$ valued (given some extra reality constraints). When the number of fields matches in each sector this allows one to interpret the first-order equation of motion as a Dirac operator and its adjoint (with respect to some norm) acting on these fields. As we will see in the nonlinear story adding the missing $3$ fields allows for the Gauss constraint to propagate which directly effects hyperbolicity.
        Therefore, we introduce another internal vector $\chi^i \in \E$, this is taken to be part of the frame sector making $16$ fields in total. This allows us to modify the equations defining the connections, \cref{eq:lin-connection-pert-in-h-and-h-hat,eq:lin-de-Donder-connection-from-h}, like so
        \begin{align}
            a^i_\mu &= \frac{1}{4} \Sigma^i_\mu{}^\nu \partial_\nu h - \partial_\mu h^i - \Sigma^{i\rho\sigma} \partial_\rho \hat{h}_{\mu\sigma} + c_1 \partial_\mu \chi^i + c_2 \epsilon^{ijk} \Sigma^j_\mu{}^\nu \partial_\nu \chi^k \\
            \xi_\mu &= -2 \partial^\nu \hat{h}_{\mu\nu} + \frac{1}{2}\partial_\mu h + c_3 \Sigma^i_\mu{}^\nu \partial_\nu \chi^i
        \end{align}
        where $c_1,c_2,c_3$ are constants. These are all the extra terms one can construct using derivatives of $\chi^i$, this can be explicitly checked using representation theory, which we will not spell out here. Similar to before we wish for Einstein's equations to be hyperbolic. Since the linearised equations are linear in their arguments and the modifications to the connection are linear, to compute the hyperbolicity of the new fields we only need to consider
        \begin{align}
            a^i_\mu &= c_1 \partial_\mu \chi^i + c_2 \epsilon^{ijk} \Sigma^j_\mu{}^\nu \partial_\nu \chi^k \\ \xi_\mu &= c_3 \Sigma^i_\mu{}^\nu \partial_\nu \chi^i.
        \end{align}
        Substituting these into the modified Einstein's equations we find 
        \begin{align}
            \Sigma^i_\mu{}^\rho \partial_\nu a^i_\rho - \Sigma^i_\mu{}^\rho \partial_\rho a^i_\nu +& \partial_{(\mu} \xi_{\nu)} = \Sigma^i_\mu{}^\rho \partial_\nu \left( c_1 \partial_\rho \chi^i + c_2 \epsilon^{ijk} \Sigma^j_\rho{}^\sigma \partial_\sigma \chi^k \right) + c_3 \partial_{(\mu} \left( \Sigma^i_{\nu)}{}^\rho \partial_\rho \chi^i \right) \nonumber \\ & \hspace{100pt} - \Sigma^i_\mu{}^\rho \partial_\rho \left( c_1 \partial_\nu \chi^i + c_2 \epsilon^{ijk} \Sigma^j_\nu{}^\sigma \partial_\sigma \chi^k \right)  \nonumber\\ & =  c_2 \Sigma^i_{\mu\nu} \partial^\rho \partial_\rho \chi^i + c_2 \Sigma^i_\nu{}^\rho \partial_\mu \partial_\rho \chi^i - (c_1 + c_2) \Sigma^i_\mu{}^\rho \partial_\nu \partial_\rho \chi^i \nonumber \\ & \hspace{100pt} + (c_1 + 2 c_2) \Sigma^i_\mu{}^\rho \partial_\nu \partial_\rho \chi^i + c_3 \partial_{(\mu} \left( \Sigma^i_{\nu)}{}^\rho \partial_\rho \chi^i \right) \nonumber\\ &= c_2 \Sigma^i_{\mu\nu} \partial^\rho \partial_\rho \chi^i + (c_3 + 2 c_2) \Sigma^i_{(\mu}{}^\rho \partial_{\nu)} \partial_\rho \chi^i.
        \end{align}
        It is then clear that $c_3 = -2 c_2$ will cause Einstein's equations to be hyperbolic in $\chi^i$.
        We also have to check that this modification keeps the hyperbolicity of the Lorenz gauge,
        \begin{align}
            \partial^\mu a^i_\mu - \frac{1}{2} \Sigma^{i\mu\nu} \partial_\mu \xi_\nu = c_1 \partial^\mu \partial_\mu \chi^i + c_2 \epsilon^{ijk} \Sigma^{j\mu\nu} \partial_\mu \partial_\nu \chi^k - \frac{c_3}{2} \Sigma^{i\mu\nu} \partial_\mu \left(\Sigma^j_\nu{}^\rho \partial_\nu \chi^j \right) \\ = \left(  c_1 + \frac{c_3}{2} \right) \partial^\mu \partial_\mu \chi^i = ( c_1 - c_2) \partial^\mu \partial_\mu \chi^i.
        \end{align}
        No new conditions are imposed by the requirement that the Lorenz gauge implies second-order hyperbolic equations. This results in a 2 parameter family of hyperbolic gauge fixing for linearised \pleb{}, assuming that both $c_1,c_2$ are not zero at the same time which would remove the gauge fixing. If $c_1 = c_2$ then it appears that the equation of motion for $\chi^i$ is removed, however, it is still provided a dynamical equation of motion through the modification to the Einstein condition.

        \subsection{Linearised Gauge Fixed \pleb{}}\label{subsec:lin-gauge-fixing-2-param-family}
        Starting with the linearised self-dual frame $(h,h^i,\hat{h}_{\mu\nu})$ we add to this sector an internal vector $\chi^i$. Then we define the modified linearised self-dual connection $a^i_\mu$ along with another connection 1-form $\xi_\mu$ as derivatives of the frame and internal vector
        \begin{align}
            a^i_\mu &= \frac{1}{4} \Sigma^i_\mu{}^\nu \partial_\nu h - \partial_\mu h^i - \Sigma^{i\rho\sigma} \partial_\rho \hat{h}_{\mu\sigma} + c_1 \partial_\mu \chi^i + c_2 \epsilon^{ijk} \Sigma^j_\mu{}^\nu \partial_\nu \chi^k \label{eq:lin-modified-12-connection-definition}\\
            \xi_\mu &= -2 \partial^\nu \hat{h}_{\mu\nu} + \frac{1}{2}\partial_\mu h - 2 c_2 \Sigma^i_\mu{}^\nu \partial_\nu \chi^i \label{eq:lin-modified-4-connection-definition}.
        \end{align}
        Such that the modified linearised Einstein equation and Lorenz condition form a hyperbolic system of equations. The modified Einstein's equations and Lorenz condition are 
        \begin{align}
            \Sigma^i_\mu{}^\rho \partial_\nu a^i_\rho - \Sigma^i_\mu{}^\rho \partial_\rho a^i_\nu + \partial_{(\mu} \xi_{\nu)} = 0 \label{eq:lin-modified-einstein-condition}\\
            \partial^\mu a^i_\mu - \frac{1}{2} \Sigma^{i\mu\nu} \partial_\mu \xi_\nu = 0.\label{eq:lin-modified-lorenz-condition}
        \end{align}
        On the removal of the gauge variables $\xi_\mu = \chi^i = 0$ this returns the usual Einstein's equations on a linearised metric $(\delta g)_{\mu\nu} = 2 \hat{h}_{\mu\nu} + \frac{1}{2} g_{\mu\nu} h$. The second order equations imposed by the above conditions are 
        \begin{align}
            - \frac{1}{4} g_{\mu\nu} \partial^\rho \partial_\rho h - \partial^\rho \partial_\rho \hat{h}_{\mu\nu} + c_2 \Sigma^i_{\mu\nu} \partial^\rho \partial_\rho \chi^i &= 0 \\
            -\partial^\mu \partial_\mu h^i + (c_1 - c_2) \partial^\mu \partial_\mu \chi^i &= 0.
        \end{align}
        We see the highest order symbols are the wave operators $\partial^\mu \partial_\mu$ for all the components. The equations (\ref{eq:lin-modified-12-connection-definition},~\ref{eq:lin-modified-4-connection-definition}) and (\ref{eq:lin-modified-einstein-condition},~\ref{eq:lin-modified-lorenz-condition}) form our linearised gauged fixed \pleb{} system.

        \subsection{Linear Separation}
        There is a coincidence, in \pleb{}'s formalism, that allows the gauge fixed Einstein's equation to be separated into two sectors. This is possible because of maps between $\Lambda^1$ and $\E \times \Lambda^1$ which are the projection and embedding of $(\C^3 \otimes \Lambda^1)_4$ into and out of $\C^3 \otimes \Lambda^1$ detailed in \cref{eq:pleb-4-embedding-into-E-1-forms,eq:pleb-4-projection-of-E-1-forms}. To show this separation we first introduce new connection fields as linear combinations of the old connections
        \begin{align}
            \Omega^i_\mu &= a^i_\mu - \frac{1}{2} \Sigma^i_\mu{}^\nu \xi_\nu \label{eq:lin-12-connection-in-a-xi}\\
            \omega_\mu &= \xi_\mu + 2 \Sigma^i_\mu{}^\nu a^i_\nu \label{eq:lin-4-connection-in-a-xi}
        \end{align}
        along with their inverse 
        \begin{align}
            a^i_\mu &= \frac{1}{2}\Omega^i_\mu - \frac{1}{2}\epsilon^{ijk} \Sigma^j_\mu{}^\nu \Omega^k_\nu - \frac{1}{4} \Sigma^i_\mu{}^\nu \omega_\nu \label{eq:lin-12-component-inverse-transformation}\\
            \xi_\mu &= \Sigma^i_\mu{}^\nu \Omega^i_\nu - \frac{1}{2}\omega_\mu \label{eq:lin-4-component-inverse-transformation}.
        \end{align}
        For our new connection fields we can look at their implied definitions, through $a^i_\mu$ and $\xi_\mu$, in terms of derivatives of the frame perturbation
        \begin{align}
            \Omega^i_\mu &= -\partial_\mu \left(  h^i - (c_1-c_2) \chi^i \right) + \Sigma^i_\mu{}^\rho \partial^\sigma \hat{h}_{\rho\sigma} - \Sigma^{i\rho\sigma} \partial_\rho \hat{h}_{\mu\sigma} \\  
            \omega_\mu &= - \partial_\mu h -2 \Sigma^i_\mu{}^\nu \partial_\nu \left( h^i - (c_1 + c_2) \chi^i \right).
        \end{align}
        If we introduce two linear combinations of the frame
        \begin{align}
        \hat{h}^i    &= h^i - (c_1 - c_2) \chi^i \label{eq:lin-hhat-i-new-linear-combiation}\\
        \hat{\chi}^i &= h^i - (c_1 + c_2) \chi^i \label{eq:lin-chihat-i-new-linear-combiation}
        \end{align}
        we find the connections become
        \begin{align}
            \Omega^i_\mu &= -\partial_\mu \hat{h}^i + \Sigma^i_\mu{}^\rho \partial^\sigma \hat{h}_{\rho\sigma} - \Sigma^{i\rho\sigma} \partial_\rho \hat{h}_{\mu\sigma} \\
            \omega_\mu &= -\partial_\mu h - \Sigma^i_\mu{}^\nu \partial_\nu \hat{\chi}^i.
        \end{align}
        Note that $\Omega^i_\mu$ only depends on $\hat{h}^i, \hat{h}_{\mu\nu}$ and $\omega_\mu$ only depends on $h,\hat{\chi}^i$. The dimension of each sector is $12$ and $4$. So far this split holds at the level of the connection definitions, but as we will see it extends to the first-order connection equations of motion. To check this we can substitute \cref{eq:lin-12-component-inverse-transformation,eq:lin-4-component-inverse-transformation} into \cref{eq:lin-modified-einstein-condition,eq:lin-modified-lorenz-condition}.
        In doing so we find 
        \begin{align}
            \Sigma^i_\mu{}^\rho F^i_{\nu\rho} + \partial_{(\mu} \xi_{\nu)} &= \frac{1}{4}g_{\mu\nu} \partial^\rho \omega_\rho -\frac{1}{4} \Sigma^i_{\mu\nu} \left( \Sigma^{i\rho\sigma} \partial_\rho \omega_\sigma + 2 \partial^\rho \Omega^i_\rho \right) \nonumber \\ & \hspace{100pt} + \Sigma^i_{\langle  \mu}{}^\rho \partial_{\nu \rangle} \Omega^i_\rho - \Sigma^i_{\langle  \mu|}{}^\rho \partial_{\rho} \Omega^i_{|\nu \rangle} = 0 \label{eq:lin-modified-Ricci-in-omega-Omega} \\
            \partial^\mu a^i_\mu - \frac{1}{2}\Sigma^{i\mu\nu} \partial_\mu \xi_\nu &= \partial^\mu \Omega^i_\mu = 0
        \end{align}
        where $T_{\langle  \mu\nu \rangle} = \frac{1}{2}\left( T_{\mu\nu} + T_{\nu\mu} - \frac{1}{2} g_{\mu\nu} g^{\rho\sigma} T_{\rho\sigma} \right)$ is the symmetric tracefree part. Clearly, the second equation is already separated as it only depends on $\Omega^i_\mu$. Now we can use the decomposition $\Lambda^1 \otimes \Lambda^1 = C^\infty(M) \oplus \Lambda^1 \otimes_{stf} \Lambda^1 \oplus \E \oplus \overline{\E}$ and compute the irreducible parts through 
        \begin{align}
            T_{\mu\nu} g^{\mu\nu} \in C^\infty(M), \quad T_{\langle \mu \nu \rangle} \in \textrm{Sym}_0^2(\Lambda^1), \quad T_{\mu\nu} \Sigma^{i\mu\nu} \in \E, \quad T_{\mu\nu} \asd^{i\mu\nu} \in \overline{\E}.
        \end{align}
        Where $\textrm{Sym}_0^2(\Lambda^1)$ is the space of symmetric tracefree tensors. For which we find the irreducible components of \cref{eq:lin-modified-Ricci-in-omega-Omega} are
        \begin{align}
            \partial^\mu \omega_\mu = 0, \quad \Sigma^{i\mu\nu} \partial_\mu \omega_\nu + 2 \partial^\mu \Omega^i_\mu = 0, \quad \Sigma^i_{\langle  \mu}{}^\rho \partial_{\nu \rangle} \Omega^i_\rho - \Sigma^i_{\langle  \mu|}{}^\rho \partial_{\rho} \Omega^i_{|\nu \rangle} = 0.
        \end{align}
        The trace and symmetric tracefree component spaces are clearly only depend on one of the two connection sectors, as for the self-dual component by taking the appropriate linear combination with modified Lorenz condition we can see that
        \begin{align}
            \partial^\mu \Omega^i_\mu = 0, \quad \Sigma^{i\mu\nu} \partial_\mu \omega_\nu = 0
        \end{align}
        are separate conditions. Therefore, the linear Einstein's equations and modified Lorenz condition together are equivalent to the separate system of equations,
        \begin{gather}
            \omega_\mu = -\partial_\mu h - \Sigma^i_\mu{}^\nu \partial_\nu \hat{\chi}^i, \quad \partial^\mu \omega_\mu = 0, \quad \Sigma^{i\mu\nu} \partial_\mu \omega_\nu = 0 \label{eq:lin-4-full-gauge-fixed-system} \\[10pt]
            \Omega^i_\mu = -\partial_\mu \hat{h}^i + \Sigma^i_\mu{}^\rho \partial^\sigma \hat{h}_{\rho\sigma} - \Sigma^{i\rho\sigma} \partial_\rho \hat{h}_{\mu\sigma}, \quad \partial^\mu \Omega^i_\mu = 0, \quad \Sigma^i_{\langle  \mu}{}^\rho \partial_{\nu \rangle} \Omega^i_\rho - \Sigma^i_{\langle  \mu|}{}^\rho \partial_{\rho} \Omega^i_{|\nu \rangle} = 0 \label{eq:lin-12-full-gauge-fixed-system}.
        \end{gather}
        This separation is quite remarkable, it is the two parameter generalisation of the separation appearing in~\cite{ChiralPerturbaKrasno2020,Plebanski_compl_Krasno_2025} and \cref{chap:Plebanski-Elliptic-Complex}. For example the splitting in~\cite{Plebanski_compl_Krasno_2025} and \cref{chap:Plebanski-Elliptic-Complex} is obtained when taking $c_1 = -\frac{1}{\sqrt{2}}$ and $c_2 = 0$, then after rescaling some fields we find the maps are equivalent. The equivalence to~\cite{ChiralPerturbaKrasno2020} is less obvious as one would need to take care when converting from spinor indices to spacetime indices. From the metric perturbation analysis earlier in this section we found that the physical degrees of freedom are in the tracefree part of the metric perturbation $\hat{h}_{\mu\nu}$, this only appears in the 12 component sector \cref{eq:lin-12-full-gauge-fixed-system}. One can then ignore the 4 component system in \cref{eq:lin-4-full-gauge-fixed-system} as they do not propagate physical degrees of freedom. This offers a more efficient route to imposing linearised Einstein's equations.

    \subsection{Separated Lagrangian}
        We note that the equations of motion for the separated system can each be obtained from a Lagrangian
        \begin{align}
            L_4 &= h \partial^\mu \omega_\mu - \hat{\chi}^i \Sigma^{i\mu\nu} \partial_\mu \omega_\nu  - \frac{1}{2} \omega^\mu \omega_\mu \label{eq:lin-4-action} \\ L_{12} &= \hat{h}^i \partial^\mu \Omega^i_\mu - 2 \hat{h}^{[\mu}{}_\rho \Sigma^{i\nu]\rho} \partial_\mu \Omega^i_\nu - \frac{1}{2} \Omega^{i\mu} \Omega^i_\mu. \label{eq:lin-12-action}
        \end{align}
        By varying with respect to each field we obtain the equations \cref{eq:lin-4-full-gauge-fixed-system,eq:lin-12-full-gauge-fixed-system}.
        There is a 1-parameter family of combined Lagrangians that describes both the 4 and 12, however, the combination that is proportional to linearised \pleb{} is given by 
        \begin{align}
            L_{pleb} = \frac{1}{4} L_4 - L_{12}.
        \end{align}
        Using the formulas for $L_4$ and $L_{12}$ and taking the transformation back to the original variables we find the total Lagrangian becomes 
        \begin{align}
            L_{pleb} = &  \left( \frac{1}{4} h \Sigma^{i\mu\nu} + \epsilon^{ijk} h^j \Sigma^{k\mu\nu} - 2 \hat{h}^{[\mu|}{}_\rho \Sigma^{i\rho|\nu]} \right) \partial_\mu a^i_\nu + \frac{1}{2} \epsilon^{ijk} \Sigma^{i\mu\nu} a^j_\mu a^k_\nu\nonumber\\
            & -\left(\hat{h}^{\mu\nu} - \frac{1}{4} g^{\mu\nu} h\right) \partial_\mu \xi_\nu - 2 c_2 \chi^i  \partial^\mu a^i_\mu - (c_1 + c_2) \chi^i \epsilon^{ijk} \Sigma^{j\mu\nu} \partial_\mu a^k_\nu \nonumber \\ & \hspace{200pt} + c_2 \chi^i \Sigma^{i\mu\nu} \partial_\mu \xi_\nu + \frac{1}{4} \xi^\mu \xi_\mu. \label{eq:lin-gauge-fixed-plebanski-action-in-original-variables}
        \end{align}
        The first line in the above is the linearisation of the \pleb{} Lagrangian, $L = \Sigma^i \wedge F^i$ and the second is the Lagrangian corresponding to the modified gauge choices. Again we see the 2 parameter family of gauge choices that produces the separation property as well as the hyperbolicity. The action in~\cite{Plebanski_compl_Krasno_2025} is also equivalent given the correct choice of $c_1$ and $c_2$. The reason any linear combination of $L_4$ and $L_{12}$ can be chosen is because $L_4$ does not propagate any physical degrees of freedom therefore the action can always be written as $L = L_{pleb} + \alpha L_4$ and as the physical theory is independent of $L_4$ the choice of $\alpha$ is irrelevant.

    \section{Dirac Operators}\label{subsec:pleb-dirac-operators}
        Since the linearised Einstein's equation are first-order and their square produces the wave operator, it should be possible to write this system as two Dirac operators on the frame and connection fields. First, we recall some basic factors about the Dirac-K\"ahler operator on differential forms. The Dirac-K\"ahler operator is a formal square root of the wave operator on the space of differential forms, and is given by
        \begin{align}
            D = -(d + d^*), \quad D : \Lambda^n \rightarrow \Lambda^{n-1} \oplus \Lambda^{n+1}.
        \end{align}
        The operators $d^*$ is the codifferential and its action is defined as
        \begin{align}
            (d^* \theta)_{\mu_1...\mu_n} = - \partial^\mu \theta_{\mu \mu_1...\mu_n} \in \Lambda^n, \quad \theta \in \Lambda^{n+1}.
        \end{align}
        It can then be shown that the Dirac squares to the wave operator $D^2 = d d^* + d^* d = \Box$. This construction is due to a correspondence between spinors and differential forms~\cite{AIHPA_1978__29_1_85_0}. The actions \cref{eq:lin-4-action,eq:lin-12-action} can then be expressed using differential forms.

        \subsection{4 Component Dirac Operator}
        For the 4 component sector we only need to identify each field with an appropriate differential form, a natural choice is
        \begin{align}
            h \in \Lambda^0, \quad \omega \in \Lambda^1, \quad \hat{\chi} = \hat{\chi}^i \Sigma^i \in \Lambda^+.
        \end{align}
        Since the Dirac-K\"ahler operator swaps between odd and even polyforms is makes sense to introduce the odd and even combinations of these fields,
        \begin{align}
            H = h + \hat{\chi} \in \Lambda^0 \oplus \Lambda^+, \qquad \omega \in \Lambda^1.
        \end{align}
        The equations of motion in \cref{eq:lin-4-full-gauge-fixed-system} are then 
        \begin{align}
            \omega = D|_{\Lambda^1} H, \quad D_+ \omega = 0
        \end{align}
        where $D|_{\Lambda^1}$ is $D$ followed by the projection onto $\Lambda^1$ and $D_+$ is $D$ followed by the projection of $\Lambda^2 \rightarrow \Lambda^+$. Despite the projections we find that the composition of the two still gives the wave operator. Explicitly the individual operators are 
        \begin{align}
            D|_{\Lambda^1} H = dh + d^* \hat{\chi}, \qquad D_+ \omega = \frac{1}{2}(1-i \star) d\omega + d^* \omega
        \end{align}
        where $P_+ = \frac{1}{2}(1-i \star) : \Lambda^2 \rightarrow \Lambda^+$ is the projector onto self-dual 2-forms. To recover the third equation in \cref{eq:lin-4-full-gauge-fixed-system} from the above system one can make use of the definition $\Sigma^i_{\mu\nu} \Sigma^{i\rho\sigma} \sim (P_+)_{\mu\nu}{}^{\rho\sigma}$. The composition of the two operators is then
        \begin{align}
            D_+ D|_{\Lambda^1} H = \frac{1}{2}(1-i\star)d(dh + d^* \hat{\chi}) + d^*(dh + d^* \hat{\chi}) = \frac{1}{2}(1-i\star)d d^* \hat{\chi} + d^* d h
        \end{align}
        We can then make use of the definition of the codifferential and self-duality to show 
        \begin{align}
            \star d d^* \hat{\chi} = \star d \star d \star \hat{\chi} = i d^* d \hat{\chi}
        \end{align}
        such that the result of the composition is 
        \begin{align}
            D_+ D|_{\Lambda^1} H = \frac{1}{2}(d d^* + d^* d) \hat{\chi} + d^*  d h = \frac{1}{2} \Box \hat{\chi} + \Box h.
        \end{align}
        From which we can easily recover the result that the second order equations on the frame are wave operators
        \begin{align}
            D_+ D|_{\Lambda^1} (h+\hat{\chi}) = 0, \quad \Rightarrow \quad \Box h = 0, \quad \Box \hat{\chi} = 0.
        \end{align}
        This sector is related to the truncated de Rham complex 
        \begin{align}
            0 \longrightarrow \Lambda^0 \overset{d}{\longrightarrow} \Lambda^1 \overset{d_+}{\longrightarrow} \Lambda^+ \longrightarrow 0
        \end{align}
        where $d_+ = P_+ \circ d$. This also plays an important role in chiral Yang-Mills theories~\cite{AtiyahHitchinSingerSelfDual1978}. Each space has an inner product defined through metric pairing, the Dirac-K\"ahler operator is then obtained by combining the adjoint operators into the maps 
        \begin{equation}
            \begin{tikzcd}
                \Lambda^0 \oplus \Lambda^+ \arrow[rr, "D|_{\Lambda^1} = d+d^*", shift left=2] &  & \Lambda^1 \arrow[ll, "D_+ =d_+ + d^*", shift left]
            \end{tikzcd}.
        \end{equation}
        Part of Einstein's equations are then equivalent to the vanishing of the composition of the top arrow followed by the bottom one.

        \subsection{12 Component Dirac Operator}
        For the 12 component sector some more work is needed in order to represent the fields as differential forms. First we tackle $\hat{h}_{\mu\nu}$ and make use of \cref{col:pleb-E-Lambda2-9-is-ASD-2-forms} to map $Sym^2_0(\Lambda^1)$ into $\Lambda^-$,
        \begin{align}
            H^i_{\mu\nu} = 2\hat{h}_{[\mu}{}^\rho \Sigma^i_{\nu]\rho} \in \E \otimes \Lambda^-
        \end{align}
        Under this map the first equation in (\ref{eq:lin-12-full-gauge-fixed-system}) becomes
        \begin{align}
            \Omega^i_\mu = - \partial_\mu \hat{h}^i + \partial^\nu H^i_{\nu\mu}.
        \end{align}
        For the last equation in (\ref{eq:lin-12-full-gauge-fixed-system}) we use the map between $Sym^2_0(\Lambda^1)$ and $3\times 3$ Hermitian matrices, \cref{eq:pleb-9-dim-basis-of-symmetric-tracefree-tensors} to find 
        \begin{align}
            \Sigma^k_\mu{}^\rho \partial_{[\nu} \Omega^k_{\rho]} \Sigma^{i\mu\sigma} \asd^{j\nu}{}_\sigma = (\delta^{ik} g^{\rho\sigma} - \epsilon^{ikl} \Sigma^{l\rho\sigma}) \partial_{[\nu} \Omega^k_{\rho]} \asd^{j\nu}{}_\sigma = \asd^{j\mu\nu} \partial_\mu \Omega^i_\nu = 0.
        \end{align}
        The above condition can then be interpreted as the anti-self-dual projection of $d\Omega^i$.
        Now the fields all have a differential form interpretation,
        \begin{align}
            \hat{h}^i \in \E \otimes \Lambda^0, \quad \Omega^i \in \E \otimes \Lambda^1, \quad H^i \in \E \otimes \Lambda^-.
        \end{align}
        Using this representation the equations of motion in \cref{eq:lin-12-full-gauge-fixed-system} become
        \begin{align}
            \Omega^i = D|_{\Lambda^1} (\hat{h}^i + H^i), \quad D_- \Omega^i = 0.
        \end{align}
        Where, similar to the 4 component case, $D_-$ is the usual Dirac-K\"ahler operator followed by projection onto $\Lambda^-$. The composition of the two operators can be computed similarly to before and results in
        \begin{align}
            D_- D|_{\Lambda^1} (\hat{h}^i + H^i) = 0, \quad \Rightarrow \quad \Box \hat{h}^i = 0, \quad \Box H^i = 0.
        \end{align}
        This sector can be placed into another truncated de Rham complex,
        \begin{align}
            0 \longrightarrow \E \otimes \Lambda^0 \overset{d}{\longrightarrow} \E \otimes \Lambda^1 \overset{d_-}{\longrightarrow} \E \otimes \Lambda^- \longrightarrow 0
        \end{align}
        this time all the fields are multiplied with $\C^3$ and the anti-self-dual 2-forms are present instead of the self-dual. As before the Dirac-K\"ahler operators are recovered using the metric inner product and constructing the following operators
        \begin{equation}
            \begin{tikzcd}
                \E \times \Lambda^0 \oplus \E \times \Lambda^- \arrow[rr, "D|_{\Lambda^1} = d+d^*", shift left=2] &  &  \E \times  \Lambda^1 \arrow[ll, "D_- =d_- + d^*", shift left]
            \end{tikzcd}.
        \end{equation}
        Their composition leads to the remaining part of Einstein's equations. We see that by parametrising the fields correctly, gauge fixed linearised gravity, through the lens of \pleb{}'s formulation, has a relatively simple interpretation using Dirac-K\"ahler operators. This is due to fact that linearised gravity around Minkowski can be thought of as a collection of truncated de Rham complexes and the gauge fixing that occurs here becomes the adjoint of the maps on the spaces of the elliptic operator. This leads to the work in the next chapter on further realisations of gravity as an elliptic complex.

    \newpage
    \chapter{Linear \pleb{} Elliptic Complex}\label{chap:Plebanski-Elliptic-Complex}
In this chapter we explore a mathematical result available through the use of self-dual 2-form formulation of differential geometry in dimension 4. Yang-Mills (YM) instantons are important equations in both physical and mathematical literature, their solutions corresponding to physical configurations of gauge fields. The YM instanton equation is 
\begin{align}
    F(A) = - \star F(A), \quad \textrm{or} \quad F_- = 0 \label{eq:ell-yang-mills-instanton-condition}
\end{align}
where $F(A)$ is the curvature of a connection $A$ and $F_-$ is its projection onto $\Lambda^-$. Instantons are critical points of the YM's action in dimension 4. The YM action, when evaluated on an instanton, is a topological invariant on the principal bundle that the connection appears on. To study the solution space of the nonlinear instanton equation one linearises around an instanton background. When also introducing the correct gauge fixing these equations form an elliptic complex. Using a result from Atiyah-Singer~\cite{The_Index_of_El_Atiyah_1968} one can relate the dimension of the space of solutions to topological data. Then through the pioneering work by Atiyah, Hitchin and Singer~\cite{AtiyahHitchinSingerSelfDual1978}, Donaldson~\cite{DonaldsonSelfDual1983} was able to obtain novel results on differential structures in dimension 4.

In this spirit we aim to enable a similar story in the case of self-dual instantons for gravity. Instantons, in the case of self-dual 2-forms, only exists in Euclidean or split signature. In any signature we have the decomposition of the self-dual curvature into self-dual and anti-self-dual 2-forms,
\begin{align}
    F^i = \Psi^{ij} \Sigma^j + \frac{R}{36} \Sigma^i + R^{ij} \asd^j.
\end{align}  
Where $\Psi^{ij}$ is the self-dual Weyl curvature, $R^{ij}$ and $R$ are the tracefree Ricci tensor and Ricci scalar. Einstein's equations, in a vacuum and with zero cosmological constant, imply,
\begin{align}
    F^i = \Psi^{ij} \Sigma^j
\end{align}
such that $R^{ij} = 0 = R$. That is the Ricci tensor vanishes. Instanton solutions to Einstein's equations are then characterised by 
\begin{align}
    \Psi^{ij} = 0, \quad \textrm{or} \quad F^i = 0.
\end{align}
In Lorentzian signature we have that $-\overline{\Psi^{ij}}^* = \Psi^{ij}$ and so the instanton condition implies that the entire of the Riemann curvature vanishes which corresponds to Minkowski space. No interesting conditions are imposed in this case. We therefore choose to focus on Euclidean signature instantons. It should be noted that a certain subset of solutions to the Euclidean instantons will correspond to Lorentzian solutions analytically continued into Euclidean ones, therefore the Euclidean case may remain of interest in physics. This chapter is heavily based on the author's paper~\cite{Plebanski_compl_Krasno_2025}.

\section{Elliptic Complex}

\begin{definition}~\label{def:ell-elliptic-complex}
    Let $E_0,E_1,\ldots,E_k$ be a sequence of vector bundles over a manifold $M$. An elliptic complex is the given by the corresponding sequence of operators $d_i$, 
    \begin{align}
        0 \longrightarrow E_0 \overset{d_1}{\longrightarrow} E_1 \overset{d_2}{\longrightarrow} \ldots \overset{d_k}{\longrightarrow} E_k \longrightarrow 0.
    \end{align}
    Such that the operators satisfy 
    \begin{enumerate}
        \item $d_{i+1} \circ d_i = 0$, which implies that $\textrm{Im}(d_i) \subseteq \textrm{Ker}(d_{i+1})$. 
        \item The associated symbols $\sigma(d_i)$ are exact, that is $\textrm{Im}(\sigma(d_i,\xi)) = \textrm{Ker}(\sigma(d_{i+1},\xi))$, when $\xi \neq 0 \in \Lambda^1$.
    \end{enumerate}
\end{definition}

The symbol of a differential operator,
\begin{align}
    d = \sum_{k=0}^N a^{\mu_1\ldots \mu_n} \partial_{\mu_1} \ldots \partial_{\mu_k}
\end{align}
for some $N \in \mathbb{N}$, is given by 
\begin{align}
    \sigma(d,\xi) = a^{\mu_1\ldots\mu_n} \xi_{\mu_1} \ldots \xi_{\mu_N}.
\end{align}
In a sense it is the Fourier transform of the highest order term of the differential operator $d$. The prototypical example of an Elliptic Complex is the de Rham complex of differential forms 
\begin{align}
    0 \longrightarrow \Lambda^0 \overset{d}{\longrightarrow} \Lambda^1 \overset{d}{\longrightarrow} \ldots \overset{d}{\longrightarrow} \Lambda^n
\end{align}
where the maps $d_i = d$ are all the exterior derivative. The symbol of the operator is then 
\begin{align}
    \sigma(d_i,\xi) = \sigma(d,\xi) = \xi \wedge : \Lambda^k \rightarrow \Lambda^{k+1}, \quad \xi \in \Lambda^1.
\end{align}
Using a standard result of linear algebra we have that for a differential form $\theta \in \Lambda^k$,
\begin{align}
    \xi \wedge \theta = 0, \quad \Rightarrow \quad \theta = \xi \wedge \alpha, \quad \alpha \in \Lambda^{k-1}.
\end{align}
Which proves that the complex is elliptic since the image of any operator is the kernel of the next.

\section{Yang-Mills Instantons}
We now show how the Euclidean Yang-Mills instanton equations appear in the context of elliptic complexes. Starting from the instanton condition \cref{eq:ell-yang-mills-instanton-condition}, we linearise the instanton equations around an instanton background and find the following elliptic complex of differential operators
\begin{align} \label{inst-complex}
    \Lambda^0(\mathfrak{g}) \overset{d^A}{\longrightarrow} \Lambda^1(\mathfrak{g}) \overset{d^A_+}{\longrightarrow} \Lambda^+(\mathfrak{g})
\end{align}
Here $d^A$ is the exterior covariant derivative with respect to the background gauge field $A$, and the second operator is the restriction of the exterior covariant derivative (of a Lie algebra valued 1-form) to the space of self-dual Lie algebra valued 2-forms. The composition of the two operators appearing above vanishes when the background gauge field is an instanton $F(A){}_+=0$. The above complex can be checked to be elliptic, this follows from the Hodge complex as the symbols are the same. Also, the full YM equations $\star d^A \star F(A)=0$ are equivalent to $\star d^A \star F_+(A)=0$, and the linearisation of the latter is then the statement that
\begin{align}\label{dd-star-YM}
    (d^A_+){}^* d^A_+ a = 0,
\end{align}
where $(d^A_+){}^*$ is the adjoint of $d^A_+ a$ with respect to the metric pairing inner product. So, the complex in \cref{inst-complex} can be said to encode the full second-order YM field equations. Adding the adjoint operators, one gets an elliptic operator
\begin{align}  \label{dirac-YM}
    \Lambda^1(\mathfrak{g}) \overset{d^A_+ + {(d^A)}^*}{\longrightarrow} \Lambda^0(\mathfrak{g}) \oplus \Lambda^+(\mathfrak{g})
\end{align}
This operator is known to be the (twisted by $A$) Dirac operator $D: S_+\otimes S_- \otimes \mathfrak{g} \to S_+\otimes S_+ \otimes \mathfrak{g}$, where $S_\pm$ are the two spinor representations of ${\rm Spin}(4)$. The instanton complex plays the key role in understanding the (local) properties of the moduli space of YM instantons, see~\cite{Donaldson:1983wm}.

\section{\pleb{} Complex}
As mentioned the aim of this chapter is to build an analogous elliptic differential complex, but this time relevant for Einstein equations rather than YM equations. A similar construction in the metric language is also considered in~\cite{The_linearised_Eastwo_2022}. Our construction became possible due to the recent re-interpretation~\cite{Su2StructureBhoja2024} of the \pleb{} description~\cite{OnTheSeparatiPleban1977} of Einstein equations.

We have seen in \cref{chap:plebanskis-formulation} that Einstein's equation can be encoded into a triple of complex 2-forms. Here we give a brief description of the changes that occur when working in Euclidean signature. Firstly, the Hodge star squares to $\star^2 = +\id$, this means that (anti-) self-dual 2-forms are real valued and no extension to complex numbers is needed. As such the reality conditions can also be dropped due to the metric being automatically real-valued. Let $\Sigma^i, i=1,2,3 \in \Lambda^2$ be a triple of 2-forms satisfying the metricity condition $\Sigma^i \wedge \Sigma^j \sim \delta^{ij}$, note the proportionality factor here should be real-valued unlike in the Lorentzian case. Some literature refers to such a set of 2-forms as perfect triples. The ${\rm GL}(4,\R)$ stabiliser of a perfect triple can be shown to be ${\rm SU}(2)$, and so perfect triples give an example of a $G$-structure, with $G={\rm SU}(2)$. Given an ${\rm SU}(2)$-structure $\Sigma^i$, there exists a triple of 1-forms $A^i$ satisfying the equation
\begin{align}\label{pleb-1}
d\Sigma^i + \epsilon^{ijk} A^j\wedge \Sigma^k = 0.
\end{align}
This triple of 1-forms can be shown~\cite{Su2StructureBhoja2024} to coincide with the so-called intrinsic torsion of the ${\rm SU}(2)$-structure $\Sigma^i$. When the intrinsic torsion vanishes the ${\rm SU}(2)$-structure is integrable. It can be shown that in this case the metric defined by $\Sigma^i$ is Ricci-flat and there are 3 integrable complex structures satisfying the algebra of the imaginary quaternions. So, when $d\Sigma^i=0$ the metric defined by $\Sigma^i$ is hyper-K\"ahler. For more information about this description of hyper-K\"ahler metrics see e.g.~\cite{The_Kummer_Cons_Jiang_2025}, Section 4.1. More generally, Einstein equations $R_{\mu\nu}=0$ can also be encoded in the same language. Indeed, as is well-known from the general context of $G$-structures, the intrinsic torsion of a $G$-structure, when $G\subset{\rm SO}(n)$, where $\dim(M)=n$, together with its first derivatives, encode some part of the Riemann curvature tensor of the Riemannian metric defined by the $G$-structure. This is why it may be possible to impose some equations on the curvature, for example Einstein equations, as a first-order in derivatives condition on the intrinsic torsion. In the case at hand, one shows that Einstein equations $R_{\mu\nu} = 0$ become the statement that 
\begin{align}\label{pleb-2}
F^i(A) = \Psi^{ij} \Sigma^j, \qquad F^i(A) = dA^i + \frac{1}{2} \epsilon^{ijk} A^j\wedge A^k.
\end{align}
Here $\Psi^{ij}$ is an arbitrary symmetric tracefree $3\times 3$ matrix. The equations (\ref{pleb-1}), (\ref{pleb-2}) are the two equations appearing in the first-order \pleb{} formalism for general relativity. Their linearisation around the hyper-K\"ahler background $d\Sigma^i=0$, and thus $A^i=0$, is what gives rise to the \pleb{} complex. Define $E=\mathfrak{su}(2) = \mathfrak{so}(3)=\R^3$ and let $S\subset E\otimes \Lambda^2$ be the tangent space to the space of perfect triples at $\Sigma^i$. We will characterise $S$ more explicitly below. Similarly, let $E \otimes \Lambda^1$ be the tangent space to the space of triples of 1-forms $A^i$. The linearisation of the equation (\ref{pleb-1}) at $A^i=0$ is then
\begin{align}\label{lin-pleb-1}
d \sigma^i + \epsilon^{ijk} a^j \wedge \Sigma^k=0, \qquad \sigma^i\in S, \quad a^i \in E\otimes \Lambda^1.
\end{align}
The solution of this equation can be stated in terms of a certain linear operator $J_1: E\otimes \Lambda^1 \to E\otimes \Lambda^1$ defined in (\ref{J-1}). This operator has two eigenspaces in $E\otimes \Lambda^1$, decomposing it $E\otimes \Lambda^1 = (E\otimes \Lambda^1){}_4 \oplus (E\otimes \Lambda^1){}_8$. The equation (\ref{lin-pleb-1}) becomes a linear equation for $a^i$, whose solution is given by $a^i = d_2 \sigma^i$, where 
\begin{align}\label{map-2}
    d_2:S\to E\otimes \Lambda^1, \qquad d_2 \sigma^i = \frac{1}{2} J_1^{-1}( \star d \sigma^i).
\end{align}
Here $\star$ is the Hodge star with respect to the background metric defined by $\Sigma^i$. This is the operator that appears as the second arrow of our to-be-constructed complex. To construct the first arrow we note that the objects $a^i$ that appear as the result of the map $d_2: S\to E\otimes \Lambda^1$ are diffeomorphism invariant. Linearised diffeomorphisms act on the space $S$ via
\begin{align}\label{map-1}
    d_1: TM \to S, \qquad d_1 \xi = d (\iota_\xi  \Sigma^i).
\end{align}
We will see in the main exposition that the object $d (\iota_\xi  \Sigma^i)$ is in $S$, the tangent space to the space of perfect triples. It is then immediate to see that the composition $d_2 d_1=0$. The final arrow of the \pleb{} complex arises by considering the map 
\begin{align}\label{map-3}
    d_3: E\otimes \Lambda^1 \to E, \qquad d_3 a^i = da^i \Big|_3 \equiv \epsilon^{ijk} \Sigma^j \wedge da^k / v_\Sigma.
\end{align}
Here $da^i\in E\otimes \Lambda^2$ is the exterior derivative of $a^i$, and the projection is taken onto the copy of $E$ inside $E\otimes \Lambda^2$. The last equality describes this projection explicitly, here $v_\Sigma$ is the volume form for the metric defined by $\Sigma^i$, which we can take to be $v_\Sigma = (1/6) \Sigma^i\wedge \Sigma^i$. The vanishing of the composition $d_3 d_2=0$ is the consequence of the equation (\ref{lin-pleb-1}). Indeed, taking the exterior derivative of this equation one gets $d_3 a^i=0$ whenever $a^i$ satisfies (\ref{lin-pleb-1}). 

Assembling the spaces and the differential operators we get what we propose to call the \pleb{} complex
\begin{align} \label{pleb-compl-intr}
    TM \overset{d_1}{\longrightarrow} S \overset{d_2}{\longrightarrow} E \otimes \Lambda^1 \overset{d_3}{\longrightarrow} E    
\end{align}
We then have the following
\begin{theorem}\label{thrm:pleb-complex-is-elliptic}
    The complex (\ref{pleb-compl-intr}) is an elliptic complex of differential operators, as in \cref{def:ell-elliptic-complex}.
\end{theorem}
We remark that the dimensions of the spaces appearing in (\ref{pleb-compl-intr}) are
\begin{equation}
\begin{tikzcd}\label{pleb-compl-intr-dims}
    4  \arrow{r}{d_1} & 13  \arrow{r}{d_2} & 12  \arrow{r}{d_3} & 3
\end{tikzcd}
\end{equation}
and thus the alternating sum of the dimensions vanishes, as it should for an elliptic complex. 

Our second statement gives an Einstein analogue of the characterisation (\ref{dd-star-YM}) of the linearised YM equations.
\begin{theorem}\label{thrm:pleb-complex-adjoint-gives-einsteins-equations} Define the inner product on $E\otimes \Lambda^1$ to be given by
\begin{align}\label{inner-prod-a}
\langle a,a\rangle = \int_M \epsilon^{ijk} \Sigma^i \wedge a^j \wedge a^k.
\end{align}
The linearisation of the Einstein's equations (\ref{pleb-2}) is the statement 
\begin{align}
d_2^* d_2 \sigma^i =0.
\end{align}
\end{theorem}
The operator $d_2^*$ also depends on the inner product in $S$, and this is not unique. But the above theorem holds for any choice of the inner product in $S$. 

Our final statement is related to a construction of an elliptic operator for (\ref{pleb-compl-intr}) that squares to the Laplacian, giving a gravity analogue of (\ref{dirac-YM}). The most natural such construction would consist in simply adding the adjoints. Indeed, let us consider the operator
\begin{align}
D: S\oplus E \to TM \oplus E\otimes\Lambda^1, \qquad D(\sigma,\chi) = (d_1^*\sigma, d_2 \sigma + d_3^*\chi).
\end{align}
The operator so constructed is guaranteed to be an elliptic operator. However, somewhat to our surprise, we have the following negative result: $D^* D$ is not a multiple of the Laplacian on $S\oplus E$, if we demand that the inner product on $E\otimes\Lambda^1$ is given by (\ref{inner-prod-a}). However, $D^* D$ is a multiple of the Laplacian if instead one uses the following inner product
\begin{align}
\langle a,a\rangle' = \int_M (a^i,a^i),
\end{align}
where $(\cdot,\cdot)$ is the metric pairing on $\Lambda^1$, defined with respect to the metric determine by $\Sigma^i$. While this choice of the inner product leads to the desired result for $D^* D$, this is not the choice that leads to $d_2^*$ being the operator appearing in the linearised Einstein equations. This makes us look for a construction of $D$ that has the property $D^* D\sim \Delta$ and that will involve the operators $d_2, d_2^*$, the latter being defined with respect to the inner product (\ref{inner-prod-a}). Such a construction exists, but is non-trivial. 

Let us consider the more general elliptic differential operator 
\begin{align}
\tilde{D}: S\oplus E \to TM \oplus E\otimes\Lambda^1, \qquad   \tilde{D}(\sigma,\chi) = (\tilde{d}_1^* \sigma + \tilde{d}_4 \chi, d_2 \sigma + \tilde{d}_3^*\chi),
\end{align}
where we introduced $\tilde{d}_4: E\to TM$. We require that $d_2$ that appears here is the one of the \pleb{} complex (\ref{pleb-compl-intr}), and that the inner product on $E\otimes \Lambda^1$ is as in (\ref{inner-prod-a}) so that $d_2^*$ is the one that appears in the linearised Einstein equations. All other operators are allowed to be completely general first order differential operators as allowed by the relevant representation theory. In particular, we do not require that $\tilde{d}_1, d_2, \tilde{d}_3, \tilde{d}_4$ form a differential complex (i.e.\ there is no demand that the compositions are zero). We have the following
\begin{theorem} There is a choice of the inner product on $S$, and there is a choice of the operators $\tilde{d}_1, \tilde{d}_3, \tilde{d}_4$ such that all the following hold:
\begin{itemize}
\item $\tilde{D}^* \tilde{D} =\pm \Delta$ on $S\oplus E$
\item The operators $\tilde{d}_1,\tilde{d}_3,\tilde{d}_4$ are linear combinations of $d_1, d_2, d_3$ of the \pleb{} complex. 
\item There exist linear maps
\begin{align}
T_1: S\oplus E\to S\oplus E, \qquad T_2 : TM\oplus E\otimes \Lambda^1 \to TM\oplus E\otimes \Lambda^1
\end{align}
such that 
\begin{align}\label{split}
T_2 \tilde{D} T_1 = D_4 \oplus D_{12},
\end{align}
with $D_4, D_{12}$ being operators
\begin{align}
D_4: \Lambda^0 \oplus E \to \Lambda^1, \qquad D_{12}: E \oplus {\rm Sym}^2_0(\Lambda^1) \to E\otimes \Lambda^1
\end{align}
explicitly given by (\ref{d4-d12}). Both of these operators are versions of the Dirac operator
\begin{align}
D_4: S_+\otimes S_+ \to S_-\otimes S_+, \qquad D_{12}: S_-\otimes S_-\otimes S_+^2 \to S_+\otimes S_-\otimes S_+^2.
\end{align}
\end{itemize}
\end{theorem}
Moreover, the choices we need to make for the above to hold are essentially unique, modulo signs that can be absorbed into field redefinitions. The operator $\tilde{D}$ that is the result of construction in this theorem is explicitly given by
\begin{align}
    \tilde{D} = \begin{pmatrix}\frac{1}{\sqrt{2}}(d^*_1 - \Phi d_2) && -\Phi d^*_3 \\ d_2 && \frac{1}{\sqrt{2}} J_1 d^*_3 \end{pmatrix}.
\end{align}
Here all operators are those of the \pleb{} complex, the linear map $\Phi: E\otimes \Lambda^1\to \Lambda^1$ is given by (\ref{Phi}) and $J_1: E\otimes \Lambda^1\to E\otimes \Lambda^1$ is the linear map given by (\ref{J-1}). This gives us a non-trivial gauge-fixing of the \pleb{} complex, by appropriately assembling the operators $d_1, d_2, d_3$ and their adjoints into an elliptic operator that can be written as (twisted by $T_1, T_2$) direct sum of two Dirac operators. 

If one does not demand that $\tilde{d}_1, \tilde{d}_3,\tilde{d}_4$ can be expressed in terms of $d_1, d_2, d_3$ of the \pleb{} complex then there is some more freedom in the choice of $\tilde{d}_1, \tilde{d}_3, \tilde{d}_4$ that leads to $\tilde{D}^* \tilde{D} =\pm \Delta$. The operators $D_4, D_{12}$ already appeared in the work~\cite{ChiralPerturbaKrasno2020}, but this reference was using spinor techniques heavily, which makes it difficult to establish any uniqueness statement. More generally, the operators are those appearing in \cref{subsec:pleb-dirac-operators} containing a 2 parameter family of Dirac operators for the linearised Einstein's equation. However, no consideration is taken to the ellipticity of the complex formed by such operators. In contrast, the present work makes all the assumptions that go into the construction of $\tilde{D}$ manifest, leading to the statement in the above theorem.

\section{Decomposition of \texorpdfstring{$E$}{E}-valued differential forms}\label{sec:decomp}

The material in this preparatory section is from~\cite{Su2StructureBhoja2024} and is a translation of \cref{subsec:pleb-C3-valued-forms-decomp} in Euclidean signature. The rotation group in four dimensions is ${\rm SO}(4)={\rm SU}(2)\times{\rm SU}(2)/\Z_2$. One of the two ${\rm SU}(2)$ factors stabilises the 2-forms $\Sigma^i$. The other ${\rm SU}(2)$ factor rotates the triple $\Sigma^i$ by an ${\rm SO}(3)$ orthogonal transformation. This means that it is natural to treat $\Sigma^i$ as a map $\Sigma: \R^3 \to \Lambda^2$. The image of this map is the space $\Lambda^2_+$ of self-dual 2-forms. The map $\Sigma$ thus identifies $\R^3$ with $\Lambda^2_+$. This makes it natural to consider how the spaces of $E$-valued differential forms on $M$ transform with respect to all of the rotation group ${\rm SO}(4)={\rm SU}(2)\times{\rm SU}(2)/\Z_2$. 

Thus, we need to understand the decomposition of the spaces $\Lambda^1(M)\otimes E, \Lambda^2(M)\otimes E$, into irreducible representations of ${\rm SU}(2)\times{\rm SU}(2)/Z_2$. Irreducible representations of ${\rm SU}(2)$ are the spin $k/2$ representations that we denote by $S^k$. They are of dimension $\dim(S^k) = k+1$. As we have already discussed, there are two different ${\rm SU}(2)$'s in the game. One ${\rm SU}(2)$ is the group with respect to which the 2-forms $\Sigma^i$ are invariant. We will choose to denote this copy of ${\rm SU}(2)$ by ${\rm SU}_-(2)$, and the corresponding representations by $S^k_-$. The other ${\rm SU}(2)$ is one that acts non-trivially on $\Sigma^i$ by mixing them, with the map $\Sigma:E\to \Lambda^2$ being equivariant with respect to this copy of ${\rm SU}(2)$. We will denote it by ${\rm SU}_+(2)$, and the corresponding representations by $S_+^k$. We then have
\begin{align}
\Lambda^1(M) = S_+\otimes S_-, \qquad \Lambda^2(M) = S_+^2 \oplus S_-^2, \qquad E = S_+^2,
\end{align}
and the decomposition of $\Lambda^1(M)\otimes E, \Lambda^2(M)\otimes E$ into irreducibles is
\begin{align}\label{E-forms-decomp}
\Lambda^1(M) \otimes E= (S_+\otimes S_-) \oplus (S_+^3\otimes S_-), \\ \nonumber
\Lambda^2(M) \otimes E = C^\infty(M)  \oplus S_+^2 \oplus S_+^4 \oplus (S_+^2 \otimes S_-^2).
\end{align}

We now go on to state the algebra for the Euclidean $\Sigma^i$'s and the component formulae for the irreducible components.

\subsection{\texorpdfstring{Algebra of $\Sigma$'s}{}}

The components of the 2-forms we call $\Sigma^i_{\mu\nu}$, where $\mu,\nu =1,2,3,4$. The Riemannian metric defined by $\Sigma^i$ can be written explicitly as
\begin{align}\label{metric}
g_{\mu\nu} v_g = \frac{1}{6} \epsilon^{ijk} \Sigma^i_{\mu\alpha} \Sigma^j_{\nu\beta} \Sigma^k_{\gamma\delta} \tilde{\epsilon}^{\alpha\beta\gamma\delta}.
\end{align}
Here $\tilde{\epsilon}^{\alpha\beta\gamma\delta}$ is the densitiesed completely antisymmetric tensor, which has components $\pm 1$ in any coordinate system and does not need a metric for its definition. Both sides of this formula are densitiesed $\mu\nu$ symmetric tensors, and $v_g=\sqrt{\det(g_{\mu\nu})}$ is the volume form for $g_{\mu\nu}$. 

One of the two indices of $\Sigma^i$ can be raised with the metric (\ref{metric}), to convert these objects into those in ${\rm End}(TM)$. We then have a triple of such endomorphisms of the tangent bundle, satisfying the same algebra of the imaginary quaternions
\begin{align}\label{algebra}
\Sigma^i_{\mu}{}^\alpha \Sigma^j_\alpha{}^\nu = - \delta^{ij} \delta_\mu{}^\nu + \epsilon^{ijk} \Sigma^k_\mu{}^\nu.
\end{align}
It can be shown that the 2-forms also satisfy the following algebra
\begin{align}\label{sigma-sigma}
\Sigma^i_{\mu\nu} \Sigma^i_{\rho\sigma}= g_{\mu\rho} g_{\nu\sigma} - g_{\mu\sigma} g_{\nu\rho} + \epsilon_{\mu\nu\rho\sigma}, \\ 
\label{sigma-sigma-epsilon}
\epsilon^{ijk} \Sigma^j_{\mu\nu} \Sigma^k_{\rho\sigma}= -2\Sigma^i_{[\mu|\rho|} g_{\nu]\sigma} + 2\Sigma^i_{[\mu|\sigma|} g_{\nu]\rho} , \\ \nonumber
\epsilon^{\mu\nu\rho\alpha} \Sigma^i_{\sigma\alpha} = \delta_\sigma^\rho \Sigma^{i\mu\nu} +  \delta_\sigma^\mu \Sigma^{i\nu\rho} + \delta_\sigma^\nu \Sigma^{i\rho\mu} .
\end{align}
These relations, as well as \cref{algebra}, will be used on numerous occasions below, without explicit mention. 

\subsection{\texorpdfstring{Decomposition of $\Lambda^1(M) \otimes E$}{}}

Given an ${\rm SU}(2)$ structure $\Sigma^i$, we have the following operator acting on $\Lambda^1(M) \otimes E$
\begin{align}\label{J-1}
\Lambda^1(M) \otimes E\ni a^i_\mu \to J_1(A){}_\mu^i := \epsilon^{ijk} \Sigma^j_\mu{}^\alpha a^k_\alpha.
\end{align}
As the algebra, \cref{algebra}, is signature independent we recover the same polynomial identity for $J_1$,
\begin{align}
J_1^2 = 2\mathbb{I} + J_1.
\end{align}
This means that the eigenvalues of $J_1$ are $2, -1$. It is easy to check that 
\begin{align}
    E \otimes \Lambda^1(M) = (\Lambda^1(M) \otimes E){}_4 \oplus (\Lambda^1(M) \otimes E){}_8
\end{align}
and that
\begin{align}\label{lambda-E-4}
    \xi^\nu \Sigma^i_{\nu\mu} \in (\Lambda^1(M) \otimes E){}_4 = (S_+ \otimes S_-)
\end{align}
and that $J_1$ has eigenvalue $+2$ on such forms. The space 
\begin{align}\label{lambda-E-8}
(\Lambda^1(M)\otimes E){}_8 = (S^3_+\otimes S_-) 
\end{align}
can then be characterised as the orthogonal complement of (\ref{lambda-E-4}) in $\Lambda^1(M)\otimes E$.

\subsection{Decomposition of \texorpdfstring{$\Lambda^2(M) \otimes E$}{}}\label{subsub:ell-decomposition-of-E-2-forms}
To characterise some of the spaces appearing in the decomposition of $\Lambda^2(M) \otimes E$ we first consider the ${\rm GL}(4)$ orbit of the 2-forms $\Sigma^i$. The tangent space to this orbit is precisely the space $S$ of tangent vectors to perfect triples that we introduced above. This tangent space is the space of $E$-valued 2-forms of the form $h_{[\mu}{}^\alpha \Sigma^i_{ |\alpha|\nu]}$. Decomposing $h_{\mu\nu}\in{\rm GL}(4)$ into its symmetric and antisymmetric parts, and noting that the antisymmetric part is valued in $\Lambda^2(M)=S_+^2 \oplus S_-^2$, we get the following list of irreducibles appearing  
\begin{align}\label{S-space-param}
h_{[\mu}{}^\alpha \Sigma^i_{ |\alpha|\nu]} \in S = C^\infty(M) \oplus S_+^2 \oplus (S_+^2 \otimes S_-^2) \subset \Lambda^2(M) \otimes E,
\end{align}
which is all spaces in the second line of (\ref{E-forms-decomp}) apart from $S_+^4$. These irreducibles in $\Lambda^2(M) \otimes E$ can then be characterised as the images of the map $h_{\mu\nu}\to h_{[\mu}{}^\alpha \Sigma^i_{ |\alpha|\nu]} $. 

One can also act on the index $i$ of $\Sigma^i$ 2-forms with a ${\rm GL}(3)$ transformation, i.e., consider the orbit of $E$-valued 2-forms of the form $h^{ij} \Sigma^j_{\mu\nu}$. Decomposing the matrix $h^{ij}$ into symmetric and antisymmetric parts, one finds the following list of irreducibles
\begin{align}
h^{ij} \Sigma^j_{\mu\nu} \in C^\infty(M) \oplus S_+^2 \oplus S_+^4
\end{align}
where $C^\infty(M), S_+^2, S_+^4$ correspond to the trace, antisymmetric and symmetric tracefree parts respectively. In the opposite direction, given an object $B^i_{\mu\nu}\in \Lambda^2(M) \otimes E$, its irreducible parts can be extracted as follows
\begin{align}\label{irreducibles}
B^{k}_{\alpha\beta} \Sigma^{k\alpha\beta} &\in C^\infty(M), \\ \nonumber
\epsilon^{ijk} B^j_{\alpha\beta} \Sigma^{k\alpha\beta} &\in S_+^2, \\ \nonumber
B^{\langle i}_{\alpha\beta} \Sigma^{j\rangle\alpha\beta} &\in S_+^4, \\ \nonumber
B^i_{\langle\mu|\alpha|} \Sigma^{i\alpha}{}_{\nu\rangle} &\in S_+^2 \otimes S_-^2 .
\end{align}
Here $T_{\langle \mu\nu \rangle} = (1/2)(T_{\mu\nu} + T_{\nu\mu} - (1/4) g_{\mu\nu} g^{\alpha\beta} T_{\alpha\beta})$, which denotes the symmetric tracefree part. 

As in the Lorentzian case let us introduce the following operator
\begin{align}\label{J2}
J_2: \Lambda^2\otimes E\to \Lambda^2\otimes E, \qquad J_2(B){}_{\mu\nu}^i = \epsilon^{ijk} \Sigma^j_{[\mu}{}^\alpha B_{|\alpha|\nu]}^k, \qquad B_{\mu\nu}^i\in \Lambda^2\otimes E.
\end{align}
For which we find the following polynomial relation
\begin{align}
J_2^4 - 2J_2^3-J_2^2 + 2 J_2=0 \qquad \text{or} \qquad J_2 (J_2-2)(J_2-1)(J_2+1)=0,
\end{align}
which implies that the eigenvalues of $J_2$ are $0,1,-1,+2$ respectively for the second line in \cref{E-forms-decomp}. We are able to identify $(\Lambda^2\otimes E){}_9 = \Lambda^-\otimes E$.~To obtain the parametrisation that we will use through this section we split the $GL(4)$ orbit partially into its irreducible parts,
\begin{align}
    h_{\mu\nu} \rightarrow h_{(\mu\nu)} +  \Sigma^i_{\mu\nu} h^i + \asd^i_{\mu\nu} \bar{h}^i.
\end{align}
Where we have used $h_{[\mu\nu]} \in \Lambda^2 = \Lambda^+ \oplus \Lambda^-$ to separate, the antisymmetric part into its self-dual and anti-self-dual components, each of these are decomposed into a basis using $\Sigma^i$ and $\asd^i$. Substituting this into the $GL(4)$ actions, and making use of the quaternion algebra, we find
\begin{align}
    \sigma^i_{\mu\nu} = 2 h_{[\mu}{}^\alpha \Sigma^i_{|\alpha|\nu]} \rightarrow 2 h_{[\mu}{}^\alpha \Sigma^i_{|\alpha|\nu]} + 2 \epsilon^{ijk} h^k \Sigma^j_{\mu\nu} + 2 \bar{h}^j \asd^j_{[\mu}{}^\alpha \Sigma^i_{|\alpha|\nu]}.
\end{align}
The last term in the above is zero as $\asd^j_{\mu}{}^\alpha \Sigma^i_{\alpha\nu} \in \rm Sym^2(\Lambda^1)$ is a basis for symmetric tracefree tensors. Next we need to check that no components of $h_{(\mu\nu)}$ are in the kernel of the above parametrisation. To check this we introduce a basis $e^0_\mu, e^i_\mu \in \Lambda^1$ and decompose the following objects in this basis,
\begin{align}
    h_{\mu\nu} & = h_{00} e^0_\mu e^0_\nu + h_{0i} (e^0_\mu e^i_\nu + e^0_\nu e^i_\mu) + h_{ij} e^i_\mu e^j_\nu \\
    \Sigma^i_{\mu\nu} & = i e^0_\mu e^i_\nu - i e^0_\nu e^i_\mu - \epsilon^{ijk} e^j_\mu e^k_\nu
\end{align}
where $h_{ij} = h_{ji}$. Substituting this into the definitions of $\sigma^i$ we find
\begin{align}
    \sigma^i_{\mu\nu} = \left( i \delta^{ij} h_{00} - \epsilon^{ijk}(2i h^k + h^k_0) - i h^{i j} \right) e^0_{[\mu} e^j_{\nu]} + \left( h^{ij} - \delta^{ij} h^{kk} - \epsilon^{ijk}(i h^k_0 + 2 h^k)\right) \epsilon^{jlm} e^l_\mu e^m_\nu.
\end{align}
For $\sigma^i = 0$ we find that $h_{\mu\nu} = 0$ and $h^i = 0$, therefore the kernel is trivial. The parametrisation of $S$ that we will use throughout this chapter is then
\begin{align}\label{param-sigma}
    \sigma^i_{\mu\nu} =  2 \epsilon^{ijk} \Sigma^j_{\mu\nu} h^k + 2 h_{[\mu}{}^\alpha \Sigma^i_{|\alpha|\nu]} \in S
\end{align}
here $h_{\mu\nu}$ is a symmetric tensor and $h^i$ is an internal vector. Another notation that we will frequently use below is to represent objects in $S$ as the following two lists
\begin{align}
\left( h^i,\ h_{\mu\nu} \right) \in& E \oplus \rm Sym^2(\Lambda^1) \\
\left( h,\ h^i,\ \tilde{h}_{\mu\nu} \right) \in& C^\infty \oplus E \oplus \rm Sym^2_0(\Lambda^1)
\end{align}
where $\rm Sym^2_0(\Lambda^1)$ denotes the tracefree part and $h_{\mu\nu} = \frac{1}{4} g_{\mu\nu} h + \tilde{h}_{\mu\nu}$. So, the second representation splits the trace and the tracefree parts. 

\section{Proof of ellipticity}

We are now equipped to study the complex (\ref{pleb-compl-intr}) in more detail. We prove \cref{thrm:pleb-complex-is-elliptic} by showing that the sequence of symbols 
\begin{equation}
\begin{tikzcd}\label{exact-sequence}
0 \arrow{r}{} & TM  \arrow{r}{\sigma(d_1)} & S  \arrow{r}{\sigma(d_2)} & E\otimes \Lambda^1  \arrow{r}{\sigma(d_3)} & E \arrow{r}{} & 0
\end{tikzcd}
\end{equation}
is exact, which means that \pleb{} complex is an elliptic complex of differential operators. 

\subsection{Proof of the \cref{thrm:pleb-complex-is-elliptic}}

We consider the symbols of all the differential operators. Symbols are obtained by replacing the operators of partial derivative $\partial_\mu$ with a factor of an arbitrary 1-form, which we denote by $k_\mu$. The symbol of the operator in the first arrow (\ref{map-1}) is
\begin{align}
\xi^\mu \to k_{[\mu} \xi^\alpha \Sigma^i_{|\alpha|\nu]}.
\end{align}
To see that this has zero kernel we use the fact that if two 1-forms wedge to zero then they necessarily point in the same direction. The kernel of the above map is then 
\begin{align}\label{map-1-kernel-condition}
\xi^\alpha \Sigma^i_{\alpha\nu} = k_\nu \phi^i
\end{align}
where $\phi^i$ is an arbitrary internal vector. We would like to show that $\phi^i$ must vanish. 
We can solve this to find $\xi_\mu$ by contracting with $\Sigma^i_\mu{}^\nu$,
\begin{align}
\xi_\mu = \frac{1}{3} \Sigma^i_\mu{}^\nu \Sigma^i_{\alpha\nu} \xi^\alpha = \frac{1}{3} \Sigma^i_\mu{}^\nu k_\nu \phi^i.
\end{align}
Substituting this $\xi_\mu$ back into the left-hand side of (\ref{map-1-kernel-condition}) and using the algebra of $\Sigma^i$'s we find 
\begin{align}
\frac{1}{3} \Sigma^{j\alpha\nu} k_\nu \phi^j \Sigma^i_{\alpha\mu} = \frac{1}{3} k_\mu \phi^i + \frac{1}{3} \epsilon^{ijk} \Sigma^j_\mu{}^\nu k_\nu \phi^k.
\end{align}
Comparing with the right-hand side of (\ref{map-1-kernel-condition}) we find 
\begin{align}
2 k_\mu \phi^i = \epsilon^{ijk} \Sigma^j_\mu{}^\nu k_\nu \phi^k.
\end{align}
Contracting the above with $k^\mu$ we find 
\begin{align}
k^2 \phi^i = 0.
\end{align}
As $k^2 = k^\mu k_\mu \neq 0$ then this requires that $\phi^i = 0$.
Which implies that kernel of the first map (\ref{map-1}) is trivial, and 
shows the exactness of the first arrow in (\ref{exact-sequence}). 

The symbol of the second arrow (\ref{map-2}) is
\begin{align}\label{symbol-2}
\sigma^i \to \frac{1}{2} J_1^{-1} (\epsilon_\mu{}^{\alpha\beta\gamma} k_\alpha \sigma^i_{\beta\gamma}).
\end{align}
We know that $J_1^{-1}$ is invertible, so we can safely ignore it and understand the kernel of 
\begin{align}
    \sigma^i \to \epsilon_\mu{}^{\alpha\beta\gamma} k_\alpha \sigma^i_{\beta\gamma}
\end{align}
Using the same linear algebra fact as before we know that $k_{[\alpha} \sigma^i_{\beta\gamma]}$ is zero when $\sigma^i_{\beta \gamma} = k_{[\beta} b^i_{\gamma]}$ for any $b^i_\mu \in E \otimes \Lambda^1$. This may make it seem that the kernel of the second map is all of $E \otimes \Lambda^1$. However, in general $k \wedge b^i \in E \otimes \Lambda^2$, which is bigger than $S$. Let us determine one-forms $b^i$ such that $k\wedge b^i \in S$. 
A 2-form from $E\otimes \Lambda^2$ belongs to $S$ if and only if 
\begin{align}\label{no-S-plus-4-in-kernel}
\Sigma^{\langle j|\mu\nu} \sigma^{i\rangle}_{\mu\nu} = \Sigma^{\langle j|\mu\nu} k_\mu b^{|i\rangle}_\nu = 0.
\end{align}
To find solutions of this equation, we decompose $b^i$ into its irreducible parts
\begin{align}
b^i_\mu = \xi^\alpha \Sigma^i_{\alpha\mu} + (b_8){}^i_\mu
\end{align}
where $\xi^\mu$ is the 4-vector part and $b^i_8\in (E\otimes\Lambda^1){}_8$.
Using the algebra of $\Sigma^i$'s it is easy to see that the 4-vector part satisfies (\ref{no-S-plus-4-in-kernel}) automatically. Let us consider the $b^i_8$ part. To analyse this further we can choose a basis for $\Sigma^i$,
\begin{align}\label{self-dual-2-forms-k-basis}
\Sigma^i_{\mu\nu} = \frac{k_\mu}{|k|} e^i_\nu - \frac{k_\nu}{|k|} e^i_\mu - \epsilon^{ijk} e^j_\mu e^k_\nu
\end{align}
where $e^i_\mu$ are orthonormal and orthogonal to $k_\mu$, i.e. $e^i_\mu k^\mu = 0$ and $e^i_\mu e^{j\mu} = \delta^{ij}$.
This allows the decomposition 
\begin{align}
(b_8){}^i_\mu = k_\mu b^i + b^{ij} e^j_\mu.
\end{align}
The eigenvalue of $b_8^i$ under the map $J_1$ is $-1$, which restricts the components of $b^{ij}$. Acting with $J_1$ gives
\begin{align}\label{J1-b}
J_1(b_8){}^i_\mu =& \epsilon^{ijk}\left(\frac{k_\mu}{|k|} e^{j\nu} - \frac{k^\nu}{|k|} e^j_\mu - \epsilon^{jmn} e^m_\mu e^{n\nu}\right)(k_\nu b^k + b^{kl} e^l_\nu) \\ =& \frac{k_\mu}{|k|} \left( \epsilon^{ijk} b^{kj} \right) + \left( -b^{ji} + \delta^{ij} b^{kk} - |k| \epsilon^{ijk} b^k \right) e^j_\mu.
\end{align}
Therefore, $J_1(b_8){}^i = -b^i_8$ implies
\begin{align}
 b^i =  \epsilon^{ijk} \frac{b^{jk}}{|k|} \quad \textrm{and} \quad b^{ij} = b^{ji} - \delta^{ij} b^{kk} + |k| \epsilon^{ijk} b^k.
\end{align}
The second equality implies $b^{ii}=0$. We have thus parameterised $b^i_8$ by a tracefree matrix $b^{ij}$ (and is not necessarily symmetric).  We can now determine which part of $b^i_8$ is actually contributing to $k\wedge b^i_8\in S$. It is clear that the $k_\mu b^i$ part does not survive in $k\wedge b^i_8$, so let us consider the other term. We want to determine if any part of $k\wedge b^{ij} e^j$ is in $S$. Substituting this into (\ref{no-S-plus-4-in-kernel}) we find 
\begin{align}
\left(\frac{k^\mu}{|k|} e^{\langle i|\nu} - \frac{k^\nu}{|k|} e^{\langle i|\mu} - \epsilon^{\langle i|lm} e^{l\mu} e^{m\nu} \right) k_\mu b^{|j\rangle n} e^n_\nu = |k| b^{\langle ij\rangle }.
\end{align}
This must vanish in order for $k\wedge b^i_8\in S$. This means that $k\wedge b^i_8$ is either zero or not in $S$. It thus cannot contribute to the non-trivial kernel of (\ref{symbol-2}). This means that $b^i_\mu = \xi^\alpha \Sigma^i_{\alpha\mu}$ generates all the non-trivial kernel, where elements in the kernel are given by $\sigma^i_{\mu\nu} = k_{[\mu|} \xi^\alpha \Sigma^i_{\alpha |\nu]}$ and its dimension is $4$.
It is also clear then that the image of $\sigma(d_1)$ and the kernel of $\sigma(d_2)$ exactly coincide. 
As the dimension of the kernel of $\sigma(d_2)$ is $4$ and the dimension of the domain is $13$ then the image will have dimension $13-4 = 9$.

We now look at the final map
\begin{align}
a^i_\mu \to \epsilon^{ijk} \Sigma^{j\mu\nu} k_\mu a^k_\nu = k^\mu J_1(a){}^i_\mu.
\end{align}
To find its kernel we decompose $a^i_\mu$
\begin{align}\label{E-valued-1-form-k-basis}
a^i_\mu = k_\mu a^i + a^{ij} e^j_\mu
\end{align}
where $a^i$ and $a^{ij}$ are an arbitrary internal vector and matrix.
Computing the action of $J_1$ is the same computation as (\ref{J1-b}), which gives
\begin{align}
J_1(a){}^i_\mu =  \frac{k_\mu}{|k|}( - \epsilon^{ijk} a^{jk}) + \left( \delta^{ij} a^{kk} - a^{ji} - \epsilon^{ijk}|k| a^k \right) e^j_\mu.
\end{align}
This implies
\begin{align}
k^\mu J_1(a){}^i_\mu = - |k| \epsilon^{ijk} a^{jk}.
\end{align}
This shows that the kernel of $\sigma(d_3)$ is all of $a^i_\mu$ apart from $a^{[ij]}$. So, the kernel is $12-3=9$ dimensional, and the image is $E$. We can also compute the image of $\sigma(d_2)$ explicitly, by using the form (\ref{d2-explicitly}) of this operator. In terms of symbols we have
\begin{align}\label{symbol-d2-explicitly}
(h^i , h_{\mu\nu}) \to - \Sigma^{i\alpha\beta} k_\alpha h_{\mu\beta} + 2 k_\mu h^i.
\end{align}
We now decompose $h_{\mu\nu}$ into its longitudinal and transverse parts
\begin{align}\label{h-decomp}
h_{\mu\nu} = \psi k_\mu k_\nu + k_{(\mu} X_{\nu)} + \tilde{h}_{\mu\nu}.
\end{align}
Here both $X_\mu, \tilde{h}_{\mu\nu}$ are transverse $k^\mu X_\mu=0, k^\mu \tilde{h}_{\mu\nu}=0$. We now want to compute the image of various part of the map (\ref{symbol-d2-explicitly}) and check that it coincides with the described above kernel of $\sigma(d_3)$. The last term in (\ref{symbol-d2-explicitly}) is precisely the first term in (\ref{E-valued-1-form-k-basis}). The image of the first term in (\ref{h-decomp}) is trivial. The image of the second term is
\begin{align}
- \frac{1}{2} \Sigma^{i\alpha\beta} k_\alpha X_\beta k_\mu,
\end{align}
which is also of the form of the first term in (\ref{E-valued-1-form-k-basis}). To compute the image of the last term in (\ref{h-decomp}) we use the explicit form of $\Sigma^{i\alpha\beta}$. We have
\begin{align}\label{image-d2-calc}
- \left(\frac{k^\alpha}{|k|} e^{i\beta} - \frac{k^\beta}{|k|} e^{i\alpha} - \epsilon^{ijk} e^{j\alpha} e^{k\beta}\right) k_\alpha \tilde{h}_{\mu\beta} = - |k| e^{i\beta} \tilde{h}_{\mu\beta}.
\end{align}
We can further decompose $\tilde{h}_{\mu\nu}$ into the basis of vectors $e^i_\mu$
\begin{align}
\tilde{h}_{\mu\nu} = h^{ij} e^i_\mu e^j_\nu,
\end{align}
where $h^{ij}=h^{(ij)}$ is a symmetric $3\times 3$ matrix. This gives for the right-hand-side of (\ref{image-d2-calc})
\begin{align}
- |k| h^{ij} e^j_\mu.
\end{align}
This is precisely the part of the second term in (\ref{E-valued-1-form-k-basis}) that is in the kernel of $\sigma(d_3)$. Thus, we have explicitly checked that the image of $\sigma(d_2)$ coincides with the kernel of $\sigma(d_3)$. This finishes the proof of ellipticity of the complex.

\section{Characterisation of Einstein equations}

The purpose of this section is to show that there exists a choice of the inner products on $E\otimes \Lambda^1$ and $S$ such that the linearisation of the Einstein equations $\delta R_{\mu\nu}=0$ becomes the statement $d_2^* d_2 \sigma^i=0$. 

\subsection{Linearisation of the Einstein condition}

As we have reviewed in the Introduction, in \pleb{} formalism the Einstein condition becomes the statement (\ref{pleb-2}). Its linearisation around $A^i=0$ is the statement
\begin{align}\label{einstein-lin}
d a^i = \psi^{ij} \Sigma^j,
\end{align}
with again $\psi^{ij}$ being an arbitrary symmetric tracefree $3\times 3$ matrix. From the discussion in \cref{subsub:ell-decomposition-of-E-2-forms} we know that the right-hand side of (\ref{einstein-lin}) is an arbitrary vector in $(\Lambda^2\otimes E){}_5\subset \Lambda^2\otimes E$. This means that (\ref{einstein-lin}) can be stated as $(da^i)\Big|_{1+3+9}=0$, i.e.\ as the statement that the projection of the exterior derivative of $a^i$ on the representations of dimensions $1,3,9$ in $\Lambda^2\otimes E$ vanishes. All these components can be recovered by computing
\begin{align}\label{einstein-lin-1}
(\partial_\mu a^i_\alpha - \partial_\alpha a^i_\mu) \Sigma^{i\alpha}{}_\nu \in \Lambda^1\otimes \Lambda^1.
\end{align}
The symmetric part of this tensor computes the $(\Lambda^2\otimes E){}_{1+9}$ parts, and the antisymmetric part only contains the $\Lambda^2_+ = (\Lambda^2\otimes E){}_3$ part. So, linearised Einstein equations can be written as the statement that (\ref{einstein-lin-1}) vanishes. Projecting (\ref{einstein-lin-1}) onto $\Sigma^{k\mu\nu}$ it is easy to see that the $(\Lambda^2\otimes E){}_3$ part of this condition reads
\begin{align}\label{einstein-lin-3}
\epsilon^{ijk} \Sigma^{j\mu\nu} \partial_\mu a_\nu^k = 0
\end{align}
The symmetric part of (\ref{einstein-lin-1}), on the other hand, gives
\begin{align}\label{einstein-lin-1+9}
\Sigma^i_{(\mu}{}^\alpha ( \partial_{\nu)} a^i_\alpha - \partial_\alpha a^i_{\nu)})=0.
\end{align}
As we already discussed, the condition (\ref{einstein-lin-3}) is an automatic consequence of the equation (\ref{lin-pleb-1}), and so holds automatically. The equations (\ref{einstein-lin-1+9}) are the ten linearised Einstein equations. 

\subsection{The adjoint of \texorpdfstring{$d_2$}{}}

We now compute the adjoint of $d_2$ given by (\ref{map-2}), with the inner product 
\begin{align}
\langle a,a\rangle = \int \epsilon^{ijk} \Sigma^{i\mu\nu} a_\mu^j a_\nu^k = - \int (a, J_1(a)),
\end{align}
where $J_1$ is the operator (\ref{J-1}) and $(a,b)=g^{\mu\nu} a_\mu^i b_\nu^i$ is the metric pairing of two objects in $\Lambda^1\otimes E$. We have
\begin{align}
\langle a, d_2 \sigma\rangle =& \langle a, \frac{1}{2} J_1^{-1} (\star d\sigma) \rangle= - \frac{1}{2} \int a_\mu^i \epsilon^{\mu\nu\rho\sigma} \partial_\nu \sigma^i_{\rho\sigma}
= -\frac{1}{2} \int \epsilon^{\mu\nu\rho\sigma} \sigma^i_{\mu\nu} \partial_\rho a_\sigma^i
\end{align}
Here we have used that $J_1^{-1} J_1=\mathbb{I}$. We already see that the adjoint of $d_2$ is related to the $da^i$. Given that $\sigma^i \in (\Lambda^2\otimes E){}_{1+3+9}$, the operator $d_2^*$ will have components in all these spaces. It is clear that these will be multiples of the left-hand sides in (\ref{einstein-lin-3}) and (\ref{einstein-lin-1+9}). This proves \cref{thrm:pleb-complex-adjoint-gives-einsteins-equations}. We will present a more detailed computation of $d_2^*$ below, after the form of the inner product on $S$ is fixed. 

\section{The \pleb{} complex case}

We will now describe the \pleb{} complex in more details, including all the adjoint maps. We will be using the parametrisation (\ref{param-sigma}) of the space $S$, so that $S$ is the sum of the following irreducible components
\begin{align}\label{S-spaces}
S = \Lambda^0 \oplus E \oplus {\rm Sym}_0^2(\Lambda^1),
\end{align}
where ${\rm Sym}_0^2(\Lambda^1)$ denotes the tracefree part. Now we establish the form of the operators $d_1, d_2, d_3$ in this parametrisation.

\subsection{\pleb{} complex in detail}

In the parametrisation (\ref{param-sigma}) of $S$ the first map of the \pleb{} complex takes the following form
\begin{align}
d_1 \xi = ( \partial^\mu \xi_\mu, \frac{1}{4} \Sigma^{i\mu\nu} \partial_\mu \xi_\nu, \frac{1}{2}( \partial_{\mu} \xi_{\nu}+ \partial_{\nu} \xi_{\mu}- \frac{1}{2} g_{\mu\nu} \partial^\alpha\xi_\alpha)).
\end{align}
Here the notation is that we list the irreducible components in $S$ in brackets, in the order they appear in (\ref{S-spaces}). The second operator is given by
\begin{align}
d_2 \sigma = \frac{1}{2} J_1^{-1}( \epsilon_\mu{}^{\alpha\beta\gamma} \partial_\alpha \sigma^i_{\beta\gamma}).
\end{align}
Let us compute this in the parametrisation (\ref{param-sigma}). We have
\begin{align}
\epsilon_\mu{}^{\alpha\beta\gamma} \partial_\alpha \sigma^i_{\beta\gamma} = 2 (\Sigma^{i\alpha\beta} \partial_\alpha h_{\mu\beta} - \Sigma^i_\mu{}^\beta \partial^\alpha h_{\alpha\beta} + \Sigma^i_\mu{}^\alpha \partial_\alpha h) + 4 \epsilon^{ijk} \Sigma^j_\mu{}^\alpha \partial_\alpha h^k.
\end{align}
Using $J_1^{-1} = (1/2)(J_1-\mathbb{I})$ we have
\begin{align}\nonumber
d_2 \sigma = \frac{1}{2} \epsilon^{ijk}\Sigma^j_\mu{}^\nu (\Sigma^{k\alpha\beta} \partial_\alpha h_{\nu\beta} - \Sigma^k_\nu{}^\beta \partial^\alpha h_{\alpha\beta} + \Sigma^k_\nu{}^\alpha \partial_\alpha h) - \frac{1}{2} (\Sigma^{i\alpha\beta} \partial_\alpha h_{\mu\beta} - \Sigma^i_\mu{}^\beta \partial^\alpha h_{\alpha\beta} + \Sigma^i_\mu{}^\alpha \partial_\alpha h)
\\ \nonumber 
+ \epsilon^{ijk}\Sigma^j_\mu{}^\nu \epsilon^{klm} \Sigma^l_\nu{}^\alpha \partial_\alpha h^m - \epsilon^{ijk} \Sigma^j_\mu{}^\alpha \partial_\alpha h^k.
\end{align}
The first term here is
\begin{align}
\frac{1}{2} \epsilon^{ijk}\Sigma^{j\mu\nu} \Sigma^{k\rho\sigma} \partial_\rho h_{\nu\sigma} = \frac{1}{2}  (- g^{\mu\rho} \Sigma^{i\nu\sigma} + g^{\nu\rho} \Sigma^{i\mu\sigma} + g^{\mu\sigma} \Sigma^{i\nu\rho} - g^{\nu\sigma} \Sigma^{i\mu\rho})  \partial_\rho h_{\nu\sigma}=\\ \nonumber
\frac{1}{2} \Sigma^{i\mu\sigma} \partial^\nu h_{\nu\sigma}+ \frac{1}{2} \Sigma^{i\nu\rho} \partial_\rho h_{\nu\mu} - \frac{1}{2} \Sigma^{i\mu\rho} \partial_\rho h.
\end{align}
The second term
\begin{align}
- \frac{1}{2} \epsilon^{ijk}\Sigma^j_\mu{}^\nu  \Sigma^k_\nu{}^\beta \partial^\alpha h_{\alpha\beta} =- \Sigma^i_\mu{}^\beta  \partial^\alpha h_{\alpha\beta} .
\end{align}
The third is
\begin{align}
\frac{1}{2} \epsilon^{ijk}\Sigma^j_\mu{}^\nu \Sigma^k_\nu{}^\alpha \partial_\alpha h =  \Sigma^i_\mu{}^\alpha  \partial_\alpha h.
\end{align}
The first term in the second line is
\begin{align}\nonumber
\epsilon^{ijk}\Sigma^j_\mu{}^\nu \epsilon^{klm} \Sigma^l_\nu{}^\alpha \partial_\alpha h^m = ( \delta^{il}\delta^{jm} - \delta^{im}\delta^{jl}) ( - \delta^{jl} \delta_\mu^\alpha + \epsilon^{jls}\Sigma^s_\mu{}^\alpha) \partial_\alpha h^m = 2 \partial_\mu h^i + \epsilon^{ijk} \Sigma^j_\mu{}^\alpha \partial_\alpha h^k.
\end{align}
Combining everything we observe multiple cancellations, with the final result being
\begin{align}\label{d2-explicitly}
d_2 \sigma = -  \Sigma^{i\alpha\beta} \partial_\alpha h_{\mu\beta} +2 \partial_\mu h^i.
\end{align}
Let us now exhibit the tracefree part and the trace of $h_{\mu\nu}$ explicitly. We parametrise
\begin{align}\label{h-tilde}
h_{\mu\nu} = \tilde{h}_{\mu\nu} + \frac{1}{4} g_{\mu\nu} h.
\end{align}
In this parametrisation 
\begin{align}
d_2 \sigma = \frac{1}{4} \Sigma^i_\mu{}^\alpha \partial_\alpha h +2 \partial_\mu h^i -  \Sigma^{i\alpha\beta} \partial_\alpha \tilde{h}_{\mu\beta} .
\end{align}
The last operator of the complex is
\begin{align}
d_3 a = \epsilon^{ijk} \Sigma^{j\mu\nu} \partial_\mu a_\nu^k.
\end{align}

\subsection{Inner products}

We will now write the general form of the inner products in all the spaces. We will be identifying $TM\sim \Lambda^1$ using the metric. The inner products on $\Lambda^1$ and $E$ are unique (up to an overall multiple which we choose to be unity). The spaces $S, E\otimes \Lambda^1$ each contains several irreducible components, and the inner product is no longer unique. We will introduce arbitrary (at this stage) constants to parametrise the arising inner products. The inner product in $\Lambda^1$ is
\begin{align}
\langle \xi,\xi\rangle = \int_M (\xi_\mu)^2
\end{align}
where $(\xi_\mu)^2=\xi^\mu \xi_\mu$ denotes the metric contraction. The inner product in $S$ an arbitrary linear combination of inner products on all 3 irreducible pieces composing $S$
\begin{align}
\langle \sigma,\sigma \rangle = \int_M \beta_1 h^2 + \beta_2 (h^i)^2 + \beta_3 (\tilde{h}_{\mu\nu})^2.
\end{align}
The inner product in $E\otimes \Lambda^1$ is similarly a linear combination of two terms
\begin{align}
\langle a,a\rangle = \int_M \gamma_1 (a_\mu^i)^2 + \gamma_2 \epsilon^{ijk} \Sigma^{i\mu\nu} a^j_\mu a^k_\nu.
\end{align}
The inner product in $E$ is
\begin{align}
\langle \chi,\chi\rangle = \int_M (\chi^i)^2.
\end{align}

\subsection{General form of the adjoints}

We will first write down the most general form that the adjoint operators can take (based on the relevant representation theory). For the adjoint of the first operator we have
\begin{align}\label{d1-adj-gen}
d_1^* \sigma = a_1' \partial_\mu h + a_2' \Sigma^i_\mu{}^\nu \partial_\nu h^i + a_3' \partial^\nu \tilde{h}_{\mu\nu}.
\end{align}
For the second operator we have
\begin{align}\label{d2-adj-gen}
d_2^* a = (b_1' \Sigma^{i\mu\nu} \partial_\mu a_\nu^i,\ b_2' \partial^\mu a_\mu^i + b_3' \epsilon^{ijk} \Sigma^{j\mu\nu} \partial_\mu a_\nu^k,\ 
2b_4' \Sigma^i_{\langle \mu}{}^\alpha \partial_{\nu\rangle }a_\alpha^i +  2b_5' \Sigma^i_{\langle \mu}{}^\alpha \partial_\alpha a_{\nu\rangle }^i).
\end{align}
The third adjoint has the form
\begin{align}\label{d3-adj-gen}
d_3^* \chi = c_1' \partial_\mu \chi^i + c_2' \epsilon^{ijk} \Sigma^j_\mu{}^\alpha \partial_\alpha \chi^k.
\end{align}

\subsection{\pleb{} case adjoints}
We now compute the \pleb{} complex adjoints. For $\langle \sigma, d_1 \xi\rangle = \langle d_1^* \sigma, \xi\rangle$ we get
\begin{align}
\int_M \beta_1 h \partial^\mu \xi_\mu + \frac{\beta_2}{4}  h^i \Sigma^{i\mu\nu} \partial_\mu \xi_\nu + \beta_3  \tilde{h}^{\mu\nu} \partial_\mu \xi_\nu =\\ \nonumber
\int_M \xi^\mu (a_1' \partial_\mu h + a_2' \Sigma^i_\mu{}^\nu \partial_\nu h^i + a_3' \partial^\nu \tilde{h}_{\mu\nu}).
\end{align}
So, we have
\begin{align}
a_1'= -\beta_1 , \qquad  a_2'= \frac{\beta_2}{4} , \qquad a_3'= - \beta_3.
\end{align}

For $\langle a, d_2 \sigma \rangle = \langle d_2^* a, \sigma \rangle$ we have
\begin{align}\nonumber
& \int_M \gamma_1 a^{i\mu} (  \frac{1}{4} \Sigma^i_\mu{}^\alpha \partial_\alpha h +2 \partial_\mu h^i -  \Sigma^{i\alpha\beta} \partial_\alpha \tilde{h}_{\mu\beta} )
 \\ \nonumber & \hspace{100pt} + \gamma_2 \epsilon^{ijk} \Sigma^{i\mu\nu} a^j_\mu ( \frac{1}{4} \Sigma^k_\nu{}^\alpha \partial_\alpha h +2 \partial_\nu h^k -  \Sigma^{k\alpha\beta} \partial_\alpha \tilde{h}_{\nu\beta}  ) \\ \nonumber & = 
\int_M \beta_1 b_1' h \Sigma^{i\mu\nu} \partial_\mu a_\nu^i + \beta_2 h^i (b_2' \partial^\mu a_\mu^i \\ \nonumber & \vspace{100pt} + b_3' \epsilon^{ijk} \Sigma^{j\mu\nu} \partial_\mu a_\nu^k) + \beta_3 \tilde{h}^{\mu\nu} (2b_4' \Sigma^i_{(\mu}{}^\alpha \partial_{\nu)}a_\alpha^i +  2b_5' \Sigma^i_{(\mu}{}^\alpha \partial_\alpha a_{\nu)}^i).
\end{align}
The last term in the first line gives
\begin{align}
\epsilon^{ijk} \Sigma^{i\mu\nu}   \Sigma^{k\rho\sigma} a^j_\mu \partial_\rho \tilde{h}_{\nu\sigma} & = ( g^{\mu\rho} \Sigma^{i\nu\sigma} - g^{\nu\rho} \Sigma^{i\mu\sigma} - g^{\mu\sigma} \Sigma^{i\nu\rho} + g^{\nu\sigma} \Sigma^{i\mu\rho})a^i_\mu \partial_\rho \tilde{h}_{\nu\sigma} \\ \nonumber & = 
- \Sigma^{i\mu\sigma}a^i_\mu \partial^\nu \tilde{h}_{\nu\sigma} - \Sigma^{i\nu\rho} a^{i\mu} \partial_\rho \tilde{h}_{\nu\mu},
\end{align}
where we used the fact that $\tilde{h}_{\mu\nu}$ is traceless. This means we have
\begin{align}\nonumber
\int_M \frac{\gamma_1- 2\gamma_2}{4} \Sigma^{i\mu\rho}a^i_\mu \partial_\rho h +2\gamma_1 a^{i\mu} \partial_\mu h^i + 2\gamma_2 \epsilon^{ijk} \Sigma^{i\mu\nu} a^j_\mu  \partial_\nu h^k 
\\ \nonumber - (\gamma_1 +\gamma_2)a^{i\mu} \Sigma^{i\alpha\beta} \partial_\alpha \tilde{h}_{\mu\beta} +\gamma_2 \Sigma^{i\mu\sigma}a^i_\mu \partial^\nu \tilde{h}_{\nu\sigma} 
 \\ \nonumber =
\int_M \beta_1 b_1' h \Sigma^{i\mu\nu} \partial_\mu a_\nu^i + \beta_2 h^i (b_2' \partial^\mu a_\mu^i + b_3' \epsilon^{ijk} \Sigma^{j\mu\nu} \partial_\mu a_\nu^k) \\ \nonumber  + \beta_3 \tilde{h}^{\mu\nu} (2b_4' \Sigma^i_{(\mu}{}^\alpha \partial_{\nu)}a_\alpha^i +  2b_5' \Sigma^i_{(\mu}{}^\alpha \partial_\alpha a_{\nu)}^i).
\end{align}
This gives
\begin{align}
b_1' =  \frac{\gamma_1-2\gamma_2}{4\beta_1}, \quad b_2' = - \frac{2\gamma_1}{\beta_2}, \quad b_3'= \frac{2\gamma_2}{\beta_2}, \quad b_4'= \frac{\gamma_2}{2\beta_3}, \quad b_5'= - \frac{\gamma_1+\gamma_2}{2\beta_3}.
\end{align}

For $\langle \chi, d_3 a\rangle = \langle d_3^* \chi, a\rangle$ we have
\begin{align}
& \int_M  \chi^i  \epsilon^{ijk} \Sigma^{j\mu\nu} \partial_\mu a_\nu^k  \\ \nonumber & =\int_M 
\gamma_1 a^{i\mu} (c_1' \partial_\mu \chi^i + c_2' \epsilon^{ijk} \Sigma^j_\mu{}^\alpha \partial_\alpha \chi^k) + \gamma_2 \epsilon^{ijk} \Sigma^{i\mu\nu} a_\mu^j (c_1' \partial_\nu \chi^k + c_2' \epsilon^{klm} \Sigma^l_\nu{}^\alpha \partial_\alpha \chi^m), 
\end{align}
or 
\begin{align}
& \int_M  \chi^i  \epsilon^{ijk} \Sigma^{j\mu\nu} \partial_\mu a_\nu^k \\ \nonumber & = \int_M 
(\gamma_1 c_1' - 2 \gamma_2 c_2' ) a^{i\mu}  \partial_\mu \chi^i +  (\gamma_1 c_2'  - \gamma_2 c_1' - \gamma_2 c_2' ) \epsilon^{ijk} a^{i\mu} \Sigma^j_\mu{}^\alpha \partial_\alpha \chi^k ,
\end{align}
and so
\begin{align}
\gamma_1 c_1' - 2 \gamma_2 c_2' =0, \qquad 1  = -\gamma_1 c_2'  + \gamma_2 c_1' + \gamma_2 c_2' .
\end{align}
This gives
\begin{align}
c_1' = \frac{2\gamma_2}{\gamma_1(\gamma_2-\gamma_1) + 2\gamma_2^2}, \quad c_2' = \frac{\gamma_1}{\gamma_1(\gamma_2-\gamma_1) + 2\gamma_2^2}.
\end{align}
We have explicitly checked that the compositions of the adjoint operators (with these coefficients) vanish, as they should. 

 \subsection{Imposing the \texorpdfstring{$\Delta$}{} condition}
 
 We now form the operators $D$ and $D^*$, see (\ref{D}) and (\ref{D*}). Requiring $D^*D$ to be a multiple of the $\Delta=\partial^\mu\partial_\mu$ operator, we are led to equations (\ref{D2-eqs}). Substituting the values of all the coefficients as they are (\ref{Pleb-coeffs}) in the \pleb{} complex case, and expressing the coefficients of the adjoint operators via those in the inner products, we can view the equations  (\ref{D2-eqs}) as those on the inner product coefficients. The solution is as follows
 \begin{align}
\langle \sigma,\sigma \rangle = \beta \int_M \frac{1}{4} h^2 + 8 (h^i)^2 +  (\tilde{h}_{\mu\nu})^2, \qquad
\langle a,a\rangle = \beta^2 \int_M  (a_\mu^i)^2 .
\end{align}
It also makes sense to choose $\beta=1$. We note that the relative coefficient in the $h^2, (\tilde{h}_{\mu\nu})^2$ terms is the natural one. We can pass to treating $h_{\mu\nu}$ as an object that contains the trace part. Indeed, using (\ref{h-tilde}) we have
\begin{align}\label{trace-tracefree}
(h_{\mu\nu})^2 = (\tilde{h}_{\mu\nu})^2 + \frac{1}{4} h^2, 
\end{align}
which is precisely what we have in the above inner product. The inner products that appear in this construction are then
\begin{align}\label{inner-prod-1}
\langle \xi,\xi \rangle =\int_M (\xi_\mu)^2, \quad
\langle \sigma,\sigma \rangle = \int_M 8 (h^i)^2 +  (h_{\mu\nu})^2, \quad
\langle a,a\rangle =  \int_M  (a_\mu^i)^2 , \quad
\langle \xi,\xi\rangle =  \int_M  (\xi^i)^2.
\end{align}

\subsection{Inner product on \texorpdfstring{$S$ in terms of $\sigma^i_{\mu\nu}$}{}} 

It is interesting to rewrite the inner product on $S$ in (\ref{inner-prod-1}) in terms of the object $\sigma^i_{\mu\nu}\in S$. Using the parametrisation (\ref{param-sigma}) we have
\begin{align}\nonumber
(\sigma^i_{\mu\nu})^2 = 2 (h_\mu{}^\alpha \Sigma^i_{\alpha\nu} - h_\nu{}^\alpha \Sigma^i_{\alpha\mu}) h^{\mu\beta} \Sigma^i_\beta{}^\nu + 8 h_\mu{}^\alpha \Sigma^i_{\alpha\nu} \epsilon^{ijk} \Sigma^{j\mu\nu} \chi^k + 4 \epsilon^{ijk} \Sigma^j_{\mu\nu} \chi^k \epsilon^{imn} \Sigma^{m\mu\nu} \chi^n= \\ \nonumber
2( 3 (h_{\mu\nu})^2 - h_\nu{}^\alpha h^{\mu\beta} ( g_{\alpha\beta} \delta_\mu{}^\nu - \delta_\alpha{}^\nu g_{\mu\beta} + \epsilon_{\alpha\mu\beta}{}^\nu)) + 32 (\chi^i)^2 = \\ \nonumber
2( 2(h_{\mu\nu})^2 + h^2) + 32 (\chi^i)^2.
\end{align}
In the parametrisation (\ref{h-tilde}) this becomes
\begin{align}
\frac{1}{4} (\sigma^i_{\mu\nu})^2 = (\tilde{h}_{\mu\nu})^2 + \frac{3}{4} h^2 + 8 (\chi^i)^2.
\end{align}
We also have
\begin{align}
\Sigma^{i\mu\nu} \sigma^i_{\mu\nu} = 2 \Sigma^{i\mu\nu} h_\mu{}^\alpha \Sigma^i_{\alpha\nu} = 6h.
\end{align}
This means that 
\begin{align}
\frac{1}{4} h^2 + 8 (h^i)^2 +  (\tilde{h}_{\mu\nu})^2 = \frac{1}{4} (\sigma^i_{\mu\nu})^2 - \frac{1}{72} ( \Sigma^{i\mu\nu} \sigma^i_{\mu\nu} )^2.
\end{align}

\subsection{Explicit form of the adjoints}

We now write the \pleb{} complex operators without splitting the tracefree and the trace parts of $h_{\mu\nu}$
\begin{align}
d_1 \xi = ( \frac{1}{4} \Sigma^{i\mu\nu} \partial_\mu \xi_\nu, \partial_{(\mu} \xi_{\nu)}) \in (E,S^2 T^*M) , \\ \nonumber
d_2 \sigma = -  \Sigma^{i\alpha\beta} \partial_\alpha h_{\mu\beta} +2 \partial_\mu h^i, \\ \nonumber
d_3 a = \epsilon^{ijk} \Sigma^{j\mu\nu} \partial_\mu a_\nu^k.
\end{align}
The adjoints, in the same notation are given by 
\begin{align}
d_1^* \sigma = 2 \Sigma^i_\mu{}^\nu \partial_\nu h^i - \partial^\nu h_{\mu\nu}, \\ \nonumber
d_2^* a =( -\frac{1}{4} \partial^\mu a_\mu^i , - \Sigma^i_{(\mu}{}^\alpha \partial_\alpha a_{\nu)}^i) \in (E,S^2 T^*M), \\ \nonumber
d_3^* \chi =  - \epsilon^{ijk} \Sigma^j_\mu{}^\alpha \partial_\alpha \chi^k.
\end{align}
As a check, we recompute the operators that appear in $D^* D$. The operator $(d_1 d_1^* +d_2^* d_2 ) \sigma$ becomes, in the $h^i$ component
\begin{align}
 =  \frac{1}{4} \Sigma^{i\mu\nu} \partial_\mu (2 \Sigma^j_\nu{}^\alpha \partial_\alpha h^j - \partial^\alpha h_{\nu\alpha})
 -\frac{1}{4} \partial^\mu (-  \Sigma^{i\alpha\beta} \partial_\alpha h_{\mu\beta} +2 \partial_\mu h^i) = \\ \nonumber
 -\frac{1}{2}   \partial^\mu \partial_\mu h^i -  \frac{1}{4} \Sigma^{i\mu\nu} \partial_\mu\partial^\alpha h_{\nu\alpha}
 + \frac{1}{4}   \Sigma^{i\alpha\beta} \partial^\mu \partial_\alpha h_{\mu\beta} - \frac{1}{2} \partial^\mu\partial_\mu h^i = -  \partial^\mu\partial_\mu h^i .
 \end{align}
 In the $h_{\mu\nu}$ component we get
 \begin{align}
 \partial_{(\mu} (2 \Sigma^i_{\nu)}{}^\alpha \partial_\alpha h^i - \partial^\alpha h_{\nu)\alpha})- \Sigma^i_{(\mu}{}^\alpha \partial_\alpha  (-  \Sigma^{i\rho\sigma} \partial_\rho h_{\nu)\sigma} +2 \partial_{\nu)} h^i)= \\ \nonumber
 2 \Sigma^i_{(\mu}{}^\alpha \partial_{\nu)} \partial_\alpha h^i -  \partial_{(\mu} \partial^\alpha h_{\nu)\alpha} + ( \delta_{(\mu}{}^\rho g^{\alpha\sigma} - \delta_{(\mu}{}^\sigma g^{\alpha\rho}) \partial_\alpha \partial_\rho h_{\nu)\sigma} - 2\Sigma^i_{(\mu}{}^\alpha \partial_\alpha \partial_{\nu)} h^i = - \partial^\alpha \partial_\alpha h_{\mu\nu}.
 \end{align}
 For the operator $d_3 d_3^* \chi$ we have
 \begin{align}
 d_3 d_3^* \chi= \epsilon^{ijk} \Sigma^{j\mu\nu} \partial_\mu (- \epsilon^{klm} \Sigma^l_\nu{}^\alpha \partial_\alpha \chi^m)=
 - \epsilon^{ijk}\epsilon^{klm} (- \delta^{jl} g^{\mu\alpha}) \partial_\mu\partial_\alpha \chi^m = - 2 \partial^\mu \partial_\mu \chi^i,
 \end{align}
 which confirms that $D^* D\sim \Delta$.  
 
\subsection{Linearised Einstein equations}

As we have seen previously, see (\ref{einstein-lin-1+9}), the linearisation of the Einstein equations is 
\begin{align}\label{correct-d2-star}
\Sigma^i_{(\mu}{}^\alpha (\partial_{\nu)} a_\alpha^i - \partial_\alpha a_{\nu)}^i) =0.
\end{align}
We would like to realise the operator that appears in this equation as $d_2^*$, but this is not the operator $d_2^*$ that arises if the inner products are chosen as in (\ref{inner-prod-1}) to satisfy $D^*D\sim \Delta$ condition. We now change the inner product to obtain the desired $d_2^*$, but this will then be in conflict with the $D^*D\sim \Delta$ condition. Still, we develop this option as the one that is relevant for the Einstein equations. The construction of an elliptic operator $\tilde{D}$ with $\tilde{D}^*\tilde{D}\sim \Delta$ will occupy us in the next sections. 

\subsection{\pleb{} case different inner product}

We now consider the case of \pleb{} complex with the inner product $\gamma_1=0$. In this case $b_3=b_4=0$, and it is not hard to see that this gives $b_4'+b_5'=0, b_2'=0$. The other coefficients are
\begin{align}
b_1'= -\frac{\gamma_2}{2\beta_1}, \quad b_3'= \frac{2\gamma_2}{\beta_2}, \quad b_4'=-b_5'= \frac{\gamma_2}{2\beta_3}
\end{align}
so that 
\begin{align}\nonumber
d_2^* a = \gamma_2 (-\frac{1}{2\beta_1} \Sigma^{i\mu\nu} \partial_\mu a_\nu^i,  \frac{2}{\beta_2} \epsilon^{ijk} \Sigma^{j\mu\nu} \partial_\mu a_\nu^k, \frac{1}{\beta_3}(\Sigma^i_{\langle\mu}{}^\alpha \partial_{\nu\rangle}a_\alpha^i -\Sigma^i_{\langle\mu}{}^\alpha \partial_\alpha a_{\nu\rangle}^i)).
\end{align}
Note that the desired operator (\ref{correct-d2-star}) appears here. We now choose 
\begin{align}
\gamma_2=1, \qquad \beta_1=\frac{1}{4}, \qquad \beta_2 = 8, \qquad \beta_3=1,
\end{align}
which corresponds to the following inner products
\begin{align}\label{inner-prod-pleb}
\langle \xi,\xi \rangle =\int_M (\xi_\mu)^2, \quad
\langle \sigma,\sigma \rangle = \int_M \frac{1}{4} h^2 +8 (h^i)^2 +  (\tilde{h}_{\mu\nu})^2, \\ \nonumber
\langle a,a\rangle =  \int_M  \epsilon^{ijk} \Sigma^{i\mu\nu} a_\mu^j a_\nu^k, \quad
\langle \xi,\xi\rangle =  \int_M  (\xi^i)^2.
\end{align}
These are the same inner products in all the spaces as in (\ref{inner-prod-1}), apart from that in $E\otimes\Lambda^1$, where we instead take the inner product that contains the $J_1$ operator. This makes the inner product on $E \otimes \Lambda^1$ not positive-definite. An important consequence of this is that the composition of maps with their adjoints will not be elliptic and therefore also not equal to $\Delta$. With these inner products, the full complex becomes
\begin{align}
d_1 \xi = ( \partial^\mu \xi_\mu, \frac{1}{4} \Sigma^{i\mu\nu} \partial_\mu \xi_\nu, \partial_{\langle\mu} \xi_{\nu\rangle}) \in (E,S^2 T^*M) , \\ \nonumber
d_2 \sigma = \frac{1}{4} \Sigma^i_\mu{}^\nu \partial_\nu h +2 \partial_\mu h^i -  \Sigma^{i\alpha\beta} \partial_\alpha \tilde{h}_{\mu\beta} , \\ \nonumber
d_3 a = \epsilon^{ijk} \Sigma^{j\mu\nu} \partial_\mu a_\nu^k,
\end{align}
and
\begin{align}\nonumber
d_1^* \sigma =   -\frac{1}{4}\partial_\mu h + 2\Sigma^i_\mu{}^\nu \partial_\nu h^i -  \partial^\nu \tilde{h}_{\mu\nu}, \\ \nonumber
d_2^* a = ( -2 \Sigma^{i\mu\nu} \partial_\mu a_\nu^i,\ \frac{1}{4} \epsilon^{ijk} \Sigma^{j\mu\nu} \partial_\mu a_\nu^k,\ \Sigma^i_{\langle\mu}{}^\alpha \partial_{\nu\rangle}a_\alpha^i-\Sigma^i_{\langle\mu}{}^\alpha \partial_\alpha a_{\nu\rangle}^i), \\ \nonumber
d_3^* \chi =   \partial_\mu \chi^i.
\end{align}
We will refer to the operators appearing above as those of the \pleb{} complex. Here $\langle \rangle$ denotes the symmetric tracefree part. 

\subsection{Computation of \texorpdfstring{$d_2^* d_2$}{}}

As a check, we compute the operator $d_2^* d_2$ for the operators of the \pleb{} complex as detailed above. We will need the result of this computation below. For the trace part we have
\begin{align}
-2 \Sigma^{i\mu\nu} \partial_\mu (\frac{1}{4} \Sigma^i_\nu{}^\alpha \partial_\alpha h +2 \partial_\nu h^i -  \Sigma^{i\rho\sigma} \partial_\rho \tilde{h}_{\nu\sigma} )=
\frac{3}{2} \partial^\mu \partial_\mu h - 2 \partial^\mu \partial^\nu \tilde{h}_{\mu\nu}.
\end{align}
For the vector part we have
\begin{align}
\frac{1}{4} \epsilon^{ijk} \Sigma^{j\mu\nu} \partial_\mu (\frac{1}{4} \Sigma^k_\nu{}^\alpha \partial_\alpha h +2 \partial_\nu h^k -  \Sigma^{k\rho\sigma} \partial_\rho \tilde{h}_{\nu\sigma} )=
\\ \nonumber
- \frac{1}{4}( g^{\mu\rho} \Sigma^{i\nu\sigma} - g^{\nu\rho} \Sigma^{i\mu\sigma} - g^{\mu\sigma} \Sigma^{i\nu\rho} + g^{\nu\sigma} \Sigma^{i\mu\rho}) \partial_\mu\partial_\rho \tilde{h}_{\nu\sigma} =0.
\end{align}
For the tracefree part we have
\begin{align}\nonumber
\Sigma^i_{\langle\mu}{}^\alpha \partial_{\nu\rangle}(\frac{1}{4} \Sigma^i_\alpha{}^\beta \partial_\beta h +2 \partial_\alpha h^i -  \Sigma^{i\rho\sigma} \partial_\rho \tilde{h}_{\alpha\sigma} )
-\Sigma^i_{\langle\mu}{}^\alpha \partial_\alpha (\frac{1}{4} \Sigma^i_{\nu\rangle}{}^\beta \partial_\beta h +2 \partial_{\nu\rangle} h^i -  \Sigma^{i\rho\sigma} \partial_\rho \tilde{h}_{\nu\rangle\sigma} )= \\ \nonumber
- \frac{3}{4} \partial_{\langle\mu} \partial_{\nu\rangle} h + 2 \Sigma^i_{\langle\mu}{}^\alpha \partial_{\nu\rangle}\partial_\alpha h^i + \partial_{\langle\mu}\partial^\rho \tilde{h}_{\nu\rangle \alpha}
+ \frac{1}{4} \partial_{\langle\mu} \partial_{\nu\rangle} h - 2 \Sigma^i_{\langle\mu}{}^\alpha \partial_{\nu\rangle}\partial_\alpha h^i + \partial_{\langle\mu}\partial^\rho \tilde{h}_{\nu\rangle \alpha}- \partial^\alpha \partial_\alpha \tilde{h}_{\mu\nu}= \\ \nonumber
-  \frac{1}{2} \partial_{\langle\mu} \partial_{\nu\rangle} h+ 2\partial_{\langle\mu}\partial^\rho \tilde{h}_{\nu\rangle \alpha}- \partial^\alpha \partial_\alpha \tilde{h}_{\mu\nu}.
\end{align}
Collecting the results we have
\begin{align}\label{d2}
d_2^* d_2 \sigma = ( \frac{3}{2} \partial^\mu \partial_\mu h - 2 \partial^\mu \partial^\nu \tilde{h}_{\mu\nu}, 0, -  \frac{1}{2} \partial_{\langle\mu} \partial_{\nu\rangle} h+ 2\partial_{\langle\mu}\partial^\rho \tilde{h}_{\nu\rangle \alpha}- \partial^\alpha \partial_\alpha \tilde{h}_{\mu\nu}).
\end{align}

\section{General complex}

We now change the game, and start by defining a general set of differential operators between the same spaces as appear in the \pleb{} complex. We will write down the operators most general as allowed by the relevant representation theory. The representation theory exercise is straightforward, for example the first-order differential maps from $\Lambda^1$ to $S$ can be computed by first associating the spaces and operators with
\begin{align}
    \partial_\mu = S_+ \otimes S_-, \quad \Lambda^1 = S_+ \otimes S_-, \quad S = C^\infty \oplus S_+^2 \oplus S_-^2 \oplus S_+^2 \otimes S_-^2. 
\end{align}
Then the first-order differential operators on $\xi \in \Lambda^1$ are decomposed into
\begin{align}
    \partial_\mu \xi_\nu &= S_+ \otimes S_- \otimes S_+ \otimes S_- = C^\infty \oplus S_+^2 \oplus S_-^2 \oplus S_+^2 \otimes S^2_-\nonumber \\ &= \frac{1}{4} g_{\mu\nu} \partial^\rho \xi_\rho + \frac{1}{2} \Sigma^i_{\mu\nu} \Sigma^{i\rho\sigma} \partial_\rho \xi_\sigma + \frac{1}{2} \asd^i_{\mu\nu} \asd^{i\rho\sigma} \partial_\rho \xi_\sigma + \partial_{\langle \mu} \xi_{\nu \rangle}.
\end{align}
Where $\partial_{\langle \mu} \xi_{\nu \rangle} = \frac{1}{2} \left( \partial_\mu \xi_\nu + \partial_\nu \xi_\mu - \frac{1}{2} g_{\mu\nu} \partial^\rho \xi_\rho \right)$ is the symmetric tracefree part. More generally one calculates the different number of copies of one of the spaces in (\ref{complex-spaces}) in the tensor product of another space in (\ref{complex-spaces}) and the space $\Lambda^1$. One can then write down the relevant differential operators without difficulty. 

\subsection{Differential operators defined}

We now attempt to find the most general complex of the type
\begin{equation}
\begin{tikzcd}\label{complex-spaces}
\Lambda^1 \arrow{r}{d_1} & S \arrow{r}{d_2} & E\otimes \Lambda^1 \arrow{r}{d_3} & E
\end{tikzcd}
\end{equation}
The possible maps are dictated by the representation theory. They can be parameterised as
\begin{align}\label{d1-gen}
d_1 \xi = ( a_1 \partial^\mu \xi_\mu, a_2 \Sigma^{i\mu\nu} \partial_\mu \xi_\nu, 2 a_3 \partial_{\langle\mu} \xi_{\nu\rangle}) \in ( \Lambda^0, E, {\rm Sym}_0^2 T^*M) ,
\end{align}
where $a_1, a_2, a_3$ are arbitrary coefficients and
\begin{align}
\partial_{\langle\mu} \xi_{\nu\rangle}=
\frac{1}{2}\partial_{\mu} \xi_{\nu} + \frac{1}{2}\partial_{\nu} \xi_{\mu} - \frac{1}{4} g_{\mu\nu} \partial^\alpha \xi_\alpha
\end{align}
is the symmetric tracefree part. 
Here ${\rm Sym}_0^2 T^*M$ is the space of symmetric tracefree tensors. For the second map we have
\begin{align}\label{d2-gen}
d_2 \sigma = b_1 \Sigma^i_\mu{}^\alpha \partial_\alpha h + b_2 \partial_\mu h^i + b_3 \epsilon^{ijk} \Sigma^j_\mu{}^\alpha \partial_\alpha h^k + b_4 \Sigma^i_\mu{}^\rho \partial^\sigma \tilde{h}_{\rho\sigma} + b_5 \Sigma^{i\rho\sigma}  \partial_\rho \tilde{h}_{\mu\sigma} \in E\otimes\Lambda^1.
\end{align}
Again, $b_{1,2,3,4,5}$ are arbitrary coefficients. The last map is
\begin{align}\label{d3-gen}
d_3 a = c_1 \partial^\mu a_\mu^i + c_2 \epsilon^{ijk} \Sigma^{j\mu\nu} \partial_\mu a_\nu^k \in E.
\end{align}

\subsection{Compositions}

We want to the above to be a differential complex, so we want the compositions of the arrows to give zero. We have
\begin{align}
d_2 d_1 \xi =\ & b_1 a_1 \Sigma^i_\mu{}^\rho \partial_\rho \partial^\sigma \xi_\sigma + b_2 a_2 \Sigma^{i\rho\sigma} \partial_\mu \partial_\rho \xi_\sigma + b_3 a_2 \epsilon^{ijk} \Sigma^j_\mu{}^\nu \partial_\nu \Sigma^{k\rho\sigma} \partial_\rho \xi_\sigma \\ \nonumber &
+ b_4 a_3 \Sigma^i_\mu{}^\rho \partial^\sigma (\partial_\rho \xi_\sigma + \partial_\sigma \xi_\rho- \frac{1}{2} g_{\rho\sigma}\partial^\alpha \xi_\alpha ) 
+ b_5 a_3 \Sigma^{i\rho\sigma} \partial_\rho (\partial_\mu \xi_\sigma + \partial_\sigma \xi_\mu- \frac{1}{2} g_{\mu\sigma}\partial^\alpha \xi_\alpha).
\end{align}
Next we use
\begin{align}
\epsilon^{ijk} \Sigma^j_{\mu\nu}  \Sigma^{k}_{\rho\sigma} = - g_{\mu\rho} \Sigma^i_{\nu\sigma} + g_{\nu\rho} \Sigma^i_{\mu\sigma} + g_{\mu\sigma} \Sigma^i_{\nu\rho} - g_{\nu\sigma} \Sigma^i_{\mu\rho}
\end{align}
to get
\begin{align}
d_2 d_1 \xi =\ & (b_1 a_1+b_4 a_3-b_3 a_2 + \frac{b_5 a_3 - b_4 a_3}{2}) \Sigma^i_\mu{}^\rho \partial_\rho \partial^\sigma \xi_\sigma + (b_2 a_2+b_5 a_3-b_3 a_2) \Sigma^{i\rho\sigma} \partial_\mu \partial_\rho \xi_\sigma \\ \nonumber &
+ (b_4 a_3+b_3 a_2) \Sigma^i_\mu{}^\rho \partial^\sigma \partial_\sigma \xi_\rho.
\end{align}
This gives
\begin{align}\label{eqs-b1}
b_1 a_1+b_4 a_3-b_3 a_2+ \frac{b_5 a_3 - b_4 a_3}{2}=0, \qquad b_2 a_2+b_5 a_3-b_3 a_2=0, \qquad b_4 a_3+b_3 a_2=0.
\end{align}

The other composition is as follows
\begin{align}
d_3 d_2 \sigma = & c_1 b_2 \partial^\mu \partial_\mu h^i + c_1 b_4 \Sigma^{i\mu\rho} \partial_\mu \partial^\sigma \tilde{h}_{\rho\sigma} + c_1 b_5 \Sigma^{i\rho\sigma} \partial_\rho \partial^\mu \tilde{h}_{\mu\sigma} \\ \nonumber \ &
+ c_2 \epsilon^{ijk} \Sigma^{j\mu\nu} \partial_\mu ( b_1 \Sigma^k_\nu{}^\alpha \partial_\alpha h + b_2 \partial_\nu h^k + b_3 \epsilon^{klm} \Sigma^l_\nu{}^\alpha \partial_\alpha h^m 
+ b_4 \Sigma^k_\nu{}^\rho \partial^\sigma \tilde{h}_{\rho\sigma} + b_5 \Sigma^{k\rho\sigma} \partial_\rho \tilde{h}_{\nu\sigma}).
\end{align}
The first two terms in the second line do not contribute. The third term gives
\begin{align}
c_2 b_3 \epsilon^{ijk} \Sigma^{j\mu\nu} \partial_\mu \epsilon^{klm} \Sigma^l_\nu{}^\alpha \partial_\alpha h^m = & c_2 b_3 (\delta^{il} \delta^{jm} - \delta^{im}\delta^{jl}) ( - \delta^{jl} g^{\mu\alpha} + \epsilon^{jls}\Sigma^{s\mu\alpha}) \partial_\mu\partial_\alpha h^m  \\ \nonumber
= & 2 c_2 b_3 \partial^\mu \partial_\mu h^i.
\end{align}
The fourth term gives
\begin{align}
c_2 b_4 \epsilon^{ijk} \Sigma^{j\mu\nu}\Sigma^k_\nu{}^\rho  \partial_\mu \partial^\sigma \tilde{h}_{\rho\sigma}= 2 c_2 b_4 \Sigma^{i\mu\rho} \partial_\mu \partial^\sigma \tilde{h}_{\rho\sigma}.
\end{align}
The last term in the second line gives
\begin{align}\nonumber
c_2 b_5  \epsilon^{ijk} \Sigma^{j\mu\nu}  \Sigma^{k\rho\sigma} \partial_\mu \partial_\rho \tilde{h}_{\nu\sigma}=c_2 b_5 (- g^{\mu\rho} \Sigma^{i\nu\sigma} + g^{\nu\rho} \Sigma^{i\mu\sigma} + g^{\mu\sigma} \Sigma^{i\nu\rho} - g^{\nu\sigma} \Sigma^{i\mu\rho}) \partial_\mu \partial_\rho \tilde{h}_{\nu\sigma}= 0.
\end{align}
So, overall
\begin{align}
d_3 d_2 \sigma = (c_1 b_2 +2 c_2 b_3) \partial^\mu \partial_\mu h^i + (c_1 b_4 + c_1 b_5 + 2 c_2 b_4)\Sigma^{i\mu\rho} \partial_\mu \partial^\sigma \tilde{h}_{\rho\sigma}.
\end{align}
Requiring this to vanish gives
\begin{align}\label{eqs-b2}
c_1 b_2 +2 c_2 b_3=0, \qquad c_1 b_4 + c_1 b_5 + 2 c_2 b_4=0.
\end{align}
The system of equations (\ref{eqs-b1}), (\ref{eqs-b2}) can be interpreted as that for the constants $b_{1,2,3,4,5}$ in terms of $a_{1,2,3}, c_{1,2}$. The equations are not all independent, and not all 5 $b$'s can be solved for. Taking one of them as a parameter we have
\begin{align}
 b_2 = - \frac{2 a_1 b_1 c_2 }{a_2(c_1-c_2)}, \quad b_3 = \frac{ a_1 b_1 c_1 }{a_2(c_1-c_2)}, \quad b_4 = - \frac{a_1 b_1 c_1}{a_3(c_1-c_2)}, \quad b_5 = \frac{a_1 b_1 (c_1+2c_2)}{a_3 (c_1-c_2)}.
\end{align}
This means we can make the operator $d_2$ more explicit
\begin{align}\nonumber
d_2 \sigma = b (  \Sigma^i_\mu{}^\alpha \partial_\alpha h    - \frac{2 a_1 c_2 }{a_2(c_1-c_2)} \partial_\mu h^i +  \frac{ a_1 c_1 }{a_2(c_1-c_2)} \epsilon^{ijk} \Sigma^i_\mu{}^\alpha \partial_\alpha h^k \\ \nonumber
- \frac{a_1 c_1}{a_3(c_1-c_2)} \Sigma^i_\mu{}^\rho \partial^\sigma \tilde{h}_{\rho\sigma} + \frac{a_1 (c_1+2c_2)}{a_3 (c_1-c_2)} \Sigma^{i\rho\sigma}  \partial_\rho \tilde{h}_{\mu\sigma}),
\end{align}
where $b$ is an arbitrary parameter.

\subsection{\pleb{} case}

The \pleb{} case coefficients are
\begin{align}\label{Pleb-coeffs}
a_1=1, \quad a_2 = \frac{1}{4}, \quad a_3=\frac{1}{2}, \quad c_1=0, \quad c_2=1, \\ \nonumber
b_1=\frac{1}{4}, \quad b_2=2, \quad b_3=0, \quad b_4=0, \quad b_5=-1.
\end{align}
It can be checked that the compositions vanish with these coefficients, as they should. 

\subsection{Matching adjoints with inner products}

We now parametrise the adjoint operators as in (\ref{d1-adj-gen}), (\ref{d2-adj-gen}), (\ref{d3-adj-gen}). We will keep the inner product to be general, as in Section 5.2.
Let us relate the constants appearing in the adjoint operators to those in the inner products. We start with $d_1^*$. We have
\begin{align}
\langle \sigma, d_1 \xi\rangle = \langle d_1^* \sigma, \xi\rangle,
\end{align}
which gives
\begin{align}
\int_M \beta_1 a_1 h \partial^\mu \xi_\mu + \beta_2 a_2 h^i \Sigma^{i\mu\nu} \partial_\mu \xi_\nu + 2\beta_3 a_3 \tilde{h}^{\mu\nu} \partial_\mu \xi_\nu =\\ \nonumber
\int_M  \xi^\mu (a_1' \partial_\mu h + a_2' \Sigma^i_\mu{}^\nu \partial_\nu h^i + a_3' \partial^\nu \tilde{h}_{\mu\nu}).
\end{align}
So, we have
\begin{align}\label{a-prime}
\beta_1 a_1 = -  a_1', \qquad \beta_2 a_2 =  a_2', \qquad 2\beta_3 a_3 = -  a_3'.
\end{align}

Let us proceed similarly with $d_2^*$. We have
\begin{align}
\langle a, d_2 \sigma \rangle = \langle d_2^* a, \sigma \rangle,
\end{align}
which gives
\begin{align}\nonumber
\int_M \gamma_1 a^{i\mu} ( b_1 \Sigma^i_\mu{}^\alpha \partial_\alpha h + b_2 \partial_\mu h^i + b_3 \epsilon^{ijk} \Sigma^j_\mu{}^\alpha \partial_\alpha h^k + b_4 \Sigma^i_\mu{}^\rho \partial^\sigma \tilde{h}_{\rho\sigma} + b_5 \Sigma^{i\rho\sigma}  \partial_\rho \tilde{h}_{\mu\sigma}) \\ \nonumber
+ \gamma_2 \epsilon^{ijk} \Sigma^{i\mu\nu} a^j_\mu ( b_1 \Sigma^k_\nu{}^\alpha \partial_\alpha h + b_2 \partial_\nu h^k+ b_3 \epsilon^{klm} \Sigma^l_\nu{}^\alpha \partial_\alpha h^m + b_4 \Sigma^k_\nu{}^\rho \partial^\sigma \tilde{h}_{\rho\sigma} + b_5 \Sigma^{k\rho\sigma}  \partial_\rho \tilde{h}_{\nu\sigma}) = \\ \nonumber
\int_M \beta_1 b_1' h \Sigma^{i\mu\nu} \partial_\mu a_\nu^i + \beta_2 h^i (b_2' \partial^\mu a_\mu^i + b_3' \epsilon^{ijk} \Sigma^{j\mu\nu} \partial_\mu a_\nu^k) + \beta_3 \tilde{h}^{\mu\nu} (2b_4' \Sigma^i_{(\mu}{}^\alpha \partial_{\nu)}a_\alpha^i +  2b_5' \Sigma^i_{(\mu}{}^\alpha \partial_\alpha a_{\nu)}^i).
\end{align}

We first match the $h^i$ containing terms. Using the algebra of $\Sigma$'s gives
\begin{align}
\int_M ( \gamma_1 b_2 - 2 \gamma_2 b_3)a^{i\mu} \partial_\mu h^i + (\gamma_1 b_3 - \gamma_2 b_2 - \gamma_2 b_3) \epsilon^{ijk} a_\mu^i \Sigma^j_\mu{}^\alpha \partial_\alpha h^k  = \\ \nonumber
\int_M \beta_2 h^i (b_2' \partial^\mu a_\mu^i + b_3' \epsilon^{ijk} \Sigma^{j\mu\nu} \partial_\mu a_\nu^k) ,
\end{align}
which means
\begin{align}\label{b-prime-1}
\gamma_1 b_2 - 2 \gamma_2 b_3 = - \beta_2 b_2', \qquad \gamma_1 b_3 - \gamma_2 b_2 - \gamma_2 b_3 = - \beta_2 b_3'.
\end{align}

Let us now work out the $\tilde{h}_{\mu\nu}$ terms. The fourth term in the second line is
\begin{align}
\epsilon^{ijk} \Sigma^{i\mu\nu} a^j_\mu \Sigma^k_\nu{}^\rho \partial^\sigma \tilde{h}_{\rho\sigma}=\epsilon^{ijk}  \epsilon^{iks} \Sigma^{s\mu\rho} a^j_\mu\partial^\sigma \tilde{h}_{\rho\sigma}= - 2 \Sigma^{i\mu\rho} a^i_\mu\partial^\sigma \tilde{h}_{\rho\sigma}.
\end{align}
The last term in the second line
\begin{align}
\epsilon^{ijk} \Sigma^{i\mu\nu} a^j_\mu\Sigma^{k\rho\sigma}  \partial_\rho \tilde{h}_{\nu\sigma} & = ( g^{\mu\rho} \Sigma^{i\nu\sigma} - g^{\nu\rho} \Sigma^{i\mu\sigma} - g^{\mu\sigma} \Sigma^{i\nu\rho} + g^{\nu\sigma} \Sigma^{i\mu\rho}) a^i_\mu \partial_\rho \tilde{h}_{\nu\sigma} \\ \nonumber & = 
-\Sigma^{i\mu\sigma}  a^i_\mu \partial^\nu \tilde{h}_{\nu\sigma} - \Sigma^{i\nu\rho} a^i_\mu \partial_\rho \tilde{h}_{\nu\mu},
\end{align}
where we have used that $\tilde{h}_{\mu\nu}$ is tracefree. This means that we have
\begin{align}
\int_M (\gamma_1 b_4 - 2  \gamma_2 b_4-\gamma_2 b_5 ) a^{i\mu} \Sigma^i_\mu{}^\rho \partial^\sigma \tilde{h}_{\rho\sigma} + (\gamma_1 b_5 + \gamma_2 b_5) \Sigma^{i\rho\sigma} a^{i\mu} \partial_\rho \tilde{h}_{\mu\sigma}  = \\ \nonumber
\int_M \beta_3 \tilde{h}^{\mu\nu} (2b_4' \Sigma^i_{(\mu}{}^\alpha \partial_{\nu)}a_\alpha^i +  2b_5' \Sigma^i_{(\mu}{}^\alpha \partial_\alpha a_{\nu)}^i).
\end{align}
This means that
\begin{align}\label{b-prime-2}
\gamma_1 b_4 - 2  \gamma_2 b_4-\gamma_2 b_5  = 2\beta_3 b_4', \qquad \gamma_1 b_5 + \gamma_2 b_5 = 2 \beta_3 b_5'.
\end{align}

We now match the $h$ terms
\begin{align}\nonumber
\int_M \gamma_1 b_1 a^{i\mu}  \Sigma^i_\mu{}^\alpha \partial_\alpha h    
+ \gamma_2 b_1 \epsilon^{ijk} \Sigma^{i\mu\nu} a^j_\mu \Sigma^k_\nu{}^\alpha \partial_\alpha h  = 
\int_M \beta_1 b_1' h \Sigma^{i\mu\nu} \partial_\mu a_\nu^i ,
\end{align}
which gives 
\begin{align}\label{b-prime-3}
b_1 (\gamma_1  - 2\gamma_2 ) = \beta_1 b_1' .
\end{align}

The calculation for the final adjoint is
\begin{align}
\langle \chi, d_3 a\rangle = \langle d_3^* \chi, a\rangle,
\end{align}
which reads
\begin{align}
\int_M  \chi^i (c_1 \partial^\mu a_\mu^i + c_2 \epsilon^{ijk} \Sigma^{j\mu\nu} \partial_\mu a_\nu^k ) =\\ \nonumber \int_M 
\gamma_1 a^{i\mu} (c_1' \partial_\mu \chi^i + c_2' \epsilon^{ijk} \Sigma^j_\mu{}^\alpha \partial_\alpha \chi^k) + \gamma_2 \epsilon^{ijk} \Sigma^{i\mu\nu} a_\mu^j (c_1' \partial_\nu \chi^k + c_2' \epsilon^{klm} \Sigma^l_\nu{}^\alpha \partial_\alpha \chi^m).
\end{align}
The last term in the second line simplifies
\begin{align}
\epsilon^{ijk} \Sigma^{i\mu\nu} a_\mu^j \epsilon^{klm} \Sigma^l_\nu{}^\alpha \partial_\alpha \chi^m & = ( \delta^{il} \delta^{jm} - \delta^{im}\delta^{jl}) (-\delta^{il} g^{\mu\alpha} + \epsilon^{ils}\Sigma^{s\mu\alpha}) a_\mu^j \partial_\alpha \chi^m \\ \nonumber & = 
- 2 a^{i\mu} \partial_\mu \chi^i - \epsilon^{ijk} a_\mu^i \Sigma^{j\mu\alpha}  \partial_\alpha\chi^k.
\end{align}
This means that we have
\begin{align}
\int_M \chi^i (c_1 \partial^\mu a_\mu^i + c_2 \epsilon^{ijk} \Sigma^{j\mu\nu} \partial_\mu a_\nu^k ) =\\ \nonumber \int_M 
(\gamma_1 c_1' - 2 \gamma_2 c_2' ) a^{i\mu}  \partial_\mu \chi^i + (\gamma_1 c_2'  - \gamma_2 c_1' - \gamma_2 c_2' ) \epsilon^{ijk} a^{i\mu} \Sigma^j_\mu{}^\alpha \partial_\alpha \chi^k ,
\end{align}
and so
\begin{align}\label{c-prime}
- c_1 = \gamma_1 c_1' - 2 \gamma_2 c_2' , \qquad - c_2 = \gamma_1 c_2'  - \gamma_2 c_1' - \gamma_2 c_2' .
\end{align}
The equations we obtained give the coefficients appearing in the adjoints in terms of those in the operators and those in the inner products. 

\subsection{Composition of the adjoints}

Another useful calculation is that of the composition of the adjoints. We will compute the equations that guarantee that the compositions vanish. 
The object $d_2^* d_3^* \chi$ has three irreducible parts. The $h$ part equals
\begin{align}
b_1' \Sigma^{i\mu\nu} \partial_\mu ( c_1' \partial_\nu \chi^i + c_2' \epsilon^{ijk} \Sigma^j_\nu{}^\alpha \partial_\alpha\chi^k) =0.
\end{align}
The $h^i$ part is
\begin{align}
b_2' \partial^\mu (c_1' \partial_\mu \chi^i + c_2' \epsilon^{ijk} \Sigma^j_\mu{}^\alpha \partial_\alpha \chi^k) + b_3' \epsilon^{ijk} \Sigma^{j\mu\nu} \partial_\mu (c_1' \partial_\nu \chi^k + c_2' \epsilon^{klm} \Sigma^l_\nu{}^\alpha \partial_\alpha \chi^m) = \\ \nonumber
(b_2' c_1' + 2b_3' c_2' ) \partial^\mu  \partial_\mu \chi^i .
\end{align}
We thus want
\begin{align}
b_2' c_1' + 2b_3' c_2' =0.
\end{align}
The $\tilde{h}_{\mu\nu}$ part is the tracefree part of 
\begin{align}
2b_4' \Sigma^i_{(\mu}{}^\alpha \partial_{\nu)}( c_1' \partial_\alpha \chi^i + c_2' \epsilon^{ijk} \Sigma^j_\alpha{}^\beta \partial_\beta \chi^k) 
 +  2b_5' \Sigma^i_{(\mu}{}^\alpha \partial_\alpha ( c_1' \partial_{\nu)} \chi^i + c_2' \epsilon^{ijk} \Sigma^j_{\nu)}{}^\beta \partial_\beta \chi^k) .
 \end{align}
 The second term is
 \begin{align}
 \Sigma^i_{(\mu}{}^\alpha \partial_{\nu)}\epsilon^{ijk} \Sigma^j_\alpha{}^\beta \partial_\beta \chi^k = 2 \Sigma^i_{(\mu}{}^\alpha \partial_{\nu)} \partial_\alpha\chi^i.
 \end{align}
 The last term is
 \begin{align}
 \Sigma^i_{(\mu}{}^\alpha \partial_\alpha \epsilon^{ijk} \Sigma^j_{\nu)}{}^\beta \partial_\beta \chi^k = ( \delta^\alpha_{(\nu} \Sigma^i_{\mu)}{}^\beta + \delta^\beta_{(\mu} \Sigma^{i\alpha}{}_{\nu)}) \partial_\alpha\partial_\beta \chi^k=0.
 \end{align}
 So, overall we have
 \begin{align}
 (2b_4' c_1' + 4b_4' c_2' +2b_5' c_1') \Sigma^i_{(\mu}{}^\alpha \partial_{\nu)} \partial_\alpha \chi^i ,
 \end{align}
 and thus we want
 \begin{align}
 b_4' c_1' + 2b_4' c_2' +b_5' c_1'=0.
 \end{align}
 
 We now compute the other composition. We have
 \begin{align}
 d_1^* d_2^* a = a_1' b_1' \partial_\mu \Sigma^{i\rho\sigma} \partial_\rho a_\sigma^i + a_2' \Sigma^i_\mu{}^\nu \partial_\nu (b_2' \partial^\alpha a_\alpha^i + b_3' \epsilon^{ijk} \Sigma^{j\rho\sigma} \partial_\rho a_\sigma^k) \\ \nonumber
 + a_3' \partial^\nu ( b_4' \Sigma^i_{\mu}{}^\alpha \partial_{\nu}a_\alpha^i + b_4' \Sigma^i_{\nu}{}^\alpha \partial_{\mu}a_\alpha^i +  b_5' \Sigma^i_{\mu}{}^\alpha \partial_\alpha a_{\nu}^i+  b_5' \Sigma^i_{\nu}{}^\alpha \partial_\alpha a_{\mu}^i - \frac{b_4'-b_5'}{2} g_{\mu\nu} \Sigma^{i\rho\sigma}\partial_\rho a^i_\sigma).
 \end{align}
 The last term in the first line gives
 \begin{align}
 \Sigma^{i\mu\nu} \epsilon^{ijk} \Sigma^{j\rho\sigma} \partial_\nu \partial_\rho a_\sigma^k & = ( - g^{\mu\rho} \Sigma^{i\nu\sigma} + g^{\nu\rho} \Sigma^{i\mu\sigma} + g^{\mu\sigma} \Sigma^{i\nu\rho} - g^{\nu\sigma} \Sigma^{i\mu\rho})  \partial_\nu \partial_\rho a_\sigma^i \\ \nonumber & = 
 \Sigma^{i\mu\sigma} \partial^\alpha \partial_\alpha   a_\sigma^i
 -  \Sigma^{i\rho\sigma}  \partial^\mu \partial_\rho a_\sigma^i -   \Sigma^{i\mu\rho}  \partial_\rho  \partial^\sigma a_\sigma^i.
 \end{align}
 This gives
 \begin{align}\nonumber
 d_1^* d_2^* a = (a_1' b_1' + a_3' b_4' -  a_2' b_3'  - \frac{a_3'(b_4'-b_5')}{2}) \partial_\mu \Sigma^{i\rho\sigma} \partial_\rho a_\sigma^i + (a_2' b_2'  +  a_3' b_5' -   a_2' b_3' ) \Sigma^i_\mu{}^\nu \partial_\nu  \partial^\alpha a_\alpha^i \\ \nonumber
+ (a_2' b_3' + a_3' b_4' ) \Sigma^{i}_\mu{}^{\sigma} \partial^\alpha \partial_\alpha   a_\sigma^i .
 \end{align}
We thus want
\begin{align}
a_1' b_1' + a_3' b_4' -  a_2' b_3'  - \frac{a_3'(b_4'-b_5')}{2} =0, \qquad a_2' b_2'  +  a_3' b_5' -   a_2' b_3' =0, \qquad a_2' b_3' + a_3' b_4' =0.
\end{align}
We have checked that all these equations hold provided the primed coefficients are as for the adjoint operators, and provided that squaring relations (\ref{eqs-b1}), (\ref{eqs-b2})  hold.

\subsection{Computation of the square}\label{sec:square}

We now form an elliptic operator
\begin{align}\label{D}
(S,E)\ni (\sigma,\chi) \to D(\sigma,\chi) = (d_1^*\sigma, d_2 \sigma + d_3^*\chi)\in (\Lambda^1,E\otimes\Lambda^1).
\end{align}
Its adjoint is given by 
\begin{align}\label{D*}
(\Lambda^1,E\otimes\Lambda^1) \ni (\xi,a) \to D^*(\xi,a) = ( d_1\xi + d_2^* a, d_3 a) \in (S,E).
\end{align}
The composition of these two operators is
\begin{align}
D^* D (\sigma,\chi) = ( d_1 d_1^*\sigma + d_2^* (d_2 \sigma + d_3^*\chi), d_3 (d_2 \sigma + d_3^*\chi))=( (d_1 d_1^* +d_2^* d_2 ) \sigma, d_3  d_3^*\chi).
\end{align}
We want to compute all the operators appearing here, starting with $d_1 d_1^*$. The $h$ component of $d_1 d_1^*\sigma$ is
\begin{align}
 a_1 \partial^\mu (a_1' \partial_\mu h + a_2' \Sigma^i_\mu{}^\nu \partial_\nu h^i + a_3' \partial^\nu \tilde{h}_{\mu\nu})= a_1 a_1' \partial^\mu \partial_\mu h + a_1 a_3' \partial^\mu \partial^\nu \tilde{h}_{\mu\nu}.
 \end{align}
 For the $h^i$ component we have
 \begin{align}
 a_2 \Sigma^{i\mu\nu} \partial_\mu (a_1' \partial_\nu h + a_2' \Sigma^j_\nu{}^\alpha \partial_\alpha h^j + a_3' \partial^\alpha \tilde{h}_{\nu\alpha})=
 - a_2 a_2' \partial^\mu \partial_\mu h^i+ a_2 a_3' \Sigma^{i\mu\nu} \partial_\mu \partial^\alpha \tilde{h}_{\nu\alpha}.
 \end{align}
 For the $\tilde{h}_{\mu\nu}$ component we have
 \begin{align}
 2 a_3 \partial_{(\mu} (a_1' \partial_{\nu)} h + a_2' \Sigma^i_{\nu)}{}^\alpha \partial_\alpha h^i + a_3' \partial^\alpha \tilde{h}_{\nu)\alpha})=\\ \nonumber
 2a_3 a_1'  \partial_{\mu} \partial_{\nu} h + 2a_3 a_2'  \Sigma^i_{(\mu}{}^\alpha \partial_{\nu)} \partial_\alpha h^i + 2 a_3 a_3' \partial_{(\mu}\partial^\alpha \tilde{h}_{\nu)\alpha},
 \end{align}
 where it is understood that the tracefree part of the result is taken. 
 We now compute $d_2^* d_2  \sigma$. For the $h$ component we have
 \begin{align}
  b_1' \Sigma^{i\mu\nu} \partial_\mu (b_1 \Sigma^i_\nu{}^\alpha \partial_\alpha h + b_2 \partial_\nu h^i + b_3 \epsilon^{ijk} \Sigma^j_\nu{}^\alpha \partial_\alpha h^k + b_4 \Sigma^i_\nu{}^\rho \partial^\sigma \tilde{h}_{\rho\sigma} + b_5 \Sigma^{i\rho\sigma}  \partial_\rho \tilde{h}_{\nu\sigma})= \\ \nonumber
  - 3 b_1 b_1' \partial^\mu \partial_\mu h - 3 b_4 b_1' \partial^\mu \partial^\nu \tilde{h}_{\mu\nu} + b_5 b_1' ( g^{\mu\rho} g^{\nu\sigma}- g^{\mu\sigma} g^{\nu\rho} + \epsilon^{\mu\nu\rho\sigma})
  \partial_\mu \partial_\rho \tilde{h}_{\nu\sigma} = \\ \nonumber
  ( - 3 b_1 b_1' + b_5 b_1' ) \partial^\mu \partial_\mu h - (3 b_4 b_1'  + b_5 b_1' )\partial^\mu \partial^\nu \tilde{h}_{\mu\nu}.
  \end{align}
  For the $h^i$ component we have
  \begin{align}\nonumber
  b_2' \partial^\mu (b_1 \Sigma^i_\mu{}^\alpha \partial_\alpha h + b_2 \partial_\mu h^i + b_3 \epsilon^{ijk} \Sigma^j_\mu{}^\alpha \partial_\alpha h^k + b_4 \Sigma^i_\mu{}^\rho \partial^\sigma \tilde{h}_{\rho\sigma} + b_5 \Sigma^{i\rho\sigma}  \partial_\rho \tilde{h}_{\mu\sigma}) \\ \nonumber
  + b_3' \epsilon^{ijk} \Sigma^{j\mu\nu} \partial_\mu (b_1 \Sigma^k_\nu{}^\alpha \partial_\alpha h + b_2 \partial_\nu h^k + b_3 \epsilon^{klm} \Sigma^l_\nu{}^\alpha \partial_\alpha h^m + b_4 \Sigma^k_\nu{}^\rho \partial^\sigma \tilde{h}_{\rho\sigma} + b_5 \Sigma^{k\rho\sigma}  \partial_\rho \tilde{h}_{\nu\sigma}) = \\ \nonumber
  b_2  b_2' \partial^\mu \partial_\mu h^i+  (b_4 b_2' +  b_5 b_2' )  \Sigma^i_\mu{}^\rho \partial^\mu \partial^\sigma \tilde{h}_{\rho\sigma} 
  + b_3 b_3' (\delta^{il}\delta^{jm} - \delta^{im}\delta^{jl}) ( - \delta^{jl} g^{\mu\alpha} + \epsilon^{jls}\Sigma^{s\mu\alpha}) 
   \partial_\mu \partial_\alpha h^m \\ \nonumber
   + 2b_4 b_3' \Sigma^{i\mu\rho} \partial_\mu \partial^\sigma \tilde{h}_{\rho\sigma}
     + b_5 b_3' (- g^{\mu\rho} \Sigma^{i\nu\sigma} + g^{\nu\rho} \Sigma^{i\mu\sigma} + g^{\mu\sigma} \Sigma^{i\nu\rho} - g^{\nu\sigma} \Sigma^{i\mu\rho})   \partial_\mu \partial_\rho \tilde{h}_{\nu\sigma} = \\ \nonumber
      (b_2  b_2' + 2 b_3 b_3' )\partial^\mu \partial_\mu h^i+  (b_4 b_2' +  b_5 b_2' + 2b_4 b_3' )  \Sigma^i_\mu{}^\rho \partial^\mu \partial^\sigma \tilde{h}_{\rho\sigma} .
    \end{align}
  For the $\tilde{h}_{\mu\nu}$ component we need the tracefree part of
  \begin{align}\nonumber
  2b_4' \Sigma^i_{(\mu}{}^\alpha \partial_{\nu)}(b_1 \Sigma^i_\alpha{}^\beta \partial_\beta h + b_2 \partial_\alpha h^i + b_3 \epsilon^{ijk} \Sigma^j_\alpha{}^\beta \partial_\beta h^k + b_4 \Sigma^i_\alpha{}^\rho \partial^\sigma \tilde{h}_{\rho\sigma} + b_5 \Sigma^{i\rho\sigma}  \partial_\rho \tilde{h}_{\alpha\sigma}) \\ \nonumber
  +  2b_5' \Sigma^i_{(\mu}{}^\alpha \partial_\alpha (b_1 \Sigma^i_{\nu)}{}^\beta \partial_\beta h + b_2 \partial_{\nu)} h^i + b_3 \epsilon^{ijk} \Sigma^j_{\nu)}{}^\beta \partial_\beta h^k + b_4 \Sigma^i_{\nu)}{}^\rho \partial^\sigma \tilde{h}_{\rho\sigma} + b_5 \Sigma^{i\rho\sigma}  \partial_\rho \tilde{h}_{\nu)\sigma}) = \\ \nonumber
 -6 b_1 b_4' \partial_{(\mu} \partial_{\nu)} h + (2b_2 b_4' + 4b_3 b_4' ) \Sigma^i_{(\mu}{}^\alpha \partial_{\nu)} \partial_\alpha h^i
  -(6 b_4 b_4'  + 2b_5 b_4' ) \partial_{(\mu} \partial^\sigma \tilde{h}_{\nu)\sigma} \\ \nonumber
  +2 b_1b_5' ( g_{\mu\nu} \partial^\alpha\partial_\alpha h - \partial_{(\mu} \partial_{\nu)} h) 
  +  2b_2 b_5' \Sigma^i_{(\mu}{}^\alpha \partial_\alpha \partial_{\nu)} h^i 
  +  2b_3 b_5'  ( - g_{\mu\nu} \Sigma^{i\alpha\beta} + g_{(\mu}^\alpha  \Sigma^{i}_{\nu)}{}^{\beta} - g_{(\mu}^\beta \Sigma^{i}_{\nu)}{}^\alpha) \partial_\alpha  \partial_\beta h^i
  \\ \nonumber
  +  2b_4 b_5' ( g_{\mu\nu} \partial^\rho\partial^\sigma \tilde{h}_{\rho\sigma} - \partial_{(\mu} \partial^\rho \tilde{h}_{\nu)\rho}) 
  +  2 b_5 b_5' (\partial_{(\mu} \partial^\rho \tilde{h}_{\nu)\rho} - \partial^\alpha \partial_\alpha \tilde{h}_{\mu\nu}) =  \\ \nonumber
  +( -6 b_1 b_4'  - 2 b_1b_5' ) \partial_{(\mu} \partial_{\nu)} h  +2 b_1b_5' g_{\mu\nu} \partial^\alpha\partial_\alpha h + (2b_2 b_4' + 4b_3 b_4' +  2b_2 b_5' ) \Sigma^i_{(\mu}{}^\alpha \partial_{\nu)} \partial_\alpha h^i \\ \nonumber
  +(-6 b_4 b_4'  - 2b_5 b_4' -  2b_4 b_5'+  2 b_5 b_5') \partial_{(\mu} \partial^\sigma \tilde{h}_{\nu)\sigma} 
  +  2b_4 b_5'  g_{\mu\nu} \partial^\rho\partial^\sigma \tilde{h}_{\rho\sigma} 
  -  2 b_5 b_5'  \partial^\alpha \partial_\alpha \tilde{h}_{\mu\nu}.
     \end{align}
Here we have used that $\tilde{h}_{\mu\nu}$ is tracefree. Dropping the trace parts, and collecting all the terms, we have the following expression for the $\tilde{h}_{\mu\nu}$ part of $(d_1 d_1^* +d_2^* d_2 ) \sigma$
 \begin{align}
 ( -6 b_1 b_4'  - 2 b_1b_5' + 2a_3 a_1' ) \partial_{\langle\mu} \partial_{\nu\rangle} h  + (2b_2 b_4' + 4b_3 b_4' +  2b_2 b_5' + 2a_3 a_2' ) \Sigma^i_{\langle\mu}{}^\alpha \partial_{\nu\rangle} \partial_\alpha h^i \\ \nonumber
  +(-6 b_4 b_4'  - 2b_5 b_4' -  2b_4 b_5'+  2 b_5 b_5'+ 2 a_3 a_3' ) \partial_{\langle\mu} \partial^\sigma \tilde{h}_{\nu\rangle\sigma} 
  -  2 b_5 b_5'  \partial^\alpha \partial_\alpha \tilde{h}_{\mu\nu}.
  \end{align}
  For the $h^i$ part we have
  \begin{align}
     (b_2  b_2' + 2 b_3 b_3'  - a_2 a_2' )\partial^\mu \partial_\mu h^i+  (b_4 b_2' +  b_5 b_2' + 2b_4 b_3' + a_2 a_3'  ) \Sigma^i_\mu{}^\rho \partial^\mu \partial^\sigma \tilde{h}_{\rho\sigma}.
    \end{align}
For the trace part $h$ we have 
    \begin{align}
     ( - 3 b_1 b_1' + b_5 b_1' + a_1 a_1' ) \partial^\mu \partial_\mu h + (-3 b_4 b_1'  - b_5 b_1' + a_1 a_3' )\partial^\mu \partial^\nu \tilde{h}_{\mu\nu}.  
      \end{align}
      
  We also need to compute the operator $d_3 d_3^* \chi$. We have
  \begin{align}
  d_3 d_3^* \chi & = c_1 \partial^\mu (c_1' \partial_\mu \chi^i + c_2' \epsilon^{ijk} \Sigma^j_\mu{}^\alpha \partial_\alpha \chi^k) 
  + c_2 \epsilon^{ijk} \Sigma^{j\mu\nu} \partial_\mu (c_1' \partial_\nu \chi^k + c_2' \epsilon^{klm} \Sigma^l_\nu{}^\alpha \partial_\alpha \chi^m) \\ & =  \nonumber
 ( c_1 c_1' + 2 c_2 c_2')\partial^\mu \partial_\mu \chi^i.
  \end{align}
      
\subsection{Imposing the \texorpdfstring{$\Delta$}{} Conditions}
      
 We want the $D^* D$ operator to be a multiple of the $\Delta=\partial^\alpha \partial_\alpha$ operator. This means we want to have the following quantities vanishing
 \begin{align}\label{D2-eqs}
 -6 b_1 b_4' - 2 b_1b_5' + 2a_3 a_1' =0, \quad 2b_2 b_4' + 4b_3 b_4' +  2b_2 b_5' + 2a_3 a_2' =0, \\ \nonumber
  -6 b_4 b_4'  - 2b_5 b_4' -  2b_4 b_5'+  2 b_5 b_5'+ 2 a_3 a_3' =0, 
 \\ \nonumber
 b_4 b_2' +  b_5 b_2' + 2b_4 b_3' + a_2 a_3' =0, \quad -3 b_4 b_1'  - b_5 b_1' + a_1 a_3'=0.
 \end{align}
 When these conditions are satisfied, the parts of $D^* D$ become
 \begin{align}\nonumber
 (d_1 d_1^* +d_2^* d_2 ) \sigma &=  ( - 3 b_1 b_1' + b_5 b_1' + a_1 a_1' ) \partial^\mu \partial_\mu h +(b_2  b_2' + 2 b_3 b_3'  - a_2 a_2' )\partial^\mu \partial_\mu h^i -  2 b_5 b_5'  \partial^\alpha \partial_\alpha h_{\mu\nu}, \\ \nonumber
 d_3 d_3^* \chi &= ( c_1 c_1' + 2 c_2 c_2')\partial^\mu \partial_\mu \chi^i.
 \end{align}
Solving the conditions (\ref{D2-eqs}) for the operators $d_1, d_2, d_3$ of the \pleb{} complex leads to the choice of the inner products as in (\ref{inner-prod-1}). As we have already described, the resulting in this case operator $d_2^*$ is not the one relevant for the linearised Einstein equations. This means we need a different operator $D$. We will build it in the next section, based on the general computations that were carried out in this section.  
 
\section{Twisted operator}

\subsection{Construction of a more general operator \texorpdfstring{$\tilde{D}$}{}}

We know that the operators $D,D^*$  (\ref{D}), (\ref{D*}) constructed from the \pleb{} complex do not satisfy $D^* D\sim\Delta$, due to the fact they come from an complex that is not elliptic. We would now like to use the operators of the \pleb{} complex as building blocks of more general operators $\tilde{D},\tilde{D}^*$. Thus, we consider the twisted operators
\begin{align}\label{tilde-D}
\tilde{D}(\sigma,\chi) = (\tilde{d}_1^* \sigma + \tilde{d}_4 \chi, \tilde{d}_2 \sigma + \tilde{d}_3^*\chi),
\end{align}
Here all operators are general, of the type considered in the previous section, see (\ref{d1-gen}), (\ref{d2-gen}), (\ref{d3-gen}), and we have introduced a new operator $d_4:E\to \Lambda^1$
\begin{align}
\tilde{d}_4 \chi=f \Sigma^i_\mu{}^\alpha \partial_\alpha \chi^i .
 \end{align}
 Let us also introduce
\begin{align}
\tilde{D}^*(\xi, a) = (\tilde{d}_1 \xi + \tilde{d}_2^* a, \tilde{d}_4^* \xi + \tilde{d}_3 a).
\end{align}
We do not yet assume that the operators appearing here are adjoints of those in (\ref{tilde-D}) with respect to some inner product. For now these are just the most general operators of the type (\ref{d1-adj-gen}), (\ref{d2-adj-gen}), (\ref{d3-adj-gen}) and we introduce
\begin{align}
\tilde{d}_4^* \xi = f' \Sigma^{i\mu\nu} \partial_\mu\xi_\nu.
\end{align}

The composition $\tilde{D}^* \tilde{D}$ is 
\begin{align}
\tilde{D}^* \tilde{D}(\sigma,\chi) = (\tilde{d}_1 (\tilde{d}_1^* \sigma + \tilde{d}_4 \chi) + \tilde{d}_2^* (\tilde{d}_2 \sigma + \tilde{d}_3^*\chi), \tilde{d}_4^* (\tilde{d}_1^* \sigma + \tilde{d}_4 \chi)+ \tilde{d}_3 (\tilde{d}_2 \sigma + \tilde{d}_3^*\chi)).
\end{align}
Most of the operators appearing here were already evaluated in section\ \ref{sec:square}. Let us compute the parts that have not been previously computed.

\subsection{\texorpdfstring{Computation of $\tilde{d}_4^* \tilde{d}_1^* +\tilde{d}_3 \tilde{d}_2$}{}}

We have
\begin{align}\nonumber
\tilde{d}_4^* \tilde{d}_1^* +\tilde{d}_3 \tilde{d}_2 =f' \Sigma^{i\mu\nu} \partial_\mu(a_1'  \partial_\nu h + a_2' \Sigma^j_\nu{}^\alpha \partial_\alpha h^j +a_3' \partial^\alpha \tilde{h}_{\nu\alpha}) \\ \nonumber
+ c_1 \partial^\mu ( b_1 \Sigma^i_\mu{}^\alpha \partial_\alpha h + b_2 \partial_\mu h^i + b_3 \epsilon^{ijk} \Sigma^j_\mu{}^\alpha \partial_\alpha h^k + b_4 \Sigma^i_\mu{}^\rho \partial^\sigma \tilde{h}_{\rho\sigma} + b_5 \Sigma^{i\rho\sigma}  \partial_\rho \tilde{h}_{\mu\sigma} ) \\ \nonumber
+  c_2 \epsilon^{ijk} \Sigma^{j\mu\nu} \partial_\mu ( b_1 \Sigma^k_\nu{}^\alpha \partial_\alpha h + b_2 \partial_\nu h^k + b_3 \epsilon^{klm} \Sigma^l_\nu{}^\alpha \partial_\alpha h^m + b_4 \Sigma^k_\nu{}^\rho \partial^\sigma \tilde{h}_{\rho\sigma} + b_5 \Sigma^{k\rho\sigma}  \partial_\rho \tilde{h}_{\nu\sigma} ) =  \\ \nonumber
- f' a_2' \partial^\alpha \partial_\alpha h^i
+ f' a_3'\Sigma^{i\mu\nu} \partial_\mu \partial^\alpha \tilde{h}_{\nu\alpha}  + c_1 b_2 \partial^\mu \partial_\mu h^i + c_1 (b_4 +b_5) \Sigma^{i\mu\rho} \partial_\mu \partial^\sigma \tilde{h}_{\rho\sigma} 
\\ \nonumber
+ c_2 b_3 \epsilon^{ijk} \Sigma^{j\mu\nu}  \epsilon^{klm} \Sigma^l_\nu{}^\alpha \partial_\mu \partial_\alpha h^m + c_2 b_4 \epsilon^{ijk} \Sigma^{j\mu\nu}
\Sigma^k_\nu{}^\rho \partial_\mu \partial^\sigma \tilde{h}_{\rho\sigma} + c_2 b_5 \epsilon^{ijk} \Sigma^{j\mu\nu} \Sigma^{k\rho\sigma}  \partial_\mu \partial_\rho \tilde{h}_{\nu\sigma}.
\end{align}
The terms in the last line compute as follows
\begin{align}
\epsilon^{ijk} \Sigma^{j\mu\nu}  \epsilon^{klm} \Sigma^l_\nu{}^\alpha \partial_\mu \partial_\alpha h^m= (\delta^{il} \delta^{jm} - \delta^{im}\delta^{jl}) ( - \delta^{jl} g^{\mu\alpha} + \epsilon^{jls}\Sigma^{s\mu\alpha})  \partial_\mu \partial_\alpha h^m= 2 \partial^\mu \partial_\mu h^i, \\ \nonumber
\epsilon^{ijk} \Sigma^{j\mu\nu}\Sigma^k_\nu{}^\rho \partial_\mu \partial^\sigma \tilde{h}_{\rho\sigma}= 2 \Sigma^{i\mu\rho} \partial_\mu \partial^\sigma \tilde{h}_{\rho\sigma}, \\ \nonumber
\epsilon^{ijk} \Sigma^{j\mu\nu} \Sigma^{k\rho\sigma}  \partial_\mu \partial_\rho \tilde{h}_{\nu\sigma} = (- g^{\mu\rho} \Sigma^{i\nu\sigma} + g^{\nu\rho} \Sigma^{i\mu\sigma} + g^{\mu\sigma} \Sigma^{i\nu\rho} - g^{\nu\sigma} \Sigma^{i\mu\rho})\partial_\mu \partial_\rho \tilde{h}_{\nu\sigma}= 0.
\end{align}
This gives
\begin{align}\nonumber
\tilde{d}_4^* \tilde{d}_1^* +\tilde{d}_3 \tilde{d}_2 =(- f' a_2' +2 c_2 b_3+ c_1 b_2) \partial^\alpha \partial_\alpha h^i
+ (f' a_3' + c_1 (b_4 +b_5) + 2 c_2 b_4)\Sigma^{i\mu\nu} \partial_\mu \partial^\alpha \tilde{h}_{\nu\alpha}.
\end{align}
In order for our operator $\tilde{D}$ to square to a multiple of the box operator the second term here must vanish, and so we impose
\begin{align}\label{f-eqn}
f' a_3' + c_1 (b_4 +b_5) + 2 c_2 b_4=0.
\end{align}
This can be viewed as an equation giving $f'$ provided all other parameters are known.

\subsection{Finding \texorpdfstring{$\tilde{D},\tilde{D}^*$}{} that encode Einstein equations mod gauge}

We now look for the solution of the system of equations $\tilde{D}^* \tilde{D}\sim\Delta$ assuming that $\tilde{d}_2=d_2$ and $\tilde{d}_2^* = d_2^*$ of the \pleb{} complex. These are the two operators whose composition $d_2^* d_2$ encodes the linearised Einstein equations, and we would like to keep this part of $\tilde{D}^*\tilde{D}$ intact. All other entries in $\tilde{D}^*\tilde{D}$ are related to gauge, and we are at will to choose them as convenient. Choosing the coefficients $b,b'$ as they are in the \pleb{} complex case, and substituting these into (\ref{D2-eqs}), we get 
\begin{align}
a_2= a_2'=0, \quad a_1=-2a_3, \quad a_1'= \frac{1}{4a_3}, \quad a_3'=-\frac{1}{a_3}, \quad f' = - c_1 a_3.
\end{align}
This gives
\begin{align}
\tilde{d}_1 \xi = a_3 ( - 2\partial^\mu \xi_\mu, 0, 2\partial_{\langle\mu} \xi_{\nu\rangle} ), \\ \nonumber
\tilde{d}_1^* \sigma =  \frac{1}{4a_3} \partial_\mu h - \frac{1}{a_3} \partial^\nu \tilde{h}_{\mu\nu}.
\end{align}
We can choose $a_3$ in such a way that $\tilde{d}_1^*$ is the adjoint of $\tilde{d}_1$ with respect to the inner products (\ref{inner-prod-pleb}). We have
\begin{align}
(\tilde{d}_1\xi,\sigma) = \int -\frac{a_3}{2} (\partial^\mu \xi_\mu) h + 2a_3 \tilde{h}^{\mu\nu} \partial_\mu \xi_\nu = \int ( \frac{a_3}{2} \partial_\mu h - 2a_3 \partial^\nu \tilde{h}_{\mu\nu}) \xi^\mu.
\end{align}
So, the adjoint of $\tilde{d}_1$ matches $\tilde{d}_1^*$ if 
\begin{align}
\frac{a_3}{2}= \frac{1}{4a_3}, \qquad - 2a_3 = - \frac{1}{a_3} \qquad \Rightarrow \qquad a_3=\frac{1}{\sqrt{2}}.
\end{align}
Thus, finally
\begin{align}\label{tilde-d1-d1*}
\tilde{d}_1 \xi = \sqrt{2} ( - \partial^\mu \xi_\mu, 0, \partial_{\langle\mu} \xi_{\nu\rangle} ), \\ \nonumber
\tilde{d}_1^* \sigma =  \sqrt{2}( \frac{1}{4} \partial_\mu h -  \partial^\nu \tilde{h}_{\mu\nu}).
\end{align}

As a check we compute
\begin{align}
\tilde{d}_1 \tilde{d}_1^* \sigma = 2( - \frac{1}{4}\partial^\mu \partial_\mu h + \partial^\mu \partial^\nu \tilde{h}_{\mu\nu}, 0, \frac{1}{4} \partial_{\langle\mu} \partial_{\nu\rangle} h -  \partial_{\langle \mu} \partial^\alpha \tilde{h}_{\nu\rangle\alpha}).
\end{align}
Recalling (\ref{d2}) we have
\begin{align}
(d_2^* d_2 + \tilde{d}_1 \tilde{d}_1^*)\sigma = (\partial^\mu \partial_\mu h, 0, - \partial^\alpha \partial_\alpha \tilde{h}_{\mu\nu}),
\end{align}
which is as desired. Note that the composition $d_2 \tilde{d}_1$ does not vanish, so these operators do not form a complex. However, as we will now see, the tilded operators can be rewritten in terms of those of \pleb{} complex. 

\subsection{Rewriting} 

We would like to rewrite the operators $\tilde{d}_1, \tilde{d}_1^*$ in terms of those appearing in the \pleb{} complex. For this, we introduce two linear operators 
\begin{align}\label{Phi}
\Phi: E\otimes \Lambda^1 \to \Lambda^1, \qquad \Phi(a)_\mu = \Sigma^i_{\mu}{}^\alpha a^i_\alpha,
\end{align}
and
\begin{align}
\Phi^*: \Lambda^1 \to E\otimes \Lambda^1, \qquad \Phi^*(\xi) = \frac{1}{2 \gamma_2 - \gamma_1} \Sigma^i_\mu{}^\alpha \xi_\alpha
\end{align}
such that
\begin{align}
\langle \xi, \Phi(a) \rangle = \langle a^i, \Phi^*(\xi)^i \rangle
\end{align}
where $\Phi^*$ is the adjoint of $\Phi$.
Note that for the case of \pleb{} $\gamma_1 = 0, \gamma_2 = 1$ so
\begin{align}
\Phi^*(\xi)^i_\mu = \frac{1}{2} \Sigma^i_\mu{}^\alpha \xi_\alpha.
\end{align}
We then compute
\begin{align}\label{Phi-d2}
\Phi (d_2 \sigma)= \Sigma^i_\mu{}^\alpha( \frac{1}{4} \Sigma^i_\alpha{}^\nu \partial_\nu h +2 \partial_\alpha h^i -  \Sigma^{i\rho\sigma} \partial_\rho \tilde{h}_{\alpha\sigma}) =
 \\ \nonumber
-\frac{3}{4} \partial_\mu h + 2 \Sigma^i_\mu{}^\alpha\partial_\alpha h^i - ( \delta_\mu{}^\rho g^{\alpha\sigma} - \delta_\mu{}^\sigma g^{\alpha\rho}) \partial_\rho \tilde{h}_{\alpha\sigma} =
- \frac{3}{4} \partial_\mu h + \partial^\nu \tilde{h}_{\mu\nu} + 2 \Sigma^i_\mu{}^\alpha\partial_\alpha h^i .
\end{align}
It is now clear that there is a linear combination of $d_1^*\sigma$ and $\Sigma^i_\mu{}^\alpha (d_2 \sigma)_\alpha^i$ in which the $h^i$ term cancels
\begin{align}\label{tilde-d1-star}
(d_1^*\sigma)_\mu - \Phi (d_2 \sigma)_\mu = -\frac{1}{4}\partial_\mu h -  \partial^\nu \tilde{h}_{\mu\nu} + \frac{3}{4} \partial_\mu h - \partial^\nu \tilde{h}_{\mu\nu}
=  \frac{1}{2} \partial_\mu h-2 \partial^\nu \tilde{h}_{\mu\nu} =  \sqrt{2} \tilde{d}_1^* \sigma.
\end{align}
Notably, we were able to rewrite the operator  $\tilde{d}_1^*$ that we deduced is necessary for $\tilde{D}^* \tilde{D}\sim\Delta$ in terms of the original \pleb{} operators. In fact, the inner products (\ref{inner-prod-pleb}) were chosen as they are so that this becomes possible. 

We also have
\begin{align}
2 d_2^*( \Phi^*(\xi)) = (-2\Sigma^{i\mu\nu}  \Sigma^i_\nu{}^\alpha \partial_\mu \xi_\alpha, \frac{1}{4} \epsilon^{ijk} \Sigma^{j\mu\nu}  \Sigma^k_\nu{}^\alpha \partial_\mu \xi_\alpha, \Sigma^i_{\langle\mu|}{}^\alpha \Sigma^i_\alpha{}^\beta \partial_{|\nu\rangle} \xi_\beta-\Sigma^i_{\langle\mu|}{}^\alpha \Sigma^i_{|\nu\rangle}{}^\beta \partial_\alpha  \xi_\beta ) \\ \nonumber = 
 (  6 \partial^\mu \xi_\mu, \frac{1}{2} \Sigma^{i\mu\nu} \partial_\mu  \xi_\nu, -2 \partial_{\langle\mu}\xi_{\nu\rangle} ).
\end{align}
This means we have
\begin{align}
d_1\xi - d_2^*( \Phi^*(\xi)) = (-2 \partial^\mu \xi_\mu , 0, 2 \partial_{\langle\mu}\xi_{\nu\rangle} ) = \sqrt{2} \tilde{d}_1 \xi.
\end{align}
Again, we were able to rewrite the tilded operator in terms of the original \pleb{} complex untilded ones. This means that the operators $\tilde{D},\tilde{D}^*$ start to take a more concrete form
\begin{align}
\tilde{D}(\sigma,\chi) = (\frac{1}{\sqrt{2}}(d_1^*\sigma - \Phi(d_2\sigma)) + \tilde{d}_4 \chi, d_2\sigma + \tilde{d}_3^* \chi), \\ \nonumber
\tilde{D}^*(\xi,a) = ( \frac{1}{\sqrt{2}} (d_1 \xi -  d_2^* (\Phi^*(\xi))) + d_2^* a, \tilde{d}_3 a + \tilde{d}_4^* \xi).
\end{align}
It remains to choose the operators $\tilde{d}_3, \tilde{d}_4$ and $\tilde{d}_3^*, \tilde{d}_4^*$ in the most convenient way.

\subsection{Requiring the adjoints}

We have imposed the condition that $\tilde{d}_1, \tilde{d}_1^*$ are adjoints of each other. Let us demand that the same holds true for the operators $\tilde{d}_3, \tilde{d}_4$ and $\tilde{d}_3^*, \tilde{d}_4^*$. The general operator $\tilde{d}_3$ is parameterised
\begin{align}
\tilde{d}_3 a = c_1 \partial^\mu a_\mu^i + c_2 \epsilon^{ijk} \Sigma^{j\mu\nu} \partial_\mu a_\nu^k.
\end{align}
In the inner products (\ref{inner-prod-pleb}) the coefficients of its adjoint are $c_1'=c_2 - c_1/2, c_2'   = c_1/2$, so that 
\begin{align}
\tilde{d}_3^* \chi = ( c_2 - \frac{c_1}{2}) \partial_\mu \chi^i + \frac{c_1}{2} \epsilon^{ijk} \Sigma^j_\mu{}^\alpha \partial_\alpha \chi^k.
\end{align}
For the operators $\tilde{d}_4, \tilde{d}_4^*$ we have
\begin{align}
(\xi, \tilde{d}_4 \chi) = \int f \xi^\mu \Sigma^i_\mu{}^\alpha \partial_\alpha \chi^i = (\tilde{d}_4^* \xi, \chi),
\end{align}
where 
\begin{align}
\tilde{d}_4^* \xi = f \Sigma^{i\mu\nu} \partial_\mu \xi_\nu,
\end{align}
so that $f'=f$. 

\subsection{Additional computations}

We have
\begin{align}\nonumber
(\tilde{d}_4^* \tilde{d}_1^* +\tilde{d}_3 d_2)\sigma =f' \Sigma^{i\mu\nu} \partial_\mu \sqrt{2} (\frac{1}{4} \partial_\nu h -  \partial^\alpha \tilde{h}_{\nu\alpha}) 
+ c_1 \partial^\mu ( \frac{1}{4} \Sigma^i_\mu{}^\nu \partial_\nu h +2 \partial_\mu h^i -  \Sigma^{i\alpha\beta} \partial_\alpha \tilde{h}_{\mu\beta} ) \\ \nonumber
+  c_2 \epsilon^{ijk} \Sigma^{j\mu\nu} \partial_\mu ( \frac{1}{4} \Sigma^k_\nu{}^\alpha \partial_\alpha h +2 \partial_\nu h^k -  \Sigma^{k\rho\sigma} \partial_\rho \tilde{h}_{\nu\sigma}) = 
-(\sqrt{2} f' +c_1) \Sigma^{i\mu\nu} \partial_\mu \partial^\alpha \tilde{h}_{\nu\alpha}  + 2 c_1 \partial^\mu \partial_\mu h^i. 
\end{align}
The absence of the first term demands $f'=- c_1/\sqrt{2}$.

We also have, for the operator $(\tilde{d}_1 \tilde{d}_4 + d_2^* \tilde{d}_3^*)\chi$, for the trace part
\begin{align}
=  - \sqrt{2} f\Sigma_\mu^i{}^\alpha \partial^\mu \partial_\alpha \chi^i    -2 \Sigma^{i\mu\nu} \partial_\mu (c_1' \partial_\nu \chi^i + c_2' \epsilon^{ijk} \Sigma^j_\nu{}^\alpha \partial_\alpha \chi^k)=0.
\end{align}
For the vector part
\begin{align}
\frac{1}{4} \epsilon^{ijk} \Sigma^{j\mu\nu} \partial_\mu (c_1' \partial_\nu \chi^k + c_2' \epsilon^{klm} \Sigma^l_\nu{}^\alpha \partial_\alpha \chi^m)= \frac{c_2'}{2} \partial^\alpha \partial_\alpha \chi^i.
\end{align}
For the tracefree part
\begin{align}\nonumber
\sqrt{2} \partial_{\langle\mu} f\Sigma_{\nu\rangle}^i{}^\alpha \partial_\alpha \chi^i + \Sigma^i_{\langle\mu}{}^\alpha \partial_{\nu\rangle}(c_1' \partial_\alpha \chi^i + c_2' \epsilon^{ijk} \Sigma^j_\alpha{}^\beta \partial_\beta \chi^k)
-\Sigma^i_{\langle\mu}{}^\alpha \partial_\alpha (c_1' \partial_{\nu\rangle} \chi^i + c_2' \epsilon^{ijk} \Sigma^j_{\nu\rangle}{}^\beta \partial_\beta \chi^k)=\\ \nonumber
(\sqrt{2} f+2c_2') \Sigma^i_{\langle\mu}{}^\alpha \partial_{\nu\rangle} \partial_\alpha \chi^i.
\end{align}
The vanishing of this requires $f=-\sqrt{2} c_2'$. Note that this is compatible with $2 c_2' = c_1$ and $f=f'$. This fixes all the operators, apart from the operator $\tilde{d}_3 a$ which can still be chosen arbitrarily. We also need to compute
\begin{align}
(\tilde{d}_4^* \tilde{d}_4 + \tilde{d}_3 \tilde{d}_3^*) \chi = f' \Sigma^{i\mu\nu} \partial_\mu ( f \Sigma^j_\nu{}^\alpha \partial_\alpha \chi^j) + c_1 \partial^\mu ( c_1' \partial_\mu \chi^i + c_2' \epsilon^{ijk} \Sigma^j_\mu{}^\alpha \partial_\alpha \chi^k) \\ \nonumber
+ c_2 \epsilon^{ijk} \Sigma^{j\mu\nu} \partial_\mu ( c_1' \partial_\nu \chi^k + c_2' \epsilon^{klm} \Sigma^l_\nu{}^\alpha \partial_\alpha \chi^m)
= (-ff' +c_1 c_1' + 2c_2 c_2') \partial^\mu \partial_\mu \chi^i.
\end{align}

\subsection{The operator \texorpdfstring{$\tilde{D}^* \tilde{D}$}{}}

With the choices we have made, the operator $\tilde{D}^* \tilde{D}$ takes the following form
\begin{align}
\tilde{D}^* \tilde{D} \left( \begin{array}{c} h \\ h^i \\ \tilde{h}_{\mu\nu} \\ \chi^i \end{array}\right) = - \partial^\mu \partial_\mu \left( \begin{array}{c} - h \\ - \frac{c_1}{4} \chi^i \\ \tilde{h}_{\mu\nu} \\ -2c_1 h^i + (c_1^2-2c_1 c_2)\chi^i\end{array}\right).
\end{align}
We have parameterised all the appearing coefficients in terms of $c_1, c_2$. We see that the operator is not diagonal in the space $h^i,\chi^i$, but rather a matrix
\begin{align}
\left( \begin{array}{cc} 0 & - \frac{c_1}{4} \\ -2c_1 & c_1^2-2c_1 c_2 \end{array}\right)
\end{align}
appears. The additional requirement that the eigenvalues of this matrix are $\pm 1$ is chosen arbitrarily, which requires the diagonal elements to be both zero, and the product of the off-diagonal elements to be unity. This means
\begin{align}
c_1= \sqrt{2}, \qquad c_2 = \frac{1}{\sqrt{2}}.
\end{align}
Note that this means $f=-1, c_1'= 0, c_2'= 1/\sqrt{2}$. With these choices the matrix mixing $h^i, \chi^i$ becomes
\begin{align}
\left( \begin{array}{cc} 0 & - \frac{1}{2\sqrt{2}} \\ -2\sqrt{2} & 0 \end{array}\right)
\end{align}

\subsection{Final result for the operator \texorpdfstring{$\tilde{D}$}{D}}
We now note that, with the choices we have made, the twisted operator $\tilde{D}$, written as a matrix acting on $S \oplus E$ in terms of the original \pleb{} maps becomes 
\begin{align}
\tilde{D} = \begin{pmatrix}\frac{1}{\sqrt{2}}(d^*_1 - \Phi d_2) && -\Phi d^*_3 \\ d_2 && \frac{1}{\sqrt{2}} J_1 d^*_3 \end{pmatrix}.
\end{align}
This should be contrasted with the ``naive'' map (\ref{D}), which in the same notations is represented by the following matrix
\begin{align}
D = \begin{pmatrix}d^*_1 && 0 \\ d_2 &&  d^*_3 \end{pmatrix}.
\end{align}

\subsection{Lagrangian}

We can obtain the operators $\tilde{D}, \tilde{D}^*$ as arising from an action
\begin{align}
S = \int  \xi^\mu  (\tilde{d}_1^* \sigma+ \tilde{d}_4 \chi)_\mu + \epsilon^{ijk} \Sigma^{i\mu\nu} a^j_\mu( d_2\sigma+ \tilde{d}_3^* \chi)_\nu^k - \frac{1}{2} (\xi_\mu)^2 - \frac{1}{2} \epsilon^{ijk} \Sigma^{i\mu\nu} a^j_\mu a^k_\nu.
\end{align}
Explicitly, substituting the operators, we get
\begin{align}
S = \int  \xi^\mu  (\sqrt{2}( \frac{1}{4} \partial_\mu h -  \partial^\nu \tilde{h}_{\mu\nu}) - \Sigma^i_\mu{}^\alpha \partial_\alpha \chi^i) \\ \nonumber
+ \epsilon^{ijk} \Sigma^{i\mu\nu} a^j_\mu( \frac{1}{4} \Sigma^k_\nu{}^\alpha \partial_\alpha h +2 \partial_\nu h^k -  \Sigma^{k\alpha\beta} \partial_\alpha \tilde{h}_{\nu\beta} + \frac{1}{\sqrt{2}} \epsilon^{klm} \Sigma^l_\nu{}^\alpha \partial_\alpha \chi^m) \\ \nonumber
- \frac{1}{2} (\xi_\mu)^2 - \frac{1}{2} \epsilon^{ijk} \Sigma^{i\mu\nu} a^j_\mu a^k_\nu.
\end{align}
This is very similar to the Lagrangian that appears as (8.174) with (8.178) in~\cite{ChiralPerturbaKrasno2020}, apart from the last term in the second line, which would need to be chosen differently in order to get the Lagrangian in~\cite{ChiralPerturbaKrasno2020}. This is also a specific choice of $c_1,c_2$ in \cref{eq:lin-gauge-fixed-plebanski-action-in-original-variables} such that the equations of motion are related to the non gauge fixed maps through the twisting described here.

Varying this with respect to $\xi, a$ we get
\begin{align}
\xi = \tilde{d}_1^* \sigma+ \tilde{d}_4 \chi, \qquad a = d_2\sigma+ \tilde{d}_3^* \chi. 
\end{align}
Substituting this back into the action we get a second order in derivatives action functional that depends only on $\sigma, \chi$
\begin{align}\nonumber
S= \frac{1}{2} \int \left(\sqrt{2}( \frac{1}{4} \partial_\mu h -  \partial^\nu \tilde{h}_{\mu\nu})- \Sigma^i_\mu{}^\alpha \partial_\alpha \chi^i \right)^2 \\ \nonumber
+ \epsilon^{ijk} \Sigma^{i\mu\nu} ( \frac{1}{4} \Sigma^j_\mu{}^\alpha \partial_\alpha h +2 \partial_\mu h^j -  \Sigma^{j\alpha\beta} \partial_\alpha \tilde{h}_{\mu\beta} + \frac{1}{\sqrt{2}} \epsilon^{jlm} \Sigma^l_\mu{}^\alpha \partial_\alpha \chi^m)\times \\ \nonumber
( \frac{1}{4} \Sigma^k_\nu{}^\rho \partial_\rho h +2 \partial_\nu h^k -  \Sigma^{k\rho\sigma} \partial_\rho \tilde{h}_{\nu\sigma} + \frac{1}{\sqrt{2}} \epsilon^{kpq} \Sigma^p_\nu{}^\rho \partial_\rho \chi^q).
\end{align}
Evaluating this gives
\begin{align}\label{sec-order-action}
S=\frac{1}{2} \int - \frac{1}{4} (\partial_\mu h)^2 - 4\sqrt{2} \partial_\mu h^i \partial^\mu \chi^i + ( \partial_\alpha \tilde{h}_{\mu\nu})^2.
\end{align}
This computation of the second order action is just another way to state that the introduced operator $\tilde{D}$ has the property $\tilde{D}^* \tilde{D}\sim \Delta$.

\section{Splitting the elliptic operator \texorpdfstring{$\tilde{D}$}{}}

We now want to understand the arising elliptic operator $\tilde{D}$ better. 

\subsection{Changing the parametrisation of the \texorpdfstring{$(\xi,a)$}{} space}

Let us introduce new coordinates on the space $E\oplus E\otimes\Lambda^1$
\begin{align}
\Omega_\mu^i = a_\mu^i - \frac{1}{\sqrt{2}} \Sigma^i_\mu{}^\alpha \xi_\alpha, \qquad \omega_\mu = \xi_\mu + \sqrt{2} \Sigma^i_\mu{}^\alpha a_\alpha^i.
\end{align}
The inverse transformation is
\begin{align}
\xi_\mu = \frac{1}{2}\omega_\mu + \frac{1}{\sqrt{2}} \Sigma_\mu^i{}^\alpha \Omega^i_\alpha, \qquad a_\mu^i = \frac{1}{2}( \Omega_\mu^i + \epsilon^{ijk}\Sigma^j_\mu{}^\alpha \Omega_\alpha^k) - \frac{1}{2\sqrt{2}} \Sigma^i_\mu{}^\alpha \omega_\alpha.
\end{align}
Using the previously introduced operators $\Phi,\Phi^*$ we have
\begin{align}
\Omega^i = a^i - \sqrt{2} \Phi^*(\xi)^i, \quad \omega = \xi + \sqrt{2} \Phi(a)
\end{align}
\begin{align}
\xi = \frac{1}{2}\omega + \frac{1}{\sqrt{2}} \Phi(\Omega), \quad a^i = \frac{1}{2}(\Omega^i + J_1(\Omega)^i) - \frac{1}{\sqrt{2}} \Phi^*(\omega).
\end{align}
The transformation is selected so that
\begin{align}\label{inner-omega}
\epsilon^{ijk} \Sigma^{i\mu\nu} a_\mu^j a_\nu^k + (\xi_\mu)^2 = (\Omega_\mu^i)^2 - \frac{1}{2} (\omega_\mu)^2. 
\end{align}
We have thus found the variables that diagonalise the indefinite quadratic form that appears on the left-hand side of (\ref{inner-omega}). This makes it clear that it is more interesting to consider the operator $\tilde{D}$ composed with the transformation to $\omega, \Omega$ variables. 

\subsection{\texorpdfstring{Splitting of $\tilde{D}$}{}}

Let $T_2$ be the linear map on $(E,E\otimes\Lambda^1)$ giving $(\omega,\Omega)$ from $(\xi,a)$
\begin{align}
\left( \begin{array}{c} \omega \\ \Omega \end{array}\right) = \left( \begin{array}{cc} \mathbb{I} & \sqrt{2}\Phi \\ - \sqrt{2} \Phi^* & \mathbb{I}\end{array}\right) \left( \begin{array}{c} \xi \\ a \end{array}\right).
\end{align}
We then have
\begin{align}
T_2 \tilde{D} = \begin{pmatrix} 1 && \sqrt{2} \Phi \\ - \sqrt{2} \Phi^* && 1 \end{pmatrix} \begin{pmatrix} \frac{1}{\sqrt{2}}(d_1^* - \Phi d_2) && - \Phi d_3^* \\ d_2 && \frac{1}{\sqrt{2}} J_1 d_3^* \end{pmatrix} =\\ \nonumber \begin{pmatrix} \frac{1}{\sqrt{2}} d_1^* + (\sqrt{2}-\frac{1}{\sqrt{2}})  \Phi d_2  && - \Phi d_3^* + \Phi J_1 d_3^* \\ d_2 - \Phi^*(d_1^* - \Phi d_2) && \sqrt{2} \Phi^* \Phi d_3^* + \frac{1}{\sqrt{2}} J_1 d_3^* \end{pmatrix}
\end{align}
To simplify this further we will need the following identities 
\begin{align}
(\Phi J_1)(a)_\mu = \Sigma^i_\mu{}^\nu \epsilon^{ijk} \Sigma^j_\nu{}^\rho a^k_\rho = 2 \Sigma^i_\mu{}^\nu a^i_\nu = 2 \Phi(a)_\mu,
\end{align}
\begin{align}
(\Phi^* \Phi)(a)^i_\mu = \frac{1}{2} \Sigma^i_\mu{}^\nu \Sigma^j_\nu{}^\rho a^j_\rho = -\frac{1}{2} a^i_\mu - \frac{1}{2} \epsilon^{ijk}\Sigma^j_\mu{}^\nu a^k_\nu = -\frac{1}{2}(1+J_1)(a)^i_\mu.
\end{align}
Using the language of operators the above is simply 
\begin{align}
\Phi J_1 = 2 \Phi, \qquad \Phi^* \Phi = -\frac{1}{2}(1+J_1).
\end{align}
This gives
\begin{align}
T_2 \tilde{D} =\begin{pmatrix} \frac{1}{\sqrt{2}} (d_1^* +  \Phi d_2)  &&  \Phi d_3^*  \\ \frac{1}{2}(1-J_1) d_2 - \Phi^*d_1^*  &&- \frac{1}{\sqrt{2}}  d_3^* \end{pmatrix}
\end{align}
The operators appearing in the first column can be further simplified. We have
\begin{align}
(d_1^* +  \Phi d_2)\sigma =  -\frac{1}{4}\partial_\mu h + 2\Sigma^i_\mu{}^\nu \partial_\nu h^i -  \partial^\nu \tilde{h}_{\mu\nu}  - \frac{3}{4} \partial_\mu h + \partial^\nu \tilde{h}_{\mu\nu} + 2 \Sigma^i_\mu{}^\alpha\partial_\alpha h^i = \\ \nonumber
- \partial_\mu h + 4\Sigma^i_\mu{}^\nu \partial_\nu h^i,
\end{align}
where we took $\Phi d_2$ from (\ref{Phi-d2}). Note cancellation of the $\tilde{h}_{\mu\nu}$ terms. We also have
\begin{align}
\Phi^* d_1^* = \frac{1}{2} \Sigma^i_\mu{}^\alpha ( -\frac{1}{4}\partial_\alpha h + 2\Sigma^j_\alpha{}^\nu \partial_\nu h^j -  \partial^\nu \tilde{h}_{\alpha\nu} ) = \\ \nonumber
- \frac{1}{8} \Sigma^i_\mu{}^\alpha \partial_\alpha h - \partial_\mu h^i - \epsilon^{ijk} \Sigma^j_\mu{}^\alpha \partial_\alpha h^k - \frac{1}{2}  \Sigma^i_\mu{}^\alpha \partial^\nu \tilde{h}_{\alpha\nu} , 
\end{align}
and
\begin{align}
J_1 d_2 \sigma = \epsilon^{ijk} \Sigma^j_\mu{}^\alpha ( \frac{1}{4} \Sigma^k_\alpha{}^\nu \partial_\nu h +2 \partial_\alpha h^k -  \Sigma^{k\rho\sigma} \partial_\rho \tilde{h}_{\alpha\sigma} )= \\ \nonumber
\frac{1}{2} \Sigma^i_\mu{}^\nu \partial_\nu h + 2 \epsilon^{ijk} \Sigma^j_\mu{}^\alpha \partial_\alpha h^k + \Sigma^{i\alpha\beta} \partial_\alpha \tilde{h}_{\beta\mu}-
\Sigma^{i}_\mu{}^\alpha \partial^\beta \tilde{h}_{\alpha\beta},
\end{align}
so that
\begin{align}
(1-J_1)d_2\sigma =- \frac{1}{4} \Sigma^i_\mu{}^\nu \partial_\nu h +2 \partial_\mu h^i  - 2 \epsilon^{ijk} \Sigma^j_\mu{}^\alpha \partial_\alpha h^k - 2\Sigma^{i\alpha\beta} \partial_\alpha \tilde{h}_{\beta\mu}+
\Sigma^{i}_\mu{}^\alpha \partial^\beta \tilde{h}_{\alpha\beta}.
\end{align}
This gives, finally
\begin{align}\nonumber
\left(\frac{1}{2}(1-J_1) d_2 - \Phi^*d_1^*\right)\sigma = - \frac{1}{8} \Sigma^i_\mu{}^\nu \partial_\nu h + \partial_\mu h^i  -  \epsilon^{ijk} \Sigma^j_\mu{}^\alpha \partial_\alpha h^k - \Sigma^{i\alpha\beta} \partial_\alpha \tilde{h}_{\beta\mu}+\frac{1}{2} 
\Sigma^{i}_\mu{}^\alpha \partial^\beta \tilde{h}_{\alpha\beta} \\ \nonumber
+ \frac{1}{8} \Sigma^i_\mu{}^\alpha \partial_\alpha h + \partial_\mu h^i + \epsilon^{ijk} \Sigma^j_\mu{}^\alpha \partial_\alpha h^k + \frac{1}{2}  \Sigma^i_\mu{}^\alpha \partial^\nu \tilde{h}_{\alpha\nu} = \\ \nonumber
2 \partial_\mu h^i+\Sigma^{i}_\mu{}^\alpha \partial^\beta \tilde{h}_{\alpha\beta}- \Sigma^{i\alpha\beta} \partial_\alpha \tilde{h}_{\beta\mu}.
\end{align}
Note the cancellation of the $h$-dependent terms here. We now rewrite the operator $T_2 \tilde{D}$ as one acting on a multiplet of fields $(h,h^i,\chi^i,\tilde{h}_{\mu\nu})$. Using matrix notation we have
\begin{align}
\left( \begin{array}{c} \omega \\ \Omega\end{array}\right)=
T_2 \tilde{D} \left( \begin{array}{c} h \\ h^i \\ \chi^i \\ \tilde{h}_{\mu\nu}\end{array}\right) = \left(\begin{array}{cccc} - \frac{1}{\sqrt{2}} d & 2\sqrt{2} \Phi d & \Phi d & 0 \\
0 & 2 d & - \frac{1}{\sqrt{2}} d & \tilde{d} \end{array}\right) \left( \begin{array}{c} h \\ h^i \\ \chi^i \\ \tilde{h}_{\mu\nu}\end{array}\right),
\end{align}
where we have defined new operators 
\begin{align}
d:\Lambda^0\to \Lambda^1, \qquad & (dh)_\mu := \partial_\mu h, \\ \nonumber
d: E\to E\otimes \Lambda^1, \qquad & (dh^i)_\mu := \partial_\mu h^i, \\ \nonumber
\tilde{d}:{\rm Sym}^2_0(\Lambda^1)\to E\otimes \Lambda^1, \qquad & (\tilde{d}\tilde{h})_\mu^i := \Sigma^{i}_\mu{}^\alpha \partial^\beta \tilde{h}_{\alpha\beta}- \Sigma^{i\alpha\beta} \partial_\alpha \tilde{h}_{\beta\mu}.
\end{align}
We note that we can rewrite the operator $\tilde{d}$ differently using the $\diamond$ operator
\begin{align}
(\tilde{h}\diamond \Sigma^i)_{\mu\nu}:= \tilde{h}_\mu{}^\alpha \Sigma^i_{\alpha\nu} -  \tilde{h}_\nu{}^\alpha \Sigma^i_{\alpha\mu}.
\end{align}
Then
\begin{align}
 (\tilde{d}\tilde{h})_\nu^i = - \partial^\mu (\tilde{h}\diamond \Sigma^i)_{\mu\nu} = d^* (\tilde{h} \diamond \Sigma^i)_\mu.
 \end{align}
Let also write the arising objects $\omega,\Omega$ explicitly
\begin{align}
\omega_\mu = - \frac{1}{\sqrt{2}} \partial_\mu h  + \Sigma^i_\mu{}^\alpha \partial_\alpha ( 2\sqrt{2} h^i + \chi^i), \\ \nonumber
\Omega_\mu^i =\Sigma^i_\mu{}^\alpha\partial^\beta \tilde{h}_{\alpha\beta} -  \Sigma^{i\alpha\beta} \partial_\alpha \tilde{h}_{\mu\beta} + \frac{1}{\sqrt{2}} \partial_\mu (2 \sqrt{2} h^i - \chi^i).
\end{align}

We now introduce
\begin{align}
h_\pm^i := 2 h^i \pm \frac{1}{\sqrt{2}} \chi^i.
\end{align}
Note that these are precisely the variables that diagonalise the mixed term in (\ref{sec-order-action}). 
\begin{align}
4 \sqrt{2} \partial^\mu h^i \partial_\mu \chi_i =  ( ( \partial_\mu h_+)^2 -  ( \partial_\mu h_-)^2).
\end{align}
We also have
\begin{align}
8 (h^i)^2 + (\chi^i)^2 = (h_+^i)^2 + (h_-^i)^2,
\end{align}
so that these are also the correct variables to diagonalise the inner product quadratic form. We can then write
\begin{align}
\left( \begin{array}{c} h\\ h^i \\ \chi^i \\ \tilde{h}_{\mu\nu} \end{array}\right) = 
 \left( \begin{array}{cccc}  1& 0 & 0 & 0 \\ 0 & \frac{1}{4} & \frac{1}{4} & 0 \\ 0&  \frac{1}{\sqrt{2}} & - \frac{1}{\sqrt{2}} &0 \\ 0 & 0 & 0 & 1 \end{array} \right) \left( \begin{array}{c} h\\ h^i_+ \\ h^i_- \\ \tilde{h}_{\mu\nu} \end{array} \right) = T_1  \left( \begin{array}{c} h\\ h^i_+ \\ h^i_- \\ \tilde{h}_{\mu\nu} \end{array} \right) .
\end{align}
The result is then
\begin{align}
T_2 \tilde{D} T_1 = \left(\begin{array}{cccc} - \frac{1}{\sqrt{2}} d & \sqrt{2} \Phi d & 0 & 0 \\
0 & 0 &d & \tilde{d} \end{array}\right).
\end{align}
This shows that the operator $T_2 \tilde{D} T_1$ splits into the direct sum of two operators. 

Let us describe the arising operators more explicitly. We define
\begin{align}
D_4: \Lambda^0 \oplus E \to \Lambda^1, \qquad D_{12}: E \oplus {\rm Sym}^2_0(\Lambda^1) \to E\otimes \Lambda^1,
\end{align}
given by
\begin{align}
D_4 (h,h_+) = \left( \begin{array}{cc} - \frac{1}{\sqrt{2}} d & \sqrt{2} \Phi d\end{array}\right) \left( \begin{array}{c} h \\ h_+\end{array}\right), \\ \nonumber
D_{12} (h_-, \tilde{h}) = \left( \begin{array}{cc} d & \tilde{d} \end{array}\right) \left( \begin{array}{c} h_- \\ \tilde{h} \end{array}\right).
\end{align}
Explicitly, these operators are given by
\begin{align}\label{d4-d12}
D_4 (h,h^i_+) = 
- \frac{1}{\sqrt{2}} \left( \partial_\mu h - 2\Sigma^i_\mu{}^\alpha \partial_\alpha h^i_+ \right), \\ \nonumber
 D_{12} (h^i_-, \tilde{h}_{\mu\nu})  =  \partial_\mu h^i_- + \Sigma^i_\mu{}^\alpha\partial^\beta \tilde{h}_{\alpha\beta} -  \Sigma^{i\alpha\beta} \partial_\alpha \tilde{h}_{\mu\beta}.
\end{align}

Each of these two operators times its adjoint gives a multiple of $\Delta$. This is easiest to confirm by performing the action computation. We have 
\begin{align}
\int (D_4 (h,h_+) )^2 = \frac{1}{2} \int  (\partial_\mu h)^2 + 4(\partial_\alpha h^i_+)^2, \\ \nonumber
\int (D_{12} (h_-, \tilde{h}))^2 = \int (\partial_\alpha h^i_-)^2 + (\partial_\alpha \tilde{h}_{\mu\nu})^2.
\end{align}
Here we used
\begin{align}\nonumber
\int ( \Sigma^i_\mu{}^\alpha\partial^\beta \tilde{h}_{\alpha\beta} -  \Sigma^{i\alpha\beta} \partial_\alpha \tilde{h}_{\mu\beta})^2=\int 3 (\partial^\nu \tilde{h}_{\mu\nu})^2 - 2 (\partial^\nu \tilde{h}_{\mu\nu})^2+ (\partial_\alpha \tilde{h}_{\mu\nu})^2 - (\partial^\nu \tilde{h}_{\mu\nu})^2 = \int (\partial_\alpha \tilde{h}_{\mu\nu})^2.
\end{align}
We then have
\begin{align}
\int (D_{12} (h_-, \tilde{h}))^2 - \frac{1}{2} (D_4 (h,h_+) )^2 = \int (\partial_\alpha \tilde{h}_{\mu\nu})^2 + (\partial_\alpha h^i_-)^2 - (\partial_\alpha h^i_+)^2 - \frac{1}{4} (\partial_\mu h)^2 = 2S,
\end{align}
where $S$ is the action (\ref{sec-order-action}).

    \newpage
    \part{Nonlinear Gravity}\label{part:nonlinear-gravity}

    \newpage
    \chapter{Type D Lorentzian Spacetimes}\label{chap:type-D-conformal-to-kahler}

Type D spacetimes are class of solutions to Einstein's equations characterised by the structure of their Weyl tensor, where type D means that the self-dual part of the Weyl tensor has two repeated eigenvalues and is of the form $\Psi^{ij} = {\rm diag}(2\alpha,-\alpha,-\alpha)$. In general type D regions contain massive objects such as stars or black holes. One of the more famous type D spacetimes is the Kerr metric~\cite{GravitationalFKerr1963}. Its original derivation was done by brute force, assuming algebraic speciality and using the null tetrad formalism~\cite{An_Approach_to_Newman_1962}. Derivations of the Kerr metric that are found in the literature are either quite involved~\cite{TheMathematicaChandr1998} or they use a somewhat unexplained trick, for example the Newman-Janis shift~\cite{NoteOnTheKerNewman1965,The_Kerr_Newman_Adamo_2014}, and the Kerr-Schild form~\cite{RelativityInMDeruel2018}. Another reference surrounding the question of simplifying the derivation of the Kerr metric can be found in~\cite{PhysicallyMotiBaines2022}.

The Kerr family of solutions is a member of a more general \pleb{}-\demi{} family~\cite{RotatingChargPleban1976}. All metrics of this family can be analytically continued to produce Euclidean signature metrics in 4D. Being an analytic continuation of a Lorentzian algebraically special metric of type D, the resulting Euclidean metrics are type D with respect to both halves of their Weyl curvature. A remarkable result in~\cite{SelfDualKahleDerdzi1983} states that a 4D Ricci-flat metric, with a one-sided type D Weyl curvature, is conformal to a K\"ahler metric. In particular, there is an integrable almost complex structure. This theorem can be applied independently to both halves of the Weyl curvature, and tells us that the Euclidean analytic continuation of any Lorentzian metric that solves the vacuum Einstein equations and is type D has two distinct integrable almost complex structures. These are of two different orientations, and complex structures in 4D that are of different orientations commute. This implies that for the Euclidean Kerr metric there exists two commuting complex structures, meaning there are two K\"ahler metrics in the conformal class. By assumption, the Kerr metric also has two commuting Killing vector fields. This is a very rigid setup and the Ricci-flat Euclidean metrics of this type can be described completely. This is the viewpoint taken in~\cite{AmbitoricGeomeAposto2013} where the \pleb{}-\demi{} family of metrics are derived as examples of ambitoric 4D geometries, i.e.\ metrics that are toric with respect to two different complex structures. A symplectic toric manifold of dimension $2n$ is a symplectic manifold endowed with a hamiltonian action of the torus ${(S^1)}^n$.

This chapter is a presentation of the author's paper~\cite{Kerr_metric_fro_Krasno_2024}, which itself is a translation of the paper~\cite{AmbitoricGeomeAposto2013} into the language of~\pleb{}'s formalism. The aim is to streamline and simplify their argument as much as possible and present an elementary derivation of the Kerr metric. In \cref{sec:type-D-historical-remarks} we present a review of the works leading to~\cite{AmbitoricGeomeAposto2013}. Sections~\ref{sec:type-D-euclidea-plebanski-formalism} and~\ref{sec:SU2_Killing_Vectors} briefly describe \pleb{}'s formalism and a characterisation of Killing vectors adapted to this formalism. In \cref{sec:ConformalToKahler} we present an alternative proof to that of \derd{}, that type D metrics are conformal to K\"ahler. Section~\ref{sec:Kerr} reviews the Kerr metric as a metric that is conformal to K\"ahler metrics in two ways and makes explicit the two commuting complex structures. Which motivates the setup for the remainder of the chapter. In \cref{sec:DerivationOfTypeDSpacetimes} we present our derivation of type D metrics following~\cite{AmbitoricGeomeAposto2013}, but using \pleb{}'s formalism to impose Einstein's equations.

\section{Historical Remarks}\label{sec:type-D-historical-remarks}
    Since the discovery of the Kerr metric~\cite{GravitationalFKerr1963} in 1963 much effort has been put towards understanding the geometry and properties that it carries. The Kerr metric is a 2 parameter family of type D solutions to Einstein's equations. In 1969 Kinnersley parameterised all Lorentzian type D solutions~\cite{TypeDVacuumMKinner1969}. This was not given in a compact form and many different metrics were presented. Later, in 1976 \pleb{} and \demi{} found a single expression for the most general black hole solution~\cite{RotatingChargPleban1976,ANewLookAtTGriffi2005}. The \pleb{}-\demi{} (PD) family of metrics are a 7 parameter family of solutions to the Einstein-Maxwell equations. When removing the electric and magnetic charge parameters from the PD metric we are left with a 5 parameter family of solutions to Einstein's equations. All the type D spacetimes have Euclidean analogues, obtained by taking an analytic continuation through a complex transformation. Doing so to the PD metric results in a double-sided type D Euclidean metric. Performing the analytic continuation to the 4-parameter subfamily with no cosmological constant gives rise to the asymptotically locally Euclidean (ALE) metric. More recently, Chen and Teo generalised the Ricci-flat PD metric to an asymptotically locally flat (ALF) metric~\cite{AFiveParameteChen2015}, which becomes ALE in the appropriate limit. However, not all the Chen-Teo metrics contain Lorentzian sections. Recently, Biquard and Gauduchon showed that the ALF, Ricci-flat, Hermitian, toric metrics fall into one of 4 types~\cite{OnToricHermitBiquar2021}. These 4 types are Kerr, Chen-Teo, Taub-NUT and Taub-bolt.
    
    One of the first realisations of complex properties of type D was through the Newman-Janis shift~\cite{NoteOnTheKerNewman1965}. The Newman-Janis shift links the Kerr and Schwarzschild solutions of Einstein's field equations through a complex diffeomorphism transformation. It was noticed by Flaherty~\cite{AnIntegrableSFlaher1974,HermitianAndK1976} that this shift was related to the existence of an integrable almost complex structure that all type D spacetimes are equipped with. Type D spacetimes were known to be locally Hermitian because of this. Later, \derd{} proved, in Euclidean signature, that all one-sided type D metrics (with an extra condition implied by Einstein's equations) are conformal to K\"ahler~\cite{SelfDualKahleDerdzi1983}. Soon after \derd{}'s result Przanowski and Baka derived a condition using this locally Hermitian condition to reduce the vanishing of the Ricci tensor for type D to a second order PDE of a single function~\cite{OneSidedTypePrzano1984}. However, the PDE was written using the complex coordinates making solutions difficult to find. This PDE is a transformation of \cref{eq:SU_Toda_Lattice} to complex coordinates. In 1991 LeBrun derived a related result, LeBrun was able to obtain a simple form of K\"ahler metrics with a $U(1)$ symmetry~\cite{ExplicitSelfDClaude1991}. The metric is parameterised using 2 functions $u = u(x,y,z)$, $w = w(x,y,z)$ and is given by 
\begin{align}
    g = e^u w(dx^2 + dy^2) + w dz^2 + w^{-1} \alpha^2, \quad \alpha = dt + \theta, \quad \alpha = J(w dz)
\end{align}
where $\theta \in \Lambda^1(\mathbb{R}^3)$ and $J$ is the complex structure. The Killing vector is $\partial_t$, which generates the $U(1)$ symmetry. Vanishing scalar curvature for LeBrun's ansatz becomes the $SU(\infty)$ Toda-Lattice equation which is another second order PDE for the function $u$\hypertarget{add:fixed-toda-equation},
\begin{align}
    u_{xx} + u_{yy} + {(e^u)}_{zz}\  = 0  \label{eq:SU_Toda_Lattice}
\end{align}
and is Ricci flat when $w = c u_z$ for some constant $c$. This ansatz covers a large class of K\"ahler metrics in dimension 4, however, the Toda equation is difficult to solve in general. More recently, Tod~\cite{OneSidedTypeTodP2020} analysed the case of one-sided type D metrics that have an additional commuting Killing vector field. An even more recent related work is~\cite{Rod_Structures_Tod_P_2024}. In this case the $SU(\infty)$ Toda equation linearises, as shown by Ward in a different context~\cite{EinsteinWeylSWard1990}. The solutions can then be obtained from axisymmetric solutions of the flat three-dimensional Laplacian. However, even after the linearisation, deriving a general solution from \cref{eq:SU_Toda_Lattice} remains difficult. Apostolov, Calderbank and Gauduchon took a different approach in~\cite{Aposto2003,AmbitoricGeomeAposto2013} building upon the fact that many interesting type D metrics are type D with respect to both halves of the Weyl curvature. This is also the approach we follow in the present work.

This chapter proceeds to derive useful results for \pleb{}'s formulation involving Killing vectors. We then compute the Kerr example to show that it satisfies the properties we have stated. In particular the two commuting complex structures are presented explicitly. Then we move to a translation of the results in~\cite{AmbitoricGeomeAposto2013} to a language more familiar to physicists. An explicit calculation is shown rather than a more involved algebraic topology argument

\section{Euclidean \pleb{} Formalism}\label{sec:type-D-euclidea-plebanski-formalism}
In this section we briefly review the Euclidean version of the \pleb{} formulation. For an introduction to \pleb{}'s formulation see \cref{chap:plebanskis-formulation} or~\cite{PlebanskiFormuKrasno2009,FormGenRelGravity2020}.

The Hodge star splits the space of 2-forms into $\Lambda^2 = \Lambda^+ \oplus \Lambda^-$, with eigenvalues $\pm 1$ respectively. Self-dual 2-forms have an eigenvalue of $+1$ and anti-self-dual have $-1$. Each eigenspace has dimension 3 (as $\dim(\Lambda^2) = 6$), meaning we can prescribe bases $\Sigma^i$ and $\asd^i$ for $\Lambda^+$ and $\Lambda^-$. The index $i = 1,2,3$ is an $\mathfrak{so}(3) = \mathfrak{su}(2)$ index, and as such $\Sigma^i$ are called $SU(2)$ structures. The Riemann curvature can be viewed as an endomorphism acting on the space of 2-forms. It can then be decomposed with respect to the decomposition of the space of 2-forms, as seen in \cref{eq:pleb-riemann-curvature-sd-asd-decomposition}. To impose Einstein's condition only the self-dual projection of the Riemann curvature is needed and in turn the self-dual projection is encoded in the curvature of the self-dual connection. The self-dual connection, $A^i \in \mathfrak{su}(2) \otimes \Lambda^1$, is defined by 
\begin{align}
    d\Sigma^i + \epsilon^{ijk} A^j \wedge \Sigma^k = 0.
\end{align}
Where the 2-forms satisfy the metricity condition
\begin{align}
    \Sigma^i \wedge \Sigma^j \sim \delta^{ij}.
\end{align}
Einstein's equations are then
\begin{align}
    F^i = \left( \Psi^{ij} + \frac{\Lambda}{3} \delta^{ij} \right) \Sigma^j, \quad {\rm with} \quad F^i = dA^i + \frac{1}{2} \epsilon^{ijk} A^j \wedge A^k \label{eq:kerr-self-dual-curvature-decomposition}
\end{align}
where $\Psi^{ij}$ is the symmetric tracefree matrix encoding the self-dual Weyl tensor and $\Lambda$ is the cosmological constant. Given a frame $e^I \in \Lambda^1$ the self-dual and anti-self-dual 2-forms are given by 
\begin{align}
    \Sigma^i & = e^0 \wedge e^i - \frac{1}{2} \epsilon^{ijk} e^j \wedge e^k \label{eq:kerr-self-dual-2-forms-in-frame} \\
    \asd^i & = e^0 \wedge e^i + \frac{1}{2} \epsilon^{ijk} e^j \wedge e^k. \label{eq:kerr-anti-self-dual-2-forms-in-frame}
\end{align}
The metric can then be recovered by \urb{} formula 
\begin{align}
    g_{\mu\nu} = \frac{1}{12} \epsilon^{ijk} \teps^{\rho\sigma\alpha\beta} \Sigma^i_{\mu\rho} \Sigma^j_{\nu\sigma} \Sigma^k_{\alpha\beta}. \label{eq:kerr-euclidean-urbantke-formula}
\end{align}
The self-dual 2-forms also satisfy the same quaternion algebra,
\begin{align}
    \Sigma^i_\mu{}^\rho \Sigma^j_\rho{}^\nu = -\delta^{ij} \delta_\mu^\nu + \epsilon^{ijk} \Sigma^k_\mu{}^\nu. \label{eq:kerr-sigma-quaternion-algebra}
\end{align}
All the fields in this case are real-valued which means we do not need to impose any reality conditions.

\section{$SU(2)$ Structures and Killing Vectors}\label{sec:SU2_Killing_Vectors}

Killing vectors fields are infinitesimal symmetries of the metric $\mathcal{L}_X g_{\mu \nu} = 0$, where $\mathcal{L}$ is the Lie derivative. For our purposes below we need to characterise Killing vectors from the point of view of $SU(2)$ structures. Given that the metric (\ref{eq:pleb-urbantke-metric}) defined by a triple of 2-forms $\Sigma^i$ is invariant with respect to the ${\rm SO}(3)$ action on $\Sigma^i$, it is suggestive that in order for a vector field $X$ to be infinitesimal isometries at the level of $\Sigma^i$, the Lie derivative of $\Sigma^i$ should be an ${\rm SO}(3)$ transformation
 \begin{align}\label{def:KillingVectorDef}
        \mathcal{L}_X \Sigma^i = \epsilon^{ijk} \tilde{\theta}^j \Sigma^k.
    \end{align}
We will call a vector field $X$ satisfying this property a $\Sigma$-Killing vector field. The following lemma describes an equivalent way of stating the $\Sigma$-Killing condition
\begin{lemma}
    $\Sigma$-Killing vectors $X \in TM$ can be equivalently described as those satisfying 
    \begin{align}
       d^A \iota_X \Sigma^i = \epsilon^{ijk} \theta^j \Sigma^k,
    \end{align}
    where $d^A$ is the exterior covariant derivative with respect to the $\Sigma$-compatible $d^A \Sigma^i=0$ connection $A^i$. 
    \end{lemma}
\begin{proof}
    To see this we use the Cartan's magic formula
    \begin{align}
        \mathcal{L}_X \Sigma^i = \iota_X d\Sigma^i + d \iota_X \Sigma^i.\label{eq:Cartans_Magic_Formula}
    \end{align}
   We now require this to be a gauge transformation    
       \begin{align}
        \mathcal{L}_X \Sigma^i = \epsilon^{ijk} \tilde{\theta}^j \Sigma^k \label{eq:SU2_LieD_KillingEqn}.
    \end{align}
    Using $d^A \Sigma^i = 0$ and \cref{eq:Cartans_Magic_Formula} we find 
    \begin{align}
        \mathcal{L}_X \Sigma^i = d^A \iota_X \Sigma^i + \epsilon^{ijk} {(\iota_X A)}^j \Sigma^k = \epsilon^{ijk} \tilde{\theta}^j \Sigma^k  \quad \Rightarrow \quad d^A \iota_X \Sigma^i = \epsilon^{ijk} \theta^j \Sigma^k
    \end{align}
    where $\theta^i = \tilde{\theta}^i - \iota_X A^i$ is the shifted $SO(3)$ gauge transformation parameter.
\end{proof}

Next we link $\Sigma$-Killing vector fields with the usual Killing vector fields for the metric.
\begin{proposition}\label{prop:Sigma-killing-vector-is-metric-killing-vector}
    A $\Sigma$-Killing vector field is a Killing vector field for the metric (\ref{eq:kerr-euclidean-urbantke-formula}) defined by the triple of 2-forms $\Sigma^i$. 
\end{proposition}

Before we prove this statement, we need to remind the reader some facts about the decomposition of $\mathfrak{su}(2)$-valued 2-forms. This construction is from~\cite{Su2StructureBhoja2024} and details are given in \cref{subsec:pleb-C3-valued-forms-decomp}, here we will only summarise the main points. Let us denote $E=\mathfrak{su}(2)$, and consider the space of $E$-valued 2-forms. The space $\Lambda^2 \otimes E$ decomposes into the eigenspaces of this operator
\begin{align}
    \Lambda^2 \otimes E = {(\Lambda^2 \otimes E)}_1 \oplus {(\Lambda^2 \otimes E)}_3 \oplus {(\Lambda^2 \otimes E)}_5 \oplus {(\Lambda^2 \otimes E)}_9.
\end{align}
The subscript in ${(\Lambda^2 \otimes E)}_k$ indicates the dimension of the corresponding space. It can also be shown that ${(\Lambda^2 \otimes E)}_9 = \Lambda^- \otimes E$. Here $\Lambda^-$ is the space of anti-self dual 2-forms, with $\Lambda^2=\Lambda^+ \oplus \Lambda^-$. We also have the following explicit projections to each of the eigenspaces
\begin{align}\nonumber
B^{(i}_{\alpha\beta} \Sigma^{j)\alpha\beta} \in S^2(E) = (\Lambda^2 \otimes E)_1 \oplus (\Lambda^2 \otimes E)_5, \\
\epsilon^{ijk} B^j_{\alpha\beta} \Sigma^{k\alpha\beta} \in E = (\Lambda^2 \otimes E)_3, \\ \nonumber
B^i_{(\mu|\alpha|} \Sigma^{i\alpha}{}_{\nu)} \in S^2(T^*M) = (\Lambda^2 \otimes E)_1 \oplus (\Lambda^2 \otimes E)_9.
\end{align}
The first and last maps share a common trace and therefore contain the same $(\Lambda^2 \otimes E)_1$ component. The tracefree part of the first map gives the $5$ dimensional components as a symmetric tracefree internal matrix. The second map provides the $3$ components as an internal vector. Lastly, the final map returns the $9$ components as a symmetric tracefree spacetime tensor. We see that the different irreducible components can be packaged into other objects which will be useful for the rest of this chapter.

\begin{proof}
We now prove the \cref{prop:Sigma-killing-vector-is-metric-killing-vector}. To this end, we consider the $E$-valued 2-form $d^A \iota_X \Sigma^i$, for an arbitrary vector field $X$, and compute its irreducible parts. We have
\begin{align}
    (d^A \iota_X \Sigma^i)_{\mu \nu} = \partial_{[\mu} ( X^{\rho} \Sigma^i_{|\rho| \nu]} )  + \epsilon^{ijk} A^j_{[\mu} X^{\rho} \Sigma^i_{|\rho| \nu]}.
   \end{align}
We can replace the partial derivative in the first term by the covariant derivative with respect to the metric, because it is applied to a 1-form, and the result is $\mu\nu$ anti-symmetrised. Replacing the derivative we have
  \begin{align}\label{d-i-Sigma}
    (d^A \iota_X \Sigma^i)_{\mu \nu}   = \nabla_{[\mu} ( X^{\rho} \Sigma^i_{|\rho| \nu]} )  + \epsilon^{ijk} A^j_{[\mu} X^{\rho} \Sigma^i_{|\rho| \nu]}.
     \end{align} 
 Now it is possible to use the fact that the connection 1-forms $A^i_\mu$ are the ``intrinsic torsion'' components of the ${\rm SU}(2)$-structure, that is we have the following equation
  \begin{align}\label{torsion}
 \nabla_\mu \Sigma^i_{\rho\nu} + \epsilon^{ijk} A^j_\mu \Sigma^k_{\rho\nu}=0.
  \end{align} 
 This fact is proven in ~\cite{Su2StructureBhoja2024}, by considering the $E$-valued 2-forms $X^\mu \nabla_\mu \Sigma^i_{\rho\sigma}$, and showing that only the $(\Lambda^2 \otimes E)_3$ projection is non-vanishing. This is then parametrised by $X^\mu A_\mu^i$. Contracting (\ref{torsion}) with $X^\rho$ and $\mu\nu$ anti-symmetrising we see that we can rewrite (\ref{d-i-Sigma}) as
 \begin{align}   
    (d^A \iota_X \Sigma^i)_{\mu \nu}  = \Sigma^i_{[\mu}{}^\rho \nabla_{\nu]} X_\rho.
 \end{align}
We can now use the algebra \cref{eq:kerr-sigma-quaternion-algebra} of $\Sigma$'s to compute the projections of 
 $d^A \iota_X \Sigma^i$ onto the irreducible components. The result is
\begin{align}
    (\Lambda^2 \otimes E)_1&:  \quad \Sigma^{i\mu\nu} (d^A \iota_X \Sigma^i)_{\mu \nu} = 3 \nabla_\mu X^\mu \nonumber\\
    (\Lambda^2 \otimes E)_3&: \quad \epsilon^{ijk} \Sigma^{j\mu \nu} (d^A \iota_X \Sigma^k)_{\mu \nu} = -2\Sigma^{i \mu \nu} \nabla_{[\mu} X_{\nu]} \\
    (\Lambda^2 \otimes E)_5&: \quad \Sigma^{\langle i | \mu \nu} (d^A \iota_X \Sigma^{|j\rangle})_{\mu \nu} = 0 \nonumber\\
    (\Lambda^2 \otimes E)_9&: \quad \Sigma^i_{\langle \mu|}{}^\alpha (d^A \iota_X \Sigma^i)_{\alpha | \nu \rangle} = -3 \nabla_{\langle \mu} X_{\nu \rangle} \nonumber
\end{align}
Here $\langle \mu \nu \rangle$ denotes the symmetric traceless part.
We now assume that the $\Sigma$-Killing equation is satisfied and so $d^A \iota_X \Sigma^i= \epsilon^{ijk} \theta^j \Sigma^k$. The right-hand side here has the only non-vanishing projection 
\begin{align}
    (\Lambda^2 \otimes E)_3: \quad \epsilon^{ijk} \Sigma^{j\mu\nu} (\epsilon^{klm} \theta^l \Sigma^m)_{\mu\nu} = 8 \theta^i.
\end{align}
It is then clear that the $\Sigma$-Killing equation is equivalent to 
\begin{align}
    \nabla_{(\mu} X_{\nu)} = 0, \quad \theta^i = -\frac{1}{4}\Sigma^{i\mu\nu} \nabla_{[\mu} X_{\nu]}.
\end{align}
The first is the usual Killing equation, proving the proposition. The last relation gives $\theta^i$ in terms of the vector field $X$. In words, $\theta^i$ is the vector giving the self-dual projection of $dX^\flat$, where $(X^\flat)_\mu = g_{\mu\nu} X^\nu$ is the one-form corresponding to the vector field $X$ using the musical isomorphism $\flat : TM \rightarrow \Lambda^1$. 
\end{proof}

\section{Type D metrics are conformal to K\"ahler}\label{sec:ConformalToKahler}

The aim of this section is to present an alternative (and new) proof of the theorem by \derd{}~\cite{SelfDualKahleDerdzi1983}. We use the chiral formalism, which makes the statement almost manifest. We also prove another statement from~\cite{SelfDualKahleDerdzi1983}, namely that all one-sided type D Einstein manifolds have a Killing vector. Einstein manifolds being smooth manifolds with a metric that satisfies Einstein's equation.

\subsection{\derd{} theorem}

We start by using the chiral formalism to prove the following theorem 
\begin{theorem}\label{thrm:DerdziConfKahler}
    Let $(M,g)$ be a smooth 4-manifold that is one-sided type D, i.e. with one of the two halves of the Weyl curvature, say SD, having two coinciding eigenvalues. Assume in addition that the SD part of the Weyl curvature is divergence-free $\nabla \cdot W^+ = 0$. Then $g$ is conformal to a K\"ahler metric. 
\end{theorem}

This was first proved by \derd{} in~\cite{SelfDualKahleDerdzi1983}. Note that there is no assumption here that the metric is Einstein, one only assumes that the SD Weyl tensor is divergence-free. This is true for Einstein manifolds (as we will also see later), but is a weaker condition. The proof we present here is an elementary consequence of the chiral \pleb{} formalism. In fact, one can motivate the usefulness of the chiral formalism by the fact that this, reasonably non-trivial statement of Riemannian geometry, becomes apparent in this formalism. 

We start by translating the divergence-free condition as a condition on objects in the \pleb{} formalism. As is clear from \cref{eq:kerr-self-dual-curvature-decomposition}, the self-dual part of the Weyl curvature is directly related to the matrix $\Psi^{ij}$, and we have $W^+_{\mu \nu \rho \sigma} = \Psi^{ij} \Sigma^i_{\mu\nu} \Sigma^j_{\rho\sigma}$.
The condition that the self-dual Weyl curvature is divergence-free translates to
\begin{align}\label{div-W}
    0 = (\nabla \cdot W^+)_{\nu \rho \sigma} = \nabla^\mu W^+_{\mu \nu \rho \sigma} = \nabla^{A\mu} \left( \Psi^{ij} \Sigma^i_{\mu \nu} \Sigma^j_{\rho\sigma} \right) = \nabla^{A\mu} \Psi^{ij} \Sigma^i_{\mu \nu} \Sigma^j_{\rho \sigma},
\end{align}
where $\nabla^A_\mu$ is the total gauge covariant derivative $\nabla^A_\mu \theta^i_\nu = \nabla_\mu \theta^i_\nu + \epsilon^{ijk} A^j_\mu \theta^k_\nu$ and $A^i$ is the self-dual connection defined by $d^A \Sigma^i = 0$. As we have already remarked in (\ref{torsion}), by construction, this covariant derivative has the property $\nabla_\mu^A \Sigma^i_{\rho\sigma}=0$, which justifies the last equality. 

The object on the right-hand side of (\ref{div-W}) is self-dual with respect to the index pair $\rho\sigma$. It can therefore be converted to an $E$-valued one-form 
\begin{align}\label{div-W-1} 
    (\nabla \cdot W^+)_{\nu \rho \sigma} \Sigma^{i\rho\sigma} = 4\nabla^{A\mu}(\Psi^{ij} \Sigma^j_{\mu \nu}) =4 \nabla^{A\mu}(\Psi^{ij}) \Sigma^j_{\mu \nu}.
\end{align}
This can be rewritten in form notations. Considering the 3-form $(\nabla^A \Psi^{ij}) \Sigma^j$, and taking the Hodge star we obtain a multiple of (\ref{div-W-1}). Thus, we see that the following two statements are equivalent
\begin{align}
    (\nabla \cdot W^+) = 0 \quad \Longleftrightarrow   \quad d^A (\Psi^{ij} \Sigma^j) = 0. \label{eq:Divergence_Free_SD_Weyl}
\end{align}

We now have the following simple proposition of the \pleb{} formalism
\begin{proposition}
    Every Einstein 4-Manifold has divergence free self-dual Weyl curvature, $\nabla \cdot W^+ = 0$. \label{rem:Einstein_Divergence_Free}
\end{proposition}
\begin{proof}[Proof of \cref{rem:Einstein_Divergence_Free}]
    The proof follows from the Bianchi identity $d^A F^i= 0$. In \pleb{}'s formalism, the Einstein equation takes the following form
    \begin{align}\label{Einstein}
        F^i = \left( \Psi^{ij} + \frac{\Lambda}{3} \delta^{ij} \right)\Sigma^j.
    \end{align}
    Applying the exterior covariant derivative on the left, and using the Bianchi identity, we have
    \begin{align}
        0 = d^A F^i = d^A \left( \Psi^{ij} \Sigma^j + \frac{\Lambda}{3} \Sigma^i \right) = d^A (\Psi^{ij} \Sigma^j ) = 0
    \end{align}
    where we have used $d^A \Sigma^i = 0$ and that $\Lambda$ is constant.
    Using \cref{eq:Divergence_Free_SD_Weyl} we see that Einstein 4-manifolds have divergence free self-dual Weyl curvature. 
\end{proof}

\begin{proof}[Proof of \cref{thrm:DerdziConfKahler}]
    Let us denote the eigenvalues of $\Psi^{ij}$ as $\alpha,\beta,\gamma$. We can assume, without loss of generality that $\alpha \geq \beta \geq \gamma$. 
    The traceless condition on $\Psi^{ij}$ imposes $\alpha + \beta + \gamma = 0$.
    This means that $\alpha > 0$ and $\gamma < 0$.
    There are then two possibilities for two eigenvalues to coincide. One is when the middle eigenvalue coincides with the negative one $\beta = \gamma$. The other is when the two positive eigenvalues coincide. Thus, the matrix $\Psi^{ij}$ is of the two possible forms
    \begin{align}
        \Psi^{ij} = \begin{pmatrix}
            2\alpha & 0 & 0 \\
            0 & -\alpha & 0 \\
            0 & 0 & -\alpha
        \end{pmatrix} \label{eq:TypeD_Weyl_Curvature_Matrix}, \qquad 
         \Psi^{ij} = \begin{pmatrix}
            \alpha & 0 & 0 \\
            0 & \alpha & 0 \\
            0 & 0 & -2\alpha
        \end{pmatrix}
    \end{align}
    with $\alpha$ being a positive real function. The two cases differ by the sign of ${\rm det}(\Psi)$. See section 2 of~\cite{Gravitational_I_Biquar_2023} for a related discussion. We consider the first case in details. The second case is treated analogously, by observing that it can be mapped to the first by allowing $\alpha$ to be negative and relabelling the basis vectors. In the proof below, we never need to assume that $\alpha>0$, and so the proof covers both of the cases. 
    
    We conclude that for every one-sided type D Riemannian metric, one can always pick a basis $\Sigma^i$ in the space of SD 2-forms, so that $\Psi^{ij}$ is diagonal and of the form of the first matrix in \cref{eq:TypeD_Weyl_Curvature_Matrix}. With this form of $\Psi^{ij}$ the equations (\ref{eq:Divergence_Free_SD_Weyl}) become
    \begin{align}
        d\left( 2\alpha \Sigma^1 \right) + A^2 \left( -\alpha \Sigma^3 \right) - A^3 \left( -\alpha \Sigma^2 \right) &= 0 \nonumber\\
        d\left( -\alpha \Sigma^2 \right) + A^3 \left( 2\alpha \Sigma^1 \right) - A^1 \left( -\alpha \Sigma^3 \right) &= 0 \label{eq:Explicit_Divergence_Equations}\\
        d\left( -\alpha \Sigma^3 \right) + A^1 \left( -\alpha \Sigma^2 \right) - A^2 \left( 2\alpha \Sigma^1 \right) &= 0. \nonumber
    \end{align}
    Using also $d^A \Sigma^i = 0$, these equations can be rewritten as follows 
    \begin{gather}
        2 d\alpha \Sigma^1 + 3\alpha d\Sigma^1 = 0\\
        d\alpha \Sigma^2 + 3\alpha (d\Sigma^2 - A^1 \Sigma^3) = 0, \quad d\alpha \Sigma^3 + 3\alpha (d\Sigma^3 + A^1 \Sigma^2) = 0. \label{eq:dSigma_Equations_123}
    \end{gather}
    The first equation in the above can be written as 
    \begin{align}
        d\omega = 0, \quad \omega:= \alpha^\frac{2}{3} \Sigma^1 \label{eq:ConfKahler_Closed_2form_Equation}.
    \end{align}
    We now introduce a scaled complex linear combinations of the remaining 2-forms
    \begin{align}
        \Omega := \alpha^\frac{2}{3} (\Sigma^2 + i \Sigma^3) \quad (resp. \quad \bar{\Omega} := \alpha^{\frac{2}{3}} (\Sigma^2 - i\Sigma^3)),
    \end{align}
    where the scaling factor in front is the same as for $\omega$. The last two equations in \cref{eq:Explicit_Divergence_Equations} can be rewritten as
    \begin{align}\label{d-Sigma-23}
    d\alpha (\Sigma^2+i \Sigma^3) + 3\alpha d(\Sigma^2+i \Sigma^3) + 3i \alpha A^1 (\Sigma^2+i \Sigma^3)=0.
    \end{align}
    For later purposes, we note that using $d^A \Sigma^i=0$ we can also rewrite the last two equations as
    \begin{align}
    d\alpha \Sigma^2 - 3\alpha A^3\Sigma^1=0, \qquad  d\alpha \Sigma^3 + 3\alpha A^2\Sigma^1=0,
   \end{align}
    and thus their complex linear combination gives
    \begin{align}\label{d-Omega-Sigma-1}
    d\alpha (\Sigma^2+i \Sigma^3) + 3i \alpha (A^2+i A^3) \Sigma^1=0.
    \end{align}
    This will be useful later. 
    
  We now use (\ref{d-Sigma-23}) to get an equation for $\Omega$
    \begin{align}\label{d-Omega}
        d\Omega +\left( i A^1 - \frac{d\alpha}{3\alpha}\right) \Omega = 0.
        \end{align}
We will rewrite this equation in a more suggestive way below.    
Finally, the algebraic equations $\Sigma^i\Sigma^j\sim \delta^{ij}$ imply the following equations
    \begin{align}
        \omega \Omega = 0, \quad \Omega \Omega = 0, \quad \omega^2 = \frac{1}{2}\Omega \bar{\Omega} \label{eq:ConfKahler_metricity}.
    \end{align}
    
    We can now understand where the conformal to K\"ahler comes from. The second equation in (\ref{eq:ConfKahler_metricity}) implies that the complex 2-form $\Omega$ is decomposable. Its two 1-form factors can be declared to be the $(1,0)$ forms of an almost complex structure, which defines this almost complex structure. The equation in (\ref{d-Omega}) can then be rewritten in a better way. It is clear that it states
    \begin{align}
        d\Omega +\left( i A^1 - \frac{1}{3\alpha} d\alpha \Big|_{(0,1)} \right) \Omega = 0,
        \end{align}
        because only the projection of $d\alpha$ to the space $\Lambda^{0,1}$ survives the wedge product with $\Omega$. We can then add another term that is valued in $\Lambda^{1,0}$ and rewrite this equation as 
          \begin{align}
        d\Omega +i a \Omega = 0, \qquad a = A^1 + \frac{1}{3i \alpha} ( d\alpha \Big|_{(1,0)}  - d\alpha \Big|_{(0,1)} ).\label{eq:ConfKahler_IntegrableComplexStrcture}
        \end{align}
        The connection 1-form $a$ is now real-valued. We note that it can also be written as       
        \begin{align}
        \quad a_\mu = A^1_\mu + \omega_\mu{}^\nu \partial_\nu \ln\left(\alpha^\frac{1}{3}\right). 
    \end{align}
    
    The first equation (\ref{eq:ConfKahler_IntegrableComplexStrcture}) implies that this complex structure is integrable. Indeed, it implies that $d\Omega$ has only a $\Lambda^{2,1}$ part, which is equivalent to integrability~\cite{Besse1987}. The first of the equations in (\ref{eq:ConfKahler_metricity}) implies that $\omega$ is in $\Lambda^{1,1}$ with respect to this complex structure. It can thus be declared to be the K\"ahler form. The last equation in  (\ref{eq:ConfKahler_metricity}) is the correct normalisation condition linking the volume form as defined by the K\"ahler form $\omega$ and the volume form defined by $\Omega$. The equation (\ref{eq:ConfKahler_Closed_2form_Equation}) states that the K\"ahler form is closed. Altogether, we deduce that the metric defined by $\omega,\Omega$ is K\"ahler, with Ricci form $\rho = da$. It is also clear that the original metric, defined by the $\Sigma^i$'s, is conformal to this K\"ahler metric 
    \begin{align}
        g_K = \alpha^\frac{2}{3} g_\Sigma
    \end{align}
    This proves \cref{thrm:DerdziConfKahler}. 
\end{proof}

\subsection{Killing Vectors in Einstein one-sided type D spaces}

Another statement proven in~\cite{SelfDualKahleDerdzi1983} is about the existence of Killing vectors in every Einstein manifold satisfying the assumptions of \cref{thrm:DerdziConfKahler}. 
\begin{proposition} \label{prop:ConfKahlerKillingVector}
    Let $(M,\Sigma^i)$ be an Einstein manifold with type D self-dual Weyl tensor. Let $\Sigma^i$ be a basis for the triple of self-dual 2-forms chosen so that the SD part of the Weyl curvature is diagonal, and $\Sigma^1$ is in the direction of the special (non-repeated) eigenvalue. Then $X^\mu = \Sigma^{1\mu \nu} \nabla_\nu \alpha^{-\frac{1}{3}}$ is a Killing vector field. Here $\alpha$ with $\alpha^2 = \frac{1}{6}\Tr{\Psi^2}$ is the repeated eigenvalue of the self-dual Weyl tensor.
\end{proposition}
\begin{proof}
    Inserting $X$ into $\Sigma^1$ we find
    \begin{align}
        (\iota_X \Sigma^1)_\mu = X^\nu \Sigma^1_{\nu \mu} = \Sigma^{1 \nu \rho} \nabla_\rho \alpha^{-\frac{1}{3}} \Sigma^1_{\nu \mu} = \nabla_\mu \alpha^{-\frac{1}{3}} \quad \Rightarrow \quad \iota_X \Sigma^1 = d\alpha^{-\frac{1}{3}}
    \end{align}
    where we have used the quaternion algebra of the 2-forms. The other two 1-forms are computed as
    \begin{align}
    (\iota_X \Sigma^2)_\mu = \Sigma^{1\sigma\rho} \nabla_\rho \alpha^{-\frac{1}{3}} \Sigma^2_{\sigma\mu}= \Sigma^1_\mu{}^\sigma \Sigma^2_\sigma{}^\rho \nabla_\rho \alpha^{-\frac{1}{3}},
      \end{align}
      where we have used the algebra of $\Sigma$'s to anti-commute $\Sigma^{1,2}$. We similarly have
     \begin{align}
     (\iota_X \Sigma^3)_\mu = \Sigma^1_\mu{}^\sigma \Sigma^3_\sigma{}^\rho \nabla_\rho \alpha^{-\frac{1}{3}}.
     \end{align}
     Introducing $\Sigma^+ = \Sigma^2+ i \Sigma^3$ we have
      \begin{align}\label{i-X-Sigma-plus}
      (\iota_X \Sigma^+)_\mu = \Sigma^1_\mu{}^\sigma \Sigma^+_\sigma{}^\rho \nabla_\rho \alpha^{-\frac{1}{3}}.
      \end{align}
      We now use (\ref{d-Omega-Sigma-1}), which we rewrite as 
      \begin{align}
      d\alpha^{-1/3} \Sigma^+ = i \alpha^{-1/3} A^+ \Sigma^1,
      \end{align}
      where we introduced $A^+=A^2+ iA^3$. Taking the Hodge dual we get
       \begin{align}
      \Sigma^+_\mu{}^\rho \nabla_\rho \alpha^{-1/3}  = i \alpha^{-1/3} \Sigma^1_\mu{}^\rho A^+_\rho.
      \end{align}
      Using this in (\ref{i-X-Sigma-plus}) we get
       \begin{align}
      (\iota_X \Sigma^+)_\mu =  i \alpha^{-1/3} \Sigma^1_\mu{}^\sigma  \Sigma^1_\sigma{}^\rho A^+_\rho = -i \alpha^{-1/3}A^+_\mu.
      \end{align}
      We now summarise the above results
      \begin{align}\label{i-X-Sigma-plus*}
      \iota_X \Sigma^1 = d \alpha^{-1/3}, \qquad \iota_X \Sigma^+= -i \alpha^{-1/3}A^+, \qquad  \qquad \iota_X \Sigma^-= i \alpha^{-1/3}A^-.
 \end{align}
 
  Having computed the components of $\iota_X \Sigma^i$, we can compute its exterior covariant derivative and use the characterisation of the $\Sigma$-Killing vectors to show that $X$ is a Killing vector field. It will be convenient to rewrite the formulas for the exterior covariant derivative of $\iota_X \Sigma^i$ in terms of the introduced objects $\Sigma^\pm, A^\pm$. We have
   \begin{align}
  d_A \iota_X \Sigma^1 = d \iota_X \Sigma^1 + \frac{1}{2i} (A^- \iota_X \Sigma^+ - A^+ \iota_X \Sigma^-), \\ \nonumber
  d_A \iota_X\Sigma^+ = d  \iota_X \Sigma^+ - i A^+ \iota_X \Sigma^1 + i A^1 \iota_X \Sigma^+.
    \end{align}
    Substituting (\ref{i-X-Sigma-plus*}) into the first line we see it vanishes $d_A \iota_X \Sigma^1=0$. For the first term this is manifest from $d^2=0$. In the second term there is a cancellation of the two $A^+ A^-$ terms. On the second line we have
    \begin{align}
     d_A \iota_X\Sigma^+ = -i d( \alpha^{-1/3} A^+)  - i A^+ d \alpha^{-1/3} +  A^1 \alpha^{-1/3}A^+ = - i \alpha^{-1/3} (dA^+ +i A^1 A^+).
    \end{align}
What arises here is the curvature 
 \begin{align}
 F^+ = F^2 +i  F^3 = dA^+ +i A^1 A^+.
 \end{align}
 Using the Einstein equations in the form (\ref{Einstein}) we have
 \begin{align}
 F^+ = (-\alpha + \frac{\Lambda}{3}) \Sigma^+.
\end{align}
 This means that all in all
 \begin{align}
 d_A \iota_X\Sigma^1=0, \qquad d_A \iota_X\Sigma^+ = -i \alpha^{-1/3}(-\alpha + \frac{\Lambda}{3}) \Sigma^+.
 \end{align}
    
    On the other hand, the $\Sigma$-Killing condition with the vector $\theta^i=(\theta,0,0)$ becomes
    \begin{align}
    d_A \iota_X \Sigma^1 = 0, \qquad d_A \iota_X\Sigma^+= i \theta \Sigma^+.
     \end{align}
     We see that this is indeed satisfied, with
     \begin{align}
     \theta = \alpha^{-1/3}(\alpha - \frac{\Lambda}{3}).
     \end{align}
     This finishes the proof. 
   \end{proof}
   
\section{Two complex structures of Euclidean Kerr}\label{sec:Kerr}
   
The aim of this section is to exhibit the structures described in the previous section for the example of the Euclidean Kerr metric. This section serves only to motivate the considerations that follow. 

\subsection{Metric and Chiral objects}

The Euclidean Kerr metric is given by
\begin{equation}
\begin{aligned}\label{eq:EuclideanKerrMetric} 
    g = & \left( 1- \frac{2Mr}{\rho^2} \right) dt^2 + \frac{\rho^2}{\Delta} dr^2 + \rho^2 d\theta^2 + \left( r^2 -a^2 - \frac{2Mr a^2}{\rho^2} \sin(\theta)^2 \right) \sin(\theta)^2 d\phi^2 \\ & \qquad + \frac{4Mra\sin(\theta)^2}{\rho^2} dt d\phi,
\end{aligned}
\end{equation}
where 
\begin{align}\label{rho-delta}
\rho^2 = r^2 - a^2 \cos(\theta)^2, \qquad \Delta = r^2 - 2Mr - a^2.
\end{align}
The constants $a$ and $M$ are the Euclidean versions of the angular momentum and mass of the black hole~\cite{The_Kerr_spacet_Visser_2007}. This is obtained from the usual Lorentzian Kerr metric in Boyer-Lindquist coordinates through a transformation $t \rightarrow it$, $a \rightarrow ia$.
A useful tetrad basis is given by
\begin{align}
    e^0 = \frac{\sqrt{\Delta}}{\rho}(dt - a \sin(\theta)^2 d\phi), \quad e^1 = \frac{\rho}{\sqrt{\Delta}}dr,\quad e^2 = \rho d\theta,\quad e^3 = \frac{\sin(\theta)}{\rho}(a dt + (r^2 - a^2) d\phi).\label{eq:KerrTetradBasis}
\end{align}

\subsection{Self-dual objects}

The self-dual 2-forms are readily calculated from the \cref{eq:kerr-self-dual-2-forms-in-frame} using the tetrad basis. We obtain 
\begin{align}
    \Sigma^1 &= (dt-a\sin(\theta)^2 d\phi) \wedge dr - \sin(\theta) d\theta \wedge (a dt + (r^2-a^2) d\phi)\\
    \Sigma^2 &= \sqrt{\Delta}(dt-a\sin(\theta)^2 d\phi) \wedge d\theta -\frac{\sin(\theta)}{\sqrt{\Delta}}(a dt + (r^2-a^2) d\phi) \wedge dr \\
    \Sigma^3 &= \sqrt{\Delta} \sin(\theta) dt \wedge d\phi - \frac{\rho^2}{\sqrt{\Delta}} dr \wedge d\theta.
\end{align}
The self-dual connections obtained by solving $d^A \Sigma^i = d\Sigma^i + \epsilon^{ijk} A^j \wedge \Sigma^k = 0$ are the following 
\begin{align}
    A^1 &= \frac{M}{z_+^2} (dt - a\sin(\theta)^2 d\phi) + \frac{r\cos(\theta) - a}{z_+} d\phi \\
    A^2 &= -\frac{\sqrt{\Delta}\sin(\theta)}{z_+} d\phi \\
    A^3 &= -\frac{a\sin(\theta)}{\sqrt{\Delta} z_+} dr + \frac{\sqrt{\Delta}}{z_+} d\theta
\end{align}
where $z_+ = r-a \cos(\theta)$.
The self-dual curvatures can be computed (by hand!) and are given by
\begin{align}
    F^1 &= \frac{2M}{z_+^3}  \Sigma^1,\quad F^2 = -\frac{M}{z_+^3} \Sigma^2, \quad F^3 = -\frac{M}{z_+^3} \Sigma^3.
\end{align}
This verifies that Kerr is Ricci-flat, and also gives the components of the self-dual Weyl tensor $F^i = \Psi^{ij} \Sigma^j$ 
\begin{align}
    \Psi^{ij}_+ = \begin{pmatrix}
        2\alpha_+ & 0 & 0 \\
        0 & -\alpha_+ & 0 \\
        0 & 0 & -\alpha_+
    \end{pmatrix}, \quad \alpha_+ = \frac{M}{z_+^3}.
\end{align}
The matrix $\Psi^{ij}$ of the self-dual Weyl is tracefree as expected. Note that we have just used \pleb{} formalism to verify that the Kerr metric is Ricci flat. It is worth noting that this computation is easy to do, even by hand. This once again demonstrates the usefulness and power of the \pleb{} formalism. 

\subsection{Anti-self-dual objects}

The anti-self-dual 2-forms can also be calculated using \cref{eq:kerr-anti-self-dual-2-forms-in-frame} to find
\begin{align}
    \bar{\Sigma}^1 &= (dt - a\sin(\theta)^2 d\phi) \wedge dr + \sin(\theta) d\theta \wedge ( a dt + (r^2-a^2) d\phi ) \\
    \bar{\Sigma}^2 &= \sqrt{\Delta}(dt-a\sin(\theta)^2 d\phi) \wedge d\theta +\frac{\sin(\theta)}{\sqrt{\Delta}}(a dt + (r^2-a^2) d\phi) \wedge dr \\
    \bar{\Sigma}^3 &= \sqrt{\Delta} \sin(\theta) dt \wedge d\phi + \frac{\rho^2}{\sqrt{\Delta}} dr \wedge d\theta.
\end{align}
The corresponding anti-self-dual connections are 
\begin{align}
    \bar{A}^1 &= -\frac{M}{z_-^2} (dt-a\sin(\theta)^2 d\phi) + \frac{r\cos(\theta) + a}{z_-} d\phi \\
    \bar{A}^2 &= -\frac{\sqrt{\Delta} \sin(\theta)}{z_-} d\phi\\
    \bar{A}^3 &= \frac{a \sin(\theta)}{\sqrt{\Delta} z} dr + \frac{\sqrt{\Delta}}{z_-} d\theta.
\end{align}
Here $z_-=r+a\cos(\theta)$.
The anti-self-dual part of the curvature is 
\begin{align}
    \bar{F}^1 = -\frac{2M}{z_-^3} \bar{\Sigma}^1, \quad \bar{F}^2 = \frac{M}{z_-^3} \bar{\Sigma}^2, \quad \bar{F}^3 = \frac{M}{z_-^3} \bar{\Sigma}^3.
\end{align}
Clearly the anti-self-dual Weyl tensor is 
\begin{align}
    \Psi_-^{ij} = \begin{pmatrix}
        2\alpha_- & 0 & 0 \\
        0 & -\alpha_- & 0 \\
        0 & 0 & -\alpha_-
    \end{pmatrix}, \quad \alpha_- = -\frac{M}{z_-^3}
\end{align}
Note that the anti-self-dual and self-dual objects are related by the transformation $t\rightarrow -t$, $a \rightarrow -a$, which is the Euclidean analogue of the complex conjugation in the Lorentzian signature.

\subsection{Conformal to K\"ahler}

Looking at both halves of the Weyl curvature we can see that Euclidean Kerr is double-sided type D. Then by \derd{}'s theorem we know Euclidean Kerr is conformal to two different K\"ahler metrics. The two conformal factors are 
\begin{align}
    \lambda_\pm = \frac{1}{z_\pm} \simeq \alpha_\pm^{\frac{1}{3}} \label{eq:KerrConformalFactors}
\end{align}
the factor of $M$ can be ignored as this is a trivial conformal transformation. Conformally transforming the 2-forms $\Sigma^1$ and $\bar{\Sigma}^1$ we find a pair of closed 2-forms 
\begin{align}
    \omega_+ &= \frac{1}{z_+^2} \Sigma^1 = -dt \wedge d\left( \frac{1}{r-a\cos(\theta)} \right) + d\phi \wedge d\left( \frac{a-r\cos(\theta)}{r-a\cos(\theta)} \right) \\[10pt]
    \omega_- &= \frac{1}{z_-^2} \bar{\Sigma}^1 = dt \wedge d\left( \frac{1}{r+a\cos(\theta)} \right) - d\phi \wedge d \left( \frac{a + r\cos(\theta)}{r+a\cos(\theta)} \right).
\end{align}
The 2-forms $\omega_\pm$ are the K\"ahler forms of the two K\"ahler metrics.

\subsection{Complex Structures}

We can extract the complex structures from the metric and the corresponding K\"ahler forms.
It is useful to introduce a basis for $TM$ dual to the tetrad basis in \cref{eq:KerrTetradBasis}, these are given by
\begin{align}
    e_0 = \frac{r^2-a^2}{\rho \sqrt{\Delta}} \partial_t - \frac{a}{\rho \sqrt{\Delta}}  \partial_\phi, \quad e_1 = \frac{\sqrt{\Delta}}{\rho} \partial_r, \quad e_2 = \frac{1}{\rho}\partial_\theta, \quad e_3 = \frac{1}{\rho} \left( a \sin(\theta) \partial_t + \frac{1}{\sin(\theta)} \partial_\phi \right).
\end{align} 
These are dual to the tetrads in the sense that $e_I (e^J) = e^\mu_I e^J_\mu = \delta_I^J$.
The complex structures can be calculated at the level of the Einstein metric as they are conformally invariant.
Hence, we look for operators $J_\pm$ that satisfy $g(\cdot,J_\pm \cdot) = \lambda^{-2}_\pm \omega_\pm$.
Written in terms of tetrads and their duals this is simply
\begin{align}
    J_\pm &= e^1 \otimes e_0 - e^0 \otimes e_1 \mp e^3 \otimes e_2 \pm e^2 \otimes e_3.
\end{align}
It is not difficult to see that $J^2_\pm = - \id$. 
Substituting the expression in Boyer-Lindquist coordinates we find
\begin{align}
    J_\pm = &\frac{1}{\Delta} dr \otimes \left( (r^2-a^2) \partial_t  - a \partial_\phi \right) - \frac{\Delta}{\rho^2}(dt-a \sin(\theta)^2 d\phi) \otimes \partial_r \\
    & \mp \frac{\sin(\theta)}{\rho^2}(adt + (r^2-a^2)d\phi) \otimes \partial_\theta \pm d\theta \otimes \left( a \sin(\theta) \partial_t + \frac{1}{\sin(\theta)} \partial_\phi \right).
\end{align}
We have explicitly checked that these complex structures are integrable (by computing and checking that their Nijenhuis tensors, defined in \cref{eq:nijenhuis-tensor}, vanishes) and that they commute. 

\subsection{Killing Vectors}

The Euclidean Kerr metric is conformal to two different K\"ahler metrics with opposite orientations. The general statements of the previous section imply that there should be two Killing vector fields, one coming from each of the two orientations. These are given by 
\begin{align}
    X_\pm^\mu = J_\pm^\mu{}_\nu (d\lambda_\pm^{-1})^\nu
\end{align}
Plugging in the conformal factors, \cref{eq:KerrConformalFactors}, we calculate the intermediate objects
\begin{align}
    d\lambda_\pm^{-1} = dr \pm a \cos(\theta) d\theta.
\end{align}
The vector fields dual to the 1-forms are given by 
\begin{align}
    (d\lambda_\pm^{-1})^\sharp = \frac{\Delta}{\rho^2} \partial_r \pm \frac{a \sin(\theta)}{\rho^2} \partial_\theta
\end{align}
We get the Killing vectors by applying the complex structures
\begin{align}
    X_\pm = J_\pm (d\lambda_\pm^{-1})^\sharp = -\partial_t.
\end{align}
Thus, we find that the two Killing vectors coming from the two orientations coincide in the case of Kerr. Thus, the Euclidean Kerr is covered by the case (iii) of Proposition 11 from~\cite{AmbitoricGeomeAposto2013}. We note that the property that these two Killing vector fields coincide is related to the fact that the Kerr Laplace operator is separable. Indeed, this property holds only when one of the parameters of the \pleb{}-\demi{} solution is zero $\epsilon=0$, where $\epsilon$ is the constant appearing in the solutions \cref{eq:HR_solutions_cases}. This is also when the Laplace operator is separable. For a discussion related to the separability and Killing forms see e.g.~\cite{Killing2FormsGauduc2017}.

Even though we see that for the Kerr metric there is just one Killing vector field coming from the general construction of Proposition \ref{prop:ConfKahlerKillingVector}, there is still another Killing vector field $\partial_\phi$ unrelated to the conformal structure. The two Killing vector fields $\xi_1=\partial_t, \xi_2=\partial_\phi$ commute, and they also span an isotropic subspace $\omega_\pm(\xi_1,\xi_2)=0$ with respect to either of the K\"ahler forms.
It is also obvious that each of these two Killing vector fields is $J_\pm$-holomorphic, in the sense that ${\mathcal L}_{\xi_{1,2}} J_\pm=0$. This is manifest in the case of the Kerr metric, because all the geometric objects constructed from it are $t,\phi$ independent. We will take these properties of $\xi_1,\xi_2$ as the starting point of the derivation of the Kerr metric in the next section. 

For completeness, let us also compute the result of the action of the complex structures $J_\pm$ on $\xi_{1,2}$. We have 
\begin{gather}
    J_\pm(\partial_t) = \frac{\Delta}{\rho^2}\partial_r \pm \frac{a \sin(\theta)}{\rho^2} \partial_\theta \\
    J_\pm(\partial_\phi) = -\frac{a \sin(\theta)^2\Delta}{\rho^2} \partial_r \pm \frac{(r^2-a^2)\sin(\theta)}{\rho^2} \partial_\theta
\end{gather}
It is clear that $\xi_1,\xi_2, J(\xi_1),J(\xi_2)$ span all of $TM$, for each of the two complex structures. It can also be checked that all four of these vector fields are mutually commuting, for either $J$. The resulting frame will play an important role in the derivation of the next section.

\section{Toric double-sided type D metrics}\label{sec:DerivationOfTypeDSpacetimes}

We now proceed with a derivation of the Euclidean Kerr metric. In fact, our discussion will be more general, and include the more general \pleb{}-\demi{} family of metrics. 
The \pleb{}-\demi{} (PD) spacetimes~\cite{RotatingChargPleban1976,ANewLookAtTGriffi2005} are Lorentzian algebraically special (type D) spacetimes that generalise many
important GR solutions. In general the PD spacetimes are solutions to the Einstein-Maxwell equations, however, here we will only consider the subclass of metrics that satisfy the vacuum Einstein equations. Our aim is to provide a ``simple" derivation of this class of metrics by building on the conformal to K\"ahler ideas described in the previous section. 
Trying to set up Einstein's equation and solve them by brute force is doable but is far from the best strategy, see~\cite{TheMathematicaChandr1998, PhysicallyMotiBaines2022, APossibleIntuNikoli2012} for examples of such methods. Instead, we follow the approach of~\cite{Aposto2003,AmbitoricGeomeAposto2013} that is based on the availability of two different commuting complex structures in these spaces. 

\subsection{Properties of PD Spacetimes}

We begin by detailing properties of the spacetimes we aim to derive. We are interested in Lorentzian signature spacetimes that are algebraically special (type D), although we will solve for their Euclidean analytic continuations.
For example, to obtain a Euclidean metric from the Kerr metric one sends $t \rightarrow it$ and $a \rightarrow ia$, where $t$ and $a$ are the time coordinate and the angular momentum parameter respectively. The self-dual (SD) and anti-self-dual (ASD) Weyl tensor in Lorentzian spacetimes are complex conjugates of each other, i.e. $W^+ = (W^-)^*$.
This is a consequence of the fact that SD 2-forms are complex-valued objects and the ASD 2-forms are complex conjugates of their SD counterparts.
Unlike the Euclidean signature case, there are no one-sided type D metrics in Lorentzian signature. Thus, analytically continuing a type D Lorentzian metric to Euclidean signature we get a Riemannian metric that is type D with respect to both $W^\pm$. 

We are looking for Einstein metrics, for which we know both $W^\pm$ are divergence free. This lets us use \derd{}'s theorem, see \cref{thrm:DerdziConfKahler}, which in this case implies that the metric is conformal to two K\"ahler metrics, in general not coinciding. We denote these two K\"ahler metrics by $g_\pm$, so that $g = \lambda^2_\pm g_\pm$ where $g$ is the Einstein metric of interest. Being conformal to two K\"ahler metrics implies the existence of two integrable almost complex structures $J_\pm$. Each complex structure is associated with the SD and ASD orientations, hence the complex structures have different orientations and are not coinciding. Moreover, in 4D, two complex structures of different orientations commute. Having two commuting integrable complex structures is an extremely strong property.

By the discussion in the previous section, one-sided type D Einstein metrics are equipped with a Killing vector, see \cref{prop:ConfKahlerKillingVector}. In our situation of double-sided type D metrics we have two Killing vector fields of this type. However, they may coincide, which, as we have seen in the previous section, is indeed the situation in Kerr. However, the metric we want to reproduce has two commuting Killing vector fields, one related to its stationary property, the other related to it being axisymmetric. So, we will assume that there are two commuting Killing vectors $\xi_1,\xi_2$. In the Euclidean setting both of these Killing vectors have compact orbits. This means that we have the two-dimensional torus $\mathbb{T}^2$ acting on our space by isometries. Such spaces are called toric. Moreover, in our setup with two different K\"ahler metrics $g_\pm$, we have the action of $\mathbb{T}^2$ on both $g_\pm$. This is the reason why these spaces are called ambitoric in~\cite{AmbitoricGeomeAposto2013}.

The final condition we impose is that the Killing vectors $\xi_1,\xi_2$ are $J$-holomorphic with respect to both $J_\pm$, that is ${\mathcal L}_{\xi_{1,2}} J_\pm=0$, and that $\omega_\pm(\xi_1,\xi_2)=0$. We have seen that this is true in the case of Kerr, and so this is a geometrically well-motivated assumption. Having motivated the geometry of the problem, we are ready to convert the assumptions into a concrete metric ansatz. 

\subsection{A frame of commuting vector fields}

Let us consider just one of the two complex structures for the moment. The first statement that we would like to prove is that the collection of vector fields $\xi_1,\xi_2,J(\xi_1),J(\xi_2)$ spans $TM$ and mutually commutes. The first half of the statement follows from the fact that $\xi_1,\xi_2$ span an isotropic subspace $\omega(\xi_1,\xi_2)=0$. This means that $J(\xi_1),J(\xi_2)$ is the complementary subspace. 

To prove commutativity we need to use both J-holomorphicity, and the integrability of $J$. Let us start by giving an equivalent way of stating J-holomorphicity. This is a standard statement in complex geometry.
\begin{lemma} A vector field $X$ is $J$-holomorphic, ${\mathcal L}_X J=0$, if and only if
     \begin{align}
        [X,JY]=J[X,Y], \forall Y\in TM.
     \end{align}
\end{lemma}
\begin{proof} We have
\begin{align}
{\mathcal L}_X JY = ({\mathcal L}_X J)(Y) + J({\mathcal L}_X Y) \qquad \Rightarrow \qquad ({\mathcal L}_X J)(Y) = [X,JY]- J[X,Y].
\end{align}
This shows that $({\mathcal L}_X J)=0$ if and only if $[X,JY]- J[X,Y]=0$ for any $Y$.
\end{proof}

We can now use this to show that most of the vector fields in $\xi_1,\xi_2,J(\xi_1),J(\xi_2)$ commute. 
\begin{lemma} Let $[\xi_1,\xi_2]=0$ and $\xi_{1,2}$ be $J$-holomorphic. Then 
\begin{align} 
[\xi_1, J(\xi_1)] =0, \quad [\xi_1, J(\xi_2)] =0 \quad [\xi_2, J(\xi_1)] =0  \quad [\xi_2, J(\xi_2)] =0.
\end{align}
\end{lemma}
\begin{proof} This is obvious from the $J$-holomorphicity of $\xi_{1,2}$ and the previous lemma. 
\end{proof}

To prove that $J(\xi_1),J(\xi_2)$ commute, we need to show that the integrability of $J$ implies that they are also $J$-holomorphic. 
\begin{lemma} If $J$ is integrable, and $X$ is a $J$-holomorphic vector field, then $J(X)$ is also $J$-holomorphic.
\end{lemma} 
\begin{proof}
$J$ is integrable if and only if the Nijenhuis tensor 
\begin{align}
N_J(X,Y) = [JX,JY]-J[X,JY]-J[JX,Y]-[X,Y] \label{eq:nijenhuis-tensor}
\end{align}
vanishes. Let us assume that $X$ here is $J$-holomorphic. Then the second and the fourth terms annihilate each other, and the vanishing of $N_J$ implies
\begin{align}
[JX,JY]=J[JX,Y],
\end{align}
which is precisely the statement that $JX$ is $J$-holomorphic.
\end{proof}

Now, with both $J(\xi_1),J(\xi_2)$ being $J$-holomorphic, it is clear that $[J(\xi_1),J(\xi_2)]=J^2 [\xi_1,\xi_2]=0$. We have established that $\xi_1,\xi_2,J(\xi_1),J(\xi_2)$ are all mutually commuting. We end this subsection with another simple lemma.
\begin{lemma}
    Let $\xi_{1,2,3,4}$ span $TM$ and all commute. Let $\theta_{1,2,3,4}$ be the dual 1-forms. Then all these 1-forms are closed. 
\end{lemma}
\begin{proof}This follows from the standard formula
    \begin{align}
        d\theta(X,Y) = X(\theta(Y)) - Y(\theta(X)) - \theta([X,Y]), \quad \theta \in \Lambda^1, \quad X,Y \in TM.
    \end{align}
    Indeed, evaluating $d\theta$ for any of the 1-forms, on any pair of the vector fields $\xi_{1,2,3,4}$, we see that all the terms on the right-hand side are zero.  
   \end{proof}
   
\subsection{Construction of the frame}

This subsection contains a central part of the construction for the derivation of the Kerr metric. Here we use the availability of the two commuting complex structures to find a convenient parametrisation of the 1-forms dual to $\xi_1,\xi_2,J(\xi_1),J(\xi_2)$. After such a parametrisation is obtained, one can characterise the two commuting complex structures explicitly. Here we follow~\cite{AmbitoricGeomeAposto2013} rather closely, and even keep some notations of these authors. However, unlike this reference, we try to be as elementary as possible, which means without using any algebraic geometry.

First, we note that we can always choose the frame dual to $\xi_1,\xi_2,J(\xi_1),J(\xi_2)$ to be of the form $\theta_1,\theta_2,J(\theta_1),J(\theta_2)$. Since the original vector fields commute, these 1-forms are all closed. This means that there are local coordinates such that 
   \begin{align} 
   \theta_1 = d\tau, \qquad \theta_2= d\varphi.
   \end{align}
   The condition that the other two 1-forms are closed becomes
    \begin{align} 
   dJd\tau = 0, \qquad dJ d\varphi =0.
   \end{align}

We now come to the main statement of this subsection.
\begin{proposition} There exists a choice of coordinates $x,y$, and of two functions $F=F(x), G=G(y)$, such that the two commuting complex structures $J_\pm$ are realised as follows
\begin{align}
 J_\pm d\varphi &= \frac{1}{F}dx \pm \frac{1}{G} dy, \nonumber \\
    J_\pm d\tau &= \frac{x}{F}dx  \pm \frac{y}{G} dy, \nonumber\\
    J_\pm dx &= -\frac{F}{y-x}(d\tau - y d\varphi), \nonumber\\
    J_\pm dy &= \pm \frac{G}{y-x}(d\tau - x d\varphi). \label{eq:J_pm_action_on_txyf_basis}
\end{align}
\end{proposition}

\begin{proof}
Since $dJ_\pm d\tau = 0, dJ_\pm d\varphi = 0$, there exist coordinates $\xi,\eta$ such that 
\begin{gather}
    d\xi = \frac{1}{2}(J_++J_-)d\varphi, \quad d\eta = \frac{1}{2}(J_+-J_-)d\varphi. \nonumber \\
    J_\pm d\varphi= d\xi \pm d\eta
\end{gather}
The two coordinates introduced $(\xi,\eta)$ form a good coordinate system together with $\tau,\varphi$ because both complex structures map $(d\tau,d\varphi)$ to a complementary subspace, which we have now parametrised by $(d\xi,d\eta)$. The action of the complex structures on $(d\xi,d\eta)$ gives a linear combination of $(d\tau,d\varphi)$, and can be parametrised as follows. 
\begin{gather}
    J_\pm d\xi = \alpha_\pm d\tau + \beta_\pm d\varphi, \quad J_\pm d\eta = \chi_\pm d\tau + \delta_\pm d\varphi.
\end{gather}
Here $\alpha_\pm,\beta_\pm,\chi_\pm,\delta_\pm$ are eight at this stage arbitrary functions. However, with $\partial_\tau,\partial_\varphi$ being Killing vectors, we can assume that these functions depend on $\xi,\eta$ only. Next we impose that $J_\pm^2 = -1$ on $d\varphi$
\begin{align}\label{J-pm-squared-eq}
    J_\pm^2 d\varphi = J_\pm d\xi \pm J_\pm d\eta = (\alpha_\pm \pm \chi_\pm) d\tau + (\beta_\pm \pm \delta_\pm) d\varphi = -d\varphi \nonumber \\
    \Rightarrow \quad \alpha_\pm \pm \chi_\pm = 0, \quad \beta_\pm \pm \delta_\pm =-1.
\end{align}
Next we enforce that $J_+$ and $J_-$ commute. We have
\begin{align}
    J_- J_+ d\varphi = J_- d\xi + J_- d\eta = (\alpha_- +\chi_-) d\tau +(\beta_- +\delta_-)  d\varphi, \nonumber \\
    J_+ J_- d\varphi = J_+ d\xi - J_+ d\eta = (\alpha_+ -\chi_+) d\tau +(\beta_+ -\delta_+)  d\varphi.
\end{align}
The commutativity $[J_+,J_-]=0$, together with the conditions in \cref{J-pm-squared-eq}, implies that the action of $J_\pm$ on $d\xi,d\eta$ can be parametrised by only two functions
\begin{gather}
    J_\pm d\xi = -\chi d\tau - (1+\delta) d\varphi, \qquad J_\pm d\eta = \pm (\chi d\tau + \delta d\varphi).
\end{gather}
Lastly, we look at the action on $d\tau$, in general it will have the form 
\begin{align}
    J_\pm d\tau = a_\pm d\xi + b_\pm d\eta \nonumber
\end{align}
with 4 arbitrary functions $a_\pm, b_\pm$. Again we check
\begin{gather}
    J_\pm^2 d\tau = a_\pm J_\pm d\xi + b_\pm J_\pm d\eta = \left( -a_\pm  \pm b_\pm \right) \chi d\tau +  \left( - a_\pm (1+ \delta) \pm b_\pm  \delta\right) d\varphi= -d\tau \nonumber \\
    \Rightarrow \quad a_\pm = -\frac{\delta}{\chi}, \quad b_\pm = \mp \frac{1+\delta}{\chi}
\end{gather}
Hence the action on $d\tau$ is 
\begin{gather}
    J_\pm d\tau = - \frac{\delta}{\chi} d\xi \mp \frac{1+\delta}{\chi} d\eta.
\end{gather}
We have already imposed the closure $d J_\pm d\varphi=0$. It remains to impose $d J_\pm d\tau=0$. For this we find
\begin{gather}
  0=  d(J_\pm d\tau) = \left[ \left( \frac{\delta}{\chi} \right)_\eta \mp \left( \frac{1+\delta}{\chi} \right)_\xi \right] d\xi \wedge d\eta \nonumber \\
    \Rightarrow \quad \left( \frac{\delta}{\chi} \right)_\eta = 0, \quad \left( \frac{1+\delta}{\chi} \right)_\xi = 0
\end{gather}
These equations are easily solved by introducing two functions $a = a(\xi)$, $b = b(\eta)$ such that $\delta = -\frac{a}{a-b}, \chi = \frac{1}{a-b}$.
Summarising the results so far we have
\begin{align}
    J_\pm d\varphi =& d\xi \pm d\eta \nonumber \\
    J_\pm d\tau =& a d\xi \pm b d\eta \nonumber \\
    J_\pm d\xi =& \frac{1}{a-b} \left( - d\tau + b d\varphi \right) \nonumber \\
    J_\pm d\eta =& \pm\frac{1}{a-b}\left(  d\tau -a d\varphi \right)
\end{align}
From here we make a coordinate transformation $x = a(\xi), y = b(\eta)$. Writing $dx = a' d\xi \equiv F d\xi, dy = b' d\eta= G d\eta$, 
we rewrite the action of the complex structures as
\begin{align}
    J_\pm d\varphi =& \frac{dx}{F} \pm \frac{dy}{G} \nonumber \\
    J_\pm d\tau =& \frac{x dx}{F} \pm \frac{y dy}{G} \nonumber \\
    J_\pm dx =& - \frac{F}{x-y}\left(  d\tau - y d\varphi \right) \nonumber \\
    J_\pm dy =& \pm \frac{G}{x-y}\left(  d\tau - x d\varphi\right),
\end{align}
where $F=F(x), G=G(y)$. Therefore, proving the original statement.
\end{proof}

\subsection{Product structure}
We now consider the operator $j:= -J_+ J_-$. This squares to the identity $j^2 = \id$. It is also clear that, because both $J_\pm$ are orthogonal (i.e. metric-compatible) complex structures, the operator $j$ is also metric-compatible. Such an operator $j:TM\to TM: j^2 = +\id$ is known as an orthogonal product structure. The reason for this terminology is that it decomposes the (tangent and) cotangent space into eigenspaces of eigenvalue $\pm 1$. It is easy to see that the subspaces of opposite eigenvalue are metric orthogonal. Indeed, let $\eta_+, \eta_-\in \Lambda^1$ be arbitrary eigenforms of $j$ belonging to different eigenspaces 
    \begin{align}
        j \eta_+ = +\eta_+, \quad j \eta_- = - \eta_-.
    \end{align}
    Such that
    \begin{align}
        (\eta_-,\eta_+) = 0,
    \end{align}
    where the bracket denotes the metric pairing. Now, using one of the complex structures, say $J_+$, we can introduce two more 1-forms
    \begin{align}
        J_+ \eta_-, J_+ \eta_+ \in \Lambda^1, \quad s.t. \quad j(J_+ \eta_-) = - J_+ \eta_-, \quad j(J_+ \eta_+) = +J_+ \eta_+.
    \end{align}
    It is easy to see that $(J_+ \eta_-,\eta_-)=0, (J_+ \eta_+,\eta_+)=0$ and so the basis of 1-forms 
        \begin{align}
        \eta_+,\quad J_+ \eta_+,\quad \eta_-,\quad J_+ \eta_-
    \end{align}
   is metric-orthogonal. 

\subsection{Metric ansatz}

Using the complex structures defined in \cref{eq:J_pm_action_on_txyf_basis}, it is now an easy exercise to check that
\begin{align}
j(dx)= dx, \qquad j(dy)=- dy.
\end{align}
This means that the 1-forms 
\begin{align}
dx, dy, (d\tau - y d\varphi), (d\tau- x d\varphi),
\end{align}
where the last two 1-forms are obtained by applying e.g. $J_+$ to $dx,dy$, form a metric orthogonal basis. This means that the metric we are considering is of \hypertarget{rem:eq-full-stop} the form
\begin{align}
    g = A(d\tau-y d\varphi)^2 + B dx^2 + C dy^2 + D(d\tau- x d\varphi)^2 \label{eq:General_Ambitoric_Metric}
\end{align}
where $A,B,C,D$ are functions of $x,y$ only. However, metrics of this form are not always compatible with the 2 complex structures.
To require this we impose that $g(\cdot ,J_\pm \cdot) \in \Lambda^2$.
Computing the tensor $g(\cdot, J_\pm \cdot)$ we find
\begin{align}
    g(\cdot, J_\pm \cdot) = \frac{A(x-y)}{F}dx \otimes (dt - y d\varphi) - BF\frac{(dt-y d\varphi)}{x-y} \otimes dx \\ \nonumber
    \mp \frac{D(x-y)}{G}dy \otimes (dt - x d\varphi) \pm CG \frac{(dt - x d\varphi)}{x-y} \otimes dy.
\end{align}
Imposing that this is a 2-form results in the following restrictions to the components of $g$,
\begin{align}
    \frac{A(x-y)}{F} = \frac{BF}{x-y}, \quad \frac{D(x-y)}{G} = \frac{CG}{x-y}.
\end{align}
Which are most conveniently solved by introducing two new functions $U = U(x,y)$ and $V = V(x,y)$ so that the solutions become
\begin{align}
    A = \frac{FU}{(x-y)^2}, \quad B = \frac{U}{F}, \quad C = \frac{V}{G}, \quad D = \frac{GV}{(x-y)^2}.
\end{align}
This leads us directly to the following proposition.
\begin{proposition}
    A 4-dimensional Riemannian metric, and its corresponding compatible 2-forms, with two distinct commuting Killing vectors that are J-holomorphic with two complex structures are given locally by
    \begin{align}
    g &= FU \left( \frac{d\tau - y d\varphi}{x-y} \right)^2 + \frac{U}{F} dx^2 + \frac{V}{G} dy^2 + GV \left( \frac{d\tau-xd\varphi}{x-y} \right)^2 \\
    \Sigma^1_\pm &= \frac{U}{x-y} dx \wedge (dt - yd\varphi) \mp \frac{V}{x-y} dy \wedge (d\tau - x d\varphi)\label{eq:Conformal2Form_and_ConformalMetric}.
\end{align}
\end{proposition}
It should be appreciated how close this metric ansatz is to the Euclidean Kerr metric. We emphasise that the only information that went into the construction of this ansatz is that there are two commuting Killing vector fields that are J-holomorphic with the two commuting integrable complex structures. No field equations have yet been imposed.

\subsection{Conformal to two K\"ahler metrics}

We now want to impose the requirement that $g$ is conformal to two different K\"ahler metrics. That is, we demand that $\Sigma^1_\pm$ are conformal to closed 2-forms 
\begin{align}
    d( \lambda^2_\pm \Sigma^1_\pm) = 0.
\end{align}
Where $\lambda_\pm = \lambda_\pm(x,y)$ are the conformal factors for each K\"ahler metric.
The Lee forms of $\Sigma^1_\pm$ are 1-forms, $\theta_\pm$, defined by 
\begin{align}
    d\Sigma^1_\pm = \theta_\pm \wedge \Sigma^1_\pm.
\end{align}
When $\Sigma^1_\pm$ are conformal to closed 2-forms the Lee forms themselves are closed.
In terms of the conformal factors we find that $\theta_\pm = - d\ln(\lambda^2_\pm)$, and hence closed.
Using the 2-forms in \cref{eq:Conformal2Form_and_ConformalMetric} and solving for $\theta_\pm$ we find them to be
\begin{align}
    \theta_\pm = \left( \frac{V_x}{V} \pm \frac{U}{V(x-y)} \right) dx + \left( \frac{U_y}{U} \mp \frac{V}{U(x-y)} \right)dy,
\end{align}
where we write $V_x = \partial_x V$ and similarly for $y$. We want to impose the condition that both of these are closed 1-forms. 
It is more efficient to take linear combinations. Firstly, we demand 
\begin{align}
    d(\theta_+ + \theta_-) = 0 \quad \Rightarrow \quad \ln(U)_{xy} - \ln(V)_{xy} = 0
\end{align}
which can be solved by reparameterising by new functions $A = A(x),\ B = B(y)$ and $H = H(x,y)$, such that
\begin{align}
    U = \frac{x-y}{H^2A}, \quad V = \frac{x-y}{H^2B}.
\end{align}
The $A,B$ here are unrelated to those used to parametrise the metric earlier. The introduction of $x-y$ factor here is to help simplify later formulae.
The other linear combination is 
\begin{align}
    d(\theta_+ - \theta_-) = 0 \quad \Rightarrow \quad -(B^2)_y (x-y) - 2B^2 = (A^2)_x (x-y) - 2A^2. \label{eq:AB_Quadratic_Equation}
\end{align}
Taking derivatives twice with respect to $x$ we get $(A^2)_{xxx} = 0$. Similarly, taking the derivative with respect to $y$ twice we get  $(B^2)_{yyy}=0$. This means that 
these equations can be solved by $A^2$ and $B^2$ being quadratic polynomials in their respective variables, with arbitrary coefficients. Further, 
substituting the quadratic ansatz for $A^2,B^2$ back into \cref{eq:AB_Quadratic_Equation} we find that their coefficients coincide, that is 
\begin{align}
    A^2 = R(x), \quad B^2 = R(y),\quad R(z) = r_0 + r_1 z + r_2 z^2. \label{eq:AmbiToric_AB_R_Solution}
\end{align}
We thus find that, given the two functions $F(x),G(y)$ and the 3 constants $r_0,r_1,r_2$ defining $A(x),B(y)$, the metric 
\begin{align}
    g = \frac{1}{H^2} \left[ \frac{F}{A(x-y)} (d\tau-yd\varphi)^2 + \frac{x-y}{FA} dx^2 + \frac{x-y}{BG} dy^2 + \frac{G}{B(x-y)} (d\tau-x d\varphi)^2 \right] \label{eq:GeneralAmbitoric_Metric}
\end{align}
is an ambitoric metric that is conformal to two K\"ahler metrics with opposite orientations. This ansatz for the metric is half-way to determining the Kerr metric.

\subsection{Solving for the conformal factor}

We are interested in a metric that is Einstein, so we now impose Einstein's equations. 
The easiest way to do this is using \pleb{}'s formalism, the details of which are in \cref{chap:plebanskis-formulation}. The derivation we present here is different from that in~\cite{AmbitoricGeomeAposto2013}, and, we hope, more straightforward. From \cref{eq:GeneralAmbitoric_Metric} we choose a convenient basis for the frame 
\begin{equation}
\begin{aligned}
    &e^0 = \frac{1}{H}\sqrt{\frac{F}{A(x-y)}}(d\tau-yd\varphi), && e^1 = \frac{1}{H}\sqrt{\frac{x-y}{FA}}dx, \\  & e^2 = \frac{1}{H}\sqrt{\frac{x-y}{BG}}dy, && e^3 = \frac{1}{H}\sqrt{\frac{G}{B(x-y)}} (d\tau - x d\varphi).
\end{aligned}
\end{equation}
Using \cref{eq:kerr-self-dual-2-forms-in-frame} we can build the SD 2-forms
\begin{equation}
\begin{aligned}
    \Sigma^1 &= \frac{1}{H^2} \left( \frac{1}{A} (d\tau-yd\varphi) \wedge dx - \frac{1}{B} dy \wedge (d\tau-xd\varphi) \right) \\
    \Sigma^2 &= \frac{1}{H^2} \left( \sqrt{\frac{F}{ABG}} (d\tau-y d\varphi) \wedge dy - \sqrt{\frac{G}{ABF}} (d\tau-x d\varphi) \wedge dx \right) \\
    \Sigma^3 &= \frac{1}{H^2} \left( \sqrt{\frac{FG}{AB}} d\varphi\wedge d\tau - \frac{x-y}{\sqrt{ABFG}} dx \wedge dy \right).
\end{aligned}
\end{equation}
The self-dual connection, obtained by solving $d^A \Sigma^i = d\Sigma^i + \epsilon^{ijk} A^j \wedge \Sigma^k = 0$, in this basis becomes
\begin{align}
    A^1 &= \left( \frac{BF}{A} - \frac{F^2}{AH^2} \left( \frac{(x-y)AH^2}{F} \right)_x \right) \frac{d\tau-y d\varphi}{2(x-y)^2} \nonumber \\ & \hspace{150pt} - \left( \frac{AG}{B} + \frac{G^2}{BH^2} \left( \frac{(x-y)BH^2}{G} \right)_y \right) \frac{d\tau - x d\varphi}{2(x-y)^2} \nonumber\\
    A^2 &= \sqrt{\frac{FG}{AB}} \left[  \frac{A H_x - B H_y}{H(x-y)} d\tau + \frac{H(A-B) + 2yB H_y - 2xA H_x}{H(x-y)} d\varphi \right] \label{eq:AmbiToric_Connection_xy}\\
    A^3 &= \frac{1}{\sqrt{AB}} \left[ \sqrt{\frac{F}{G}} \frac{H(B-A) + 2B(x-y)H_y}{2H(x-y)} dx + \sqrt{\frac{G}{F}} \frac{H(A-B)-2A(x-y) H_x}{2H(x-y)} dy \right].\nonumber
\end{align}

The curvatures are given by rather long expressions. This is the only place where we needed to resort to algebraic manipulation software. We state the curvature 2-forms by decomposing them into their self-dual and anti-self dual components. We write 
\begin{align}
    F^i = M^{ij} \Sigma^j + R^{ij} \bar{\Sigma}^j.
\end{align}
The non-zero components of $M^{ij}$ are 
\begin{flalign}
    &\begin{aligned}
    M^{11} =&\frac{AH^2}{24}\left( \frac{F\log(A)_x}{x-y} \right)_x - \frac{AH^3}{12(x-y)} \left( F\left( \frac{1}{H} \right)_x \right)_x - \frac{AH^2}{24}\left( \frac{F}{x-y} \right)_{xx} \\
    &+\frac{BH^2}{24}\left( \frac{G\log(B)_x}{x-y} \right)_y - \frac{BH^3}{12(x-y)} \left( G\left( \frac{1}{H} \right)_y \right)_y\\ & - \frac{BH^2}{24}\left( \frac{G}{x-y} \right)_{yy} - \frac{BGH^2}{48(x-y)^3}  - \frac{AFH^2}{48(x-y)^3} \\
    &+ \frac{ABH^2}{12}\left[ \left( \frac{G}{B(x-y)^2} \right)_y - \left( \frac{F}{A(x-y)^2} \right)_x \right] - \frac{H^2}{16(x-y)^3}\left( \frac{A^3 G + B^3 F}{AB} \right)
    \end{aligned} \\
    &\begin{aligned}
    M^{22} = M^{33} =& \frac{ABH^2}{24}\left( \frac{F}{A(x-y)^2} \right)_x - \frac{ABH^2}{24}\left( \frac{G}{B(x-y)^2} \right)_y \\ & - \frac{AH^3}{12(x-y)}\left( F\left( \frac{1}{H} \right)_x \right)_x - \frac{AFH^2}{48(x-y)^3} - \frac{AH^2}{24} \left( \frac{F}{(x-y)^2} \right)_x \\ & - \frac{BH^3}{12(x-y)}\left( G\left( \frac{1}{H} \right)_y \right)_y - \frac{BGH^2}{48(x-y)^3} + \frac{BH^2}{24} \left( \frac{G}{(x-y)^2} \right)_y  \\ & + \frac{H^2}{48(x-y)^3} \left( \frac{A^3G+B^3F}{AB} \right)
    \end{aligned} \\
    &\begin{aligned}
    M^{12} = M^{21} =& -\sqrt{\frac{FG}{AB}}\frac{H^2}{24(x-y)^3} \left( \frac{x-y}{2}(A^2)_x - A^2 +\frac{x-y}{2}(B^2)_y + B^2 \right)
    \end{aligned}
\end{flalign}
We remind that one of the Einstein equations states that the trace of the matrix $M^{ij}$ coincides with the cosmological constant $M^{11}+M^{22}+M^{33}=\Lambda$. The components of $R^{ij}$ are
\begin{flalign}
&\begin{aligned}
    R^{11} =& \frac{AFH}{12}\left( \frac{H}{x-y} \right)_{xx} - \frac{AH^2}{24}\left( \frac{F}{x-y} \right)_{xx} + \frac{AHF_x H_x}{12(x-y)} +\frac{AH^2}{24}\left( \frac{F\log(A)_x}{x-y} \right)_x \\ & - \frac{7AFH^2}{48(x-y)^3} -\frac{BGH}{12}\left( \frac{H}{x-y} \right)_{yy} - \frac{BH^2}{24}\left( \frac{G}{x-y} \right)_{yy} + \frac{BHG_y H_y}{12(x-y)} \\ & + \frac{BH^2}{24}\left( \frac{G\log(B)_y}{x-y} \right)_y - \frac{7BGH^2}{48(x-y)^3} + \frac{H^2}{16(x-y)^3}\left( \frac{B^3 F - A^3 G}{AB} \right)
\end{aligned} \\
&\begin{aligned}
    R^{22} =& -\frac{FH(AH_x)_x}{24(x-y)} + \frac{FH^2A_x}{24(x-y)^2} - \frac{AFH^2}{48(x-y)^3} +\frac{GH(BH_y)_y}{24(x-y)} + \frac{GH^2B_y}{24(x-y)^2} \\
    & - \frac{BGH^2}{48(x-y)^3} + \frac{H^2}{48(x-y)^3} \left( \frac{B^3 F - A^3 G}{AB} \right)
\end{aligned} \\
&\begin{aligned}
    R^{33} =& -\frac{FH(AH_x)_x}{24(x-y)} + \frac{FH^2A_x}{24(x-y)^2} - \frac{AFH^2}{48(x-y)^3} -\frac{GH(BH_y)_y}{24(x-y)} - \frac{GH^2B_y}{24(x-y)^2}\\
    &  + \frac{BGH^2}{48(x-y)^3} - \frac{H^2}{48(x-y)^3} \left( \frac{B^3 F - A^3 G}{AB} \right)
\end{aligned} \\
&\begin{aligned}
    R^{12} =& \sqrt{\frac{FG}{AB}}\frac{H}{12(x-y)^2} \left[ 2AB\left( 2(x-y)H_{xy} + H_x - H_y \right) \right. \\ & \quad + \left.\left( 2B^2 H_y - \frac{H}{2}(B^2)_y - 2A^2 H_x + \frac{H}{2}(A^2)_x \right) \right]
\end{aligned} \\
&\begin{aligned}
    R^{21} =& \sqrt{\frac{FG}{AB}}\frac{H}{12(x-y)^2} \left[ 2AB\left( 2(x-y)H_{xy} + H_x - H_y \right) \right. \\ & \quad - \left. \left( 2B^2 H_y - \frac{H}{2}(B^2)_y - 2A^2 H_x + \frac{H}{2}(A^2)_x \right) \right]
\end{aligned}
\end{flalign}

We find it useful to first analyse the equations imposed by $R^{12} = 0, R^{21}=0$. Taking the sum and difference of these equations we get
\begin{align}
    2(x-y)H_{xy} + H_x - H_y = 0 \label{eq:H_equation_xy} \\
    2A^2 H_x - \frac{H}{2}(A^2)_x -2B^2 H_y + \frac{H}{2}(B^2)_y = 0. \label{eq:HAB_equation_xy}
\end{align}
To solve the first equation we introduce a coordinate substitution $t,r = (x+y,x-y)$ such that is becomes 
\begin{align}
    H_{tt} = r\left( \frac{1}{r}H_r \right)_r.
\end{align}
Looking for separable solutions of the form $H = T(t) + R(r)$ we obtain the following class of solutions
\begin{align}
    H(t,r) = c_0 r^2 \ln(r) + c_3 r^2 + c_0 t^2  + c_1 t + c_2.
\end{align}
Another class of solutions is obtained by taking $H = H(z)$ and $z = \sqrt{t^2-r^2}$ the equation reduces to 
\begin{align}
    H_{zz} = 0
\end{align}
which has a solution of the form 
\begin{align}
    H(t,r) = c_4 \sqrt{t^2-r^2} + const.
\end{align}
As the original PDE is linear in $H$, we can add these solutions.
We note that other separable solutions exist in terms of Bessel functions, but these will not be relevant.
Substituting the solution into \cref{eq:HAB_equation_xy} we can restrict the allowed values of the constants $c_i$ and $r_i$, see \cref{eq:AmbiToric_AB_R_Solution} for the definition of the latter. In doing so we find 3 different solutions for $H(x,y)$ and $R(z)$
\begin{alignat}{2}
    1) \quad H(x,y) &= 1 + \varepsilon \sqrt{xy}, \quad &&R = z \nonumber\\
    2) \quad H(x,y) &= \sqrt{xy}, \quad &&R = z^2 + 2\varepsilon z \label{eq:HR_solutions_cases} \\
    3) \quad H(x,y) &= 1 + \varepsilon (x+y), \quad &&R = 1 \nonumber
\end{alignat}
where we have renamed some constants in such a way that $\varepsilon$ is the only constant left. In the main text we will focus on case $1)$ as this gives the family that contains the PD metrics. The other cases are described in~\cite{AmbitoricGeomeAposto2013} and briefly in the appendix of~\cite{Kerr_metric_fro_Krasno_2024}. It happens that the physically relevant case is $1)$, we have that the solutions for $H,A,B$ are
\begin{align}
    H = 1+\varepsilon \sqrt{xy}, \quad A(x) = \sqrt{x}, \quad B(y) = \sqrt{y}.
\end{align}
We now perform another change of variables to remove the square roots. We take $x = r^2,\ y = q^2$, taking the positive branch of the square root.
The metric in these variables becomes 
\begin{align}
    g = \frac{1}{(1+\varepsilon rq)^2} \left( \frac{C}{r^2-q^2}(d\tau-q^2 d\varphi)^2 + \frac{r^2-q^2}{C} dr^2 + \frac{r^2-q^2}{D}dq^2 + \frac{D}{r^2-q^2} (d\tau-r^2 d\varphi)^2 \right) \label{eq:PlebanskiDemianskiMetric}
\end{align}
where we have introduced $C(r) = F/r$, $D(q) = G/q$.

\subsection{\pleb{}-\demi{} family of solutions}

Having chosen a solution for the conformal factor function $H^2$, we now calculate the Ricci scalar condition $M^{11}+M^{22}+M^{33} = \Lambda$. The components of the matrix $M^{ij}$ for the metric \cref{eq:PlebanskiDemianskiMetric} are as follows
\begin{flalign}
    &\begin{aligned}
    M^{11} =& \frac{1}{6}\frac{(1+\varepsilon r q)^2(r+q)^3}{q^2-r^2}\left[ \left( \frac{C}{(r+q)^3} \right)_{rr} \hspace{-10pt}+ \left( \frac{D}{(r+q)^3} \right)_{qq} \right] \\
    +& \frac{(1+\varepsilon r q)^5}{12(q^2-r^2)} \left[ \left( \frac{C}{(1+\varepsilon rq)^3} \right)_{rr} \hspace{-10pt} + \left( \frac{D}{(1+\varepsilon rq)^3} \right)_{qq} \right],
    \end{aligned} \\
    &\begin{aligned}
    M^{22} = M^{33} =& -\frac{1}{12}\frac{(1+\varepsilon r q)^2(r+q)^3}{q^2-r^2}\left[ \left( \frac{C}{(r+q)^3} \right)_{rr} \hspace{-10pt}+ \left( \frac{D}{(r+q)^3} \right)_{qq} \right] \\
    + & \frac{(1+\varepsilon r q)^5}{12(q^2-r^2)} \left[ \left( \frac{C}{(1+\varepsilon rq)^3} \right)_{rr} \hspace{-10pt} + \left( \frac{D}{(1+\varepsilon rq)^3} \right)_{qq} \right].
    \end{aligned}
\end{flalign}
Taking the trace results in the following equation
\begin{align}
    12 \varepsilon^2 (q^2 C + r^2 D) - 6\varepsilon(1+\varepsilon rq)(qC_r + r D_q) + (1+\varepsilon rq)^2 (C_{rr} + D_{qq}) = 4 (q^2-r^2) \Lambda.
\end{align}
By taking derivatives of both sides 3 times with respect to $r$ and $q$ gives 
\begin{align} \label{eq:pleb-demi-ode-equations}
    C_{rrrrr} = 0, \quad D_{qqqqq} = 0.
\end{align}
This shows that $C$ and $D$ are now 4th order polynomials of their respective variables.
The only remaining nonzero component of the traceless Ricci tensor is then
\begin{equation}
\begin{aligned}
    R^{11} =& (r^2-q^2)(1+\varepsilon rq) D_{qq} + (4q+6\varepsilon rq^2 - 2\varepsilon r^3) D_q -4(1+3\varepsilon rq) D \nonumber\\
            & -(r^2-q^2)(1+\varepsilon rq) C_{rr} + (4r+6\varepsilon r^2q - 2\varepsilon q^3) C_r -4(1+3\varepsilon rq) C.
\end{aligned}
\end{equation}
This acts to further restrict $C,D$. We substitute a generic quartic polynomial ansatz for C, D into $M^{11}+M^{22}+M^{33} = \Lambda$ and $R^{11} = 0$.
This gives the following solution
\begin{equation} \label{eq:PlebanskiDemianskiFinalSolution}
\begin{aligned}
    C = -b + 2mr +  r^2 + 2n\varepsilon r^3 - (\varepsilon^2 b + \Lambda/3) r^4 \\
    D = b + 2n q - q^2 + 2m\varepsilon q^3 + (\varepsilon^2 b + \Lambda/3) q^4.
\end{aligned}
\end{equation}
We have used the freedom of rescaling $r,q \rightarrow \lambda r, \lambda q$ to set the coefficients in front of $r^2, -q^2$ to unity.

We have thus derived the Euclidean version of the \pleb{}-\demi{} metrics with their 5 parameters, $b,m,n,\varepsilon,\Lambda$~\cite{RotatingChargPleban1976,ANewLookAtTGriffi2005}. This is the same result that was obtained in~\cite{AmbitoricGeomeAposto2013}, however the approach here was, we believe, more elementary, without any need for algebraic geometry considerations. 

To check that the metric we obtained is indeed double-sided type D we can calculate both parts of the Weyl curvature. In doing so we find
\begin{align}
    \Psi = \begin{pmatrix}
        2\alpha & 0 & 0 \\
        0 & -\alpha & 0  \\
        0 & 0 & -\alpha
    \end{pmatrix}, \quad \alpha = \frac{n-m}{6} \left( \frac{1+\varepsilon rq}{r+q} \right)^3, \\ \nonumber
    \bar{\Psi} = \begin{pmatrix}
        2\bar{\alpha} & 0 & 0 \\
        0 & -\bar{\alpha} & 0 \\
        0 & 0 & -\bar{\alpha}
    \end{pmatrix}, \quad \bar{\alpha} = \frac{n+m}{6} \left( \frac{1+\varepsilon rq}{r-q} \right)^3.
\end{align}
Clearly both halves of the Weyl tensor are of type D.
When $m = \pm n$ one of the sides of the Weyl curvature vanishes, in this case the spacetimes are called self-dual \pleb{}-\demi{}~\cite{KillingYanoTeNozawa2015}.

\subsection{Kerr Metric}\label{sec:PDMetricToKerrMetric}

To obtain the Kerr metric from the more general \pleb{}-\demi{} family of metrics we set $\Lambda = 0$. We want the metric to be asymptotically flat, requiring that $g_{rr} \rightarrow 1$ as $r \rightarrow \infty$, where $g_{rr} = g_{\mu\nu} (\partial_r)^\mu (\partial_r)^\nu$ is the radial component of the metric. For this to be the case the polynomial $C(r)$ must be at most quadratic, which necessitates $\varepsilon = 0$.
Now both $C,D$ are simple quadratic functions
\begin{align}
    C(r) = r^2 - 2Mr -a^2, \qquad 
    D(q) = a^2 - q^2,
\end{align}
where we renamed the constant terms suggestively, and also set the NUT charge (vanishing in Kerr) $n=0$. Now $C(r)=\Delta$ of the Kerr metric, and setting $q=a\cos\theta$ makes $D=a^2\sin^2\theta$. These choices can be motivated using the requirement that the metric is an appropriate analytic continuation of a Lorentzian signature metric, and that the Lorentzian metric is axisymmetric, but we will not attempt this here. The main claim we are making is that the Euclidean Kerr metric is a member of a large family of metrics whose determination reduces to an exercise in linear algebra, determining the metric ansatz, following by a straightforward computation of curvatures using the \pleb{} formalism. 

With these choices the metric \cref{eq:PlebanskiDemianskiMetric} becomes
\begin{align}
    g = \frac{\Delta}{\rho^2}(d\tau- a^2\cos^2\theta d\varphi)^2 + \frac{\rho^2}{\Delta} dr^2 + \rho^2 d\theta^2 + \frac{\sin^2\theta}{\rho^2} (a d\tau-a r^2 d\varphi)^2.
 \end{align}
 Now further changing the coordinates
 \begin{align}
 d\varphi = -\frac{1}{a} d\phi, \qquad d\tau = dt - a d\phi
 \end{align}
 gives the usual form of the Kerr metric in Boyer-Lindquist coordinates, with the frame given by \cref{eq:KerrTetradBasis}. The $\Lambda\not=0$ Kerr metric is another member of the \pleb{}-\demi{} family, but we have not attempted to exhibit it explicitly.

\section{Discussion}
We have presented an alternative derivation of the (Euclidean) Kerr metric, one that proceeds by deriving the more general \pleb{}-\demi{} family. The derivation we described is ``elementary" in the sense that the most non-trivial step of the construction, which is establishing the ansatz (\ref{eq:Conformal2Form_and_ConformalMetric}), proceeds by a calculation in linear algebra. Thus, one gets rather far without imposing any differential equations whatsoever. Some differential equations are imposed at the next step, which requires that the metric is conformal to two different K\"ahler metrics. This leads to the metric ansatz (\ref{eq:GeneralAmbitoric_Metric}). The rest of the analysis consists in imposing Einstein equations. This is seen to reduce the arbitrariness to two quartic polynomials, each of one variable. One can then recognise the Euclidean Kerr metric in the obtained family of solutions without difficulty. It corresponds to the case of quadratic polynomials. 

Our main reason to be interested in the geometry we described is the fact that the (Euclidean) Kerr metric exhibits the beautiful and rich geometry if viewed as a complex 4D manifold.  This is of course also one of the motivations of~\cite{AmbitoricGeomeAposto2013}. It can also be seen from the derivation we presented that the \pleb{} chiral formalism is best-suited for viewing 4D Euclidean geometry via the prism of complex geometry. Most of what we described is not new for mathematicians, apart from our new characterisation of Killing vector fields in \cref{sec:SU2_Killing_Vectors} and the new proof of \derd{} theorem in \cref{sec:ConformalToKahler}. Some reasoning in section 6, in particular in the part where we impose Einstein equations, is also different from~\cite{AmbitoricGeomeAposto2013}. But our main motivation in describing these results is our hope that the adopted here point of view of the \pleb{} formalism will make these beautiful ideas more digestible for the gravitational physicists.

    \newpage
       \chapter{Nonlinear Gauge Fixing} \label{chap:nonlinear-gauge-fixing}
        \pleb{}'s formulation of gravity places Einstein's equations (EEs) in a form that resembles a gauge theory. Locally, two different groups act on this theory, the first is the diffeomorphism group which contains all coordinate transformations on the manifold. The second group is particular to the \pleb{} formulation and is the $SO(3,\C)$ group of rotations of the frame. To solve Einstein's equations one has to fix a representative in both of these groups, this is known as gauge fixing. How exactly one should gauge fix is arbitrary, but there are some gauges with more desirable properties. In this Thesis we are ultimately interested in obtaining stable numerical schemes for Einstein's equations, as such we consider first-order hyperbolic gauge fixings. See \cref{chap:numerical-relativity} for definitions of stable, well-posed and hyperbolic. Only the highest order in derivatives term contribute to the hyperbolicity, and therefore we can ignore the lower order terms when checking hyperbolicity. We develop the required modifications to \pleb{}'s equations of motion that produce the second-order wave operators for fundamental fields.

        In \cref{chap:Linearised-Gravity} we considered the linearised version of this problem, and developed a first-order hyperbolic gauge fixing for \pleb{}'s formulation of gravity. There we had to introduce extra fields to control the diffeomorphism and rotation gauge, they arose as a 1-form and internal vector, that is $\xi \in \Lambda^1$ and $\chi^i \in \C^3$. The linearisation of the self-dual 2-forms $\Sigma^i$ could be parametrised by an internal vector and symmetric tensor $h^i, h_{\mu\nu}$, and its self-dual connection was $a^i \in \C^3 \otimes \Lambda^1$. Using a modified de Donder and Lorenz gauge it was possible to create a system whose second-order dynamics was the wave equation. Furthermore, these equations conformally separated into spaces of dimension $12$ and $4$, each evolving independently. Here we present the nonlinear version of the same gauge fixing and produce as much as possible from the conformal separation. The linearised parametrisation of the self-dual 2-forms is no longer possible as there is no background with which to expand around, we instead must use the full 2-form. 
        
        We begin with a short overview of the Einstein's equations in \pleb{}'s formulation. Then in \cref{sec:nonlin-hyperbolicity-check} we compute the highest order structure of these equations to show that they are indeed not hyperbolic in the sense that the second-order wave operator structure is spoilt by the presence of other terms. After this, in \cref{sec:nonlin-harmonic-gauge,sec:nonlin-lorenz-gauge}, we develop the minimum gauge fixing required to obtain the wave equation for all the present fields. This, however, does not fix the $SO(3,\C)$ gauge which is done by introducing a modification to the de Donder and Lorenz gauge. The form of this modification is derived in \cref{sec:nonlin-modified-gauge-fixing} and the resulting system is presented in \cref{sec:nonlin-pleb-gauge-fixed-system}. Some results pertaining to the conformal separation are available at the nonlinear level, these are explored in \cref{sec:nonlin-pleb-conformal-separation} and an evolution system based on the conformal separation is given in \cref{sec:nonlin-pleb-conformal-system}.
        
        \section{First-Order \pleb{} Equations of Motion}
        We very briefly recap Einstein's equations as imposed using the \pleb{} formulation. The main fields are the self-dual 2-forms and the connections,
        \begin{align}
            \Sigma^i \in \Lambda^+, \quad A^i \in \C^3 \otimes \Lambda^1.
        \end{align}
        The 2-forms satisfy the metric and reality conditions \cref{eq:pleb-metricity-condition,eq:pleb-reality-condition}. Einstein's equation, in a vacuum, are then neatly encoded into the following first order system
        \begin{align}
            \Sigma^i_\mu{}^\rho F^i_{\nu\rho} = 0, \quad d^A \Sigma^i = 0. \label{eq:nonlinear-plebanski-eom}
        \end{align}
        where $F^i = dA^i + \frac{1}{2} \epsilon^{ijk} A^j \wedge A^k$ is the curvature 2-form. The first equation in the above is Ricci flat condition on the self-dual curvature and the second defines $A^i$ as the self-dual connection for $\Sigma^i$. This way of writing the equations is not at its most useful for the gauge fixing we consider here. Instead, we solve the second equation in the above for the self-dual connection.  First by moving to components and taking the Hodge star of the second equation in \cref{eq:nonlinear-plebanski-eom} we find 
        \begin{align}
            2 (\star d^A \Sigma)^i_\mu = \epsilon_\mu{}^{\nu\rho\sigma} \partial_\nu \Sigma^i_{\rho\sigma} + \epsilon^{ijk} \epsilon_\mu{}^{\nu\rho\sigma} A^j_\nu \Sigma^k_{\rho\sigma} = 0.
        \end{align}
        Then by using the self-duality we can write this as 
        \begin{align}
            \epsilon_\mu{}^{\nu\rho\sigma} \partial_\nu \Sigma^i_{\rho\sigma} - 2 i \epsilon^{ijk} \Sigma^j_\mu{}^\nu A^k_\nu = 0.
        \end{align}
        Using the previously introduced operator $J_1: E \times \Lambda^1 \rightarrow E \times \Lambda^1$ we can write this as 
        \begin{align}
            \star d \Sigma^i = i J_1(A)^i.
        \end{align}
        Then by applying $-i J_1^{-1}$ to both sides 
        \begin{align}
            A^i = -i J_1^{-1}(\star d \Sigma)^i
        \end{align}
        expanding into components results in
        \begin{align}
            A^i_\mu &=  -\frac{i}{4}( \epsilon^{ijk} \Sigma^j_\mu{}^\nu - \delta^{ik} \delta_\mu{}^\nu ) \epsilon_\nu{}^{\alpha\rho\sigma} \partial_\alpha \Sigma^k_{\rho\sigma} \\
            &= \frac{i}{4} \epsilon_\mu{}^{\nu\rho\sigma} \partial_\nu \Sigma^i_{\rho\sigma} + \frac{1}{2} \epsilon^{ijk} \Sigma^{j\rho\sigma} \partial_\rho \Sigma^k_{\mu\sigma} - \frac{1}{4} \epsilon^{ijk} \Sigma^{j\rho\sigma} \partial_\mu \Sigma^k_{\rho\sigma} \label{eq:nonlinear-self-dual-connection-from-Sigma}
        \end{align}
        We see that this is defined in a very similar way to the Christoffel symbols in that it is a linear combination of the partial derivatives of the metric variable, in this case the metric-like variable is $\Sigma^i_{\mu\nu}$. The first-order system that imposes Einstein's equations can then be written as
        \begin{align}
            A^i = -i J_1^{-1}(\star d \Sigma)^i, \quad \Sigma^i_\mu{}^\rho F^i_{\nu\rho} = 0. \label{eq:nonlinear-original-plebanski-system}
        \end{align}
        From here we can start considering gauge fixings. Before that we compute the highest order symbol of the second-order dynamical equations implied by this system.

        \section{Hyperbolicity of Einstein's Equations} \label{sec:nonlin-hyperbolicity-check}
        We can now take our original \pleb{} system and compute the highest order terms arising to check that the equations are not hyperbolic already. The notion of hyperbolicity here refers to the second order equation containing a leading order term that resembles the wave equation. Similarly to the linear story we will see there is a tensor, arising as a nonlinear version of the de Donder 4-vector, that destroys the hyperbolicity.

        To begin we compute the components of the curvature using \cref{eq:nonlinear-self-dual-connection-from-Sigma} and keeping only the highest order terms 
        \begin{align}
            F^i_{\mu\nu} = 2\partial_{[\mu} A^i_{\nu]} = \epsilon^{ijk} \Sigma^{j\rho\sigma}\partial_{\rho[\mu} \Sigma^k_{\nu]\sigma} - \frac{i}{2} \epsilon_{[\mu}{}^{\nu\rho\sigma} \partial_{\nu]\alpha} \Sigma^i_{\rho\sigma}.
        \end{align}
        Not that we have introduced the notation $\partial_{\mu\nu} = \partial_\mu \partial_\nu = \partial_{(\mu} \partial_{\nu)}$ and later $\partial^2 = \partial^\mu \partial_\mu$. The reason being is that at the level of the leading order terms, $\partial_{\mu\nu}$ behaves as a single derivative and satisfies the same product rule
        \begin{align}
            \partial_{\mu\nu}(fg) = g \partial_{\mu\nu} f  + f \partial_{\mu\nu} g + {\rm lower\ order\ terms}.
        \end{align}
        The metric can then be used to raise and lower either index as one would for the single derivative. Substituting the curvature into the Einstein condition we find 
        \begin{align}
            \Sigma^i_{\mu}{}^\rho F^i_{\nu\rho} = & g_{\mu\nu} \left( \frac{1}{4} \Sigma^{i\rho\sigma} \partial^2 \Sigma^i_{\rho\sigma} + \frac{1}{2} \Sigma^{i\rho\sigma} \partial_{\rho}{}^\alpha \Sigma^i_{\sigma\alpha}\right) \\ & + \partial_\mu \left( -\frac{1}{2} \Sigma^i_\nu{}^\rho \partial^\sigma \Sigma^i_{\rho\sigma} + \frac{1}{2} \Sigma^{i\rho\sigma} \partial_\rho \Sigma^i_{\nu\sigma} - \frac{1}{4} \Sigma^{i\rho\sigma} \partial_\nu \Sigma^i_{\rho\sigma} \right) \\ & + \partial_\nu \left( -\frac{1}{2} \Sigma^i_\mu{}^\rho \partial^\sigma \Sigma^i_{\rho\sigma} + \frac{1}{2} \Sigma^{i\rho\sigma} \partial_\rho \Sigma^i_{\mu\sigma} - \frac{1}{4} \Sigma^{i\rho\sigma} \partial_\mu \Sigma^i_{\rho\sigma} \right) \\ & - \Sigma^i_{(\mu}{}^\rho \partial^2 \Sigma^i_{\nu)\rho} - \Sigma^i_{\rho(\mu} \partial^{\rho\sigma} \Sigma^i_{\nu)\sigma}.
        \end{align}
        In this form it is difficult to extract anything meaningful. The result is symmetric in $\mu\nu$ as the antisymmetric part is equal to $\epsilon^{ijk} \Sigma^j \wedge F^k$ and this vanishes whenever $A^i$ is coming from its definition in \cref{eq:nonlinear-original-plebanski-system}. However, there are a collection of identities derived in \cref{apx:second-order-formulas-for-Sigma} which can be used to simplify the above. The identities are derived by considering only the highest order structure, here we will state the identities without derivation. To begin we use the following 
        \begin{align}
            \Sigma^{i\rho\sigma} \partial^2 \Sigma^i_{\rho\sigma} = 6 \partial^2 \ln(\sqrt{-g}) \\ \Sigma^{i\rho\sigma} \partial_\rho{}^\alpha \Sigma^i_{\sigma\alpha} = - \partial^{\rho\sigma} g_{\rho\sigma} - \partial^2 \ln(\sqrt{-g})
        \end{align}
        which allows us to simplify the terms in brackets on the first line
        \begin{align}
            \frac{1}{4} \Sigma^{i\rho\sigma} \partial^2 \Sigma^i_{\rho\sigma} + \frac{1}{2} \Sigma^{i\rho\sigma} \partial_{\rho}{}^\alpha \Sigma^i_{\sigma\alpha} &= \frac{3}{2} \partial^2 \ln(\sqrt{-g}) - \frac{1}{2} \partial^2 \ln(\sqrt{-g}) - \frac{1}{2} \partial^{\rho\sigma} g_{\rho\sigma}  \\ &= \partial^2 \ln(\sqrt{-g}) + \frac{1}{2} \partial_{\rho\sigma} g^{\rho\sigma}.
        \end{align}
        For the second and third lines we can use 
        \begin{align}
            -\frac{1}{2} \Sigma^i_\mu{}^\rho \partial^\sigma \Sigma^i_{\rho\sigma} + \frac{1}{2} \Sigma^{i\rho\sigma} \partial_\rho \Sigma^i_{\mu\sigma} &= \partial_\mu \ln(\sqrt{-g}) - g_{\mu\rho} \partial_\sigma g^{\rho\sigma} \\ \frac{1}{4} \Sigma^{i\rho\sigma} \partial_\mu \Sigma^i_{\rho\sigma} &= \frac{3}{2} \partial_\mu \ln(\sqrt{-g})
        \end{align}
        to find 
        \begin{align}
            -\frac{1}{2} \Sigma^i_\mu{}^\rho \partial^\sigma \Sigma^i_{\rho\sigma} + \frac{1}{2} \Sigma^{i\rho\sigma} \partial_\rho \Sigma^i_{\mu\sigma} - \frac{1}{4} \Sigma^{i\rho\sigma} \partial_\mu \Sigma^i_{\rho\sigma} = -\frac{1}{2} \partial_\mu \ln(\sqrt{-g}) - g_{\mu\rho} \partial_\sigma g^{\rho\sigma}.
        \end{align}
        Using these in Einstein's equations gives
        \begin{align}
            \Sigma^i_{\mu}{}^\rho F^i_{\nu\rho} = & g_{\mu\nu} \left( \partial^2 \ln(\sqrt{-g}) + \frac{1}{2} \partial_{\rho\sigma} g^{\rho\sigma} \right)  \\ & + \partial_{\mu} \left( -\frac{1}{2} \partial_\nu \ln(\sqrt{-g}) - g_{\nu\rho} \partial_\sigma g^{\rho\sigma} \right) \\ & + \partial_{\nu} \left( -\frac{1}{2} \partial_\mu \ln(\sqrt{-g}) - g_{\mu\rho} \partial_\sigma g^{\rho\sigma} \right) \\ & - \Sigma^i_{(\mu}{}^\rho \partial^2 \Sigma^i_{\nu)\rho} - \Sigma^i_{\rho(\mu} \partial^{\rho\sigma} \Sigma^i_{\nu)\sigma}.
        \end{align}
        The final line in the above can be simplified using the following identities
        \begin{align}
            \Sigma^i_{(\mu}{}^\rho \partial^2 \Sigma^i_{\nu)\rho} &= \partial^2 g_{\mu\nu} + g_{\mu\nu} \partial^2 \ln(\sqrt{-g}) \\ \Sigma^i_{\rho(\mu} \partial^{\rho\sigma} \Sigma^i_{\nu)\sigma} &= -\frac{1}{2} \partial_\mu (g_{\nu\rho} \partial_\sigma g^{\rho\sigma}) -\frac{1}{2} \partial_\nu (g_{\mu\rho} \partial_\sigma g^{\rho\sigma}) + \frac{1}{2} g_{\mu\nu} \partial_{\rho\sigma} g^{\rho\sigma} - \frac{1}{2} \partial^2 g_{\mu\nu}.
        \end{align}
        Substituting these into the original equation, and after many cancellations, the result greatly simplifies to
        \begin{align}
            \Sigma^i_{\mu}{}^\rho F^i_{\nu\rho} = -\frac{1}{2} \partial^2 g_{\mu\nu} + \partial_{(\mu} \Gamma_{\nu)}, \quad \Gamma_\mu = -\partial_\mu \ln(\sqrt{-g}) - g_{\mu\sigma} \partial_\rho g^{\rho\sigma} = - \utilde{g}_{\mu\rho} \partial_\sigma \tg^{\rho\sigma}. \label{eq:nonlinear-einstein-equations-as-box-g-plus-H}
        \end{align}
        Where $\Gamma^\mu = \Gamma^\mu_{\rho\sigma} g^{\rho\sigma}$ and $\Gamma^\mu = \Box x^\mu = 0$ is the Harmonic condition. The Harmonic condition is the nonlinear version of the de Donder gauge. We have also introduced the notation $\tg^{\mu\nu} = \sqrt{-g} g^{\mu\nu}$ and $\utilde{g}_{\mu\nu} = g_{\mu\nu}/\sqrt{-g}$. The 2-forms $\Sigma^i$ contain the metric plus a frame, it is clear that Einstein's equation are then equations on the metric part of this 2-form. Using the fact that, in this notation, $(\delta g)_{\mu\nu} = 2 \hat{h}_{\mu\nu} + \frac{1}{2} h g_{\mu\nu}$ and linearising the Harmonic condition we see that it reproduces the linear gauge fixing \cref{eq:lin-de-Donder-connection-from-h}. From this we see that the highest order structure of the equations is closely related to the linearisation, which is perhaps unsurprising as they are very similar calculations. We can now suggest nonlinear equations of motion to remove the unwanted terms to the highest order, keeping in mind that the linearisations should match those in the previous section. In fact, due to the similarities, the nonlinear gauge can almost be read directly from the linear gauges.

        \section{Highest Order Harmonic Gauge} \label{sec:nonlin-harmonic-gauge}
        From \cref{eq:nonlinear-einstein-equations-as-box-g-plus-H} we can see that $\Gamma^\mu = 0$ would make the equations hyperbolic. To impose this, as in the linear case, we introduce a first-order field to impose the condition dynamically. Let $\xi \in \Lambda^1$ be a connection 1-form, we can modify the equations of motion such that 
        \begin{align}
            \Sigma^i_\mu{}^\rho F^i_{\nu\rho} + \partial_{(\mu} \xi_{\nu)} = 0 \label{eq:nonlinear-modified-einsteins-conditions-on-connections}.
        \end{align}
        When supplemented with 
        \begin{align}
            \xi_\mu = - \Gamma_\mu
        \end{align}
        and the original equation for $A^i$ we obtain a hyperbolic second order partial differential equation for $g_{\mu\nu}$ to highest order
        \begin{align}
            \Sigma^i_\mu{}^\rho F^i_{\nu\rho} + \partial_{(\mu} \xi_{\nu)} = -\frac{1}{2} \partial^2 g_{\mu\nu}.
        \end{align}
        When $\xi_\mu = 0$ we recover the Harmonic conditions and the modifications to the Ricci flat condition are removed. Hence, use the gauge fixed Einstein's equations along with $\xi = - \Gamma$, as long as $\xi = 0$ is imposed somehow. This provides a hyperbolic second-order equation for the metric components in $\Sigma^i$, however, for the $SO(3,\C)$ frame we require a different gauge fixing.
        
        \section{Highest Order Modified Lorenz Gauge} \label{sec:nonlin-lorenz-gauge}
        Taking inspiration from the linear case we can fix the $SO(3,\C)$ freedom by imposing the Lorenz condition non-linearly
        \begin{align}
            \partial^\mu A^i_\mu = g^{\mu\nu} \partial_\mu A^i_\nu = 0
        \end{align}
        Using our nonlinear definition for $A^i$, \cref{eq:nonlinear-self-dual-connection-from-Sigma}, we can compute the Lorenz condition to leading order and find 
        \begin{align}
            \partial^\mu A^i_\mu = -\frac{1}{4}\epsilon^{ijk} \Sigma^{j\mu\nu} \partial^2 \Sigma^k_{\mu\nu} - \frac{1}{2} \epsilon^{ijk} \Sigma^{j\rho\sigma} \partial^\alpha{}_\rho \Sigma^k_{\sigma\alpha}.
        \end{align}
        The last term in the above can be manipulated to reveal
        \begin{align}
            \epsilon^{ijk} \Sigma^{j\rho\sigma} \partial^\alpha{}_\rho \Sigma^k_{\sigma\alpha} = &- \Sigma^{i\mu\nu} \partial_\mu \left( \partial_\nu \ln(\sqrt{-g}) + g_{\nu\rho} \partial_\sigma g^{\rho\sigma} \right) - \frac{1}{4} \epsilon^{ijk} \Sigma^{j\mu\nu} \partial^2 \Sigma^k_{\mu\nu} \\ = & \Sigma^{i\mu\nu} \partial_\mu \Gamma_\nu - \frac{1}{4} \epsilon^{ijk} \Sigma^{j\mu\nu} \partial^2 \Sigma^k_{\mu\nu}.
        \end{align}
        In which case the Lorenz condition becomes
        \begin{align}
            \partial^\mu A^i_\mu = -\frac{1}{2} \Sigma^{i\mu\nu} \partial_\mu \Gamma_\nu - \frac{1}{8} \epsilon^{ijk} \Sigma^{j\mu\nu} \partial^2 \Sigma^k_{\mu\nu}.
        \end{align}
        This suggests a modification to the gauge choice, whose linear structure matches that gauge fixing found in \cref{chap:Linearised-Gravity}, which we call the modified Lorenz gauge
        \begin{align}
            \partial^\mu A^i_\mu - \frac{1}{2} \Sigma^{i\mu\nu} \partial_\mu \xi_\nu = 0.
        \end{align}
        Then clearly with $\xi_\mu = - \Gamma_\mu$ we have 
        \begin{align}
            \partial^\mu A^i_\mu - \frac{1}{2} \Sigma^{i\mu\nu} \partial_\mu \xi_\nu = -\frac{1}{8}\epsilon^{ijk} \Sigma^{j\mu\nu} \partial^2 \Sigma^k_{\mu\nu}.
        \end{align}
        Which has the wave operator as its leading order term. Computing the linearisation of the right-hand side also confirms that this is the wave equation on the perturbation of the $SO(3,\C)$ frame. This condition combined with the Harmonic gauge converts the leading order symbol for Einstein's equations to be that of the wave operator acting on all the components in the 2-form $\Sigma^i$.
        
        \section{Modified Harmonic and Connections} \label{sec:nonlin-modified-gauge-fixing}
        As in the linear case we can count the same number of components in the frame and connection sector to find $13$ in $\Sigma^i$ and $16$ in the connections $A^i$ and $\xi$. Since the 16 connections are defined by 13 frame variables we will find that in the numerical scheme that some equations become constraints instead of evolution equations. Specifically in this case we will find 3 Gauss constraints related to the $SO(3,\C)$ gauge. The desired numerical scheme will evolve the constraints instead of imposing them as additional conditions. To achieve this at the covariant level we introduce an internal vector $\chi^i$ which we interpret as being part of the frame. This matches the number of components in the frame and connection sectors, at the linear level this allowed for the construction of the hyperbolic Dirac operator. We will see that a similar construction is available at the nonlinear level, and modifications can be made to the previously described gauges that give $\chi^i$ a hyperbolic second-order equations of motion.
        
        The first modification is to the definition of the self-dual connection, we add all the possible terms that contains a derivative of $\chi^i$ to find
        \begin{align}
            A^i &= -i J_1^{-1}(\star d \Sigma)^i + c_1 d\chi^i + c_2 J_1(d\chi)^i \label{eq:nonlinear-self-dual-connection-modified-definition}
        \end{align}
        which in components is 
        \begin{align}
            A^i_\mu = \frac{i}{4} \epsilon_\mu{}^{\nu\rho\sigma} \partial_\nu \Sigma^i_{\rho\sigma} + \frac{1}{2} \epsilon^{ijk} \Sigma^{j\rho\sigma} \partial_\rho \Sigma^k_{\mu\sigma} - \frac{1}{4} \epsilon^{ijk} \Sigma^{j\rho\sigma} \partial_\mu \Sigma^k_{\rho\sigma} + c_1 \partial_\mu \chi^i + c_2 \epsilon^{ijk} \Sigma^j_\mu{}^\nu \partial_\nu \chi^k.
        \end{align}
        The intrinsic torsion of the self-dual 2-form has changed due to this gauge fixing, performing a computation similar to that in the proof \cref{eq:plebanski-unique-intrinsic-torsion-proof} we find that the intrinsic torsion is now 
        \begin{align}
            -T^i = A^i - c_1 d\chi^i - c_2 J_1(d\chi)^i.
        \end{align}
        This means that the metric covariant derivative of the self-dual 2-forms is then 
        \begin{align}
            \nabla_\mu \Sigma^i_{\rho\sigma} = -\epsilon^{ijk} \left( A^j_\mu - c_1 \partial_\mu \chi^j - c_2 J_1(d\chi)^j_\mu \right) \Sigma^k_{\rho\sigma}.
        \end{align}
        Or that the total covariant derivative is 
        \begin{align}
            \nabla_\mu^A \Sigma^i_{\rho\sigma} = \epsilon^{ijk}(c_1 \partial_\mu \chi^j + c_2 J_1(d\chi)^j_\mu) \Sigma^k_{\rho\sigma} \label{eq:nonlinear-plebanski-modified-total-covariant-deriative-on-Sigma}.
        \end{align}
        The only possible modification to the connection 1-form is
        \begin{align}
            \xi_\mu = -\Gamma_\mu + c_3 \Sigma^i_\mu{}^\nu \partial_\nu \chi^i.
        \end{align}
        Here, $c_1,c_2,c_3 \in \C$ are constants. This is the simplest nonlinear modification that reproduces the linear modification. The structure is exactly the same as the linear case, however, the $\Sigma^i$ here are assumed to be dynamical instead of the flat Minkowski 2-forms. Contributions from $\partial \Sigma^i$ terms have already been computed, and the highest order terms are linear in their arguments. Therefore, we can consider
        \begin{align}
            A^i_\mu & = c_1 \partial_\mu \chi^i + c_2 \epsilon^{ijk} \Sigma^j_\mu{}^\nu \chi^k \\ \xi_\mu & = c_3 \Sigma^i_\mu{}^\nu \partial_\nu \chi^i.
        \end{align}
        We are able to separate the solutions this way as the highest order terms are linear in their arguments. Substituting the above into the gauge fixed Einstein's equations, \cref{eq:nonlinear-modified-einsteins-conditions-on-connections}, we find each term to be 
        \begin{align}
            \Sigma^i_\mu{}^\rho F^i_{\nu \rho} & = c_2 \Sigma^i_{\mu\nu} \partial^2 \chi^i + 2c_2 \Sigma^i_{(\mu}{}^\rho \partial_{\nu)\rho} \chi^i \\
            \partial_{(\mu} \xi_{\nu)} & = c_3 \Sigma^i_{(\mu}{}^\rho \partial_{\nu)\rho} \chi^i.
        \end{align}
        Combining we find 
        \begin{align}
            \Sigma^i_\mu{}^\rho F^i_{\nu \rho} + \partial_{(\mu} \xi_{\nu)} = c_2 \Sigma^i_{\mu\nu} \partial^2 \chi^i + (c_3 + 2c_2) \Sigma^i_{(\mu}{}^\rho \partial_{\nu)\rho} \chi^i.
        \end{align}
        The correct choice of constants is then $c_3 = -2c_2$, which again matches the linear story. The modified Lorenz condition in this case remains hyperbolic
        \begin{align}
            \partial^\mu A^i_\mu - \frac{1}{2}\Sigma^{i\mu\nu} \partial_\mu \xi_\nu = (c_2 - c_1) \partial^2 \chi^i.
        \end{align}
    
        \section{Gauge Fixed \pleb{} System} \label{sec:nonlin-pleb-gauge-fixed-system}
        Collecting all the gauge fixing results together we find our first-order hyperbolic equations of motion for \pleb{}'s formulation of gravity are
        \begin{gather}
            A^i = -i J_1^{-1}(\star d\Sigma)^i + c_1 d\chi^i + c_2 J_1(d\chi)^i, \quad \xi_\mu = - \Gamma_\mu - 2c_2 \Sigma^i_\mu{}^\nu \partial_\nu \chi^i \label{eq:nonlinear-modified-connection-definitions}\\
            \Sigma^i_\mu{}^\rho F^i_{\nu\rho} + \partial_{(\mu} \xi_{\nu)} = 0, \quad \partial^\mu A^i_\mu - \frac{1}{2} \Sigma^{i\mu\nu} \partial_\mu \xi_\nu = 0 \label{eq:nonlinear-modified-einstein-and-lorenz-conditions}.
        \end{gather}
        The dynamical fields are then $\Sigma^i, \chi^i$ which we call the frame variables, along with the connection variables $A^i, \xi$. The connections are defined as linear combinations of the derivatives of the frame fields. Einstein's equations on these connections are then first-order differential equations. The second-order evolution equations for the frame, implied by the above system, have the wave operator as their leading order symbols. The fields $\chi^i$ and $\xi$ are 3 and 4 components that are introduced to dynamically control the $SO(3,\C)$ and diffeomorphism gauges. The first equation is a modified definition for the self-dual connection, the second is a modification to the Harmonic condition involving $\chi^i$. Einstein's equations is adjusted with derivatives of $\xi$ and the final equation is a modified Lorenz condition on the self-dual connection. This kind of modification to Einstein's equations has appeared in the literature before and has been used to prove that Einstein's equations are well-posed in \cite{Theoreme_d_exis_Foures_1952,Existence_of_me_DeTurc_1981} and in other numerical works, see \cite{General_covaria_Bona_2003,A_new_approach_Headri_2009}. As for the Lorenz condition this appears in the context of Yang-Mills theories where it is well known that it produces wave operator, e.g. see chapter 8 of \cite{FormGenRelGravity2020}. In numerical relativity the Lorenz condition has appeared in implementations of the tetrad formalism \cite{HyperbolicTetrBuchma2003}.
        
        It is clear that in order to recover the original system we must also impose $\xi = 0 = \chi^i$ such that the changes to the original equations of motion are undone. At the level of the Hamiltonian these new fields arise as conjugate momentum for the Lagrange multipliers, therefore, the primary constraints of the theory require them to vanish. It is also the reason why we can add these terms to the equations of motion without any worry, as the modifications keep the first-class structure of the primary and secondary constraints. The reason we have not already set the new fields to zero is because as an evolution system removing $\xi$ or $\chi^i$ would destroy the hyperbolicity and therefore cause the system to be ill-posed. Instead, we notice that since every field satisfies the wave equation we can impose $\partial_t \xi = \xi = 0 = \chi^i = \partial_t \chi^i$ for the initial data, and analytically it will remain true for all times. This is exactly the same mechanism that is used in the chiral Maxwell gauge fixing. This setup does require that the initial data be a solution to the original Einstein equations, and this is not always possible to find for every physical setup. Instead, as we will discuss in more detail in \cref{chap:numerical-relativity}, it may be useful to add damping terms to achieve this. Damping terms can be added to this gauge fixing as they occur at a lower order and can be added such that Einstein's equations are recovered as the additional fields tend to zero. In practice, we can add any terms containing $\xi$ and $\chi^i$ as long it can be shown that numerically these quantities tend to zero, or in some cases to a constant, sufficiently quickly such that true solutions to Einstein's equations are recovered in a reasonable time interval.

        One can also notice that this system no longer transforms covariantly under diffeomorphism transformations. This is most notably a problem in spherical polar coordinates as some objects have angular dependence that clearly breaks the spherical symmetry. In some cases the lack of covariance will introduce equations of motion that destroy the system and force the metric to be degenerate. We briefly describe how each field transforms under diffeomorphisms, to see where the problem occurs. Under diffeomorphism transformations we know that the frame variables all transform as tensors by definition, the self-dual connection also transforms covariantly. However, the Harmonic 1-form $\Gamma_\mu$ is not a tensor under diffeomorphisms as it is constructed from the Christoffel symbols which are themselves not tensors. While this is not necessarily an issue as this term exists to fix the diffeomorphism gauge, as mentioned in spherical symmetry this condition is not well-defined. To resolve this and restore covariance we follow~\cite{Covariant_formu_Brown_2009,Formulations_of_Alcubi_2010} and introduce a non-dynamical reference metric $\bar{g}_{\mu\nu}$. This reference metric is fixed to be the Minkowski metric in whichever coordinate system is being used. We can then compute its Christoffel symbols $\bar{\Gamma}^\mu_{\rho\sigma}$ which are in general not zero since they have different values for different choices of coordinates, and there may exist a coordinate system where they are not zero. This can then be used to construct the tensor,
        \begin{align}
            \Delta \Gamma_\mu = \Gamma_\mu - g^{\rho\sigma} g_{\mu\nu} \bar{\Gamma}^\nu_{\rho\sigma}.
        \end{align}
        Then whenever $\Gamma_\mu$ appears we can replace it with $\Delta \Gamma_\mu$ without changing the physical dynamics. In the usual Cartesian coordinates, $x^\mu = (t,x,y,z)$, it is easy to see that $\bar{\Gamma} = 0$ and so the original Harmonic condition is recovered. We call this the covariant Harmonic 1-form.

        The other noncovariant term is in the modification to the Einstein condition, $\partial_{(\mu} \xi_{\nu)}$, the use of a partial derivative destroys the covariance of this term. It is possible to promote this to covariant derivatives, this can be recognised as the Z4 equations \cite{General_covaria_Bona_2003} which have been widely successful in numerical relativity. However, as we will see in later chapters, promoting the derivative in this way can destroy the polynomial structure that occurs in the 3+1 decomposition of this system. We argue that as this term should vanish for suitable schemes then it should not produce a problem during the numerical evolution. Therefore, we choose to leave this term as partial derivative. The other noncovariant equation is the Lorenz gauge where partial derivatives are employed instead of metric covariant derivatives. For the same reason we leave this term as a partial derivative as the structure of the evolution system is simpler in this choice. A brief look into the numerical performance of these choices will be presented in \cref{chap:plebanski-numerical-relativity}.
    
        \section{Nonlinear Gauge Conformal Separation} \label{sec:nonlin-pleb-conformal-separation}
        In \cref{chap:Linearised-Gravity} we presented a splitting on the field equations into first-order sectors of size 4 and 12. In this section we show here that a slightly weaker result is obtainable with minimal modification to \cref{eq:nonlinear-modified-einstein-and-lorenz-conditions}. It is possible to obtain a first order system of 12 frame and connection variables that is separate from the remaining 4 component sector. However, the 4 sector still depends on 12, this is due to the fact that the full metric is a mix of the 12 and 4 component of the full frame. It was seen that the conformal transformations were stored in the 4 component part and the rest remained in the 12 component sector. Using this knowledge we develop a conformal transformation for the nonlinear fields that separates the gauge fixed equations of motion.

        \subsection{Connection Definition Conformal Transformation}
        We begin by performing a conformal decomposition of the frame variables. We denote the metric related to the 2-forms $\Sigma^i$ by $g$, then by introducing a scalar, $\lambda = 1/\sqrt{\sqrt{-\det(g)}}$,  we can define a new metric with unit determinant,
        \begin{align}
            g_H = \lambda g \label{eq:nonlinear-conformal-metric}
        \end{align}
        such that $\det(g_H) = -1$. The 2-forms that correspond to this conformal metric are
        \begin{align}
            H^i = \lambda \Sigma^i \in \Lambda^+ \subset \Lambda^2
        \end{align}
        and $g_H$ is recovered from $H^i$ using the \urb{} metric. Due to the conformal invariance of the Hodge star on middle degree forms, $H^i$ and $\Sigma^i$ are self-dual under both metrics. We note that the spacetime indices on $H^i_{\mu\nu}$ are raised and lowered using $g_H$, such that it satisfies a quaternion algebra with this metric 
        \begin{align}
            H^i_{\mu\rho} H^{j\rho}{}_\nu = -\delta^{ij}g_H{}_{\mu\nu} + \epsilon^{ijk} H^k_{\mu\nu}. \label{eq:conformal-H-quaternion-algebra}
        \end{align}
        All other identities follow from the replacement of $(\Sigma^i,g) \rightarrow (H^i,g_H)$.

        Next we can look at the effect of this conformal transformation on the definitions of the connections in \cref{eq:nonlinear-modified-connection-definitions}. We note that $\Sigma^i_\mu{}^\nu = H^i_\mu{}^\nu$ as they are conformally invariant, therefore $J_1$ is also unchanged by any conformal transformation. The modified self-dual connection becomes 
        \begin{align}
            A^i & = -i J_1^{-1}\left( \star d\left(\frac{1}{\lambda} H^i\right) \right) + c_1 d\chi^i + c_2 J_1(d\chi)^i \\ 
                & = i J_1^{-1}\left( \frac{1}{\lambda} \star \left( d\log(\lambda) \wedge H^i \right) \right) -i J_1^{-1} \left( \frac{1}{\lambda} \star dH^i \right) + c_1 d\chi^i + c_2 J_1(d\chi)^i.
        \end{align}
        Where $\star$ is the Hodge star for $g$. Using the conformal transformation we can relate the conformal Hodge star $\star_H$ to usual Hodge star $\star$. Indeed, when $\star$ is acting on a 3-form we find 
        \begin{align}
            (\star T)_\mu = \epsilon{}_\mu{}^{\nu\rho\sigma} T_{\nu\rho\sigma} = g{}_{\mu\alpha} \epsilon^{\alpha\nu\rho\sigma} T_{\nu\rho\sigma} = \frac{1}{\lambda} g_H{}_{\mu\nu} \sqrt{-g}\epsilon^{\alpha\nu\rho\sigma}_H T_{\nu\rho\sigma} = \lambda \epsilon_H{}_\mu{}^{\nu\rho\sigma} T_{\nu\rho\sigma} = \lambda (\star_H T)_\mu
        \end{align}
        where we have used the definition of $\lambda$. The self-dual connection simplifies slightly to become 
        \begin{align}
            A^i & = i J_1^{-1}\left( \star_H \left( d\log(\lambda) \wedge H^i \right) \right) -i J_1^{-1} \left( \star_H dH^i \right) + c_1 d\chi^i + c_2 J_1(d\chi)^i.
        \end{align}
        For the first term we find the following 
        \begin{align}
            \star_H (d\log(\lambda) \wedge H^i)_\mu = \frac{1}{2} \epsilon_{H\mu}{}^{\nu\rho\sigma} \partial_\nu \log(\lambda) H^i_{\rho\sigma} = i H^i_\mu{}^\nu \partial_\nu \log(\lambda) = i (H^i \cdot d \log(\lambda))_\mu
        \end{align}
        In the above we have used the notation $(H^i \cdot \theta)_\mu = H^i_\mu{}^\nu \theta_\nu$, and later we will use the same notation with $\Sigma^i$.  Next, we recall that $J_1^{-1} = \frac{1}{2}(J_1 - 1)$ and $J_1( H^i \cdot X ) = 2 H^i \cdot X$, which can be checked using the quaternion algebra and that $\Sigma^i_\mu{}^\nu = H^i_\mu{}^\nu$, to show that $J_1^{-1}(H^i \cdot X) = \frac{1}{2} H^i \cdot X$. The conformal decomposition of the self-dual connection is then
        \begin{align}
            A^i = -\frac{1}{2} H^i \cdot d\log(\lambda) -i J_1^{-1}( \star_H d H^i) + c_1 d\chi^i + c_2 J_1(d\chi)^i.
        \end{align}

        The same conformal transformation can be applied to the connection 1-form, $\xi$. Using the standard formula for the conformal transformation of the Christoffel symbols reveals 
        \begin{align}
            \Gamma_{\mu} = g^{\rho\sigma} g_{\mu\nu} \Gamma^\nu{}_{\rho\sigma} = g_H^{\rho\sigma} g_{H\mu\nu} \Gamma_H^\nu{}_{\rho\sigma} + \partial_\mu \log(\lambda) = \Gamma_{H\mu} + \partial_\mu \log(\lambda).
        \end{align}
        Applying this to the connection 1-form becomes 
        \begin{align}
            \xi = -\Gamma_H - d\log(\lambda) - 2c_2 H^i \cdot d\chi^i.
        \end{align}
        As in the linear case we introduce two special combinations of the self-dual and 1-form connections,
        \begin{align}
            \Omega^i = A^i - \frac{1}{2} H^i \cdot \xi, \quad \omega = \xi + 2 H^i \cdot A^i. \label{eq:nonlinear-Omega-in-terms-of-A-and-xi}
        \end{align}
        Recall that $H^i \cdot = \Sigma^i \cdot$ as they are conformally invariant, so this is equivalent to the definitions in \cref{chap:Linearised-Gravity}.
        These can be inverted to find 
        \begin{align}
            A^i = \frac{1}{2} \Omega^i + \frac{1}{2}J_1(\Omega)^i - \frac{1}{4} H^i \cdot \omega, \quad \xi = -\frac{1}{2} \omega + H^i \cdot \Omega^i \label{eq:nonlinear-A-xi-to-Omega-connection}
        \end{align}
        which gives a way to convert between the two connections. Computing the first connection in \cref{eq:nonlinear-Omega-in-terms-of-A-and-xi} using the conformal transformation we find
        \begin{align}
            \Omega^i & = -\frac{1}{2} H^i \cdot d \log(\lambda) - i J_1^{-1}(\star_H dH^i) + c_1 d\chi^i + c_2 J_1(d\chi)^i \\ & \qquad -\frac{1}{2} \left( - H^i \cdot \Gamma_H - H^i \cdot d \log(\lambda) + 2 c_2 d\chi^i + 2 c_2 J_1(d\chi)^i \right) \nonumber \\
            & = -i J_1^{-1}(\star_H dH^i) + \frac{1}{2} H^i \cdot \Gamma_H + (c_1 - c_2) d\chi^i.
        \end{align}
        The second of the new connections can be derived using $H^i \cdot J_1^{-1}(A)^i = \frac{1}{2} H^i \cdot A^i$ to give
        \begin{align}
            \omega & = - \Gamma_H - d \log(\lambda) - 2 c_2 H^i \cdot d\chi^i + 2 \left( \frac{3}{2} d \log(\lambda) - \frac{i}{2} H^i \cdot \star_H dH^i + (c_1 + 2 c_2) H^i \cdot d\chi^i \right) \nonumber \\
            & = - \Gamma_H + 2 H^i \cdot d \log(\lambda) -i H^i \cdot \star_H dH^i + 2(c_1 + c_2) H^i \cdot d\chi^i
        \end{align}

        \subsection{Conformal Connection}
        To summarise, the conformal connections are defined using the conformal frame variables $H^i,\lambda,\chi^i$ in the following way 
        \begin{align}
            \Omega^i & = -i J_1^{-1}(\star_H dH^i) + \frac{1}{2} H^i \cdot \Gamma_H + (c_1 - c_2) d\chi^i \\
            \omega & = -\Gamma_H + 2 d\log(\lambda) - i H^i \cdot \star_H dH^i + 2(c_1 + c_2) H^i \cdot d\chi^i
        \end{align}

        One specific choice of $c_1,c_2$ deserves some more attention. Similarly to the linear case we see that when $c_2 = c_1 = c$ then the 12 connection, $\Omega^i$, moreover, as we require $c \neq 0$ such that not all the gauge fixing is removed, we can scale $\chi^i$ such that $c = 1$ can always be chosen and the connection equations become
        \begin{align}
            \Omega^i & = -i J_1^{-1}(\star_H dH^i) + \frac{1}{2} H^i \cdot \Gamma_H \\
            \omega & = -\Gamma_H + 2 d\log(\lambda) - i H^i \cdot \star_H dH^i + 4 H^i \cdot d\chi^i.
        \end{align}
        Counting the number of components in $H^i$ we find it contains a metric with determinant 1 and the $SO(3,\C)$ frame which constitutes $10-1+3=12$ components The 4 remaining components are accounted for by $\lambda, \chi^i$. This suggests that the 12,4 split is also available at the nonlinear level, albeit a weaker version where the 12 components are independent of the 4 components, but the reverse is not true.
        
        \subsection{Connection Evolution Conformal Transformation}
        The conformal separation of the new connection variables is expected to be preserved when rewriting Einstein's equations and the Lorenz gauge with the new connections.  Here, we will show that, by changing the lower order terms in \cref{eq:nonlinear-modified-einstein-and-lorenz-conditions} and slightly modifying the higher order terms, the separation of the 12 dimensional sector is achievable at the nonlinear level.

        We start by making an educated guess as to the equations of motion for $\Omega^i, \omega$ we then show that this is indeed a gauge fixing of Einstein's equations. Looking at \cref{eq:lin-12-full-gauge-fixed-system} we define the equations of motion for the conformal connection to be
        \begin{gather}
            H^i_{\langle \mu}{}^\rho F(\Omega)^i_{\nu\rangle\rho} = 0, \quad g_H^{\mu\nu} \partial_\mu \Omega^i_\nu = 0 \label{eq:nonlinear-pleb-12-connection-eoms}\\
            H^{i\mu\nu} \partial_\mu \omega_\nu = 0, \quad g_H^{\mu\nu} \partial_\mu \omega_\nu = 0 \label{eq:nonlinear-pleb-4-connection-eoms}.
        \end{gather}
        Where $F(\Omega)^i = d\Omega^i + \frac{1}{2} \epsilon^{ijk} \Omega^j \wedge \Omega^k$, however we expect to modify these equations of motion with potential terms later. All that needs to be checked is that these equations are equivalent to Einstein's equations plus terms involving the primary constraints $\xi, \chi^i$, then this system is equivalent to the original equations of motion  on the constraint surface.

        \subsection{$\Omega^i$ Connection Equations}
        Beginning with the first equation in \cref{eq:nonlinear-pleb-12-connection-eoms} we use \cref{eq:nonlinear-Omega-in-terms-of-A-and-xi} to expand the curvature,
        \begin{align}
            F(\Omega)^i_{\mu\nu} & = F(A)^i_{\mu\nu} + \partial^A_{[\mu}\left( \Sigma^i_{\nu]}{}^\rho \xi_\rho \right) + \frac{1}{4}\epsilon^{ijk} \Sigma^j_{[\mu}{}^\rho \Sigma^k_{\nu]}{}^\sigma \xi_\rho \xi_\sigma \\
            & = F(A)^i_{\mu\nu} + \partial^A_{[\mu}\left( \Sigma^i_{\nu]}{}^\rho \xi_\rho \right) + \frac{1}{2} \Sigma^i_{[\mu}{}^\rho \xi_{\nu]} \xi_\rho
        \end{align}
        which when substituting into the equation of motion results in 
        \begin{align}
            H^i_{\langle \mu}{}^\rho F(\Omega)^i_{\nu\rangle\rho} & = \Sigma^i_{\langle \mu}{}^\rho F(\Omega)^i_{\nu\rangle\rho} = \Sigma^i_{\langle \mu}{}^\rho F(A)^i_{\nu\rangle\rho} -\frac{1}{2}\Sigma^i_{\langle \mu}{}^\rho \partial^A_{\nu\rangle} \left( \Sigma^i_\rho{}^\sigma \xi_\sigma \right) \nonumber \\ & \hspace{100pt} + \frac{1}{2} \Sigma^i_{\langle \mu}{}^\rho \partial^A_\rho \left( \Sigma^i_{\nu \rangle}{}^\sigma \xi_\sigma \right) + \frac{1}{2} \xi_{\langle \mu} \xi_{\nu \rangle} \nonumber\\ 
            & = \Sigma^i_{\langle \mu}{}^\rho F(A)^i_{\nu\rangle\rho} + \partial_{\langle \mu} \xi_{\nu \rangle} + \frac{1}{2} \xi_{\langle\mu} \xi_{\nu\rangle} - \frac{1}{2}\Sigma^i_{\langle \mu|}{}^\rho \xi_\sigma \partial^A_{|\nu\rangle} \Sigma^i_{\rho}{}^{\sigma} + \frac{1}{2} \Sigma^i_{\langle \mu|}{}^\rho \xi_\sigma \partial^A_\rho \Sigma^i_{|\nu \rangle}{}^{\sigma}.
        \end{align}
        Performing a similar expansion for the second equation in \cref{eq:nonlinear-pleb-12-connection-eoms} gives
        \begin{align}
            \frac{1}{\lambda} g_H^{\mu\nu} \partial_\mu \Omega^i_\nu = g^{\mu\nu} \partial_\mu \Omega^i_\nu = g^{\mu\nu} \partial_\mu A^i_\nu - \frac{1}{2} g^{\mu\nu} \partial_\mu \left( \Sigma^i_\nu{}^\rho \xi_\rho \right) =  g^{\mu\nu} \partial_\mu A^i_\nu - \frac{1}{2} \Sigma^{i\mu\nu} \partial_\mu \xi_\nu - \frac{1}{2} g^{\mu\nu} \xi_\rho \partial_\mu \Sigma^i_\nu{}^\rho = 0.
        \end{align}
        We see that in both cases we obtain the previously derived gauge condition with an extra term proportional to $\xi_\mu$. This means that when on the constraint surface, that is $\xi_\mu = 0$, these additional terms vanish and Einstein's equations are recovered. The presence of the derivatives of $\Sigma^i$ do not change the hyperbolicity as these will only enter at the first-order level in the second-order equations of motion for the frame variables. For the new Lorenz condition the same is structure is present and upon removing the constraint terms we are left with the usual Lorenz gauge.
        
        In conclusion for the $\Omega^i$ variable we have seen that the proposed evolution equations are indeed modifications to the Einstein's equations (and Lorenz condition) that return Einstein solutions when the constraints are satisfied.

        \subsection{$\omega$ Connection Equations}
        We now turn to the 4 dimensional part of the connection. Starting with the equation of motion $g_H^{\mu\nu} \partial_\mu \omega_\nu = 0$, we show that it is a modification of part of Einstein's equations.
        \begin{align}
            \frac{1}{\lambda} g_H^{\mu\nu} \partial_\mu \omega_\nu & = g^{\mu\nu} \partial_\mu \omega_\nu = g^{\mu\nu} \partial_\mu \xi_\nu + 2 g^{\mu\nu} \partial_\mu \left( \Sigma^i_\nu{}^\rho A^i_\rho \right) \nonumber \\ & = g^{\mu\nu}\partial_\mu \xi_\nu + 2 \Sigma^{i\mu\nu} \partial_\mu A^i_\nu + 2 A^i_\rho g^{\mu\nu} \partial_\mu \Sigma^i_\nu{}^\rho.
        \end{align}
        Using the previous result for the last term we can write this as 
        \begin{align}
            \frac{1}{\lambda} g_H^{\mu\nu} \partial_\mu \omega_\nu & = g^{\mu\nu}\partial_\mu \xi_\nu + 2 \Sigma^{i\mu\nu} \partial_\mu A^i_\nu -2 A^i_\mu \epsilon^{ijk} \Sigma^{j\mu\nu} \Omega^k_\nu \\ & = \Sigma^{i\mu\nu} F(A)^i_{\mu\nu} + g^{\mu\nu} \partial_\mu \xi_\nu - \epsilon^{ijk} \Sigma^{i\mu\nu} A^j_\mu A^k_\nu - 2 A^i_\mu \epsilon^{ijk} \Sigma^{j\mu\nu} \Omega^k_\nu \nonumber \\ & = \Sigma^{i\mu\nu} F(A)^i_{\mu\nu} + g^{\mu\nu} \partial_\mu \xi_\nu + \frac{1}{2} g^{\mu\nu} \xi_\mu \xi_\nu + g^{\mu\nu} \left( \frac{1}{4} \omega_\mu \omega_\nu + \Omega^i_\mu \Omega^i_\nu \right) - \Sigma^{i\mu\nu} \omega_\mu \Omega^i_\nu.
        \end{align}
        Going from the second to the third line in the above we have used the definitions of the new connections. The potential part contain terms of the form $A^i A^j$, these will change the dynamics from being Einstein. To remove them we simply use the last line in the above to change the equation of motion in the connection variables to be
        \begin{align}
            g_H^{\mu\nu} \partial_\mu \omega_\nu - g_H^{\mu\nu} \left( \frac{1}{4} \omega_\mu \omega_\nu + \Omega^i_\mu \Omega^i_\nu \right) + H^{i\mu\nu} \omega_\mu \Omega^i_\nu = 0.
        \end{align}
        Hence, the unwanted potential terms are no longer there at the cost of a slightly more complicated equation of motion. This involves both of the 12 and 4 component connections, this mixing of sectors is not unsurprising as the same occurs for the definition of the connections. Lastly, the final evolution equation for the connections is 
        \begin{align}
            \frac{1}{\lambda} H^{i\mu\nu} \partial_\mu \omega_\nu & = \Sigma^{i\mu\nu} \partial_\mu \omega_\nu = \Sigma^{i\mu\nu} \partial_\mu \xi_\nu + 2 \Sigma^{i\mu\nu} \partial_\mu \left( \Sigma^j_\nu{}^\rho A^j_\rho \right) = \Sigma^{i\mu\nu} \partial_\mu \xi_\nu + 2 \Sigma^{i\mu\nu} \partial^A_\mu \left( \Sigma^j_\nu{}^\rho A^j_\rho \right) \nonumber \\ & = \Sigma^{i\mu\nu} \partial_\mu \xi_\nu - 2 g^{\mu\nu} \partial_\mu A^i_\nu - 2\epsilon^{ijk} \Sigma^{j\mu\nu} \partial^A_\mu A^k_\nu + 2 A^j_\rho \Sigma^{i\mu\nu} \partial^A_\mu \Sigma^j_\nu{}^\rho \nonumber \\ & = \Sigma^{i\mu\nu} \partial_\mu \xi_\nu - 2g^{\mu\nu} \partial_\mu A^i_\nu - \epsilon^{ijk}\Sigma^{j\mu\nu} F^k_{\mu\nu} - \epsilon^{ijk} \Sigma^{j\mu\nu} \epsilon^{klm} A^l_\mu A^m_\nu + 2 A^j_\rho \Sigma^{i\mu\nu} \partial^A_\mu \Sigma^j_\nu{}^\rho.
        \end{align}
        Here we can derive another useful identity, using \cref{eq:nonlinear-plebanski-modified-total-covariant-deriative-on-Sigma,eq:nonlinear-modified-connection-definitions}, to find
        \begin{align}
            \Sigma^{i\mu\nu} \partial^A_\mu \Sigma^j_\nu{}^\rho & = \Sigma^{i\mu\nu} \nabla^A_\mu \Sigma^j_\nu{}^\rho + \delta^{ij} g^{\mu\sigma} \Gamma_{\mu\sigma}^\rho \nonumber \\ & = \Sigma^{i\mu\nu} \epsilon^{jlm} \left( c \partial_\mu \chi^l + c J_1(d\chi)^l_\mu \right) \Sigma^m_\nu{}^\rho + \delta^{ij} \Gamma^\rho \nonumber \\ & = \Sigma^{i\mu\nu} \epsilon^{jlm} \left( c \partial_\mu \chi^l + c J_1(d\chi)^l_\mu \right) \Sigma^m_\nu{}^\rho - \delta^{ij} (\xi^\rho + 2c \Sigma^{k\rho\nu} \partial_\nu \chi^k) \nonumber \\ & = -\delta^{ij} \xi^\rho
        \end{align}
        Which means that the last term in the equation of motion contains $\xi_\mu$ and as such does not contribute on the constraint surface. Another way of seeing the same result is to note that $\nabla^A_\mu \Sigma^i_{\rho\sigma} = 0$ on the constraint surface as all the equations return to their unmodified form. We can then safely ignore that term and write the connection squared terms as
        \begin{align}
            \frac{1}{\lambda} H^{i\mu\nu} \partial_\mu \omega_\nu & \simeq -2 \left( g^{\mu\nu} \partial_\mu A^i_\nu - \frac{1}{2} \Sigma^{i\mu\nu} \partial_\mu \xi_\nu \right) - \epsilon^{ijk} \Sigma^{j\mu\nu} F^k_{\mu\nu} - \epsilon^{ijk} \Sigma^{j\mu\nu} \epsilon^{klm} A^l_\mu A^m_\nu \nonumber \\ 
            & = -2 g^{\mu\nu} \partial_\mu \Omega^i_\nu - \epsilon^{ijk}\Sigma^{j\mu\nu} F^k_{\mu\nu} - \epsilon^{ijk} \Sigma^{j\mu\nu} \epsilon^{klm} A^l_\mu A^m_\nu \nonumber \\ &= - \epsilon^{ijk}\Sigma^{j\mu\nu} F^k_{\mu\nu} - g^{\mu\nu} \omega_\mu \Omega^i_\nu - \frac{1}{2} \epsilon^{ijk} \Sigma^{k\mu\nu} \omega_\mu \Omega^j_\nu + \Sigma^{j\mu\nu} \Omega^i_\mu \Omega^j_\nu.
        \end{align}
        The first term in the last line is proportional to $d^A d^A \Sigma^i$, and as $d^A \Sigma^i = ... d\chi^i$ where the right-hand side contains $\chi^i$ which vanishes on the constraint surface. Knowing this implies that $\epsilon^{ijk}\Sigma^{j\mu\nu} F^k_{\mu\nu}$ also vanishes on the constraint surface. Therefore, the equation of motion becomes
        \begin{align}
            H^{i\mu\nu} \partial_\mu \omega_\nu + g^{\mu\nu} \omega_\mu \Omega^i_\nu + \frac{1}{2} \epsilon^{ijk} H^{k\mu\nu} \omega_\mu \Omega^j_\nu - H^{j\mu\nu} \Omega^i_\mu \Omega^j_\nu = 0
        \end{align}
        which is proportional to the algebraic constraint $d^A d^A \Sigma^i = \epsilon^{ijk} F^j \wedge \Sigma^k = 0$ plus additional terms proportional to the constraints $\xi_\mu, \chi^i$.

        Summarising the connection equations of motion we have 
        \begin{gather}
            g_H^{\mu\nu} \partial_\mu \omega_\nu = g_H^{\mu\nu} \left( \frac{1}{4} \omega_\mu \omega_\nu + \Omega^i_\mu \Omega^i_\nu \right) - H^{i\mu\nu} \omega_\mu \Omega^i_\nu \label{eq:nonlinear-plebanski-conformal-1-of-4}\\
            H^{i\mu\nu} \partial_\mu \omega_\nu + g_H^{\mu\nu} \omega_\mu \Omega^i_\nu + \frac{1}{2} \epsilon^{ijk} H^{k\mu\nu} \omega_\mu \Omega^j_\nu - H^{j\mu\nu} \Omega^i_\mu \Omega^j_\nu \label{eq:nonlinear-plebanski-conformal-3-of-4}\\
            g_H^{\mu\nu} \partial_\mu \Omega^i_\nu = 0, \quad H^i_{\langle \mu}{}^\rho F(\Omega)^i_{\nu\rangle \rho} = 0 \label{eq:nonlinear-plebanski-conformal-12}.
        \end{gather}
        By comparing these equations with the linearisations in \cref{chap:Linearised-Gravity} we see that the highest order terms match, the only difference is the addition of some connection squared terms in the 4 dimensional sector.

        \section{Conformal System} \label{sec:nonlin-pleb-conformal-system}
        Here we gather the equations of motion for the conformal \pleb{} system. The basic  variables for the frame are the triple of 2-forms $H^i \in \Lambda^+ \subset \Lambda^2$, a scalar function $\lambda$ and an internal vector $\chi^i$. The triple of 2-forms satisfies $H^i \wedge H^j = 2i\delta^{ij} d^4x$ such that the metric they describe has constant determinant. The physical metric is then $g = \frac{1}{\lambda} g_H$ where $g_H$ is the metric defined by $H^i$. By taking derivatives of the frame variables we are able to define the conformal connections, 
        \begin{equation} \label{eq:nonlinear-conformal-connection-definitions}
            \begin{aligned}
                \Omega^i & = -i J_1^{-1}(\star_H dH^i) + \frac{1}{2} H^i \cdot \Gamma_H \\
                \omega & = -\Gamma_H + 2 d\log(\lambda) - i H^i \cdot \star_H dH^i + 4 c H^i \cdot d\chi^i.
            \end{aligned}
        \end{equation}
        Where $(\Gamma_H)_\mu = g_{H\mu\nu} \partial_\rho g_H^{\rho\nu}$ is the conformal Harmonic 1-form. These connections are related to the self-dual connection of the physical spacetime via \cref{eq:nonlinear-A-xi-to-Omega-connection}. The gauge fixed Einstein's equations in terms of the conformal connections $\Omega^i$ and $\omega$ are then imposed through 
        \begin{equation} \label{eq:nonlinear-conformal-einstens-and-gauge-equations}
            \begin{gathered}
                g_H^{\mu\nu} \partial_\mu \omega_\nu = g_H^{\mu\nu} \left( \frac{1}{4} \omega_\mu \omega_\nu + \Omega^i_\mu \Omega^i_\nu \right) - H^{i\mu\nu} \omega_\mu \Omega^i_\nu \\
                H^{i\mu\nu} \partial_\mu \omega_\nu + g_H^{\mu\nu} \omega_\mu \Omega^i_\nu + \frac{1}{2} \epsilon^{ijk} H^{k\mu\nu} \omega_\mu \Omega^j_\nu - H^{j\mu\nu} \Omega^i_\mu \Omega^j_\nu \\
                g_H^{\mu\nu} \partial_\mu \Omega^i_\nu = 0, \quad H^i_{\langle \mu}{}^\rho F(\Omega)^i_{\nu\rangle \rho} = 0.
            \end{gathered}
        \end{equation}
        We have seen that the gauge fixing produces a surprisingly simply system of equations for the connection. Moreover, the 12 dimensional conformally flat sector $(H^i,\Omega^i)$ can be evolved completely independently of the remaining $4$ dimensional sector $(\lambda,\chi^i,\omega)$. Unlike the linear version the $4$ dimensional sector still depends on the $12$ sector to evolve, this is a slightly weaker result, but it is still remarkable that it occurs. The $12$ dimensional sector has $9\R + 3\C + 12 \C = 39\R$ components that one needs to evolve, this is smaller than the full system which has $39 + 3\C + 1 \R + 4 \C = 54\R$ components. The linearisation of this system produces two decoupled Dirac equations, which are known to be strongly hyperbolic (as their characteristic matrices are Hermitian), therefore the full system is said to be hyperbolic. This property along with it being first-order make it a useful starting point for considering numerical relativity.

        \subsection{Weyl Curvature Extraction}
        For strongly gravitating systems one often expects gravitational waves to be produced during the evolution. These are contained in the Weyl curvature of this system. We briefly detail two ways  how one can extract the Weyl curvature components from these new connection variables. First one can reconstruct the original connection variable 
        \begin{align}
            A^i_\mu = \frac{1}{2} \Omega^i_\mu - \frac{1}{2} \epsilon^{ijk} H^j_\mu{}^\nu \Omega^k_\nu - \frac{1}{4} H^i_\mu{}^\nu \omega_\nu.
        \end{align}
        From this one can compute its curvature $F(A)^i = dA^i + \frac{1}{2} \epsilon^{ijk} A^j \wedge A^k$. The self-dual Weyl tensor is then given by the $3 \times 3$ symmetric tracefree matrix 
        \begin{align}
            \Psi^{ij} = \frac{1}{\lambda} H^{\langle i|\mu\nu|} F(A)^{j\rangle}_{\mu\nu}.
        \end{align}
        The 5 Weyl-NP scalars are then packaged into $\Psi^{ij}$. The exact linear combinations depend on the basis chosen when computing $H^i$. Assuming that during the evolution the primary constraints $\xi$ vanish, then we find that $A^i = \Omega^i$ and in this case the Weyl curvature can be computed as 
        \begin{align}
            \Psi^{ij} = \frac{1}{\lambda} H^{\langle i|\mu\nu} F(\Omega)^{j\rangle}_{\mu\nu}.
        \end{align}
        We see that both the 12 and the 4 sectors are still needed and the Weyl curvature cannot be computed with only the 12.
    
    \section{Discussion}

    We have seen that the full nonlinear Einstein's equations, when written using \pleb{}'s formulation, can be gauge fixed into a 2 parameter family of first-order hyperbolic equations. The fundamental fields are the self-dual 2-forms and connection, $\Sigma^i \in \Lambda^+, A^i \in \Lambda^1$, we also introduce a connection 1-form and internal vector $\xi \in \Lambda^1, \chi^i \in \C^3$. The additional fields are used to impose the diffeomorphism and $SO(3,\C)$ gauge freedom in the form of dynamical equations of motion rather than fixing components. The resulting family of equations of motion shown in \cref{eq:nonlinear-modified-connection-definitions,eq:nonlinear-modified-einstein-and-lorenz-conditions}. It is seen that $\xi$ encodes the Harmonic 1-form and is itself used to modify Einstein's equations on the self-dual connection. The definition of both the self-dual and Harmonic connections are modified by the introduction of derivatives of $\chi^i$ in such a way that it satisfies a second-order wave equation to leading order. The resulting system has $16 + 16 = 32$ equations of motion for the $16 + 16 = 32$ fields. The nature of this gauge fixing is to promote every field to be dynamical and as such there is no gauge freedom in choosing a fixed value for a field, only its initial and boundary conditions need to be specified. It is very similar to the covariant form of the Z4 gauge fixing, there the 1-form $\xi$ is used to modify Einstein's equations and is defined through the Harmonic condition much in the same way as here. This hyperbolic gauge fixing can be seen as a generalisation of this to the situation where there is a complex spatial frame and hence an $SO(3,\C)$ symmetry is available. The role of $\chi^i$ is then exactly the same as $\xi$ in the Z4 case, it is used to give evolution equations and therefore dynamically control the constraints that arise due to the extra $SO(3,\C)$ symmetry. While this appears like an increase in the number of fields needed as compared to the metric and Z4 formulations, we find here that the first-order structure is much simpler, and later it will be seen that the equations of motion are polynomial in all the basic fields. 

    Further benefits are available, that can only be seen in the chiral formulation, in the form of the conformal \pleb{} system. We have derived in \cref{sec:nonlin-pleb-conformal-system} what we call the conformal \pleb{} system. By choosing a specific member of the 2 parameter gauge fixing one can decouple a unimodular frame along with its conformal connection from the rest of the fields. The conformal frame is described by a triple of self-dual 2-forms $H^i \in \Lambda^+$ and its self-dual connections are $\Omega^i$ which appear as a linear combination of $A^i$ and $\xi$. The definition of $\Omega^i$ only  involves $H^i$, and it is shown that 9 components of Einstein's equations and the 3 Lorenz conditions can be imposed through 12 differential equations on $H^i$ and $\Omega^i$ only. This distinguishes a 12 components sector that evolves completely independently of the remaining components. The missing components are the conformal factor and the internal vector $\lambda \in C^\infty, \chi^i \in \C^3$ and the remaining 4 components of the connections are packaged into $\omega$, which again is a linear combination of $A^i$ and $\xi$. The equations of motion that encode the connection $\omega$ and the leftover components of Einstein's equation involve both the 12 and 4 sector components. Only the 12 sector is completely separate and the 4 requires all the fields to be evolved. This system opens up the possibility of different evolution schemes where the two sectors are evolved separately, and it may be that there is some benefit in doing so. One example this would be to evolve the 12 dimensional system on its own and instead of evolving the $4$ part one can simply find a ``static" solution for them that implies that the constraints are satisfied. This line of reasoning is not explored here, however, schemes similar to this could be developed that are more efficient than standard techniques. It should be noted that the gauge fixing required to obtain this system is a slight modification to that proposed in the normal \pleb{} gauge fixing, however they are equivalent on the constraint surface and so given a sufficient method to set $\xi$ and $\chi$ to zero these two systems describe Einstein's equations.

    At the level of the Hamiltonian this gauge fixing promotes the Lagrange multipliers, that impose the constraints, to be dynamical by adding the terms proportional to $\xi, \chi^i$, which are their conjugate momentum. This modification is done to keep the first-class structure of the constraints, meaning that any dynamics on the constraint surface preserves the constraints exactly. For numerical purposes this is the desired situation as theoretically this means that constraints are exactly preserved during evolution, although numerical artefacts cause deviations from the constraint surface that need to be dealt with. One way that the systems displayed here could resolve this issue is by noticing that the constraints themselves satisfy wave equations to leading order, as such any violation in the constraints will be propagated away and eventually leave the system. This idea is called constraint sweeping and is the preferred situation to gauge fixing that does not affect the constraint propagation at all. Modifications can be made, even at the covariant level, that cause constraints to be damped, these will be mentioned more in the numerical chapters.

    Some mention of the reality conditions is needed as in general the 2-forms $\Sigma^i$ and $H^i$ can encode a complex-valued metric. At the covariant level one can impose the reality conditions in the usual way, $\Sigma^i \wedge \asd^j = 0$ and $H^i \wedge \bar{H}^j = 0$ where $\bar{H}^i = -(H^*)^i$. In the conformal system the extra reality condition is needed, ${\rm Im}(\lambda) = 0$, such that the conformally related metric is real-valued also. When the reality conditions are satisfied, and on the constraint surface, the connections then encode the correct derivatives of the metric and frame such that their equations of motion impose Einstein's equations correctly. Therefore, the reality condition can be considered at the level of the frame rather than imposing any conditions on the connection.

    \newpage
    \part{Numerical Relativity}\label{part:numerical-relativity}

    \newpage
    \chapter{Metric and ADM Formalism}\label{chap:numerical-relativity}
General relativity is compromised of 10 nonlinear partial differential equations (PDEs) on the metric tensor $g_{\mu\nu}$ on a manifold $M$, and general solutions to them are not known. To obtain a solution one usually has to assume an underlying symmetry is present, for example the Kerr metric~\cite{GravitationalFKerr1963} has an axial symmetry. The lack of more general solutions led to development of numerical methods of solving Einstein's equations, called numerical relativity (NR). Most modern methods require that Einstein's equation be formulated as an initial value problem or Cauchy problem.

The covariant nature of Einstein's equations does not distinguish a time coordinate with which to integrate the Cauchy problem, therefore one must choose a foliation of the underlying manifold adapted to the timelike direction. General relativity is invariant under the action of diffeomorphisms, this means that changing the coordinates does not affect the physical solutions. Many ways exist to choose a representative for the coordinates and foliations, this choice is known as gauge-fixing. The first successful numerical simulation of binary black hole spacetime was completed in 2005, see~\cite{Evolution_of_Bi_Pretor_2005}, and by now NR is a mature field~\cite{NumericalRelatBaumga2010}. However, the problem of finding a good numerical scheme is by no means solved, much work has gone towards understanding the requirements for gauge fixing (amongst other properties) that result in long term numerical stability. One such exploration is based on the formulation of Einstein's equation, as different formulations possess different numerical properties, see~\cite{Formulations_of_Shinka_2008} for a review. It is then possible that one formulation produces numerical results that are more stable than the others.

It is known that robust numerical simulations require well-posed PDEs to guarantee the existence of solutions. Here, a well-posed partial differential equations has a unique solutions for a given initial data that depends continuously on the initial data, when formulated as a cauchy problems. Hyperbolicity is a condition on the symbol of a PDE that can be used to compute if a system is well-posed.  The different representations of Einstein's equations, in the context of NR, are usually restricted to those that are hyperbolic PDEs and as such carry useful numerical properties. The existence of such a system was first proven by Choquet-Bruhat in~\cite{Hyperb1+3YvonneTommaso1983}, where a particular gauge was assumed such that the PDEs resembled the wave equation, which is the prototypical second-order hyperbolic equation. Since the PDEs are well-posed we assume that there exists a numerical scheme that is able to integrate them, given sufficient initial data, to give solutions at arbitrary later times.

This is an introductory chapter that summarises the relevant theory required to place general relativity and Einstein's equation into a well-posed system of partial differential equations suitable for numerical techniques. It is loosely based on the books~\cite{AlcubierreMiguelIntro3+12008,NumericalRelatBaumga2010} where a more in depth description can be found.

In \cref{sec:numerical-method} we begin describing our chosen numerical method that we will use to integrate Einstein's equations, these are the finite difference schemes. Hyperbolicity for first-order systems of equations is introduced and its relation to the finite difference methods are discussed. In \cref{sec:num-asympytotically-constrained-system} we explain the different methods available that are able to control the constraint violation that occurs in our chosen numerical method. Then the separation of the temporal and spatial components of the metric formulation of GR into its Hamiltonian form, known as the ADM formulations, is described in \cref{sec:numiercal-adm-formulation}, there the evolution and constraint equations are written explicitly. The standard methods of gauge fixing the ADM formulation are discussed in \cref{sec:numerical-gauge-fixing}. The Z4 system is introduced in \cref{sec:numerical-Z4-system} and used as a motivational example for the gauge fixing considered for the \pleb{} formulation. Finally, chiral Maxwell theory is used as a numerical example, due its close relationship to chiral gravity. Constraint damping and propagation are considered, and the results are displayed in \cref{sec:numerical-chiral-maxwell}.

\section{Numerical Methods} \label{sec:numerical-method}
Gravity is an example of a field theory on a 4-manifold $M$. In general any field theory can be described by a set of partial differential equations, if we package all the fields into a solution vector $u \in C^\infty(M)^N$ where $N$ is the number of fields, these differential equations can be written as
\begin{align}
    L(u,\partial_\mu u,...) = 0
\end{align}
where $L : C^\infty(M)^N \rightarrow C^\infty(M)^{N'}$ where $N'$ is a positive integer not necessarily equal to $N$. This allows for a universal treatment of all field theories. The main problem for numerical methods is then how to convert the above system into something that can be computed by a computer. There are many ways of achieving this, the most common methods are finite difference, finite elements(volume) and spectral methods.

Before discussing the main method used in this Thesis it is worth mentioning another noteworthy method. When a field theory can be constructed using only differential forms and the exterior algebra then one can make use of discrete exterior calculus (DEC). One can then construct a variational integrator based on discretising the action on a lattice which preserves the constraint exactly through evolution. Importantly the chiral formulations of Yang-Mills and gravity can be described efficiently by differential forms, allowing for these types of integrators to be developed. In~\cite{NumericalRelatFarr2010} this was performed for BF formalism of gravity and showed promising results. The constraints were preserved exactly through evolution, which circumvents most of the problems in current techniques to integrate Einstein's equations. However, it was noted that the choice of boundary conditions is important, and a bad choice would lead to unstable results. Despite this it would seem that the most useful integrator one could develop would be through this route, however, the development is involved and out of the scope of this Thesis.

\subsection{Finite Difference Method}
The main idea behind the finite difference method (FDM) is to discretise the spacetime into a set of points. First, we assume that the spacetime can be foliated, that is $M \cong \R \times \spasurf$, with coordinates $x^\mu = t,x^a$ where $t \in \R$ is the time coordinate and $x^a \in \spasurf$ ($a = 1,2,3$) are the 3 spatial coordinates for the spatial hypersurface $\spasurf$. A manifold that can be foliated in this way is called globally hyperbolic~\cite{Leray1953}. Both the time and the spatial coordinates are discretised into intervals which we represent with the end points. The discrete spacetime we call the grid (or numerical grid), the spacing we denote $\Delta t$ and $\Delta x^a$ and a point on the grid is given by the tuple of integers $(n,i,j,k)$ corresponding to the point $x^\mu = (t_0 + n \Delta t,x^1_0 + i \Delta x^1,x^2_0 + j \Delta x^2,x^3_0 + k \Delta x^3)$. In this was the spatial hypersurface becomes a primitive cubic lattice of points. Smooth functions, or more generally tensors, are mapped onto discrete tensors that are evaluated at each point on the grid. Derivatives can then be computed by taking the appropriate linear combinations of the field at different points. The time derivative is given by
\begin{align}
    (\partial_t u)^n_{i,j,k} = \frac{1}{\Delta t}(u^{n+1}_{i,j,k} - u^n_{i,j,k}). \label{eq:num-time-finite-difference}
\end{align}
This simply computes the gradient between $u$ at the points $t_n = n \Delta t$ and $t_{n+1} = t_n + \Delta t$. Using the Taylor expansion of $u$  around $t_n$ we see that
\begin{align}
    u^{n+1} = u(t_n + \Delta t) = u^n + \Delta t \partial_t u |_{t = t_n} + \mathcal{O}(\Delta t^2).
\end{align}
It is clear then that \cref{eq:num-time-finite-difference} computes the time derivative up to order $\mathcal{O}(\Delta t^2)$, for this reason we say that has an accuracy order of 1. For the spatial derivatives we can use a similar method called the central difference operator 
\begin{align}
    (\partial_{x^1} u)^n_{i,j,k} = \frac{1}{2 \Delta x} \left( u^n_{i+1,j,k} - u^n_{i-1,j,k} \right) \label{eq:num-space-finit-difference}.
\end{align}
The analogues for $x^2,x^3$ are obvious. Computing the Taylor expansion around $x^1_i$ we see that this has an accuracy order of 2. For most of the numerical work this order of spatial derivative will suffice, if higher accuracy is needed one can take linear combinations of more points in the grid. At the boundary of the grid we are not able to use the central difference approximation, instead we can use one-sided finite differences to approximate. Another method is to extend the boundary with ghost points that are populated by some interpolation algorithm, in this case the central difference operator can be used. A more advanced finite difference operator is available through the SBP-SAT (summation by parts - simultaneous approximation term) operators, these are finite difference operators that exhibit a discrete summation-by-parts property, see~\cite{Review_of_summa_Del_Re_2014,Review_of_summa_Svard_2014} for reviews. In essence, they are still linear combinations of the points on the grid, however, they have been shown to be well-posed and stable.

An important theorem in the study of numerical methods is the Lax Equivalence Theorem~\cite{LaxRichtmyer1956},
\begin{theorem}
    For a consistent finite difference scheme, it being stable implies that it is also convergent.
\end{theorem}
In words a finite difference scheme is consistent if the numerical approximations to the differential operators are equal to differential operator in the analytical limit, e.g. as the grid separation goes to zero. For example, this is true of \cref{eq:num-time-finite-difference} because in the limit as $\Delta t \rightarrow 0$ this is equal to the time derivative. A finite difference scheme is called stable, for a particular choice of norm, if there exists a constant independent of the grid or time step spacing that bounds solutions at a later time by the solutions value at a previous time step. Lastly, a scheme is convergent if for any solution to the numerical system, it is also a solution to the continuous PDE in the analytic limit. All operators we consider are consistent and stable implying that they will be convergent. However, this theorem only holds when the original PDE is well-posed~\cite{Well_Posed_Prob_Wendro_1968}, the definition of well-posed we discuss below.

Another important topic for finite difference methods is boundary conditions. As the numerical grid is finite one must consider what happens to the numerical solution at the boundaries of the grid. Depending on the physical scenario we consider three different methods of imposing boundary conditions,
\begin{enumerate}
    \item Robin Boundary Conditions: $a u + b \partial_n u = f$,
    \item Periodic Boundary Conditions $u(x+L) = u(x)$,
    \item Absorbing Boundary Condition: $\partial_t u = \pm v \partial_n u$.
\end{enumerate}
The Robin boundary condition imposes that on the boundary the solution and its first derivative are fixed by a function $f$, this is useful when the boundary value of a function is known analytically or one wishes to fix the gradient at a boundary. An example use case for this is in spherical symmetry where at $r = 0$ there is a boundary due to coordinate artefacts, and a Robin type boundary condition has been shown to perform well for the Schwarzschild metric in the ADM formulation~\cite{AlcubierreMiguelIntro3+12008}. As for periodic boundary conditions these depend on the topology of the physical manifold, this case is primarily used for testing as most physical situations are not periodic. In this case any error accumulated during a simulation will remain on the grid, this represents a worse case scenario from an error accumulation point of view and hence it is useful to see how the system behaves under such a consideration. Finally, the most reasonable boundary condition is the absorbing boundary condition, where outgoing solutions are absorbed at the boundary. Numerical errors and physical waves can both exit the grid in this case which is both physically accurate and useful for controlling errors. Without absorbing boundary conditions often the solution, along with numerical errors, would reflect and radiate back into the system causing unwanted instabilities. Usual methods of constructing absorbing boundary conditions are based on decomposing the solution into outgoing waves at the boundaries and removing this component.

\subsection{Method of Lines}
To make use of the finite difference method it is easiest to assume that the differential operator $L$ can be written as a Cauchy initial value problem. More specifically, we assume it can be put in the form
\begin{align}
    \partial_t u = S(u)
\end{align}
where $S$ is a differential operator that depends only on the $u$ and its spatial derivatives. This method is a specific example of an abstract Cauchy problems~\cite{Pazy1983}, which generalises this type of evolution to the language of operators. The method of lines form is useful for numerical works because different approximation techniques can be used in the spatial and temporal parts of the problem. Given some numerical method for approximating $S(u)$, one can then use a separate time integrator to obtain a solution at a later time. For the purposes of this Thesis we will be using the standard Runge-Kutta 4-th order time integrator, which is known to be stable and converge for use with the method of lines~\cite{Time_Dependent_Gustaf_2013}. As for evaluating $S(u)$ we will use second the central difference finite scheme which are suitable enough for our tests.

\subsection{Hyperbolicity and Well-posedness} \label{sec:num-hyperbolicity-and-well-posedness}
Hyperbolicity is an important concept for initial value problems. A partial differential equation that is hyperbolic has a sufficiently  ``wave-like" structure and for it there exists a finite domain of dependence for each point in the solution. This means that there is a notion of causality and two distinct points on a particular hypersurface can only influence each other at a later time. For the purposes of this text we only consider the hyperbolicity of first-order partial differential equations, that is differential equations of the form
\begin{align}
    \partial_t u = M^a \partial_a u + Q \label{eq:num-first-order-hyperbolic-system}
\end{align}
where $M^a(u,x)$ are $N \times N$ matrices that depend on the solution vector and the position, $Q(u,x)$ is called the source term that also depends on $u$ and $x$. All the equations we discuss can be or are already presented in this form. If one has a second-order system of equations, by introducing the appropriate first-order variables one can reduce the system to a first-order system. The hyperbolicity can then be characterised through the characteristic matrix $P = M^a n_a$, where $n_a$ is an arbitrary unit vector. The different levels of hyperbolicity are
\begin{enumerate}
    \item Weakly Hyperbolic: Eigenvalues of $P$ are all real
    \item Strongly Hyperbolic: $P$ is diagonalisable with all real eigenvalues
    \item Symmetric Hyperbolic: $P$ is hermitian
\end{enumerate}
The three types are such that symmetric hyperbolicity implies strong hyperbolicity and both imply weak~\cite{Kreiss1989-cm}. If the characteristic matrix is at least strongly hyperbolic then one finds that the solutions are well-posed~\cite{An_Introduction_Hildit_2013}. A partial differential equation (PDE) is well-posed when the following conditions hold,
\begin{enumerate}
    \item The PDE has a solution given an initial condition
    \item The solution is unique for each initial condition
    \item The solution is changes continuously with the initial condition.
\end{enumerate}
These are the Hadamard criteria for well-posed PDEs~\cite{SurLesProblemsHadamard1920}.

To see how strong hyperbolicity implies well-posedness we can consider the example where $M^a$ is constant and does not depend on the solution vector $u$. In this case we can diagonalise $M^a = R \Lambda^a R^{-1}$, where $\Lambda^a$ is a diagonal real matrix and the matrix $R$ is independent of the index $a$ since $P = M^a n_a$ can be diagonalised for any $n_a$. In which case we can introduce the solution eigenvector $\omega = R^{-1}u$ such that the partial differential equations become
\begin{align}
    \partial_t \omega = \Lambda^a \partial_a \omega
\end{align}
where we have limited ourselves to the case when $Q(u) = 0$. For this we can construct a norm,
\begin{align}
    ||\omega(t)||^2 = \int_\spasurf \omega^\dagger \omega.
\end{align}
Computing the time derivative of this norm we find that
\begin{align}
    \partial_t ||\omega(t)||^2 = \int_\spasurf \partial_a \omega^\dagger \left( \Lambda^a \right)^\dagger \omega + \omega^\dagger \Lambda^a \partial_a \omega = \int_\spasurf \partial_a \left( \omega^\dagger \Lambda^a \omega \right) = \int_{\partial \spasurf} \omega^\dagger \Lambda^a \omega.
\end{align}
Furthermore, if the fields vanish on the boundary, or there is no boundary, then the r.h.s.~of the above integral vanishes. What we are left with is the result that
\begin{align}
    \partial_t ||\omega(t)||^2 = 0,
\end{align}
and integrating this in time reveals that
\begin{align}
    ||\omega(t)||^2 = ||\omega(0)||^2.
\end{align}
This is an example of a conserved energy norm, which is named this way as in some systems this norm corresponds with the physical energy. To prove uniqueness we can consider two solutions $\omega_1, \omega_2$ such that $\omega_1|_{t=0} = \omega_2|_{t=0}$, and consider the evolution of their difference $w = \omega_1 - \omega_2$. Since $w|_{t=0} = 0$ it is clear that $||w(0)||^2 = 0$ and therefore $||w(t)||^2 = 0$ for all later times $t$. This implies that $\omega_1 = \omega_2$ is true throughout the evolution and are hence the same solution. As for continuity consider again $\omega_1, \omega_2$, this time with different initial conditions $\omega_1|_{t=0} = f(x)$ and $\omega_2|_{t=0} = g(x)$. The difference between the solutions, $w = \omega_1 - \omega_2$, has an initial value $w|_{t=0} = f(x) - g(x)$ and satisfies $||w(t)||^2 = ||f(x)-g(x)||^2$. In the limit as $f \rightarrow g$ we find that $||w(t)||^2 \rightarrow 0$ meaning that the solution varies continuously on the initial data. Clearly the behaviour of the energy norm depends on the boundary conditions, meaning one must take care to choose them appropriately.

More generally the matrix $M^a$ will not be constant and depend on the solution $u$, also $Q(u) \neq 0$ in general. To analyse the more general case one needs to linearise around a known solution, then the above method of checking the hyperbolicity can be used. In which case we can only prove well-posedness locally around a given solution, also called the local well-posedness, but often this is enough to consider the system worth exploring over other systems where this is not the case. The linearisation and hyperbolicity check does not account for long time singularities, or other discontinuities, that may appear in the solutions.

Therefore, given that our initial condition is suitably smooth we expect that, for strongly hyperbolic PDE's, that the numerical solution to be well-behaved for small time deviations from the initial condition. Hyperbolicity is a key concept for numerical simulations, any modern formulation of Einstein's equations should be checked to be strongly hyperbolic (at least locally) before attempting any numerical study. Further tests of a system's stability can either be performed using some symmetry to simplify the PDEs or empirically.

\section{Asymptotically Constrained Systems}\label{sec:num-asympytotically-constrained-system}
Most field theories and formulations of gravity (and other field theories) come with some constraint equations. Constraint equations, in this context, are PDEs that only depend on the spatial derivative and the fields, and they are required to be satisfied on every spatial hypersurface. It is known that simply being at least strongly hyperbolic is not enough to ensure that numerics are well-behaved (i.e.\ satisfy the constraints and do not diverge). All the theories we are considering have first-class constraints, meaning that their time derivatives are proportional to constraints. This implies that if one were to numerically integrate the constraint system independently for some prescribed solution vector $u$, with the initial condition being zero, then the constraints remain zero throughout the evolution. This leads to the concept of free evolution, where the constraints are imposed on the solution $u$ at the initial hypersurface and then the system is evolved using the evolution equations only. While analytically this seems reasonable, numerical errors cause the system to diverge from the constraint surface (the surface in phase space where the constraints are satisfied). Small errors can build up this way and cause larger divergences that spoil the numerical solutions and often causing codes to crash.

Any first-order system can be modelled by introducing a solution vector $u$, along with the evolution and constraint equations,
\begin{align} \label{eq:numerical-generic-evolution-system}
    \partial_t u = M^a \partial_a u + Q(u), \qquad C(u) = 0 
\end{align}
where $M^a$ are matrix functions of the solution vector, $Q(u)$ are the source terms that contain no derivatives and $C(u)$ is a vector of the constraints. Since the constraints are first-class this means they satisfy an evolution equation of the type 
\begin{align} \label{eq:numerical-generaic-constraint-system}
    \partial_t C = N^a \partial_a C + B C 
\end{align}
where $N^a$ and $B$ are matrix valued functions of $u$. Most of the common evolution systems for gravity can be placed in the form of the above equations. One of the first systems was the ADM formulation, see \cref{sec:numiercal-adm-formulation} for a brief overview, for which it was realised that it was not hyperbolic and so behaved poorly during numerical simulations. In an attempt to remedy this it was noticed that by adding constraints to the equations on could place the ADM equations in a hyperbolic form. This search gave rise to the BSSN formulation~\cite{On_the_Numerica_Baumga_1998}, which is strongly hyperbolic and was able to achieve long term stable numerical results. It was noticed that the numerical advantage could be attributed to the addition of the constraints which appeared to damp their growth~\cite{Formulations_of_Shinka_2008}. The BSSN approach also employs a conformal transformation where the conformal factor is evolved separately from the remaining part of the metric, which also contributed to its success. Although these reasons have been suggested to be the cause of the improved performance, no rigorous proof exists for long term stability without direct numerical computation.

Due to the success of the BSSN formulation and the realisation that adding constraints improves the performance, this led to the development of two related methods for adapting a set of evolution equations of the type \cref{eq:numerical-generic-evolution-system}, to improve the convergence of the solution towards the constraint hypersurface. These are called the adjusted system~\cite{Hyperbolic_form_Yoneda_2001} and the $\lambda$-system~\cite{Einstein_s_Equa_Brodbe_1998}. These two methods are not the only ways of doing this, but they do allow for some useful quantitative analysis.

\subsection{Adjusted System}
The adjusted system starts with a first-order system, where the constraints are assumed to be first class. Therefore, we have the same setup as in \cref{eq:numerical-generaic-constraint-system,eq:numerical-generic-evolution-system}. We now ``adjust" this system by adding constraints 
\begin{align}
    \partial_t u = M^a \partial_a u + Q +K C + L^a \partial_a C
\end{align}
for some matrices $K, L^a$. This modification is justified as the constraints are weakly vanishing and so this is equivalent to adding a zero to the right-hand side of the system. Here, the constraints that are added to the equations of motion are written as their definitions in terms of the solution vector $u$, that is  $C = C(u)$ rather than independent variables. The modification to the evolution equations produces a modification to the constraint evolution equations 
\begin{align} \label{eq:numerical-adjusted-constraint-evolution}
    \partial_t C = N^a \partial_a C + B C + F^{ab} \partial_a \partial_b C + G^a \partial_a C + H C 
\end{align}
where $F^{ab}, G^a$ and $H$ are matrix functions of $u$. The next step is to determine the functions $K, L^a$ by analysing the constraint evolution equation. Treating the constraints $C$ as new independent fields we are able to linearise \cref{eq:numerical-adjusted-constraint-evolution} around a known background for $u$. Evidently the equations are now linear and one can perform a spatial Fourier transformation
\begin{align}
    C = \int \hat{C}(k) e^{i k_a x^a} d^3k
\end{align}
to find that 
\begin{align}
    \partial_t \hat{C} = (i N^a k_a + B - F^{ab} k_a k_b + i G^a k_a + H) \hat{C} = P \hat{C}.
\end{align}
We can then restrict the modification, that is the form of $K$ and $L^a$, through restriction of the eigenvalues of $P$. The strategy is to obtain as many negative eigenvalues for $P$ as possible, this causes most of the constraint modes to damp to zero during their evolution. If this is not possible then one can try to obtain as many pure imaginary eigenvalues as possible, this leads to propagation of the constraints and while it does not cause them vanish, they do not grow either and the simulations can remain stable. Of course this analysis is local and is only suitable for small deviations in time, however, it is preferred that the constraint do not grow exponentially in time even locally. At the same time it may also be possible to choose $K$ and $L^a$ such that the system changes from weakly to strongly (or even symmetric) hyperbolic, if possible this should be taken into consideration such that the equations are also locally well-posed.

\subsection{\texorpdfstring{$\lambda$}{λ}-systems} \label{subsec:lambda-system}
For the $\lambda$-system we begin with the same system \cref{eq:numerical-generic-evolution-system,eq:numerical-generaic-constraint-system}, but we now assume that it is at least strongly hyperbolic as the following modifications do not change the hyperbolicity. This modification follows by introducing a vector $\lambda$ with the same number of components as the constraints and imposing a dissipative equation of motion on it 
\begin{align}
    \partial_t \lambda = A C - E \lambda \label{eq:num-lambda-system-dissipative-equations}
\end{align}
where $A,E$ are diagonal matrices with constant coefficients and $C = C(u)$ is the constraint in terms of the solution $u$. The components of $\lambda$ measure the failure of the numerical approximation to impose the constraints, indeed we see that when $\lambda$ is zero then the constraints must vanish. The dissipative system should be constructed such that $\lambda$ is damped towards zero. We then modify the highest order system of equations, the system before the modification has the leading order structure
\begin{align}
    \partial_t \binom{u}{\lambda} = \begin{pmatrix}
        M^a && 0 \\ \mathcal{A}^a && 0
    \end{pmatrix} \partial_a \binom{u}{\lambda}
\end{align}
where we use the notation $A C(u) = \mathcal{A}^a \partial_a u$ to leading order. To make this strongly hyperbolic, using our assumption that $M^a$ is already strongly hyperbolic, we add the corresponding off-diagonal terms
\begin{align}
    \partial_t \binom{u}{\lambda} = \begin{pmatrix}
        M^a && \bar{\mathcal{A}}^a \\ \mathcal{A}^a && 0
    \end{pmatrix} \partial_a \binom{u}{\lambda}. \label{eq:num-lambda-system-general}
\end{align}
where $\bar{\mathcal{A}}^a$ is the Hermitian conjugate of $\mathcal{A}^a$. By taking the Hermitian conjugate of the full system we see that the total characteristic matrix is strongly hyperbolic, furthermore, if $M^a$ is symmetric hyperbolic then so is the full system. Next one has to find suitable values for $A,E$ such that dissipative terms control the  constraints. Other than empirical guidance, one can look at the modification that occurs to the constraint evolution equation given \cref{eq:num-lambda-system-general}. In general this modification will be of the form 
\begin{align}
    \partial_t C = D^a \partial_a C + B C + G^{ab} \partial_{ab} \lambda + H^{a} \partial_a \lambda + I \lambda
\end{align}
with matrices $G^{ab}, H^a, I$. Together with \cref{eq:num-lambda-system-dissipative-equations} this forms a closed system with the above equation. Linearising around an arbitrary background and performing a spatial Fourier transform,
\begin{align}
    \lambda = \int \hat{\lambda}(k) e^{ik_a x^a} d^3k, \quad C = \int \hat{C}(k) e^{ik_a x^a} d^3 k,
\end{align}
the two equations can be written as 
\begin{align}
    \partial_t \binom{\hat{\lambda}}{\hat{C}} = \begin{pmatrix}
        -E && A \\ -G^{ab} k_a k_b + i H^a k_a + I && i D^a k_a + B
    \end{pmatrix} \binom{\hat{\lambda}}{\hat{C}} = P \binom{\hat{\lambda}}{\hat{C}}.
\end{align}
The matrix $P$ is known as the amplification matrix. The goal is then to choose $A,E$ such that the eigenvalues of the amplification matrix have no positive real part as this would indicate exponential growth. The imaginary parts indicate propagation and should affect only the speed at which the constraints travel in the spatial hypersurface. If it is not possible to make all the eigenvalues negative then in~\cite{Formulations_of_Shinka_2008} it is suggested to make as many of the eigenvalues as possible non-positive. This idea is very similar to the adjusted system where a similar analysis occurs.

It should be noted that the idea of adding constraints on the equations of motion does not originate in these two modifications, and it has been used since the early days of numerical analysis. The above two systems give a systematic and quantitative way to analyse the modification and as such can help guide numerical simulations. Otherwise, one has to modify any evolution equations and check empirically that convergence is obtained and this can be an arduous task. In general these types of systems are called asymptotically constrained systems as during the evolution the numerical solution converges to the true solution instead of it being required at each time step.

For a generic evolution system that contains first-class constraints then there exists guidelines on how to check if the constraints will diverge or decay, given in~\cite{Diagonalizabili_Yoneda_2002}. The authors there describe analysing the constraint propagation matrix in exactly the same way as mentioned here. This can be done for any modification that retains the first-class structure of the constraints. If any of the eigenvalues are positive then a constraint mode will grow exponentially, this is the situation we wish to avoid. It also suggests that imaginary eigenvalues are more beneficial than zero eigenvalues as this causes the constraints to propagate instead of accumulating in position. We will follow these guidelines when considering our gauge fixed system.

\section{ADM Formulation} \label{sec:numiercal-adm-formulation}

In this section we describe the decomposition of the metric and related variables onto and perpendicular to the spatial hypersurfaces. It will allow us to produce a 3+1 system of equations, of the form \cref{eq:num-first-order-hyperbolic-system}. We do this in the usual metric formulation, but the technology used can and will be applied to other formalisms. This was first done by  Arnowitt, Deser and Misner in~\cite{Canonical_Varia_Arnowi_1960}, in an attempt to find a quantum theory for gravity, and as such was named the ADM formalism. Soon after it was discovered that it placed Einstein's equations in a useful form for numerical relativity. The seminal paper by York~\cite{York:1978gql}, is an early example of a summary of the formulation of gravity as a cauchy problem.

\subsection{Metric Decomposition and Extrinsic Curvature}
The first step in deriving the ADM formulation is to distinguish a timelike direction, in doing so we choose a function $t : \R \rightarrow M$, to which we associate the time coordinate. For later purposes it will be useful to introduce coordinates on the spatial hypersurfaces, $x^a \in \spasurf$, where $a = 1,2,3$. By taking the exterior derivative we find, $dt \in \Lambda^1$, that is the 1-form (or covector) that points in the direction normal to the constant $t$ surfaces. Next, we introduce a vector $n^\mu$, that also points in the direction of $dt$, i.e. such that 
\begin{align}
    n^\mu (dt)_\mu = 1
\end{align}
in components we have that $(dt)_\mu = (1,0)$ and $n^\mu = (1,-N^a)$ for an arbitrary 3-vector $N^a$. The above condition implies that $n^\mu$ is timelike as it points in the same direction as $dt$. We can now begin to decompose the metric, $g_{\mu\nu}$, for which it is easiest to do so to the inverse metric. Contracting with $dt$ twice we define 
\begin{align}
    g^{\mu\nu} (dt)_\mu (dt)_\nu = (dt)^\mu (dt)_\mu = -\frac{1}{N^2}
\end{align}
where $N$ is the lapse function. Fixing the meaning of the 3-vector we can define 
\begin{align}
    (dt)^\mu = g^{\mu\nu} (dt)_\nu = -\frac{1}{N^2} n^\mu
\end{align}
where $N^a$ is now the shift vector. For the metric formalism is it convenient to define $\N^\mu = \frac{1}{N} n^\mu$, such that 
\begin{align}
    \N_\mu = -N (dt)_\mu, \quad \N^\mu \N_\mu = -1.
\end{align}
The vector $\N^\mu$ is then normal to the hypersurfaces $t = constant$ and as $\N^0 > 0$ it is future pointing. The lapse and shift define the changes for in the coordinates between the hypersurfaces for normal observers (i.e. observers travelling along the integral curves of the normal vector), 
\begin{align}
    \frac{dx^\mu}{d\tau} = \N^\mu.
\end{align}
Splitting into 3+1 coordinates we find 
\begin{align}
    d\tau = N dt, \quad dx^a = -N^a dt,
\end{align}
and we see concretely that lapse defines the change in the proper time and the shift the change in the spatial coordinates. They specify how the coordinates evolve between spatial hypersurfaces. The specification of $t,x^a,N$ and $N^a$ is enough information to locally decompose the manifold into the foliations $M = \R \times \spasurf$.

The projector onto the spatial hypersurfaces is then 
\begin{align}
    P_\mu^\nu = \delta_\mu^\nu + \N_\mu \N^\nu, \quad P_\mu^\rho P_\rho^\nu = P_\mu^\nu.
\end{align}
It can be checked that, by contracting with $\N^\mu$, the timelike components are removed with this projection. We can then use the projectors to define the spatial metric as
\begin{align}
    \gamma_{\mu\nu} = P_\mu^\rho P_\nu^\sigma g_{\rho\sigma}
\end{align}
or equivalently 
\begin{align}
    \gamma_{\mu\nu} = g_{\mu\nu} + \N_\mu \N_\nu.
\end{align}
We see that it is purely spatial as $\N^\nu \gamma_{\nu\mu} = 0$ along with the symmetry implies there are no timelike components. In 3+1 coordinates we find that 
\begin{align}
    g_{\mu\nu} = \begin{pmatrix}
        -N^2 + N_c N^c && N_a \\ N_b && \gamma_{ab}
    \end{pmatrix}, \quad \N_\mu = (-N,0), \quad N_a = \gamma_{ab} N^b.
\end{align}
The inverse metric is given by 
\begin{align}
    g^{\mu\nu} = \begin{pmatrix}
        -\frac{1}{N^2} && \frac{N^a}{N^2} \\ \frac{N^b}{N^2} && \gamma^{ab} - \frac{N^a N^b}{N^2}
    \end{pmatrix}.
\end{align}
We define $\gamma_{ab}$ as the $6$ spatial metric components, and they define the proper distance in the hypersurface $ds^2 = \gamma_{ab} dx^a dx^b$. We use the name spatial metric for $\gamma_{\mu\nu}$ and $\gamma_{ab}$ interchangeably, their specific meaning is distinguished by their indices. 

Our first equation of motion will be related to the time derivatives of $\gamma_{\mu\nu}$, as such we introduce the extrinsic curvature 
\begin{align}
    K_{\mu\nu} := -\frac{1}{2} \mathcal{L}_{\N} \gamma_{\mu\nu} = -D_\mu \N_\nu
\end{align}
where we have introduced
\begin{align}
    D_\mu T_{\nu_1 ... \nu_q} = P_\mu^\nu P_{\nu_1}^{\rho_1} ... P_{\mu_q}^{\rho_q} \nabla_\nu T_{\rho_1 ... \rho_q}
\end{align}
as the spatial covariant derivative. Using the linearity of the Lie derivative we find 
\begin{align}
    \mathcal{L}_n \gamma_{\mu\nu} = -2N K_{\mu\nu}.
\end{align}
One can show, by contraction with $n^\mu$ that $K_{\mu\nu}$ is purely spatial. Splitting this into 3+1 coordinates we find 
\begin{align}
    \partial_t \gamma_{ab} = 2 \left( D_{(a} N_{b)} - N K_{ab} \right),
\end{align}
which is an evolution equation for the spatial part of the metric. The extrinsic curvature is the measure of deviation in the normal vector, $\mathcal{N}^\mu$, as it is parallel transported across a hypersurface.

\subsection{ADM Evolution Equations}

Einstein's equations (EEs) are encoded into a symmetric $4$ by $4$ tensor, these are comprised of $10$ unique equations. We recall EEs in a vacuum are 
\begin{align}
    G_{\mu\nu} = R_{\mu\nu} - \frac{1}{2}R g_{\mu\nu} = 0. \label{eq:num-vacuum-einsteins-equations}
\end{align}
Now we make use of the standard result in 3+1 decompositions to decompose the Riemann curvature into timelike and spacelike parts, 
\begin{align}
    R_{\mu\nu\rho\sigma} = &\  {}^{(3)}R_{\mu\nu\rho\sigma} - 2 K_{\rho [\mu} K_{\nu]\sigma} + 4 D_{[\rho} K_{\sigma][\mu} \N_{\nu]} + 4 D_{[\mu} K_{\nu][\rho} \N_{\sigma]} \nonumber \\ & - 4 \N_{[\mu} K_{\nu]}{}^\alpha K_{\alpha[\rho} \N_{\sigma]} - \frac{4}{N} \N_{[\rho} D_{\sigma]} D_{[\mu} N \N_{\nu]} - 4 \N_{[\mu} \mathcal{L}_{\N} K_{\nu][\rho} \N_{\sigma]}
\end{align} 
where ${}^{(3)}R_{\mu\nu\rho\sigma}$ is the Riemann curvature tensor of the spatial hypersurface defined by $2 D_{[\mu} D_{\nu]} X_\rho = {}^{(3)} R_{\mu\nu\rho}{}^\sigma X_\sigma$ with $\N^\mu X_\mu = 0$. For a proof of this see~\cite{NumericalRelatBaumga2010}. Contracting with the metric gives the 3+1 decomposition of the Ricci tensor 
\begin{align}
    R_{\mu\nu} = &\ {}^{(3)}R_{\mu\nu} + K_{\mu\nu} K - 2K_\mu{}^\rho K_{\nu\rho} - \mathcal{L}_\N K_{\mu\nu} - \frac{1}{N} D_\nu D_\mu N  \nonumber \\ & + 2 \N_{(\mu} D^\rho K_{\nu)\rho} - 2 \N_{(\mu} D_{\nu)} K + \N_\mu \N_\nu \left( K_{\rho\sigma} K^{\rho\sigma} + \mathcal{L}_{\N} K + \Delta N \right)
\end{align}
where ${}^{(3)} R_{\mu\nu} = {}^{(3)}R_{\mu\rho\nu}{}^{\rho}$ is the spatial Ricci tensor, $K = K_{\mu\nu} g^{\mu\nu}$ and $\Delta = \lambda^{\mu\nu} D_\mu D_\nu$ is the spatial Laplace operator. Substituting this into \cref{eq:num-vacuum-einsteins-equations} and taking a contraction with $\N^\nu$ we find 
\begin{align}
    \N^\nu G_{\mu\nu} = - D^\nu K_{\mu\nu} + D_\mu K - \frac{1}{2} N_\mu \left( {}^{(3)} R + K^2 - K_{\rho\sigma} K^{\rho\sigma} \right) = 0.
\end{align}
From here we can read of the spatial and timelike parts to obtain 
\begin{align}
    \mathcal{H} &= {}^{(3)} R + K^2 - K_{\rho\sigma} K^{\rho\sigma} = 0 \\
    \mathcal{M}_\mu &= - D^\nu K_{\mu\nu} + D_\mu K = 0.
\end{align}
These are the Hamiltonian and Momentum constraints respectively. They are purely spatial conditions on the extrinsic curvature and spatial metric, no time derivatives are present meaning they are imposed on each hypersurface. They arise due to the diffeomorphism freedom in the theory. At the Hamiltonian field theory level the lapse and shift become the Lagrange multipliers that impose the Hamiltonian and Momentum constraints.

Lastly, for the spatial part $P_\mu{}^\rho P_\nu{}^\sigma G_{\mu\nu}$ we compute its trace reversed form, which is simply 
\begin{align}
    P_\mu{}^\rho P_\nu{}^\sigma R_{\rho\sigma} = - \mathcal{L}_\N K_{\mu\nu} + K_{\mu\nu} K - 2 K_\mu{}^\rho K_{\rho\nu} + {}^{(3)} R_{\mu\nu} - \frac{1}{2} D_{\mu} D_{\nu} N = 0.
\end{align}
Then by rearranging and solving for $\mathcal{L}_{\N} K_{\mu\nu}$ we find the evolution equation for the extrinsic curvature 
\begin{align}
    \mathcal{L}_{\N} K_{\mu\nu} = - \frac{1}{N} D_\mu D_\nu N + {}^{(3)} R_{\mu\nu}  - 2 K_{\mu\rho} K^\rho{}_\nu + K K_{\mu\nu}.
\end{align}
Using that $\mathcal{L}_{\N} = \frac{1}{N} \mathcal{L}_{N \N} = \frac{1}{N} \mathcal{L}_{n}$ and expanding the Lie derivative in terms of the covariant derivative we have that 
\begin{align}
    n^\rho \nabla_\rho K_{\mu\nu} + K_{\mu\rho} \nabla_\nu n^\rho + K_{\nu\rho} \nabla_\mu n^\rho = - D_\mu D_\nu N + {}^{(3)} R_{\mu\nu}  - 2 K_{\mu\rho} K^\rho{}_\nu + K K_{\mu\nu}.
\end{align}
Now we are in a position to write down the ADM equations of motion, by expanding into 3+1 coordinates the result is
\begin{align}
    \left( \partial_t - \mathcal{L}_{N} \right) \gamma_{ab} &= -2 N K_{ab} \\
    \left( \partial_t - \mathcal{L}_{N} \right) K_{ab} &= - D_a D_b N + N \left( {}^{(3)} R_{ab} - 2 K_{ac} K^c{}_b + K K_{ab} \right) \\[10pt]
    \mathcal{H} & = {}^{(3)} R + K^2 - K_{ab} K^{ab} = 0 \label{eq:num-metric-hamiltonian-constraint} \\
    \mathcal{M}_a & = - D^b K_{ab} + D_a K = 0. \label{eq:num-metric-momentum-constraint}
\end{align}
where $K = K_{ab} \gamma^{ab}$ and ${}^{(3)} R = {}^{(3)} R_{ab} \gamma^{ab}$. We see that there are two evolution equations for the spatial metric and extrinsic curvature. There are also 4 constraint equations, 1 Hamiltonian and 3 momentum constraints. The lapse and shift do not have equations of motion as they are gauge variables (or equivalently they are Lagrange multipliers), and they are equivalent to the choice of a 3+1 foliation. We will return to this point later when we discuss gauge choices.

To evolve the data $(\gamma_{ab}, K_{ab})$ we first need to specify it on an initial slice. Together they contain $6+6=12$ components which need to be determined. The constraint conditions are 4 equations that bring the undetermined components down to $8$, $4$ of these $8$ are related to the coordinate choice, $3$ for the choice of spatial coordinates and $1$ for the choice of hypersurface (i.e. choice of $t$). This leaves $4 = 2+2$ components which are the two polarisations of the gravitational wave, and their momentums. To check the hyperbolicity of the ADM equations one needs to introduce first order variables for the spatial derivative for all of $N, N^a$ and $\gamma_{ab}$, then the system can be written as in \cref{eq:num-first-order-hyperbolic-system} excluding the constraints. Computing the characteristic matrix can be done for a singe coordinates $x = x^1$ as the equations are written 3-variant way, so there is no preferred direction. In doing so one finds that the ADM equations are only weakly hyperbolic. This is clearly a problem as we cannot say that the solutions are well-posed even locally for small time deviations. It has been shown empirically that the ADM formulation suffered from long term unstable dynamics and solutions were found to blow up~\cite{Formulations_of_Shinka_2008}.

\section{Gauge Fixing}\label{sec:numerical-gauge-fixing}
\hypertarget{add:citations-for-gauge-fixing}
In the previous section we have seen that the lapse and shift appear in the ADM evolution equations for the spatial metric and the extrinsic curvature, however, there is no equation that determines these objects. As mentioned previously they describe how the coordinates evolve between the spatial slices. They arise as there is a diffeomorphism invariance in the theory, that is under a change of coordinates the physical properties like the motion of a particle remain unchanged. There are many ways to fix the lapse and shift, the simplest example is the geodesic slicing,
\begin{align}
    N = 1,\quad N^a = 0.
\end{align}
This implies that a normal observer is moving along geodesics as the acceleration $a_\mu = \N^\nu \nabla_\nu \N_\mu = 0$ vanishes. However, it was quickly noted that this gauge choice causes the numerical grid to fall into a black hole solution and thus making the grid singular~\cite{Numerical_Relat_Lehner_2001}. This condition may still be used for much simpler setups, but is ruled out for more complicated systems like those of binary black holes. Another choice is to impose 
\begin{align}
    K= \partial_t K = 0
\end{align}
which, through the equation of motion for $K$ gives an elliptic equation for the lapse
\begin{align}
    \Delta N = N K_{ab} K^{ab}. \label{eq:num-maximal-slicing-conditions}
\end{align}

It can be shown that this condition maximises the volume of the hypersurface for normal observers and as such is called maximal slicing~\cite{Maximally_Slici_Estabr_1973}. This slicing is such that it keeps volume elements fixed on the numerical grid and coordinates no longer collide and become singular during evolution. The other important property for maximal slicing is that it is singularity avoiding, meaning that the lapse shrinks to zero in the regions close to the black hole effectively freezing the numerical simulation at those points. Outside the singularity the lapse remains non-zero, and the simulation can still evolve the remaining spacetime. There are two major drawbacks to maximal slicing, the first is that as the slices near the singularity are frozen and the ones outside continue to evolve, the slices between can become stretched and distorted which causes a rapid growth in the radial components of the metric. The second downside is that solving \cref{eq:num-maximal-slicing-conditions} at each time step is computationally expensive and sometimes can require more compute time than the evolution. It is therefore useful to consider gauge fixings that are not second-order elliptic differential equations. Which leads directly into considering evolutionary gauge fixing where the lapse and shift are promoted to dynamical fields and the gauge prescribes dynamical equations for them.

\paragraph{Harmonic Gauge}
The harmonic gauge is motivated by the want for Einstein's equations to be hyperbolic as in \cref{sec:num-hyperbolicity-and-well-posedness}. Historically it has been used in contexts to form a hyperbolic system from Einstein's equations, see~\cite{Hyperb1+3YvonneTommaso1983,Numerical_Relat_Pretor_2004}. It is well known in the metric formulation that to obtain second-order hyperbolic equations from Einstein's equations one can impose 
\begin{align}
    \Gamma^\mu = g^{\rho\sigma} \Gamma^\mu_{\rho\sigma} = 0. \label{eq:numerical-harmonic-gauge}
\end{align}
It is called Harmonic since in this gauge the coordinate functions are Harmonic functions~\cite{Besse1987}, that is $\Box x^\mu = 0$. This was also seen to be the case in \cref{chap:nonlinear-gauge-fixing}, where the harmonic condition changes the highest order symbol of Einstein's equations to be the box operator. Expanding the Christoffel symbols in terms of the 3+1 decomposition of the metric we find that the harmonic condition gives the following evolution equation for the lapse and shift
\begin{align}
    (\partial_t - N^a \partial_a) N &= -N^2 K. \label{eq:num-lapse-harmonic-slicing} \\
    (\partial_t - N^b \partial_b) N^a &= - N \partial^a N + N^2 {}^{(3)} \Gamma^a
\end{align}
It was quickly found that this slicing condition also approached the singularity very quickly, as well as it leading to shock pathologies~\cite{Pathologies_of_Alcubi_1997}. Also, the gauge fixing in \cref{eq:numerical-harmonic-gauge} is not covariant, which can be seen due to it being linear combinations of the Christoffel symbols which are not tensors. In particular this causes problems when moving to a different coordinate system, for example in spherical coordinates there exists a $\frac{1}{r}$ term in ${}^{(3)} \Gamma^a$ that causes divergences near the origin.  At the same time there are angular terms that destroy the spherical symmetry. To solve both of these issues a modification to the equations of motion was introduced, called the generalised harmonic/hyperbolic gauge
\begin{align}
    (g^{\mu\nu} - h(N) \N^\mu \N^\nu) \left(\Gamma^\rho_{\mu\nu} - \bar{\Gamma}^\rho_{\mu\nu}\right) = 0
\end{align}
where $h(N)$ is a function of the lapse and $\bar{\Gamma}^\rho_{\mu\nu}$ are the Christoffel symbols for a flat reference metric $\bar{g}_{\mu\nu}$, which is not dynamical~\cite{Generalized_har_Alcubi_2005}. With the correct choice of $h(N) = \frac{N}{2} - 1$ one finds that the slicing becomes singularity avoiding, in particular the evolution equation imposed on the lapse is called Bona-Masso slicing or 1+log slicing, and has been found to be well-behaved for most spacetimes. A similar property occurs for the shift equation, where given this choice of $h(N)$ the evolution of the shift can counteract the dragging and stretching of the coordinate grid for rotating bodies. For suitably simple simulations imposing $N^a = 0$ is usually acceptable. Finally, the presence of $\bar{\Gamma}^\mu$ is to restore covariance to the gauge fixing. It is known that the difference of two connections is a tensor and therefore $\Gamma - \bar{\Gamma}$ transforms covariantly under changes in coordinates. Depending on the coordinate system the non-covariance is not always a problem, for example in Cartesian coordinates $\bar{\Gamma}^\mu = 0$ and the original condition can be implemented, alternatively in spherical coordinates we find $\bar{\Gamma}^\mu \neq 0$  and this term is needed such that problematic terms are not present~\cite{Generalized_har_Sorkin_2009}.

\section{Z4 Evolution System} \label{sec:numerical-Z4-system}
Here we present the Z4 system~\cite{General_covaria_Bona_2003}, which is close in spirit to the \pleb{} evolution system that we will describe in \cref{chap:plebanski-numerical-relativity}. It begins by introducing the 1-form $Z_\mu \in \Lambda^1$ which we can take to be the conjugate momentum for the lapse and shift. As $Z_\mu$ are conjugate to Lagrange multipliers this implies that they should vanish, so we take $Z_\mu = 0$ as 4 primary constraints. Since these constraints should be satisfied everywhere, we can use $Z_\mu$ to provide covariant modification to Einstein's equations,
\begin{align}
    R_{\mu\nu} + \nabla_{(\mu} Z_{\nu)} = 0. \label{eq:num-Z4-equations}
\end{align}
Since $Z_\mu = 0$ on the constraint surface this is equivalent to the usual Einstein equations. However, we are considering the free evolution of this system, and therefore it needs to be checked how $Z_\mu$ evolves under this system. By taking the contraction with the covariant derivative of the trace reversed form of \cref{eq:num-Z4-equations} we find,
\begin{align}
    \Box Z_\mu + R_{\mu}{}^\nu Z_\nu = 0,
\end{align}
that $Z_\mu$ satisfies a homogeneous wave equation in $Z_\mu$. This means that if we impose $Z_\mu |_{t=0}=0$ and $\partial_t Z_{\mu}|_{t=0} = 0$ this implies that $Z_\mu$ will remain zero for all time. Analytically this is true, however, numerically this rarely holds and deviations from $Z_\mu = 0$ occur and damping terms are often added to \cref{eq:num-Z4-equations} to counteract this deviation~\cite{Constraint_damp_Gundla_2005}. By decomposing the equations of motion into 3+1 using the ADM formalism one finds that the evolution equations are 
\begin{align}
    \left( \partial_t - \mathcal{L}_{N} \right) \gamma_{ab} &= -2 K_{ab} \\
    \left( \partial_t - \mathcal{L}_{N} \right) K_{ab} &= - D_a D_b N + N \left( {}^{(3)} R_{ab} - 2 K_{ac} K^c{}_b + (K - 2 \Theta) K_{ab} + D_a Z_b + D_b Z_a \right) \\
    \left( \partial_t - \mathcal{L}_{N} \right) \Theta &= \frac{N}{2} \left( \mathcal{H} + 2 D_a Z^a - \frac{2}{N} Z^a D_a N  \right) \\
    \left( \partial_t - \mathcal{L}_{N} \right) Z_a &= N \left( \mathcal{M}_a + D_a \Theta - \frac{\Theta}{N} D_a N - 2 K_a{}^b Z_b \right)
\end{align}
where $\Theta = \mathcal{N}^\mu Z_\mu$ and $Z_i$ are the spatial components. A natural gauge to impose that produces a hyperbolic system is the first-order Harmonic gauge,
\begin{align}
    Z_\mu = \Gamma_\mu
\end{align}
where $\Gamma_\mu$ is the Harmonic 1-form. As $Z_\mu$ tends to zero this imposes the usual Harmonic condition that causes the leading order symbol for the Ricci flat condition to be the wave equation. This provides evolution equations for the lapse and shift, which can then be modified to obtain the Bona-Masso slicing while retaining hyperbolicity. Tests of this system were done in the Harmonic gauge, and it was found that there were improvements over the ADM system~\cite{A_symmetry_brea_Bona_2003}. In a series of standardised test developed in~\cite{Apples_with_App_Daveri_2018}, the constraint damped version of the Z4 system was shown to improve constraint preservation over the undamped cases. Cementing the idea that damping terms, while perhaps not as mathematically beautiful, are needed in this type of discretisation procedure.  Similar to the jump from the ADM to the BSSN systems, a conformal decomposition that effectively separates the conformal modes, was developed in~\cite{Conformal_and_c_Alic_2012}. Again the separation of the conformal structure appears to perform better than no separation, it was also shown in their specific example that it performed better than the BSSN system in the same situation.

In~\cite{ActionPrinciplBona2011} it was shown that the Z4 can be derived from a gauge fixed Lagrangian for the metric formulation of gravity. The development of which can be summarised into the following algorithm. First, compute the gauge fixing that makes the equations of motion hyperbolic in the sense that the wave equation is the leader order contribution to the second-order dynamics. Then impose that gauge by introducing first-order parameters, which when set to zero recover the gauge fixing as required. The 3+1 decomposition of this system can then be performed and modifications can be made assuming that the hyperbolicity and constraint conservation properties are kept. Following this rough guideline with the metric formulation results in the various Z4 system that are commonplace in the literature. The plan for this Thesis is to perform the same steps in the \pleb{} formulation of GR and explore the numerical properties that the resulting system has. The leading order contributions to the gauge fixing have already been computed in \cref{chap:nonlinear-gauge-fixing}, along with a similar conformal decomposition. All that remains is to find a suitable 3+1 decomposition that is useful for numerics. Before doing so we quickly reproduce these steps for the chiral Maxwell system as it contains many parallels with the chiral GR theories.

\section{Numerical Chiral Maxwell} \label{sec:numerical-chiral-maxwell}
With the numerical techniques described in the previous parts of this chapter, we can apply them on a chiral theory different from gravity that is much simpler but contains a similar structure. Chiral Maxwell or $U(1)$ Yang-Mills is a simple example that still keeps many of the properties from \pleb{}'s formulation of gravity. We recall that, in the 3+1 language with coordinates $t,x^i$, the main variables are the Riemann-Silberstein vector and the connection 1-form are
\begin{align}
    \phi^i \in \C^3, \quad A = A_0 dt + A^i dx^i \in \Lambda^1.
\end{align}
On a Minkowski background the original equations of motion become
\begin{align}
    \phi^i = -i(\partial_t A^i - \partial^i A_0) - \epsilon^{ijk} \partial^j A^k, \quad \partial_t \phi^i = -i \epsilon^{ijk} \partial^j \phi^k, \quad \partial^i \phi^i = 0.
\end{align}
We see that the evolution equations for $\phi^i$ are completely separate from the equations involving $A^i$, allowing one to consider only the evolution and constraint equations for $\phi^i$. The physical fields are the electric and magnetic components, and we have seen that their complex combination is in $\phi^i$. This gives us what we call the Maxwell system
\begin{align}
    \partial_t \phi^i = -i\epsilon^{ijk} \partial^j \phi^k, \quad C_\phi = \partial^i \phi^i = 0.
\end{align}
Now we can apply the $\lambda$-system modification to this. The first check is that the principal symbol for the original system is symmetric hyperbolic, as the fields are complex we need to check that they are Hermitian. It is clear that the principal symbol is $P^{ij} = i \epsilon^{ijk}n_k$ and by taking the adjoint we see it is Hermitian. We are then free to continue with the modification, introducing the field $\lambda \in \C$ we can give it a dissipative equation for the constraint,
\begin{align}
    \partial_t \lambda = \alpha \partial^i \phi^i + \beta \lambda.
\end{align}
The total system becomes
\begin{align}
    \partial_t \binom{\phi^i}{\lambda} = \begin{pmatrix}
        -i \epsilon^{ijk} && 0 \\ \alpha \delta^{jk} && 0
    \end{pmatrix} \partial^j \binom{\phi^k}{\lambda} + \binom{0}{\beta \lambda}.
\end{align}
To make the principal part symmetric hyperbolic we simply introduce  $\bar{\alpha} \delta^{ij}$, where $\bar{\alpha}$ is the complex conjugate, into the top-right block, such that the total $\lambda$-system modification becomes 
\begin{align}
    \partial_t \binom{\phi^i}{\lambda} = \begin{pmatrix}
        -i \epsilon^{ijk} && \bar{\alpha} \delta^{ij} \\ \alpha \delta^{jk} && 0
    \end{pmatrix} \partial^j \binom{\phi^k}{\lambda} + \binom{0}{\beta \lambda}.
\end{align}
Writing these equations of motion out individually gives 
\begin{align}
    \partial_t \phi^i = -i\epsilon^{ijk} \partial^j \phi^k + \bar{\alpha} \partial^i \lambda, \quad \partial_t \lambda = \alpha \partial^i \phi^i + \beta \lambda.
\end{align}
To continue the analysis we can compute the equation of motion for the constraint $C_\phi$, 
\begin{align}
    \partial_t C_\phi = \partial_t \partial^i \phi^i = \partial^i \left( -i\epsilon^{ijk} \partial^j \phi^k + \bar{\alpha} \partial^i \lambda \right) = \bar{\alpha} \partial^i \partial^i \lambda.
\end{align}
Therefore, $\lambda$ and $C_\phi$ form a closed system. Next we can take the Fourier transform of the coupled $\lambda,C_\phi$ system,
\begin{align}
    \lambda = \int \hat{\lambda}(k) e^{ik_a x^a} d^3k, \quad C_\phi = \int \hat{C_\phi}(k) e^{ik_a x^a} d^3k
\end{align}
to find 
\begin{align}
    \partial_t \binom{\hat{\lambda}}{\hat{C_\phi}} = \begin{pmatrix}
        \beta && \alpha \\ -\bar{\alpha} k^2 && 0
    \end{pmatrix} \binom{\hat{\lambda}}{\hat{C_\phi}}.
\end{align}
The propagation matrix has eigenvalues $\lambda_\pm = \frac{1}{2} \left( \beta \pm \sqrt{\beta^2 - 4 |\alpha|^2 k^2} \right)$, it can be seen that the real part of the eigenvalues are negative when ${\rm Re}(\beta) <  0$ and $\alpha \neq 0$. 

It is worth comparing this result to the gauge fixing described in \cref{chap:Linearised-Gravity}. There the variable $\lambda$ is called $\theta$ and appears as the conjugate momentum for $A_0$. The hyperbolic gauge fixing presented previously also fixes $\alpha = i, \beta = 0$, however, $\beta \neq 0$ would still be an acceptable range for the desired properties of the gauge fixing. The highest order structure is not effected by $\beta$, so it can take any values and the wave operator is still recovered. The combined constraint and hyperbolicity analysis reveal $\alpha = i, \beta \leq 0$ as optimal choices for the gauge fixed system. The follows the guideline set out in the previous section, the gauge fixing suggested by the $\lambda$-system is the same as the hyperbolic gauge fixing. The fields introduced in order to place the gauge fixing in a first-order formulation are $\lambda$ (or $\theta$). The original chiral Maxwell equations are recovered as $\lambda \rightarrow 0$. And as we can see from the eigenvalues of the propagation matrix the constraints are damped throughout evolution, given the correct modification. This system has all the same analytic properties that appear in the Z4 system and that we wish to apply to the \pleb{} system. As a confirmation of this result we perform some basic numerical tests of the system.

\subsection{Numerical Results and Discussion}
We present some numerical results for the first-order hyperbolic chiral Maxwell system with damping. We use the method of lines and the finite difference method (FDM) as described in this chapter. The 3 dimensional grid contains $40\times 40\times 40$ points and the spatial coordinates are periodic with ranges $-2.5 <  x,y,z <  2.5$ which gives $dx=dy=dz=0.125$. Runge-Kutta 4th order was used as the time integrator and 2nd order central differences were used for the spatial derivatives. We choose to simulate from $t = 0$ to $t = 10$ with time steps $dt = 0.25 dx = 0.03125$. The system is initialised with
\begin{align}
    \phi^i(0,x) &= \varphi^i e^{i k^j x^j}, \quad \varphi^i = i E^i - B^i \\ 
    \theta(0,x) &= 0,
\end{align}
the wave vector is chosen in the $xy$ direction, $k^i = \frac{2\pi}{5} (1,1,0)$, the electric and magnetic parts are then $E^i = (0,1,0)$ and $B^i = -\epsilon^{ijk} k^j E^k/k^2$. This choice of direction for $E^i$ means that the initial condition does not satisfy the constraints, this is in order to check the constraint conserving property of the system.

\begin{figure}[H]
    \begin{subfigure}{0.5\textwidth}
        \includegraphics[width=\textwidth]{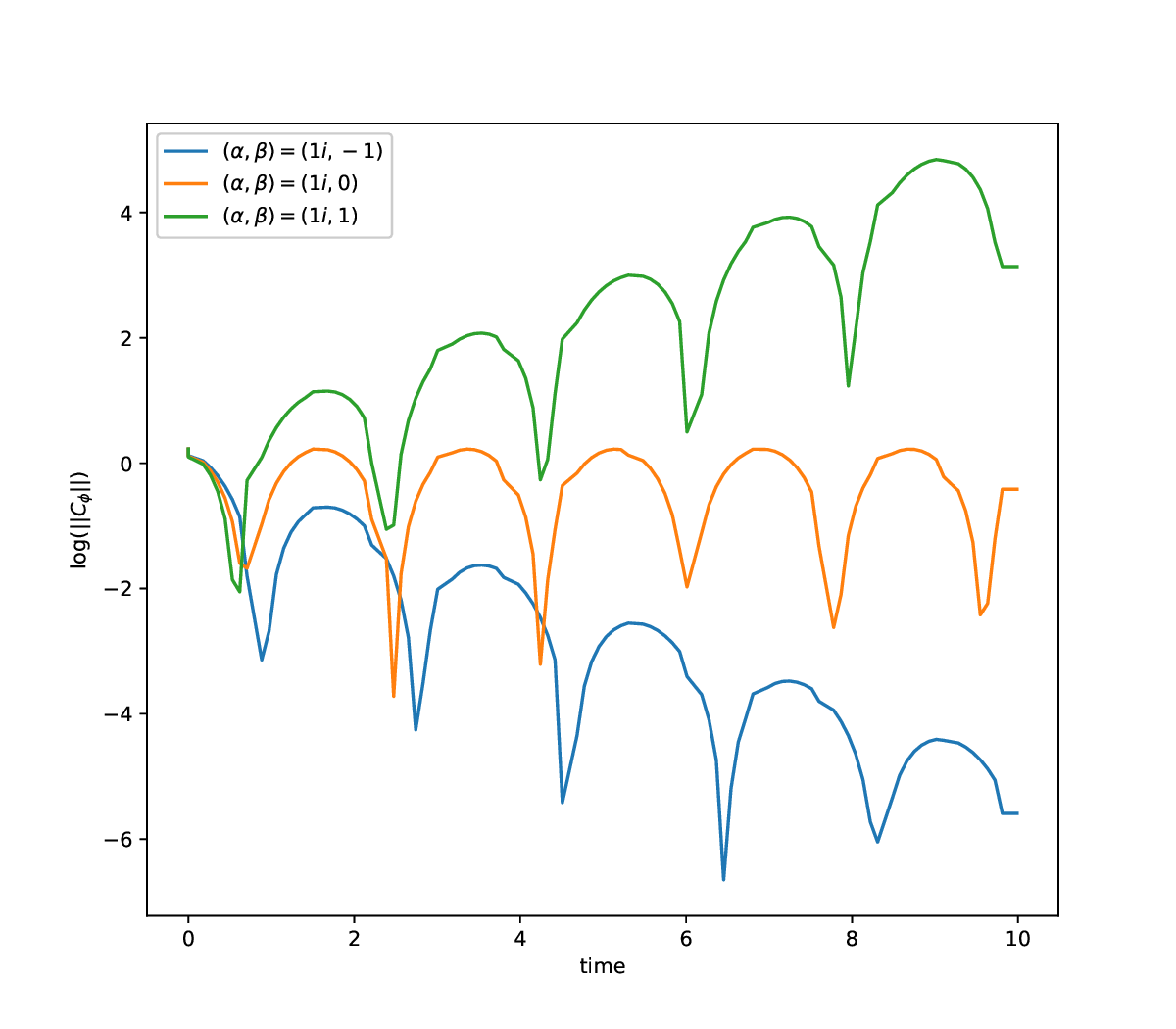}
        \caption{Parameter Search} 
        \label{fig:chiral-maxwell-parameter-search}
    \end{subfigure}
    \hfill
    \begin{subfigure}{0.44\textwidth}
        \includegraphics[width=\textwidth]{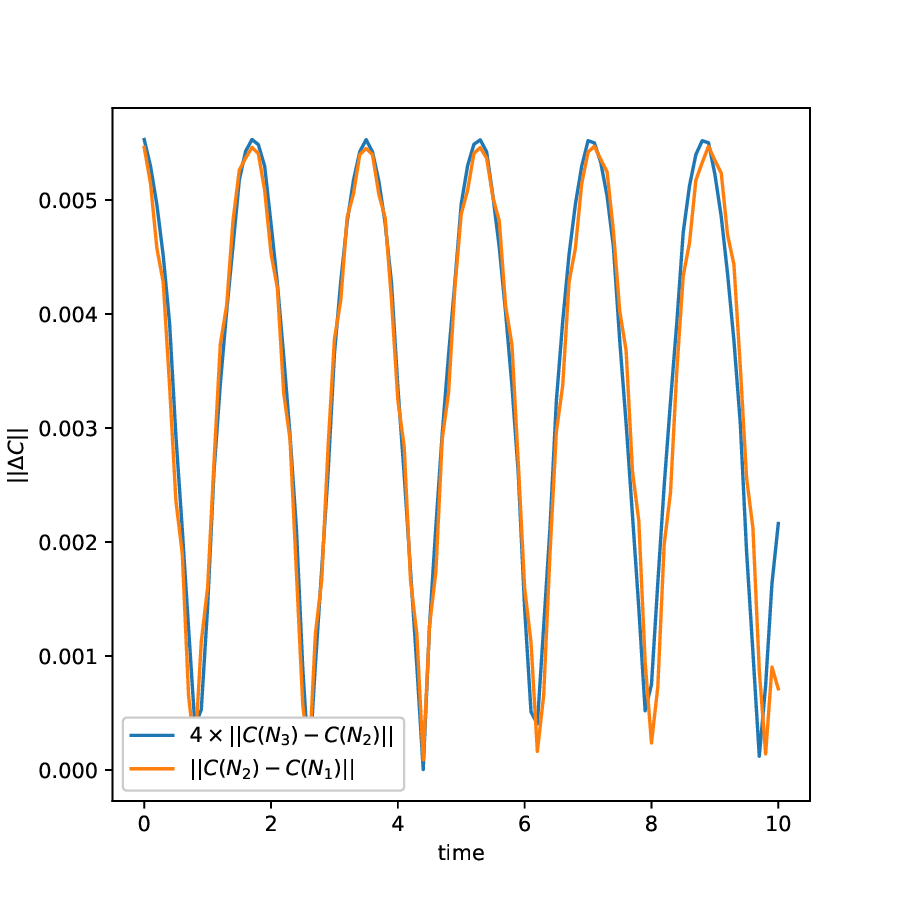}
        \caption{Convergence Test}
        \label{fig:chiral-maxwell-convergence-test}
    \end{subfigure}

    \caption{ Figure~\ref{fig:chiral-maxwell-parameter-search} contains the results of a parameter search for optimal values of $\alpha, \beta$ displaying that the sign of $\beta$ affected the convergence of the constraint as expected from the discussion in the main text. Figure~\ref{fig:chiral-maxwell-convergence-test} the difference between the Gauss constraints for different resolutions, $N = 20,40,80$. The error converges at a rate of $4$ with is correct given the order of accuracy used was $2$.}
    \label{fig:chiral-maxwell-numerical-results}
\end{figure}

From the plots in \cref{fig:chiral-maxwell-numerical-results} we see that with a non-zero damping parameter the system is able to remove the artificially introduced constraint. We see that the combination of the hyperbolic gauge fixing and damping (or equivalently the $\lambda$-system as they obtain the same system here) removes the constraint as expected by the Fourier analysis. It is worth noting that with the damping removed the constraint neither grows nor diminishes, this could be due to the propagation of the constraint effectively diffusing its violating modes removing the chance increase at a specific point. It should be noted that simulation time is relatively low and that over a longer period of time the simulations without damping are expected to diverge and crash, considering the domain is periodic and the constraints, while able to propagate, are not able to leave. Nevertheless, with damping we see positive results. This encourages our exploration into applying these techniques to the chiral formulations of gravity.

As a final note, in this chapter, we comment on how the plots in \cref{fig:chiral-maxwell-parameter-search,fig:chiral-maxwell-convergence-test} are computed. In the first plot, we compute the mean, over all points on the grid, of the absolute value of the constraint. The constraint itself is computed from the solution at each time step. The mean of the absolute values is denoted in the graphs as $||\cdot||$. The natural logarithm of this value is plotted versus time for a resolution of $N=40$. Similarly, the second plot instead computes the constraints for three different resolutions. To compute their difference at different resolutions we interpolate the solution at the higher resolution (greater value of $N$) to the lower resolution. The solutions are also interpolated in time so that their difference can be computed at the same time value. Since the order of accuracy used was $2$ and the resolution doubles for each resolution we find that the two differences should be a factor of $2^2 = 4$, as displayed by the plot.

    \newpage
    \chapter{Hyperbolic First-Order Chiral Formulation} \label{chap:plebanski-numerical-relativity}

    Stable evolutions of a binary black hole configuration were achieved in 2005, see~\cite{Evolution_of_Bi_Pretor_2005,Gravitational_w_Baker_2005}. Numerical relativity is by now a mature field, see e.g.~\cite{NumericalRelatBaumga2010}. At the same time, existing numerical schemes take a very long time to run, with up to one month to complete a single simulation. Geometrodynamics in its standard form is difficult and computationally expensive business. It is known that general relativity can be reformulated in many ways, see e.g.~\cite{FormGenRelGravity2020}, and it is not impossible that one of such reformulations may hide the potential to be significantly more efficient for numerical simulations. For a review of different formulations from the perspective of numerical relativity see e.g.~\cite{Formulations_of_Shinka_2008}. The goal of this chapter is to explore one of the alternative formulations of GR in the hope to find a more economic evolution system than the one provided by the standard metric formulation. 

    It has been shown in \cref{chap:Linearised-Gravity} that the \pleb{} formulation has some remarkable linear properties. Firstly, there exists a ``nice'' gauge fixing in the sense that the system can be written in terms of Dirac operators acting on a collection of differential forms. It is well-known that Dirac operators are strongly hyperbolic, due to the fact that they are the formal square roots of the wave operator. Furthermore, it was seen that after being gauge fixed the resulting set equations could be separated into sectors of size 12 and 4. At the linear level each of these sectors could be treated independently. The 4 component sector contained the information about the linear conformal factor and the remaining metric components were contained in the 12 sector. Being strongly hyperbolic and naturally separating the conformal factor out are useful properties that are used in modern numerical relativity algorithms such as the BSSN~\cite{NumericalRelatBaumga2010} and CCZ4 formulations~\cite{Conformal_and_c_Alic_2012}. This leads directly to the work done in \cref{chap:nonlinear-gauge-fixing} where a particular lift of the linear gauge fixing to the nonlinear equations of motion was shown. At the nonlinear level these equations are naturally locally strongly hyperbolic, this is because the local hyperbolicity is a linear property and therefore the linear hyperbolicity can be lifted directly. In the same chapter it was also shown that a conformally separated system of equations exists for the nonlinear gauge fixing also. This is a noteworthy characteristic that, due to the success of other conformal separations, should also be explored numerically. It is therefore not impossible that this chiral reformulation may led to improvements in numerical simulations.
    
    This chapter begins with an introduction to Ashtekar's formulation which arises as the Hamiltonian version of \pleb{}'s formalism. Previous numerical studies of Ashtekar's variables have been performed in~\cite{Hyperbolic_form_Yoneda_2001,Formulations_of_Shinka_2008} and the numerical results have been promising. It is known that Ashtekar variables are useful as their equations of motion become polynomial, the price one has to pay is using complex-valued fields. To solve this complex-valued problem one has to impose reality-conditions, there is no unique way to impose them but once they are satisfied the resulting metric will be real-valued. The aim of this chapter is then to covert the nonlinear hyperbolic gauge fixing for \pleb{} into a 3+1 evolution system, which should reveal itself as a gauge fixing of Ashtekar's equations of motion.

    We begin this chapter with a derivation of the 3+1 decomposition of the fields appearing in \pleb{}'s formulation, this is detailed in \cref{sec:pleb-num-1+3-decomposition} where the metricity and reality conditions are also solved in a similar manner. Then in \cref{sec:pleb-num-ashtekar-formalism} we describe the 3+1 decomposed fields in the context of Ashtekar's Hamiltonian formulation. In \cref{sec:pleb-num-gauge-fixed-1+3-decomposition} we perform the main computation, which involves decomposing the gauge fixed equations of motion into the Ashtekar fields. The resulting system is presented in \cref{sec:pleb-num-gauge-fixed-evolution-system} where additional damping terms are considered in order to control the constraints. Next we present some results about the conformally separated system, along with some partially conformally separated systems that arise as a change of variables as compared to the original gauge fixed system. A derivation and exposition of these systems is shown in \cref{sec:pleb-num-conformal-evolution-systems}. Finally, a standard numerical test is performed on the partially conformal system and the results are shown in \cref{sec:pleb-num-numerical-results}.

    \section{3+1 Decomposition} \label{sec:pleb-num-1+3-decomposition}

    Here we describe the process of decomposing \pleb{}'s formulation into temporal and spatial components with respect to a foliation $M = \R \times \spasurf$. In doing so we solve the metricity and reality conditions in order to correctly parametrise the fields. The fundamental variables for \pleb{}'s formalism are the self-dual 2-forms $\Sigma^i \in \Lambda^+(M)$ and the self-dual connections $A^i$. The self-dual 2-forms describe a real-valued metric when they satisfy the metricity and reality conditions 
    \begin{align}
        \Sigma^i \wedge \Sigma^j \sim \delta^{ij}, \quad {\rm Re}(\Sigma^i \wedge \Sigma^i) = 0, \quad \Sigma^i \wedge \asd^j = 0.
    \end{align}
    A triple of complex-valued 2-forms has $18\C$ components and the reality and metricity constraints are $5\C + 1\R + 9 \R = 10\C$ conditions. This leaves $10\R + 3\C$ components in $\Sigma^i$ which is enough for a real-valued metric and an $SO(3,\C)$ frame. The definition of the self-dual connection in terms of $\Sigma^i$ and the Einstein equations on said connection are given by
    \begin{align}
        A^i = -i J_1^{-1}(\star d\Sigma^i), \quad \Sigma^i_\mu{}^\rho F^i_{\nu\rho} = 0.
    \end{align}

    The gauge fixing shown in \cref{chap:nonlinear-gauge-fixing} introduces a connection 1-form $\xi_\mu$ and an internal vector $\chi^i$, we will that see these are introduced to control the constraints. We repeat the gauge fixed equations here, 
    \begin{align} \label{eq:num-covariant-hyperbolic-gauge-fixing}
        A^i = -i J_1^{-1}(\star d\Sigma^i) + c_1 d\chi^i + c_2 J_1(d\chi)^i, \quad \xi_\mu = - \Gamma_\mu - 2 c_2 \Sigma^i_\mu{}^\nu \partial_\nu \chi^i, \\
        \Sigma^i_\mu{}^\rho F^i_{\nu\rho} + \partial_{(\mu} \xi_{\nu)} = 0, \quad \partial^\mu A^i_\mu - \frac{1}{2} \Sigma^{i\mu\nu} \partial_\mu \xi_\nu = 0.
    \end{align}
    This gauge fixing is only for the highest order terms as to obtain the wave equation when substituting the connection definition into its equations of motion. Lower order modifications are possible, but they are best done in the 3+1 language.

    We being the decomposition onto $\R$ and $\spasurf$ by first considering the self-dual 2-forms $\Sigma^i$. The space $\R$ is parametrised by the time coordinate $t : \R \rightarrow M$, and the coordinates on the spatial hypersurface we denote $x^a \in \spasurf$. As before we take the exterior derivative of the time coordinate to obtain a 1-form in the timelike direction, $dt \in \Lambda^1(M)$, to which we define an arbitrary vector $n^\mu$ such that 
    \begin{align}
        n^\mu (dt)_\mu = 1. \label{eq:num-n-and-dt-are-dot-product}
    \end{align}
    In the coordinates $x^\mu = (t,x^a)$, the vector is $n^\mu = (1,-N^a)$ where $N^a$ is an arbitrary 3-vector. Instead of using the metric as the basic variables we instead introduce the following bivector 
    \begin{align}
        \tE^{i\mu\nu} = \frac{1}{2}\teps^{\mu\nu\rho\sigma} \Sigma^i_{\rho\sigma} \in \Lambda^2(TM) \label{eq:num-tilde-E-in-eps-S}
    \end{align}
    which is dual to the 2-forms $\Sigma^i$. The tilde over a field represents a density weight of $+1$, a tilde under a field conveys a weight $-1$. The triple of 2-forms only defines a metric up to conformal factor, as such we fix the physical metric by fixing conformal factor in the \urb{} formula for the metric, i.e. \cref{eq:pleb-urbantke-metric}. When a physical metric is chosen we can equate the density weight to the determinant of the metric. Keeping track of the number of upper and lower tildes becomes useful for error correction, as the number of up tildes minus the number of down should remain constant in any equation. The parametrisation in terms of the bivectors are chosen because they are in a sense the conjugate momentums for the connections $A^i$, meaning that the Lagrangian has the leading order kinetic term $L = \tE^{i\mu\nu} \partial_\mu A^i_\nu$. The Levi-civita symbol takes values in $\teps^{\mu\nu\rho\sigma} = \pm 1$, and is defined without need for the metric. In this way we can take $\tE^{i\mu\nu}$ to be the basic variables of the theory without the need to introduce a metric. To decompose the bivectors we insert the 1-form into this bivector 
    \begin{align}
        \tE^{i\mu} = (dt)_\nu \tE^{i\nu\mu} \label{eq:num-4-variant-triad-definition}
    \end{align}
    to give a triple of vectors. By computing $(dt)_\mu \tE^{i\mu} = (dt)_{[\mu} (dt)_{\nu]} \tE^{i\mu\nu} = 0$ we see that these are tangent to the hypersurfaces. In 3+1 coordinates this means we can parametrise them with $E^{i\mu} = (0,\tE^{ia})$ where we call $\tE^{ia}$ the triads (even though they are really densitised inverse triads, we call them triads and leave for ease of writing). We see that we now have a vector basis $n^\mu, \tE^{ia} \in TM$, which we can use to construct a basis for $\Lambda^2(TM)$
    \begin{align}
        n^{[\mu} \tE^{i\nu]} \in \Lambda^2(TM), \quad \epsilon^{ijk} \tE^{j\mu} \tE^{k\nu} \in \Lambda^2(TM).
    \end{align}
    This basis is $6$ dimensional as expected for the space of bivector in dimension 4. The full bivector can now be decomposed into this basis 
    \begin{align}
        \tE^{i\mu\nu} = \utilde{X}^{ij} \epsilon^{jkl} \tE^{k\mu} \tE^{l\nu} + M^{ij} n^{[\mu} \tE^{i\nu]}
    \end{align}
    for arbitrary matrices $\utilde{X}^{ij}$ and $M^{ij}$. By contracting this with $(dt)_\mu$ and making use of \cref{eq:num-n-and-dt-are-dot-product,eq:num-4-variant-triad-definition} we find that $M^{ij} = 2\delta^{ij}$ and therefore 
    \begin{align}
        \tE^{i\mu\nu} = \utilde{X}^{ij} \epsilon^{jkl} \tE^{k\mu} \tE^{l\nu} + 2n^{[\mu} \tE^{i\nu]}
    \end{align}
    is now the decomposition. For the next step we use a result in linear algebra that any $3\times 3$ matrix can be written as, $\utilde{X}^{ij} = \utilde{X}^{(ij)} + \epsilon^{ijk} \utilde{X}^k$, the sum of a symmetric and antisymmetric part where in dimension 3 the antisymmetric part can be parametrised by an internal 3-vector $\tilde{X}^i$. Plugging this into the above decomposition we find that 
    \begin{align}
        \tE^{i\mu\nu} & = \utilde{X}^{(ij)} \epsilon^{jkl} \tE^{k\mu} \tE^{l\nu} + \epsilon^{ijm} \utilde{X}^m \epsilon^{jkl} \tE^{k\mu} \tE^{l\nu} + 2 n^{[\mu} \tE^{i\nu]} \\
        & = \utilde{X}^{(ij)} \epsilon^{jkl} \tE^{k\mu} \tE^{l\nu} + 2(n^{[\mu} + \utilde{X}^j \tE^{j[\mu}) \tE^{i\nu]}
    \end{align}
    Defining a new complex vector $N^\mu = n^{\mu} + \utilde{X}^j \tE^{j\mu}$ which in components we can write it as $N^\mu = (1,-X^a)$ where $X^a = N^a - \utilde{X}^j \tE^{ja}$ is a complex 3-vector.  With this in mind the decomposition is then 
    \begin{align}
        \tE^{i\mu\nu} = \utilde{X}^{(ij)} \epsilon^{jkl} \tE^{k\mu} \tE^{l\nu} + 2 N^{[\mu} \tE^{i\nu]}. \label{eq:num-bivector-complex-decomposition}
    \end{align}
    Counting the number of components at this point gives $6\C + 3\C + 9\C = 18\C$ for $\utilde{X}^{(ij)}, X^a$ and $\tE^{ia}$ respectively. This is equal to the number of components in a generic triple of complex 2-forms as expected. We wish for the bivectors to encode information about a real-valued metric and $SO(3,\C)$ frame only, as we have seen to reduce the number of components we need to satisfy the metricity and reality conditions. 
    \subsection{Decomposing the Metricity Condition}
    First we focus on the metricity condition,
    \begin{align}
        \Sigma^i \wedge \Sigma^j \sim \delta^{ij}.
    \end{align}
    Writing this out in components gives 
    \begin{align}
        \Sigma^i \wedge \Sigma^j = \frac{1}{4}\Sigma^i_{\mu\nu} \Sigma^j_{\rho\sigma} \teps^{\mu\nu\rho\sigma} d^4x
    \end{align}
    Using that $\Sigma^i_{\mu\nu} = -\frac{1}{2} \uteps_{\mu\nu\rho\sigma} \tE^{i\rho\sigma}$ we can write this condition as 
    \begin{align}
        \tE^{i\mu\nu} \tE^{j\rho\sigma} \uteps_{\mu\nu\rho\sigma} \sim \delta^{ij}.
    \end{align}
    We have introduced $\uteps_{\mu\nu\rho\sigma} = \pm 1$ being the totally antisymmetric symbol with indices down. Substituting \cref{eq:num-bivector-complex-decomposition} into this and expanding we find 
    \begin{align}
        \tE^{i\mu\nu} \tE^{j\rho\sigma} \uteps_{\mu\nu\rho\sigma} = 4\utilde{X}^{(i|k} \epsilon^{klm} \tE^{l\mu} \tE^{m\nu} N^\rho \tE^{|j)\sigma} \epsilon_{\mu\nu\rho\sigma}
    \end{align}
    where we now assume that $\utilde{X}^{ij}$ is symmetric. To continue we note the identity 
    \begin{align}
        \uteps_{\mu\nu\rho\sigma} N^\mu \tE^{i\nu} \tE^{j\rho} \tE^{k\sigma} = \epsilon^{ijk} \det(\tE) \label{eq:num-epsilon-N-E-E-to-eps-det}
    \end{align}
    where $\det(\tE) = \frac{1}{6} \epsilon^{ijk} \uteps_{abc} \tE^{ia} \tE^{jb} \tE^{kc}$, and we have fixed the orientation to be $\teps^{0abc} = -\teps^{abc}$ or $\uteps_{0abc} = +\uteps_{abc}$. To see this is true we notice the left-hand side is totally antisymmetric in $ijk$ and therefore must be equal to a multiple of $\epsilon^{ijk}$. Contracting both sides with $\epsilon_{ijk}$ and separating into 3+1 coordinates we find the left and right sides agree. Applying this identity to the metricity conditions reveals
    \begin{align}
        \tE^{i\mu\nu} \tE^{j\rho\sigma} \uteps_{\mu\nu\rho\sigma} \sim 4 \utilde{X}^{(i|k} \epsilon^{klm} \epsilon^{|j)lm} = 8 \utilde{X}^{(ij)} \sim \delta^{ij}.
    \end{align}
    As $\utilde{X}^{ij}$ is symmetric this implies that there is a scalar function $\utilde{X}$ such that $\utilde{X}^{ij} = \utilde{X} \delta^{ij}$. This leaves us with the decomposition 
    \begin{align}
        \tE^{i\mu\nu} = \utilde{X} \epsilon^{ijk} \tE^{j\mu} \tE^{k\nu} + 2 N^{[\mu} \tE^{i\nu]}.
    \end{align}
    Counting the remaining components gives $13 \C = 10\C + 3\C$, enough to form a complex-valued metric and an $SO(3,\C)$ frame. We see that the metricity condition removes the unwanted components that do not belong in the usual theory of gravity.

    \subsection{Decomposing the Reality Conditions}
    To restrict ourselves to a real-valued metric we need to impose the reality conditions,
    \begin{align}
        \Sigma^i \wedge \asd^j = 0, \quad {\rm Re}(\Sigma^i \wedge \Sigma^i) = 0.
    \end{align}
    The latter of the above we see imposes that 
    \begin{align}
        \utilde{X} \det(\tE) \sim i V \label{eq:num-plebanski-1R-reality-condition}
    \end{align}
    where $V$ is some real-valued function which we will determine later. The first reality condition requires that we define the anti-self-dual 2-forms, as we are in Lorentzian signature they are related to the complex conjugates of the self-dual 2-forms and their decomposition can be adopted from the self-dual 2-forms as 
    \begin{align}
        \bar{\tE}^{i\mu\nu} = \bar{\utilde{X}}  \epsilon^{ijk} \bar{\tE}^{j\mu} \bar{\tE}^{k\nu} - 2 \bar{N}^{[\mu} \bar{\tE}^{i\nu]}.
    \end{align}
    Here the bar above a term denotes the negative complex conjugate, i.e. $\bar{X} = -X^*$ where ${}^*$ denotes complex conjugation. Inserting this into the first reality condition and expanding we find 
    \begin{equation} \label{eq:pleb-num-reality-condition-step-1}
        \begin{aligned}
                \utilde{X} \bar{\utilde{X}} \epsilon^{ijk} \epsilon^{jmn} \tE^{k\mu} \tE^{l\nu} \bar{\tE}^{m\rho} \bar{\tE}^{n\sigma} \uteps_{\mu\nu\rho\sigma} - 2\utilde{X} \epsilon^{ikl} \tE^{k\mu} \tE^{l\nu} \bar{N}^\rho \bar{\tE}^{j\sigma} \uteps_{\mu\nu\rho\sigma} \\ +  2 \bar{\utilde{X}} \epsilon^{jmn} \bar{\tE}^{m\rho} \bar{\tE}^{n\sigma} N^\mu \tE^{i\nu} \uteps_{\mu\nu\rho\sigma} - 4 N^\mu \bar{N}^\rho \tE^{i\nu} \bar{\tE}^{j\sigma} \uteps_{\mu\nu\rho\sigma} = 0.
        \end{aligned}
    \end{equation}
    To simplify this we note that the complex conjugates $\bar{\tE}^{i\mu}$ are still purely spatial with respect to $dt$, therefore they span the same space as $\tE^{i\mu}$, the first term is the antisymmetric product of 4 3-vectors which will contain repeats and is therefore zero due to symmetry. As for the middle two terms we introduce $\utE^i_\mu$ as the inverse of the triad $\tE^{i\mu}$, i.e.
    \begin{align}
        \tE^{i\mu} \utE^j_\mu = \delta^{ij}, \quad \tE^{i\mu} \utE^i_\nu = \delta^\mu_\nu - N^\mu (dt)_\nu.
    \end{align}
    Which implies that its components are $\utE^i_\mu = (-N^b \utE^i_b, \utE^i_a)$. Which we use to simplify the second term in appearing in the first line of \cref{eq:pleb-num-reality-condition-step-1}, it can be shown that
    \begin{align}
        \utilde{X} \epsilon^{ikl} \tE^{k\mu} \tE^{l\nu} \bar{N}^\rho \bar{\tE}^{j\sigma} \uteps_{\mu\nu\rho\sigma} = - 2 \utE^i_\mu \bar{\tE}^{j\mu} \utilde{X} \det(\tE).
    \end{align}
    By separating both sides into 3+1 and using the component definitions we see that both sides match. By taking the complex conjugate of this and using the definition of the barred variables we obtain
    \begin{align}
        \bar{\utilde{X}} \epsilon^{ikl} \bar{\tE}^{k\mu} \bar{\tE}^{l\nu} N^\rho \tE^{j\sigma} \uteps_{\mu\nu\rho\sigma} & = 2 \bar{\utE}^i_\mu \tE^{j\mu} \bar{\utilde{X}} \det(\bar{\tE}) \\
        & = - 2 \bar{\utE}^i_\mu \tE^{j\mu} \utilde{X} \det(\tE).
    \end{align}
    Where in the last equality we have used \cref{eq:num-plebanski-1R-reality-condition} to reveal that $\bar{\utilde{X}} \det(\bar{\tE}) = - \utilde{X} \det(\tE)$. The reality condition then becomes
    \begin{align}
        \utE^i_\mu \bar{\tE}^{j\mu} \utilde{X} \det(\tE) - \bar{\utE}^j_\mu \tE^{i\mu} \utilde{X} \det(\tE) - N^\mu \bar{N}^\rho \tE^{i\nu} \bar{\tE}^{j\sigma} \uteps_{\mu\nu\rho\sigma} = 0.
    \end{align}
    To compute the last term we split the vector $N^\mu$ into $N^\mu = t^\mu - \utilde{N}^i \tE^{i\mu}$ where $t^\mu = (1,0)$ is purely timelike. A similar split is available for $\bar{N}^\mu = -t^\mu + \bar{\utilde{N}}^i \bar{\tE}^{i\mu}$ where the bar operator is defined the same way with respect to the unbarred fields. The only terms that survive in the last term of the equation above are 
    \begin{align}
        N^\mu \bar{N}^\rho \tE^{i\nu} \bar{\tE}^{j\sigma} \uteps_{\mu\nu\rho\sigma} = -\utilde{N}^k \tE^{k\mu} \bar{N}^\rho \tE^{i\nu} \bar{\tE}^{j\sigma} \uteps_{\mu\nu\rho\sigma} + \bar{\utilde{N}}^k \bar{\tE}^{k\mu} N^\rho \tE^{i\nu} \bar{\tE}^{j\sigma} \uteps_{\mu\nu\rho\sigma}.
    \end{align}
    Using the previous results this becomes 
    \begin{align}
        N^\mu \bar{N}^\rho \tE^{i\nu} \bar{\tE}^{j\sigma} \uteps_{\mu\nu\rho\sigma} = -\epsilon^{ikl} \utilde{N}^k \utE^l_\mu  \bar{\tE}^{j\mu} \det(\tE) - \epsilon^{jkl}  \bar{\utilde{N}}^k \bar{\utE}^l_\mu \tE^{i\mu} \det(\bar{\tE}).
    \end{align}
    Collecting all the results the reality condition becomes 
    \begin{align}
        \utE^i_\mu \bar{\tE}^{j\mu} \utilde{X} \det(\tE) - \bar{\utE}^j_\mu \tE^{i\mu} \utilde{X} \det(\tE) + \epsilon^{ikl} \utilde{N}^k \utE^l_\mu  \bar{\tE}^{j\mu} \det(\tE) + \epsilon^{jkl}  \bar{\utilde{N}}^k \bar{\utE}^l_\mu \tE^{i\mu} \det(\bar{\tE}) = 0.
    \end{align}
    To understand this condition we introduce the matrix $H^{ij} = \utE^i_\mu \bar{\tE}^{j\mu}$ along with the vector $H^i = \frac{\utilde{N}^i}{\utilde{X}}$ such that it becomes 
    \begin{align}
        H^{ij} + \bar{H}^{ji} + \epsilon^{ikl} H^k H^{lj} - \epsilon^{jkl} \bar{H}^k \bar{H}^{li} = 0
    \end{align}
    where we have made use of \cref{eq:num-plebanski-1R-reality-condition} and removed the factors of $\utilde{X} \det(\tE)$. From here we can notice that $-\bar{H}^{ij}$ is the inverse of $H^{ij}$, indeed by computing 
    \begin{align}
        - H^{ik} \bar{H}^{kj} = \utE^i_\mu \bar{\tE}^{k\mu} \bar{\utE}^k_\nu \tE^{j\nu} = \utE^i_\mu (\delta^\mu_\nu - N^\mu (dt)_\nu) \tE^{j\nu} = \utE^i_\mu \tE^{j\mu} = \delta^{ij}
    \end{align}
    it is clear that they are indeed inverses of each other. We can then multiply the reality condition with the appropriate factor of $-\bar{H}$ to find 
    \begin{align}
        \delta^{ij} - \bar{H}^{ki} \bar{H}^{kj} + \epsilon^{ijk} H^k + \epsilon^{kml} \bar{H}^l \bar{H}^{mi} \bar{H}^{kj} = 0.
    \end{align}
    The first two terms in the above are symmetric in the $ij$ indices and the last two are antisymmetric and as such we can treat these terms separately. The first two terms then imply that $\bar{H}^T = \bar{H}^{-1}$, or equivalently 
    \begin{align}
        H^{ij} = -\bar{H}^{ji}
    \end{align}
    which by multiplying by $\tE^{i\mu}$ and $\bar{\tE}^{j\nu}$ suggests that 
    \begin{align}
        \tE^{i\mu} \tE^{i\nu} = \bar{\tE}^{i\mu} \bar{\tE}^{i\nu}.
    \end{align}
    Meaning that the tensor
    \begin{align}
        \tilde{\tilde{\gamma}}^{\mu\nu} = \tE^{i\mu} \tE^{i\nu}
    \end{align}
    is real-valued. An immediate consequence of this is that $\det(\tE) = \pm \det(\bar{\tE})$ and therefore $\det(\bar{H}) = \pm 1$. Without loss of generality we choose the branch $\det(\tE) = -\det(\bar{\tE})$ which implies that $\det(\tE) = \det(\tE^*)$, $\det(\bar{H}) = 1$ and $\utilde{X} = \bar{\utilde{X}}$. This allows us to identify $\bar{H} \in SO(3,\C)$ and make use of the identity
    \begin{align}
        \epsilon^{kml} \bar{H}^{mi} \bar{H}^{kj}= \epsilon^{jik} \bar{H}^{lk}
    \end{align}
    such that the antisymmetric terms in the reality condition become
    \begin{align}
        \epsilon^{ijk} H^k + \epsilon^{kml} \bar{H}^l \bar{H}^{mi} \bar{H}^{kj} = \epsilon^{ijk} H^k - \epsilon^{ijk} \bar{H}^{lk} \bar{H}^l = 0.
    \end{align}
    It is clear then at the only independent condition is 
    \begin{align}
        H^i - \bar{H}^{ji} \bar{H}^j = 0.
    \end{align}
    Substituting in the definitions of the H fields we find 
    \begin{align}
        \frac{\utilde{N}^i}{\utilde{X}} - \bar{\utE}^j_\mu \tE^{i\mu} \frac{\bar{\utilde{N}}^j}{\bar{\utilde{X}}} = 0.
    \end{align}
    Now we can make use of $\utilde{X} = \bar{\utilde{X}}$ and multiply by $\tE^{i\mu}$ to find 
    \begin{align}
        \bar{\utilde{N}}^i \bar{\tE}^{i\mu} - \utilde{N}^i \tE^{i\mu} = 0.
    \end{align}
    Finally, using the definition of $N^\mu$ and $\bar{N}^\mu$ this is equivalent to 
    \begin{align}
        N^\mu + \bar{N}^\mu = 0, \quad \textrm{ or } \quad N^\mu = (N^\mu)^*.
    \end{align}
    Collecting all the results of the reality conditions we find 
    \begin{align}
        \utilde{X} = \pm i \dlap \in i \R, \quad N^\mu \in \R, \quad \tE^{i\mu} \tE^{i\nu} \in \R.
    \end{align}
    Since $N^\mu$ is real this also means that the 3-vector $N^a$ is real-valued. The decomposition of the bivector that satisfies the reality and metricity conditions is then
    \begin{align}
        \tE^{i\mu\nu} = \pm i \dlap \epsilon^{ijk} \tE^{j\mu} \tE^{k\nu} + 2 N^{[\mu} \tE^{i\nu]} \label{eq:num-plebanski-bivector-real-decomposition}
    \end{align}
    Counting the number of components we see that there are $1\R + 3\R + 6\R + 3\C = 10\R + 3\C$ which make up the real metric and choice of $SO(3,\C)$ frame.

    \subsection{Recovering the Metric}
    Now that we have the split of the variables for $\Sigma^i$, we can use this and the \urb{} formula for the metric to compare the \pleb{} story with the usual ADM metric language. To do so we make use of the inverse \urb{} formula in \cref{eq:pleb-urbantke-inverse-metric} and fix the conformal factor such that it is related to the inverse of the physical metric,
    \begin{align}
        \sqrt{-g} \tg^{\mu\nu} = -\frac{1}{12} \epsilon^{ijk} \tE^{i\mu\rho} \tE^{j\alpha\beta} \tE^{k\sigma\nu} \uteps_{\rho\sigma\alpha\beta}.
    \end{align}
    Here $g_{\mu\nu}$ is the physical metric, $\sqrt{-g} = \sqrt{-\det(g)}$ and $\tg^{\mu\nu} = \sqrt{-g} g^{\mu\nu}$. Substituting \cref{eq:num-plebanski-bivector-real-decomposition} into this, expanding all the terms, and cancelling expressions due to symmetry we find 
    \begin{align}
        \sqrt{-g} \tg^{\mu\nu} = & \left(\frac{1}{12} \dlap^2 \tE^{l\mu} \tE^{l\nu} - \frac{1}{6} N^\mu N^\nu \right) \epsilon^{ijk} \uteps_{\alpha\beta\rho\sigma} N^\alpha \tE^{i\beta} \tE^{j\rho} \tE^{k\rho} \\ & + \dlap^2 \tE^{i\mu} \tE^{j\nu} \left(\frac{1}{3} \epsilon^{jkl} \tE^{j\beta} - \frac{1}{12} \epsilon^{ikl} \tE^{i\beta} \right) \uteps_{\alpha\beta\rho\sigma} N^\alpha \tE^{k\rho} \tE^{l\sigma}.
    \end{align}
    Making use of \cref{eq:num-epsilon-N-E-E-to-eps-det} and expanding the product of two epsilons we find 
    \begin{align}
        \sqrt{-g} \tg^{\mu\nu} = \dlap \det(\tE) \left( \dlap \tE^{i\mu} \tE^{i\nu} - \frac{N^\mu N^\nu}{\dlap} \right).
    \end{align}
    We note that the sign of in front of the first term in \cref{eq:num-plebanski-bivector-real-decomposition} does not change the definition of the metric. To remove the factors of the determinant of the metric we compute the determinant of $\sqrt{-g} \tg^{\mu\nu}$ to find 
    \begin{align}
        \det(\sqrt{-g} \tg) = \det(g)^3 = - \dlap^6 \det(\tE)^6, \quad \Rightarrow \sqrt{-g} = \pm \dlap \det(\tE).
    \end{align}
    Here we can fix the sign by requiring that $\dlap \det(\tE) > 0$ such that the square root of the determinant of the metric is 
    \begin{align}
        \sqrt{-g} = \dlap \det(\tE).
    \end{align}
    Therefore, the inverse densitised metric is 
    \begin{align}
        \tg^{\mu\nu} = \dlap \tE^{i\mu} \tE^{i\nu} - \frac{N^\mu N^\nu}{\dlap},
    \end{align}
    and for the inverse metric we find 
    \begin{align}
        g^{\mu\nu} = \frac{\tE^{i\mu} \tE^{i\nu}}{\det(\tE)} - \frac{N^\mu N^\mu}{\dlap^2 \det(\tE)}.
    \end{align}
    Computing the lapse we see that, 
    \begin{align}
        N = (-(dt)_\mu (dt)_\nu g^{\mu\nu})^{-\frac{1}{2}} = \dlap \sqrt{\det(\tE)}.
    \end{align}
    The shift vector, $N^a$, is identified with the 3-vector inside $N^\mu = (1,-N^a)$. The inverse spatial metric is 
    \begin{align}
        \gamma^{\mu\nu} = \frac{\tE^{i\mu} \tE^{i\nu}}{\det(\tE)}.
    \end{align}
    This gives a dictionary between the metric and \pleb{} components.

    \subsection{Enforcing Self-Duality}
    The final step in this decomposition is to fix the sign in \cref{eq:num-plebanski-bivector-real-decomposition}; we check that these bivectors are self-dual. To do so we compute
    \begin{align}
        \frac{1}{2} \tg^{\mu\rho} \tg^{\nu\sigma} \uteps_{\rho\sigma\alpha\beta} \tE^{i\alpha\beta} = \mp i \dlap \det(\tE) \left( \pm i \dlap \epsilon^{ijk} \tE^{j\mu} \tE^{k\nu} + 2 N^{[\mu} \tE^{i\nu]} \right) = \mp i \dlap \det(\tE) \tE^{i\mu\nu}
    \end{align}
    where the $\pm$ relative to the $\pm$ in \cref{eq:num-plebanski-bivector-real-decomposition}. We have used the decomposition of $\tg^{\mu\nu}$ and $\tE^{i\mu\nu}$ into $\tE^{i\mu}, N^\mu, \dlap$ and then used the previous identities to obtain the right-hand side. It is a lengthy calculation so only the result is stated. We see that the self-dual bivectors are then given by
    \begin{align}
        \tE^{i\mu\nu} = -i \dlap \epsilon^{ijk} \tE^{j\mu} \tE^{k\nu} + 2 N^{[\mu} \tE^{i\nu]}.
    \end{align}

    \subsection{3+1 Coordinate Decomposition} \label{subsec:num-1+3-decomposition-complete}
    We have computed the decomposition of the bivector into $\tE^{i\mu}, N^\mu$ and $\dlap$. It will be useful later to have these objects separated into temporal and spatial coordinates $x^\mu = (t,x^a)$. In doing so we find 
    \begin{align}
        \tE^{i\mu} = (0,\tE^{ia}), \quad N^\mu = (1,-N^a)
    \end{align}
    as such the basic variables are the triad $\tE^{ia}$, the shift vector $N^a$ and the inverse densitised lapse $\dlap$. The decomposition of the bivector is then 
    \begin{align}
        \tE^{i0a} = \tE^{ia}, \quad \tE^{iab} = -i\dlap \epsilon^{ijk} \tE^{ja} \tE^{kb} - N^a \tE^{ib} + N^b \tE^{ia}. \label{eq:num-bivector-1+3-decomposition}
    \end{align}
    The metric in 3+1 is given by 
    \begin{align}
        \tg^{00} = -\frac{1}{\dlap},\quad \tg^{0a} = \frac{N^a}{\dlap},\quad \tg^{ab} = \dlap \tE^{ia} \tE^{ib} - \frac{N^a N^b}{\dlap}
    \end{align}
    and the inverse metric is given by 
    \begin{align}
        \utilde{g}_{00} = -\dlap + \utilde{g}_{ab} N^a N^b, \quad \utilde{g}_{0a} = \utilde{g}_{ab} N^b, \quad \utilde{g}_{ab} = \frac{1}{\dlap} \utE^i_a \utE^i_b.
    \end{align}
    For later purposes it will be useful to compute the decomposition of the following endomorphisms
    \begin{align}
        \Sigma^i_\mu{}^\nu = -i \utilde{g}_{\mu\rho} \tE^{i\rho\nu},
    \end{align}
    which we find gives
    \begin{equation}
    \begin{gathered}
        \Sigma^i_0{}^0 = i \utE^i_a \frac{N^a}{\dlap}, \quad \Sigma^i_a{}^0 = \frac{i}{\dlap} \utE^i_a, \quad \Sigma^i_0{}^a = i\dlap \tE^{ia} - i \utE^i_b \frac{N^a N^b}{\dlap} - \epsilon^{ijk} \utE^j_b N^b \tE^{ka} \\ \Sigma^i_a{}^b = -i \utE^i_a \frac{N^a}{\dlap} - \epsilon^{ijk} \utE^j_a \tE^{kb}
    \end{gathered} \label{eq:num-conformal-2-form-1+3-decomposition}
    \end{equation}
    This completes the decomposition of the metric variables into temporal and spatial components. For the connection variables the story is much simpler, we simply introduce
    \begin{align}
        A^i_\mu dx^\mu = A^i_0 dt + A^i_a dx^a
    \end{align}
    and the curvatures are 
    \begin{align}
        F^i_{0a} = \partial_t A^i_a - \partial^A_a A^i_0, \quad F^i_{ab} = \partial_a A^i_b - \partial_b A^i_a + \epsilon^{ijk} A^j_a A^k_b
    \end{align}
    where $\partial^A_a \xi^i_b = \partial_a \xi^i_b + \epsilon^{ijk} A^j_a \xi^k_b$ is the spatial gauge covariant derivative. We call $A^i_0$ the temporal connection and $A^i_a$ the spatial connection and no extra conditions need to be satisfied by them. This completes the decomposition of the frame and connection fields in the \pleb{} formulation of gravity. Now we can start to apply it to the equations of motion.

    \section{Ashtekar's Formalism} \label{sec:pleb-num-ashtekar-formalism}
    Ashtekar's formalism was first introduced in $1985$ by Abhay Ashtekar~\cite{NewVariablesFAshtek1986}, where they introduced a canonical pair of position and momentum variables for gravity. The idea was to use densitised inverse triads as the main position variables, $\tE^{ia} =  \det(e) e^{ia}$ where $e^I_\mu$ is the tetrad and $e^{ia}$ is the inverse of the spatial part. When the action for gravity is modified by the addition of the purely imaginary Holst term then the canonical momentum variables appear as a complex linear combination of the spin connection and extrinsic curvature $A^i_a = \epsilon^{ijk} \omega^{jk}_a + i e^{ib} K_{ab}$. When written in these variables the Hamiltonian becomes polynomial in the variables, moreover there are no second-class constraints as compared to the frame formalism. This is ideal for numericals as the constraints satisfy a closed set of evolution equations, as previously discussed this allows for modification to the equations of motion that are attractive towards the constraint surface. We show how the action for \pleb{} becomes polynomial and write down the equations of motion and constraints that define the theory.

    \subsection{Ashtekar Lagrangian}
    We now write the \pleb{} Lagrangian in terms of the new fundamental variables, $\tE^{ia}$,$N^a$,$\dlap$,$A^i_0$ and $A^i_a$. We start with the \pleb{} Lagrangian and move to components to find
    \begin{align}
        L = \Sigma^i \wedge F^i /d^4x = \frac{1}{4} \teps^{\mu\nu\rho\sigma} \Sigma^i_{\mu\nu} F^i_{\rho\sigma} = \frac{1}{2} \tE^{i\mu\nu} F^i_{\mu\nu}.
    \end{align}
    Then by separating the coordinates into time and space, $x^\mu = (t,x^a)$, and using the 3+1 split in \cref{subsec:num-1+3-decomposition-complete} we find 
    \begin{align}
        L = \tE^{ia} \left( \partial_t A^i_a - \partial^A_a A^i_0 \right) -\frac{i}{2} \dlap \epsilon^{ijk} \tE^{ja} \tE^{kb} F^i_{ab} + \tE^{ia} N^b F^i_{ab}.
    \end{align}
    This is the Lagrangian for gravity in Ashtekar variables. It is now easy to see that $\tE^{ia}$ is conjugate to $A^i_a$ and as the time derivative only appears between them these are the only dynamical variables. The variation with respect to these two fields gives
    \begin{align}
        \frac{\delta L}{\delta \tE^{ia}} &= \partial_t A^i_a - \partial^A_a A^i_0 - i \dlap \epsilon^{ijk} \tE^{jb} F^k_{ab} + N^b F^i_{ab} = 0 \\
        \frac{\delta L}{\delta A^i_a} &= -\partial_t \tE^{ia} - \epsilon^{ijk} A^j_0 \tE^{ka} - i \epsilon^{ijk} \partial_b^A \left( \dlap \tE^{ja} \tE^{kb} \right) + 2 \partial_b^A \left( \tE^{i[a} N^{b]} \right) = 0.
    \end{align}
    Where we have made use of the variational formula $\delta F^i_{ab} = 2\partial^A_{[a} \delta A^i_{b]}$. We see that from their variations we obtain two evolution equations for the triad and spatial connection,
    \begin{align}
        \partial_t A^i_a  &= \partial^A_a A^i_0 + i \dlap \epsilon^{ijk} \tE^{jb} F^k_{ab} + N^b F^i_{ba} \\
        \partial_t \tE^{ia} &= - \epsilon^{ijk} A^j_0 \tE^{ka} - i \epsilon^{ijk} \partial_b^A \left( \dlap \tE^{ja} \tE^{kb} \right) + 2 \partial_b^A \left( \tE^{i[a} N^{b]} \right). \label{eq:num-ashtekar-triad-evolution-equation}
    \end{align}
    Varying with the remaining fields we obtain the following constraints
    \begin{align}
        \frac{\delta L}{\delta \dlap} & = -\frac{i}{2} \epsilon^{ijk} \tE^{ja} \tE^{kb} F^i_{ab} = 0 \\
        \frac{\delta L}{\delta N^a} & = \tE^{ib} F^i_{ba} = 0 \\
        \frac{\delta L}{\delta A^i_0} & = \partial_a^A \tE^{ia} = 0.
    \end{align}
    Variation with respect to the lapse and shift give the Hamiltonian and momentum constraints 
    \begin{align}
        \mathcal{H} & = \epsilon^{ijk} \tE^{ja} \tE^{kb} F^i_{ab} = 0, \\
        \mathcal{M}_a & = \tE^{ib} F^i_{ba} = 0.
    \end{align}
    These generate the time and space diffeomorphism gauge transformations. The Ashtekar formalism comes with an extra constraint known as the Gauss constraint 
    \begin{align}
        \mathcal{G}^i = \partial^A_a \tE^{ia}
    \end{align}
    these generate the $SO(3,\C)$ gauge transformations. Due to the extra $SO(3,\C)$ gauge symmetry in Ashtekar and \pleb{}'s formulations, as compared to the metric formulation, we obtain the additional gauss constraint. It can be shown that the constraints are first-class and as such form a closed system under the time derivative~\cite{Hamiltonian_Ana_Buffen_2004}. Moreover, every equation of motion and constraint is polynomial in the fields or derivatives of them, this is surprisingly simple for GR and the highest order degree polynomial appearing is degree $4$ in the equations of motion (or degree $5$ at the level of the Lagrangian). The polynomial property can have important consequences for numerical simulations, for certain initial conditions some fields may pass through degenerate points and in the standard formalisms one would have to divide by zero, this causes equations to diverge and codes to crash, see section 6.6 of~\cite{AlcubierreMiguelIntro3+12008}. In this formulation as no division is required this problem does not occur.

    It has been shown that the Ashtekar system is weakly hyperbolic~\cite{Constructing_hy_Yoneda_1999}. It was then shown in~\cite{Einstein_s_equa_Iriond_1997} that by adding the appropriate constraint to the evolution equations Ashtekar's evolution system becomes a symmetric hyperbolic system. This property was taken further in a series of papers by Shinkai and Yoneda~\cite{Constructing_hy_Yoneda_1999,Formulations_of_Shinka_2008,Hyperbolic_form_Shinka_2000,Hyperbolic_form_Yoneda_2001,Will_hyperbolic_Shinka_2001} where they explored the numerical properties of Ashtekar's variables. They applied the adjusted and $\lambda$-systems to the evolution and constraint equations to form a strongly hyperbolic evolution system. It was shown that modifying the equations of motion with the constraints worked as expected and the constraint growth was tamed. Interestingly they found no striking difference between the weakly, strongly and symmetric hyperbolic systems and found that the constraint was under control for all systems~\cite{Will_hyperbolic_Shinka_2001}. This is in agreement with~\cite{Numerical_Relat_Hern_2000} where a similar result was found for a different system of equations for gravity. It was suggested that the requirement on the constraint propagation eigenvalues is more important than the hyperbolicity, that is constraint damping is more important than hyperbolicity. With this result in mind we require that the system by strongly hyperbolic and have as many eigenvalues of the constraint amplification matrix be non-positive as possible.

    \subsection{Wave Equation Analogy}  \label{subsec:wave-equation-analogy}
    The difference between the ADM and Ashtekar formalism can be nicely summed in an analogy with the wave equation. In dimension 4 the hyperbolic wave equation is 
    \begin{align}
        \partial^2_t u = \partial^i \partial^i u
    \end{align}
    where $u(t,x^i)$ is a real function on a Minkowski manifold. The solution $u$ represents Einstein's equation on the metric, in the Harmonic gauge it is a collect of wave like equations on $g_{\mu\nu}$. The goal of the ADM and Ashtekar systems is then put Einstein's equation in a form where the method of lines can be applied. To do this for the wave equation we can introduce a first-order variable for the time and spatial derivatives of $u$
    \begin{align}
        v = \partial_t u, \quad  a^i = \partial^i u
    \end{align}
    Then the wave equation becomes the following first order in time system 
    \begin{align}
        \partial_t \begin{pmatrix}
            u \\ v \\ a^i
        \end{pmatrix} = \begin{pmatrix}
            0 & 0 & 0 \\ 0 & 0 & \delta^{jk} \\ 0 & \delta^{ij} & 0 
        \end{pmatrix} \partial^j \begin{pmatrix}
            u \\ v \\ a^k
        \end{pmatrix} + \begin{pmatrix}
            v \\ 0 \\ 0
        \end{pmatrix}
    \end{align}
    This is the usual way of producing a first-order system from a second-order system. It is hyperbolic as this is a first-order reduction of the wave equation, furthermore, in this representation it is symmetric hyperbolic. The first row and column in the above matrix do not contribute the characteristic matrix as there are no spatial differential operators involved. This is effectively the ADM way of reducing the Einstein equations, where the extrinsic curvature is the time derivative, $v$, of the metric and the Levi-Civita connections are the spatial derivatives, $a^i$. 

    Another way of producing a first-order system for the wave equation is to use the technology of spinors and the gamma matrices. The gamma matrices in 4d are a set of four matrices $\gamma^\mu$ such that $\gamma^{(\mu} \gamma^{\nu)} = g^{\mu\nu}$. More formally the gamma matrices, $\gamma^\mu$, generate the matrix representation of the Clifford algebra $Cl_{1,3}(\R)$. The smallest real representation of the gamma matrices is the $4 \times 4$ Majorana representation, for which we can take the solution vector, $U$, to be a real 4 component vector. The first-order reduction of the wave equation is then
    \begin{align}
        \gamma^\mu \partial_\mu U = 0.
    \end{align}
    To write this in a similar form to the ADM version we split the matrices into $\gamma^\mu = \gamma^0, \gamma^i$ and using that $-\gamma^0$ is the inverse of $\gamma^0$ we find 
    \begin{align}
        \partial_t U = \Gamma^i \partial^i U
    \end{align}
    where $\Gamma^i = \gamma^0 \gamma^i$. This first-order system is strongly hyperbolic as the gamma matrices are diagonalisable in any representation and $\Gamma^{(i} \Gamma^{j)} = \delta^{ij}$, so the eigenvalues are all real. Moreover, in dimension 4 there exists a basis such that $\gamma^0$ is also anti-Hermitian and $\gamma^i$ are Hermitian, in this basis the wave equation becomes symmetric hyperbolic and real-valued. The spinor version of the first-order decomposition is analogous to the Ashtekar variable formalism, where the variables $(\tE^{ia}, A^i_a)$ are equivalent to the spinor $U$. For the wave equation in the ADM form, only the original variable $u$ satisfies the wave equation to second order, whereas the functions $a^i$ do not. This differs from the spinor formulation where all the fields in the 4 component vector $U$ satisfy the wave equation.

    As for other systems, like gravity, the spinor formulation usually involves complex variables as spinors are naturally complex-valued. In the wave equation example, the existence of a Majorana basis allows the spinor to be real-valued, whereas for gravity the issue of restricting to a real-valued field is done through imposing the reality conditions. The aim of the rest of this chapter is then to further explore the spinor-like first-order equations of motion for gravity, and in particular develop an evolution system ready for use with the numerical approaches mentioned previously.

    \section{Gauge Fixed 3+1 Decomposition} \label{sec:pleb-num-gauge-fixed-1+3-decomposition}
    Now we have seen how the space-time decomposition works for the usual \pleb{} formalism and Lagrangian we can apply the same techniques to our gauge fixed system. As there is no Lagrangian for the gauge fixing found in \cref{chap:nonlinear-gauge-fixing} we must decompose the equations of motion directly into 3+1. We repeat the system here for the reader
    \begin{align}
        A^i = -i J_1^{-1}(\star d\Sigma)^i + c_1 d\chi^i + c_2 J_1(d\chi)^i, \quad \xi_\mu = -\Gamma_\mu - 2c_2 \Sigma^i_\mu{}^\nu \partial_\nu \chi^i \\
        \Sigma^i_\mu{}^\rho F^i_{\nu\rho} + \partial_{(\mu} \xi_{\nu)} = 0, \quad \partial^\mu A^i_\mu - \frac{1}{2} \Sigma^{i\mu\nu} \partial_\mu \xi_\nu = 0.
    \end{align}
    Before performing the 3+1 decomposition we need to manipulate these equations into a form that best suits this process. 
    \subsection{Component Formulas}
    The basic metric variables in the \pleb{}  formalism are efficiently encoded into the bivector $\tE^{i\mu\nu}$, hence, we will try to place $\Sigma^i$ in this form where possible. As shown in the previous section the inverse densitised metric $\tg^{\mu\nu}$ naturally decomposes into the components of $\tE^{i\mu\nu}$, and manipulating terms such that they involve $\tg^{\mu\nu}$ will therefore be useful for the decomposition. Starting with the definition for $A^i$ we apply $J_1$ to both sides and use $J^2_1 = J_1 + 2 \mathbb{I}$ to find 
    \begin{align}
        J_1(A)^i = -i \star d\Sigma^i + (c_1 + c_2) J_1(d\chi)^i + 2c_2 d\chi^i
    \end{align}
    which in components reads 
    \begin{align}
            \epsilon^{ijk} \Sigma^j_\mu{}^\nu A^k_\nu = -\frac{i}{2} \epsilon_\mu{}^{\nu\rho\sigma} \partial_\rho \Sigma^i_{\rho\sigma} + (c_1 + c_2) \epsilon^{ijk} \Sigma^j_\mu{}^\nu \partial_\nu \chi^k + 2c_2 \partial_\mu \chi^i.
        \end{align}
    The raising the index with the densitised inverse metric, $\tg^{\mu\nu}$, we find
    \begin{align}
            \epsilon^{ijk} \tsd^{j\mu\nu} A^k_\nu = -\frac{i}{2} \teps^{\mu\nu\rho\sigma} \partial_\rho \Sigma^i_{\rho\sigma} + (c_1 + c_2) \epsilon^{ijk} \tsd^{j\mu\nu} \partial_\nu \chi^k + 2c_2 \tg^{\mu\nu} \partial_\nu \chi^i.
    \end{align}
    Where we have used that $\tsd^{i\mu\nu} = \sqrt{-g} \Sigma^{i\mu\nu}$. From here we can use that $\teps^{\mu\nu\rho\sigma} = \pm 1$ to commute it with the partial derivative and then use the self-duality $\teps^{\mu\nu\rho\sigma} \Sigma^i_{\rho\sigma} = 2i \tsd^{i\mu\nu}$. After which the equation becomes 
    \begin{align}
            \epsilon^{ijk} \tsd^{j\mu\nu} A^k_\nu = \partial_\nu \tsd^{i\mu\nu} + (c_1 + c_2) \epsilon^{ijk} \tsd^{j\mu\nu} \partial_\nu \chi^k + 2c_2 \tg^{\mu\nu} \partial_\nu \chi^i.
    \end{align}
    The term involving $A^i$ can be packaged into the gauge covariant derivative, i.e. $\partial^A_\mu \chi^i = \partial_\mu \chi^i + \epsilon^{ijk} A^j_\mu \chi^k$, such that
    \begin{align}
        \partial^A_\nu \tsd^{i\mu\nu} + (c_1 + c_2) \epsilon^{ijk} \tsd^{j\mu\nu} \partial_\nu \chi^k + 2c_2 \tg^{\mu\nu} \partial_\nu \chi^i = 0.
    \end{align}
    Using the definition $\tE^{i\mu\nu} = \frac{1}{2} \teps^{\mu\nu\rho\sigma} \Sigma^i_{\rho\sigma} = i \tsd^{i\mu\nu}$ we rewrite the above equation as 
    \begin{align}
            \partial^A_\nu \tE^{i\mu\nu} + (c_1 + c_2) \epsilon^{ijk} \tE^{j\mu\nu} \partial_\nu \chi^k + 2 i c_2 \tg^{\mu\nu} \partial_\nu \chi^i = 0.
    \end{align}
    This form of the first equation will be easier to decompose into 3+1 components.

    Next we tackle the equation defining the new connection 1-form, $\xi_\mu$. The Harmonic 1-form takes a simple form when using the inverse densitiesed metric,
    \begin{align}
        \Gamma_\mu = \Gamma_{\mu\rho\sigma} g^{\rho\sigma} = g^{\rho\sigma} \partial_\rho g_{\mu\sigma} - \frac{1}{2} g^{\rho\sigma} \partial_\mu g_{\rho\sigma} = - \utilde{g}_{\mu\nu} \partial_\rho \tg^{\nu\rho}.
    \end{align}
    Similarly to before we raise the index with $\tg^{\mu\nu}$ and using this form of the Harmonic 1-form we have 
    \begin{align}
        \tg^{\mu\nu} \xi_\nu = \partial_\nu \tg^{\mu\nu} - 2c_2 \tsd^{i\mu\nu} \partial_\nu \chi^i.
    \end{align}
    Converting to $\tE^{i\mu\nu}$ reveals 
    \begin{align}
        \partial_\nu \tg^{\mu\nu} + 2ic_2 \tE^{i\mu\nu} \partial_\nu \chi^i - \tg^{\mu\nu} \xi_\nu = 0.
    \end{align}

    It remains to express the two equations on the connections using these variables, for the modified Einstein condition we simply raise the $\mu$ index, densitise and convert to $\tE^{i\mu\nu}$ resulting in 
    \begin{align}
        \tE^{i\mu\rho} F^i_{\nu\rho} + i \tg^{\mu\rho} \partial_{(\nu} \xi_{\rho)} = 0.
    \end{align}
    The Lorenz condition we can multiply by $\sqrt{-g}$ to give the appropriate density weight, 
    \begin{align}
        \tg^{\mu\nu} \partial_\mu A^i_\nu + \frac{i}{2} \tE^{i\mu\nu} \partial_\mu \xi_\nu = 0.
    \end{align}
    
    Collecting the new equations of motion we find 
    \begin{align}
        \partial^A_\nu \tE^{i\mu\nu} + (c_1 + c_2) \epsilon^{ijk} \tE^{j\mu\nu} \partial_\nu \chi^k + 2 i c_2 \tg^{\mu\nu} \partial_\nu \chi^i & = 0 \label{eq:num-plebanski-connection-definition-components}\\
        \partial_\nu \tg^{\mu\nu} + 2ic_2 \tE^{i\mu\nu} \partial_\nu \chi^i - \tg^{\mu\nu} \xi_\nu & = 0 \label{eq:num-plebanski-harmonic-definition-components}\\
        \tE^{i\mu\rho} F^i_{\nu\rho} + i \tg^{\mu\rho} \partial_{(\nu} \xi_{\rho)} & = 0\label{eq:num-plebanski-einstein-condition-components}\\
        \tg^{\mu\nu} \partial_\mu A^i_\nu + \frac{i}{2} \tE^{i\mu\nu} \partial_\mu \xi_\nu & = 0. \label{eq:num-plebanski-lorenz-gauge-components}
    \end{align}
    When written this way the metric variables only appear in $\tE^{i\mu\nu}$ or $\tg^{\mu\nu}$ and as such they can be decomposed using the technology from the previous section. The connection variables also appear with their indices in fixed positions, therefore, the 3+1 decomposition is simple to perform.

    \subsection{3+1 Frame Evolution Equations}
    Recalling the space-time split from the previous section we defined a time coordinate $t$ and an induced 1-form $dt$. Using these we found the following decomposition for our main variables 
    \begin{align}
        \tE^{i\mu\nu} & = -i\dlap \epsilon^{ijk} \tE^{j\mu} \tE^{k\nu} + N^\mu \tE^{i\nu} - N^\nu \tE^{i\mu} \\
        \tg^{\mu\nu} & = \dlap \tE^{i\mu} \tE^{i\nu} - \frac{N^\mu N^\nu}{\dlap}
    \end{align}
    where
    \begin{align}
        (dt)_\nu \tE^{i\nu\mu} = \tE^{i\mu}, \quad \textrm{and} \quad (dt)_\nu \tg^{\mu\nu} = -\frac{N^\mu}{\dlap}.
    \end{align}
    In the adapted spacetime coordinates, $x^\mu = (t,x^a)$, the above decomposition becomes
    \begin{align}
        \tE^{i0a} &= \tE^{ia}, \quad \tE^{iab} = -i \epsilon^{ijk} \dlap \tE^{ja} \tE^{kb} - N^a \tE^{ib} + N^b \tE^{ia} \\
        \tg^{00} &= -\frac{1}{\dlap}, \quad \tg^{0a} = \frac{N^a}{\dlap}, \quad \tg^{ab} = \dlap \tE^{ia} \tE^{ib} - \frac{N^a N^b}{\dlap}.
    \end{align}

    With this mind we can start decomposing \cref{eq:num-plebanski-connection-definition-components,eq:num-plebanski-harmonic-definition-components,eq:num-plebanski-einstein-condition-components,eq:num-plebanski-lorenz-gauge-components} into 3+1 components. 
    First we tackle \cref{eq:num-plebanski-connection-definition-components} and look at the $\mu =0$ component,
    \begin{align}
        \partial_a^A \tE^{i0a} + (c_1 + c_2) \epsilon^{ijk} \tE^{j0a} \partial_a \chi^i + 2i c_2 \tg^{00} \partial_t \chi^i + 2ic_2 \tg^{0a} \partial_a \chi^i = 0
    \end{align}
    then using the decomposition we find
    \begin{align}
        \partial_a^A \tE^{ia} + (c_1 + c_2) \epsilon^{ijk} \tE^{ja} \partial_a \chi^i - 2ic_2\frac{1}{\dlap} ( \partial_t - N^a \partial_a) \chi^i = 0.
    \end{align}
    Solving for $(\partial_t - N^a \partial_a) \chi^i$ gives
    \begin{align}
        (\partial_t - N^a \partial_a) \chi^i = \frac{1}{2 i c_2} \dlap \partial^A_a \tE^{ia} + \frac{1}{2 i c_2} (c_1 + c_2) \epsilon^{ijk} \dlap \tE^{ja} \partial_a \chi^k.
    \end{align}
    Which is our equation of motion for $\chi^i$. Next we take the $\mu = a$ component of the same equation 
    \begin{align}
        \partial_t^A \tE^{ia0} + \partial_b^A \tE^{iab} + (c_1 + c_2) \epsilon^{ijk} \tE^{ja0} \partial_t \chi^k + (c_1 + c_2) \epsilon^{ijk} \tE^{jab} \partial_b \chi^k \nonumber \\ + 2i c_2 \tg^{a0} \partial_t \chi^i + 2i c_2 \tg^{ab} \partial_b \chi^i = 0.
    \end{align}
    Which becomes 
    Applying the decomposition expands to 
    \begin{align}
        - \partial_t^A \tE^{ia} + \partial^A_b \left( -i \epsilon^{ijk} \dlap \tE^{ja} \tE^{kb} + \tE^{ia} N^b - \tE^{ib} N^a \right) - (c_1 + c_2) \epsilon^{ijk} \tE^{ja} \partial_t \chi^k  \nonumber\\ + (c_1 + c_2) \epsilon^{ijk} (-i \epsilon^{jlm} \dlap \tE^{la} \tE^{mb} + \tE^{ja} N^b - \tE^{jb} N^a) \partial_b \chi^k \nonumber \\ + 2i c_2 \frac{N^a}{\dlap} (\partial_t - N^b \partial_b) \chi^i + 2ic_2 \dlap \tE^{ja} \tE^{jb} \partial_b \chi^i = 0.
    \end{align}
    After some rearranging we can factor this into 
    \begin{align}
        (\partial_t^A - N^b \partial_b) \tE^{ia} = -i \epsilon^{ijk} \tE^{kb} \partial^A_b \left( \dlap \tE^{ja} \right) -i \epsilon^{ijk} \dlap \tE^{ja} \partial^A_b \tE^{kb} - \tE^{ib} \partial_b N^a + \tE^{ia} \partial^A_b N^b \nonumber\\ - (c_1 + c_2) \epsilon^{ijk} \tE^{ja} (\partial_t - N^b \partial_b) \chi^k  - i (c_1 + c_2) \epsilon^{ijk} \epsilon^{jlm} \dlap \tE^{la} \tE^{mb} \partial_b \chi^k  \nonumber \\ + 2i c_2 \frac{N^a}{\dlap} \left( (\partial_t - N^b \partial_b) \chi^i - \frac{1}{2 i c_2} \dlap \partial^A_b \tE^{ib} - \frac{1}{2 i c_2} (c_1+c_2) \epsilon^{ijk} \dlap \tE^{jb} \partial_b \chi^k \right) \nonumber\\ + 2ic_2 \dlap \tE^{ja} \tE^{jb} \partial_b \chi^i
    \end{align}
    Substituting the evolution equation for $\chi^i$ where it appears simplifies this to be
    \begin{align}
        (\partial^A_t - N^a \partial^A_a) \tE^{ia} = -i \epsilon^{ijk} \tE^{kb} \partial_b^A \left(\dlap \tE^{ja}\right) - \tE^{ib} \partial_b N^a + \tE^{ia} \partial_b N^b + i(c_1 + c_2) \dlap \tE^{ia} \tE^{jb} \partial_b \chi^j \nonumber\\ + \frac{c_2-c_1}{2 i c_2}\left( \epsilon^{ijk} \dlap \tE^{ja} \partial_b^A \tE^{kb} + (c_1+c_2) \dlap \tE^{ja} \tE^{ib} \partial_b \chi^j + (3c_2 + c_1) \dlap \tE^{ja} \tE^{jb} \partial_b \chi^i \right).
    \end{align}
    Revealing an evolution equation for the triad. Moving to the next equation we look at the $\mu = 0$ component of \cref{eq:num-plebanski-harmonic-definition-components},
    \begin{align}
        \partial_0 \tg^{00} + \partial_a \tg^{0a} + 2i c_2 \tE^{i0a} \partial_a \chi^i - \tg^{00} \xi_0 - \tg^{0a} \xi_a = 0
    \end{align}
    which decomposes to be
    \begin{align}
        \frac{1}{\dlap^2} \partial_t \dlap + \partial_a \left( \frac{N^a}{\dlap} \right) + 2i c_2 \tE^{ia} \partial_a \chi^i + \frac{1}{\dlap} ( \xi_0 - N^a \xi_a) = 0.
    \end{align}
    Solving for $(\partial_t - N^a \partial_a) \dlap$ gives
    \begin{align}
        (\partial_t - N^a \partial_a)\dlap = - \dlap \partial_a N^a + 2ic_2 \dlap^2 \tE^{ia} \partial_a \chi^i - \dlap ( \xi_0 - N^a \xi_a)
    \end{align}
    which is an evolution equation for the densitiesed lapse. Taking the $\mu = a$ component of the same equation gives us 
    \begin{align}
        \partial_t \tg^{a0} + \partial_b \tg^{ab} + 2i c_2 \tE^{ia0} \partial_t \chi^i + 2i c_2 \tE^{iab} \partial_b \chi^i - \tg^{a0} \xi_0 - \tg^{ab} \xi_b = 0.
    \end{align}
    Which becomes 
    \begin{align}
        \partial_t \left( \frac{N^a}{\dlap} \right) + \partial_b \left( \dlap \tE^{ia} \tE^{ib} - \frac{N^a N^b}{\dlap} \right) - 2i c_2 \tE^{ia} \partial_t \chi^i \nonumber \\ + 2ic_2 \left( -i \epsilon^{ijk} \dlap \tE^{ja} \tE^{kb} + \tE^{ia} N^b - \tE^{ib} N^a \right) \partial_b \chi^i 
        - \frac{N^a}{\dlap} \left( \xi_0 - N^b \xi_b \right) - \dlap \tE^{ia} \tE^{ib} \xi_b = 0.
    \end{align}
    Making use of the following property for covariant derivatives 
    \begin{align}
        \partial_b ( \dlap \tE^{ia} \tE^{ib} ) = \partial_b^A (\dlap \tE^{ia} \tE^{ib}) = \dlap \tE^{ia} \partial_b^A \tE^{ib} + \tE^{ib} \partial_b^A (\dlap \tE^{ia})
    \end{align}
    and collecting terms suggestively we obtain 
    \begin{align}
        -\frac{N^a}{\dlap^2} \left( (\partial_t - N^b\partial_b) \dlap + \dlap \partial_b N^b + 2ic_2 \dlap^2 \tE^{ib} \partial_b \chi^i + \dlap (\xi_0 - N^b \xi_b) \right) + \tE^{ib} \partial_b^A (\dlap \tE^{ia}) \nonumber \\ - 2ic_2 \tE^{ia} \left( (\partial_t - N^b \partial_b) \chi^i + \frac{i}{2 c_2} \dlap \partial_b^A \tE^{ib} \right) + \frac{1}{\dlap} (\partial_t - N^b \partial_b) N^a \nonumber \\ + 2 c_2 \epsilon^{ijk} \dlap \tE^{ja} \tE^{kb} \partial_b \chi^i - \dlap \tE^{ia} \tE^{ib} \xi_b = 0.
    \end{align}
    Into which we can use the evolution equation for $\dlap$ to kill the first and the evolution for $\chi^i$ to replace the first term on the second line. The result is
    \begin{align}
        \frac{1}{\dlap} (\partial_t - N^b \partial_b) N^a + \tE^{ib} \partial_b^A (\dlap \tE^{ia})  + (c_2- c_1) \epsilon^{ijk} \dlap \tE^{ja} \tE^{kb} \partial_b \chi^i  - \dlap \tE^{ia} \tE^{ib} \xi_b = 0.
    \end{align}
    Solving for $(\partial_t - N^b \partial_b) N^a$ we have another equation of motion
    \begin{align}
        (\partial_t - N^b \partial_b) N^a = - \dlap \tE^{ib} \partial_b^A( \dlap \tE^{ia}) + (c_1 - c_2) \epsilon^{ijk} \dlap \tE^{ja} \tE^{kb} \partial_b \chi^i + \dlap^2 \tE^{ia} \tE^{ib} \xi_b
    \end{align}
    this time for the shift, $N^a$. Combining these results we find an evolution system for the frame variables.

    The evolution equations for the frame variables are
    \begin{align}
        (\partial^A_t - N^a \partial^A_a) \tE^{ia} = & -i \epsilon^{ijk} \tE^{kb} \partial_b^A \left(\dlap \tE^{ja}\right) - \tE^{ib} \partial_b N^a + \tE^{ia} \partial_b N^b + i(c_1 + c_2) \dlap \tE^{ia} \tE^{jb} \partial_b \chi^j \nonumber\\ & + \frac{c_2-c_1}{2 i c_2}\left( \epsilon^{ijk} \dlap \tE^{ja} \partial_b^A \tE^{kb} + (c_1+c_2) \dlap \tE^{ja} \tE^{ib} \partial_b \chi^j \right. \nonumber \\ & \left. \hspace{150pt} + (3c_2 + c_1) \dlap \tE^{ja} \tE^{jb} \partial_b \chi^i \right). \label{eq:num-plebanski-triad-evolution-equation} \\
        (\partial_t - N^b \partial_b) N^a = & - \dlap \tE^{ib} \partial_b^A( \dlap \tE^{ia}) + (c_1 - c_2) \epsilon^{ijk} \dlap \tE^{ja} \tE^{kb} \partial_b \chi^i + \dlap^2 \tE^{ia} \tE^{ib} \xi_b \label{eq:num-plebanski-shift-evolution-equation} \\
        (\partial_t - N^a \partial_a)\dlap = & - \dlap \partial_a N^a + 2ic_2 \dlap^2 \tE^{ia} \partial_a \chi^i - \dlap ( \xi_0 - N^a \xi_a) \label{eq:num-plebanski-lapse-evolution-equation} \\
        (\partial_t - N^a \partial_a) \chi^i = & \frac{1}{2 i c_2} \dlap \partial^A_a \tE^{ia} + \frac{1}{2 i c_2} (c_1 + c_2) \epsilon^{ijk} \dlap \tE^{ja} \partial_a \chi^k. \label{eq:num-plebanski-internal-vector-evolution}
    \end{align}
    We now discuss the structure of the above system of evolution equations for the frame as compared to the original Ashtekar system \cref{eq:num-ashtekar-triad-evolution-equation}. It is still wished for the gauge fixed system to describe the original evolution system as such we require that the Gauss constraint,
    \begin{align}
        \mathcal{G}^i = \partial^A_a \tE^{ia} \simeq 0,
    \end{align}
    to vanish during evolution. As in the Maxwell example \cref{sec:first-order-maxwell-gauge-fixing} we also require the new gauge variables $\chi^i = 0$ to be or close to zero during evolution. This constraint is due to the fact that $\chi^i$ plays the role of the conjugate momentum for $A^i_0$, which is a Lagrange multiplier for the Gauss constraint, therefore, its conjugate momentum should vanish or be constant as it is a primary constraint (it being zero means the contribution to the Lagrangian, $\chi^i \partial_t A^i_0$, vanishes). Requiring this means that the modifications to the evolution equation for the triad vanishes and the result is the original equation of motion from Ashtekar's formalism. As for the evolution equations for the lapse and shift, these components are pure gauge variables and as such their equations of motion can be freely specified. The above choice is a modification of Harmonic slicing which has no physical meaning and is simply a change in description. Finally, the evolution of the internal gauge vector $\chi^i$ evolves proportional to the Gauss constraint and derivatives of itself. Therefore, if the Gauss constraint and $\chi^i$ are zero initially, then the $\partial_t \chi^i |_{t=0}$ is true also and locally $\chi^i$ will remain constant. For system where the constraint cannot be imposed initially then one needs to check the constraint amplification matrix to make sure that the violating modes are damped. An analysis of this will be performed later once all the equations of motion have been found. Next we consider the equations of motion for the constraints.

    \subsection{3+1 Connection Evolution Equations}
    Having derived the evolution system for the frame sector fields, we can now perform the same for the connection sector. We begin with $\mu,\nu = 0,a$ component of \cref{eq:num-plebanski-einstein-condition-components} to find 
    \begin{align}
     \tE^{i0b} F^i_{ab} + \frac{i}{2} \tg^{00} \left(\partial_t \xi_a + \partial_a \xi_0\right) + \frac{i}{2} \tg^{0b} \left(\partial_a \xi_b + \partial_b \xi_a\right) = 0.
    \end{align}
    Using the decomposition and solving for $(\partial_t - N^b \partial_b) \xi_a$ we find 
    \begin{align}
        (\partial_t - N^b \partial_b) \xi_a = -\partial_a \xi_0 + N^b \partial_a \xi_b - 2i \dlap \tE^{ib} F^i_{ab}.
    \end{align}
    Next we can look at the $\mu,\nu = a,b$ components of the same equation,
    \begin{align}
     \tE^{ia0} F^i_{b0} + \tE^{iac} F^i_{bc} + \frac{i}{2} \tg^{a0} \left(\partial_b \xi_0 + \partial_t \xi_a \right) + \frac{i}{2} \tg^{ac} \left(\partial_b \xi_c + \partial_c \xi_b\right) = 0.
    \end{align}
    Using the decomposition gives 
    \begin{align}
        \tE^{ia} F^i_{0b} + \left(-i \epsilon^{ijk} \tE^{ja} \tE^{kc} + \tE^{ia} N^c - \tE^{ic} N^a\right) F^i_{bc} \nonumber \\ + \frac{i}{2} N^a (\partial_b \xi_0 + \partial_t \xi_b) + \frac{i}{2}(\tE^{ia} \tE^{ic} - N^a N^c)(\partial_b \xi_c + \partial_c \xi_b) = 0.
    \end{align}
    Collecting into useful factors we have
    \begin{align}
        \tE^{ia}( F^i_{0b} + N^c F^i_{bc} - i \epsilon^{ijk} \dlap \tE^{jc} F^k_{bc} + i \dlap \tE^{ic} \partial_{(b} \xi_{c)}) \nonumber \\ + \frac{i}{2} N^a \left( (\partial_t - N^c \partial_c) \xi_b + \partial_b \xi_0 - N^c \partial_b \xi_c - 2i \dlap \tE^{ic} F^i_{bc} \right) = 0.
    \end{align}
    The term in the brackets on the second line is zero via the evolution equation for $\xi_a$, and multiplying by $\utE^i_a/\dlap$ we find 
    \begin{align}
        F^i_{0a} = - N^b F^i_{ab} + i\epsilon^{ijk} \dlap \tE^{jb} F^k_{ab} - i \dlap \tE^{ib} \partial_{(a} \xi_{b)}.
    \end{align}
    To put this in the form of an evolution equation we decompose the first two curvatures appearing using $F^i_{0a} = \partial_t A^i_a - \partial_a A^i_0 + \epsilon^{ijk} A^j_0 A^k_a = \partial_t A^i_a - \partial^A_a A^i_0$ and similarly $F^i_{ab} = \partial^A_a A^i_b - \partial_b A^i_a$ such that 
    \begin{align}
        (\partial_t - N^b \partial_b) A^i_a = \partial^A_a A^i_0 - N^b \partial^A_a A^i_b + i \epsilon^{ijk} \dlap \tE^{jb} F^k_{ab} - i \dlap \tE^{ib} \partial_{(a} \xi_{b)}.
    \end{align}
    The only component remaining is the $\mu,\nu = 0,0$ components, for which we find 
    \begin{align}
        \dlap \tE^{i0a} F^i_{0a} + i \dlap \tg^{00} \partial_t \xi_0 + \frac{i}{2} \dlap \tg^{0a} (\partial_t \xi_a + \partial_a \xi_0) = 0
    \end{align}
    and with the decomposition becomes 
    \begin{align}
        \dlap \tE^{ia} F^i_{0a} - i (\partial_t - N^a \partial_a) \xi_0 + \frac{i}{2} N^a ( \partial_t \xi_a - \partial_a \xi_0) = 0.
    \end{align}
    Rearranging this for $(\partial_t - N^a \partial_a) \xi_0$ reveals
    \begin{align}
        (\partial_t - N^a \partial_a)\xi_0 = - N^a \partial_a \xi_0 - (\dlap^2 \tE^{ia} \tE^{ib} - N^a N^b) \partial_a \xi_b \nonumber \\ -2i \dlap N^a \tE^{ib} F^i_{ab} + \epsilon^{ijk} \dlap^2 \tE^{ia} \tE^{jb} F^k_{ab}.
    \end{align}

    The final evolution equation is in the Lorenz condition, \cref{eq:num-plebanski-lorenz-gauge-components}. Expanding this into 3+1 components gives 
    \begin{align}
        \tg^{00} \partial_t A^i_0 + \tg^{0a} (\partial_t A^i_a + \partial_a A^i_0) + \tg^{ab} \partial_a A^i_b + \frac{i}{2} \tE^{i0a} ( \partial_t \xi_a - \partial_a \xi_0) + \frac{i}{2} \tE^{iab} \partial_a \xi_b = 0.
    \end{align}
    Using the decomposition and collecting terms we find 
    \begin{align}
        -\partial_t A^i_0 + N^a (\partial_t A^i_a + \partial_a A^i_0) + \dlap^2 \tE^{ja} \tE^{jb} \partial_a A^i_b + \frac{i}{2} \tE^{ia}( \partial_t \xi_a - \partial_a \xi_0) \nonumber \\ + \frac{1}{2} \epsilon^{ijk} \dlap \tE^{ja} \tE^{kb} \partial_a \xi_b + \frac{i}{2} \tE^{ia} N^b \partial_a \xi_b + \frac{i}{2} \tE^{ib} N^a \partial_b \xi_a = 0.
    \end{align}
    Substituting the previous evolution equations for $A^i_a$ and $\xi_a$ and solving for $(\partial_t -N^a \partial_a) A^i_0$ gives our final connection evolution equation 
    \begin{align}
        (\partial_t - N^a \partial_a) A^i_0 = & \ (\dlap^2 \tE^{ja} \tE^{jb} - N^a N^b) \partial_a A^i_b + N^a \partial^A_a A^i_0 - i \epsilon^{ijk} \dlap \tE^{ja} N^b F^k_{ab} \nonumber \\ &\ + \dlap^2 \tE^{ia} \tE^{jb} F^j_{ab} - i \dlap \tE^{ia} \partial_a \xi_0 - i N^a \dlap \tE^{ib} \partial_{(a} \xi_{b)} \nonumber \\ &\  + i N^a \dlap \tE^{ib} \partial_b \xi_a + \frac{1}{2} \epsilon^{ijk} \dlap^2 \tE^{ja} \tE^{kb} \partial_a \xi_b
    \end{align}

    The result of this decomposition is an evolution equation for the self-dual connection and the connection 1-form
    \begin{equation}
    \begin{aligned}
        (\partial_t - N^b \partial_b) A^i_a & = \partial^A_a A^i_0 - N^b \partial^A_a A^i_b + i \epsilon^{ijk} \dlap \tE^{jb} F^k_{ab} - i \dlap \tE^{ib} \partial_{(a} \xi_{b)} \\
        (\partial_t - N^a \partial_a) A^i_0 & = \ (\dlap^2 \tE^{ja} \tE^{jb} - N^a N^b) \partial_a A^i_b + N^a \partial^A_a A^i_0 - i \epsilon^{ijk} \dlap \tE^{ja} N^b F^k_{ab} \\ &\ + \dlap^2 \tE^{ia} \tE^{jb} F^j_{ab} - i \dlap \tE^{ia} \partial_a \xi_0 - i N^a \dlap \tE^{ib} \partial_{(a} \xi_{b)} \\ &\  + i N^a \dlap \tE^{ib} \partial_b \xi_a + \frac{1}{2} \epsilon^{ijk} \dlap^2 \tE^{ja} \tE^{kb} \partial_a \xi_b \\
        (\partial_t - N^a \partial_a)\xi_0 & = - N^a \partial_a \xi_0 - (\dlap^2 \tE^{ia} \tE^{ib} - N^a N^b) \partial_a \xi_b \\ &\  -2i \dlap N^a \tE^{ib} F^i_{ab} + \epsilon^{ijk} \dlap^2 \tE^{ia} \tE^{jb} F^k_{ab} \\
        (\partial_t - N^b \partial_b)\xi_a & = 2i \dlap \tE^{ib} F^i_{ba} - \partial_a \xi_0 + N^b \partial_a \xi_b
    \end{aligned}
    \end{equation}
    As with the evolution of the frame we want for this system to describe original Ashtekar formulation of Einstein's equations, but now the focus is on the evolution equation for the connection. The system before gauge fixing had 4 constraints which were found to be the Hamiltonian and Momentum constraints
        \begin{align}
        \mathcal{H} = \epsilon^{ijk} \tE^{ia} \tE^{jb} F^k_{ab}, \quad \mathcal{M}_a = \tE^{ib} F^i_{ba}.
    \end{align}
    The components of the 1-form, $\xi_0$ and $\xi_a$, appear as conjugate momentum for the lapse and shift Lagrange multipliers and can thus be taken as primary constraints that should vanish. The equation of motion for the spatial connection is only modified by the derivative of the primary constraints $\xi_a$ and as such it recovers the original dynamics on the constraint surface. The equation of motion for $A^i_0$ is pure gauge, due to $A^i_0$ being a Lagrange multiplier for the Gauss constraint, and the evolution simply changes the gauge orbit of the full connection $A^i_\mu$ without changing the physical solution. Finally, for $\xi_0$ and $\xi_a$, they evolve proportional to constraints and their derivatives. The same argument that was applied to the frame evolution system can be applied here, since the modification to the equations of motion are either pure gauge or are proportional to constraints the dynamics are unchanged on the constraint surface. If one prepares the system with all the constraints satisfied initially we see that their time derivatives are also satisfied and so should remain zero throughout evolution. It remains to be checked that this system is attractive towards the constraint surface during evolution.

    \section{Gauge Fixed Evolution System} \label{sec:pleb-num-gauge-fixed-evolution-system}
    Collecting both the frame and connection evolution equations we find our full evolution system that corresponds to the hyperbolic gauge fixing 
    \begin{align}
    (\partial^A_t - N^a \partial^A_a) \tE^{ia} & = -i \epsilon^{ijk} \tE^{kb} \partial_b^A \left(\dlap \tE^{ja}\right) - \tE^{ib} \partial_b N^a + \tE^{ia} \partial_b N^b + i(c_1 + c_2) \dlap \tE^{ia} \tE^{jb} \partial_b \chi^j \nonumber\\ + & \frac{c_2-c_1}{2 i c_2}\left( \epsilon^{ijk} \dlap \tE^{ja} \partial_b^A \tE^{kb} + (c_1+c_2) \dlap \tE^{ja} \tE^{ib} \partial_b \chi^j + (3c_2 + c_1) \dlap \tE^{ja} \tE^{jb} \partial_b \chi^i \right). \\
    (\partial_t - N^b \partial_b) N^a & = - \dlap \tE^{ib} \partial_b^A( \dlap \tE^{ia}) + (c_1 - c_2) \epsilon^{ijk} \dlap \tE^{ja} \tE^{kb} \partial_b \chi^i + \dlap^2 \tE^{ia} \tE^{ib} \xi_b  \\
    (\partial_t - N^a \partial_a)\dlap & = - \dlap \partial_a N^a + 2ic_2 \dlap^2 \tE^{ia} \partial_a \chi^i - \dlap ( \xi_0 - N^a \xi_a) \\
    (\partial_t - N^a \partial_a) \chi^i & = \frac{1}{2 i c_2} \dlap \partial^A_a \tE^{ia} + \frac{1}{2 i c_2} (c_1 + c_2) \epsilon^{ijk} \dlap \tE^{ja} \partial_a \chi^k. \\[20pt]
    (\partial_t - N^a \partial_a)\xi_0 & = - N^a \partial_a \xi_0 - (\dlap^2 \tE^{ia} \tE^{ib} - N^a N^b) \partial_a \xi_b \nonumber \\ &\  -2i \dlap N^a \tE^{ib} F^i_{ab} + \epsilon^{ijk} \dlap^2 \tE^{ia} \tE^{jb} F^k_{ab} \label{eq:num-plebanski-xi0-evolution-equation} \\
    (\partial_t - N^b \partial_b)\xi_a & = 2i \dlap \tE^{ib} F^i_{ba} - \partial_a \xi_0 + N^b \partial_a \xi_b \label{eq:num-plebanski-xia-evolution-equation}  \\
    (\partial_t - N^b \partial_b) A^i_a & = \partial^A_a A^i_0 - N^b \partial^A_a A^i_b + i \epsilon^{ijk} \dlap \tE^{jb} F^k_{ab} - i \dlap \tE^{ib} \partial_{(a} \xi_{b)} \label{eq:num-plebanski-Aia-evolution-equation}  \\
    (\partial_t - N^a \partial_a) A^i_0 & = \ (\dlap^2 \tE^{ja} \tE^{jb} - N^a N^b) \partial_a A^i_b + N^a \partial^A_a A^i_0 - i \epsilon^{ijk} \dlap \tE^{ja} N^b F^k_{ab} \nonumber \\ &\ + \dlap^2 \tE^{ia} \tE^{jb} F^j_{ab} - i \dlap \tE^{ia} \partial_a \xi_0 - i N^a \dlap \tE^{ib} \partial_{(a} \xi_{b)} \nonumber \\ &\  + i N^a \dlap \tE^{ib} \partial_b \xi_a + \frac{1}{2} \epsilon^{ijk} \dlap^2 \tE^{ja} \tE^{kb} \partial_a \xi_b. \label{eq:num-plebanski-Ai0-evolution-equation}
    \end{align}

    To check that this system is strongly hyperbolic one has to diagonalise the characteristic matrix and check that the eigenvalues are real. As this is a lengthy calculation that does not reveal any insight to the equations, we simply state that the characteristic matrix is strongly hyperbolic. The eigenvalues all take values in $- N^a k_a \pm \sqrt{k^2}$ where $k^2 = k_a k^a$ and $k_a$ is the arbitrary wave vector used to construct $P = M^a k_a$. This means that none of the fields have zero propagation speed modes, in~\cite{Towards_an_unde_Alcubi_1999} this was argued to be a useful property for avoiding instabilities. The strong hyperbolicity also means that we can trust our evolution system to be well-posed, at least locally. The strong hyperbolicity of this system is somewhat expected, and as we have seen in \cref{chap:nonlinear-gauge-fixing} this system obeys the wave equation at second order which is strongly hyperbolic.

    \subsection{Constraint Amplification Analysis}
    Now we check that our evolution system is indeed attractive towards the constraint surface, this is so that any constraint violation is quickly damped and removed rather than growing. To check we expand around the Minkowski background and perform a Fourier transform, as shown in \cref{sec:num-asympytotically-constrained-system}, the eigenvalues of the amplification matrix will indicate whether the constraint will grow or shrink. On the Minkowski background the fields have components $\dlap = 1, N^a = 0, \tE^{ia} = \delta^{ia}, A^i_\mu =0, \xi_\mu = 0$ and $\chi^i = 0$. Around this background the equation of motion for $\tE^{ia}$ becomes 
    \begin{align}
        \partial_t \tE^{ia} = -i \epsilon^{ijk} \partial^k \tE^{ja} -i\epsilon^{iak} \partial^k \dlap - \partial^i N^a + \delta^{ia} \partial_b N^b + i(c_1 +c_2) \delta^{ia} \partial^j \chi^j + \nonumber \\ \frac{c_2 - c_1}{2ic_2} \left( \epsilon^{iak} \mathcal{G}^k + (c_1+c_2) \partial^i \chi^a + (3c_2 + c_1) \partial^a \chi^i \right)
    \end{align}
    where we have used $\tE^{ia} = \delta^{ia}$ on the background to convert indices between internal and spatial. With this we can compute the time derivative of the linearised Gauss constraint 
    \begin{align}
        \partial_t \mathcal{G}^i = \frac{c_1 + c_2}{2ic_2} \epsilon^{ijk} \partial^k \mathcal{G}^j - \frac{(c_1+c_2)^2}{2ic_2} \partial^i \partial^j \chi^j + \frac{(c_2-c_1)(3c_2 + c_1)}{2ic_2} \partial^j \partial^j \chi^i.
    \end{align}
    The linearised equation for the internal vector, $\chi^i$, is then 
    \begin{align}
        \partial_t \chi^i = \frac{1}{2ic_2} \mathcal{G}^i - \frac{c_1+c_2}{2ic_2} \epsilon^{ijk} \partial^k \chi^j.
    \end{align}
    We see that $(\chi^i,\mathcal{G}^i)$ form a closed system. Taking the Fourier transform of these equations and writing them in matrix form gives us 
    \begin{align}
        \partial_t \binom{\hat{\mathcal{G}}^i}{\hat{\chi}^i} = \begin{pmatrix}
            \frac{c_1 + c_2}{2ic_2} \epsilon^{ijk} k^k && \frac{(c_1+c_2)^2}{2ic_2} k^i k^j - \frac{(c_2 - c_1)(3c_2+c_1)}{2ic_2} k^2 \delta^{ij} \\ \frac{1}{2ic_2} \delta^{ij} && -\frac{c_1+c_2}{2c_2} \epsilon^{ijk} k^k
        \end{pmatrix} \binom{\hat{\mathcal{G}}^i}{\hat{\chi}^i}.
    \end{align}
    Where $k^i$ is the wave vector for the Fourier transformation and $k^2 = k^i k^i$. Computing the eigenvalues for the characteristic matrix we find 
    \begin{align}
        \lambda_\pm = \pm \sqrt{\frac{k^2}{2}} \sqrt{1-2\frac{c_1}{c_2}-\left(\frac{c_1}{c_2}\right)^2}
    \end{align}
    with multiplicity 3 for each sign. The condition for stability is that all the eigenvalues should have a nonpositive real part for any $k^i$. For this to be the case we must require that $c_1 = \alpha c_2$ where $\alpha \leq -1 -\sqrt{2}$ or $\alpha \geq -1 + \sqrt{2}$. In which case
    \begin{align}
        1-2\alpha-\alpha^2 \leq 0 \label{eq:pleb-num-alpha-constraint-range}
    \end{align}
    and hence the eigenvalues are purely imaginary. On the connection side of the evolution system we can compute the equations of motion for the Hamiltonian and Momentum constraints, skipping the details, to find 
    \begin{align}
        \partial_t \mathcal{H} & = 2i \partial^i \mathcal{M}^i, \\
        \partial_t \mathcal{M}^i & = \frac{i}{2} \partial^i \mathcal{H} + \frac{i}{2} \partial^i \partial^j \xi^j - \frac{i}{2} \partial^j \partial^j \xi^i.
    \end{align}
    The linearised equations of motion for $\xi_0$ and $\xi^i$ are 
    \begin{align}
        \partial_t \xi_0 = \mathcal{H} - \partial^i \xi^i \\
        \partial_t \xi^i = 2i \mathcal{M}^i - \partial^i \xi_0.
    \end{align}
    Then by taking the Fourier transform and taking the solution vector to be $u = (\hat{\xi_0},\hat{\xi^i},\hat{\mathcal{H}},\hat{\mathcal{M}}^i)$ the amplification matrix is 
    \begin{align}
        \begin{pmatrix}
            0 & -i k^j & 1 & 0 \\ -ik^i & 0 & 0 & 2i \delta^{ij} \\ 0 & 0 & 0 & -2k^j \\ 0 & \frac{i}{2} k^2 \delta^{ij} - \frac{i}{2} k^i k^j & -\frac{k^i}{2} & 0 
        \end{pmatrix}.
    \end{align}
    The eigenvalues of the amplification matrix are 
    \begin{align}
        \pm \sqrt{k^2}, \quad \pm i\sqrt{k^2}
    \end{align}
    with multiplicity $2$ for the first and $4$ for the second. Only the real part of one of the eigenvalues is positive, while this means that there exists a constraint mode that will grow during evolution it says nothing about the rate of growth and the simulation could still converge before the constraint grows to overthrow the true solution. To check this one would need to perform numerical simulations.

    \subsection{Constraint Damping}
    So far we have seen that the eigenvalues for the constraints are purely imaginary, this implies that the constraints are propagated instead of growing or damped. To induce damping we can add linear terms involving the momentum that are conjugate to the Lagrange multipliers, i.e. $\chi^i, \xi_0$ and $\xi_i$, to their respective equations of motion. One of the simpler modifications that can introduce damping is
    \begin{equation} \label{eq:num-damped-lambda-system}
        \begin{aligned}
            (\partial_t -N^a \partial_a) \chi^i & = \frac{1}{2ic_2} \dlap \partial_a^A \tE^{ia} + \frac{c_1 + c_2}{2ic_2} \epsilon^{ijk} \dlap \tE^{ja} \partial_a \chi^k - \beta \chi^i \\
            (\partial_t - N^a \partial_a)\xi_0 = & 2i \dlap N^a \tE^{ib}  F^i_{ba} + \epsilon^{ijk} \dlap^2 \tE^{ia} \tE^{jb} F^k_{ab} - (\dlap^2 \tE^{ia} \tE^{ib} - N^a N^b) \partial_a \xi_b \\&\  - N^a \partial_a \xi_0 - \rho \xi_0\\
        (\partial_t - N^b \partial_b)\xi_a = & 2i \dlap \tE^{ib} F^i_{ba} - \partial_a \xi_0 + N^b \partial_a \xi_b - \sigma \xi_a\\
        \end{aligned}
    \end{equation}
    where $\beta,\rho,\sigma$ are damping coefficients. This modification is allowed as $\chi^i, \xi_0$ and $\xi_a$ are all weakly zero and during the evolution they should converge to $0$. The eigenvalues for the $(\chi^i,\mathcal{G}^i)$ amplification matrix becomes 
    \begin{align}
        \lambda_\pm & = -\frac{\beta}{2} \pm \frac{1}{2} \sqrt{\beta^2 + 2(1-2\alpha-\alpha^2)k^2} \\
        \lambda_\pm & = -\frac{\beta}{2} \pm \frac{1}{2} \sqrt{\beta^2 + 2i\beta(\alpha+1)\sqrt{k^2} + 2(1-2\alpha-\alpha^2)k^2} \\
        \lambda_\pm & = -\frac{\beta}{2} \pm \frac{1}{2} \sqrt{\beta^2 - 2i\beta(\alpha+1)\sqrt{k^2} + 2(1-2\alpha-\alpha^2)k^2}
    \end{align}
    For the real part of these eigenvalues to be nonpositive we notice that when $\alpha=1$ they reduce to 
    \begin{align}
        \lambda_\pm & = -\frac{\beta}{2} \pm \frac{1}{2} \sqrt{\beta^2 -4k^2} \\
        \lambda_\pm & = -\frac{\beta}{2} \pm \frac{1}{2} (\beta + 2i\sqrt{k^2}) \\
        \lambda_\pm & = -\frac{\beta}{2} \pm \frac{1}{2} (\beta - 2i\sqrt{k^2}).
    \end{align}
    Then choosing $\beta \geq 0$ we find that 4 of the eigenvalues have negative real parts and 2 have zero. Doing the same calculation for the connection amplification matrix we find the eigenvalues to be 
    \begin{align}
        \lambda_\pm & = -\frac{\rho+\sigma}{2} \pm \frac{1}{2} \sqrt{(\rho-\sigma)^2 -4k^2} \\
        \lambda_\pm & = \pm \sqrt{k^2} \\
        \lambda_\pm & = -\frac{\sigma}{2} \pm \frac{1}{2} \sqrt{\sigma^2 - 4k^2}.
    \end{align}
    The last eigenvalue requires $\sigma \geq 0$ for no growth to occur, and for the first eigenvalue we need $\rho > 0$ to ensure the real part is $\leq 0$. However, the damping has not remedied the middle eigenvalue for which the positive sign still produces a growing eigenvalue. When $\beta,\rho,\sigma$ are all greater than zero then some constraint modes should be damped, all but one of the modes of the constraint will be damped. The presence of this positive eigenvalue suggests that there is perhaps a modification to the gauge fixing that will remove this positive eigenvalue and perform better numerically. Since the search space for such a gauge fixing is effectively infinite we do not consider such modifications. It is still possible that the system is well-behaved numerically as it could be that the corresponding components are never active. To check this more thoroughly one would need to perform numerical test.

    The choice $\alpha=1$ implies that $c_1 = c_2 = c$ for some non-zero constant $c$, we will see in the next section that this choice also leads to the development of a new evolution system with simplified algebraic properties and that makes use of the conformal separation property that was noted at the linear level.

    \section{Conformal Evolution System} \label{sec:pleb-num-conformal-evolution-systems}

    In the linear analysis of the hyperbolic gauge fixing, \cref{chap:Linearised-Gravity}, we saw that the fields separated conformally into sectors of dimension 12 and 4. This surprising property allowed one to ignore the $4$ component sector as the physical content was contained in the $12$ dimensional part. A similar story occurred for the nonlinear gauge fixing in \cref{chap:nonlinear-gauge-fixing} where the 12 dimensional system could be defined independently of the 4, however, the 4 dimensional system was still dependent on the 12. As the 12 contains the physical degrees of freedom this is not a problem and one can still ignore the $4$ dimensional system. The equations of motion for the 12 component connection involved combinations of the original connections and frame variables, this mixing of sectors introduces more complicated dynamics for the constraint propagation, for this reason we also consider a partially conformal system where only the frame variables are separated into their conformal parts. The partially conformal system offers some economy over the system in the previous section which we explore.

    \subsection{Conformal Frame Evolution System}
    We begin by finding the nonlinear analogues of the linearised separated variables. The metric linearises to $(\delta g)_{\mu\nu} = 2 h_{\mu\nu} = 2 \hat{h}_{\mu\nu} + \frac{1}{2} h g_{\mu\nu}$, where $(\delta e)^I{}_\mu = h^I{}_\mu$. From this we wish to separate $\hat{h}_{\mu\nu}$ and $h$, that is the tracefree and trace parts. For the trace part we can make use of a formula from linear algebra
    \begin{align}
        \delta \sqrt{-g} = \sqrt{-g} g^{\mu\nu} h_{\mu\nu} = h
    \end{align}
    where we have used that $\sqrt{-g}=1$ on the background. It shows that the nonlinear version of $h$ is then $\sqrt{-g}$, in practice we can use any function of $\sqrt{-g}$ without disrupting the splitting. The tracefree part can then be found by multiplying the densitiesed inverse metric (we choose the densitised inverse metric as it is more conveniently written using the components of $\tE^{i\mu\nu}$) by the appropriate power of $\sqrt{-g}$, that is
    \begin{align}
        \delta( \sqrt{-g}^n \tg^{\mu\nu} ) = \delta( \sqrt{-g}^{n+1} g^{\mu\nu} ) = (n+1) g^{\mu\nu} h - 2 h^{\mu\nu} \nonumber \\ \quad = (n+1) g^{\mu\nu} h - \frac{1}{2} g^{\mu\nu} h - \hat{h}^{\mu\nu} = \left( n + \frac{1}{2} \right) g^{\mu\nu} h - \hat{h}^{\mu\nu}.
    \end{align}
    We can see that to remove $h$ and leave $\hat{h}^{\mu\nu}$ as the only irreducible component we can pick $n = -\frac{1}{2}$. In which case the nonlinear analogue of $\hat{h}^{\mu\nu}$ is $\frac{1}{\sqrt{\sqrt{-g}}} \tg^{\mu\nu}$ which matches the conformal metric appearing in \cref{eq:nonlinear-conformal-metric}. Moving to the 3+1 components we recall that $\sqrt{-g} = \dlap \det(\tE)$, and we can see that 
    \begin{align}
        \frac{1}{\sqrt{\sqrt{-g}}} \tg^{\mu\nu} = \frac{1}{\sqrt{\dlap \det(\tE)}} \begin{pmatrix}
            -\frac{1}{\dlap} && \frac{N^a}{\dlap} \\ \frac{N^b}{\dlap} && \dlap \tE^{ia} \tE^{ib} - \frac{N^a N^b}{\dlap}
        \end{pmatrix} \nonumber \\ = \frac{1}{\sqrt{\dlap^3 \det(\tE) }} \begin{pmatrix}
            -1 && N^a \\ N^b && \dlap^2 \tE^{ia} \tE^{ib} - N^a N^b
        \end{pmatrix}.
    \end{align}
    If we introduce a new conformal triad $H^{ia} = \dlap \tE^{ia}$ then the above simplifies slightly 
    \begin{align}
        \frac{1}{\sqrt{\sqrt{-g}}} \tg^{\mu\nu} = \frac{1}{\sqrt{\det(H)}} \begin{pmatrix}
            -1 && N^a \\ N^b && H^{ia} H^{ib} - N^a N^b
        \end{pmatrix}.
    \end{align}
    Which shows that the conformal metric is a function of $\hat{\gamma}^{ab} = H^{ia} H^{ib}$ and $N^a$. These are $6+3=9$ components which matches the number of components $\hat{h}_{\mu\nu}$. As for the $SO(3,\C)$ frame perturbation, $h^i$, this only possible to write down in relation to another frame, as the nonlinear version of $h^i$ would be an $SO(3,\C)$ transformation. Instead, we know that it will exist in the choice of frame $H^{ia}$ as this contains $6+3 = 9$ components where the $6$ count the components of the spatial metric and the remaining $3$ belong to the Euler angles needed to rotate to the chosen frame. Therefore, we can say that ($H^{ia},N^a$) contains the nonlinear fields that correspond to the linear components ($h^i, \hat{h}_{\mu\nu}$) which parametrises the 12 components sector. As for the remaining $4$ components we know that they will be functions of all the fields, as $h = \delta\sqrt{-g} = \delta(\dlap \det(\tE))$ and $\hat{\chi}^i = h^i - (c_1+c_2) \chi^i$, so instead of creating the nonlinear combination we find the resulting equations simpler when leaving $\dlap$ and $\chi^i$ separate. We can now write the frame sector, \cref{eq:num-plebanski-triad-evolution-equation,eq:num-plebanski-shift-evolution-equation,eq:num-plebanski-lapse-evolution-equation,eq:num-plebanski-internal-vector-evolution}, using the conformal triad $H^{ia}$. We start with evolution of the conformal triad and find that it is

    \begin{align}
        (\partial^A_t - N^b \partial^A_b) H^{ia} = & -i \epsilon^{ijk} H^{kb} \partial_b^A H^{ja} - H^{ib} \partial_b N^a - H^{ia} (\xi_0 - N^b \xi_b) \nonumber\\  & \quad + \frac{i}{2c_2}(c_1-c_2) \left( \epsilon^{ijk} H^{ja} \partial^A_b H^{kb} - \epsilon^{ijk} H^{ja} H^{kb} \partial_b n  + (c_1+c_2) H^{ja} H^{ib} \partial_b \chi^j \right. \nonumber\\ & \left. \hspace{100pt} + (3c_2 + c_1) H^{ja} H^{jb} \partial_b \chi^i + 2c_2 H^{ia} H^{jb} \partial_b \chi^j  \right).
    \end{align}
    Here we have introduced a new variable $n = \log(\dlap)$ with which it will be useful to write some equations of motion. The equation for $H^{ia}$ also contains $\chi^i$ components, and from the analysis in \cref{chap:Linearised-Gravity} we know that the linearised $12$ sector is parameterised by ($\hat{h}_{\mu\nu}, \hat{h}^i$), with, $\hat{h}^i = h^i - (c_1 - c_2) \chi^i$. This means that, in general, we will have to form a nonlinear combination out of $\chi^i$, $H^{ia}$ and $N^a$ in order to find the splitting. Instead of constructing more complicated combinations of these fields we choose an easier tactic, by fixing $c_2 = c_1 = c$ we see that $\hat{h}^i = h^i$ and the nonlinear evolution equation for $H^{ia}$ loses all of its terms containing $\chi^i$. This choice of $c_1,c_2$ implies that $\alpha = 1$ which is within the correct range for the constraints to not exponentially grow as per \cref{eq:pleb-num-alpha-constraint-range}, this value also agrees with the damped eigenvalue analysis. The evolution equation for the conformal triad is then
    \begin{align}
        (\partial^A_t - N^b \partial^A_b) H^{ia} = -i \epsilon^{ijk} H^{kb} \partial^A_b H^{ja} - H^{ib} \partial_b N^a - H^{ia} (\xi_0 - N^b \xi_b).
    \end{align}
    As for the shift evolution equation, when $c_2 = c_1 = c$ we find 
    \begin{align}
        (\partial_t - N^b \partial_b) N^a = - H^{ib} \partial_b^A H^{ia} + H^{ia} H^{ib} \xi_b.
    \end{align}
    It is clear then that the only frame variables required for the evolution of ($H^{ia},N^a$) are themselves. The other $4$ equations of motion, when using $H^{ia}$ become 
    \begin{align}
        (\partial_t - N^a \partial_a) n &= - \partial_a N^a - 2ic H^{ia} \partial_a \chi^i - \xi_0 + N^a \xi_a \\
        (\partial_t - N^a \partial_a) \chi^i &= -\frac{i}{2 c} \partial^A_a H^{ia} + \frac{i}{2c} H^{ia} \partial_a n - i \epsilon^{ijk} H^{ja} \partial_a \chi^k.
    \end{align}
    At the nonlinear level we do not see a separation of the two sectors entirely, the $4$ component sector depends on the $12$ but the $12$ is still independent.

    \paragraph{Conformal Frame Evolution System} With the conformal frame, $H^{ia}$, we find that the system simplifies greatly if supported with the correct choice of $c_1,c_2$. We call the choice of $c_1 = c_2 = c$ our conformal evolution system, for which the evolution of the frame variables is.
    \begin{equation}
        \begin{aligned}
            (\partial^A_t - N^b \partial^A_b) H^{ia} &= -i \epsilon^{ijk} H^{kb} \partial^A_b H^{ja} - H^{ib} \partial_b N^a - H^{ia} (\xi_0 - N^b \xi_b) \\
            (\partial_t - N^b \partial_b) N^a &= - H^{ib} \partial_b^A H^{ia} + H^{ia} H^{ib} \xi_b \\
            (\partial_t - N^a \partial_a) n &= - \partial_a N^a - 2ic H^{ia} \partial_a \chi^i - \xi_0 + N^a \xi_a \\
            (\partial_t - N^a \partial_a) \chi^i &= -\frac{i}{2 c} \partial^A_a H^{ia} + \frac{i}{2c} H^{ia} \partial_a n - i \epsilon^{ijk} H^{ja} \partial_a \chi^k.
        \end{aligned}\label{eq:num-plebanski-conformally-seperated-frame-evolution-system}
    \end{equation}

    Some discussion of the hyperbolicity of this system, as compared to the original system, is needed. We see that $c$ is now a rescaling parameter for $\chi^i$, fixing it to a specific value does not change the hyperbolicity. Choosing $c = \frac{i}{2}$ we can write the characteristic matrix in block matrix form, choosing the solution vector to be $((H^{ia},N^a),(\chi^i,n))$ we find
    \begin{align}
        P(n) = \begin{pmatrix}
            A(n)^I{}_J \otimes \delta^a_b && 0 \\ -\delta^I_J n_b && -A(n)^I{}_J
        \end{pmatrix}
    \end{align}
    where $A^I{}_J(n)$ acts on objects of the type $(X^i,X)$ with $4$ components. Checking the structure of $A^I{}_J$ we see that is Hermitian,
    \begin{align}
        A^I{}_J = \begin{pmatrix}
            -i\epsilon^{ijk} n^k && - n^i \\ -n^j && 0
        \end{pmatrix}
    \end{align}
    where $n^i = H^{ia} n_a$. This means that $A^I{}_J$ has a full set of eigenvectors $v_A$, and corresponding eigenvalues $\lambda_A = \pm \sqrt{n^2}$, that span the solution space. Due to the reality conditions the eigenvalues are real-valued and as such $A$ is strongly hyperbolic. We can then try to construct eigenvectors for $P(n)$, which amounts to solving 
    \begin{align}
        \begin{pmatrix}
            A(n)^I{}_J \otimes \delta^a_b && 0 \\ -\delta^I_J n_b && -A(n)^I{}_J
        \end{pmatrix} \begin{pmatrix}
            u^{Jb} \\ v^J
        \end{pmatrix} = \lambda \begin{pmatrix}
            u^{Jb} \\ v^J
        \end{pmatrix}
    \end{align}
    or as a system of equations 
    \begin{align}
        A^I{}_J u^{Ja} = \lambda u^{Ia} \\
        -n_a u^{Ia} - A^I{}_J v^J = \lambda v^I.
    \end{align}
    If $u^{Ia} = 0$ then we find there are 4 eigenvectors inherited from $A$,
    \begin{align}
        \begin{pmatrix}
            0 \\ v_A^I
        \end{pmatrix}
    \end{align}
    with eigenvalues $\lambda_A$. If $u^{Ia} = 0$ then we see that $u^{Ja} = v_A^J u^a$ and $\lambda = \lambda_A$ for some spatial 1-form $u_a$. The second eigenvalue equations will become
    \begin{align}
        -n_a u^a v_A^I - A^I{}_J v^J = \lambda_A v^I
    \end{align}
    To solve this we introduce an ansatz $v^I = \alpha v_A^I$ and see that this equation becomes 
    \begin{align}
        -n_a u^a - \alpha \lambda_A = \alpha \lambda_A, \quad \Rightarrow \alpha = -\frac{n_a u^a}{2 \lambda_A}.
    \end{align}
    Which gives the remaining 3 eigenvectors 
    \begin{align}
        \begin{pmatrix}
            u^a v_A^I \\ -\frac{n_a u^a}{2\lambda_A}
        \end{pmatrix}
    \end{align}
    with eigenvalues $\lambda_A$. We see that $P(n)$ has a full set of eigenvectors and real values and is hence strongly hyperbolic. It is also worth mentioning that the matrix, $A^I{}_J$, also appears in the context of Chiral Yang-Mills. In fact, it is the characteristic matrix there as well. We see that gravity is two copies of Maxwell with some mixing terms, this has the flavour of the double copy structure (e.g. \cref{sec:self-dual-YM-and-gravity}) but the link in its current form is tenuous at best.

    \subsection{Conformal Connection Evolution System}
        For the connection evolution equations we explore two options, one related to the original system and where the constraint analysis directly carries across which we call the partially conformal system. The first system is simply the connection evolution equations using the conformal frame. However, the second system completely decouples the 12 components using the modification to the nonlinear gauge fixing presented in \cref{chap:nonlinear-gauge-fixing}.

        \subsubsection{Partially Conformal System}
        For the partially conformal system we simply replace the frame variables in the equations of motion for the frame with the conformally transformed variables. Interestingly, looking at \cref{eq:num-plebanski-xi0-evolution-equation,eq:num-plebanski-xia-evolution-equation,eq:num-plebanski-Aia-evolution-equation,eq:num-plebanski-Ai0-evolution-equation} we find that the variables $H^{ia}$ appear naturally and one can immediately write this system with only $H^{ia}$ and $N^a$ as the frame variables,
        \begin{equation} \label{eq:num-nonlinear-partial-conformal-connection-system}
        \begin{aligned}
            (\partial_t - N^a \partial_a) \xi_0 & = 2i N^a H^{ib} F^i_{ab} - \epsilon^{ijk} H^{ia} H^{jb} F^k_{ab} - (H^{ia} H^{ib} - N^a N^b) \partial_a \xi_b \\ & \quad\quad  - N^a \partial_a \xi_0 \\
            (\partial_t - N^b \partial_b) \xi_a & = 2i H^{ib} F^i_{ab} - \partial_a \xi_0 + N^b \partial_a \xi_b \\
            (\partial_t - N^b \partial_b) A^i_a & = \partial^A_a A^i_0 - N^b \partial_a^A A^i_b + i \epsilon^{ijk} H^{jb} F^k_{ab} - i H^{ib} \partial_{(a} \xi_{b)} \\
            (\partial_t - N^a \partial_a) A^i_0 & = (H^{ja} H^{jb} - N^a N^b) \partial_a A^i_b + N^a \partial^A_a A^i_0 - i \epsilon^{ijk} H^{ja} N^b F^k_{ab} \\ & \quad\quad -H^{ia} H^{jb} F^j_{ab} - i H^{ia} \partial_a \xi_0 + i N^a H^{ib} \partial_{(a} \xi_{b)} \\  & \quad\quad +i N^a H^{ib} \partial_b \xi_a + \frac{1}{2} \epsilon^{ijk} H^{ja} H^{kb} \partial_a \xi_b
        \end{aligned} 
        \end{equation}

        This along with \cref{eq:num-plebanski-conformally-seperated-frame-evolution-system} constitutes the partially conformal evolution system. As compared to the system involving $\tE^{ia}$, this system has a lower polynomial order meaning that the algebraic structure is slightly simpler. Since it is simply a change of variables, the hyperbolicity and constraint analysis give the same results.

        \subsubsection{Full Conformal System}
        Up to the addition of constraint terms, the gauge fixed Einstein's equations can be placed into an evolution system involving only the modified connection, the covariant form of these equations is shown in \cref{eq:nonlinear-plebanski-conformal-1-of-4,eq:nonlinear-plebanski-conformal-3-of-4,eq:nonlinear-plebanski-conformal-12}. We decompose these evolution equations with respect to the 3+1 coordinates and obtain a new set of evolution equations for gravity. 
        
        First we compare the new connections to the original ones, the covariant definitions of the new connections in terms of the originals are
        \begin{align}
            \Omega^i_\mu = A^i_\mu - \frac{1}{2} \Sigma^i_\mu{}^\nu \xi_\nu, \quad \omega_\mu = \xi_\mu + 2 \Sigma^i_\mu{}^\nu A^i_\nu.
        \end{align}
        In the 3+1 coordinates, we can make use of \cref{eq:num-conformal-2-form-1+3-decomposition}, for which we find the above definitions become
        \begin{align}
            \Omega^i_0 & = A^i_0 - \frac{i}{2\dlap} \utE^i_a N^a (\xi_0 - N^b \xi_b) - \frac{i}{2} \dlap \tE^{ia} \xi_a + \frac{1}{2} \epsilon^{ijk} \utE^j_a N^b \tE^{ka} \xi_a \nonumber \\ & = A^i_0 - \frac{i}{2} H^i_a N^a ( \xi_0 - N^b \xi_b) - \frac{i}{2} H^{ia} \xi_a + \frac{1}{2} \epsilon^{ijk} H^j_a N^a H^{kb} \xi_b \\
            \Omega^i_a & = A^i_a - \frac{i}{2\dlap} \utE^i_a (\xi_0 - N^b \xi_b) + \frac{1}{2} \epsilon^{ijk} \utE^j_a \tE^{kb} \xi_b \nonumber \\ & = A^i_a - \frac{i}{2} H^i_a (\xi_0 - N^b \xi_b) + \frac{1}{2} \epsilon^{ijk} H^j_a H^{kb} \xi_b
        \end{align}
        and
        \begin{align}
            \omega_0 & = \xi_0 + \frac{2i}{\dlap} \utE^i_a N^a (A^i_0 - N^b A^i_b) + 2i\dlap \tE^{ia} A^i_a - 2 \epsilon^{ijk} \utE^j_b N^b \tE^{kb} A^i_b \nonumber \\ & = \xi_0 + 2iH^i_a N^a (A^i_0 - N^b A^i_b) + 2iH^{ia} A^i_a - 2 \epsilon^{ijk} H^j_b N^b H^{ka} A^i_a \\
            \omega_a & = \xi_a + \frac{2i}{\dlap} \utE^i_a (A^i_0 - N^b A^i_b) - 2 \epsilon^{ijk} \utE^j_a \tE^{kb} A^i_b \nonumber \\ & = \xi_a + 2i H^i_a (A^i_0 - N^b A^i_b) - 2 \epsilon^{ijk} H^j_a H^{kb} A^i_b.
        \end{align}
        The covariant equations,  \cref{eq:nonlinear-plebanski-conformal-1-of-4,eq:nonlinear-plebanski-conformal-3-of-4,eq:nonlinear-plebanski-conformal-12}, are then
        \begin{gather}
        \Sigma^i_{\langle\mu}{}^\rho F(\Omega)^i_{\nu\rangle\rho} = 0, \quad g^{\mu\nu} \partial_\mu \Omega^i_\nu = 0. \label{eq:num-conformal-covariant-connection-system} \\
        \Sigma^{i\mu\nu} \partial_\mu \omega_\nu + g^{\mu\nu} \omega_\mu \Omega^i_\nu + \frac{1}{2} \epsilon^{ijk} \Sigma^{k\mu\nu} \omega_\mu \Omega^j_\nu - \Sigma^{j\mu\nu} \Omega^j_\mu \Omega^i_\nu = 0 \label{eq:num-conformal-3-of-4-covariant-connection-system} \\
        g^{\mu\nu} \partial_\mu \omega_\nu - g^{\mu\nu} \left(\frac{1}{4} \omega_\mu \omega_\nu + \Omega^i_\mu \Omega^i_\nu\right) + \Sigma^{i\mu\nu} \omega_\mu \Omega^i_\nu = 0 \label{eq:num-conformal-1-of-4-covariant-connection-system}
        \end{gather}

        To decompose the first equation in \cref{eq:num-conformal-covariant-connection-system} we first contract the $\mu,\nu$ indices with $\Sigma^{i\mu\sigma} \asd^j_\sigma{}^\nu$ and using the algebra of $\Sigma^i$'s we find that this is equivalent to 
        \begin{align}
            \asd^{j\mu\nu} F(\Omega)^i_{\mu\nu} = 0.
        \end{align}
        This keeps all the equations of motion as $\Sigma^{i\mu\sigma} \asd^j_\sigma{}^\nu$ is a complete basis for the symmetric tracefree $4\times 4$ tensors. Moving from $\asd^i \rightarrow \bar{\tE}^i$ and using the using that its decomposition is 
        \begin{align}
            \bar{\tE}^{i0a} = \bar{\tE}^{ia}, \quad \bar{\tE}^{iab} = -i \epsilon^{ijk} \dlap \bar{\tE}^{ja} \bar{\tE}^{kb} + \bar{\tE}^{ia} N^b - \bar{\tE}^{ib} N^a
        \end{align}
        the decomposition of the connection equation is 
        \begin{align}
            \bar{\tE}^{j\mu\nu} F(\Omega)^i_{\mu\nu} = 2 \bar{\tE}^{jb} F(\Omega)^i_{0b} - i \epsilon^{jkl} \dlap \bar{\tE}^{kb} \bar{\tE}^{lc} F(\Omega)^i_{bc} + 2 \bar{\tE}^{jb} N^c F(\Omega)^i_{bc} = 0.
        \end{align}
        Using the inverse of the anti-self-dual triad, $\bar{\utE}^i_a$, we can convert the index $j$ to $a$,
        \begin{align}
            \frac{1}{2} \bar{\utE}^j_a \bar{\tE}^{j\mu\nu} F(\Omega)^i_{\mu\nu} = F(\Omega)^i_{0a} - i \dlap \bar{\utE}^j_a \epsilon^{jkl} \dlap \bar{\tE}^{kb} \bar{\tE}^{lc} F(\Omega)^i_{bc} + 2 N^b F(\Omega)^i_{ab} = 0.
        \end{align}
        From here we can use the following identity,
        \begin{align}
            \epsilon^{jkl} \bar{\tE}^{kb} \bar{\tE}^{lb} = \det(\bar{\tE}) \teps^{dbc} \bar{\utE}^j_d \label{eq:num-matrix-identity-on-bartE}
        \end{align}
        and solve for $F(\Omega)^i_{0a}$ to obtain 
        \begin{align}
            F(\Omega)^i_{0a} = i \det(\bar{\tE}) \bar{\utE}^j_a \bar{\utE}^j_d  \teps^{dbc} F(\Omega)^i_{bc} + 2 N^b F(\Omega)^i_{ba}.
        \end{align}
        We see that the spatial metric has appeared $\gamma_{ab} = -\det(\bar{\tE}) \bar{\utE}^j_a \bar{\utE}^j_d$ to replace it with the usual self-dual triads, $\gamma_{ab} = \det(\tE) \utE^j_a \utE^j_b$ (the minus sign appears because the bar operation is the negative of the complex conjugate). Equivalently it is a consequence of the reality conditions. The evolution equation can be written in terms of purely self-dual objects 
        \begin{align}
            F(\Omega)^i_{0a} = -i \det(\tE) \utE^j_a \utE^j_d  \teps^{dbc} F(\Omega)^i_{bc} + 2 N^b F(\Omega)^i_{ba}.
        \end{align}
        Using the inverse of \cref{eq:num-matrix-identity-on-bartE} but instead on $\tE$ and expanding the first and last curvatures into derivatives of $\Omega^i$ gives
        \begin{align}
            (\partial_t - N^b \partial_b) \Omega^i_a = \partial^\Omega_a \Omega^i_0 - N^b \partial^\Omega_a \Omega^i_b - \frac{i}{2} \utE^j_a \epsilon^{jkl} \tE^{kb} \tE^{lc} F(\Omega)^i_{bc}.
        \end{align}
        Finally, by using $H^{ia} = \dlap\tE^{ia}$ and $H^i_a = \utE^i_a/\dlap$ we have the equation of motion for $\Omega^i_a$ written using only objects in the 12 dimensional sector,
        \begin{align}
            (\partial_t - N^b \partial_b) \Omega^i_a = \partial^\Omega_a \Omega^i_0 - N^b \partial^\Omega_a \Omega^i_b - \frac{i}{2} H^j_a \epsilon^{jkl} H^{kb} H^{lc} F(\Omega)^i_{bc}.
        \end{align}

        The second equation in \cref{eq:num-conformal-covariant-connection-system} we can densitise the inverse metric and then decompose into 3+1 like so 
        \begin{align}
            \dlap \tg^{\mu\nu} \partial_\mu \Omega^i_\nu = -\partial_t \Omega^i_0 + N^a \left( \partial_t \Omega^i_a + \partial_a \Omega^i_0 \right) + \left(\dlap^2 \tE^{ja} \tE^{jb} - N^a N^b\right) \partial_a \Omega^i_b = 0
        \end{align}
        Substituting in the equation of motion for $\Omega^i_a$ and replacing $\tE^{ia}$ with $H^{ia}$ gives
        \begin{align}
            (\partial_t - N^a \partial_a) \Omega^i_0 & = N^a \partial_a \Omega^i_0 + ( H^{ja} H^{jb} - N^a N^b ) \partial_a \Omega^i_b \nonumber \\ & - \frac{i}{2} N^a H^j_a \epsilon^{jkl} H^{kb} H^{lc} F(\Omega)^i_{bc} + \epsilon^{ijk} N^a \Omega^j_a \Omega^k_0 = 0.
        \end{align}

        The connection terms in the evolution equations for $H^{ia}$ and $N^a$ in \cref{eq:num-plebanski-conformally-seperated-frame-evolution-system} can be written using only $\Omega^i$. To see this we make explicit the connection terms, 
        \begin{align}
            (\partial_t - N^b \partial_b) H^{ia} &= -i \epsilon^{ijk} H^{kb} \partial_b H^{ja} - H^{ib} \partial_b N^a \nonumber\\ & \quad +i H^{ja} H^{jb} A^i_b - i H^{ia} H^{jb} A^j_b - \epsilon^{ijk}(A^j_0 - N^b A^j_b) H^{ka} - H^{ia} (\xi_0 - N^b \xi_b) \\
            (\partial_t - N^b \partial_b) N^a &= - H^{ib} \partial_b^A H^{ia} - \epsilon^{ijk} H^{ib} A^j_b H^{ka} + H^{ia} H^{ib} \xi_b.
        \end{align}
        Writing the connection terms using the new connections the evolution equations become 
        \begin{align}
            (\partial_t - N^b \partial_b) H^{ia} &= -i \epsilon^{ijk} H^{kb} \partial_b H^{ja} - H^{ib} \partial_b N^a \nonumber\\ & \quad +i H^{ja} H^{jb} \Omega^i_b - i H^{ia} H^{jb} \Omega^j_b - \epsilon^{ijk}(\Omega^j_0 - N^b \Omega^j_b) H^{ka} \\
            (\partial_t - N^b \partial_b) N^a &= - H^{ib} \partial_b H^{ia} - \epsilon^{ijk} H^{ib} \Omega^j_b H^{ka}.
        \end{align}
        \paragraph{Conformal 12 System}
        The evolution equations for the frame and the connections then form a separate system which we call the conformal system,
        \begin{equation}
            \begin{aligned} \label{eq:num-plebanski-12-conformal-system}
                (\partial^\Omega_t - N^b \partial^\Omega_b) H^{ia} &= -i \epsilon^{ijk} H^{kb} \partial^\Omega_b H^{ja} - H^{ib} \partial_b N^a \\
                (\partial_t - N^b \partial_b) N^a &= - H^{ib} \partial_b^\Omega H^{ia} \\
                (\partial_t - N^b \partial_b) \Omega^i_a & = \partial^\Omega_a \Omega^i_0 - N^b \partial^\Omega_a \Omega^i_b - \frac{i}{2} H^j_a \epsilon^{jkl} H^{kb} H^{lc} F(\Omega)^i_{bc} \\
                (\partial_t - N^a \partial_a) \Omega^i_0 & = N^a \partial^\Omega_a \Omega^i_0 + ( H^{ja} H^{jb} - N^a N^b ) \partial_a \Omega^i_b \nonumber \\ & - \frac{i}{2} N^a H^j_a \epsilon^{jkl} H^{kb} H^{lc} F(\Omega)^i_{bc} = 0.
            \end{aligned}
        \end{equation}
        Where $\partial^\Omega_\mu \chi^i = \partial_\mu \chi^i + \epsilon^{ijk} \Omega^j_\mu \chi^k$ is the conformal gauge covariant derivative. It is remarkable that this system is completely independent of the remaining evolution equations. The only drawback is that it is no longer polynomial in the fields, one has to invert the triad. Practically this is no different from the metric version, where the metric has to be inverted in order to compute the right-hand side of the evolution equations. A similar decomposition is available for \cref{eq:num-conformal-1-of-4-covariant-connection-system,eq:num-conformal-3-of-4-covariant-connection-system} but there all the connection are involved, and the result is not as simple. 
        
        As these equations are only defined up to addition of constraint terms, in this case terms proportional to $\xi_\mu$, there exists a large family of generalisation to these equations of motion all with different numerical properties. Here we will not explore this family numerically, but simply state it as a surprising result that such a system exists and that its numerical properties would be of interest to study.

    \section{Numerical Results} \label{sec:pleb-num-numerical-results}

    The last part of this chapter we dedicate to some minor numerical results for the systems in previous sections. We take our evolution system to be 
    \begin{equation}
    \begin{aligned}
        (\partial^A_t - N^b \partial^A_b) H^{ia} &= -i \epsilon^{ijk} H^{kb} \partial^A_b H^{ja} - H^{ib} \partial_b N^a - H^{ia} (\xi_0 - N^b \xi_b) \nonumber\\
        (\partial_t - N^b \partial_b) N^a &= - H^{ib} \partial_b^A H^{ia} + H^{ia} H^{ib} \xi_b \nonumber\\
        (\partial_t - N^a \partial_a) n &= - \partial_a N^a - 2ic H^{ia} \partial_a \chi^i - \xi_0 + N^a \xi_a \nonumber\\
        (\partial_t - N^a \partial_a) \chi^i &= -\frac{i}{2 c} \partial^A_a H^{ia} + \frac{i}{2c} H^{ia} \partial_a n - i \epsilon^{ijk} H^{ja} \partial_a \chi^k - \beta \chi^i \nonumber\\[20pt]
        (\partial_t - N^a \partial_a) \xi_0 & = 2i N^a H^{ib} F^i_{ab} - \epsilon^{ijk} H^{ia} H^{jb} F^k_{ab} - (H^{ia} H^{ib} - N^a N^b) \partial_a \xi_b \nonumber\\ & \quad\quad  - N^a \partial_a \xi_0 - \rho \xi_0 \nonumber\\
        (\partial_t - N^b \partial_b) \xi_a & = 2i H^{ib} F^i_{ab} - \partial_a \xi_0 + N^b \partial_a \xi_b - \sigma \xi_a\nonumber\\
        (\partial_t - N^b \partial_b) A^i_a & = \partial^A_a A^i_0 - N^b \partial_a^A A^i_b + i \epsilon^{ijk} H^{jb} F^k_{ab} - i H^{ib} \partial_{(a} \xi_{b)} \nonumber\\
        (\partial_t - N^a \partial_a) A^i_0 & = (H^{ja} H^{jb} - N^a N^b) \partial_a A^i_b + N^a \partial^A_a A^i_0 - i \epsilon^{ijk} H^{ja} N^b F^k_{ab} \nonumber\\ & \quad\quad -H^{ia} H^{jb} F^j_{ab} - i H^{ia} \partial_a \xi_0 + i N^a H^{ib} \partial_{(a} \xi_{b)} \nonumber\\  & \quad\quad +i N^a H^{ib} \partial_b \xi_a + \frac{1}{2} \epsilon^{ijk} H^{ja} H^{kb} \partial_a \xi_b.
    \end{aligned} \label{eq:num-plebanski-partially-conformal-system}
    \end{equation}
    which is our partially conformal system with additional damping terms. We have chosen this system as the frame variables $H^{ia},N^a,\chi^i,n$ constitute a simpler system than when written in Ashtekar's variables (simpler here means fewer terms and multiplications needed). The standard test in~\cite{Apples_with_App_Daveri_2018} are performed, where random noise is added to the Minkowski solution.

    We set up the numerical simulation in Cartesian coordinates $(t,x,y,z)$ with the grid being a simple cubic lattice of $N\times N \times N$ points. The length of each side of the grid is we choose to be $L = 5.0$. We use the Runge-Kutta 4th order time integrator with central finite differencing for the spatial derivatives. The boundaries were chosen to be periodic as this represents the ``worse" case scenario in the sense that any constraints that would be propagated out of the grid remains and contributes to the errors. Following~\cite{Apples_with_App_Daveri_2018} the initial condition chosen is 
    \begin{gather}
        H^{ia} = \delta^{ia} + \varepsilon^{ia}, \quad N^a = \varepsilon^a, \quad \dlap = \varepsilon, \quad \chi^i = 0 \\
        A^i_\mu = \varepsilon^i_\mu, \quad \xi_\mu = \varepsilon_\mu
    \end{gather}
    where $\varepsilon \in [-10^{-10}/p^2,+10^{-10}/p^2]$, with $p = N/20$, is a randomly chosen variable within the given range. Each component and position on the grid are given a different random number. We then evolve this initial condition up to time $T = 10.0$ or $100.0$ and fix $c = 1$.

    \begin{figure}[H]
        \begin{subfigure}[b]{0.49\textwidth}
        \includegraphics[width=\textwidth]{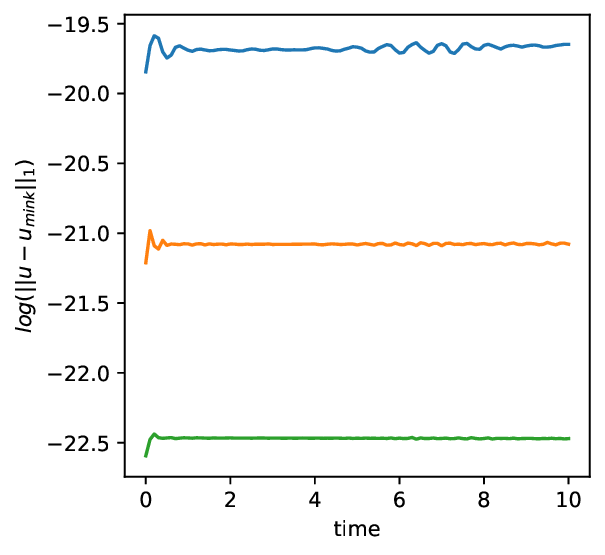}
        \caption{Convergence Testing}
        \label{fig:num-distance-minkowski}
        \end{subfigure}
        \hfill
        \begin{subfigure}[b]{0.49\textwidth}
            \includegraphics[width=\textwidth]{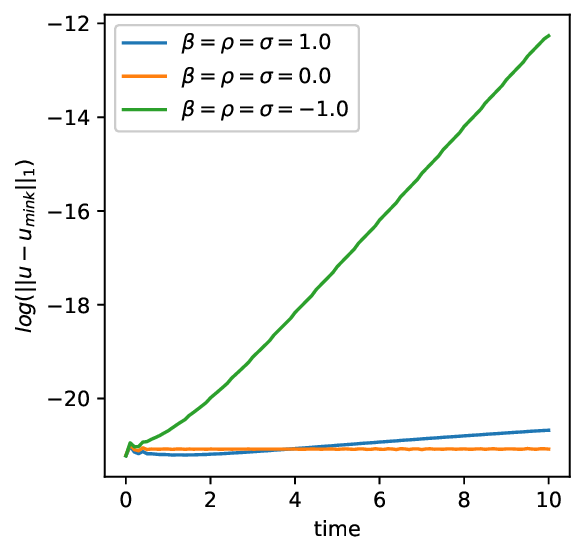}
            \caption{Distance from Minkowski}
            \label{fig:num-minkowski-distance-10}
        \end{subfigure}
        \caption{In (a) we compare the $L_1$ norm, which is defined as $||u||_1 = \frac{1}{N} \sum_i^N |u_i|$ where $|\cdot|$ is the absolute value and the index $i$ runs over all the grid points and components in $u$ (which is denoted as $N$), of the difference between the numerical solution and the Minkowski solution for $N=20,40,80$ for the top, middle and bottom lines respectively. The error in the solution decreases as the number of points, and we can take $N=40$ as being in the convergent regime. In (b) we compare the different values for the damping coefficient, with $+1,0$ performing better than $-1$, as expected from the eigenvalue analysis.}
    \end{figure}
    \hypertarget{rep:L1-norm-discussion}{}

\begin{figure}[H]
        \centering
        \begin{subfigure}{0.4\textwidth}
            \includegraphics[width=\textwidth]{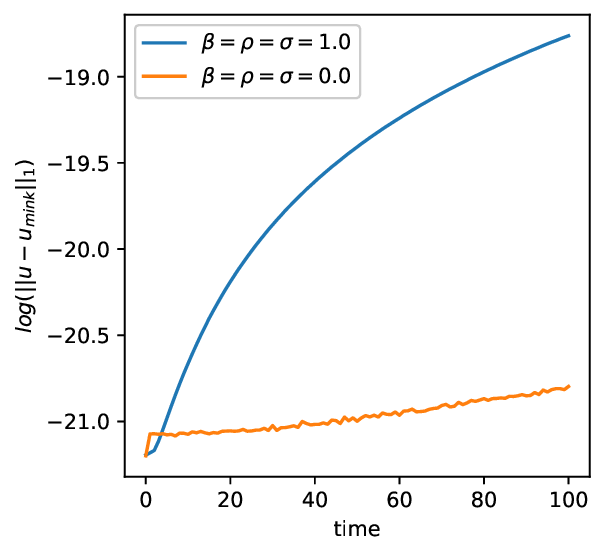}
            \caption{\cref{fig:num-minkowski-distance-10} extended to $T = 100.0$}
            \label{fig:num-minkowski-distance-100}
        \end{subfigure}
        \hfill
        \begin{subfigure}{0.4\textwidth}
            \includegraphics[width=\textwidth]{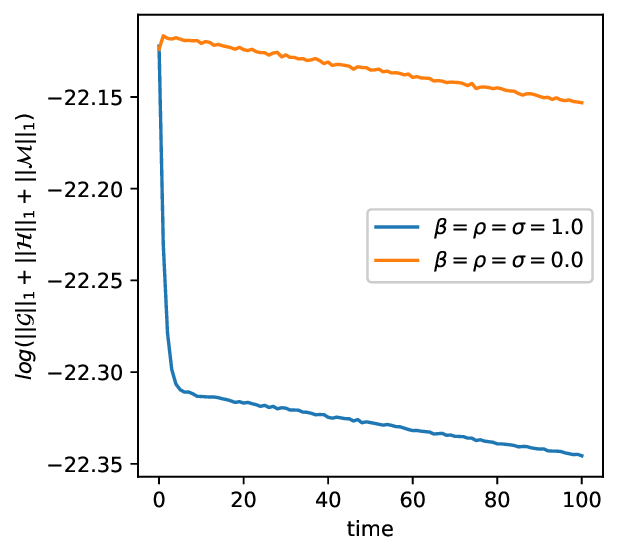}
            \caption{Sum over all constraints}
            \label{fig:num-constraint-damping}
        \end{subfigure}
        \begin{subfigure}{0.4\textwidth}
            \includegraphics[width=\textwidth]{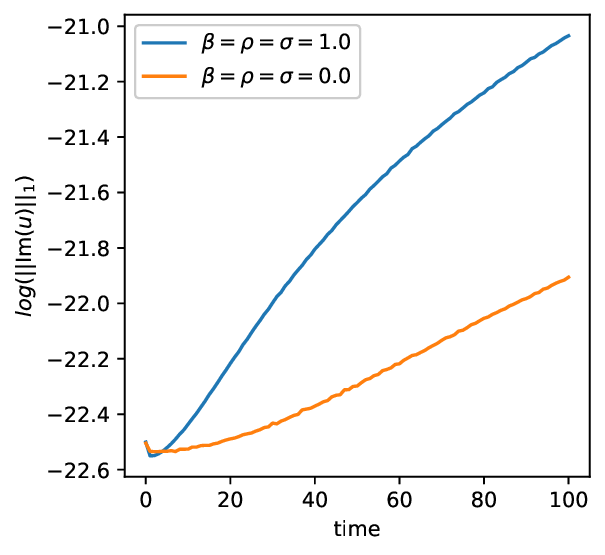}
            \caption{Sum of imaginary metric components}
            \label{fig:num-reality-condition}
        \end{subfigure}
        \caption{Plot (a) is the same as \cref{fig:num-minkowski-distance-10} but for a longer simulation time. In (b) we plot the $L_1$ norm for the sum of the absolute values of all the constraints for the damped and undamped system. In (c) we plot the $L_1$ norm of the imaginary part of the metric for both systems.}
    \end{figure}

    In \cref{fig:num-distance-minkowski} we see that the numerical code converges with respect to the increase of number of points, this means we can take $N=40$ as being in the convergent regime. The different values for the damping coefficient are compared in \cref{fig:num-minkowski-distance-10}, and it is seen that values $\geq 0$ perform better than $<  0$. In fact for damping coefficients $< 0$ the numerical simulation terminates before $T = 20$ due to divergent components. This is in agreement with the eigenvalue analysis and from now on we only compare the damped and undamped systems. We also observe, that for later times, the damped system begins to diverge from Minkowski quicker than the system without damping and the reason for this is unknown. Simulating the same but up to $T=100.0$, see \cref{fig:num-minkowski-distance-100}, we find the effect is more pronounced. It appears that the growth in the distance becomes smaller as time goes on, longer simulations would be needed to see if it eventually converges to a value or continues to grow. The undamped system is expected to diverge eventually as the boundary is periodic and there exists no way for the error to exit the grid.

    \begin{figure}[H]
    \end{figure}

    There are two important properties that need to be checked, the constraints and the reality conditions. In \cref{fig:num-constraint-damping} we see that both systems cause the constraint to decrease from its initial value, however, the damped system reduces it much further initially and then settles down to the same rate as the undamped system. A positive damping coefficient performs as expected in removing the constraint violating modes. Finally, the reality of the metric needs to checked, we see that in \cref{fig:num-reality-condition} that for both systems the imaginary part of the metric grows and interestingly the effect is worse for the damped system. The growth in the reality condition increases by roughly 1 order of magnitude from its initial position, since the initial value is small the end value is also small and within a suitable range. It is reasonable to say that for this initial condition the reality condition has been preserved, even if for some extremely large $T$ value it may cause the simulation to crash.

    \section{Discussion}
    In this chapter we have presented a variety of new evolution systems for general relativity based on the 3+1 decomposition of \pleb{}'s formulation. The evolution systems are based on the decomposition of the hyperbolic gauge fixing \cref{eq:num-covariant-hyperbolic-gauge-fixing}. The main fields the densitised triad $\tE^{ia}$, shift $N^a$, inverse densitised lapse $\dlap$ and the temporal and spatial parts of the self-dual connection $A^i_0, A^i_a$. In the original Ashtekar evolution system only $\tE^{ia}$ and $A^i_a$ are dynamical, the remaining fields are Lagrange multipliers enforcing constraints and as such are gauge fields. In this work the gauge fixing is chosen so that all the fields are promoted to be dynamical. Compared to the ADM formulation we find there are 3 extra Gauss constraints, $\mathcal{G}^i$ that arise due to the $SO(3,\C)$ symmetry present. The hyperbolic gauge fixing introduces conjugate momentum variables for each of the Lagrange multipliers, they are the internal vector $\chi^i$ and the connection 1-form $\xi_\mu$. The evolution equations for these contain the constraints and first-order derivatives of themselves at leading order, and therefore have the structure of a $\lambda$-system~\cite{Hyperbolic_form_Yoneda_2001,Einstein_s_Equa_Brodbe_1998}. Friction terms are also introduced in \cref{eq:num-damped-lambda-system} for which the eigenvalues of the amplification matrix suggest that most of the constraint modes should be damped or are caused to propagate. The resulting system is locally strongly hyperbolic independently of the choice of damping coefficients. We see that the choice of a ``nice" gauge fixing has generated many sought after numerical properties that a 3+1 system should contain. The gauge fixings introduced in the definitions of $\xi_0$  and $\xi_a$ are the Harmonic gauge for the lapse and shift, these are known to have terms that cause shocks (discontinuities in the numerical solution) which are not stable for strong gravitating bodies. The more popular 1+log slicing has shown to be more well-behaved in comparison~\cite{Gauge_condition_Alcubi_2003}, this suggests that the gauge fixing here should also be modified similarly. This modification has not been considered here.

    A further property of this gauge fixing is that it admits a conformal decomposition into two sectors, one that evolves completely on its own. By introducing a conformal triad $H^{ia} = \dlap \tE^{ia}$ the evolution equations for the frame separate into sectors of dimension 12 and 4 as seen in \cref{eq:num-plebanski-conformally-seperated-frame-evolution-system}. The evolution equations for the connection can also be written as a separated system of dimension 12 and 4, see \cref{eq:num-conformal-covariant-connection-system}. This system is equivalent to Einstein's equations in the sense that additional terms are all proportional to constraints which vanish on the constraint surface. The 12 dimensional sector $(H^{ia},N^a,\Omega^i_a,\Omega^i_0)$ evolves independently of the 4 dimensional sector $(\chi^i,n,\omega_a,\omega_0)$, but the reverse is not true. The 12 component system is found in \cref{eq:num-plebanski-12-conformal-system}. If one is only interested in certain components of the metric tensor at the end of the evolution, then evolving with this system will require fewer components than the full gauge fixed system. Usually when evolving binary compact objects one is interested in the Weyl curvature, to compute this one needs to evolve both the 12 and 4 sector. It can be argued that this system is then not useful, however, we have seen that the specific structure of the evolution equations can greatly effect the numerical performance. The idea of performing a conformal transformation is not new, in fact it is used in BSSN~\cite{On_the_Numerica_Baumga_1998,Evolution_of_th_Shibat_1995} and CCZ4~\cite{Conformal_and_c_Alic_2012} systems. Similarly to this conformal system, a metric with determinant one and a new conformally transformed connection appear. A comparison of the numerical performance of the BSSN and Z4 systems can be found in~\cite{Compact_binary_Hildit_2012}, where the Z4 system shows better performance. As our system can be viewed as a Z4-like system for Ashtekar's formulation we are motivated to believe that it may have some numerical benefits.

    The numerical simulations performed in this chapter suggest that the constraint damping terms work as expected as seen in \cref{fig:num-constraint-damping}, even without the damping terms it appears that the constraints are under control. A possible reason for this could be the non-zero wave speeds for all the fields as suggested in~\cite{Towards_an_unde_Alcubi_1999}, this causes the constraints to effectively homogenise and not build up in a single place. However, the distance from the Minkowski solution and the reality condition grow during evolution. One can argue that this is due to the periodic boundary terms, and usually absorbing boundary conditions would be employed, and the error would be able to exit the numerical grid. The growth of these two quantities is rather slow, and the simulation is stable up to $T = 100$ and likely for longer, although the initial condition is relatively simple. More numerical tests would need to be performed to check if this system is truly stable.

    The reality conditions are a unique characteristic of this formulation and as such should have some comments. Reality conditions exist because chiral objects in Lorentzian signature are necessarily complex-valued, but the physical fields constructed from them should be real-valued. In this case the triad $\tE^{ia}$ is complex but the metric $\tE^{ia} \tE^{ib}$ should be real-valued, this is our reality condition. To resolve the reality conditions, in~\cite{Asymptotically_Shinka_1999}, it was suggested to enlarge the $\lambda$-system and add new dynamical fields that impose the reality conditions asymptotically, although no tests were performed to show its usefulness. Here, we take the free evolution approach, due to its simplicity, where the reality condition and its time derivative are imposed initially and the evolution (at least analytically) should remain zero. Practically this is done by obtaining a solution in the metric formulation and the using the frame to construct the Ashtekar variables. Numerically we have shown that this is not the case and the violation in the reality conditions grows with time, however more test would be required to determine how stable the system is in more complicated setups and for longer times. Another route to ensure the metric remains real, is to split the triad into real and imaginary parts $H^{ia} = E^{ia} + i \epsilon^{ijk} \phi^k E^{ja}$ where $E^{ja}$ and $\phi^i$ are real tensors such that $H^{ia} H^{ib} = E^{ia}E^{ib} (1-\phi^2) + \phi^{i}E^{ia} \phi^{j} E^{jb}$. One would also have to separate the connections into real and imaginary components, and it is not known how to do this in a way that keeps all the nice properties of the original system. Reality conditions are also key to ensuring that the first-order PDEs remain strongly hyperbolic. Since the triads appear in the characteristic matrix and their second-order equation of motion contains $\tE^{ia} \tE^{ib}$ as the leading order terms, it is clear that for this to become the real-valued wave equation the reality conditions must be satisfied. The effect of violating the reality condition on the system is not yet fully understood and should be explored in more detail.

    Overall, the system we have introduced is a generalisation of the Z4 system to the \pleb{} formulation. We see that there is a large degree of freedom in choosing the correct variables with which to integrate. Due to this large search space we have only been able to analyse a small number of systems; without further analytic or numerical conditions to reduce this space many tests will need to be performed. Nevertheless, the ideas presented here give rise to a new way of thinking about numerical relativity in terms of the chiral formulation. This way of view gravity has led to the conformal system where degrees of freedom can be completely separated into a separate system. This is the only known formulation where this is possible; its uniqueness makes it noteworthy and worth further considerations.

    \newpage
    \chapter{Weyl Curvature Evolution System for GR}~\label{chap:pure-connection-numerical-relativity}

We have seen that general relativity (GR), in dimension four, admits a series of chiral formulations. In Lorentzian signature these are necessarily complex-valued objects, but in all signatures it elucidates many of GR's hidden features. In this chapter we explore the numerical potential of the chiral first-order pure connection formulation, see \cref{sec:connection-formalism} for a different introduction to the pure connection formulation. The main goal of this chapter is to recast the field equations of this formulation into an evolution system form, and also suitably gauge fix it to ensure the constraints are satisfied throughout numerical evolution.

The behaviour of the linearisation of the pure connection action, \cref{eq:pleb-first-order-pure-connection-aciton}, around an arbitrary (Einstein) background becomes particularly striking in the pure connection formalism, see Section 5.9.5 of~\cite{PlebanskiFormuKrasno2009} and also~\cite{Local_rigidity_Fine_2021}. It can be shown that the linearisation of the \pleb{} action around an arbitrary Einstein background, with all linearised fields apart from the connection integrated out, is of an incredibly simple form, schematically $\Psi^{-1} (\partial a)^2$, when a certain gauge is imposed. It is the availability of a certain gauge in the land of connections, together with the incredible simplicity of the arising gauge-fixed action, that suggested to us that the pure connection formalism may also hide potential as an efficient alternative to the standard metric evolution system for GR. The simplicity of the arising evolution equations in this formalism, see below, shows that this hope is at least to some extent realised. 

This chapter is heavily based on the author's paper~\cite{WeylCurvatureKrasno2022}, where it is noted that a similar evolution system was developed for the pure connection system in~\cite{A_connection_ap_Salisb_1994}. However, here we consider a different gauge fixing procedure that makes the constraint violation propagating, thus ensuring ``constraint sweeping". Another novelty of this work is that we draw a parallel between the connection description of gravity and the chiral description of Maxwell's equations.

A brief overview of what is mention in this chapter is contained in \cref{sec:num-pure-con-relativity-introduction}. A comparison of this system with Maxwell's equations of motion is given in \cref{sec:num-pure-con-maxwell}. More details on the chiral first-order pure connection formalism are given in \cref{sec:pure-connection-chiral-first-order}. A detailed derivation of the evolution system \cref{evol-equations} is found in sections \ref{sec:num-pure-con-3+1}, \ref{sec:num-pure-con-metric} and \ref{sec:num-pure-con-evol-system}. Furthermore, we proceed to describe the gauge-fixing that we expect to be most appropriate for the system \cref{evol-equations} in \cref{sec:numerical-purec-connection-gauge-fixing}.

\section{Introduction} \label{sec:num-pure-con-relativity-introduction}

The parent of all chiral formulations of GR, in four spacetime dimensions, is the \pleb{} formalism, see \cref{chap:plebanskis-formulation}. This formalism describes GR as the dynamical theory of 3 fields $A^i, \Sigma^i, \Psi^{ij}$, with the dynamics following from the following action
\begin{align}\label{action-Pleb}
S[A,\Sigma,\Psi] = \int_M \Sigma^i \wedge F^i - \frac{1}{2}\Psi^{ij} \Sigma^i\wedge \Sigma^j.
\end{align}
Here $F^i = dA^i + (1/2)\epsilon^{ijk} A^j\wedge A^k$ is the curvature of the $SO(3,\C)$ connection $A^i$, and $\epsilon^{ijk}$ is the completely antisymmetric tensor in $\R^3$. The lowercase Latin indices from the middle of the alphabet are ``internal" $\R^3$ indices, $i=1,2,3$. This is the action describing zero cosmological constant GR. The inclusion of a non-zero $\Lambda$ corresponds to adding one more term to the action, but we will not consider non-zero cosmological constant here, as it is irrelevant for black hole binaries modelling.

We repeat the equations of motion, that arise by minimising the above action, for the reader's convenience
\begin{align}
    \Sigma^i\wedge \Sigma^j & \sim \delta^{ij}, \label{eq:num-pure-con-metricity}\\
    d^A \Sigma^i & = d\Sigma^i + \epsilon^{ijk} A^j \wedge \Sigma^k=0, \\
    F^i & = \Psi^{ij} \Sigma^j. \label{eq:num-pure-con-F-sigma}
\end{align}
The first of which is the metricity condition that ensure that the triple of 2-forms encodes the metric, then the second is the definition of the self-dual connection. Lastly, the third equation is precisely that the Riemann curvature only contains the Weyl tensor which is represented by the symmetric tracefree matrix $\Psi^{ij}$. Together these equations impose Einstein's condition on the metric defined by $\Sigma^i$ through the \urb{} formula \cref{eq:pleb-urbantke-metric}. In Lorentzian signature the 2-form are complex, as the Hodge star has complex eigenvalues, therefore we also need the reality condition 
\begin{align}
    \Sigma^i \wedge \asd^j = 0, \qquad {\rm Re}(\Sigma^i \wedge \Sigma^i) = 0. \label{eq:num-pure-con-reality-conditions}
\end{align}
Without this condition the metric defined by \cref{eq:pleb-urbantke-metric} would be complex-valued. Imposing the reality condition is a subtle issue in the pure connection descriptions and will be discussed more at the end of the chapter. The 3+1 decomposition of this system has already been discussed in \cref{chap:plebanski-numerical-relativity} and is known to produce Ashtekar's Hamiltonian formulation of GR.

The formulation of GR that is the main subject of this chapter arises by solving the (algebraic) field equations \cref{eq:num-pure-con-F-sigma} for the 2-form field $\Sigma^i$, and substituting the result back into the action \cref{action-Pleb}. This results in the following action
\begin{align}\label{eq:num-pure-con-action}
S[A,\Psi] = \int \left(\Psi^{-1}\right)^{ij} F^i \wedge F^j.
\end{align}
The result is the first-order pure connection formulation of GR as described in \cref{sec:connection-formalism}, however, here there is no cosmological constant. It is a diffeomorphism invariant gauge theory, and the main fields are $SO(3,\C)$ connection $A^i$ and the self-dual Weyl matrix $\Psi^{ij}$.

One of the main results in this chapter is that the field equations resulting from \cref{eq:num-pure-con-action} can be rewritten in the form of evolution equations, resulting in a remarkably simple system. We present these equations already in this section, to hopefully motivate the reader to understand further details. The new evolution system of GR that follows from \cref{eq:num-pure-con-action} makes the field $\Psi^{ij}$, which we remind encodes the self-dual part of the Weyl curvature, the main dynamical field. It is for this reason that we call this a Weyl evolution system. However, the evolution equation for $\Psi^{ij}$ explicitly contains the spatial $SO(3,\C)$ connection $A^i_a$, where $a=1,2,3$ is the spatial 1-form index. For this reason one must also evolve the spatial connection. The evolution system is as follows
\begin{equation}
\begin{aligned}\label{evol-equations}
D_t \Psi^{ij} - N^a D_a \Psi^{ij} = \im N \epsilon^{klj} \gamma^{ka} D_a \Psi^{il}, \\
D_t A_a^i - \partial_a A_0^i - N^b F^i_{ba} = \im N \Psi^{ij} \gamma^j_a.
\end{aligned}
\end{equation}
Here $N, N^a$ are the usual lapse and shift (which are real quantities). The objects $D_t, D_a$ are the gauge covariant derivatives with respect to the $A_0^i$ and $A_a^i$ components of the connection respectively. The object $F^i_{ab}= 2\partial_{[a} A^i_{b]} + \epsilon^{ijk} A^j_a A^k_b$ is the curvature of the spatial connection. The object $\gamma^{i a}$ is the (inverse) spatial triad, which is required to be ``real" by the reality conditions that this system must be supplemented with, and which is constructed from the curvature of the spatial connection according to
\begin{align}\label{spatial-frame}
\gamma^{ia} := \sqrt{\frac{{\rm det}(\Psi)}{{\rm det}(\tilde{F})}} (\Psi^{-1})^{ij} \tilde{F}^{ja}, \qquad \tilde{F}^{ia}: = \frac{1}{2} \tilde{\epsilon}^{abc} F^i_{bc}, \qquad {\rm det}(\tilde{F}) := \frac{1}{6} \epsilon^{ijk} \utilde{\epsilon}_{abc} \tilde{F}^{ia} \tilde{F}^{jb}\tilde{F}^{kc}.
\end{align}
Here and in what follows the tilde over a symbol encodes the fact that the object has density weight one, and the tilde under a symbol denotes density weight minus one. There is a single set of constraints that the system \cref{evol-equations} must be supplemented with. These are
\begin{align}\label{constr}
\gamma^{ia} D_a \Psi^{ij}=0.
\end{align}

We note that the objects $\Psi^{ij}$ and $A^i_a$ are inherently complex fields, as is in particular manifest from the fact that their evolution equations contain an explicit factor of the imaginary unit on the right-hand side. Nevertheless, the spatial frame $\gamma^{ia}$ is required to be real (in order to produce a real metric of Lorentzian signature), in the sense that $\gamma^{ia}\gamma^{ib}$ is a real symmetric $3\times 3$ tensor. It can be shown that the evolution equations \cref{evol-equations} are compatible with these reality conditions in the sense that, if the reality conditions and their time derivatives are imposed at one moment of time, they will remain satisfied at all times. 

From the point of view of evolution equations \cref{evol-equations}, the objects that are dynamical are the Weyl curvature field $\Psi^{ij}$ and the spatial connection $A^i_a$. All other fields present in \cref{evol-equations}, namely $N,N^a, A^i_0$ are not dynamical and can be chosen to be what one wishes. Their presence in the evolution system is of course the manifestation of the diffeomorphism and gauge invariance of the theory. The quantity $A_0^i$ has been referred to as the ``triad lapse" in~\cite{Asymptotically_Shinka_1999,Hyperbolic_form_Yoneda_2001}, but we will not be using this terminology in the present chapter.

The evolution equations \cref{evol-equations} encode all the dynamics of GR. Their form is considerably simpler than that of the standard ADM evolution system, see e.g.~\cite{NumericalRelatBaumga2010}, equations (2.134), (2.135)
\begin{align}
\partial_t K_{ab} - N^c \partial_c K_{ab} - K_{ac} \partial_b N^c - K_{cb} \partial_b N^c = 
N( R_{ab} - 2K_{ac}K^c{}_b + K K_{ab}) - D_a D_b N, \\ \nonumber
\partial_t \gamma_{ab} - D_a N_b - D_b N_a = - 2N K_{ab}.
\end{align}
We have rewritten the equations from~\cite{NumericalRelatBaumga2010} using our index conventions, and placed all the terms involving the shift to the left-hand side of the equations. The second of these ADM equations is the definition of the extrinsic curvature $K_{ab}$, while the first gives its evolution. To draw an analogy with our system, the second equation in \cref{evol-equations} can be viewed as the definition of $\Psi^{ij}$, while the first one describes the evolution of this field. It is clear that the right-hand side of the first equation in \cref{evol-equations} is significantly simpler than the right-hand side of the ADM evolution equation for the extrinsic curvature. There is also a simplification as compared to the usual story in regard to the constraints. The familiar Hamiltonian and Momentum constraints of GR are absent in the formulation \cref{evol-equations} entirely, and are replaced by a single ``Gauss" constraint \cref{constr}. This fact has significant consequences for the problem of numerical constraint violation and gauge-fixing, and will be further dealt with below. Another feature of the system \cref{evol-equations} that makes it differ from the ADM evolution system is that it is first order in both the time and spatial derivatives. 

However, the price that one pays for the simplicity is the non-linearity of the spatial frame \cref{spatial-frame} as constructed from the curvature of the connection. The other price is the fact that all quantities have become complex-valued and one must impose the reality conditions to recover real GR. This last issue is not a point of concern in ``exact" general relativity, because the reality conditions can be imposed on the initial data, and it is guaranteed that they will be preserved at all times. However, the issue of the reality conditions may become a problem in numerical GR, when it is no longer guaranteed that the reality conditions are preserved by the evolution, and small errors introduced by the discretisation may make the numerical code invalid very quickly. We will come back to the discussion of these issues in the last section. 

\section{Chiral description of Maxwell's theory} \label{sec:num-pure-con-maxwell}

Most of the things we do in the case of gravity have a Maxwell analogue. We discuss the much simpler Maxwell case first, as the parallel with it helps to understand the constructions that follow. 

\subsection{3+1 Chiral Maxwell}

We recall the 3+1 decomposition of the chiral Maxwell equations shown in \cref{sec:chiral-maxwell}. The basic fields are the complex vector $\phi^i$ and the connection $A = A_0 dt + A^i dx^i$, where $i = 1,2,3$ and the coordinate decomposition is $x^\mu = t,x^i$. The evolution equations for these fields are
\begin{align}
    \phi^i  & = -\im (\partial_t A^i -\im  \partial^i A_0) - \epsilon^{ijk} \partial^j A^k = i E^i - B^i \label{evol-eqn-a} \\
    \partial_i \phi^i & = 0 \label{phi-constr} \\
    \partial_t \phi^i & = -i \epsilon^{ijk} \partial^j \phi^k. \label{evol-eqn-phi}
\end{align}
The first equation gives the definition of the Riemann-Silberstein vector in terms of the electric and magnetic fields, $E^i$ and $B^i$. The beauty of this last equation is that it is independent of the electromagnetic potentials $A_i, A_0$, and combines the two Maxwell's equations on $B^i, E^i$ into a single complex equation 
\begin{align}\label{evol-eqn-M-compl}
\im \dot{\vec{\phi}} =\nabla \times \vec{\phi}
\end{align}
on the complex vector $\vec{\phi}= \vec{B}+\im \vec{E}$. We can now interpret the equation \cref{evol-eqn-phi} as the evolution equation for $\phi^i$, and \cref{evol-eqn-a} as the evolution equation for the electromagnetic potential. In this evolution equation the quantity $A_0$ remains an arbitrary function whose presence there reflects gauge-invariance. The evolution equation for $\phi^i$ must be supplemented by the constraint \cref{phi-constr}. The interesting feature of this system is that, if one is only interested in the physical fields $E^i, B^i$, it is sufficient to evolve only $\phi^i$, as its evolution equation decouples from that for the potential. 

\subsection{Constraint sweeping}

The above system of equations puts the dynamical equations of Maxwell's theory into the form of an evolution system subject to the single Gauss constraint \cref{phi-constr}. It is not hard to see from the form of the equation \cref{evol-eqn-phi} that if the constraint is satisfied at some moment of time, it will remain satisfied always. Indeed, this follows from the fact that the divergence of the curl is zero, and so the time derivative of $\phi^i$ is divergence-free.

However, any numerical implementation of the above evolution system will unavoidably lead to constraint violation. If this is not dealt with, any numerical code will become unstable. There is a very simple way to deal with this, which is to introduce an additional field whose time evolution is controlled by the would-be constraint. If the constraint is satisfied the additional field remains zero. However, when constraint violation is introduced, the modified system can be arranged in such a way that any constraint violation propagates away with the speed of light. 

Another way to motivate the modification of \cref{evol-eqn-phi} is to consider the double time derivative of $\phi^i$. We have
\begin{align}
\ddot{\phi}^i = - \im \epsilon^{ijk} \partial^j \dot{\phi}^k = - \epsilon^{ijk} \partial^j \epsilon^{klm} \partial^l \phi^m = - \partial^i \partial^j \phi^j + \Delta \phi^i.
\end{align}
Only the last term here is the one desired to obtain the $\Box\phi^i=0$ equation. The first term vanishes when the Gauss constraint is satisfied, but in general obscures the desired $\Box$-type structure of the squared evolution equation. 

All these problems are solved in one go if one introduces a new field $\phi$, which has an evolution equation of its own, and which modifies the evolution equation for $\phi^i$. The new system of evolution equations is
\begin{align}\label{eqs-M-modified}
\partial_t \phi & = \im \partial^i \phi^i, \\ \nonumber
\partial_t \phi^i & = - \im (\epsilon^{ijk} \partial^j \phi^k + \partial^i \phi).
\end{align}
This is the same as the hyperbolic gauge fixing shown in \cref{chap:Linearised-Gravity} during the chiral Maxwell example. The squaring procedure now gives
\begin{align}
\ddot{\phi}^i =- \im \epsilon^{ijk} (\partial^j \dot{\phi}^k + \partial^i \dot{\phi}) = - \partial^i \partial^j \phi^j + \Delta \phi^i + \partial^i \partial^j \phi^j = \Delta \phi^i.
\end{align}
It is also easy to see that $\ddot{\phi} = \Delta\phi$, and so both $\phi^i,\phi$ propagate with the speed of light. The described modification \cref{eqs-M-modified} of the Maxwell evolution system are known under the name of ``constraint sweeping". The idea is that the constraint can be imposed only initially, and then any numerical violation of the constraint will propagate off the grid, preventing dangerous accumulation of the constraint violation that can render the code unstable. Such ``constraint sweeping" is known to led to improved numerical stability, see e.g. the discussion surrounding formula (11.59) in~\cite{NumericalRelatBaumga2010}. Note that ``constraint sweeping" is different from ``constraint damping". The first type of modification to the evolution equations makes the constraint violation propagate away from the point in the grid where it occurs. The second type of modifications makes the constraint violation exponentially suppressed, by adding to the equations specially adjusted friction terms. We will return to the discussion of whether ``constraint sweeping" is sufficient to deal with the issue of constraint violation in the Discussion section.

\subsection{Lagrangian for gauge-fixing}

The described modification \cref{eqs-M-modified} of the evolution system can be obtained from a simple gauge-fixed Lagrangian. Indeed, we supplement the action for chiral Maxwell by the following gauge-fixing term, which is clearly a version of the Lorentz gauge
\begin{equation}\label{gf-M}
    S_{g.f} = \int \phi \eta^{\mu\nu} \partial_\mu A_\nu - \frac{1}{2} \phi^2 = \int \phi ( - \partial_t A_0 + \partial^i A^i) - \frac{1}{2} \phi^2.
\end{equation}
It is easy to see that the modifications this addition does to the field equations of Maxwell's theory are precisely as described by \cref{eqs-M-modified}. Indeed, the field equation that result by varying the Lagrangian with respect to $A_0$ is now precisely the first equation in \cref{eqs-M-modified}. The field equation that follows by varying with respect to $A^i$ is the second equation in \cref{eqs-M-modified}. 
Finally, the equation that follows by varying the action with respect to the new field $\phi$ is
\begin{align}\label{evol-eqn-A0}
\phi = - \partial_t A_0 + \partial^i A^i.
\end{align}
This can be viewed as an evolution equation for $A_0$, given $\phi, A^i$. Thus, once our dynamical system is gauge-fixed as described there remain no free functions, and $A_0$ evolves together with $A^i$ as dictated by the corresponding equation. The full gauge-fixed evolution system is then 
\begin{align}
    \partial_t \phi^i & = - \im (\epsilon^{ijk} \partial^j \phi^k + \partial^i \phi) \label{eq:evol-phi-i}\\ 
    \partial_t \phi & = \im \partial^i \phi^i, \label{eq:evol-phi}\\
    \partial_t A^i & = \partial^i A_0 + \im \epsilon^{ijk} \partial^j A^k + \im \phi^i \label{eq:evol-A-i}\\
    \partial_t A_0 & = \partial^i A^i - \phi \label{eq:evol-A-0}
\end{align}

\subsection{Reality conditions} \label{subsec:num-pure-conn-maxwell-reality-conditions}

\begin{figure}[H]
    \centering
    \includegraphics[width=0.5\linewidth]{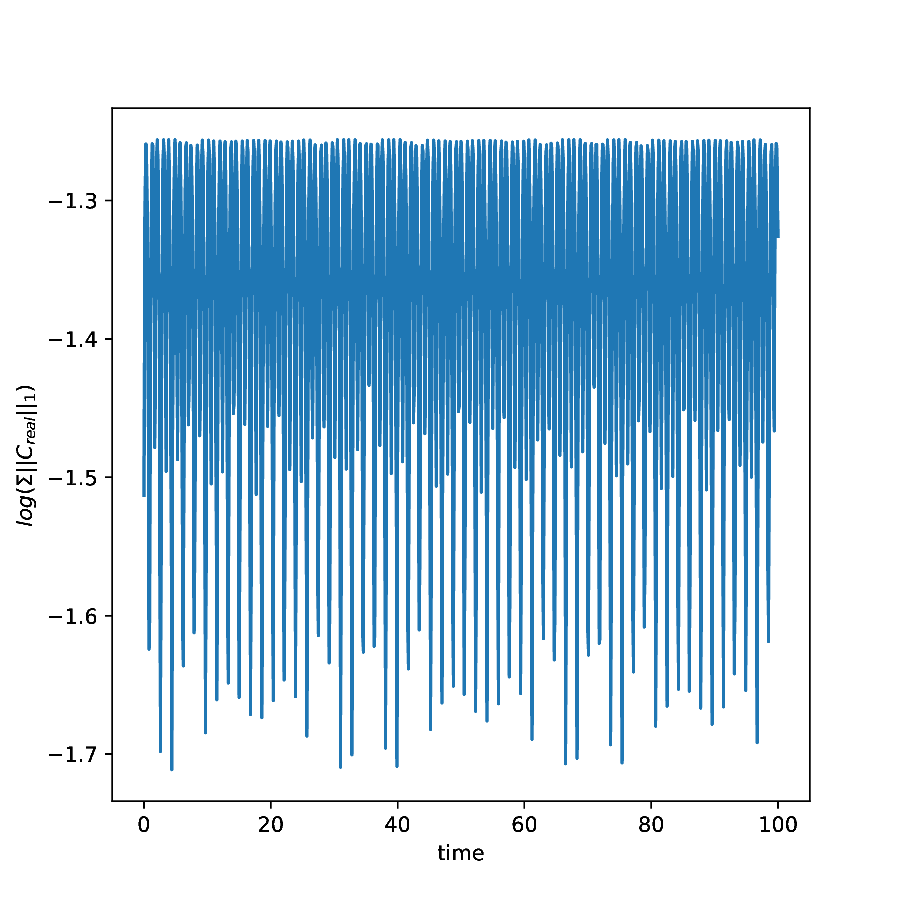}
    \label{fig:reality-condition-chiral-maxwell}
    \caption{Error in the reality conditions, \cref{eq:chiral-maxwell-reality-conditions}, versus time, for the chiral maxwell evolution system \cref{eq:evol-phi-i,eq:evol-phi,eq:evol-A-i,eq:evol-A-0}. It can be seen that the reality condition remains bounded and oscillates below a certain value.}
\end{figure}

It is interesting to discuss the question of the reality conditions that the system of equations \cref{eqs-M-modified} must be supplemented with. It is clear that we want the electromagnetic potentials $A_0, A^i$ to be real. Furthermore, it is then clear that \cref{eq:evol-A-0} implies that $\phi$ is real also. Using that $\phi^i = i E^i - B^i$ we can see that \cref{eq:evol-A-i} implies that
\begin{align}
     {\rm Re}(\phi^i + \epsilon^{ijk} \partial^j A^k) = -B^i + \epsilon^{ijk} \partial^j A^k = 0
\end{align}
and therefore $E^i = \partial^i A_0 - \partial_t A^i$. The evolution equation, \cref{eq:evol-phi-i}, then splits into real and imaginary parts, $\partial_t \vec{B} = -\nabla \times \vec{E}$ and $\partial_t \vec{E} = \nabla \times \vec{B}$. Finally, the imaginary part of \cref{eq:evol-phi} is $\partial^i B^i = 0$ which is true given that $B^i = \epsilon^{ijk} \partial^j A^k$. The real part of the last equation is $\partial_t \phi = -\partial^i E^i$. We see that the reality conditions form a set of constraints
\begin{align}
    C^i_R = {\rm Re}(\phi^i + \epsilon^{ijk} \partial^j A^k), \quad C^i = {\rm Im}(A^i), \quad C_0 = {\rm Im}(A_0), \quad C = {\rm Im}(\phi). \label{eq:chiral-maxwell-reality-conditions}
\end{align}
It is here that we encounter a problem with imposing reality conditions. Suppose that numerically the fields $A^i,A_0$ are described by real data, then \cref{eq:evol-A-i,eq:evol-phi} each split into an evolution equation and a constraint equation. This real and imaginary split returning exactly the evolution and constraint equation is only easy to compute here due to the linear nature of the equation. In the case of GR it is unlikely that the separation will be as clear. Instead, we suggest a more general strategy where one promotes all the fields to be complex, initialise them to satisfy the reality conditions and then evolve via the system in \cref{eq:evol-phi-i,eq:evol-phi,eq:evol-A-i,eq:evol-A-0}. This is also known as free evolution in the context of usual reality conditions. To see how the reality behave we look at the time derivative of them under this evolution system, in doing so we find the following closed system
\begin{align}
    \partial_t C^i_R & = \partial^i C + \partial^j \partial^j C^i - \partial^i \partial^j C^j \\
    \partial_t C^i & = \partial^i C_0 + C^i_R \\
    \partial_t C_0 & = \partial^i C^i - C_0 \\
    \partial_t C & = \partial^i C^i_R.
\end{align} 
Therefore, if one assume that the system is prepared in a state where the reality constraints are satisfied initially the constraints will be solved for all time. This only holds true analytically and numerically errors can be introduced that move the system away from the reality surface. To see how the solution grows and decays we can perform the Fourier transformation analysis done in \cref{subsec:lambda-system} and compute the eigenvalues of the amplification matrix. In doing so we find the matrix has the eigenvalues,
\begin{align}
    \lambda = \pm \sqrt{-k^2}, \quad \lambda = \pm \sqrt{ik^2}, \quad \lambda = -\frac{1}{2} \pm \frac{\sqrt{1-4k^2}}{2}
\end{align}
with multiplicity $2,1,1$ each. The real part of all but 1 eigenvalues is nonpositive, and the imaginary parts are all non-zero. This means that the reality conditions will propagate and all but 1 will be damped. Therefore, any violation of in the reality condition will be propagated off the numerical grid. To check the performance of this free evolution of the reality conditions, we perform the same numerical simulation as done in \cref{sec:numerical-chiral-maxwell} with the additional initial conditions for the potentials
\begin{align}
    A^i = -\epsilon^{ijk} \frac{k^j}{\sqrt{k^2}} E^k \cos(k^i x^i) - \frac{E^i}{\sqrt{k^2}} \sin(k^i x^i), \quad A_0 = 0.
\end{align}
We then perform the simulation up to $T = 100.0$ and without the previously introduced constraint damping terms active. In \cref{fig:reality-condition-chiral-maxwell} we see that the reality condition remains bounded, which suggest that the free evolution is capable of controlling the reality conditions despite the single positive real part of the eigenvalue.

\hypertarget{rep:Fixed-square-in-plot}{}
The reality conditions for GR we expect to handle similarly, by gauge fixing the full system in such a way that the constraints are handled. Below we will see that the action \cref{eq:num-pure-con-action} is for GR what the action \cref{eq:chiral-YM-action} is for Maxwell theory. 

\section{Chiral pure connection gravity in the first-order formalism} \label{sec:pure-connection-chiral-first-order}

Our presentation of the chiral pure connection formalism is very brief, and we only indicate the main interpretational steps. For more details the reader is referred to Section 6 in the book~\cite{FormGenRelGravity2020}.

\subsection{Action and field equations}

Chiral Gravity can be described using an $SO(3,\mathbb{C})$ connection $A^i$ and the complex symmetric tracefree matrix of quantities $\Psi^{ij}$.\footnote{ We remind the reader that the Latin indices $i,j,k,\ldots,z$ are the internal ones on which $SO(3,\mathbb{C})$ acts, Latin indices at the start of the alphabet $a,b,c,\ldots,h$ represent 3d spatial coordinates and Greek indices are the spacetime indices.} As already mentioned, the matrix of the quantities $\Psi^{ij}$ is required to be  tracefree  and symmetric 
\begin{equation}
    Tr(\Psi) = \Psi^{kk} = 0, \quad \Psi^{(ij)} = \Psi^{ij}.
\end{equation}
Varying the action \cref{eq:num-pure-con-action} with respect to the connection produces
\begin{equation}
   d^A \left( \frac{1}{\Psi^{ij}} F^j \right) = 0 
    \label{eq:CPC_Lagrangian_variation_wrt_A}
\end{equation}
where $d^A$ is the exterior covariant derivative. The notation that we have used and will continue to use in what follows is 
\begin{align}
\frac{1}{\Psi^{ij}} \equiv (\Psi^{-1})^{ij}.
\end{align}

Using the Bianchi identity on the curvature form
\begin{equation}
    d^A F^i = 0,
\end{equation}
one can rewrite \cref{eq:CPC_Lagrangian_variation_wrt_A} as
\begin{equation}
    d^A \left( \frac{1}{\Psi^{ij}} \right) F^j = 0.
    \label{eq:CPC_A_equation_of_motion}
\end{equation}
As we will see below, this contains both a constraint and an evolution equation. Varying the action \cref{eq:num-pure-con-action} with respect to $\Psi^{ij}$ leads to
\begin{equation}\label{FF-eqn}
    \frac{1}{\Psi^{ik}} F^k \wedge F^l \frac{1}{\Psi^{lj}} \sim \delta^{ij},
\end{equation}
where the Kronecker delta on the right-hand side appears because only the tracefree part of the variation is required to vanish, thus implying that there is only the (arbitrary) trace part on the right-hand side.
Multiplying this by $(\Psi^2)^{ij}$ results in
\begin{equation}
    F^i \wedge F^j \sim \Psi^{ik} \Psi^{kj}.
    \label{eq:CPC_psi_equation_of_motion}
\end{equation}
The proportionality constant is a (different from zero) scalar function that will be fixed below. 

\subsection{Interpretation of the field equations}

The two equations of motion for our system are thus \cref{eq:CPC_A_equation_of_motion} and 
 \cref{eq:CPC_psi_equation_of_motion}. The second equation receives the following interpretation. As we will discuss below, a triple of 2-forms $\Sigma^i$ satisfying $\Sigma^i \wedge \Sigma^j\sim \delta^{ij}$ defines the metric \cref{metric-Sigma}. One obtains real Lorentzian signature metrics when the 2-forms are complex and satisfy the reality conditions \cref{eq:num-pure-con-reality-conditions}. Coming back to the equation \cref{FF-eqn}, we can see that this equation suggest that we define
 \begin{align}\label{Sigma-def}
 \Sigma^i := \frac{1}{\Psi^{ij}} F^j.
 \end{align}
 The object $\Sigma^i$ so defined is constructed algebraically from the fields $\Psi^{ij}$ and the curvature of the connection $A^i$. The equation \cref{eq:CPC_psi_equation_of_motion} then says that this object satisfies $\Sigma^i \wedge \Sigma^j\sim \delta^{ij}$ and thus defines the metric via \cref{metric-Sigma}.
 
 The interpretation of the first equation \cref{eq:CPC_A_equation_of_motion} is then as follows. It is clear that using the introduced object $\Sigma^i$ it can be rewritten as $d^A \Sigma^i=0$. It can then be shown that, when \cref{eq:CPC_psi_equation_of_motion} holds, this equation implies that $A^i$ coincides with the self-dual part of the Levi-Civita connection for the metric defined by $\Sigma^i$. When this is the case the Einstein equations for the metric defined by $\Sigma^i$ are a simple consequence of \cref{Sigma-def}. Indeed, when $A^i$ is the self-dual part of the Levi-Civita connection, then $F^i$ is the self-dual part of the Riemann curvature. The definition \cref{Sigma-def} can be rewritten as $F^i =\Psi^{ij} \Sigma^j$, and thus implies that the self-dual part of the Riemann curvature is self-dual as the 2-form. This is one of the ways to state the Einstein condition. With $\Psi^{ij}$ being tracefree, the relation $F^i =\Psi^{ij} \Sigma^j$ also implies that the cosmological constant is zero. Thus, all in all, $F^i =\Psi^{ij} \Sigma^j$ is a rewriting of the Ricci-flatness condition, provided both \cref{eq:CPC_A_equation_of_motion} and  \cref{eq:CPC_psi_equation_of_motion} are satisfied. 
 
 The action \cref{eq:num-pure-con-action} thus gives a chiral, pure connection description of general relativity with zero cosmological constant. It becomes the description of Lorentzian signature GR when the reality conditions \cref{eq:num-pure-con-reality-conditions} get imposed. The field equations \cref{eq:CPC_A_equation_of_motion} and  \cref{eq:CPC_psi_equation_of_motion} can be shown to have the property that the evolution they describe commutes with the reality conditions. Thus, if reality conditions are imposed at a moment of time, they will continue to hold throughout the evolution. For this reason our strategy will be to first understand the space plus time decomposition of the equations \cref{eq:CPC_A_equation_of_motion} and  \cref{eq:CPC_psi_equation_of_motion}, as well as the problems related to the necessary gauge-fixings of this system. Only at the very end we will return to the problem of the reality conditions and comment on the most sensible ways to deal with it. 

\section{Field equations in the 3+1 split form} \label{sec:num-pure-con-3+1}

In this section we start the procedure of casting the field equations into the form of evolution equations. 

\subsection{Notations and conventions}

A 3+1 split of the connection is performed as
\begin{equation}
    A^i = A^i_0 dt + A^i_a dx^a.
\end{equation}
The curvature form can be similarly split
\begin{equation}
    F^i = \frac{1}{2}F^i_{\mu\nu} dx^\mu dx^\nu = F^i_{0a} dt dx^a + \frac{1}{2}F^i_{ab} dx^a dx^b.
\end{equation}
These two components appearing can be written using the connection
\begin{equation}\label{t-partial-A}
    F^i_{0a} = \partial_t A^i_a - D_a A^i_0, \quad F^i_{ab} = \partial_a A^i_b - \partial_b A^i_a + \epsilon^{ijk}A^j_a A^k_b
\end{equation}
where $D_a \chi^i = \partial_a \chi^i + \epsilon^{ijk} A^j_a \chi^k$ is the spatial covariant derivative. A useful description of $F^i_{ab}$ is given by its dual
\begin{equation}\label{tilde-F}
    \tilde{F}^{ia} := \frac{1}{2}\tilde{\epsilon}^{abc} F^i_{bc} \quad \Rightarrow F^i_{ab} = \utilde{\epsilon}_{abc}\tilde{F}^{ic}.
\end{equation}
The determinant of the field $\tilde{F}^{ia}$ is given by
\begin{equation}
    \det{\tilde{F}} := \frac{1}{6} \epsilon^{ijk} \utilde{\epsilon}_{abc} \tilde{F}^{ia} \tilde{F}^{jb} \tilde{F}^{kc}.
\end{equation}
The inverse of $\tilde{F}^{ia}$ can be defined as
\begin{equation}
    \utilde{P}^{i}_a = \frac{1}{2\det{\tilde{F}}}\epsilon^{ijk}\utilde{\epsilon}_{abc} \tilde{F}^{jb} \tilde{F}^{kc},
\end{equation}
and satisfies
\begin{equation}
    \tilde{F}^{ia} \utilde{P}^j_a = \delta^{ij}, \quad \tilde{F}^{ia} \utilde{P}^i_b = \delta^a_b.
\end{equation}
This new variable $\tilde{F}^{ia}$ fully defines the spatial curvature form. Using this parameterisation the equations of motion can be split into the 3+1 formalism.

\subsection{Field equations in the 3+1 form}

The equation \cref{eq:CPC_A_equation_of_motion} is equivalent to
\begin{align}
    \tilde{\epsilon}^{\mu\nu\rho\sigma} D_\nu \left(\frac{1}{\Psi^{ij}} \right) F^j_{\rho\sigma} = 0
\end{align}
To obtain these in 3+1 form, we set the index $\mu=0,a$. The first case results in 
\begin{equation}
    \tilde{\epsilon}^{0abc} D_a \left(\frac{1}{\Psi^{ij}}\right)F^j_{bc} = \tilde{\epsilon}^{abc} D_a \left(\frac{1}{\Psi^{ij}}\right)F^j_{bc}.
\end{equation}
This gives a constraint equation, which is natural to refer to as the Gauss constraint:
\begin{equation}
    C_G^i = D_a \left(  \frac{1}{\Psi^{ij}} \right) \tilde{F}^{ja} \equiv 0.
    \label{eq:Gauss_constraint}
\end{equation}
The case $\mu = a$ gives
\begin{align}
    \tilde{\epsilon}^{a \nu \rho\sigma} D_\nu \left( \frac{1}{\Psi^{ij}} \right) F^j_{\rho\sigma} = \tilde{\epsilon}^{a0bc} D_t \left( \frac{1}{\Psi^{ij}} \right) F^j_{bc} + 2 \tilde{\epsilon}^{ab0c} D_b \left( \frac{1}{\Psi^{ij}} \right) F^j_{0c}.
\end{align}
We use conventions in which $\tilde{\epsilon}^{0abc}=\tilde{\epsilon}^{abc}$, where a single tilde over the object denotes the fact that this object is a density of weight plus one. One thus gets the evolution equation for $\Psi^{ij}$
\begin{equation}
    D_t \left( \frac{1}{\Psi^{ij}} \right) \tilde{F}^{ja} = \tilde{\epsilon}^{abc} D_b \left( \frac{1}{\Psi^{ij}} \right) F^j_{0c}.
    \label{eq:Inv_Psi_Evolution_Equation}
\end{equation}
This can be rewritten in a convenient form by multiplying by $\utilde{P}^k_a$. This gives
\begin{equation}
    D_t \left( \frac{1}{\Psi^{ik}} \right) = \tilde{\epsilon}^{abc} \utilde{P}^k_a D_b \left( \frac{1}{\Psi^{ij}} \right) F^j_{0c},
    \label{eq:Inv_Psi_Equation_Of_Motion}
\end{equation}
which should be interpreted as the evolution equation for (the inverse of) $\Psi^{ij}$. 

We now perform the 3+1 split of the ``metricity" equation \cref{eq:CPC_psi_equation_of_motion}. We have
\begin{equation}
    F^i \wedge F^j = \frac{1}{4}\tilde{\epsilon}^{\mu\nu\rho\sigma} F^i_{\mu\nu}F^j_{\rho\sigma} = \frac{1}{2}\tilde{\epsilon}^{0abc} \left[ F^i_{0a}F^j_{bc} + F^i_{bc} F^j_{0a} \right] = 2 F^{(i}_{0a} \tilde{F}^{j)a} \sim \Psi^{ik}\Psi^{kj}.
\end{equation}
Both the left and right sides are symmetric in $ij$, the symmetry can be removed on both sides by introducing an antisymmetric term on the right-hand side, encoded by $X^i$. The proportionality constant can be dealt with by introducing a scalar field, $\tilde{\mu}$. The equation of motion looks like
\begin{equation}
    F^i_{0a} \tilde{F}^{ja} = \tilde{\mu} \left( \Psi^{ik} \Psi^{kj} + \epsilon^{ijk} X^k \right)
\end{equation}
again multiplying this by $\utilde{P}^j_a$ results in
\begin{equation}\label{F-0a}
    F^i_{0a} = \tilde{\mu}(\Psi^{ik}\Psi^{kj} + \epsilon^{ijk}X^k)\utilde{P}^j_a.
\end{equation}
The two evolution equations are \cref{eq:Inv_Psi_Equation_Of_Motion} and \cref{F-0a}.

\section{Self-dual 2-forms and the metric} \label{sec:num-pure-con-metric}

The goal of this section is to fix our conventions related to the self-duality, and also describe the formula for the spacetime metric as defined by a triple of 2-forms. We then explicitly compute the metric produced by the 2-forms \cref{Sigma-def}, and thus give interpretation to various objects appearing in the field equations. 

\subsection{Self-dual 2-forms}

Given a frame $e^0, e^i$, the basis in the space of self-dual 2-forms is given by
\begin{align}\label{Sigma-frame}
\Sigma^i = \im e^0 \wedge e^i + \frac{1}{2} \epsilon^{ijk} e^j\wedge e^k.
\end{align}
\footnote{The choice of orientation for the self-dual 2-forms is different here than in the rest of the text, the final result is independent of this choice, so either orientation is acceptable.} They are self-dual 
\begin{align}\label{Sigma-duality}
\frac{1}{2} \epsilon_{\mu\nu}{}^{\rho\sigma}\Sigma^i_{\rho\sigma} =  \im \Sigma^i_{\mu\nu}
\end{align}
in the orientation $\epsilon^{0123}=+1$, and the metric being of signature mostly plus
\begin{align}
ds^2 = - (e^0)^2+(e^1)^2+(e^2)^2+(e^3)^2.
\end{align}
These 2-forms satisfy the following algebra of the quaternions
\begin{align}\label{Sigma-algebra}
\Sigma^i_\mu{}^\alpha \Sigma^j_\alpha{}^\nu = - \delta^{ij} \delta_\mu{}^\nu - \epsilon^{ijk} \Sigma^k_\mu{}^\nu,
\end{align}
where our 2-form conventions are $\Sigma^i= (1/2)\Sigma^i_{\mu\nu} dx^\mu dx^\nu$. Further, the 2-forms \cref{Sigma-frame} satisfy
\begin{align}\label{Sigma-Sigma}
    \Sigma^i \wedge \Sigma^j =  2\im \delta^{ij} v_g, \quad v_g = e^0\wedge e^1\wedge e^2\wedge e^3.
\end{align}
We take our 4-manifold to be oriented by $v_g$, so that it defines the positive orientation. Using the coordinate volume form
\begin{align}
    dx^\mu dx^\nu dx^\rho dx^\sigma = \tilde{\epsilon}^{\mu\nu\rho\sigma} d^4x,
\end{align}
we can then write
\begin{align}
v_g = \sqrt{-g} d^4x,
\end{align}
where $\sqrt{-g}$ is (minus) the determinant of the frame, or the square root of (minus) the determinant of the metric. Finally, given a basis $\Sigma^i$ in the space of self-dual 2-forms that satisfies all the relations above, the metric can be recovered via the \urb{} formula
\begin{align}\label{metric-Sigma}
\im g_{\mu\nu}  \sqrt{-g} = \frac{1}{12} \tilde{\epsilon}^{\alpha\beta\gamma\delta} \epsilon^{ijk} \Sigma^i_{\mu\alpha} \Sigma^j_{\nu\beta}  \Sigma^k_{\gamma\delta}.
\end{align}
The coefficient on the right-hand side of this expression can be checked by multiplying the whole expression with $g^{\mu\nu}$, and then using \cref{Sigma-algebra} and \cref{Sigma-duality}. 

\subsection{3+1 Metric}

To compute the spacetime metric in the 3+1 decomposed form, we need to fix our orientation conventions. Given that $v_g= e^0e^1 e^2e^3$ is taken to be the positive orientation, we have $dx^0 dx^1 dx^2 dx^3 = \tilde{\epsilon}^{0123} d^4x = d^4x$. We will also take the positive spatial orientation to be $e^1 e^2 e^3$, which gives $\tilde{\epsilon}^{0abc}= \tilde{\epsilon}^{abc}$.

We now compute the metric produced by the connection. The corresponding orthonormal 2-forms are given by \cref{Sigma-def}, and we have
\begin{align}
    2i\delta^{ij} v_g =  \Sigma^i \wedge \Sigma^j  = \frac{1}{\Psi^{ik}}\frac{1}{\Psi^{jl}} (F^k_{0a} \tilde{F}^{la} + F^l_{0a} \tilde{F}^{ka}) d^4x= 2\tilde{\mu}\delta^{ij} d^4x,
\end{align}
where we have used \cref{tilde-F}, as well as (\ref{F-0a}). Now $v_g =  \sqrt{-g}  d^4x$ and we have
\begin{equation}
   \tilde{\mu}=\im \sqrt{-g},
\end{equation}
which identifies the function $\tilde{\mu}$ in \cref{F-0a} with the imaginary unit times the square root of the metric determinant. We can now compute all the metric components using \cref{metric-Sigma}. We have
\begin{align}
\im g_{00} \sqrt{-g} =  \frac{1}{6} \tilde{\epsilon}^{abc} \epsilon^{ijk} \Sigma^i_{0a} \Sigma^j_{0b} \Sigma^k_{0c}.
\end{align}
Now we can use (\ref{F-0a}) in the form
\begin{align}\label{Sigma-0}
\Sigma^i_{0a}= \frac{1}{\Psi^{ij}} F^j_{0a} = \tilde{\mu} (\Psi^{ij} \utilde{P}^j_a + \im \epsilon^{ijk} (\Psi^{jm} \utilde{P}^m_a) N^k), \qquad N^i := -\im{\rm det}(\Psi) \Psi^{ij} X^j,
\end{align}
where we have introduced a convenient combination $N^i$, to obtain
\begin{align}\label{g00}
g_{00}  =  \tilde{\mu}^2 {\rm det}(\Psi \utilde{P}) (1- N^i N^i),
\end{align}
where
\begin{align}
{\rm det}(\Psi \utilde{P}) = {\rm det}(\Psi) {\rm det}(\utilde{P}), \qquad {\rm det}(\utilde{P})= \frac{1}{6} \tilde{\epsilon}^{abc} \epsilon^{ijk} \utilde{P}^i_a \utilde{P}^j_b \utilde{P}^k_c.
\end{align}
We also have
\begin{align}
\im g_{ab} \sqrt{-g} = - \frac{1}{3{\rm det}(\Psi)} \epsilon^{ijk} F^i_{0(a} F^j_{b)c} \tilde{F}^{kc} + \frac{1}{6{\rm det}(\Psi)} \epsilon^{ijk} \tilde{\epsilon}^{pqr} F^i_{ap} F^j_{bq} F^k_{0r}.
\end{align}
Using the definition of the $\utilde{P}^i_a$, this can be transformed to
\begin{align}
\im g_{ab} \sqrt{-g} = \frac{{\rm det}(\tilde{F})}{{\rm det}(\Psi)}  F^i_{0(a} \utilde{P}^i_{b)}, 
\end{align}
which gives
\begin{align}
g_{ab} =  \frac{{\rm det}(\tilde{F})}{{\rm det}(\Psi)}  \Psi^{ki}\Psi^{kj}\utilde{P}^i_a \utilde{P}^j_b = \frac{1}{{\rm det}(\Psi \utilde{P})}  \Psi^{ki}\Psi^{kj}\utilde{P}^i_a \utilde{P}^j_b = \gamma^i_a \gamma^i_b,
\end{align}
where we have introduced the triad for the spatial metric given by
\begin{align}\label{gamma_a}
\gamma_a^i := \frac{1}{\sqrt{{\rm det}(\Psi \utilde{P})}} \Psi^{ij} \utilde{P}^j_a.
\end{align}

Finally, we can compute
\begin{align}
\im g_{0a} \sqrt{-g} = \frac{1}{6} \epsilon^{ijk}  \Sigma^i_{0b} \Sigma^j_{0a} \frac{1}{\Psi^{kl}} \tilde{F}^{lb} + \frac{1}{6} \epsilon^{ijk} \tilde{\epsilon}^{bcd} \Sigma^i_{0b} \frac{1}{\Psi^{jm}} F^m_{ac} \Sigma^k_{0d}=
 \frac{1}{2} \epsilon^{ijk}  \Sigma^i_{0b} \Sigma^j_{0a} \frac{1}{\Psi^{kl}} \tilde{F}^{lb}.
\end{align}
Using the expression (\ref{Sigma-0}) and simplifying using the algebra of $\utilde{P}$ and $\tilde{F}$ we get
\begin{align}
g_{0a}= -\im \tilde{\mu} N^i \Psi^{ij} \utilde{P}^j_a.
\end{align}
Comparing with the usual parametrisation of the metric by lapse $N$, shift $N^a$ and the spatial metric $\gamma_{ab}$
\begin{align}
g_{\mu\nu} = \left( \begin{array}{cc} - N^2 + N^a N^b \gamma_{ab} & N^b \gamma_{ab} \\ N^b \gamma_{ab} & \gamma_{ab} \end{array}\right),
\end{align}
we can identify
\begin{align}
N^a = \sqrt{-g} \, {\rm det}(\Psi \utilde{P}) N^i \frac{1}{\Psi^{ij}} \tilde{F}^{ja}, \qquad N= \sqrt{-g} \sqrt{{\rm det}(\Psi \utilde{P})}.
\end{align}
We can now see that the reality conditions saying that the 4-metric is real-valued become the conditions that the objects $\sqrt{-g}, N^i$ and $\Psi^{ij} \utilde{P}^j_a$ are real, modulo possibly an $SO(3,\C)$ rotation. Note that separately the quantities $\Psi^{ij}$ and $\utilde{P}^j_a$ are complex-valued, and it is only their product that is supposed to satisfy a reality condition and give rise to the frame of a real spatial metric. 

For future reference we mention that the inverse metric is given by
\begin{align}
g^{\mu\nu} = \left( \begin{array}{cc} - N^{-2}  & N^a N^{-2} \\ N^a N^{-2} & \gamma^{ab} - N^{-2} N^a N^b \end{array}\right).
\end{align}

\subsection{Useful formulas}

Now that the components of the curvature $F^i_{\mu\nu}$ are identified with the metric components, it is useful to rewrite the equation (\ref{Sigma-0}) in terms of the metric. The intermediate useful formulas that we need are
\begin{align}\label{useful-formulas}
\sqrt{{\rm det}(\Psi\utilde{P})} = \frac{1}{{\rm det}(\gamma)} = \frac{N}{\sqrt{-g}}, \\ \nonumber
\Psi^{ij} \utilde{P}^j_a = \sqrt{{\rm det}(\Psi\utilde{P})} \gamma^i_a = \frac{N}{\sqrt{-g}} \gamma^i_a, \\
\nonumber
\frac{1}{\Psi^{ij}} \tilde{F}^{ja} = \frac{1}{\sqrt{{\rm det}(\Psi\utilde{P})}} \gamma^{ia} = \frac{\sqrt{-g}} {N}\gamma^{ia}, \\ \nonumber
N^i = \frac{N^a \gamma^i_a}{N}.
\end{align}
Substituting all this into (\ref{Sigma-0}), and using also $\tilde{\mu}=\im \sqrt{-g}$ we get
\begin{align}\label{Sigma-0-useful}
\Sigma^i_{0a}= \frac{1}{\Psi^{ij}} F^j_{0a} = \im N \gamma^i_a - \epsilon_{abc} N^b \gamma^{ic}.
\end{align}

\section{Rewriting the evolution equations} \label{sec:num-pure-con-evol-system}

The goal of this section is to rewrite the evolution equations \cref{eq:Inv_Psi_Equation_Of_Motion} and \cref{F-0a} in the form that makes their interpretation most transparent. In particular, we will make use of the quantities $N, N^a, \gamma^i_a$ that were identified by the metric computation in the previous section. 

\subsection{Rewriting of the $\Psi^{ij}$ equation}

We start by manipulating \cref{eq:Inv_Psi_Equation_Of_Motion} so that it is $\Psi$ that evolves rather than $\Psi^{-1}$. We use
\begin{align}
d\left(\frac{1}{\Psi}\right) = - \frac{1}{\Psi}d\Psi \frac{1}{\Psi} ,
\end{align}
and then multiply the resulting equation by $\Psi$ from the left. This gives
\begin{align}
D_t \Psi^{is} \frac{1}{\Psi^{sj}} = \tilde{\epsilon}^{abc} \utilde{P}^j_a D_b \Psi^{is} \frac{1}{\Psi^{sk}} F^k_{0c}.
\end{align}
We then multiply by another copy of $\Psi$ from the right to get
\begin{align}
D_t \Psi^{ij} = \tilde{\epsilon}^{abc} \utilde{P}^s_a \Psi^{sj} D_b \Psi^{il} \frac{1}{\Psi^{lk}} F^k_{0c}.
\end{align}
The next step is to substitute the expressions \cref{useful-formulas} and (\ref{Sigma-0-useful}). This gives
\begin{align}
D_t \Psi^{ij} = \frac{N}{\sqrt{-g}} \tilde{\epsilon}^{abc} \gamma^j_a  D_b \Psi^{il} 
 (\im N \gamma^l_c - \epsilon_{cdf} N^d \gamma^{l f}).
 \end{align}
 We then use 
 \begin{align}
\tilde{\epsilon}^{abc} \gamma^i_a \gamma^j_b = {\rm det}(\gamma) \epsilon^{ijk} \gamma^{kc}
 \end{align}
in the first term, while we expand the product of two epsilons in the second term
\begin{align}
\tilde{\epsilon}^{abc} \epsilon_{cdf} = {\rm det}(\gamma) (\delta^a_d \delta^b_f - \delta^a_f\delta^b_d).
\end{align}
The term where $b$ gets contracted with $f$ vanishes by the Gauss constraint. What remains can be written as 
\begin{equation}
    (D_t - N^a D_a) \Psi^{ij} = i N \epsilon^{klj}\gamma^{ka} D_a \Psi^{il}.
    \label{eq:Psi_evolution_equation}
\end{equation}
It is manifest the right-hand side here is automatically tracefree. It is not hard to see that the antisymmetric part of the right-hand side is a multiple of the constraint  (\ref{eq:Gauss_constraint}), and so vanishes on the constraint surface. The obtained evolution equation for $\Psi^{ij}$ is very simple, and will be, after appropriate modifications related to gauge-fixing, one of our two main evolution equations. 

It is worth remarking that the obtained evolution equation for $\Psi^{ij}$ is completely analogous to the one we have previously described in the case of Maxwell theory, see \cref{evol-eqn-phi}. To make the parallel even stronger, we note that in the gauge $A_0^i=0, N^a=0, N=1$, the equation (\ref{eq:Psi_evolution_equation}) can be written as $\partial_t \Psi^{ij} = \im (\nabla \times \Psi)^{ij}$, where $(\nabla \times \Psi)^{ij}:= \epsilon^{klj} \gamma^{ka} D_a \Psi^{il}$ is the generalisation of the usual curl on vector fields to rank two symmetric tracefree tensors. We thus indeed see that the Weyl curvature, which is what encodes the spin two degrees of freedom of the gravitational field, evolves in an exact parallel to how the spin one field $\vec{\phi}=\vec{B} +\im \vec{E}$ evolves in the case of electromagnetism. The main difference in the case of gravity is that the metric has now become dynamical and determined by the potential $A_a^i$. There was no need to evolve the electromagnetic potential in the case of Maxwell, as the equation for $\vec{\phi}$ did not contain the potential. In contrast, in the case of gravity we must evolve the connection together with $\Psi^{ij}$ because the spatial triad that appears on the right-hand side of (\ref{eq:Psi_evolution_equation}) is constructed from the curvature of the connection (and $\Psi^{ij}$). 

\subsection{Rewriting the evolution equation for the connection}

The other evolution equation that we need is that for the connection, see (\ref{F-0a}). It is best to take it in the form (\ref{Sigma-0-useful}). We then multiply the equation by $\Psi$, and note that 
\begin{align}
\epsilon_{abc} \Psi^{ij} \gamma^{jc} = \frac{N}{\sqrt{-g}} \epsilon_{abc} \tilde{F}^{ic} = \utilde{\epsilon}_{abc} \tilde{F}^{ic} = F^i_{ab}.
\end{align}
It is then clear that the evolution equation for the connection can be written as
\begin{align}\label{evolution-connection}
D_t A^i_a - \partial_a A^i_0 - N^b F^i_{ba} = \im N \Psi^{ij} \gamma^j_a,
\end{align}
where we have taken the term involving the shift vector to the left-hand side. This equation clearly interprets the field $\Psi^{ij}$ as the time derivative of the spatial connection $A^i_a$, modulo terms involving $A_0^i$ and $N^a$ and related to gauge. 

We note that the right-hand side of both evolution equations explicitly contains the factor of the imaginary unit, so the evolution system obtained is intrinsically complex. We also note that the Maxwell analogue of equation \cref{evolution-connection} is \cref{evol-eqn-a}. Again, there is a direct parallel, apart from the fact that $A_a^i$ has now indices of two different types, and so the right-hand side of \cref{evolution-connection} must involve an object that relates them. 

\section{Constraint sweeping} \label{sec:numerical-purec-connection-gauge-fixing}

The evolution system we have obtained is a system of two evolution equation, for $\Psi^{ij}$ and for the connection. To understand the evolution equation for $\Psi^{ij}$ in particular, we consider a similar evolution equation in a background of Minkowski metric. 

\subsection{Spin two fields in Minkowski spacetime}

As a warm-up, and to contrast the situation in gravity with that in the case of Maxwell field, we assume that the metric is that of Minkowski space, and $\gamma^i_a=\delta^i_a$, so that there is no distinction between the internal and spatial indices. We also assume that shift $N^a=0$ and lapse $N=1$. We get the following evolution equation for $\Psi^{ij}$:
\begin{align}\label{psi-evolution-flat}
\partial_t \Psi^{ij} = \im \epsilon^{klj} \partial^k \Psi^{il},
\end{align}
which should be solved subject to the constraint that $\Psi^{ij}$ is symmetric tracefree and transverse $\partial^j \Psi^{ij}=0$, as this is what the constraint (\ref{eq:Gauss_constraint}) becomes. 

Motivated by the Maxwell example, we look for a modification of this evolution system that introduces a new field $\Phi^i$ whose time derivative is related to the Gauss constraint. We then search for an evolution system that evolves both fields $\Psi^{ij}, \Phi^i$ and is such both fields satisfy the $\Box\Psi^{ij}=\Box \Phi^i=0$ equation. It is not hard to show that there is the unique system that reduces to \cref{psi-evolution-flat} when $\Phi^i=0$ and has these properties. It is given by
\begin{align}\label{spin-two-flat}
\partial_t \Psi^{ij} = \im \left(\epsilon^{kl(j} \partial^k \Psi^{i)l} + 3 \partial^{(i} \Phi^{j)} - \delta^{ij} \partial_k \Phi^k\right), \\ \nonumber
\partial_t \Phi^i = - \frac{\im}{2} \left( \epsilon^{ijk} \partial^j \Phi^k + \partial^{j} \Psi^{ji}\right).
\end{align}
We note that the last term in the first equation is needed to make the right-hand side tracefree. Taking another time derivative of the first equation and using both equations one can obtain $\Box\Psi^{ij}=0$. Similarly, taking the time derivative of the second equation and using both equations one obtains $\Box\Phi^i=0$. This shows that the system \cref{spin-two-flat} describes a spin two fields in Minkowski spacetime, with its $5+3=8$ components. The constraint equations $\partial^j\Psi^{ij}=0$ has been replaced by an evolution equation for 3 new fields $\Phi^i$, so that now all the components of $\Psi^{ij}, \Phi^i$ propagate and satisfy the box equation. This guarantees that any constraint violation that accumulates by some numerical error will propagate away and not accumulate where it occurs. Another desirable feature of ``constraint sweeping" is that it allows for any error in the initial conditions (when these are imposed only approximately) to propagate away and not spoil the simulation. This is the desired ``constraint sweeping". 

We note that it is essential that the right-hand sides of both equations in \cref{spin-two-flat} contain the factor of the imaginary unit. It is these factors that give the right sign for the box equation in Minkowski spacetime on squaring. This is similar to the situation in the Maxwell case, see \cref{eqs-M-modified}. 

\subsection{Non-linear version}

There is an obvious non-linear version of the flat evolution system \cref{spin-two-flat} that is a modification of (\ref{eq:Psi_evolution_equation}). It is given by
\begin{align}\label{evol-eqs-GR-modified}
 ( D_t - N^a D_a) \Psi^{ij} = i N \left[ \epsilon^{kl(j} \gamma^{k a} 
 D_a \Psi^{i)l} + 3 \gamma^{a(i} D_a \Phi^{j)} - \delta^{ij}\gamma^{a k} D_a \Phi^{k} \right],
\\ \nonumber
  ( D_t - N^a D_a)  \Phi^i = -\frac{i N}{2} \left[ \epsilon^{ijk}\gamma^{aj} D_a \Phi^k + \gamma^{aj} D_a \Psi^{ij}  \right].
\end{align}
We take this to be our main evolution system for $\Psi^{ij}, \Phi^i$, while we will continue to evolve  the spatial connection according to \cref{evolution-connection}. 

\subsection{Gauge Fixing}

In the Maxwell case the new terms in the evolution equations for $\phi^i$ came from adding the Lorentz gauge term to the Lagrangian, see \cref{gf-M}. We could follow a similar strategy for gravity. Thus, we could consider adding to the Lagrangian $\Phi^i g^{\mu\nu} \partial_\mu A_\nu^i$. This is a step in the right direction. However, the difficulty that arises is that now the metric is dynamical and is a function of the connection. The variation with respect to the connection will produce terms that are not present in the most natural equations \cref{evol-eqs-GR-modified}. Thus, it appears to be a better strategy to just postulate the evolution equations \cref{evol-eqs-GR-modified} for $\Psi^{ij}, \Phi^i$, and add to them some version of the Lorentz gauge $g^{\mu\nu} \partial_\mu A_\nu^i=0$, or possibly $g^{\mu\nu} \partial_\mu A_\nu^i=\Phi^i$. This will make the $A_0^i$ component of the connection evolve, which is desirable, for one will not need to select this set of functions by hand. However, it is clear that the best form of this evolution equation for $A_0^i$ is likely to depend on how the evolution system we derived behaves under numerics. So, we refrain from trying to fix the form of this equation in this text. 

Another possible strategy, which is different to the strategy considered in the Maxwell case, is to remove 4 components from the connection. That is we set $\Sigma^i_\mu{}^\nu A^i_\nu = 0$ such that only 8 independent components remain. In terms of the 3+1 components this condition suggest that $\gamma^{ia} A^i_a = 0$ and that $A^i_0$ is a function of the frame and the remaining components in $A^i_a$. One then no longer needs an equation of motion for $A^i_0$ as it is determinant by the other variables in the theory. This matches the number of components that are in the gauge-fixed spin two fields $\Psi^{ij}$ and $\Phi^i$. It may then be that the equations of motion can be interpreted as a Dirac operator which are known to be hyperbolic and square to box, for more on this see section 8.5 of~\cite{FormGenRelGravity2020}. This gauge would be more desirable than the Lorentz gauge as it is produces a system with fewer variables and is therefore easier to compute. This does however change the leading order structure of the equations of motion due to there being derivatives of $A^i_0$ present. Further consequences of this gauge fixing are not considered here.

\section{Discussion}

In this chapter we have rewritten Einstein equations of GR as a system of two evolution equations \cref{evol-equations} for the fields $\Psi^{ij}$, which is a complex symmetric tracefree field encoding the self-dual part of the Weyl curvature, and the field $A_a^i$, which is a complex $SO(3,\C)$ spatial connection. The equations also involve the usual lapse $N$ and shift $N^a$, which are real, and the temporal component of the connection $A_0^i$, which is complex. These fields do not evolve and can be chosen to be arbitrary functions. The spatial triad $\gamma^i_a$ and its inverse $\gamma^{ia}$ that appear on the right-hand sides of the evolution equations are constructed algebraically (but non-linearly), see \cref{spatial-frame} from the field $\Psi^{ij}$ and the curvature $F^i_{ab}$ of the spatial connection. 

The evolution equations \cref{evol-equations} are incredibly simple. They directly generalise the evolution equations of the chiral formulation of Maxwell theory, see \cref{evol-eqn-a}, \cref{evol-eqn-phi}. In the case of Maxwell, one forms the complex linear combination of the electric and magnetic fields $\vec{\phi}=\vec{B}+\im \vec{E}$. In terms of this field, Maxwell's equations become a single complex evolution equation \cref{evol-eqn-M-compl}. The evolution equation \cref{evol-eqn-a} for the electromagnetic potential can be also read of as the definition of $\vec{\phi}$. The analogy between \cref{evol-equations} and the Maxwell case is that the first equation generalises \cref{evol-eqn-M-compl}, while the second gives the evolution equation for the spatial connection and thus can be read as a definition of $\Psi^{ij}$. The main difference between the gravitational and the Maxwell cases is that in the latter one does not need to evolve the electromagnetic potential as it does not appear in the evolution equation for the $\vec{\phi}$. In the gravity case the field strength of the spatial connection is what determines the metric that appears on the right-hand sides of both equations, and for this reason the spatial connection must be evolved together with $\Psi^{ij}$. At the same time, the structure of the field equations \cref{evol-equations} is completely analogous to that of \cref{evol-eqn-a}, \cref{evol-eqn-phi}. In fact, one could have guessed the system of equations \cref{evol-equations} as the most natural generalisation of the system that arises in the spin one case, with the operator $\nabla\times \Psi^{ij}:=\epsilon^{klj} \partial^k \Psi^{il}$ being the generalisation of the usual curl $\nabla\times \phi^i = \epsilon^{ijk} \partial^j \phi^k$ from spin one to spin two.

In both the Maxwell and gravity cases there are constraints. In the Maxwell case this is the Gauss constraint $\partial^i \phi^i=0$, in the gravity case this is similarly the Gauss constraint \cref{constr}, which is again the most natural generalisation of the Gauss constraint in the case of electromagnetism (and so could have been guessed as well). We find it quite satisfactory that the field equations of GR, in the form of evolution equations, could have been guessed without doing any computations, when one works in the appropriate formalism. This is of course consistent with the fact that GR is the unique theory with ``some good properties". The form of the evolution equations \cref{evol-equations} tells us that GR is the ``most natural" theory describing the dynamics of the spin two field $\Psi^{ij}$ propagating on the metric background determined by an $SO(3)$ connection.

In the case of both Maxwell and GR the evolution equations for the field $\vec{\phi}$ and $\Psi^{ij}$ can be modified to introduce the ``constraint sweeping". In both cases the idea is the same. Instead of trying to impose the Gauss constraint at every step of the evolution, one can introduce a new field ($\phi$ in the Maxwell case and $\Phi^i$ in the gravity case) whose time derivative is a multiple of the Gauss constraint. The right-hand sides of the evolution equations for $\vec{\phi}, \Psi^{ij}$ are then modified by $\phi, \Phi^i$ in such a way that all fields satisfy the box equation. This makes the constraint violation propagating, and any possible constraint error that is created by a numerical scheme will propagate away from the grid. While the currently preferred by the numerical relativity community method of dealing with constraint violation seems to be ``constraint damping" rather than ``constraint sweeping", which directly damps any possible constraint violation rather than making it propagate away, the introduced here ``constraint sweeping" is the option preferred mathematically, as one leading to the nicest right-hand sides of the evolution equations. This can be made even more prominent by rewriting the equations in spinor notations. It then turns out that the first order differential operators that arise after the gauge-fixing are versions of the Dirac operator. This is explained in Section 8 of~\cite{FormGenRelGravity2020}. Prior to numerical experiments, it only remains to hope that the form of the gauge-fixed equations preferred by mathematics is also the one that leads to good numerical stability. We hope to return to these issues in future work. 

It is very interesting that in the gravity case there are no analogues of the familiar scalar and vector constraints of the metric formalism of GR. In fact, there is a sense in which these constraints have been solved by the formalism that treats $\Psi^{ij}$ as one of the dynamical variables. This is particularly clear when one compares the formalism of this chapter to that of Ashtekar Hamiltonian formulation~\cite{New_Hamiltonian_Ashtek_1987}. The link from our formalism to that in~\cite{New_Hamiltonian_Ashtek_1987} is provided by introducing 
\begin{align}
\tilde{\gamma}^{ia} = \frac{1}{\Psi^{ij}} \tilde{F}^{ja}
\end{align}
and treating it as the main dynamical variable rather than $\Psi^{ij}$. One then has to impose the constraints which guarantee that $\Psi^{ij}$ constructed from $\tilde{\gamma}^{ia}$ and $\tilde{F}^{ia}$ is symmetric tracefree. These are precisely the vector and scalar constraints of the Hamiltonian formulation~\cite{New_Hamiltonian_Ashtek_1987}. Thus, the scalar and vector constraints are solved by the formalism that uses $\Psi^{ij}$ in a single stroke, which is one of its attractive features. Despite the fact that there are no scalar and vector constraints, diffeomorphisms are still gauge, which is signalled by the presence of the freely specifiable lapse and shift functions in the evolution equations. One can still introduce some gauge-fixing of the diffeomorphisms, and thus constrain the lapse and shift in some way.

One other issue that needs discussing is the non-linearity of the evolution equations that most strongly manifests itself in the fact that one needs to take the factors of $1/\Psi$ in constructing the spatial frame. An unpleasant feature of this formalism is therefore the fact that whenever the Weyl curvature $\Psi^{ij}$ becomes degenerate, it becomes difficult to interpret the evolution equations. And Weyl curvature can be degenerate in some physically interesting situations. In particular, Minkowski space with its vanishing Weyl curvature cannot be treated by the described pure connection formalism. Another example is that of the pp-wave spacetime, where the Weyl curvature matrix $\Psi^{ij}$  has a single non-zero eigenvalue, and so is not invertible. At the same time, the matrix $\Psi^{ij}$ for the Schwarzschild solution is invertible. Sufficiently far from the two merging black holes, the spacetime is approximately that of gravitational radiation propagating away on the background of a spherically symmetric Schwarzschild solution. The Weyl curvature is small, but the matrix $\Psi^{ij}$ remains invertible. It should be possible to read off the scalar $\psi_4$ that encodes the gravitational wave signal not too far from the merging black holes so that $\Psi^{ij}$ continues to be invertible, while at the same time as far as practically possible to get an accurate approximation of $\psi_4$ at infinity. We find it quite nice that the evolution system \cref{evol-equations} propagates directly the Weyl curvature $\Psi^{ij}$, which is where the observable curvature scalar $\psi_4$ is stored. Thus, the procedure for extraction of the observable gravitational wave template should be much less non-trivial business than it is in the case of the metric formalism. Another possible idea to deal with non-invertibility of $\Psi^{ij}$ is as follows. The only quantity that is needed for the evolution equations is the spatial triad. When some eigenvalues of the Weyl curvature are zero and $\Psi^{ij}$ stops being invertible the corresponding components of the spatial curvature are also zero. What this signals is just that the corresponding components of the spatial triad take the values they are in the case of Minkowski spacetime. So, it may be possible to give interpretation to the evolution equations \cref{evol-equations} even in the case when some eigenvalues of $\Psi^{ij}$ are zero and this matrix is not invertible.

Our further comments are on the issue of the reality conditions. We first note that the reality conditions represent in the Maxwell case were under control in the free evolution setup. All fields could be made complex and the violation of the imaginary parts remained small but bounded and did not need to be separately imposed. Our envisaged strategy of dealing with the reality conditions in the gravity case is similar. In that case the main reality condition states that the spatial frame $\gamma^i_a$ that is constructed from $\Psi^{ij}$ and $F^i_{ab}$ is real, in the sense that the spatial metric $\gamma^i_a \gamma^i_b$ is real. This is a complicated nonlinear set of equations, which is algebraic in $\Psi^{ij}$ but involves first derivatives of the spatial connection. At the same time, it is guaranteed by the link to the \pleb{} formalism and then its link to the real metric formulation of GR that this reality condition is compatible with the evolution equations in the sense that one must only preserve this condition (and its time derivative) at a single moment of time, and it will remain satisfied at all times. It is clear that the time derivative of the metric reality condition is a reality condition on the Weyl curvature $\Psi^{ij}$. It is also clear from general principles that this can be given a form of some second-order in derivatives equation on $\Psi^{ij}$, similar to how the reality condition of $\vec{\phi}$ in the Maxwell case is first-order in derivatives. The evolution equations for the constraints are then considered to be closed, or at the very least can be closed by introducing further constraints, such that the constraint amplification matrix analysis can be applied. It remains to be checked if the constraints converge to the reality surface (the surface in phase space for which the reality conditions are satisfied), this is a future calculation that we do not consider here. All in all, the hope should be that one can evolve the system \cref{evol-equations} unconstrained by the reality conditions (and thus only constrain the initial data), as in the Maxwell case, and that the numerical results are sufficiently ``real".

Finally, we would like to comment on the issue of initial data in the described pure connection formulation. We do not anticipate any difficulty with this at least for the purpose of starting the numerical experiments, because the fields that are evolved by our system \cref{evol-equations} have direct link with the usual metric fields. The object $A^i$ is the self-dual part of the spin connection for a frame constructed from the metric. The object $\Psi^{ij}$ is the self-dual part of the Weyl curvature tensor. Both can be computed (numerically if needed) for any of the initial data that one works with in the standard numerical relativity schemes. So, initial data can be determined as they are determined in the usual metric formalism, and then translated into the $A^i, \Psi^{ij}$ variables. An alternative strategy is to solve the constraints \cref{constr} directly in the described formalism. However, more work would be needed to understand how to describe e.g. a black hole binary in this language.

    \newpage
    \chapter{Conclusion and Outlooks}
    Here we provide a conclusion of each topic especially considering the overarching theme of this Thesis, that being using chiral formulations to find solutions to GR.\@ Included is a list of the novel results achieved in this work. Interesting directions for future research projects are also considered and briefly described.

    \section{Linear Structure of Chiral Gravity}
    In \cref{part:linearised-gravity}, we explored the structure of the first-order equations of motion that impose linearised Einstein equations (EEs) in \pleb{}'s formulation. A useful representation for the linearisation of self-dual 2-forms is obtained by breaking apart the perturbation of the frame into its irreducible components, as shown in \cref{eq:lin-2-form-pert-in-irreducibles-frame-comps}. Similarly, the self-dual connection has a simple definition in this representation, see \cref{eq:lin-sd-connection-pert-in-irrep-frame-comps}. The main result here is the discovery of a new 2-parameter family of hyperbolic gauge fixing, presented in \cref{subsec:lin-gauge-fixing-2-param-family}. In chiral theories, $SO(3,\mathbb{C})$ and diffeomorphism freedoms must be fixed. To achieve this, we introduce first-order Feynman forms of the modified Lorenz and de Donder gauge fixings. The new fields are internal vectors $\chi^i$ and a connection 1-form $\xi_\mu$. Their equations of motion fix the gauge orbit such that the second-order equations satisfy the wave operator, hence the name ``hyperbolic''. This is an extension of the gauge fixing presented in~\cite{ChiralPerturbaKrasno2020}. There the authors also noted a separation of components into two independent systems. Here we recover the same result, the $16+16$ components of frame and connection perturbations separate into a $4+4$ and $12+12$ system. This gauge fixing is unique, up to the choice of the 2 parameters, because it only considers linear modifications. The $4$-component and $12$-component systems can be associated with self-dual and anti-self-dual truncated de Rham complexes, respectively. This gauge fixing and separation are possible only in chiral formulations due to the correct number of components required to form a Dirac operator between spaces.

    Using this unique linear structure and interpreting $SU(2)$-structures as described in~\cite{Su2StructureBhoja2024}, we constructed what we call the \pleb{} complex \cref{pleb-compl-intr}. Within this complex, we encode the linearised version of Euclidean EEs. A twisted Dirac operator is constructed between these spaces that squares to the Laplace operator. The separation of the previous linear gauge fixed system into sectors of size $4$ and $12$ extends to the \pleb{} complex. Using an appropriate untwisting operator, it can be written as a sum of two truncated de Rham complexes. The difference between this result and more generic 2-parameter families of gauge fixings is that not every member of the family can be written as a twisting of the original \pleb{} complex. Tighter restrictions on the gauge fixing are required for this to be possible. The existence of this complex in Yang-Mills theory has led to significant developments, such as index theorems~\cite{The_Index_of_El_Atiyah_1968}, topology of four-manifolds, and classification of smooth structures~\cite{Donaldson:1983wm}. We hope that this complex will allow for similar advancements and provide further insights into these topics. It is clear that viewing linearised theory from \pleb{}'s formulation offers an interesting perspective.

    \section{Nonlinear Structure of Chiral Gravity}

    \subsection{Nonlinear Hyperbolic Gauge Fixing}
    Part~\ref{part:nonlinear-gravity} contains results pertaining to the full nonlinear Einstein equations. In \cref{chap:nonlinear-gauge-fixing}, we extend linear gauge fixing to a nonlinear hyperbolic first-order gauge fixing. The details of the gauge can be found in \cref{eq:nonlinear-modified-connection-definitions,eq:nonlinear-modified-einstein-and-lorenz-conditions}. These gauge conditions form a 2-parameter family, involving the first-order Feynman forms of modified Harmonic and Lorentz conditions. Additionally, changes are made to the definition of the self-dual connection and Einstein equations. This gauge fixing is constructed such that when the first-order gauge fields vanish, the equations revert to their original form. It is hyperbolic in nature because the second-order equations of motion have the wave operator as the leading order in derivatives term. The nonlinear theory allows for lower order modifications to these equations, making this gauge fixing non-unique. We present the simplest choice with no lower order terms. One of the most remarkable results in this section is that, with only slight modification, the nonlinear gauge fixing admits a separation into sectors of size 4 and 12, just as it does in the linear case. However, the nonlinearity causes the 4-dimensional sector to still depend on components from the 12-dimensional part. We show that this separation can be derived by performing a conformal separation that isolates a unimodular metric. The new basic fields become: a conformal factor $\lambda \in C^\infty$, three 2-forms $H^i \in \Lambda^2$ encoding a metric with determinant 1, and linear combinations of the connection components shown in \cref{eq:nonlinear-Omega-in-terms-of-A-and-xi}. This system is called the conformal gauge system and is given by \cref{eq:nonlinear-conformal-connection-definitions,eq:nonlinear-conformal-einstens-and-gauge-equations}. The 12-component system being separate means that one can find solutions to it independently. Since the number of equations matches the number of components, all 12 components should be determined by these equations.

    This gauge fixing has implications in various aspects of general relativity, with a particular focus on its numerical properties. More details about this system will be discussed in the next section. The gauge fixing presented is new and thus requires further exploration to fully understand its properties. It would be interesting to see if this gauge fixing can be derived from an action principle, which could have applications in quantum field theory, particularly as a BRST quantisation (see~\cite{BRST_antifield_Fuster_2005} for a review). Since the system seems to be in some sense separable this could have useful consequences, at the level of the action. An appropriate action for this gauge would involve modifying lower order terms in the gauge fields and should be computable, making it an interesting line of research that we have yet to explore.

    \subsection{Type D Spacetimes}

    We have seen that 4-dimensional geometry through the lens of chiral formalisms allows for a deeper insight into Einstein metrics, especially when the Weyl curvature has a type D structure. In \cref{chap:type-D-conformal-to-kahler}, an alternative derivation of the \pleb{}-\demi{} family of metrics is given using \pleb{}'s formulation of general relativity. Choosing this chiral description simplifies the proof of \derd{}'s theorem, see \cref{thrm:DerdziConfKahler}. Only basic algebra is needed to confirm this result. Other new results include proving the equivalence of $\Sigma$-Killing and metric Killing vectors \cref{prop:Sigma-killing-vector-is-metric-killing-vector} and showing that type D spacetimes have an inherent Killing vector \cref{prop:ConfKahlerKillingVector}. While this chapter is a reinterpretation of~\cite{AmbitoricGeomeAposto2013} as presented in~\cite{Kerr_metric_fro_Krasno_2024}, it serves as a showcase for the effectiveness of describing gravity in terms of self-dual 2-forms. The assumption of two Killing vectors, one of which is inherent to type D, along with additional assumptions that they are holomorphic with respect to the two complex structures provides a natural basis and ansatz for the metric and the self-dual 2-forms. Other new results include proving the equivalence of $\Sigma$-Killing and metric Killing vectors \cref{prop:Sigma-killing-vector-is-metric-killing-vector} and showing that type D spacetimes have an inherent Killing vector \cref{prop:ConfKahlerKillingVector}. While this chapter is a reinterpretation of~\cite{AmbitoricGeomeAposto2013} as presented in~\cite{Kerr_metric_fro_Krasno_2024}, it serves as a showcase for the effectiveness of describing gravity in terms of self-dual 2-forms. The assumption of two Killing vectors, one of which is inherent to type D, along with additional assumptions that they are holomorphic with respect to the two complex structures provides a natural basis and ansatz for the metric and the self-dual 2-forms. The rest of the derivation involves imposing EEs on this ansatz using \pleb{}'s formulation. The result is a pair of ordinary differential equations \cref{eq:pleb-demi-ode-equations} with quartic polynomial solutions \cref{eq:PlebanskiDemianskiFinalSolution}. Kerr can be recovered by taking the appropriate limits, following further assumptions about the spacetime.
    
    Due to the rich structure that appears for type D spacetimes, it is probable that this method will be useful for gravitational perturbations around a black hole. The problem of scalar, vector and tensor perturbations in the Kerr background are known to be separable~\cite{TheMathematicaChandr1998}. The existence of a non-trivial rank 2 Killing tensor implies that the wave operator is separable. Killing tensors can be constructed from antisymmetric Killing-Yano tensors related to 2-forms appearing in this context, as explained in~\cite{Killing2FormsGauduc2017}. The Teukolsky equations reduce the problem of gravitational perturbation of Kerr to a single scalar potential function for the metric. It is likely that the geometry described in \cref{chap:type-D-conformal-to-kahler} will lead to a useful formalism simplifying these computations, particularly using pure connection formalisms, see~\ref{sec:pure-connection-chiral-first-order} or~\cite{Gravity_as_a_di_Krasno_2011}. Steps in this direction have been made in~\cite{Pure_connection_Krasno_2024}, where self-dual Anti de Sitter nonlinear perturbations are parametrised by a single scalar. It is believed that similar results extend to all type D metrics discussed here, and it is a worthwhile avenue of research due to the increasing interest in gravitational wave analysis.

    Another aspect of this particular application is its extensions to other Petrov classification types; due to the assumptions made in the main text some degenerations of type D are not obtainable. For example, pp-wave spacetimes cannot be derived through this construction. A similar analysis, to Weyl tensors with different structure, using \pleb{}'s formulation could lead to a generalisation of the work here to Petrov types. Kundt spacetimes~\cite{Coley2009} are important examples of type N spacetimes for which this method might be useful in further understanding. Overall, we expect that this particular method of analysing geometry will be particularly powerful in understanding Einstein spacetimes with specified Weyl curvature tensors. Overall, we expect that this particular method of analysing geometry will be particularly powerful in understanding Einstein spacetimes with specified Weyl curvature tensors.

    \section{Chiral Numerical Relativity}

    \subsection{First-Order Hyperbolic \pleb{} Evolution System}

	Part~\ref{part:numerical-relativity} is the last part of the Thesis, dedicated to the numerical study of gauge fixed chiral formulations. Chapter~\ref{chap:plebanski-numerical-relativity} develops an evolution system for the first-order hyperbolic gauge fixing. The resulting evolution system is a gauge fixed version of Ashtekar's formulation, which itself is known to be the Hamiltonian formulation of \pleb{}. This system is displayed in \cref{sec:pleb-num-gauge-fixed-evolution-system}, interestingly it retains the polynomial structure of the Ashtekar's formulation. The constraint analysis of this system suggests that one mode will be undamped while the rest are damped or propagating. By adding artificial damping terms to specific equations of motion, see \cref{eq:num-damped-lambda-system}, it is found that the propagating modes become damped, but the single undamped mode remains unchanged. The evolution system for the conformal system is also presented; introducing new variables as nonlinear combinations of Ashtekar's variables and the additional gauge fields. It is known that the conformal system separates into sectors of size 4 and 12, the evolution equations for the 12 components are shown in \cref{eq:num-plebanski-conformally-seperated-frame-evolution-system}, it can be evolved completely independently of the remaining equations of motion. A partially conformal system that uses the conformal frame, but not the conformal connection variables is also present as an alternative system to both, it appears in \cref{eq:num-nonlinear-partial-conformal-connection-system} which along with \cref{eq:num-plebanski-conformally-seperated-frame-evolution-system} constitutes all the gauge fixed Einstein equations. The partially conformal system is preferred over fully conformal alternatives due to its simpler evolution system and direct link to previous constraint analysis. To test these systems numerically, random noise was added to Minkowski solutions, which were then evolved freely. The sum of the absolute value of all the constraints is under control, as shown in \cref{fig:num-constraint-damping}, however the violation in the reality conditions and the distance from the Minkowski solutions increased \cref{fig:num-minkowski-distance-100,fig:num-reality-condition}. Both the initial and the final deviations in these conditions were small enough to consider the result approximately correct. This suggests that these systems have promising applications for numerical relativity. 

    The nonlinear version of the gauge fixing is not unique due to the arbitrarily many lower-order terms that can be added, while remaining strongly hyperbolic. Both analytic and numerical explorations of this large space of evolution systems would provide valuable insights into these systems. Additionally, testing the conformal system numerically remains an interesting challenge as it offers a unique separation property not seen in other numerical relativity formulations. Another point to consider is due to the Lorentzian signature requirement the fields become complex-valued, as such the partial differential equations (PDE) also become complex. A deeper understanding of nonlinear complex differential equations would be beneficial, especially when considering the reality conditions and could be the subject of future work.
    
    A different line of study, that is not considered here but mentioned in the Introduction, is that of structure preserving integrators (SPIs). More specifically, the use of discrete exterior calculus (DEC) to construct constraint preserving integration schemes. By using DEC one can discretise the action and produce field equations from said action that accurately preserves the constraints during evolution. In this context, it is natural to want the basic fields to be described in terms of differential forms. For \pleb{}'s formulation gravity this is the case, and one such integrator was developed in~\cite{NumericalRelatFarr2010}. There only the constraints related to the Lorentz group was conserved, rather than the diffeomorphism invariance. Furthermore, in~\cite{Finite_element_Arnold_2009} it was noted that the existence of an elliptic complex for a PDE implied the existence of a well-posed DEC integrator. This along with the recent developments in discretising bundle-valued forms, see~\cite{braune2025discreteexteriorcalculusbundlevalued}, it is possible that there exists a numerical scheme that uses \pleb{}'s formulation which preserves the all the constraints and is well-posed. Such a system would be superb result for numerical relativity improving the accuracy and eliminating the need for artificial constraint preserving terms.

    \subsection{First-Order Hyperbolic Pure Connection Evolution System}
    
    In the last chapter of this Thesis, \cref{chap:pure-connection-numerical-relativity}, a numerical evolution system was developed for the pure connection descriptions of chiral gravity. The main fields are the self-dual Weyl curvature and the spatial self-dual connection. The evolution equations are given in \cref{evol-equations} and match previous results from~\cite{A_connection_ap_Salisb_1994}. Only the constraint with respect to the chiral half of the Lorentz group remains, allowing one to essentially ignore diffeomorphism constraints. In this work, we developed a gauge fixing procedure inspired by the chiral Maxwell equations, this procedure enforces constraint sweeping. Near Minkowski spacetimes, where the Weyl curvature vanishes, the presence of the inverse Weyl tensor in the equations of motion causes them to be ill-defined. Without a method of dealing with these degenerate points, only spacetimes with non-degenerate Weyl curvature can be evolved.

    Certain aspects of this system require further exploration before it can be considered robust enough for generic use. The issue of reality conditions needs to be explored both analytically and numerically, as they are crucial for ensuring hyperbolicity. A similar analysis to that done in \cref{subsec:num-pure-conn-maxwell-reality-conditions} would be needed to ensure the reality conditions are preserved. Gauge fields, such as the temporal connection are also left undetermined in the main text. Some analysis of different gauge fixing and their effect on the reality conditions and hyperbolicity is necessary. Methods for dealing with degenerate Weyl curvature are also desirable, as this can affect the system's performance. Since this description involves bundle-valued differential forms, it may be possible to construct an SPI using DEC technology described in the previous section. This would be advantageous over \pleb{}'s formalism due to the reduced number of variables and constraints. All these considerations would make promising future work.

    \newpage
    \printbibliography
    \addcontentsline{toc}{chapter}{Bibliography}

    \appendix

    \chapter{\texorpdfstring{Second Order Formulas for $\Sigma$'s}{Second Order Formulas for Σ's}} \label{apx:second-order-formulas-for-Sigma}
        Here we briefly describe the process to derive the identities used in the main text.
        We note that as we are interested in highest order derivative terms only the second derivative $\partial_{\mu\nu}$ behaves as if it was a single derivative, that is 
        \begin{align}
            \partial_{\mu\nu} (fg) = \partial_{\mu\nu}(f) g + f \partial_{\mu\nu}(g).
        \end{align}
        Throughout this section the algebraic identities in \cref{eq:pleb-epsilon-S-to-ddS}
        \paragraph{Derivation of: \texorpdfstring{$\Sigma^{i\rho\sigma} \partial^2 \Sigma^i_{\rho\sigma} = 6 \partial^2 \ln(\sqrt{-g})$}{Log}}
        \begin{align}
            \Sigma^{i\rho\sigma} \partial^2 \Sigma^i_{\rho\sigma} =& \partial^2 \left( \Sigma^{i\rho\sigma} \Sigma^i_{\rho\sigma} \right) - \Sigma^i_{\rho\sigma} \partial^2 \Sigma^{i\rho\sigma} = \partial^2 (12) - \Sigma^i_{\rho\sigma} \partial^2 \left( -\frac{i}{2}\teps^{\rho\sigma\alpha\beta} \frac{1}{\sqrt{-g}} \Sigma^i_{\alpha\beta} \right) \nonumber\\ =& \frac{i}{2} \teps^{\rho\sigma\alpha\beta} \Sigma^i_{\rho\sigma} \partial^2 \left(\frac{1}{\sqrt{-g}} \Sigma^i_{\alpha\beta} \right) = - \sqrt{-g}\Sigma^{i\rho\sigma} \partial^2 \left( \frac{1}{\sqrt{-g}} \Sigma^i_{\rho\sigma} \right) \nonumber\\ =& -\Sigma^{i\rho\sigma} \partial^2 \Sigma^i_{\rho\sigma} - \sqrt{-g} \Sigma^{i\rho\sigma} \Sigma^i_{\rho\sigma} \partial^2 \left( \frac{1}{\sqrt{-g}} \right) \nonumber\\[10pt] \Rightarrow \Sigma^{i\rho\sigma} \partial^2 \Sigma^i_{\rho\sigma} =& \frac{1}{2}\Sigma^{i\rho\sigma}\Sigma^i_{\rho\sigma}\partial^2 \ln(\sqrt{-g}) = 6 \partial^2 \ln(\sqrt{-g})
        \end{align}                                                                 
        Where the last equality in the first line uses the self-duality of $\Sigma^i$. From the first line to the second we use that fact that $\teps^{\mu\nu\rho\sigma} = \pm 1$ and can be passed through any derivative.

        \paragraph{Derivation of: \texorpdfstring{$\Sigma^{i\rho\sigma} \partial_\rho{}^\alpha \Sigma^i_{\sigma\alpha} = -\partial^{\rho\sigma} g_{\rho\sigma} - \partial^2 \ln(\sqrt{-g})$}{None}}
        \begin{align}
            \Sigma^{i\rho\sigma} \partial_\rho{}^\alpha \Sigma^i_{\sigma\alpha} =& \Sigma^i_{\rho\beta} g^{\sigma\beta} \partial^{\rho\alpha} \Sigma^i_{\sigma\alpha} = g^{\sigma\beta} \partial^{\rho\alpha}\left( \Sigma^i_{\rho\beta} \Sigma^i_{\sigma\alpha} \right) - \Sigma^{i\sigma}{}_\alpha \partial^{\alpha\rho} \Sigma^i_{\rho\sigma} \nonumber\\ =& g^{\sigma\beta} \partial^{\rho\alpha} \left( g_{\rho\sigma} g_{\beta\alpha} - g_{\rho\sigma} g_{\beta\sigma} + i \epsilon_{\rho\beta\sigma\alpha} \right) - \Sigma^{i\rho\sigma} \partial_\rho{}^\alpha \Sigma^i_{\sigma\alpha} \nonumber\\ =& -2 \partial^{\rho\sigma} g_{\rho\sigma} - 2 \partial^2 \ln(\sqrt{-g}) - \Sigma^{i\rho\sigma} \partial_\rho{}^\alpha \Sigma^i_{\sigma\alpha} \nonumber\\[10pt] \Rightarrow \Sigma^{i\rho\sigma} \partial_\rho{}^\alpha \Sigma^i_{\sigma\alpha} =& - \partial^{\rho\sigma} g_{\rho\sigma} - \partial^2 \ln(\sqrt{-g}).
        \end{align}

        \paragraph{Derivation of: \texorpdfstring{$\frac{1}{2}\Sigma^i_\mu{}^\rho \partial^\sigma \Sigma^i_{\rho\sigma} - \frac{1}{2}\Sigma^{i\rho\sigma} \partial_\rho \Sigma^i_{\mu\sigma} = g_{\mu\rho} \partial_\sigma g^{\rho\sigma} - \partial_\mu \ln(\sqrt{-g})$}{PDFstring}}
        \begin{align}
            \frac{1}{2}\Sigma^i_\mu{}^\rho \partial^\sigma \Sigma^i_{\rho\sigma} &= \frac{1}{2}\Sigma^i_{\mu\beta} g^{\rho\beta} \partial^\sigma \Sigma^i_{\rho\sigma} = \frac{1}{2} g^{\rho\beta} \partial^\sigma \left( \Sigma^i_{\mu\beta} \Sigma^i_{\rho\sigma} \right) + \frac{1}{2}\Sigma^{i\rho\sigma} \partial_\rho \Sigma^i_{\mu\sigma} \nonumber\\ &= \frac{1}{2} g^{\rho\beta} \partial^\sigma \left( g_{\mu\rho} g_{\beta\sigma} - g_{\mu\sigma} g_{\beta\rho} + i \epsilon_{\mu\beta\rho\sigma} \right) + \frac{1}{2}\Sigma^{i\rho\sigma} \partial_\rho \Sigma^i_{\mu\sigma} \nonumber\\ &= -\partial^\rho g_{\mu\rho} - \partial_\mu \ln(\sqrt{-g}) + \frac{1}{2}\Sigma^{i\rho\sigma} \partial_\rho \Sigma^i_{\mu\sigma} \nonumber\\ &= g_{\mu\rho} \partial_\sigma g^{\rho\sigma} - \partial_\mu \ln(\sqrt{-g}) + \frac{1}{2}\Sigma^{i\rho\sigma} \partial_\rho \Sigma^i_{\mu\sigma} \nonumber\\[10pt]
            \Rightarrow \frac{1}{2} \Sigma^i_\mu{}^\rho \partial^\sigma \Sigma^i_{\rho\sigma} - \frac{1}{2}\Sigma^{i\rho\sigma} \partial_\rho \Sigma^i_{\mu\sigma} &= g_{\mu\rho} \partial_\sigma g^{\rho\sigma} - \partial_\mu \ln(\sqrt{-g}).
        \end{align}

        \paragraph{Derivation of: \texorpdfstring{$\frac{1}{2}\Sigma^i_{\mu\rho} \partial^{\rho\sigma} \Sigma^i_{\nu\sigma} + \frac{1}{2}\Sigma^i_{\nu\sigma} \partial^{\rho\sigma} \Sigma^i_{\mu\rho} = \frac{1}{2}g_{\mu\nu} \partial^{\rho\sigma} g_{\rho\sigma} - \frac{1}{2}\partial_\mu \partial^\rho g_{\nu\rho} - \frac{1}{2} \partial_\nu \partial^\rho g_{\mu\rho} + \frac{1}{2}\partial^2 g_{\mu\nu}$}{PDFString}}
        \begin{align}
            \frac{1}{2}\Sigma^i_{\mu\rho} \partial^{\rho\sigma} \Sigma^i_{\nu\sigma} + \frac{1}{2}\Sigma^i_{\nu\sigma} \partial^{\rho\sigma} \Sigma^i_{\mu\rho} =& \frac{1}{2} \partial^{\rho\sigma}\left( \Sigma^i_{\mu\rho} \Sigma^i_{\nu\sigma} \right) = \frac{1}{2} \partial^{\rho\sigma} \left( g_{\mu\nu} g_{\rho\sigma} - g_{\mu\rho} g_{\nu\sigma} + i \epsilon_{\mu\rho\nu\sigma} \right) \nonumber\\ =& \frac{1}{2} g_{\mu\nu} \partial^{\rho\sigma} g_{\rho\sigma} + \frac{1}{2} \partial^2 g_{\mu\nu} - \frac{1}{2}\partial_\mu{}^\rho g_{\nu\rho} - \frac{1}{2}\partial_\nu{}^\rho g_{\mu\rho}.
        \end{align}

        \paragraph{Derivation of: \texorpdfstring{$\Sigma^i_{(\mu}{}^\rho \partial^2 \Sigma^i_{\nu)\rho} = \partial^2 g_{\mu\nu} + g_{\mu\nu} \partial^2 \ln(\sqrt{-g})$}{PDFString}}
        \begin{align}
            \Sigma^i_{(\mu}{}^\rho \partial^2 \Sigma^i_{\nu)\rho} &= \frac{1}{2}g^{\rho\sigma} \left( \Sigma^i_{\mu\sigma} \partial^2 \Sigma^i_{\nu\rho} + \Sigma^i_{\nu\sigma} \partial^2 \Sigma^i_{\mu\rho} \right) = \frac{1}{2}g^{\rho\sigma} \partial^2\left( \Sigma^i_{\mu\sigma} \Sigma^i_{\nu\rho} \right) \nonumber\\ &= \frac{1}{2} g^{\rho\sigma} \partial^2 \left( g_{\mu\nu} g_{\rho\sigma} - g_{\mu\rho} g_{\sigma\nu} + i \epsilon_{\mu\sigma\nu\rho} \right) \nonumber\\ &= \partial^2 g_{\mu\nu} + g_{\mu\nu} \partial^2 \ln(\sqrt{-g})
        \end{align}

        \paragraph{Derivation of: \texorpdfstring{$\epsilon^{ijk} \Sigma^{j\rho\sigma} \partial^\alpha{}_\rho \Sigma^k_{\sigma\alpha} = \Sigma^{i\mu\nu} \partial_\mu{}^\rho g_{\nu\rho} - \frac{1}{4} \epsilon^{ijk} \Sigma^{j\rho\sigma} \partial^2 \Sigma^k_{\rho\sigma}$}{PDFString}}
        \begin{align}
            \epsilon^{ijk} \Sigma^{j\rho\sigma} \partial^\alpha{}_\rho \Sigma^k_{\sigma\alpha} &= \epsilon^{ijk} \Sigma^{j\rho\sigma} \partial^\alpha{}_\rho \left( -\frac{i}{2}\utilde{\epsilon}_{\sigma\alpha\beta\gamma} \Sigma^{k\beta\gamma} \sqrt{-g} \right) = -\frac{i}{2}\epsilon^{ijk} \Sigma^{j\rho\sigma} \utilde{\epsilon}_{\sigma\alpha\beta\gamma} \partial^\alpha{}_\rho \left( \sqrt{-g} \Sigma^{k\beta\gamma} \right) \nonumber\nonumber\\ &= \frac{1}{2\sqrt{-g}} \epsilon^{ijk} \left( \delta^\rho_\beta \Sigma^j_{\alpha\gamma} - \delta^\rho_\gamma \Sigma^j_{\alpha\beta} - \delta^\rho_\alpha \Sigma^j_{\beta\gamma} \right) \partial^\alpha{}_\rho \left( \sqrt{-g} \Sigma^{k\beta\gamma} \right) \nonumber\nonumber\\ &= \frac{1}{\sqrt{-g}}\epsilon^{ijk} \Sigma^j_{\alpha\gamma} \partial^\alpha{}_\rho \left( \sqrt{-g} \Sigma^{k\rho\gamma} \right) - \frac{1}{2\sqrt{-g}} \epsilon^{ijk} \Sigma^j_{\rho\sigma} \partial^2 \left( \sqrt{-g} \Sigma^{k\rho\sigma} \right) \nonumber\nonumber\\ &= -\frac{2}{\sqrt{-g}} \Sigma^{i\mu\nu} \partial_{\mu\nu} \sqrt{-g} + \epsilon^{ijk} \Sigma^j_{\alpha\gamma} \partial^\alpha{}_\rho \Sigma^{k\rho\gamma} - \frac{1}{2\sqrt{-g}} \epsilon^{ijk} \Sigma^j_{\rho\sigma} \partial^2 \left( -\frac{i}{2} \teps^{\rho\sigma\alpha\beta} \Sigma^k_{\alpha\beta} \right) \nonumber\nonumber\\ &= \epsilon^{ijk} \Sigma^j_{\alpha\gamma} \partial^\alpha{}_\rho \Sigma^{k\rho\gamma} - \frac{1}{2}\epsilon^{ijk} \Sigma^{j\rho\sigma} \partial^2 \Sigma^k_{\rho\sigma} = \epsilon^{ijk} \Sigma^j_{\alpha\gamma} \partial^\alpha{}_\rho \left( \Sigma^k_{\sigma\beta} g^{\rho\sigma} g^{\gamma\beta} \right) - \frac{1}{2}\epsilon^{ijk} \Sigma^{j\rho\sigma} \partial^2 \Sigma^k_{\rho\sigma} \nonumber\nonumber\\ &= -\epsilon^{ijk} \Sigma^{j\rho\sigma} \partial^\alpha{}_\rho \Sigma^k_{\sigma\alpha} + \epsilon^{ijk} \Sigma^j_\alpha{}^\beta \Sigma^k_{\sigma\beta} \partial^\alpha{}_\rho g^{\rho\sigma} + \epsilon^{ijk}\Sigma^j_{\alpha\gamma} \Sigma^k_{\sigma\beta} \partial^{\alpha\sigma} g^{\gamma\beta} - \frac{1}{2}\epsilon^{ijk} \Sigma^{j\rho\sigma} \partial^2 \Sigma^k_{\rho\sigma} \nonumber\nonumber\\ &= -\epsilon^{ijk} \Sigma^{j\rho\sigma} \partial^\alpha{}_\rho \Sigma^k_{\sigma\alpha} - 2\Sigma^i_{\rho\sigma} \partial^\rho{}_\nu g^{\sigma\nu} + \left( g_{\alpha\sigma} \Sigma^i_{\beta\gamma} - g_{\alpha\beta} \Sigma^i_{\sigma\gamma} - g_{\gamma\sigma} \Sigma^i_{\beta\alpha} + g_{\gamma\beta} \Sigma^i_{\sigma\alpha} \right) \partial^{\alpha\sigma} g^{\gamma\beta} \nonumber\nonumber\\ &\quad\quad - \frac{1}{2}\epsilon^{ijk} \Sigma^{j\rho\sigma} \partial^2 \Sigma^k_{\rho\sigma} \nonumber\nonumber\\ &= -\epsilon^{ijk} \Sigma^{j\rho\sigma} \partial^\alpha{}_\rho \Sigma^k_{\sigma\alpha} - 2\Sigma^i_{\rho\sigma} \partial^\rho{}_\nu g^{\sigma\nu} - \Sigma^i_{\rho\sigma} \partial^\rho{}_\nu g^{\sigma\nu} + \Sigma^i_{\rho\sigma} \partial^\rho{}_\nu g^{\sigma\nu} - \frac{1}{2}\epsilon^{ijk} \Sigma^{j\rho\sigma} \partial^2 \Sigma^k_{\rho\sigma} \nonumber\nonumber\\ &= -\epsilon^{ijk} \Sigma^{j\rho\sigma} \partial^\alpha{}_\rho \Sigma^k_{\sigma\alpha} + 2\Sigma^{i\rho\sigma} \partial_\rho{}^\nu g_{\sigma\nu} - \frac{1}{2}\epsilon^{ijk} \Sigma^{j\rho\sigma} \partial^2 \Sigma^k_{\rho\sigma} \nonumber\\[10pt]
                \Rightarrow 2\epsilon^{ijk} \Sigma^{j\rho\sigma} \partial^\alpha{}_\rho \Sigma^k_{\sigma\alpha} &= 2\Sigma^{i\rho\sigma} \partial_\rho{}^\nu g_{\sigma\nu} - \frac{1}{2}\epsilon^{ijk} \Sigma^{j\rho\sigma} \partial^2 \Sigma^k_{\rho\sigma} \nonumber\nonumber\\
                \Rightarrow \epsilon^{ijk} \Sigma^{j\rho\sigma} \partial^\alpha{}_\rho \Sigma^k_{\sigma\alpha} &= \Sigma^{i\rho\sigma} \partial_\rho{}^\nu g_{\sigma\nu} - \frac{1}{4}\epsilon^{ijk} \Sigma^{j\rho\sigma} \partial^2 \Sigma^k_{\rho\sigma}
        \end{align}
   
    \newpage

\end{document}